\pdfoutput=1
\documentclass[11pt,twoside,a4paper,cmspaper,final,collab]{cms-tdr}
\def\svnVersion{36af2a0}\def\svnDate{2024/10/16}\def\cmsCernNoTag{CERN-EP-2024-095}\def\cmsCernDate{\today}\def\cmsMessage{Published in Physics Reports as \href{http://dx.doi.org/10.1016/j.physrep.2024.09.012}{\doi{10.1016/j.physrep.2024.09.012}.}}
\begin{document}\cmsNoteHeader{EXO-23-006}

\newlength\cmsTabSkip\setlength{\cmsTabSkip}{1ex}

\ifthenelse{\boolean{cms@external}}{%
    \newcommand\cmsParagraph[1]{\paragraph{#1}}
}{%
    \newcommand\cmsParagraph[1]{\textbf{#1:}}
}

\newcommand{\VLLL}{{\HepParticle{L}{}{}}\xspace}
\newcommand{\VLLE}{{\HepParticle{E}{}{}}\xspace}
\newcommand{\VLLEbar}{{\HepAntiParticle{E}{}{}}\xspace}
\newcommand{\VLLN}{{\HepParticle{N}{}{}}\xspace}
\newcommand{\VLLNbar}{{\HepAntiParticle{N}{}{}}\xspace}
\newcommand{\PAQT}{{\HepAntiParticle{T}{}{}}\xspace}
\newcommand{\PAQB}{{\HepAntiParticle{B}{}{}}\xspace}
\newcommand{\xft}{{\HepParticle{X}{5/3}{}}\xspace}
\newcommand{\xftbar}{{\HepAntiParticle{X}{5/3}{}}\xspace}
\newcommand{\yft}{{\HepParticle{Y}{4/3}{}}\xspace}
\newcommand{\yftbar}{{\HepAntiParticle{Y}{4/3}{}}\xspace}
\newcommand{\PGSmp}{{\HepParticle{\PGS}{}{\mp}}\xspace}
\newcommand{\PWst}{{\HepParticle{\PW}{}{\ast}}\xspace}
\newcommand{\PWRpm}{{\HepParticle{\PW}{R}{\pm}}\xspace}
\newcommand{\PNell}{{\HepParticle{\PN}{\Pell}{}}\xspace}
\renewcommand{\PNe}{{\HepParticle{\PN}{\Pe}{}}\xspace}
\newcommand{\PNGm}{{\HepParticle{\PN}{\PGm}{}}\xspace}
\newcommand{\PNGt}{{\HepParticle{\PN}{\PGt}{}}\xspace}
\newcommand{\PQj}{{\HepParticle{j}{}{}}\xspace}
\newcommand{\PQJ}{{\HepParticle{J}{}{}}\xspace}
\newcommand{\PQbst}{{\HepParticle{\PQb}{}{\ast}}\xspace}
\newcommand{\PU}{{\HepParticle{U}{}{}}\xspace}

\newcommand{\PellP}{{\HepParticle{\Pell}{\text{P}}{}}\xspace}
\newcommand{\Pellp}{{\HepParticle{\Pell}{}{+}}\xspace}
\newcommand{\Pellm}{{\HepParticle{\Pell}{}{-}}\xspace}
\newcommand{\Pellpm}{{\HepParticle{\Pell}{}{\pm}}\xspace}
\newcommand{\Pellmp}{{\HepParticle{\Pell}{}{\mp}}\xspace}
\newcommand{\Pellpr}{{\HepParticle{\Pell}{}{\prime}}\xspace}
\newcommand{\Pellprpr}{{\HepParticle{\Pell}{}{\prime\prime}}\xspace}
\newcommand{\PQqpr}{{\HepParticle{\PQq}{}{\prime}}\xspace}
\newcommand{\PAQqpr}{{\HepAntiParticle{\PQq}{}{\prime}}\xspace}

\newcommand{\mVLLE}{\ensuremath{m_{\VLLE}}\xspace}
\newcommand{\mZpr}{\ensuremath{m_{\PZpr}}\xspace}
\newcommand{\mWpr}{\ensuremath{m_{\PWpr}}\xspace}
\newcommand{\mWR}{\ensuremath{m_{\PWR}}\xspace}
\newcommand{\mQq}{\ensuremath{m_{\PQq}}\xspace}
\newcommand{\mQt}{\ensuremath{m_{\PQt}}\xspace}
\newcommand{\mQb}{\ensuremath{m_{\PQb}}\xspace}
\newcommand{\mQQ}{\ensuremath{m_{\PQQ}}\xspace}
\newcommand{\mQT}{\ensuremath{m_{\PQT}}\xspace}
\newcommand{\mQB}{\ensuremath{m_{\PQB}}\xspace}
\newcommand{\mhnl}{\ensuremath{m_{\PN}}\xspace}
\newcommand{\mGS}{\ensuremath{m_{\PGS}}\xspace}
\newcommand{\mNell}{\ensuremath{m_{\PNell}}\xspace}
\newcommand{\mNGt}{\ensuremath{m_{\PNGt}}\xspace}
\newcommand{\mH}{\ensuremath{m_{\PH}}\xspace}

\newcommand{\tauhnl}{\ensuremath{\tau_{\PN}}\xspace}
\newcommand{\ctauhnl}{\ensuremath{c\tau_{\PN}}\xspace}
\newcommand{\tauN}{\ensuremath{\tau_N}\xspace}
\newcommand{\tauTO}{\ensuremath{\tau_{21}}\xspace}
\newcommand{\tauTT}{\ensuremath{\tau_{32}}\xspace}
\newcommand{\rehnl}{\ensuremath{r_{\Pe}}\xspace}
\newcommand{\ruhnl}{\ensuremath{r_{\PGm}}\xspace}
\newcommand{\rthnl}{\ensuremath{r_{\PGt}}\xspace}

\newcommand{\kappaT}{\ensuremath{\kappa_{\PQT}}\xspace}
\newcommand{\kappaH}{\ensuremath{\kappa_{\PH}}\xspace}
\newcommand{\kappaW}{\ensuremath{\kappa_{\PW}}\xspace}
\newcommand{\kappaZ}{\ensuremath{\kappa_{\PZ}}\xspace}

\newcommand{\Gammatot}{\ensuremath{\Gamma_{\text{tot}}}\xspace}
\newcommand{\Gammae}{\ensuremath{\Gamma_{\Pe}}\xspace}
\newcommand{\Gammam}{\ensuremath{\Gamma_{\PGm}}\xspace}
\newcommand{\Gammat}{\ensuremath{\Gamma_{\PGt}}\xspace}
\newcommand{\GammaQ}{\ensuremath{\Gamma_{\PQQ}}\xspace}

\newcommand{\mvlq}{\ensuremath{m_{\text{VLQ}}}\xspace}
\newcommand{\mSD}{\ensuremath{m_{\text{SD}}}\xspace}
\newcommand{\rhoSD}{\ensuremath{\rho_{\text{SD}}}\xspace}
\newcommand{\minv}{\ensuremath{m_{\text{inv}}}\xspace}
\newcommand{\mjet}{\ensuremath{m_{\text{jet}}}\xspace}

\newcommand{\TTbar}{\ensuremath{\PQT\PAQT}\xspace}
\newcommand{\Tt}{\ensuremath{\PQT\PQt}\xspace}
\newcommand{\BBbar}{\ensuremath{\PQB\PAQB}\xspace}
\newcommand{\XXbar}{\ensuremath{\xft\xftbar}\xspace}
\newcommand{\YYbar}{\ensuremath{\yft\yftbar}\xspace}

\newcommand{\pp}{\ensuremath{\Pp\Pp}\xspace}
\newcommand{\qqbarpr}{\ensuremath{\PQq\PAQqpr}\xspace}
\newcommand{\ee}{\ensuremath{\Pe\Pe}\xspace}
\newcommand{\emu}{\ensuremath{\Pe\PGm}\xspace}
\newcommand{\epmmump}{\ensuremath{\Pepm\PGmmp}\xspace}
\newcommand{\mumu}{\ensuremath{\PGm\PGm}\xspace}
\newcommand{\tZ}{\ensuremath{\PQt\PZ}\xspace}
\newcommand{\tW}{\ensuremath{\PQt\PW}\xspace}
\newcommand{\tWp}{\ensuremath{\PQt\PWp}\xspace}
\newcommand{\tWm}{\ensuremath{\PQt\PWm}\xspace}
\newcommand{\tH}{\ensuremath{\PQt\PH}\xspace}
\newcommand{\bZ}{\ensuremath{\PQb\PZ}\xspace}
\newcommand{\bW}{\ensuremath{\PQb\PW}\xspace}
\newcommand{\bWp}{\ensuremath{\PQb\PWp}\xspace}
\newcommand{\bWm}{\ensuremath{\PQb\PWm}\xspace}
\newcommand{\bH}{\ensuremath{\PQb\PH}\xspace}
\newcommand{\VV}{\ensuremath{\PV\PV}\xspace}
\newcommand{\WW}{\ensuremath{\PW\PW}\xspace}
\newcommand{\WpmWpm}{\ensuremath{\PWpm\PWpm}\xspace}
\newcommand{\WZ}{\ensuremath{\PW\PZ}\xspace}
\newcommand{\WpmZ}{\ensuremath{\PWpm\PZ}\xspace}
\newcommand{\ZZ}{\ensuremath{\PZ\PZ}\xspace}
\newcommand{\Zgamma}{\ensuremath{\PZ\PGg}\xspace}
\newcommand{\Wgamma}{\ensuremath{\PW\PGg}\xspace}
\newcommand{\Vgamma}{\ensuremath{\PV\PGg}\xspace}
\newcommand{\gammagamma}{\ensuremath{\PGg\PGg}\xspace}
\newcommand{\Wb}{\ensuremath{\PW\PQb}\xspace}
\newcommand{\Zt}{\ensuremath{\PZ\PQt}\xspace}
\newcommand{\Ht}{\ensuremath{\PH\PQt}\xspace}
\newcommand{\Hb}{\ensuremath{\PH\PQb}\xspace}
\newcommand{\Zb}{\ensuremath{\PZ\PQb}\xspace}
\newcommand{\bb}{\ensuremath{\PQb\PQb}\xspace}
\newcommand{\qq}{\ensuremath{\PQq\PQq}\xspace}
\newcommand{\ellell}{\ensuremath{\Pell\Pell}\xspace}
\newcommand{\ellellpr}{\ensuremath{\Pell\Pellpr}\xspace}
\newcommand{\nunu}{\ensuremath{\PGn\PGn}\xspace}
\newcommand{\lnu}{\ensuremath{\Pell\PGn}\xspace}
\newcommand{\Bt}{\ensuremath{\PQb\PQt}\xspace}
\newcommand{\Bb}{\ensuremath{\PQb\PQb}\xspace}
\newcommand{\qW}{\ensuremath{\PQq\PW}\xspace}
\newcommand{\qH}{\ensuremath{\PQq\PH}\xspace}
\newcommand{\qZ}{\ensuremath{\PQq\PZ}\xspace}
\newcommand{\Tbbar}{\ensuremath{\PQT\PAQb}\xspace}

\newcommand{\mtZ}{\ensuremath{m_{\tZ}}\xspace}
\newcommand{\mtW}{\ensuremath{m_{\tW}}\xspace}
\newcommand{\mgg}{\ensuremath{m_{\gammagamma}}\xspace}
\newcommand{\mellell}{\ensuremath{m_{\ellell}}\xspace}
\newcommand{\mmumu}{\ensuremath{m_{\mumu}}\xspace}
\newcommand{\mlprlprpr}{\ensuremath{m_{\Pellpr\Pellprpr}}\xspace}
\newcommand{\mepmmump}{\ensuremath{m_{\epmmump}}\xspace}
\newcommand{\mellpmpimp}{\ensuremath{m_{\Pellpm\PGpmp}}\xspace}

\newcommand{\eee}{\ensuremath{\Pe\Pe\Pe}\xspace}
\newcommand{\eem}{\ensuremath{\Pe\Pe\PGm}\xspace}
\newcommand{\emm}{\ensuremath{\Pe\PGm\PGm}\xspace}
\newcommand{\mmm}{\ensuremath{\PGm\PGm\PGm}\xspace}
\newcommand{\eeX}{\ensuremath{\Pe\Pe\PX}\xspace}
\newcommand{\mmX}{\ensuremath{\PGm\PGm\PX}\xspace}
\newcommand{\ttH}{\ensuremath{\ttbar\PH}\xspace}
\newcommand{\ttW}{\ensuremath{\ttbar\PW}\xspace}
\newcommand{\ttZ}{\ensuremath{\ttbar\PZ}\xspace}
\newcommand{\ttX}{\ensuremath{\ttbar\PX}\xspace}
\newcommand{\ttV}{\ensuremath{\ttbar\PV}\xspace}
\newcommand{\Ztt}{\ensuremath{\PZ\ttbar}\xspace}
\newcommand{\Htt}{\ensuremath{\PH\ttbar}\xspace}
\newcommand{\Tbq}{\ensuremath{\PQT\PQb\PQq}\xspace}
\newcommand{\Ttq}{\ensuremath{\PQT\PQt\PQq}\xspace}
\newcommand{\WZb}{\ensuremath{\PW\PZ\PQb}\xspace}
\newcommand{\tZq}{\ensuremath{\PQt\PZ\PQq}\xspace}
\newcommand{\tHb}{\ensuremath{\PQt\PH\PQb}\xspace}
\newcommand{\tZb}{\ensuremath{\PQt\PZ\PQb}\xspace}
\newcommand{\bqq}{\ensuremath{\PQb\PQq\PQq}\xspace}
\newcommand{\blnu}{\ensuremath{\PQb\Pell\PGn}\xspace}

\newcommand{\mellellJ}{\ensuremath{m_{\Pell\Pell\PQJ}}\xspace}

\newcommand{\bHbH}{\ensuremath{\PQb\PH\PQb\PH}\xspace}
\newcommand{\bHbZ}{\ensuremath{\PQb\PH\PQb\PZ}\xspace}
\newcommand{\bZbZ}{\ensuremath{\PQb\PZ\PQb\PZ}\xspace}
\newcommand{\bHtW}{\ensuremath{\PQb\PH\PQt\PW}\xspace}
\newcommand{\bZtW}{\ensuremath{\PQb\PZ\PQt\PW}\xspace}
\newcommand{\tWtW}{\ensuremath{\PQt\PW\PQt\PW}\xspace}
\newcommand{\bWbW}{\ensuremath{\PQb\PW\PQb\PW}\xspace}
\newcommand{\ttbb}{\ensuremath{\ttbar\bbbar}\xspace}
\newcommand{\tHbq}{\ensuremath{\PQt\PH\PQb\PQq}\xspace}
\newcommand{\tZbq}{\ensuremath{\PQt\PZ\PQb\PQq}\xspace}
\newcommand{\tHtq}{\ensuremath{\PQt\PH\PQt\PQq}\xspace}
\newcommand{\tZtq}{\ensuremath{\PQt\PZ\PQt\PQq}\xspace}
\newcommand{\WbWb}{\ensuremath{\PW\PQb\PW\PQb}\xspace}

\newcommand{\BtobH}{\ensuremath{\PQB\to\bH}\xspace}
\newcommand{\BtobZ}{\ensuremath{\PQB\to\bZ}\xspace}
\newcommand{\BtotW}{\ensuremath{\PQB\to\tW}\xspace}
\newcommand{\TtobW}{\ensuremath{\PQT\to\bW}\xspace}
\newcommand{\TtotH}{\ensuremath{\PQT\to\tH}\xspace}
\newcommand{\TtotZ}{\ensuremath{\PQT\to\tZ}\xspace}
\newcommand{\Htogg}{\ensuremath{\PH\to\gammagamma}\xspace}
\newcommand{\Htobb}{\ensuremath{\PH\to\bbbar}\xspace}
\newcommand{\Ztobb}{\ensuremath{\PZ\to\bbbar}\xspace}
\newcommand{\WprtoTb}{\ensuremath{\PWpr\to\PQT\PQb}\xspace}
\newcommand{\WprtoBt}{\ensuremath{\PWpr\to\PQB\PQt}\xspace}

\newcommand{\eormu}{\ensuremath{\mkern1mu\Pe\mkern-2mu/\mkern-3mu\PGm}\xspace}

\newcommand{\ljets}{\ensuremath{\Pell\text{+jets}}\xspace}
\newcommand{\wjets}{\ensuremath{\PW\text{+jets}}\xspace}
\newcommand{\zjets}{\ensuremath{\PZ\text{+jets}}\xspace}

\newcommand{\HTlep}{\ensuremath{\HT^{\text{lep}}}\xspace}
\newcommand{\LT}{\ensuremath{L_{\mathrm{T}}}\xspace}
\newcommand{\LTmet}{\ensuremath{\LT+\ptmiss}\xspace}
\newcommand{\MT}{\ensuremath{M_{\mathrm{T}}}\xspace}
\newcommand{\ST}{\ensuremath{S_{\mathrm{T}}}\xspace}

\newcommand{\Mlb}{\ensuremath{M(\Pell,\PQb)}\xspace}
\newcommand{\minMlb}{\ensuremath{\min\Mlb}\xspace}
\newcommand{\mtilde}{\ensuremath{\widetilde{m}_{\PQT}}\xspace}

\newcommand{\BR}{\ensuremath{\mathcal{B}}\xspace}
\newcommand{\BRe}{\ensuremath{\BR_{\Pe}}\xspace}
\newcommand{\BRm}{\ensuremath{\BR_{\PGm}}\xspace}
\newcommand{\BRt}{\ensuremath{\BR_{\PGt}}\xspace}
\newcommand{\BrBbH}{\ensuremath{\BR(\PQB\to\bH)}\xspace}
\newcommand{\BrBbZ}{\ensuremath{\BR(\PQB\to\bZ)}\xspace}
\newcommand{\BrBtW}{\ensuremath{\BR(\PQB\to\tW)}\xspace}

\newcommand{\mixpar}[1]{\ensuremath{V_{#1}}\xspace}
\newcommand{\mixparsq}[1]{\ensuremath{\abs{\mixpar{#1}}^2}\xspace}
\newcommand{\mixparl}{\mixpar{\Pell}}
\newcommand{\mixpareN}{\mixpar{\Pe\PN}}
\newcommand{\mixparmN}{\mixpar{\PGm\PN}}
\newcommand{\mixpartN}{\mixpar{\PGt\PN}}
\newcommand{\mixparsqN}{\mixparsq{\PN}}
\newcommand{\mixparsqlN}{\mixparsq{\Pell\PN}}
\newcommand{\mixparsqeN}{\mixparsq{\Pe\PN}}
\newcommand{\mixparsqmN}{\mixparsq{\PGm\PN}}
\newcommand{\mixparsqtN}{\mixparsq{\PGt\PN}}
\newcommand{\mixparsqemN}{\ensuremath{\abs{\mixpar{\Pe\PN}\mixpar{\PGm\PN}^\ast}^2/(\mixparsqeN+\mixparsqmN)}\xspace}

\newcommand{\pTgone}{\ensuremath{\pt(\PGg_1)}\xspace}
\newcommand{\pTgtwo}{\ensuremath{\pt(\PGg_2)}\xspace}
\newcommand{\pTi}{\ensuremath{p_{\mathrm{T},i}}\xspace}
\newcommand{\pTj}{\ensuremath{p_{\mathrm{T},j}}\xspace}
\newcommand{\pTk}{\ensuremath{p_{\mathrm{T},k}}\xspace}
\newcommand{\pTij}{\ensuremath{p_{\mathrm{T},i+j}}\xspace}

\newcommand{\dzero}{\ensuremath{d_0}\xspace}
\newcommand{\dxy}{\ensuremath{d_{xy}}\xspace}
\newcommand{\dxysig}{\ensuremath{d_{xy}^{\text{sig}}}\xspace}
\newcommand{\DeltaTwoD}{\ensuremath{\Delta_{\text{2D}}}\xspace}
\newcommand{\Deltam}{\ensuremath{\Delta m}\xspace}
\newcommand{\Deta}{\ensuremath{\Delta\eta}\xspace}
\newcommand{\Dphi}{\ensuremath{\Delta\phi}\xspace}

\newcommand{\Nsub}[1]{\ensuremath{N_{\text{#1}}}\xspace}
\newcommand{\Ntt}{\Nsub{tt}}
\newcommand{\Ntl}{\Nsub{tl}}
\newcommand{\Nll}{\Nsub{ll}}
\newcommand{\Npp}{\Nsub{pp}}
\newcommand{\Npf}{\Nsub{pf}}
\newcommand{\Nff}{\Nsub{ff}}
\newcommand{\NSRbkg}{\ensuremath{N_{\text{SR}}^{\text{bkg}}}\xspace}
\newcommand{\NSRsim}{\ensuremath{N_{\text{SR}}^{\text{sim}}}\xspace}
\newcommand{\NCRdata}{\ensuremath{N_{\text{CR}}^{\text{data}}}\xspace}
\newcommand{\NCRsim}{\ensuremath{N_{\text{CR}}^{\text{sim}}}\xspace}

\newcommand{\chisq}{\ensuremath{\chi^2}\xspace}
\newcommand{\chimod}{\ensuremath{\chisq_{\text{mod}}}\xspace}
\newcommand{\chimodndf}{\ensuremath{\chimod/\text{ndf}}\xspace}

\newcommand{\GoM}{\ensuremath{\Gamma/\mQT}\xspace}
\newcommand{\GF}{\ensuremath{G_{\text{F}}}\xspace}
\newcommand{\Cfivell}{\ensuremath{\mathrm{C}_5^{\ellellpr}}\xspace}

\newcommand{\delphes}{\ensuremath{\textsc{delphes 3}}\xspace}
\newcommand{\CSVvtwo}{\ensuremath{\textsc{CSVv2}}\xspace}
\newcommand{\DeepCSV}{\ensuremath{\textsc{DeepCSV}}\xspace}
\newcommand{\DeepJet}{\ensuremath{\textsc{DeepJet}}\xspace}
\newcommand{\DeepTau}{\ensuremath{\textsc{DeepTau}}\xspace}
\newcommand{\DeepAKeight}{\ensuremath{\textsc{DeepAK8}}\xspace}
\newcommand{\BEST}{\ensuremath{\textsc{best}}\xspace}
\newcommand{\HOTVR}{\ensuremath{\textsc{hotvr}}\xspace}
\newcommand{\ImageTop}{\ensuremath{\textsc{ImageTop}}\xspace}

\newcommand{\TypeOne}{\mbox{Type I}\xspace}
\newcommand{\TypeThree}{\mbox{Type III}\xspace}
\newcommand{\nuMSM}{\ensuremath{\PGn\text{MSM}}\xspace}

\newcommand{\SU}[1]{\ensuremath{\mathrm{SU}(#1)}\xspace}

\hyphenation{fer-mi-ons}

\cmsNoteHeader{EXO-23-006}
\title{Review of searches for vector-like quarks, vector-like leptons, and heavy neutral leptons in proton-proton collisions at \texorpdfstring{$\sqrt{s}=13\TeV$}{sqrt(s) = 13 TeV} at the CMS experiment}
\date{\today}

\abstract{
    The LHC has provided an unprecedented amount of proton-proton collision data, bringing forth exciting opportunities to address fundamental open questions in particle physics. These questions can potentially be answered by performing searches for very rare processes predicted by models that attempt to extend the standard model of particle physics. The data collected by the CMS experiment in 2015--2018 at a center-of-mass energy of 13\TeV can be used to test the standard model with high precision and potentially uncover evidence for new particles or interactions. An interesting possibility is the existence of new fermions with masses ranging from the \MeVns to the \TeVns scale. Such new particles appear in many possible extensions of the standard model and are well motivated theoretically. New fermions may explain the appearance of three generations of leptons and quarks, the mass hierarchy across these generations, and the nonzero neutrino masses. In this report, the results of searches targeting vector-like quarks, vector-like leptons, and heavy neutral leptons at the CMS experiment are summarized. The complementarity of current searches for each type of new fermion is discussed, and combinations of several searches for vector-like quarks are presented. The discovery potential for some of these searches at the High-Luminosity LHC is also discussed.
}

\hypersetup{%
pdfauthor={CMS Collaboration},%
pdftitle={Review of searches for vector-like quarks, vector-like leptons, and heavy neutral leptons in proton-proton collisions at sqrt(s) = 13 TeV at the CMS experiment},%
pdfsubject={CMS},%
pdfkeywords={CMS, VLQ, VLL, HNL}}

\maketitle
\tableofcontents

\clearpage
\section{Introduction}

The standard model (SM) of particle physics has emerged as a highly successful theory capable of explaining a large number of observations. However, alongside its achievements, the SM is unable to provide answers to intriguing questions that invite further investigation. For instance, the SM faces the hierarchy problem~\cite{Hebecker2021}, a puzzling issue related to the vastly different strengths of the electroweak (EW) force and gravity. The Higgs boson (\PH), whose corresponding field is responsible for the mass of fundamental particles, exhibits a mass that appears to be unnaturally light compared to the Planck energy scale, at which new physics must manifest. The level of fine-tuning of the Higgs boson mass that is required to cancel large quantum corrections in the SM motivates the search for new particles and interactions that could provide a potentially more natural solution.

Furthermore, neutrinos are treated as massless particles within the framework of the SM. However, observations of neutrino oscillations~\cite{Super-Kamiokande:1998kpq,SNO:2001kpb} have since revealed that neutrinos do possess mass.
The origin of the neutrino masses and the reason for their smallness are both unknown, posing further challenges to our understanding of particle physics.

Hence, a major issue confronting the SM resides in its inability to comprehensively explain the masses of fundamental particles such as the Higgs boson and neutrinos. In an attempt to address this problem, new fermions are hypothesized, with masses ranging from the \MeVns to the \TeVns scale: vector-like quarks (VLQs), vector-like leptons (VLLs), and heavy neutral leptons (HNLs). These hypothetical particles show potential to resolve the limitations posed by the SM in characterizing particle masses.

\begin{figure}[!ht]
\centering
\includegraphics[width=0.32\textwidth]{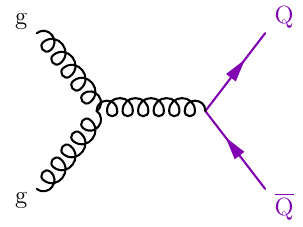}%
\hfill%
\includegraphics[width=0.32\textwidth]{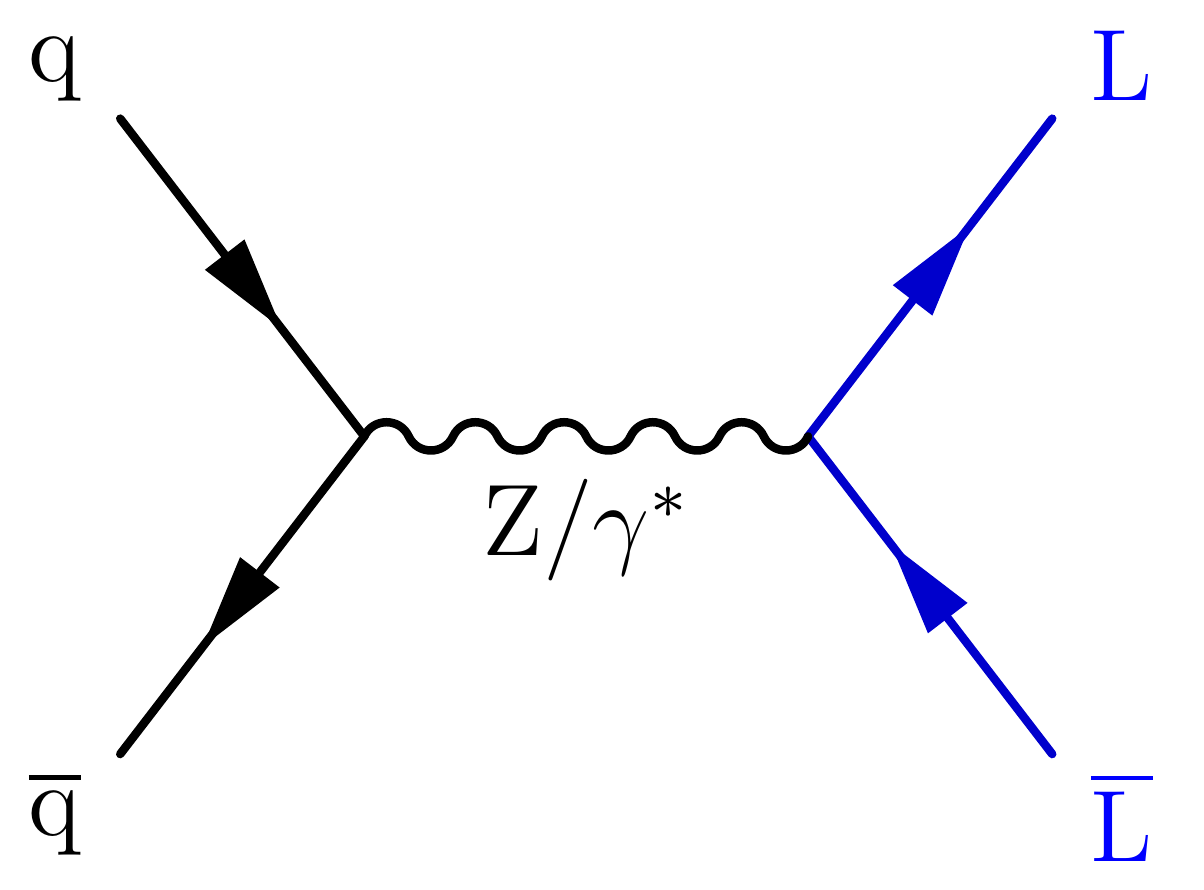}%
\hfill%
\includegraphics[width=0.32\textwidth]{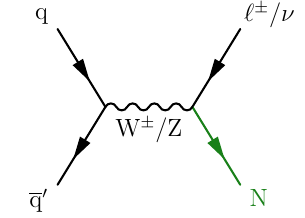}%
\caption{%
    Representative Feynman diagrams showing the production of VLQs (\PQQ, left), VLLs (\VLLL, middle), and HNLs (\PN, right) in proton-proton collisions.
}
\label{fig:introdiagrams}
\end{figure}

Beyond the SM (BSM) models in which VLQs and VLLs are introduced, such as models with extra dimensions~\cite{Randall:1999ee,Huang:2012kz} and composite Higgs models~\cite{Contino:2011np,Greco:2014aza,Falkowski:2013jya}, offer solutions to the hierarchy problem~\cite{Aguilar-Saavedra:2013qpa}, aim to explain the observed fermion flavor structure of the SM, and may provide dark matter candidates.
Meanwhile, HNLs may be the missing component in explaining the origin of neutrino masses~\cite{Mohapatra:1979ia,Schechter:1980gr,Foot:1988aq}. Additionally, HNLs could provide a solution to the observed baryon asymmetry in the universe, as well as contribute to the understanding of dark matter.
In the \GeVns up to the \TeVns range, these new fermions are typically predicted to decay into SM particles, which can be recorded in experiments such as CMS~\cite{CMS:2008xjf} at the CERN LHC.
Figure~\ref{fig:introdiagrams} shows example Feynman diagrams for the production of VLQs, VLLs, and HNLs in proton-proton (\pp) collisions.

In this report, we present the contributions of the CMS experiment to searches for VLQs, VLLs, and HNLs, using the
\pp collision data sets collected by the CMS detector during the years 2015 to 2018.
We discuss the CMS detector and event reconstruction, along with the planned CMS detector upgrade at the High-Luminosity LHC (HL-LHC), in Section~\ref{sec:detector}. This is followed by a description of the data sets and simulations in Section~\ref{sec:data_sim} and common experimental challenges faced in the searches in Section~\ref{sec:reco}.
In Sections~\ref{subsec:theory_vlq}--\ref{sec:results_hnl}, we present the theoretical basis, review the recent results from the CMS experiment, and discuss future prospects on VLQs, VLLs, and HNLs.
Finally, we summarize the review in Section~\ref{sec:summary}. All acronyms are defined in Appendix~\ref{sec:glossary}.

\section{The CMS detector and event reconstruction}
\label{sec:detector}

The central feature of the CMS apparatus is a superconducting solenoid of 6\unit{m} internal diameter, providing a magnetic field of 3.8\unit{T}. Within the solenoid volume are a silicon pixel and strip tracker, a lead tungstate crystal electromagnetic calorimeter (ECAL), and a brass and scintillator hadron calorimeter (HCAL), each composed of a barrel and two endcap sections. Forward calorimeters extend the pseudorapidity ($\eta$) coverage provided by the barrel and endcap detectors.
The forward hadron (HF) calorimeter uses steel as an absorber and quartz fibers as the sensitive material. The two halves of the HF are located 11.2\unit{m} from the interaction region, one on each end, and together they provide coverage in the range $3.0<\abs{\eta}<5.2$. They also serve as luminosity monitors.
Muons are measured in gas-ionization detectors embedded in the steel flux-return yoke outside the solenoid. A more detailed description of the CMS detector, together with a definition of the coordinate system used and the relevant kinematic variables, can be found in Ref.~\cite{CMS:2008xjf}.

Events of interest are selected using a two-tiered trigger system. The first level, composed of custom hardware processors, uses information from the calorimeters and muon detectors to select events at a rate of around 100\unit{kHz} within a fixed latency of 4\mus~\cite{CMS:2020cmk}. The second level, known as the high-level trigger (HLT), consists of a farm of processors running a version of the full event reconstruction software optimized for fast processing, and reduces the event rate to around 1\unit{kHz} before data storage~\cite{CMS:2016ngn}.

The global event reconstruction, also called particle-flow (PF) event reconstruction~\cite{CMS:2017yfk}, aims to reconstruct and identify each individual particle in an event, with an optimized combination of all subdetector information. In this process, the identification of the particle type (photon, electron, muon, charged hadron, neutral hadron) plays an important role in the determination of the particle direction and energy. Photons are identified as ECAL energy clusters not linked to the extrapolation of any charged particle trajectory to the ECAL. Electrons are identified as a primary charged particle track and potentially many ECAL energy clusters corresponding to this track extrapolation to the ECAL and to possible bremsstrahlung photons emitted along the way through the tracker material. Muons are identified as tracks in the central tracker consistent with either a track or several hits in the muon system, and associated with calorimeter deposits compatible with the muon hypothesis. Charged hadrons are identified as charged particle tracks neither identified as electrons, nor as muons. Finally, neutral hadrons are identified as HCAL energy clusters not linked to any charged particly trajectory, or as a combined ECAL and HCAL energy excess with respect to the expected charged hadron energy deposit.

The primary vertex (PV) is taken to be the vertex corresponding to the hardest scattering in the event, evaluated using tracking information alone, as described in Section 9.4.1 of Ref.~\cite{Contardo:2020886}.

The energy and direction of photons is obtained from the ECAL measurement.
In the barrel section of the ECAL, an energy resolution of about 1\% is achieved for unconverted or late-converting photons in the tens of \GeVns energy range. The energy resolution of the remaining barrel photons is about 1.3\% up to $\abs{\eta}=1$, changing to about 2.5\% at $\abs{\eta}=1.4$. In the endcaps, the energy resolution is about 2.5\% for unconverted or late-converting photons, and between 3 and 4\% for the other ones~\cite{CMS:2015myp}.

The energy of electrons is determined from a combination of the track momentum at the PV, the corresponding ECAL cluster energy, and the energy sum of all bremsstrahlung photons attached to the track.
The momentum resolution for electrons with $\pt\approx45\GeV$ from $\PZ\to\ee$ decays ranges from 1.6 to 5.0\%. It is generally better in the barrel region than in the endcaps, and also depends on the bremsstrahlung energy emitted by the electron as it traverses the material in front of the ECAL~\cite{CMS:2020uim,CMS-DP-2020-021}.

Muons are measured in the pseudorapidity range $\abs{\eta}<2.4$, with detection planes made using three technologies: drift tubes (DTs), cathode strip chambers (CSCs), and resistive-plate chambers (RPCs).
The energy of muons is obtained from the corresponding track momentum.
The single-muon trigger efficiency exceeds 90\% over the full $\eta$ range, and the efficiency to reconstruct and identify muons is greater than 96\%. Matching muons to tracks measured in the silicon tracker results in a relative transverse momentum (\pt) resolution, for muons with a \pt up to 100\GeV, of 1\% in the barrel and 3\% in the endcaps. The \pt resolution in the barrel is better than 7\% for muons with \pt up to 1\TeV~\cite{CMS:2018rym}.

The energy of charged hadrons is determined from a combination of the track momentum and the corresponding ECAL and HCAL energies, corrected for the response function of the calorimeters to hadronic showers.
Finally, the energy and direction of neutral hadrons is obtained from corrected ECAL and HCAL energies with no matching to a track.

For each event, hadronic jets are clustered from these reconstructed particles using the infrared and collinear safe anti-\kt algorithm~\cite{Cacciari:2008gp, Cacciari:2011ma} with a distance parameter of 0.4 (``small-radius'' jets) or 0.8 (``large-radius'' jets). Jet momentum is determined as the vectorial sum of all particle momenta in the jet, and is found from simulation to be, on average, within 5 to 10\% of the true momentum over the entire \pt spectrum and detector acceptance.
Jet energy corrections are derived from simulation to bring the measured response of jets to that of particle level jets on average. In situ measurements of the momentum balance in dijet, photon+jet, {\PZ}+jet, and QCD multijet events (events composed of jets produced through the strong interaction), are used to account for any residual differences in the jet energy scale between data and simulation~\cite{CMS:2016lmd}. The jet energy resolution amounts typically to 15--20\% at 30\GeV, 10\% at 100\GeV, and 5\% at 1\TeV~\cite{CMS:2016lmd}. Additional selection criteria are applied to each jet to remove jets potentially dominated by anomalous contributions from various subdetector components or reconstruction failures.

Additional \pp interactions within the same or nearby bunch crossings (pileup) can contribute additional tracks and calorimetric energy depositions to the jet momentum.
The \textit{charged hadron subtraction}~(CHS) and \textit{pileup-per-particle identification}~(PUPPI)~\cite{Bertolini:2014bba,Sirunyan:2020foa} algorithms are used to mitigate the effect of pileup.
The CHS algorithm removes charged PF candidates that are associated with a pileup vertex. The remaining pileup is removed with an offset correction after the jet clustering.
Alternatively, the PUPPI algorithm uses local shape information, event pileup properties, and tracking information to mitigate the effect of neutral pileup at the PF candidate level.
A detailed explanation of the PUPPI algorithm and the comparison to CHS can be found in Ref.~\cite{Sirunyan:2020foa}.

{\tolerance=800
Small-radius jets that originate from a \PQb quark are identified using dedicated jet tagging algorithms.
The \PQb-tagging algorithms used in the analyses presented in this article are \CSVvtwo~\cite{CMS:2017wtu}, \DeepCSV~\cite{CMS:2017wtu}, and
\DeepJet~\cite{Bols_2020,CMS-DP-2018-058}.
The working points used for these algorithms depend on the analysis. For the frequently employed ``medium'' working point of the \DeepJet (\CSVvtwo) algorithm, the tagging efficiency is typically 75 (70)\% for \PQb quark jets with percent-level misidentification rate for light quark and gluon jets, but depends on the jet \pt.
To identify hadronically decaying \PW, \PZ, and Higgs bosons as well as top quarks with high momenta,
where individual jets may be merged into a large-radius jet, dedicated identification algorithms are applied, as discussed in Section~\ref{sec:boosted}.
\par}

Hadronic \PGt lepton decays (\tauh) are reconstructed from jets, using the hadrons-plus-strips algorithm~\cite{CMS:2018jrd}, which combines 1 or 3 tracks with energy deposits in the calorimeters, to identify the tau lepton decay modes. Neutral pions are reconstructed as strips with dynamic size in $\eta$-$\phi$ (where $\phi$ is the azimuthal angle) from reconstructed electrons and photons, where the strip size varies as a function of the \pt of the electron or photon candidate.
To distinguish genuine \tauh decays from jets originating from the hadronization of quarks or gluons, and from electrons, or muons, the \DeepTau algorithm is used~\cite{CMS:2022prd}. Information from all individual reconstructed particles near the \tauh axis is combined with properties of the \tauh candidate and of the event.
The rate of a jet to be misidentified as \tauh by the \DeepTau algorithm depends on the \pt and quark flavor of the jet. In simulated events from \PW boson production in association with jets it has been estimated to be 0.43\% for a genuine \tauh identification efficiency of 70\%. The misidentification rate for electrons (muons) is 2.60 (0.03)\% for a genuine \tauh identification efficiency of 80 ($>$99)\%.

The missing transverse momentum vector \ptvecmiss is computed as the negative vector sum of the transverse momenta of all the PF candidates in an event, and its magnitude is denoted as \ptmiss~\cite{CMS:2019ctu}. The \ptvecmiss is modified to account for corrections to the energy scale of the reconstructed jets in the event.
In analyses that employ the PUPPI algorithm, the pileup dependence on the \ptvecmiss observable is removed by computing the \ptvecmiss from the PF candidates weighted by their probability to originate from the primary interaction vertex~\cite{CMS:2019ctu}.

Anomalous high-\ptmiss events can be due to a variety of reconstruction failures, detector malfunctions, or noncollision backgrounds. Such events are rejected by event filters that are designed to identify more than 85--90\% of the spurious high-\ptmiss events with a mistagging rate less than 0.1\%~\cite{CMS:2019ctu}.

The silicon pixel tracker was replaced at the year-end technical stop of the LHC after the 2016 data taking. Among other improvements, the upgraded detector features an additional barrel layer closer to the beam pipe and additional forward disks, leading to a better tracking performance and typically improved performance of charged particle reconstruction and jet flavor identification in the analysis of the data recorded in 2017 and 2018~\cite{CMSTrackerGroup:2020edz}.

The CMS detector will undergo an upgrade, called Phase 2, to take full advantage of the HL-LHC, scheduled to start in 2029.
This upgrade will involve improving the rate and latency through the first level hardware trigger upgrade to
750\unit{kHz} and 12.5\mus,
respectively, and the HLT software trigger is expected to reduce the rate by a factor of 100 to 7.5\unit{kHz}.
The silicon pixel and strip trackers will be replaced to
increase granularity, lower-density material will be used to reduce the material budget
in the tracking volume, and the geometrical coverage will extend up to $\abs{\eta}<4$.
In addition, the new strip tracker module design with two closely spaced sensors enables the deployment of a novel track trigger algorithm.
The extended tracker coverage will allow jet identification algorithms for \PQb quarks or other
massive SM particles to reach new regions of the pseudorapidity, improving sensitivity for
searches with jets. This new feature of the Phase-2 detector will be particularly important for
HL-LHC VLQ searches.

The ECAL barrel will feature updated front-end electronics that
will allow high-precision timing capabilities for photons, and the HCAL
in the barrel region will be read out by silicon photomultipliers. The endcap electromagnetic
and hadron calorimeters will be replaced with a combined sampling calorimeter that will provide
highly segmented spatial information in both transverse and longitudinal directions. The muon
system will be enhanced by upgraded electronics of the existing RPCs,
CSCs, and DTs. New muon detectors based on improved RPC and gas
electron multiplier technologies will be installed to add redundancy, increase the geometrical
coverage up to $\abs{\eta}<2.8$, and improve the reconstruction performance and trigger efficiency in the forward
region. Upgrading the endcap calorimeter and muon detectors will improve reconstruction and
identification of physics objects at high $\eta$. This will benefit searches for
heavy fermions that are accompanied by high-$\eta$ jets, such as $t$-channel single VLQ
production or vector boson fusion production of HNLs.

Additionally, a new timing detector for minimum ionizing particles will be added in both
the barrel and endcap regions, allowing for the four-dimensional reconstruction of interaction vertices,
which will disentangle the approximately 200 nearly simultaneous \pp interactions per bunch
crossing. The ability to maintain adequate pileup mitigation during HL-LHC luminosity conditions
is critical for the next generation of heavy fermion searches, as all high-level physics objects
become more difficult to reconstruct as the number of pileup interactions increases. Heavy fermion
searches at the HL-LHC will also benefit from the expected 14\TeV collision energy and
unprecedented integrated luminosity.

The detailed description of the CMS Phase-2 upgrade is presented in Refs.~\cite{Contardo:2020886, CERN-LHCC-2017-011, CERN-LHCC-2017-012, CERN-LHCC-2017-023,
CMS:2667167, CERN-LHCC-2020-004, Collaboration:2759072},
and the expected performance of the reconstruction algorithms and pileup mitigation with the
CMS Phase-2 detector is presented in Ref.~\cite{Collaboration:2650976}.

\section{Data set and simulation}
\label{sec:data_sim}

Run 2 of the LHC began in 2015 and continued through 2018, with \pp collisions at $\sqrt{s}=13\TeV$. The CMS detector recorded data corresponding to integrated luminosities of
2.3, 36.3, 41.5, and 59.8\fbinv in 2015, 2016, 2017, and 2018, respectively~\cite{LUM16,LUM17,LUM18}. Many searches presented in this report were published with partial
Run 2 data collected in 2015 or 2016. The most recent searches combine data collected in 2016--2018, corresponding to an integrated luminosity of 138\fbinv.
The 2016--2018 data set is referred to as the ``full Run 2 data set'', and data from an individual year are referred to as, for example, ``the 2016 data set''.
Sets of simulated samples are created for each year of data taking separately to match the appropriate detector operation conditions and calibrations.

The simulations for pair production of vector-like \PQT and \PQB quarks discussed in Section~\ref{subsubsec:vlq_proddec} are predominantly created at leading order (LO) in perturbative quantum chromodynamics (QCD) using \MGvATNLO~\cite{Alwall:2014hca}. 
The simulated samples for VLQ pair production cover a mass range from 0.9 to 1.8\TeV, all featuring a narrow decay width of 10\GeV. 
Decays of the \PQT and \PQB quarks to SM particles are modeled using \PYTHIA8~\cite{Sjostrand:2007gs}, with events distributed evenly across the third-generation decay modes.
The loss of chirality and spin dependence in the decays is not significant for the observables employed in the \PQT and \PQB quark pair production searches. 
Pair production of \xft quarks decaying to \tW is generated at LO in QCD using \MGvATNLO, with the decays of the \PQt quark and \PW boson performed using
\textsc{MadSpin}~\cite{Artoisenet:2012st}. This process is generated separately for left-handed and right-handed \xft quarks.
Additionally, for the EW production of single VLQs, discussed in Section~\ref{subsubsec:vlq_proddec} as well, \MGvATNLO at LO is used as the Monte Carlo (MC) event generator. These simulations consider different width hypotheses for the VLQs, including narrow-width samples with a fixed relative width of approximately 1\% and large-width samples with relative widths of 10, 20, and 30\%. Decays of the single VLQs to SM particles are modeled using \MGvATNLO.
The VLQ pair and single production cross sections are computed to next-to-next-to-leading order (NNLO) and LO in QCD, respectively, as discussed in Section~\ref{subsubsec:vlq_crosssec}.

Vector-like lepton samples in the minimal model~\cite{delAguila:1982fs,Fishbane:1985gu,Fishbane:1987tx,Montvay:1988av,Fujikawa:1994we,Kumar:2015tna,Bhattiprolu:2019vdu} and the 4321 model~\cite{4321_model,benchmark,threeSite_PS,nonuniversal4321} as discussed in Section~\ref{subsec:vll_minimal_theory} and Section~\ref{subsec:vll_4321_theory}, respectively, are created with \MGvATNLO at LO. The production cross sections used to investigate the minimal VLL model are calculated at next-to-leading order (NLO) precision~\cite{Bhattiprolu:2019vdu}. The VLL production in the 4321 model is restricted to using only SM EW couplings. In the simulated samples, the couplings of the leptoquark are set to zero for all first- and second-generation SM fermions. The masses of the leptoquarks are always taken to be 3.5\TeV and the masses of the two third-generation VLLs are always taken to be equal to each other. The simulated model parameters in the 4321 model, including the leptoquark and \PZpr boson masses, are taken from benchmark scenarios proposed to explain the lepton flavor universality~\cite{4321footprints}. However, due to the much lighter mass scale of the VLLs, the results are mostly insensitive to the values of the boson masses.

Signal events in \TypeOne seesaw models~\cite{Brdar:2019iem}, described in Section~\ref{sec:HNL_th_type-I}, are simulated at NLO precision using \MGvATNLO for the HNL particle in a mass range from 50\GeV to 25\TeV, with the mixing element \mixparmN fixed to 1. Signal events in the \TypeThree seesaw model~\cite{Fuks:2012qx,Fuks:2013vua}, described in Section~\ref{subsec:typeIIIseesaw}, are generated with \MGvATNLO at LO. The production cross sections within the \TypeThree seesaw signal model are calculated at NLO plus next-to-leading logarithmic precision, assuming that the heavy leptons are \SU2 triplet fermions.

{\tolerance=800
Signal events in left-right symmetric models (LRSMs), described in Section~\ref{subsec:LRSMtheory}, are simulated using \MGvATNLO at LO, for various right-handed (RH) \PWR boson mass hypotheses in the range from 500 to 6000\GeV with heavy neutrino (\PN) masses ranging from 100\GeV up to the \PWR mass. The production cross sections are scaled to NLO in perturbative QCD using $K$ factors obtained from the same generator. A few other LRSM searches utilize \MGvATNLO at NLO to generate signal samples assuming a \PZpr boson mass in the range from 400 to 5000\GeV with a choice of heavy neutrino mass from 100\GeV to $\mZpr/2$. In the generation, the \PWR boson mass is assumed to be 5\TeV, so that heavy neutrinos are always lighter than the \PWR boson.
\par}

The \MGvATNLO generator program is used at LO to simulate Drell--Yan and \PW boson production, production of QCD multijet events, and processes involving a single top quark and an EW or Higgs boson, two top quarks and two bosons, or three top quarks. Additionally, the \MGvATNLO generator is used at NLO to simulate \tZ, \ttW and \ttZ (\ttV), $\ttbar\ttbar$, \WW, \Zgamma, \WZ, and triboson ($\PV\PV\PV$) production, as well as single top quark $s$-channel production. Some searches utilize \POWHEG~\cite{POWHEG1,POWHEG2,POWHEG3} to simulate \WZ and \ZZ contributions from quark-antiquark annihilation production~\cite{POWHEGWZ1,POWHEGWZ2}, whereas the contribution from gluon-gluon fusion production is generated at LO using \MCFM 7.0.1~\cite{Campbell:2010ff}. The \POWHEG generator is used to simulate \ttbar~\cite{POWHEGttbar}, \ttH, and most single-top quark production processes~\cite{POWHEGsingletopST,POWHEGsingletopTW} at NLO. The top quark mass used in all simulations is 172.5\GeV. The SM processes involving Higgs boson production are generated using \POWHEG and \textsc{JHUGen} 7.0.11~\cite{Gao:2010qx,Bolognesi:2012mm,Anderson:2013afp,Gritsan:2016hjl} at NLO, for a Higgs boson mass of 125\GeV.

{\tolerance=1800
The parton distribution functions (PDFs) NNPDF3.0 NLO or LO~\cite{NNPDF30} and NNPDF3.1 NNLO~\cite{NNPDF:2017mvq} were used to generate all background and signal samples in the analyses presented in this report.
To perform the parton showering, fragmentation, and hadronization of the matrix-level events in all samples, \PYTHIA8 was used with the underlying-event tune CUETP8M1~\cite{CUETP8M1}, CUETP8M2T4~\cite{CMS-PAS-TOP-16-021}, or CP5~\cite{CMS:2019csb}, depending on the analyses.
The MLM~\cite{Hoeche:2006ps} or FxFx~\cite{Frederix:2012ps} matching schemes are used to remove double-counted partons between the matrix element calculations and parton shower, in LO and NLO setups, respectively.
The simulation of the response of the CMS detector to incoming particles is performed using the \GEANTfour toolkit~\cite{GEANT}.
Pileup collisions are simulated and incorporated in the simulated event samples, with a frequency distribution matching the one observed in collisions data (with an average of 23 collisions in 2016 and 32 in 2017--2018).
\par}

\section{Common experimental strategies}
\label{sec:reco}

Many of the searches for VLQs, VLLs, and HNLs described in this review face common challenges.
For instance, the decay products of heavy fermions may be highly Lorentz boosted, resulting in characteristic signatures in the detector. The dedicated algorithms developed and employed to identify boosted bosons and \PQt quarks experimentally are discussed in Section~\ref{sec:boosted}. Afterwards, commonly used methods for the estimation of background process contributions using simulation or from control regions (CRs) in data are outlined in Section~\ref{sec:bkgest}.
Finally, common statistical methods to quantify the presence or absence of a signal are discussed in Section~\ref{sec:statistics}.

\subsection{Boosted objects and taggers}
\label{sec:boosted}

If a resonance is much heavier than its decay products, the decay products are highly Lorentz boosted.
This results in very collimated sprays of particles from those decay products, where hadronic decays of heavy SM particles
cannot be reconstructed in individual small-radius jets, but are merged into the same large-radius jet.
In these so-called ``boosted'' final states, one may attempt to identify \PW, \PZ, and Higgs bosons as well as \PQt quarks with high transverse momenta
($\pt>200\GeV$ for \PW, \PZ, and Higgs bosons, and $\pt>400\GeV$ for \PQt quarks).
These particles have a large branching fraction into hadrons; therefore, considering hadronic final states
increases the sensitivity of searches significantly if these decays can be discriminated from QCD multijet production.

The angular distance ($\DR=\sqrt{\smash[b]{(\Deta)^2+(\Dphi)^2}}$) between the quarks from the decay of the \PW, \PZ, and Higgs bosons or the
\PQt quark depend on the mass and the momentum of the parent particle.
Hadronic decays of these particles are captured using large-radius jets.
One challenge posed by the use of large-radius jets is that the underlying-event activity and pileup contribute significantly to the jet energy,
which results in a worsening of the jet energy resolution.
Pileup mitigation, as discussed in Section~\ref{sec:detector}, and jet grooming techniques are crucial for the usage of large-radius jets.

In order to remove soft and wide-angle radiation captured by the rather large jet area of large-radius jets,
jet grooming techniques, such as \textit{pruning}~\cite{Ellis:2009me} and \textit{soft drop} (SD)~\cite{Larkoski:2014wba}, are applied.
For pruning, all jet constituents of the large-radius jet are reclustered with the
Cambridge--Aachen algorithm~\cite{Dokshitzer:1997in,Wobisch:1998wt}. In each step the two requirements
for angular separation between jet constituents $i$ and $j$ ($\DR_{ij}<m_{i+j}/\pTij$)
and soft splitting ($\min(\pTi,\pTj)\leq z_{\text{prune}}\pTij$) are checked,
where $m$ and \pt are the mass and transverse momentum of the jet constituents or their combination, and $z_{\text{prune}}=0.1$.
If both requirements are fulfilled the two constituents are combined; otherwise, the softer constituent is discarded.
The SD algorithm removes soft and wide-angle radiation from the jet by
reclustering the large-radius jet with the Cambridge-Aachen algorithm and testing the SD requirement $\min(\pTi,\pTj)>z_{\text{cut}}\pTij(\DR_{ij}/R)^{\beta}$
in each declustering step.
The standard parameters used in the CMS experiment are $z_{\text{cut}}=0.1$ and $\beta=0$.
The hardest branch is followed until the SD requirement is fulfilled, where the procedure stops.
As a consequence, at most two SD subjets are defined by this procedure.
The mass is calculated by the invariant mass of the two subjets and is called the SD mass (\mSD).

In addition to the mass of the jet, its ``prong count'' is a relevant property for boosted object identification.
The two-prong \PW/\PZ/Higgs boson decays into a pair of quarks usually results in two distinct regions of high energy density in the jet substructure.
Similarly, the three-prong \PQt quark decay into $\PQb\qqbarpr$ typically results in a distinct signature,
while jets that originate from light quarks or gluons are expected to have only a single region of high energy density, \ie,
a single prong.
The prong count of the jet is measured with the $N$-subjettiness~\cite{Thaler:2010tr},
\begin{equation}
    \tauN = \frac{1}{\sum_k\pTk R_0} \sum_k\pTk\min\left(\DR_{1,k},\DR_{2,k},\dots\DR_{N,k}\right),
\end{equation}
where $N$ is the number of axes considered, $k$ is the number of constituents of the given jet, and $R_0$ is the jet radius.
A low value of \tauN means that all radiation is aligned with the
candidate subjets, while a high value means that the radiation is not aligned.
The CMS Collaboration typically uses two ratios of $N$-subjettiness:
for two-prong tagging (\PW/\PZ/Higgs bosons), $\tauTO=\tau_2/\tau_1$ is used, and for three-prong tagging (\PQt quarks),
$\tauTT=\tau_3/\tau_2$ is used.
Requirements on the $N$-subjettiness ratios can lead to the shaping of the jet mass distribution of the QCD multijet background.
This is especially a challenge if the background estimation is based on a smoothly falling function, whereas a
selection based on \tauTO or \tauTT can introduce a peak in the jet mass distribution close to the signal mass peak.
To avoid this behavior, the \textit{designed decorrelated taggers} (DDT)~\cite{Dolen:2016kst} technique was developed,
which is a transformation of the variable \tauTO to $\tauTO^{\text{DDT}}=\tauTO-c\log(\rhoSD)$,
where $\rhoSD=\mSD/\pt$ and $c$ is a constant determined
from the \tauTO distribution as function of \rhoSD in bins of \pt.

For \Ztobb and \Htobb, as well as for \PQt quark tagging, the identification of a
subjet that originates from a \PQb quark (via the \CSVvtwo or \DeepCSV algorithms~\cite{CMS:2017wtu}) can improve the performance.
In addition, a dedicated double-\PQb tagging algorithm~\cite{CMS:2017wtu}, based on a boosted decision tree (BDT), has been developed
using observables associated with the lifetime and mass of the \PQb hadrons to identify \Htobb.
Generally, the identification of boosted objects is done by defining ``taggers'', which are a set of requirements on
the jet mass and substructure information of the jet.

The use of taggers based on neural networks (NNs), such as \DeepAKeight, \ImageTop, or the boosted event shapes tagger (\BEST)~\cite{CMS:2020poo}, offer
higher background rejection than the selection-based taggers described above.
The first technique, \DeepAKeight, is a deep NN (DNN) that uses low-level input features such as the four-momenta
of the constituents of a jet and information about secondary vertices (SVs).
It offers the possibility to classify the different large-radius jets in the event by one classifier
instead of applying different selection-based taggers that might overlap.
The \ImageTop algorithm is an image recognition technique based on a convolutional NN.
The jet energy density is displayed in a two-dimensional image where the jet energy deposit is pixelized,
with the jet axis building the center of this picture. In addition, the image
is rotated such that the major principal axis is vertical. The \ImageTop algorithm is trained
to distinguish jets originating from \PQt quarks from jets initiated by light quarks or gluons.
Both networks can be trained such that the resulting discriminators are not correlated to the jet mass.
For searches that rely on the jet mass distribution, this training method reduces any bias in the mass of light quark or gluon jets.

The \BEST algorithm is designed to identify six types of jets: light-quark/gluon, \PQb quark, \PQt quark, and \PW, \PZ, and Higgs boson jets.
The core of the algorithm targets the fundamental difference between the potential jet parent particles: the mass of the particle,
characterized by \BEST through Lorentz boosts of the jet along its centroid axis into several hypothesis-based frames. The principal idea
is that the boost hypotheses that do not correspond to the true parent particle of the jet will result in ``under-'' or ``over-boosted''
topologies. The network uses information about these event shapes to provide a class prediction for each jet.
A total of 59 input features are calculated for each jet using a mixture of Lorentz-boost invariant and frame-dependent observables.
Invariant features include the traditional jet substructure in the lab frame, such as jet \mSD, $N$-subjettiness, and subjet \PQb tagging scores.
Frame-dependent observables such as Fox--Wolfram moments and sphericity tensors aim to describe the event shape of the transformed jet constituents.

The calibration of these algorithms for \PQt quark and \PW boson jets is performed using data passing
a single-muon trigger by selecting a \ttbar-enriched region with one
\PQt quark decaying semileptonically and one \PQt quark decaying hadronically.
Pass and fail regions are defined based on the jet originating from the hadronically
decaying \PQt quark passing or failing the tagger requirement.
A template fit is done for each specific tagger and working point to extract a factor that corrects
for differences in tagging efficiency between data and simulation.
A detailed description and comparison of the performance of all the taggers
described here can be found in Ref.~\cite{CMS:2020poo}.

Especially in signatures with boosted \PQt quarks, it may happen that a lepton from the
semileptonically decaying \PQt quark overlaps a large-radius or small-radius jet in the event.
If the lepton is within $\DR<0.8$ of the large-radius jet, the large-radius jet is discarded.
However, if a semileptonically decaying \PQt quark is reconstructed by a lepton, \ptmiss,
and a \PQb jet, and the \PQt quark has high momentum, the lepton and the \PQb jet may overlap.
Requiring either an isolated lepton or discarding the small-radius jet that overlaps with the lepton
within $\DR<0.4$ may lead to significant inefficiencies.
Therefore, the lepton is removed from the small-radius jet, and its four-momentum is subtracted from the small-radius jet four-momentum.
Both objects are kept if their \pt passes a given threshold after the removal.
The lepton must pass a custom isolation criterion for boosted \PQt quarks, such that the relative transverse momentum of
the lepton to the nearest small-radius jet ($\pt^{\text{rel}}$) or the angular distance between the lepton and the
nearest small-radius jet are larger than certain thresholds, determined according to the needs of a specific analysis.

\subsection{Common strategies for background estimation}
\label{sec:bkgest}

Accurate modeling of SM background processes is essential to all searches for new fermions.
The dominant background processes depend strongly on the selected objects, particularly leptons, \ptmiss, and \PQb quark jets or jets from boosted particles. Background estimation strategies
are always tailored to an individual search, typically as a variation of one or more of the following
common methods.

\cmsParagraph{Simulation} Many SM processes are simulated at NLO, which provides a strong basis for background
estimation. For processes such as \PW/\zjets production that are often simulated at LO, ratios of cross sections
at NLO to LO can be used to weight simulated events to reproduce predictions of distributions at NLO.
In final states with charged leptons, for which QCD multijet production is unlikely to be a significant background, simulation is a common choice.
Additionally, all searches for new fermions utilize simulation to model the signal process under consideration.

Simulated events are weighted so that the efficiencies of certain selections
in simulation match those observed in data. These corrections are referred to as ``scale factors'' (SFs).
Common SFs in searches at the CMS experiment correct
for differences in the number of pileup interactions, efficiencies of trigger selections, efficiencies of charged lepton
identification and isolation selection criteria, and efficiencies of various jet identification selection criteria.
A detailed set of corrections for the jet energy scale and resolution are computed for simulated events such that the
response of the jet reconstruction algorithms is consistent between observed data and simulation. Searches may also develop
correction formulas to correct for observed mismodeling of data by simulation in certain distributions of interest.
A common correction of this type is to reweight the reconstructed \PQt quark \pt spectrum, since the NLO \PQt quark pair simulations
tend to overpredict the rate of high-\pt \PQt quark pairs~\cite{TOP-16-008,TOP-17-014}. Each correction applied to simulation carries an uncertainty that is taken into account in the statistical methods of signal extraction.

\cmsParagraph{Tight / loose or ``matrix'' methods} Searches that select multiple charged leptons often have considerable background
from events in which ``nonprompt'' leptons are selected. Nonprompt leptons are leptons not originating from the primary interaction vertex,
either because they are produced from the decay of particles with significant lifetimes, such as \PQb quarks or tau leptons,
or because their tracks are misreconstructed.
One method to estimate contributions from these events is to measure how often known prompt leptons, typically from the
decay of \PZ bosons, and known nonprompt leptons, typically from a sample of QCD multijet events, pass a certain set of
lepton selection criteria. A \PZ boson sample is created in data by selecting events with two same-flavor opposite-sign (OS)
leptons whose invariant mass lies very close to the \PZ boson mass. One lepton, known as the ``tag'', is selected using very high-purity
selection criteria, giving confidence that the other ``probe'' lepton is indeed a prompt lepton. The efficiency for the
``probe'' lepton to pass any criteria of interest can then be measured in this sample. In the context of this background
estimation method, the efficiency of the analysis selection criteria is referred to as the ``prompt rate'' $p$.
A QCD multijet sample can be created by selecting events that pass a low-momentum, low-purity, single-lepton trigger, but otherwise
exhibit no strong signs of the lepton arising from a SM boson decay. The rate at which these leptons pass the analysis selection
criteria can be measured, and is referred to as the ``nonprompt'' rate $f$. Both of these rates
describe how often either prompt or nonprompt leptons that pass some baseline ``loose'' selection also pass the ``tight''
selection criteria used in the analysis.

For searches that probe final states with two charged leptons, the probabilities for any prompt or nonprompt lepton to enter the sample must be considered
together to develop a background distribution.
The number of events with leptons passing the tight and/or loose criteria may be observed, in particular the number of events with two tight leptons, \Ntt; one tight and one loose lepton, \Ntl; and two loose leptons, \Nll.
The prompt and nonprompt rates may then be used to convert those observations into numbers of events with two prompt leptons, \Npp; one prompt and one nonprompt lepton, \Npf; and two nonprompt leptons, \Nff~\cite{Wong:2808538}:
\begin{equation}
    \begin{pmatrix}
    \Ntt \\ \Ntl \\ \Nll
    \end{pmatrix} = \left(\begin{array}{ccc}
    p^2 & pf & f^2 \\
    2p(1-p) & f(1-p) + p(1-f) & 2f(1-f) \\
    (1-p)^2 & (1-p)(1-f) & (1-f)^2
    \end{array}\right)
    \begin{pmatrix}
    \Npp \\ \Npf \\ \Nff
    \end{pmatrix}.
\end{equation}
A matrix inversion provides formulas to calculate \Npf and \Nff from the observed number of events with leptons of
varying quality. For a search selecting two tight leptons, the background from events with nonprompt leptons will be given
by $\Nsub{bkg}=pf\Npf+f^2\Nff$. This method can be extended to searches targeting final states with more than two charged leptons by expanding the probability matrix.

\cmsParagraph{Transfer factors} In many searches, one important selection criterion serves as the primary distinction between
a background-dominated CR and a region with good signal sensitivity, called the signal region (SR).
A ``transfer factor'' or ``transfer function''
that describes the efficiency of this principal selection criteria can be derived and applied to the observed data in the
CR in order to estimate the background present in the SR.

The transfer function can be computed in multiple ways. Some searches use simulation for this purpose.
The number of background events in the SR, \NSRbkg, is calculated
as $\NSRbkg=\NCRdata(\NSRsim/\NCRsim)$, where \NCRdata
is the number of observed collision events in the CR, \NSRsim is the number of simulated events in the SR,
and \NCRsim is the number of simulated events in the CR.
The transfer factor from simulation, $\NSRsim/\NCRsim$, can be computed in any bin of an observable or parameterized
with a fit, so the shape as well as the rate of background in the SR may be obtained.

\begin{figure}[!ht]
\centering
\includegraphics[width=0.444\textwidth]{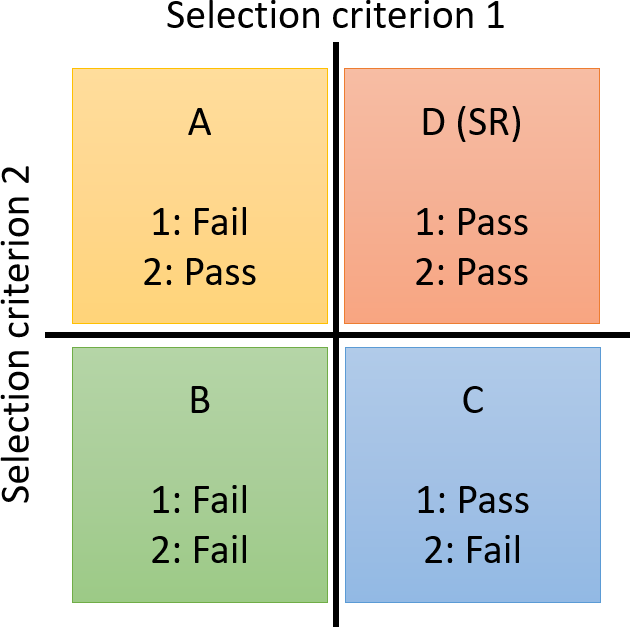}%
\hfill%
\includegraphics[width=0.506\textwidth]{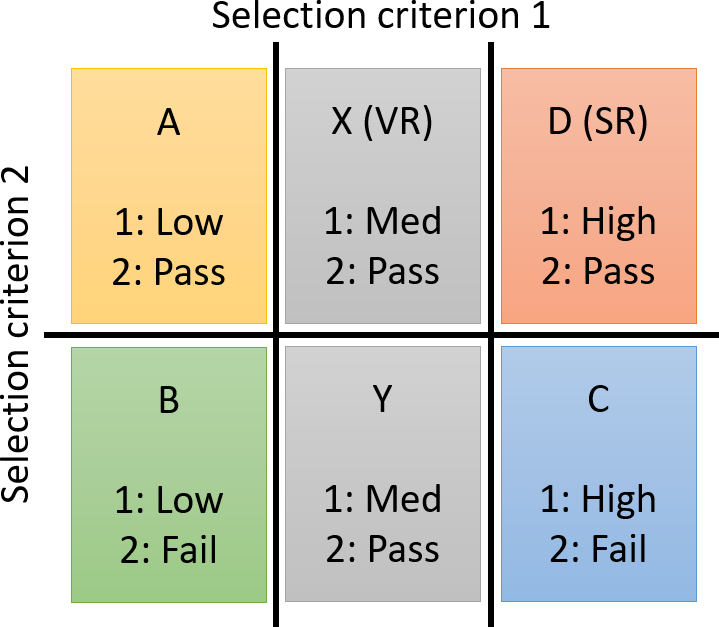}%
\caption{%
    Examples of either four (left) or six (right) selection regions used in the ABCD background estimation method.
    The region for which both criteria are satisfied is the SR.
    Expanding beyond four regions provides at least one ``validation region'' (VR).
}
\label{fig:ABCD}
\end{figure}

Other searches measure transfer factors using observed data in selection regions that are distinct from the primary SR and CR,
in which case the method might be referred to as the ``ABCD'' method. Four selection regions in the observed data are involved,
formed by events either passing or failing either of two selection criteria, as shown in Fig.~\ref{fig:ABCD} (left). The
number of background events in the SR (region D), \Nsub{D}, is calculated from observations in regions A, B, and C as
$\Nsub{A}(\Nsub{C}/\Nsub{B})$. This method may also be used in any bin of an observable to obtain a shape-based prediction for the background.
The ABCD method requires that the selection criteria are statistically independent in order to produce unbiased predictions.

If some background sources are well modelled by
simulation, these contributions may be subtracted from the observed data in each region before computing and applying the transfer function $\Nsub{C}/\Nsub{B}$.
More than four regions may be used to incorporate a method for validation into the procedure, as shown in Fig.~\ref{fig:ABCD} (right).
The number of background events in the validation region (VR) X is estimated from the observations in regions A, B, and Y as $\Nsub{A}(\Nsub{Y}/\Nsub{B})$, and if region X has a suitably low rate of
expected signal events, the observed data in this VR can be compared to the background prediction, to test the validity
of the prediction method. Some searches divide each of the two axes in three parts, resulting in a total of nine regions, and often regions close to the
SR are included in signal extraction fits to better constrain the uncertainties associated with the
background prediction.

\cmsParagraph{Sideband fits} In many searches, the observable most sensitive to the signal is a reconstructed mass
or jet mass distribution, in which the signal is expected to be resonant while the dominant background
processes are nonresonant. The shape of the background distribution may then be predicted by fitting a smooth
functional form to the observed data on either side of the region in which the signal distribution is expected to peak. This method
may be used in multiple dimensions for signals that feature more than one resonance. When multiple
functional forms offer adequate fits to the observed data, an F-statistic may be used to compare the residual sums of
squares for two formulas and determine whether a formula with more parameters provides a significantly better
fit than an alternate formula with fewer parameters (known as the Fisher $\mathcal{F}$-test~\cite{fisher_1922}).

\subsection{Common statistical techniques}
\label{sec:statistics}

All searches for new fermions determine an observable or set of observables that is used to measure the potential presence of
signal events, such as the reconstructed mass of the new fermion, a machine-learning (ML) discriminant, or
another observable that highlights unique features of the signal process. Searches prepare information
about the observed data, the signal prediction, and the background predictions using one or more of the following forms:
\begin{itemize}
\item Cut-and-count: the number of events passing the same set of final selection criteria are counted
for observed data, signal, and background.
\item Histograms: histograms of the chosen observable are created for data, signal, and background.
Each bin of the histogram serves as an independent set of information.
\item Functional forms: fits may be used to describe any distribution according to a functional form, rather
than using a histogram. This method is particularly useful for smoothing tails of distributions where few
events are predicted, and for interpolating signal predictions between simulated mass points.
\end{itemize}
Signal extraction is based on maximum likelihood fits that compare ``data'' (either collision data or pseudodata
sampled from a test distribution) to the signal ($s$) and background ($b$) predictions, with signal scaled by some unknown ratio $\mu$.
A value of $\mu=1$ represents the signal prediction according to the physics model being considered in the search. The
likelihood is assumed to follow a Poisson distribution, and all predictions are subject to various nuisance parameters, $\theta$, that
are given default values $\tilde{\theta}$ and assigned probability density functions ($p$). The likelihood function can be written as:
\begin{equation}
    \mathcal{L}(\text{data}\vert \mu,\theta) = \text{Poisson}(\text{data}\vert\mu s(\theta)+b(\theta)) p(\tilde{\theta}\vert\theta).
\end{equation}
Systematic uncertainties are incorporated into the fit as nuisance parameters. Log-normal probability distributions are assigned
to uncertainties that affect only the normalization of a histogram or rate of a predicted event yield, and Gaussian probability distributions are typically
assigned to uncertainties provided as histograms that affect the shape of a distribution. Nuisance parameters can also be assigned
a Gamma probability distribution. Where predictions are taken from simulation, the primary uncertainty that affects only the normalization is
the integrated luminosity: the integrated luminosities for the years 2015, 2016, 2017, and 2018 have 1.2--2.5\% individual uncertainties, and the total integrated luminosity of 2016--2018 has an uncertainty of 1.6\%~\cite{LUM16,LUM17,LUM18}. Uncertainties in the cross sections
of background processes may also be modeled as normalization uncertainties. Common shape-based uncertainties for simulated processes include
uncertainties in the pileup modeling, lepton and photon SF, jet flavor tagging SFs, jet energy scale and resolution corrections,
and choices of PDF and renormalization and factorization scales in simulation. Histograms are typically created for these uncertainties by
repeating the event selection process with each uncertainty source shifted up or down by one standard deviation.
Background predictions that are modeled from observed data are affected by uncertainties in the value of each parameter in a functional form,
or perhaps uncertainties due to limited event counts in the CRs used to determine a transfer function.
If a search combines multiple channels or multiple years of collision data, uncertainties
pertinant to each individual channel or data set may enter the fit as either fully correlated or as independent.
Uncertainties due to limited event counts in simulated samples are included as Poisson-distributed nuisance parameters using the Barlow--Beeston method~\cite{BBLITE1,BBLITE2}.

Observed and expected limits in the signal strength $\mu$ are extracted by comparing the compatibility
of the observed data with a background-only ($\mu=0$) hypothesis as well as with a signal+background hypothesis.
Some early searches based on 2016 data
compute Bayesian credible intervals to set 95\% confidence level (\CL) upper limits on
the signal production cross section, assuming a flat prior distribution for the signal cross section.
For most of the searches presented in this report, the \CLs method~\cite{Junk:1999kv,Read:2002hq} is used to obtain a limit at 95\% \CL using the profile
likelihood test statistic~\cite{Cowan:2010js}, often in the asymptotic approximation. The \textsc{combine}~\cite{COMBINE} software framework
used by the CMS experiment to compute limits is built on the \textsc{RooFit} and \textsc{RooStats} packages~\cite{Verkerke:2003ir} and implements statistical procedures
developed for combining ATLAS and CMS Higgs boson measurements~\cite{ATLAS:2011tau}.

\clearpage
\section{Theoretical motivation for vector-like quarks}
\label{subsec:theory_vlq}

\subsection{Motivation}

In order to address the hierarchy and naturalness problems of the SM, several extensions have been put forward that introduce the existence of new heavy quarks~\cite{Agashe:2005,Perelstein:2007,Martin:2009bg,Aguilar:2017,Zheng:2019kqu}.
These hypothetical spin-1/2 particles are vector-like in nature, which means that their left- and right-handed components transform in the same way under the EW gauge symmetry group.
The search for VLQs is strongly motivated because, unlike chiral fourth-generation quarks~\cite{Eberhardt:2012}, they are not constrained by current Higgs boson cross section measurements as their masses do not arise from Yukawa couplings~\cite{Aguilar-Saavedra:2013qpa}.

The phenomenology of VLQs is typically described in a simplified model~\cite{Aguilar-Saavedra:2013qpa}, describing interactions of VLQs via the SM gauge bosons.
In ultraviolet-complete models, however, VLQs are accompanied by new gauge bosons resulting
from symmetry requirements. These may be spin-1 resonances, as for example
in models with extra dimensions~\cite{Randall:1999ee}, where the lightest Kaluza--Klein excitation of the gluon
can couple to the lightest fermionic resonances, which are VLQs~\cite{Contino:2006nn, Bini:2011zb, Chala:2014mma}.
In minimal composite Higgs models~\cite{Contino:2011np, Greco:2014aza}, VLQs are introduced
together with new electrically neutral and charged spin-1 resonances~\cite{Giudice:2007fh, DeSimone:2012fs, Matsedonskyi:2015dns}.
In general, phenomenological models accommodating a Higgs boson with a mass of 125\GeV require fermions with masses of
$\mathcal{O}(1\TeV)$~\cite{Marzocca:2012zn, Pomarol:2012qf}, which are usually lighter
than the hypothetical spin-1 resonances.

\subsection{Production and decay modes}
\label{subsubsec:vlq_proddec}

Generally speaking, the phenomenology of VLQs depends on several parameters, such as the couplings to the SM quark generations and EW bosons, the particle mass and width, and the multiplet representation in $\SU3_{\mathrm{C}}\times\SU2_{\mathrm{L}}\times\mathrm{U}(1)_{\mathrm{Y}}$.
As singlets, the VLQs \PQT and \PQB are introduced with electrical charges of $+2/3$ and $-1/3$, respectively. Doublets and triplets incorporate two additional particles denoted by \xft (charge $+5/3$) and \yft (charge $-4/3$).

\begin{figure}[t!]
\centering
\includegraphics[width=0.4\textwidth]{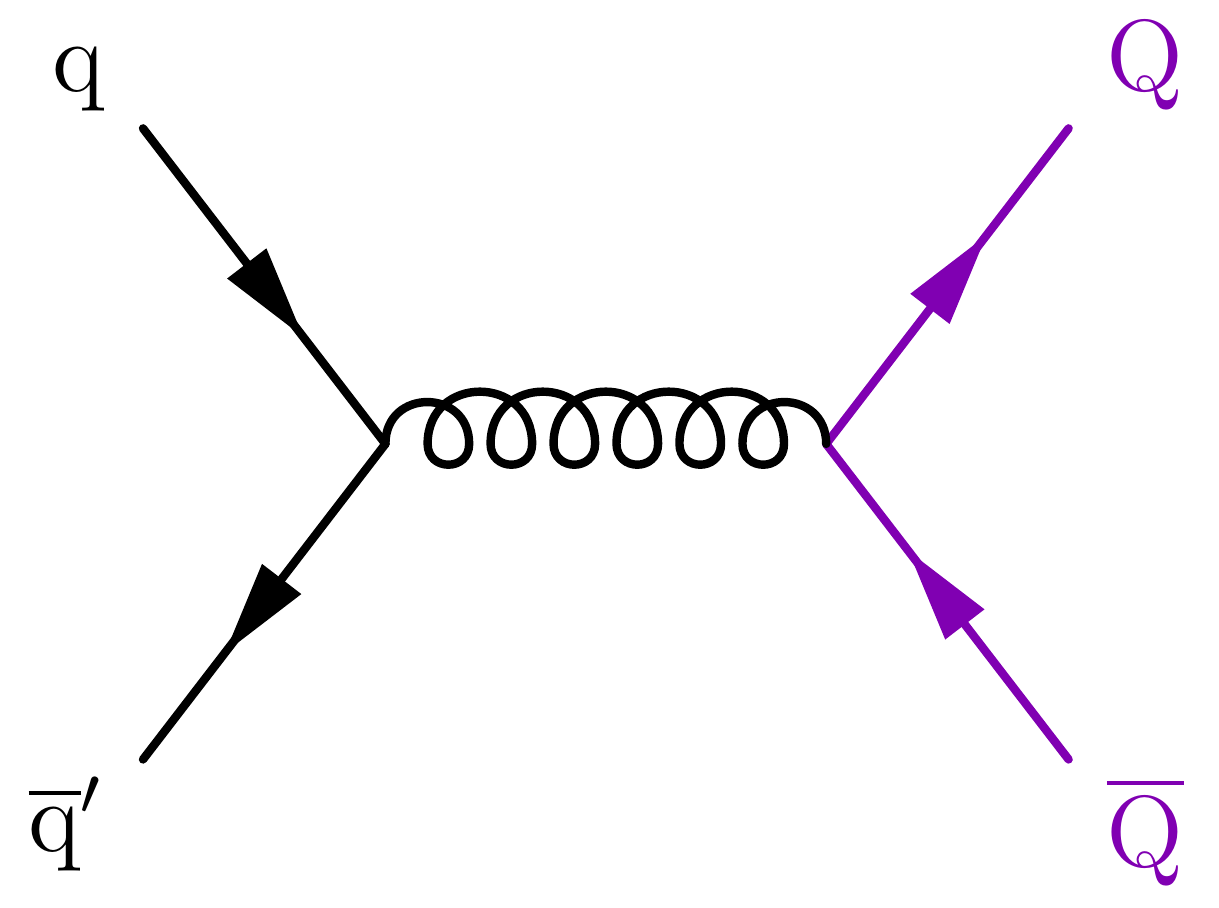}%
\hspace{0.05\textwidth}%
\includegraphics[width=0.4\textwidth]{Figure_001-a.pdf} \\
\includegraphics[width=0.4\textwidth]{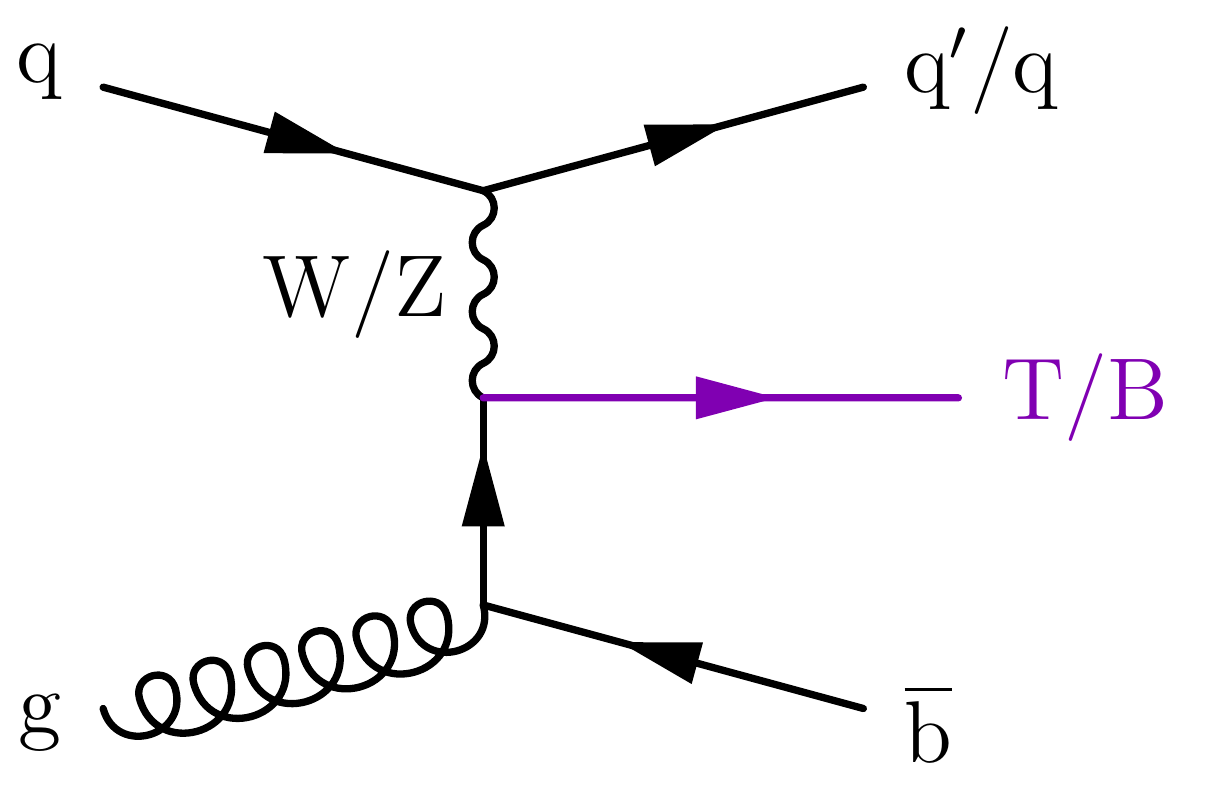}%
\hspace{0.05\textwidth}%
\includegraphics[width=0.4\textwidth]{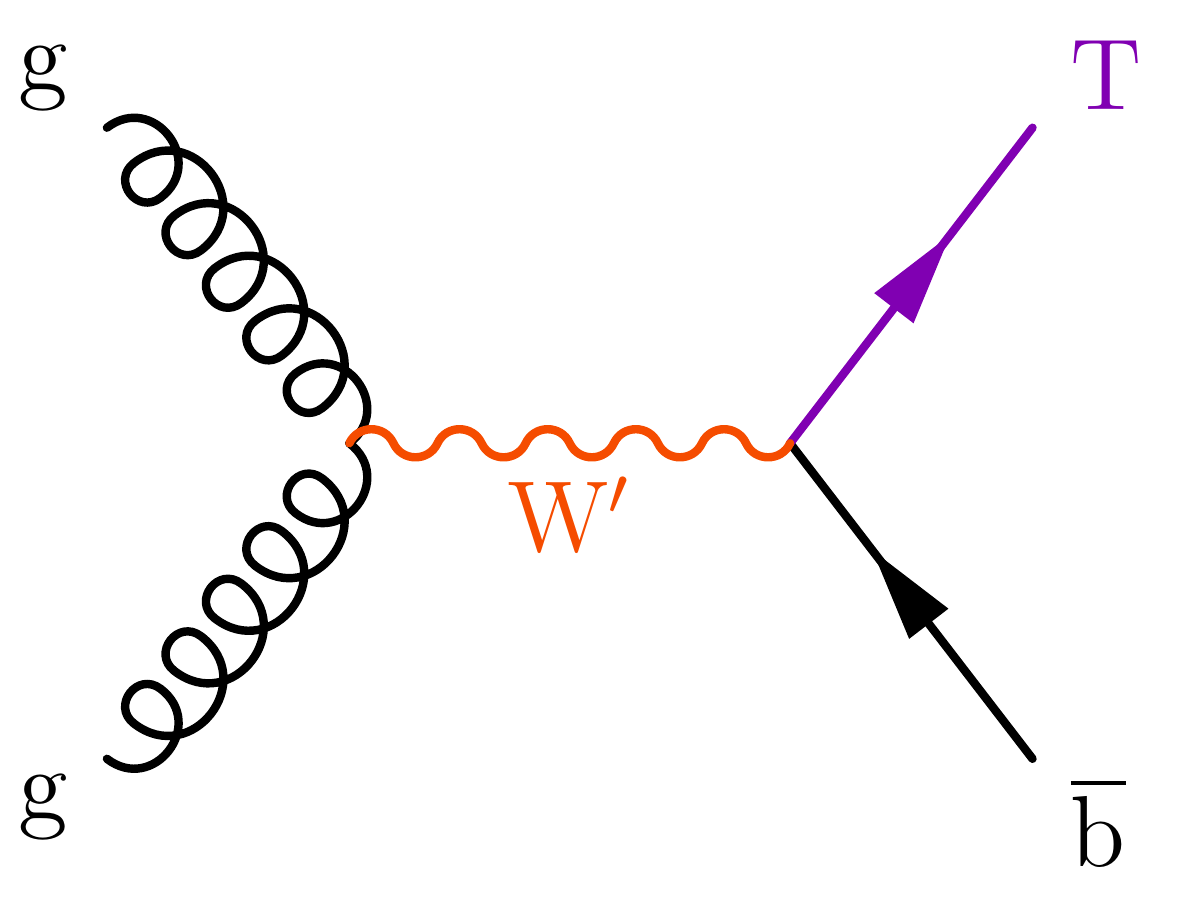}%
\caption{%
    Representative LO Feynman diagrams for pair production of VLQs via the strong interaction (upper row) and single production of VLQs via EW processes (lower left) or via new interactions (lower right).
    Here, \PQQ stands for either VLQ flavor.
}
\label{fig:vlq_diagrams}
\end{figure}

At the LHC, VLQs may be produced in pairs via the strong interaction, or singly in EW processes.
Representative LO Feynman diagrams are shown in Fig.~\ref{fig:vlq_diagrams}.
The pair production is dominant at low VLQ masses ($<$1\TeV) but the cross section decreases rapidly as a function of mass~\cite{Aguilar:2009}.
The single production cross section is larger for high VLQ masses but is more model dependent, particularly for the chosen values of the couplings to SM quarks and EW bosons and of the VLQ width~\cite{Deandrea:2021vje}.
This means that for single production, model-independent limits can only be set on the product of the cross section and branching fraction (\BR) for the different VLQ masses, as opposed to limits on the VLQ mass itself as done in pair production scenarios.
The NLO QCD corrections can have an effect on the cross sections for pair and single production, as well as on the shapes of key kinematic distributions, as discussed in Refs.~\cite{Fuks:2016ftf,Cacciapaglia:2019}.
Besides via strong and weak interactions, VLQs could also be produced via new interactions mediated by a heavy \PWpr or \PZpr boson, which will be discussed in more detail in Section~\ref{sec:resVLQ}.

In many models (\eg, Ref.~\cite{Aguilar-Saavedra:2013qpa} and references therein), VLQs decay into an SM quark plus either a \PW, \PZ, or Higgs boson.
It is usually assumed that the VLQs couple only to the third-generation SM quarks \cite{Aguilar:2013conf}---the specific branching fractions depending on the multiplet---and that they have a narrow width (this assumption is referred to as the narrow-width approximation, NWA, and remains valid up to a width-to-mass ratio of approximately 10--15\%).
The allowed decay modes for each of the VLQs are then:
\begin{align*}
&\PQT\to\bWp, \quad \TtotZ, \quad \TtotH \\
&\PQB\to\tWm, \quad \BtobZ, \quad \BtobH \\
&\xft\to\tWp \\
&\yft\to\bWm
\end{align*}
In scenarios where VLQs are introduced as singlets, their branching fractions into \qW, \qZ, and \qH are typically assumed to be 50, 25, and 25\%, respectively.
In scenarios where VLQs are introduced as doublets, a branching fraction of 50\% for \qZ and \qH is assumed for one partner in the doublet and a branching fraction of 100\% to \qW for the other partner in the doublet.
A \PQT quark most naturally forms an up-type doublet with stronger mixing between \PQt and \PQT quarks than the mixing between \PQb and \PQB quarks, resulting in $\BR(\TtotZ)=\BR(\TtotH)=50\%$ and  $\BrBtW=100\%$.
A down-type doublet with stronger mixing between the \PQb and \PQB quarks, resulting in $\BrBbZ=\BrBbH=50\%$ and $\BR(\TtobW)=100\%$, is not natural in view of the mass hierarchy $\mQt\gg\mQb$.
Exotic non-SM decays of VLQs are allowed as part of nonminimal extensions, with decay chains such as $\xft\to\PSHp\PQt\to\PWp\PH\PQt$~\cite{Banerjee:2021,Banerjee:2022}, or $\PQT\to\PQt\Pa$, where \Pa represents a new scalar or pseudoscalar particle~\cite{Cacciapaglia:2019zmj,Benbrik:2019zdp,PhysRevD.101.035015}. Exotic VLQ decays through higher-dimensional operators have also been proposed~\cite{Dobrescu:2016pda}.

\subsection{Cross sections}
\label{subsubsec:vlq_crosssec}

\begin{figure}[tp!]
\centering
\includegraphics[width=0.48\textwidth]{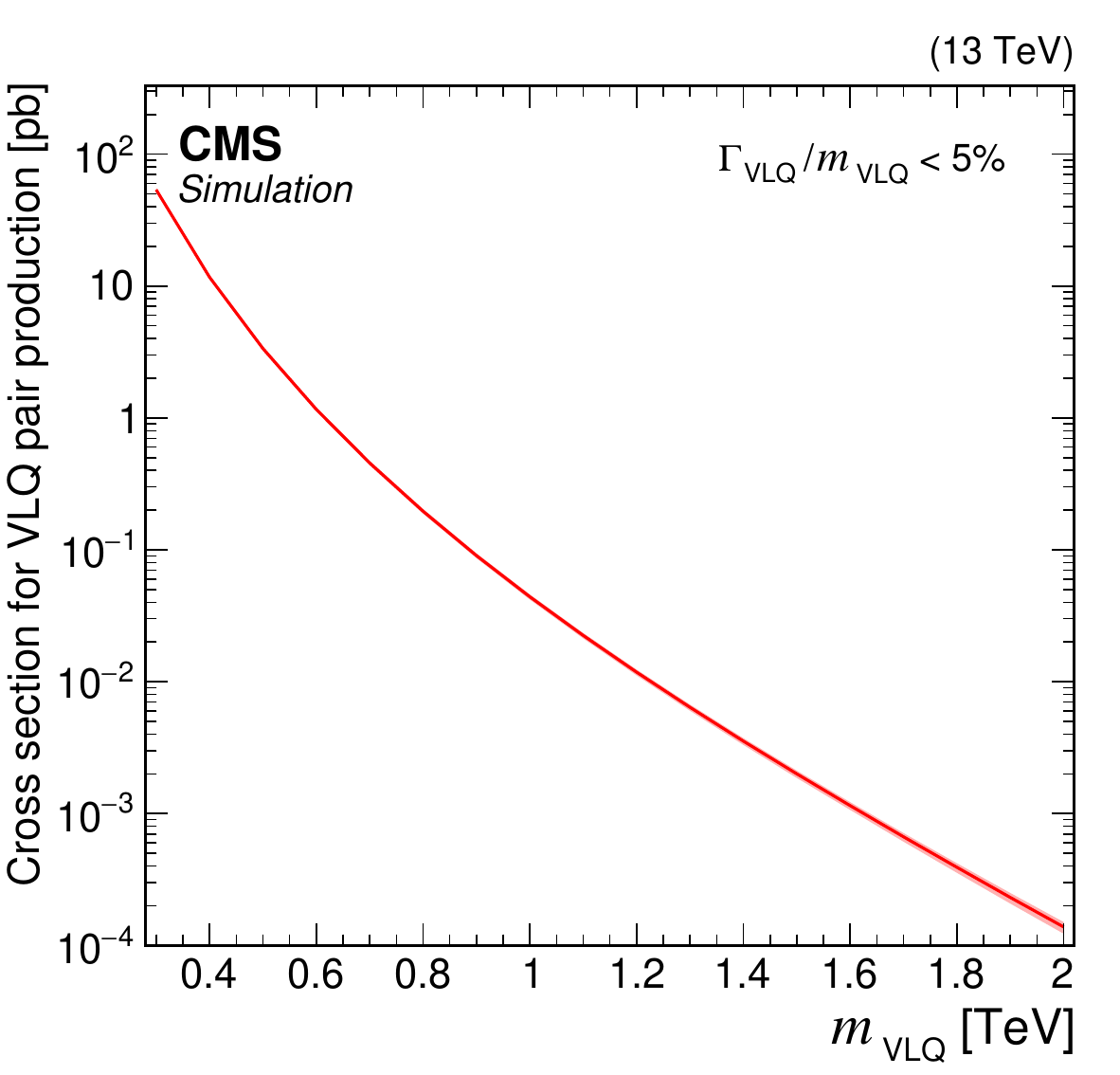}%
\hfill%
\includegraphics[width=0.48\textwidth]{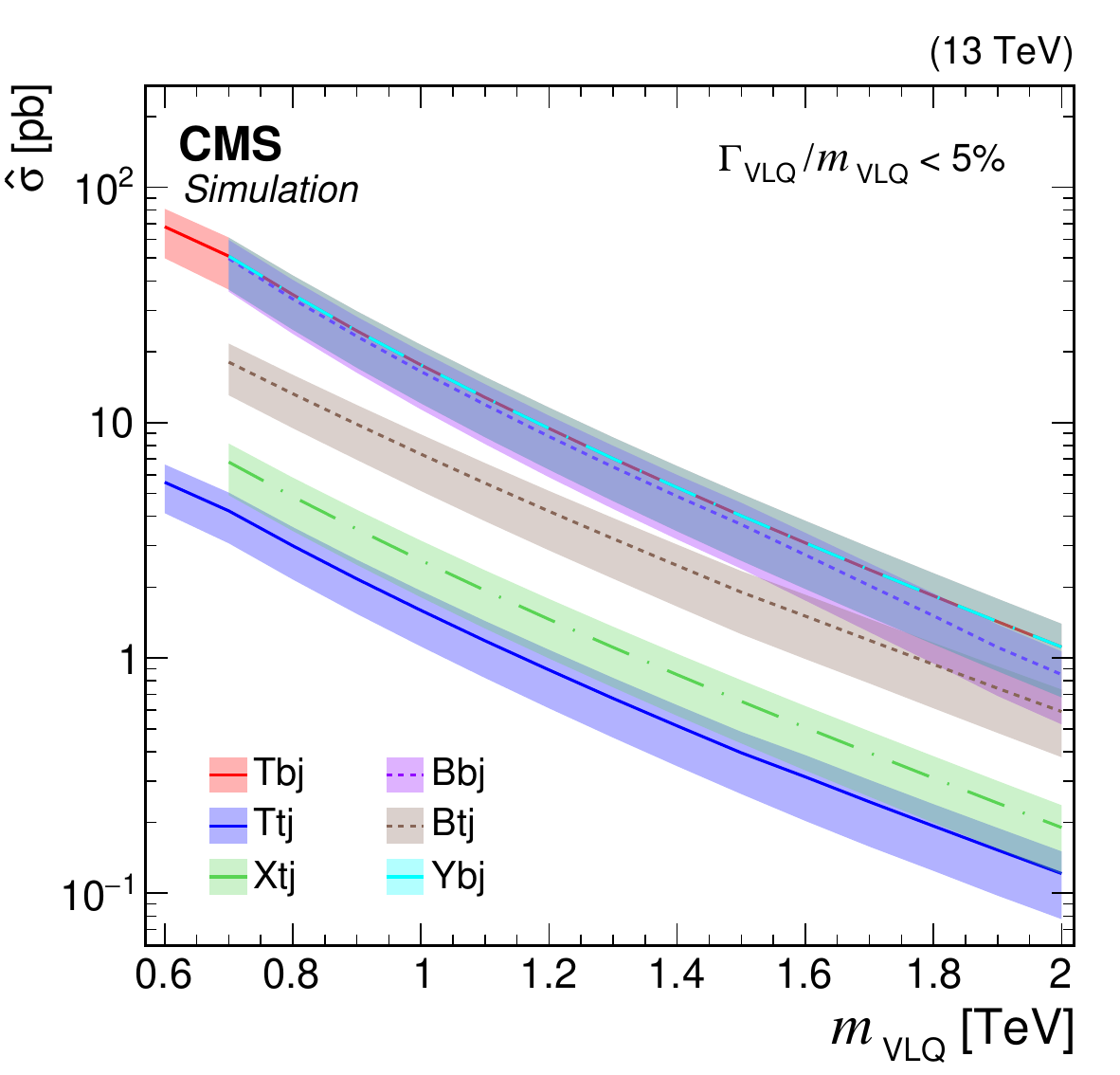}%
\caption{%
    Cross sections for the production of VLQs at $\sqrt{s}=13\TeV$ as a function of the VLQ mass.
    Pair production cross sections via the strong interaction are computed to NNLO, using the models and tools from Refs.~\cite{Czakon:2013goa,Matse:2014,Czakon:2011xx} (left).
    Reduced cross section $\hat{\sigma}$ for single production via the EW interaction is computed at LO in EW in the NWA using the models and tools from Refs.~\cite{Campbell:2004,Matse:2014,Fuks:2016ftf,Carvalho:2018} (right).
    The shaded bands indicate PDF, renormalization scale, and factorization scale uncertainties associated with the predictions.
}
\label{fig:vlq_crosssections}
\end{figure}

\begin{figure}[tp!]
\centering
\includegraphics[width=0.48\textwidth]{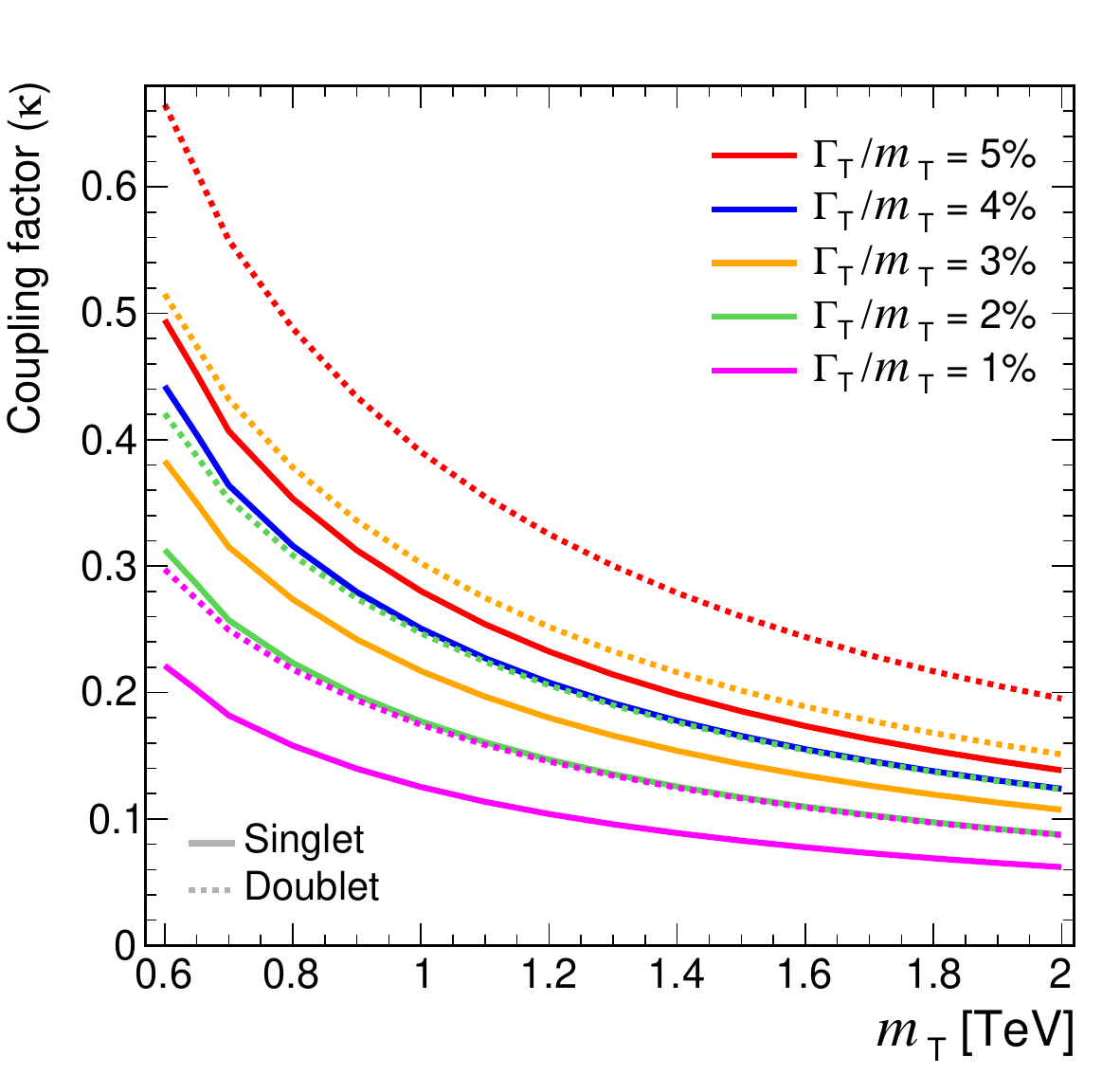}%
\hfill%
\includegraphics[width=0.48\textwidth]{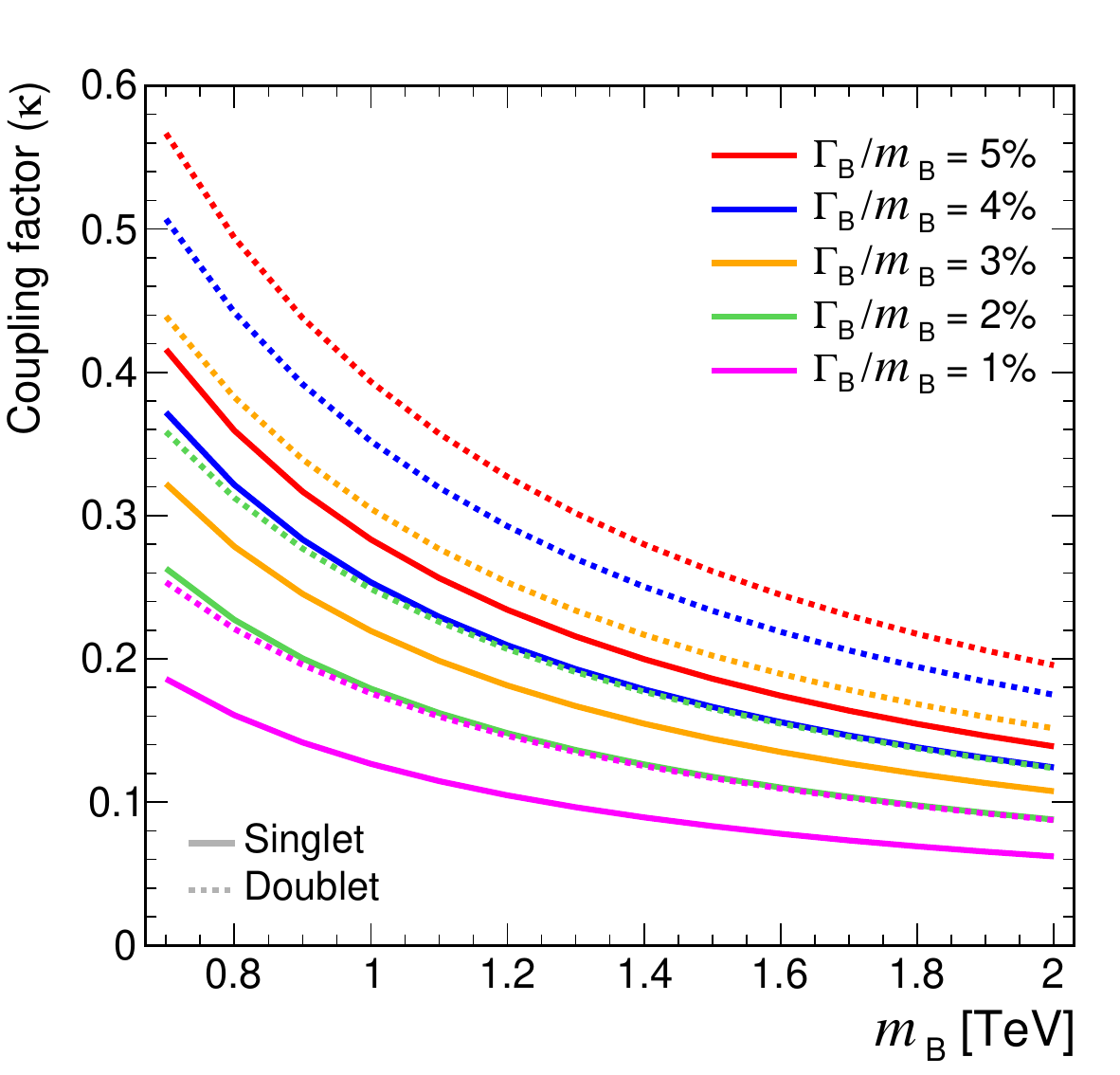} \\
\includegraphics[width=0.48\textwidth]{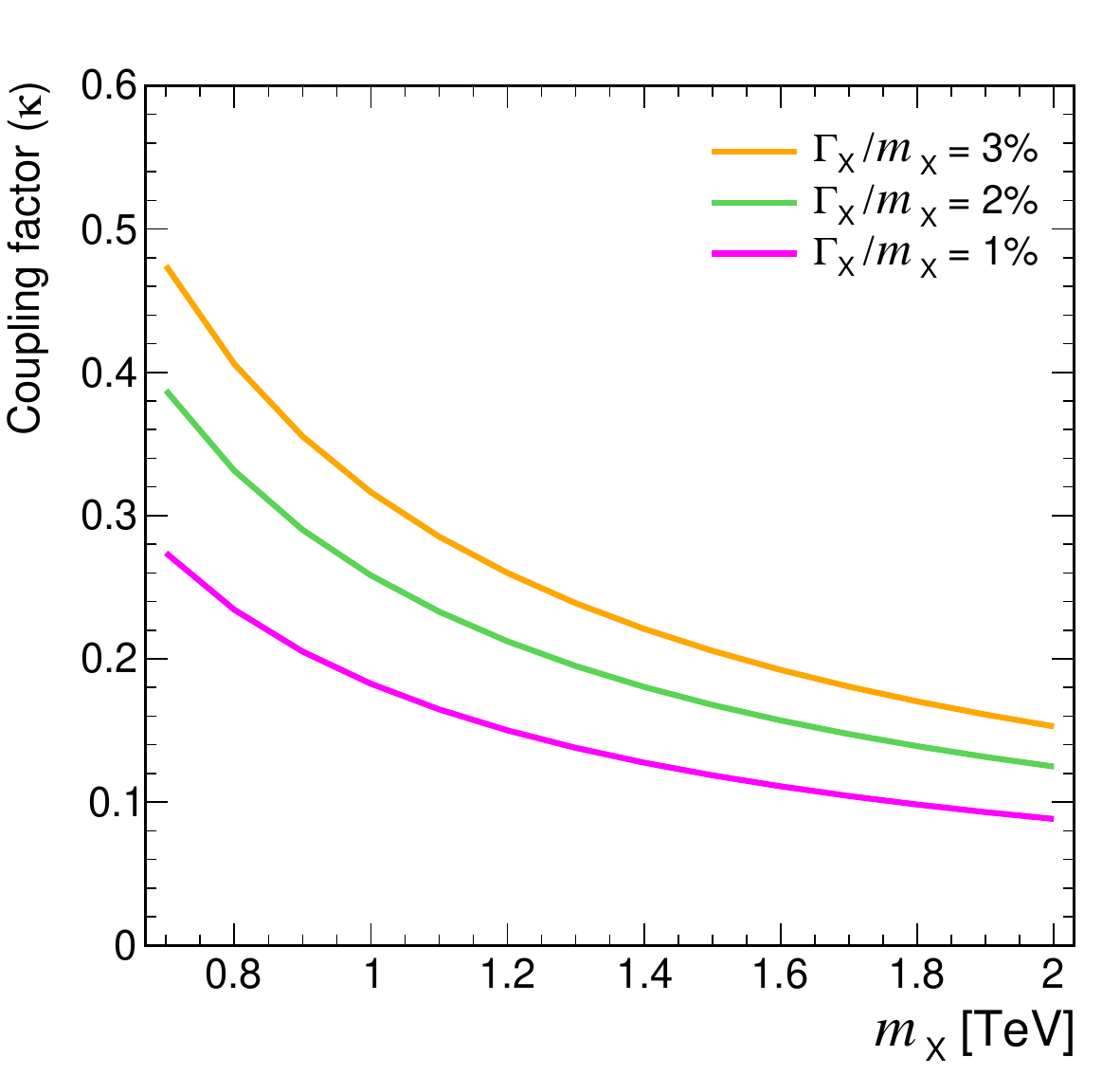}%
\hfill%
\includegraphics[width=0.48\textwidth]{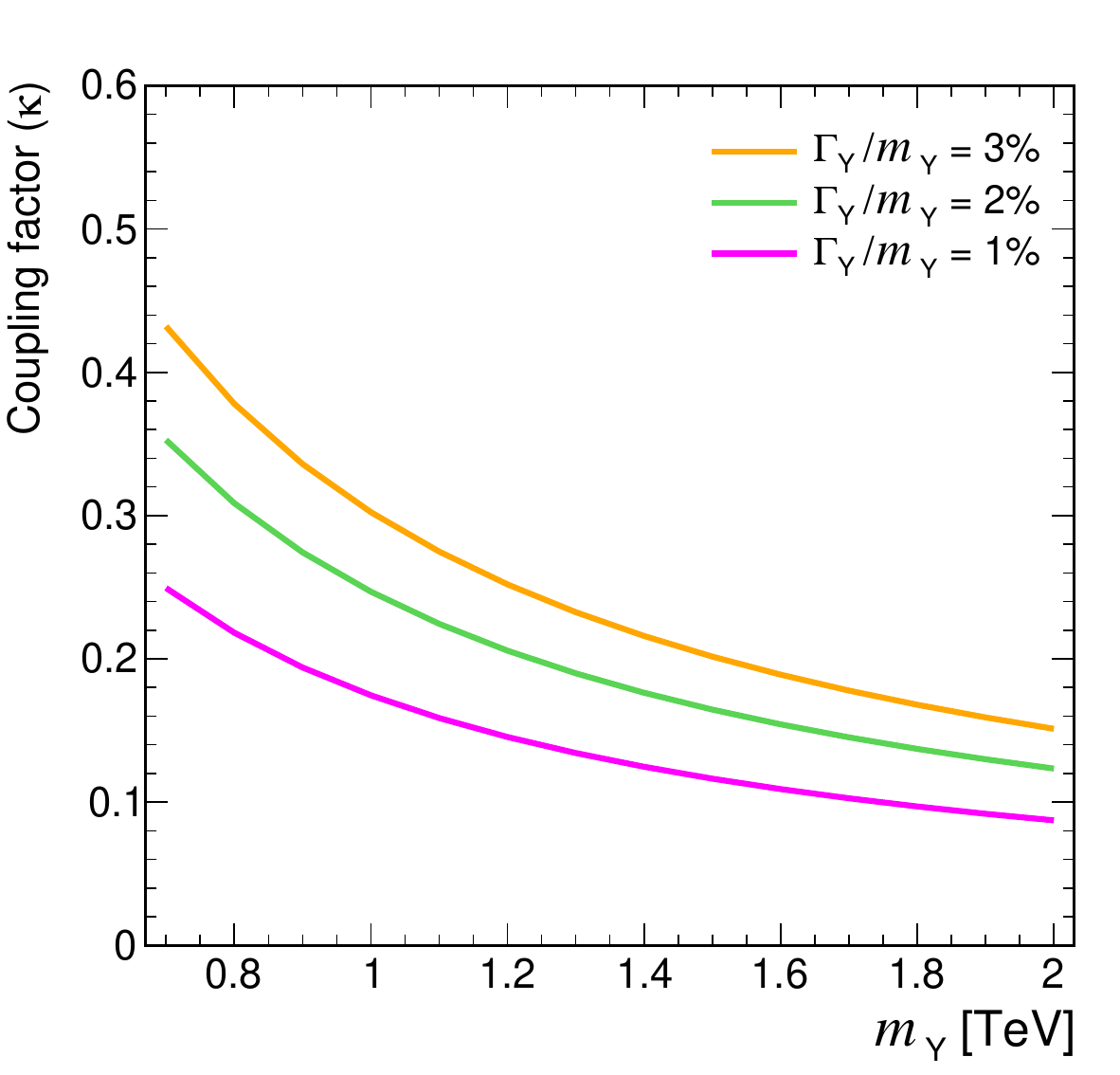}%
\caption{%
    Coupling factors for single VLQ production via the EW interaction in the narrow-width approximation as a function of the VLQ mass, using the models and tools from Refs.~\cite{Campbell:2004,Matse:2014,Fuks:2016ftf,Carvalho:2018}.
    Coupling factors in single production of \PQT (upper left), \PQB (upper right) in the singlet (solid lines) and doublet (dashed lines) scenarios.
    Coupling factors in single production of \xft (lower left), \yft (lower right) in doublet scenarios.
}
\label{fig:vlq_couplings}
\end{figure}

Predictions for the VLQ production cross sections at the LHC are given in Fig.~\ref{fig:vlq_crosssections}.
The pair production cross section is computed to NNLO in QCD~\cite{Czakon:2013goa} using the model of Ref.~\cite{Matse:2014} and the \textsc{top++2.0} program~\cite{Czakon:2011xx}, and independently of the VLQ flavor.
The single production cross section is computed at LO in the NWA using the simplified approach from Refs.~\cite{Campbell:2004,Matse:2014,Carvalho:2018}.
In this model-independent framework, the cross section does not depend on the chirality of the VLQ.
The total cross section for a VLQ \PQQ decaying to a specific final state may be written as:
\begin{equation}
    \sigma(C_1,C_2,\mQQ,\GammaQ)=C_1^2 C_2^2 \hat{\sigma}(\mQQ,\GammaQ),
\label{eq:singleTxsec}
\end{equation}
where $C_1$ indicates a production coupling parameter (\eg, $C_1=c_{\PW}$ for \PQT quarks produced in association with a \PQb quark), $C_2$ indicates a decay coupling parameter, and $\hat{\sigma}$ is the reduced cross section for an arbitrary VLQ width \GammaQ.
The VLQ width is small compared to the experimental mass resolution if the coupling between the VLQ and the relevant SM particles is $<$0.5.
As the $\GammaQ/\mQQ$ ratio drops, the production and decay contributions to the cross section can be factorized such that $\sigma=C_1^2\hat{\sigma}\BR$, where $\hat{\sigma}$ is the reduced cross section in the NWA, and \BR is the branching fraction for the VLQ to decay to the final state under consideration.
The couplings $C_1$ and $C_2$ are chosen as appropriate from $c_{\PW}$, $c_{\PZ}$, and $c_{\PH}$, which in turn depend on $\kappa$ values that are related to the mixing angles between VLQs and the corresponding SM quarks.
The computation of $\hat{\sigma}$ is carried out using a UFO-based model \cite{Degrande:2012} of Ref.~\cite{Fuks:2016ftf} with \MGvATNLO adapting the approach of Ref.~\cite{Carvalho:2018}.
The reduced cross sections $\hat{\sigma}$ and coupling factors are given in Fig.~\ref{fig:vlq_crosssections} and~\ref{fig:vlq_couplings}, respectively.
For the searches in the CMS experiment described in this report, the cross section calculation assumes $\kappaW=\kappaH=\kappaZ=\kappa$ for a singlet scenario and $\kappaH=\kappaZ$ and $\kappaW=0$ for a \PQT in an up-type doublet scenario.
The coupling factor $\kappa$ is then determined for a fixed VLQ mass and total width.
An alternative approach is to fix the branching fractions of VLQ decay modes and subsequently compute the individual coupling factors for the VLQ interactions with heavy SM bosons (\eg, used in Ref.~\cite{ATLAS:2022ozf}). The collective coupling strength is then derived using Eq.~(2.20) of Ref.~\cite{Buchkremer:2013bha}. Both approaches describe the same physics and converge to the same results when compared in the regime beyond the validity of the NWA.

\section{Review of vector-like quark searches}
\label{sec:results_vlq}

\subsection{Overview of the CMS search program}

Using data collected in 2010--2012 at $\sqrt{s}=7$ and 8\TeV, known as the Run-1 data set, the CMS Collaboration combined 10 individual searches for VLQ pair production with decays to heavy quarks into results for \TTbar production~\cite{B2G-13-005} and \BBbar production~\cite{B2G-13-006}.
Exploiting different final-state topologies, all final states for the VLQs were explored including scans over a wide range of possible branching fractions.
Another search was conducted for single or pair production of VLQs coupled only to light-flavor quarks~\cite{B2G-12-016} in events with at least one lepton (electron or muon) in the final state.

The Run-1 studies have been followed up by an extensive search program for VLQs using data collected during Run 2. The CMS experiment has carried out searches for both pair and single production of VLQs of all flavors, as well as for production of \PQT and \PQB through heavy resonance decays.
For pair production, the adopted search strategies utilize sophisticated analysis techniques, such as boosted object identification and/or multiclassifier tools to correctly identify the objects in the event. These strategies have the advantage that they are simultaneously sensitive to all decay modes of the VLQ.
In single-production searches, different widths are considered---ranging from the NWA to 30\% of the VLQ mass---as well as different settings of the coupling $\kappa$,
because for single production the cross sections depend on the VLQ flavor, mass, and width, and range from several hundred fb for low masses near 600\GeV to just fractions of a fb at masses near 1800\GeV.

A summary of all the explored channels and final states for VLQ searches by the CMS experiment is shown in Table~\ref{tab:vlqsummary}.
Specific details on the various analyses are briefly outlined in Sections~\ref{sec:PairVLQ}--\ref{sec:resVLQ}.
Their complementarity, and the statistical combination of some of the results, are discussed in Section~\ref{sec:VLQcombinationAndsummary}.
Finally an outlook for future VLQ searches is given in Section~\ref{sec:futurevlq}.

\begin{table}[!ht]
\centering
\topcaption{%
    List of VLQ searches performed by the CMS experiment grouped by production mode.
    In this table, \Pell denotes an electron or a muon.
    Additional jets in the final state are not explicitly listed in the table.
    The 0\Pell channels correspond to the all-hadronic final state.
    For the 2\Pell channels, it is indicated whether the leptons have opposite-sign (OS) or same-sign (SS) charges.
    For single VLQ searches, the channels are indicated through the decay products of the \PW, \PZ, and Higgs bosons, and \PQt quarks.
}
\label{tab:vlqsummary}
\renewcommand{\arraystretch}{1.25}
\begin{tabular}{lllll}
    Production mode & Decay mode & Channel & Section & Refs. \\
    \hline
    \TTbar & \bW, \tH, \tZ & 0\Pell, 1\Pell, OS 2\Pell, SS 2\Pell, 3\Pell & \ref{sec:TTVLQ} & \cite{B2G-17-003,B2G-17-012,B2G-18-005,B2G-20-011} \\
    \BBbar & \tW, \bH, \bZ & 0\Pell, 1\Pell, OS 2\Pell, SS 2\Pell, 3\Pell & \ref{sec:BBVLQ} & \cite{B2G-18-005,B2G-20-011,CMS:2024xbc} \\
    \XXbar & \tW & 1\Pell, SS 2\Pell & \ref{sec:XXVLQ} & \cite{B2G-17-014} \\
    \YYbar & \bW & 1\Pell & \ref{sec:TTVLQ} & \cite{B2G-17-003} \\[\cmsTabSkip]
    \PQT & \tZ & $\bqq\,\ellell$, $\bqq\,\bb$, $\bqq\,\nunu$ & \ref{sec:TZlepVLQ} & \cite{CMS:2017voh,B2G-18-003,B2G-19-004} \\
    & \tH & $\bqq\,\gammagamma$, $\bqq\,\bb$ & \ref{sec:TZHbbVLQ} & \cite{B2G-21-007,B2G-18-003,B2G-19-001} \\
    & \bW & $\PQb\,\lnu$ & \ref{sec:TYbW} & \cite{B2G-16-006} \\
    \PQB & \bH & $\PQb\,\bb$ & \ref{sec:BVLQ} & \cite{B2G-17-009} \\
    & \tW & $\bqq\,\lnu$, $\blnu\,\qq$, $\bqq\,\qq$ & \ref{sec:BVLQ} & \cite{B2G-17-018,CMS:2021iuw,B2G-20-010} \\
    \xft & \tW & $\bqq\,\lnu$, $\blnu\,\qq$, $\bqq\,\qq$ & \ref{sec:BVLQ} & \cite{B2G-17-018,B2G-20-010} \\
    \yft & \bW & $\PQb\,\lnu$ & \ref{sec:TYbW} & \cite{B2G-16-006} \\[\cmsTabSkip]
    $\PZpr\to\TTbar$ & \bW & 0\Pell & \multirow{2}{*}{\ref{sec:resVLQZprime}} & \cite{Sirunyan:2017bfa} \\
    & \tH, \tZ & 1\Pell & & \cite{Sirunyan:2018rfo} \\
    \WprtoTb & \tH, \tZ & 0\Pell & \multirow{2}{*}{\ref{sec:resVLQWprime}} & \cite{Sirunyan:2018fki, B2G-20-002} \\
    \WprtoBt & \bH, \bZ & 0\Pell & & \cite{Sirunyan:2018fki, B2G-20-002} \\
\end{tabular}
\end{table}

\subsection{Pair production}
\label{sec:PairVLQ}

Pair production of \PQT, \PQB, \xft, and \yft via gluon fusion has been studied in various searches with data collected in Run 2.
Analysis strategies for VLQ pair production typically exploit the presence of \PQt quarks and \PW, \PZ, or Higgs bosons in the decay chain, as well as the high Lorentz boost of the decay products for high VLQ masses. The searches performed in Run 1 with a data set corresponding to an integrated luminosity of 19.7\fbinv resulted in lower mass limits of 880\GeV for \xft (\tWtW decays), 920\GeV for \yft (\bWbW decays), 720--920\GeV for \PQT quarks, and 740--900\GeV for \PQB quarks~\cite{B2G-13-005,B2G-13-006}.
The Run 2 analyses use data sets corresponding to 36\fbinv or 138\fbinv, representing an increase of a factor of 1.8 or 7 with respect to Run 1, respectively. The sensitivity of the CMS experiment to VLQ production has increased dramatically as a result of the higher energy, larger data sets, and more advanced analysis techniques.
In the following, a search for \XXbar is discussed first, followed by searches for \TTbar and \BBbar.
Some of the searches for \TTbar (\BBbar) production are equally sensitive to \YYbar (\XXbar), since the selection criteria and primary observables are largely independent of the chirality
of the VLQ. These interpretations are therefore discussed in the same sections.

\subsubsection{Search for \texorpdfstring{\XXbar}{XXbar} production}
\label{sec:XXVLQ}

The search for pair produced \xft quarks in Ref.~\cite{B2G-17-014} was one of the earliest searches for VLQs at the CMS experiment in Run 2.
The search uses the 2016 data set.
It is assumed that \xft decays to a \PQt quark and a \PW boson with 100\% branching fraction. Two channels were considered:
the single-lepton channel and the same-sign dilepton channel (SSDL), where ``lepton'' refers to an
electron or a muon. In the single-lepton final state, one of the \PW bosons decays to a lepton and
neutrino, while all the other \PW bosons decay hadronically and form jets. However, in the SSDL final state, two \PW bosons
decay leptonically, giving rise to a pair of same-sign (SS) leptons, a signature that is enhanced in \tWtW decays compared to SM processes.
The \xft decays in both channels produce a large number of jets.
In the SSDL channel, background processes can be separated into three categories: SS prompt leptons (``SSP''),
opposite-sign (OS) prompt leptons (``ChargeMisID''), and SS nonprompt leptons. Here, a prompt lepton
refers to one that originates from a \PW or \PZ boson decay. The SSP background processes are modeled using simulation.
The ChargeMisID background is estimated by measuring the rate at which \PZ boson decays in a control sample are reconstructed as SS lepton events, and scaling the OS
lepton events passing all other analysis selection criteria by this rate. The nonprompt-lepton background is estimated using
the matrix method described in Section~\ref{sec:bkgest}.
In the single-lepton channel, all backgrounds are modeled using simulation.

\begin{figure}[!b]
\centering
\includegraphics[width=0.48\textwidth]{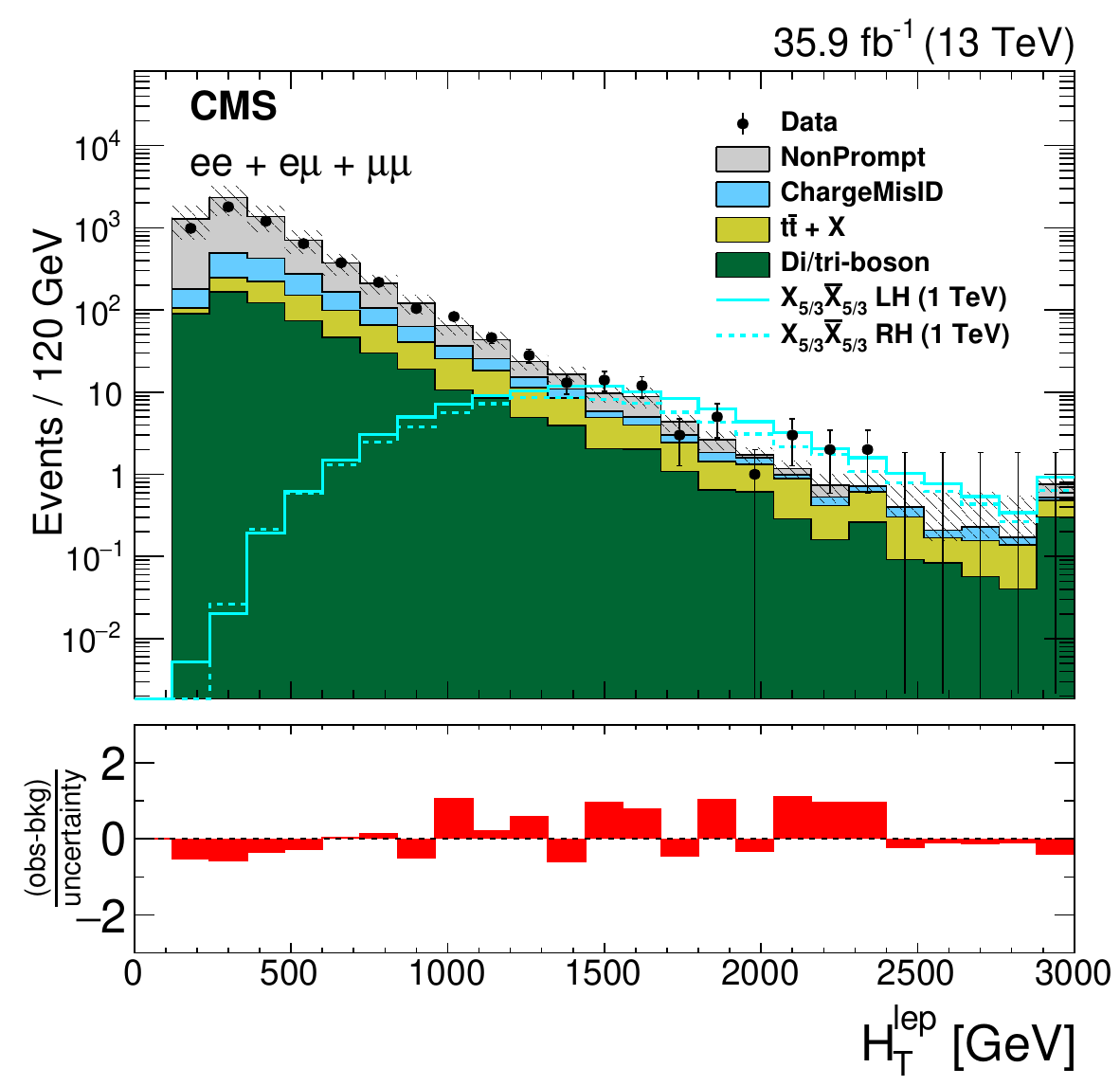}%
\hfill%
\includegraphics[width=0.48\textwidth]{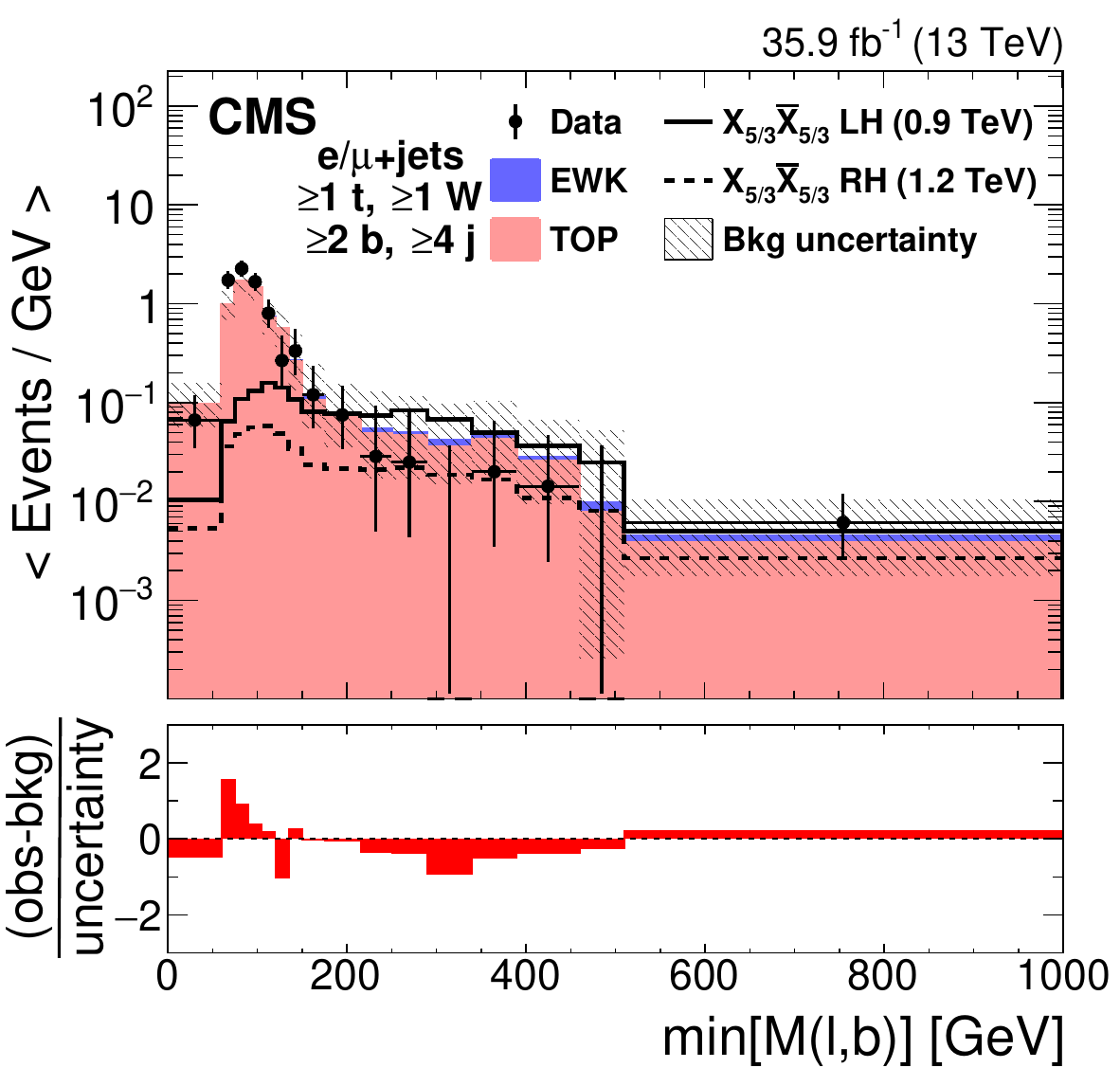}%
\caption{%
    Distributions of observables used to maximize the \XXbar signal significance for the SSDL (left) and single-lepton (right) final states.
    The left figure shows the \HTlep distribution after the SS dilepton selection, \PZ boson and quarkonia lepton invariant mass vetoes, and the requirement of at least two small-radius jets in the event, for a combination of \ee, \emu, and \mumu channels.
    The right figure shows the \minMlb distribution in events with $\geq$1 \PQt-tagged jet, $\geq$1 \PW-tagged jets,    and $\geq$2 \PQb-tagged jets for the combined electron and muon samples in the SR.
    The distribution has variable-size bins such that the statistical uncertainty in each bin is less than 30\%.
    The lower panel in each plot shows the difference between the observed and the predicted numbers of events divided by the total uncertainty.
    Figures taken from Ref.~\cite{B2G-17-014}.
}
\label{fig:Mlb1ptcats}
\end{figure}

For the SSDL final state, events are required to pass triggers based on two electrons, two muons, or one
electron and one muon. Different lepton \pt selections are used according to the trigger era corresponding
to the data-collection period.
Events are further required to have two SS tight leptons and at least two small-radius jets.
The invariant mass of the lepton pair must be greater than 20\GeV, and events are removed if they contain an
OS, same-flavor (OSSF) lepton pair having an invariant mass within 15\GeV of the mass of the \PZ boson mass.
Similarly, events with an SS electron pair having an invariant mass within 15\GeV of the \PZ boson mass are removed to eliminate Drell--Yan (DY) or charge misidentification events.
Two selection criteria are used to maximize the signal significance. First, the number of constituents
($N_{\text{const}}$), which is the number of small-radius jets plus the number of additional tight leptons beyond the SS pair, must be greater than four.
Second, \HTlep, which is defined as the scalar sum of the \pt of all constituents and the SS lepton pair, as shown in Fig.~\ref{fig:Mlb1ptcats}, must be greater than 1200\GeV.

For the single-lepton final state, events are required to pass single-electron or single-muon triggers and contain exactly one high-quality isolated lepton with $\pt>80\GeV$.
A veto on additional looser-quality leptons is applied to ensure that the channels are mutually exclusive.
Events are required to have at least four small-radius jets, with the leading jet \pt greater than 450\GeV and the subleading jet \pt greater than 150\GeV. At least one of the jets must be \PQb tagged.
Events must also have $\ptmiss>100\GeV$, representing the presence of a neutrino, and an angular separation of $\DR>1.0$ between the lepton and the subleading jet, which typically emerges from the \xft that decayed hadronically.
Large-radius jets are used to tag hadronically decaying \PQt quarks or \PW bosons,
using jet grooming techniques (\mSD and pruning, respectively) and the $N$-subjettiness observable, as described in Section~\ref{sec:boosted}.

The discriminating observable used to suppress the background contribution is the mass reconstructed from the lepton
and \PQb-tagged jet, \Mlb. If an event has more than one \PQb jet, the smallest \Mlb, \minMlb, is used.
When the lepton and \PQb quark emerge from the same \PQt quark, this distribution peaks sharply just below the \PQt quark mass, but
in signal events the lepton can emerge from the \PW boson daughter of an \xft quark, leading to a broad distribution of events above the \PQt quark mass, as shown in Fig.~\ref{fig:Mlb1ptcats} (right).
The events are separated into 16 categories based on lepton flavor (\Pe, \PGm) and the number of
\PQt-tagged, \PW-tagged, and \PQb-tagged jets.

\begin{figure}[ht!]
\centering
\includegraphics[width=0.48\textwidth]{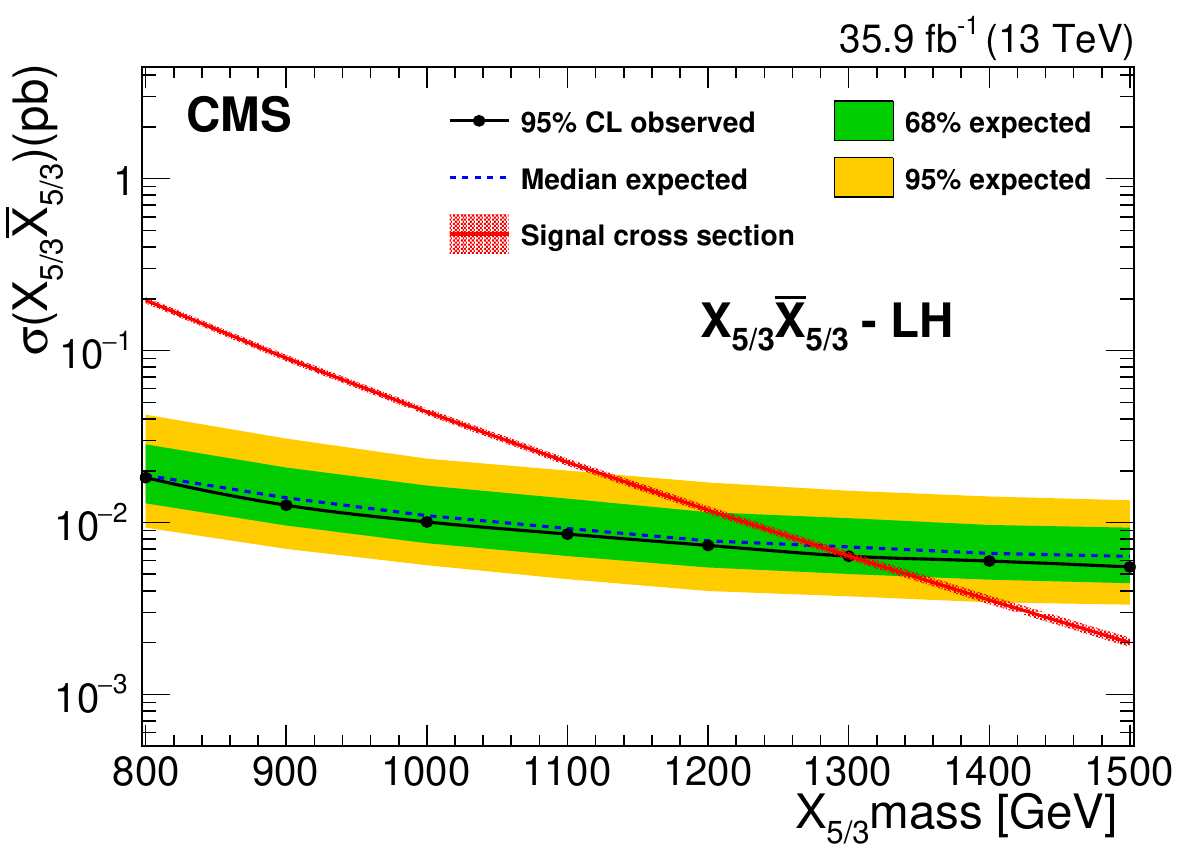}%
\hfill%
\includegraphics[width=0.48\textwidth]{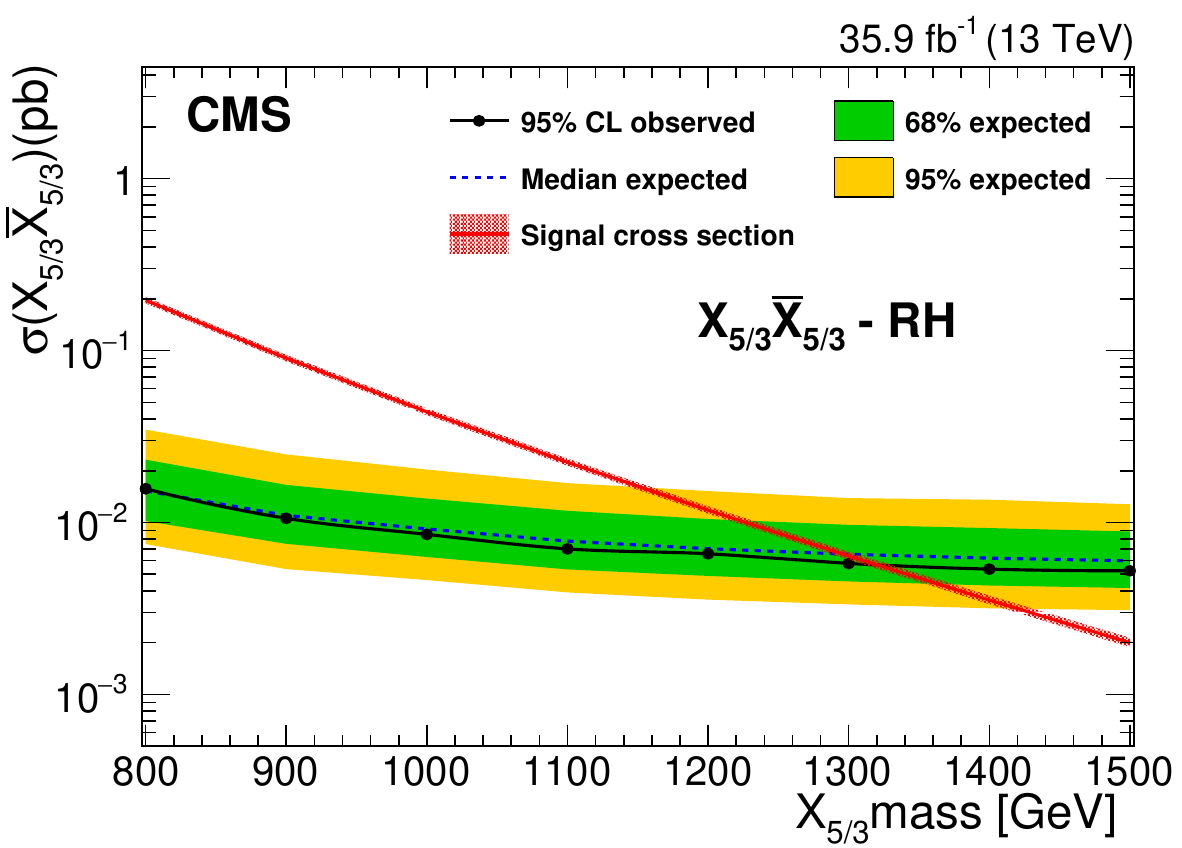}%
\caption{%
    Expected and observed cross section upper limits at 95\% \CL for an LH (left) and RH (right) \xft as a function of its mass, after combining the SS dilepton and single-lepton final states.
    The theoretical uncertainty in the signal cross section is shown with a band around the theoretical prediction.
    Figures adapted from Ref.~\cite{B2G-17-014}.
}
\label{fig:CombinedLimits}
\end{figure}

The search is performed using \minMlb histograms in the 16 lepton-flavor and jet-tag categories of the single lepton final state,
as well as the number of events in the three lepton-flavor categories of the SSDL final state.
For both final states, no statistically significant excess was observed above the SM prediction.
Upper limits are set on the production cross section at 95\% \CL.
The combination of the final states sets a lower observed (expected) limit of 1.33 (1.30)\TeV and
1.30 (1.28)\TeV on the mass of the \xft with RH or LH couplings to the \PW boson, respectively.
The limits are shown in Fig.~\ref{fig:CombinedLimits}.

The searches described below in Sections~\ref{sec:BBVLQ} and \ref{sec:VLQsummary} for \BBbar production
also provide sensitivity to \xft production when the \tWtW decay mode is considered.
With the full Run 2 data set and more advanced NN jet identification techniques, the lower mass limits for this decay mode reach up to 1.56\TeV.

\subsubsection{Searches for \texorpdfstring{\TTbar}{TTbar} production}
\label{sec:TTVLQ}

The search for pair-produced \PQT quarks is a cornerstone of the VLQ search program of the CMS experiment, and several publications were released over the course of Run 2.

\cmsParagraph{$\TTbar\to\bqq\,\PQb\ell\nu$} The first search for pair production using the 2016 data set
was a search for \TTbar or \YYbar decaying to \bWbW in the single-lepton final state~\cite{B2G-17-003}.
This search used a kinematic fit procedure to reconstruct \PQT quark candidates from one isolated charged
lepton, \ptmiss, and at least four small-radius jets, including the subjets of at least one large-radius
jet consistent with a boosted \PW boson decay ($60<\mSD<100\GeV$). The kinematic fit
constrains the lepton-neutrino pair and a quark-quark pair to be consistent with the \PW boson mass,
and constrains the two \PQT candidate masses to be equal. The fit minimizes a \chisq metric that
compares the observed kinematic properties of the lepton and jets with the fitted kinematic quantities
required to meet the constraints. This fitting technique yields many permutations for each event, since
jets may be assigned to various quarks in the decay chain. Information about jet \PW and \PQb tagging is
considered to reject certain permutations, and the fit with the highest \chisq probability is selected.
The \PQT quark mass can be reconstructed with approximately 7\% resolution using this method. No excess
beyond the simulated SM background estimate is found. A lower limit on the \PQT quark mass
is set at 1.30\TeV. Since the analysis does not discriminate between jets from \PQb and \PAQb quarks,
the signal process may be interpreted as either \TTbar or \YYbar production. Therefore, this search also
excludes \yft quarks up to a mass of 1.30\TeV.

\cmsParagraph{$\TTbar\to\tZ + \PQq\PX, \PZ\to\ell\ell$} Another search for \TTbar (or \BBbar) production using the 2016 data set that utilized 
early jet substructure identification techniques, considered an OS dilepton final state 
from the decay of one VLQ into a \PZ boson~\cite{B2G-17-012}. The \PW, \PZ, or Higgs boson 
produced by the other VLQ is reconstructed as a jet with a two-prong substructure, 
particularly for VLQs with a high mass. Two electrons with $\pt>120\GeV$ and $\pt>25\GeV$, 
or two muons with $\pt>45\GeV$ and $\pt>24\GeV$, are required. The invariant mass of the 
dilepton pair must be within 15\GeV of the \PZ boson mass, and the \pt of the pair must be 
above 100\GeV. Large-radius jets are identified as originating from \PW, \PZ, or Higgs 
bosons using the pruned mass, the $N$-subjettiness \tauTO observable, and \PQb tagging 
algorithms applied to the subjets. Small-radius jets are also included in the 
reconstruction process as \PQb quark jets, or quark jets from resolved \PW, \PZ, or Higgs 
boson decays. Events are categorized based on the jet types observed, and CRs with either 
zero \PQb-tagged jets or one \PQb-tagged jet but a total hadronic energy of less than 1\TeV 
are formed to compute a correction to the simulated \zjets background using experimental 
data. Other backgrounds are modeled using simulation. No excess of events with respect to 
the estimated background is observed, and exclusion limits were derived on \TTbar 
production for \PQT decays with at least 20\% branching fraction to \tZ. Assuming 100\% 
\tZ decays, a lower mass limit at 95\% \CL of 1.28\TeV is derived for \PQT quarks.

Two subsequent general searches for \TTbar and \BBbar production adopted deep ML algorithms 
for jet identification to enhance sensitivity to VLQ decays. In one search, using the 2016 
data set, events with a hadronic final state~\cite{B2G-18-005} are selected, and the use of 
the \BEST tagger, discussed in Section~\ref{sec:boosted}, is pioneered for identifying 
large-radius jets. In the other search, the full Run 2 data set is used, and events in a 
variety of leptonic final states~\cite{B2G-20-011} are selected. In this search, the 
flavors of large-radius jets are identified using the \DeepAKeight algorithm. Both 
searches consider the \PQT and \PQB quarks separately and scan over the various possible 
branching fractions for decays to \PW, \PZ, and Higss bosons. The analysis strategies and 
results for \TTbar production are presented here, and the \BBbar interpretations are 
presented in Section~\ref{sec:BBVLQ}.

\cmsParagraph{$\TTbar\to$ hadrons} The search of Ref.~\cite{B2G-18-005} in the all-hadronic final state is an inclusive 
search considering energetic four-jet events classified into categories with two methods. 
In the novel ``NN-based'' method, the jets are classified using the \BEST algorithm into 
one of the 126 possible multiplicities of four jets with six classes. The other, more 
traditional, ``selection-based'' method helps to verify the results of the NN-based method. 
This search targets VLQs in the \TeVns mass range, such that the SM decay products 
(\PQb/\PQt, \PW/\PZ/\PH) acquire a significant momentum. Therefore a high threshold of 
400\GeV is imposed on the \pt of each of the four jets. The scalar sum of \pt of these jets, 
\HT, is larger than 1600\GeV, hence \HT is used for triggering and for event selection, 
and as the observable of interest to search for the VLQs. In addition, a lepton veto is 
applied in these analyses, such that the event selection criteria are mutually exclusive 
with the criteria in the leptonic final state search of Ref.~\cite{B2G-20-011}, described 
later in this section.

In the NN-based method, each of the four jets is tagged as either a boson (\PW, \PZ, or 
\PH), or a quark (\PQt, \PQb, or light flavor/gluon). The hypothesis with the largest 
score from \BEST is assigned as the classification of the jet. The overall multiplicity of 
the tags in a given event determines the SR into which the event is sorted.
The dominant multijet background is estimated by measuring tagging efficiencies in 
experimental data, in a multijet-rich three-jet CR in data. The same jet selections as in 
the SR are applied, apart from the jet multiplicity ($N_{\PQj}$) requirement, which is 
modified to $N_{\PQj}=3$. In this CR, the impurity from other background processes is 
below 1\%, so \BEST classification rates in multijet events can be measured in bins of jet 
\pt. Across a range of jet \pt from 400 to 3000\GeV, the misidentification rates for \PW, 
\PZ, and Higgs bosons are 3--7\%, and for \PQt quarks this rate is 7--10\%. The 
misidentification rate for \PQb quarks is $\approx$15\% for jets with a \pt of about 
400\GeV and rises to $\approx$65\% for jets with a \pt in the range of 2000--3000\GeV. 
Jets from light-flavor quarks or gluons, which make up the vast majority of the measurement 
sample, are accurately classified by \BEST at a rate of $\approx$65\% for jets with a \pt 
of about 400\GeV, falling to $\approx$6\% for jets with a \pt in the range of 2000--3000\GeV. 
This measurement shows that the \BEST algorithm provides a very strong multijet background 
rejection for VLQ decay modes with heavy SM particles, but is not optimized for decay modes 
with \PQb quarks, such as \TtobW. The background from multijet events in each of the 126 
SRs is determined by summing up the product of the four jet classification rates over 
all relevant jet permutations in an event, and over all selected four-jet events.

\begin{figure}[!ht]
\centering
\includegraphics[width=0.48\textwidth]{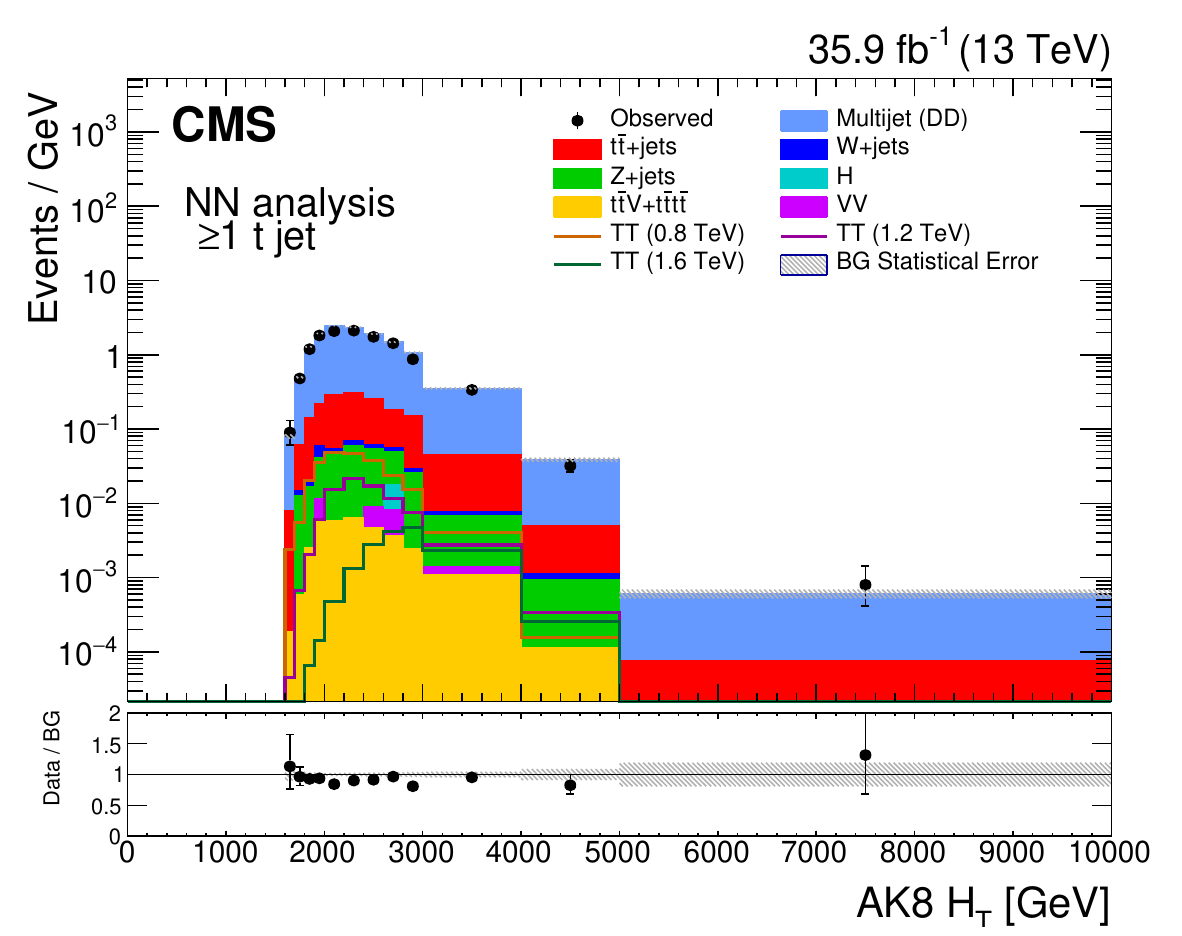}%
\hfill%
\includegraphics[width=0.48\textwidth]{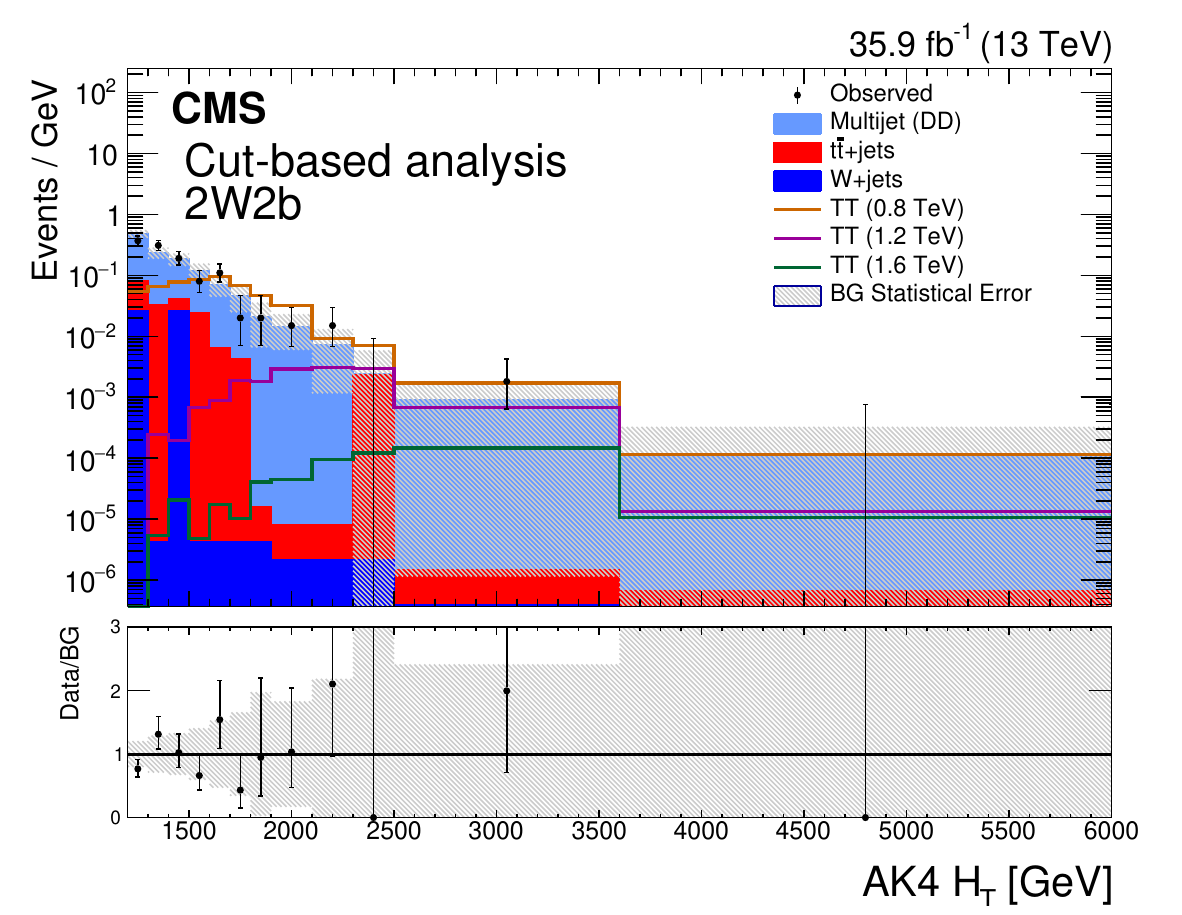}%
\caption{%
    Distributions of \HT in a combination of SRs in the NN-based approach, inclusive in $\geq$1 \PQt tags (left), and in the SR with two \PW-tagged and two \PQb-tagged jets in the selection-based approach (right).
    The lower panels show the ratio between observed data and the background estimate.
    Figures taken from Ref.~\cite{B2G-18-005}.
}
\label{fig:BESTandCBht}
\end{figure}

The observable of interest for each of the SRs is \HT, where the signal is expected to populate the tail of the distributions.
In the categories where the yield is too low to produce a meaningful distribution in \HT, a simple event counting experiment is performed.
Figure~\ref{fig:BESTandCBht} (left) shows a representative \HT distribution for events with $>$1 \PQt quark tag.

The selection-based analysis in Ref.~\cite{B2G-18-005} targets the \TtobW decay, with two \PW boson jets and two \PQb quark jets, to maximize performance. Two large-radius jets are selected first, each with $\pt>200\GeV$ and $\abs{\eta}<2.4$. In addition, two small-radius jets with $\pt>30\GeV$ and $\abs{\eta}<2.4$ are required that do not overlap the large-radius jets. Large-radius jets are taken as \PW boson candidates and are paired with a small-radius jet (\PQb quark candidate) such that the mass difference (\Deltam) between the resulting VLQ candidates is minimal. Events with $\Deltam/m<0.1$ and $\HT>1200\GeV$ are selected to reject background events and to ensure a high trigger efficiency. The events are divided into six SRs based on the \PW tag and \PQb tag multiplicity. A \PW tag is defined as a large-radius jet with $\tauTO<0.55$ and $\mSD\in [85,105]\GeV$, and a \PQb tag is defined as the operating point of the \CSVvtwo algorithm with a misidentification rate of 1\%. The background estimate for the multijet contribution is determined from data using an ABCD method fit to linear functions in each tagging category, with non-overlapping CRs obtained by inverting the \Deltam and \HT requirements. Just as in the NN-based analysis, the \HT distribution in the SRs is used to search for the presence of a \PQT quark signal. Figure~\ref{fig:BESTandCBht} (right) shows the \HT distribution in the SR with two \PW-tagged and two \PQb-tagged jets. In figures taken from Ref.~\cite{B2G-18-005}, the selection-based analysis is labeled ``cut-based analysis''.

\begin{figure}[!ht]
\centering
\includegraphics[width=0.48\textwidth]{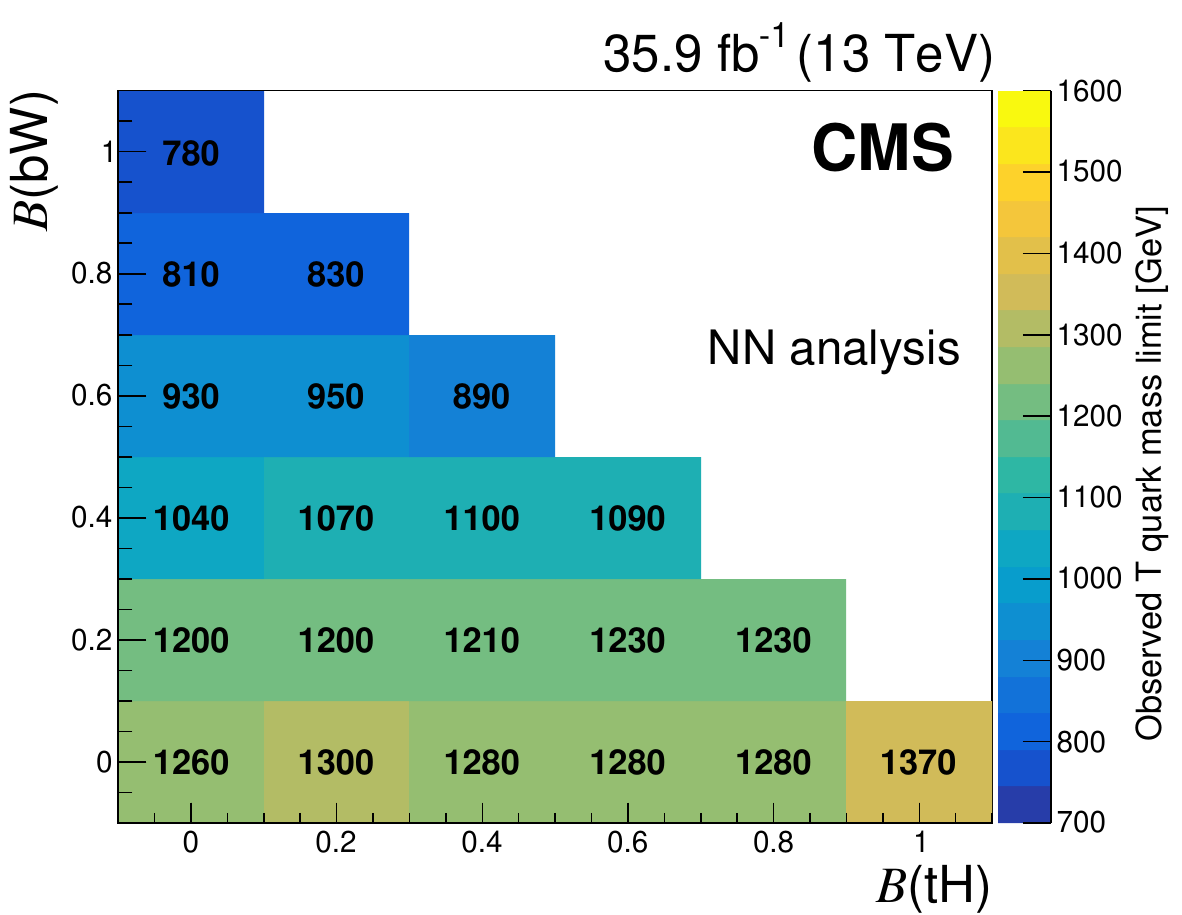}%
\hfill%
\includegraphics[width=0.48\textwidth]{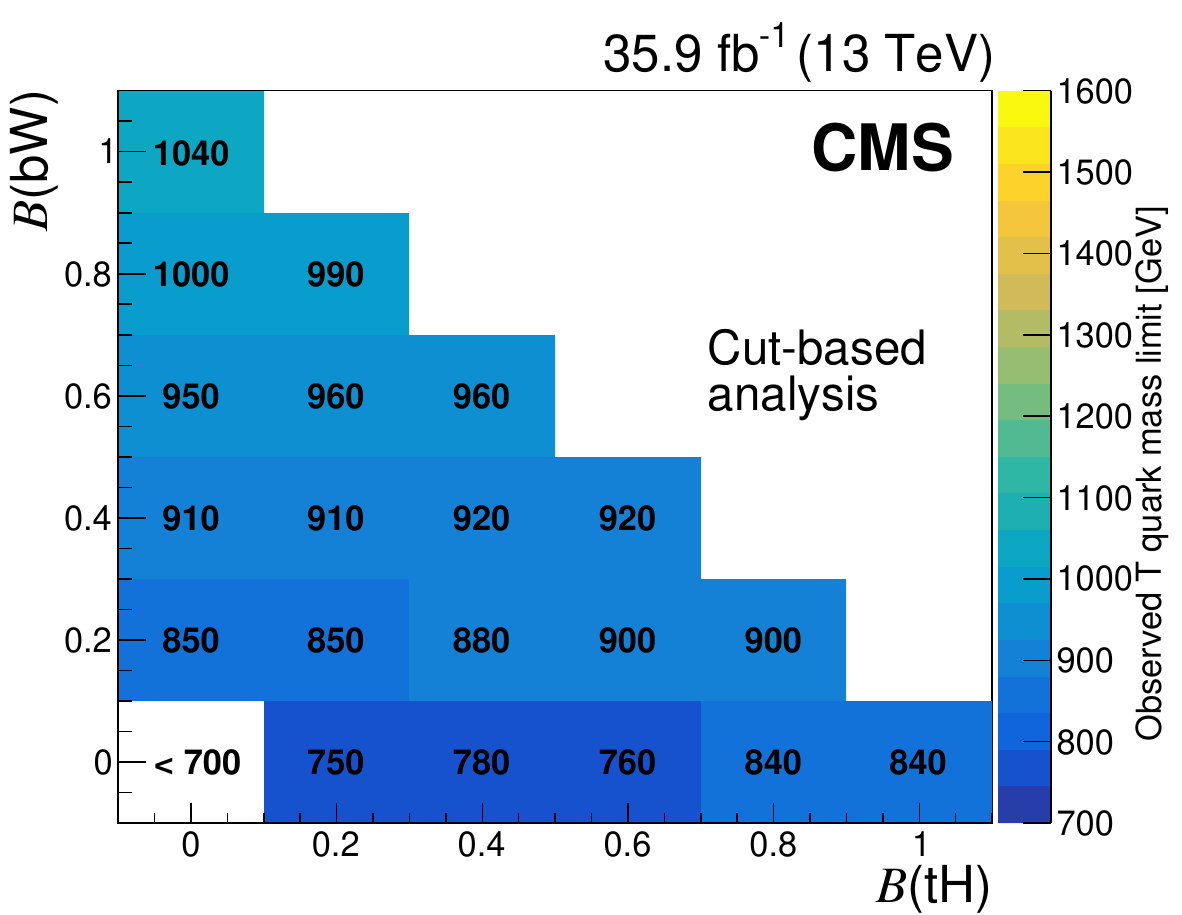}%
\caption{%
    Observed lower limits at 95\% \CL on the \PQT quark mass as functions of the \PQT quark branching fractions to \tH and \bW, using the NN-based (left) and selection-based (right) approaches.
    Figures adapted from Ref.~\cite{B2G-18-005}.
}
\label{fig:BESTscanTT}
\end{figure}

In both analyses, a simultaneous fit is performed across the 126 or six SRs by
computing Bayesian credible intervals to set 95\% \CL upper limits on the \TTbar production cross section.
No significant excess above the SM prediction is observed.
The observed lower limits on the \PQT quark mass are shown in Fig.~\ref{fig:BESTscanTT} as functions of the \PQT quark
branching fractions to \tH and \bW, for both NN-based (left) and selection-based (right) approaches.
The NN-based analysis shows more sensitivity to \TtotZ and \TtotH decay modes, excluding \PQT quark masses
below 1260 and 1370\GeV, respectively. The selection-based analysis offers stronger
sensitivity for the \TtobW decay mode, because it uses a dedicated \PQb quark tagger 
with a higher efficiency than the \BEST algorithm achieved. The NN-based analysis excludes \PQT quark
(or \yft quark) masses below 1030\GeV.

\cmsParagraph{$\TTbar\to$ leptons}
In Ref.~\cite{B2G-20-011}, a search for pair production of \PQT and \PQB quarks is presented using the full Run 2 data set. This search includes three final states containing charged electrons or muons: a single-lepton channel, an SS dilepton channel, and a ``multilepton'' channel with at least three leptons.
The three leptonic channels offer sensitivity to different potential VLQ decays. Table~\ref{tab:strategy} summarizes the main event selection criteria used to form CRs and SRs for the three channels.

\begin{table}[!ht]
\centering
\topcaption{%
    Summary of event selection criteria for the primary CRs and SRs in the three leptonic search channels.
    The phrase ``max MLP'' refers to the largest score from the single-lepton multilayer perceptron network.
    Table taken from Ref.~\cite{B2G-20-011}.
}
\label{tab:strategy}
\renewcommand{\arraystretch}{1.2}
\begin{tabular}{lc@{$\quad$}c@{$\quad$}c}
    Channel & \multicolumn{3}{c}{Event selection} \\
    & Overall & CR & SR \\
    \hline
    \multirow{7}{*}{1\Pell} & 1 tight \Pell & \NA & \NA \\
    & $\pt(\Pell)>55\GeV$ & \NA & \NA \\
    & 0 other loose \Pell, $\pt>10\GeV$ & \NA & \NA \\
    & $\ptmiss>50\GeV$ & \NA & \NA \\
    & $\geq$3 large-radius jets & \NA & \NA \\
    & \NA & max MLP not VLQ & max MLP is VLQ \\
    & \NA & & 2 VLQ candidates \\
    \hline
    \multirow{6}{*}{SS 2\Pell} & 2 tight SS \Pell & \NA & \NA \\
    & $\pt(\Pell)>40\GeV$, 30\GeV & \NA & \NA \\
    & $\geq$4 small-radius jets & \NA & \NA \\
    & $M(\ellell)>20\GeV$ & \NA & \NA \\
    & $M(\ee)$ outside 76--106\GeV & \NA & \NA \\
    & \NA & $\HTlep<1000\GeV$ & $\HTlep>1000\GeV$ \\
    \hline
    \multirow{6}{*}{3\Pell} & $\pt(\Pell)>30\GeV$ & \NA & \NA \\
    & $M(\text{OSSF }\ellell)>20\GeV$ & \NA & \NA \\
    & $\ptmiss>20\GeV$ & \NA & \NA \\
    & $\geq$1 \PQb-tagged jet & \NA & \NA \\
    & $\pt(\text{\PQb jet})>45\GeV$ & \NA & \NA \\
    & \NA & 3 loose \Pell & $\geq$3 tight $\Pell\GeV$ \\
    & \NA & 2 small-radius jets & $\geq$3 small-radius jets \\
\end{tabular}
\end{table}

\begin{figure}[!ht]
\centering
\includegraphics[width=0.48\textwidth]{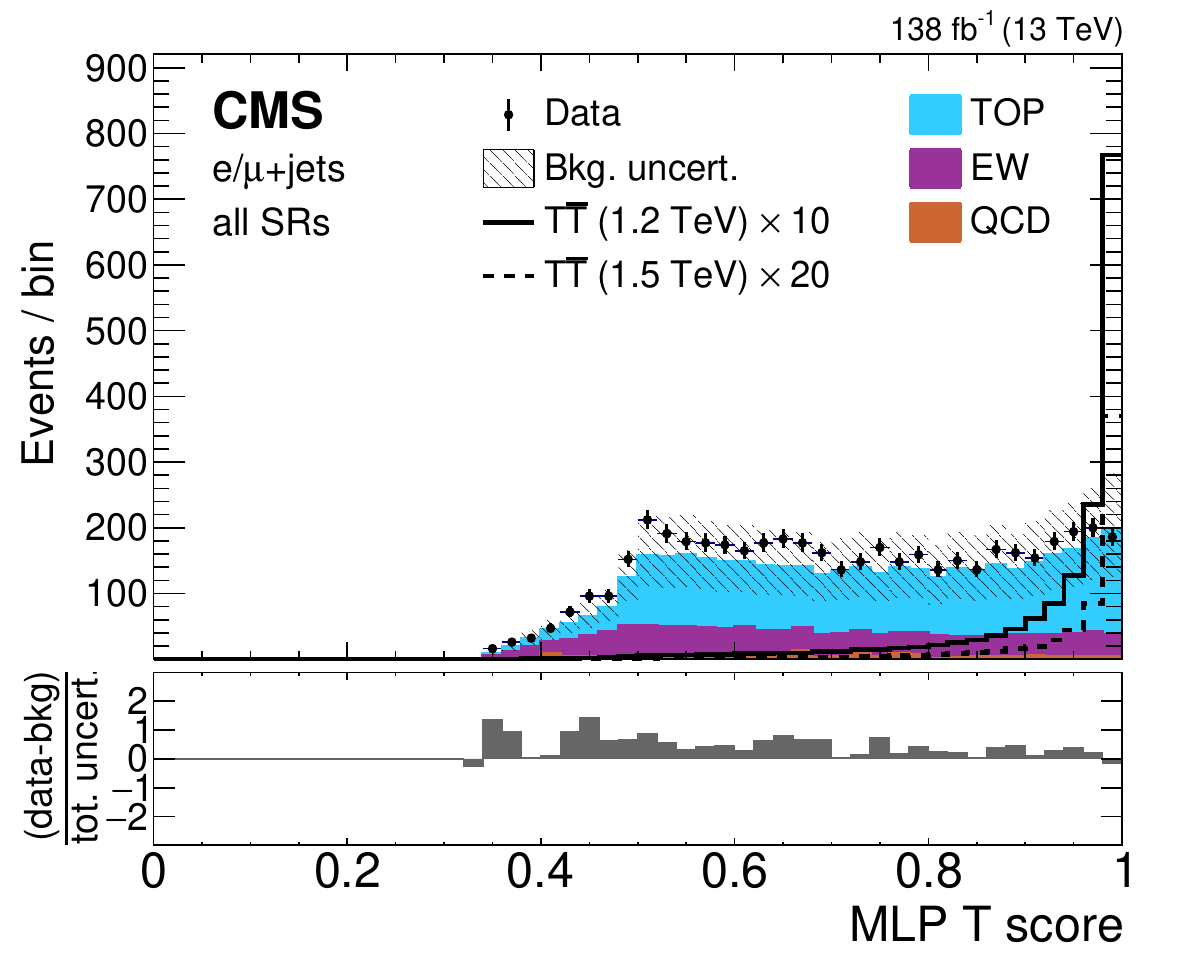}%
\hfill%
\includegraphics[width=0.48\textwidth]{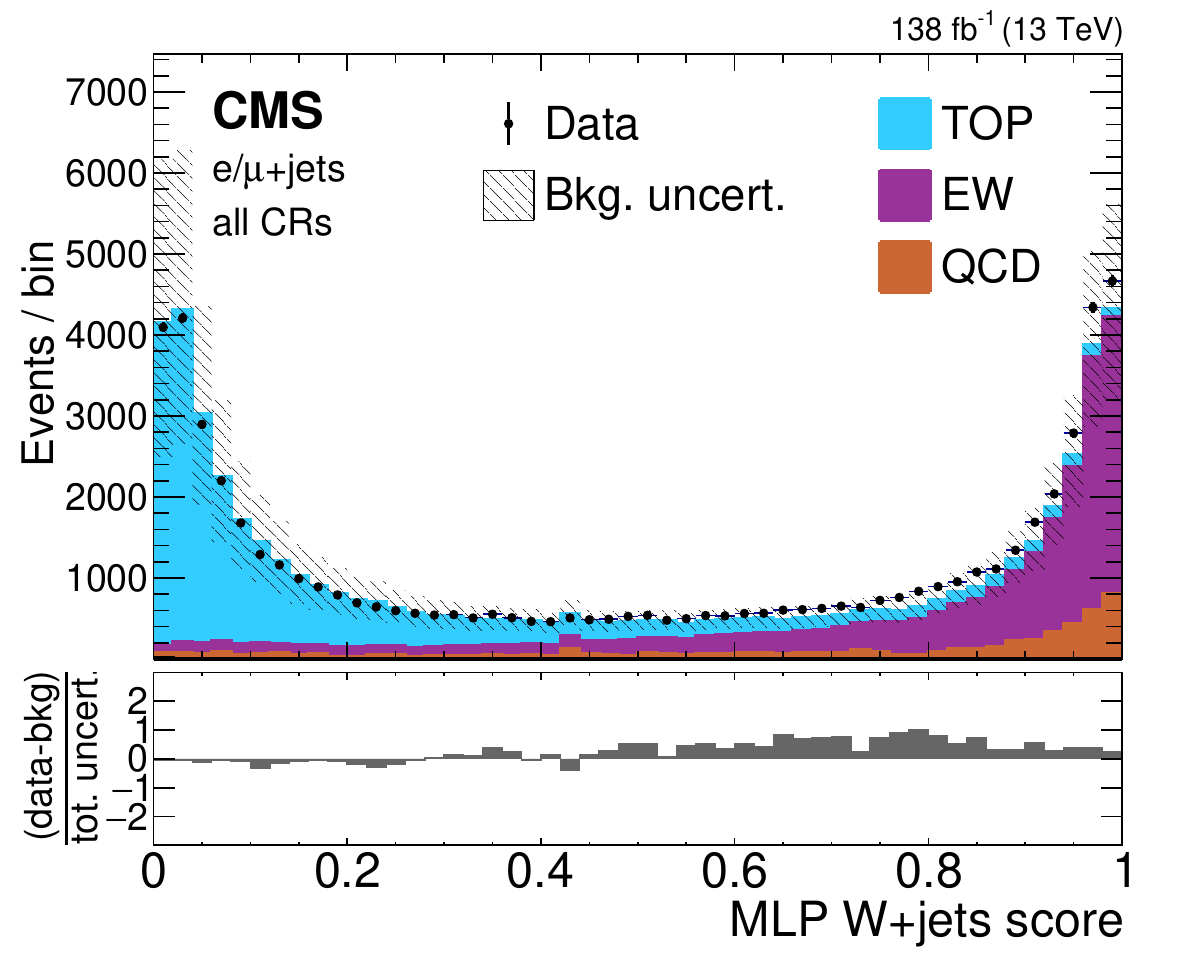}%
\caption{%
    Example single-lepton channel \TTbar NN output distributions of the \PQT quark score in the inclusive SR (left) and the \wjets score in the CRs (right).
    The observed data are shown using black markers, predicted \TTbar signals with a \PQT mass of 1.2 (1.5)\TeV in the singlet scenario using solid (dashed) lines, and backgrounds using filled histograms.
    Statistical and systematic uncertainties in the background estimate before performing the fit to data are shown by the hatched region.
    The lower panels show the difference between the observed data and the background estimate as a multiple of the total uncertainty in both sources.
    The signal predictions in the left distribution have been scaled for visibility by the factor indicated in the figure.
    Figures taken from Ref.~\cite{B2G-20-011}.
}
\label{fig:mlp}
\end{figure}

The single-lepton channel provides broad sensitivity to all \TTbar decay modes, as well as 
sensitivity to \PQB quark decays to \tW. In this channel, one of the \PQt quarks or \PW 
bosons from a VLQ decays leptonically and produces the charged lepton and a neutrino, while 
the other three initial products decay hadronically and result in large-radius jets.
The parent particles of the large-radius jets can be identified using the \DeepAKeight 
algorithm. A densely connected NN in the form of a multilayer perceptron (MLP) is trained 
to identify events as \ttbar background, \wjets background, or VLQ signal events. 
Figure~\ref{fig:mlp} shows the strong distinction between the shape of the signal and the 
background in the VLQ node score distribution in the SR, as well as the separation between 
the \ttbar and \wjets background processes in the \wjets node score in the CRs. Events are 
categorized by lepton flavor, electron or muon, and then based on the particle 
identification of the four decay products of the VLQ candidates. In this channel all background 
processes are estimated using simulation.

The SS dilepton channel is primarily sensitive to \TTbar production with \TtotH (with $\PH\to\PW\PWst$) decays.
With up to six \PW bosons produced (including those from the \PQt quark decays), two SS \PW bosons can decay leptonically to produce two SS leptons in the final state. Events are categorized by lepton flavor combinations. Three categories of background are considered: prompt lepton, nonprompt lepton, and lepton charge misidentification, as described for the \XXbar search in Section~\ref{sec:XXVLQ}. Nonprompt lepton rates used in the matrix method are extracted by fitting the predicted lepton \pt distributions in the multilepton channel CR, through the minimization of the \chisq between the observed data and the total estimated background as the nonprompt lepton rate value is varied.

The multilepton channel is primarily sensitive to contributions from \TtotZ decays.
Leptonic decays of these \PZ bosons, combined with possible leptonic decays of the \PW bosons from the decay of the \PQt quarks, may produce three or more leptons---a rare final state in SM processes. The prompt-lepton background is estimated from simulation, and the nonprompt-lepton background is again estimated via the matrix method, but extended to three leptons.
The high-energy signature of the VLQ signal is used to discriminate the signal from the 
background in the SS dilepton and multilepton channels. Following the example of the 
\XXbar search, the observable \HTlep is used in the SS dilepton channel. The observable
$\ST=\sum p_{\text{T,jets}}+\sum p_{\text{T,leptons}}+\ptmiss$ is used in the multilepton channels. Events are categorized by lepton flavor combinations.

\begin{figure}[!ht]
\centering
\includegraphics[width=0.48\textwidth]{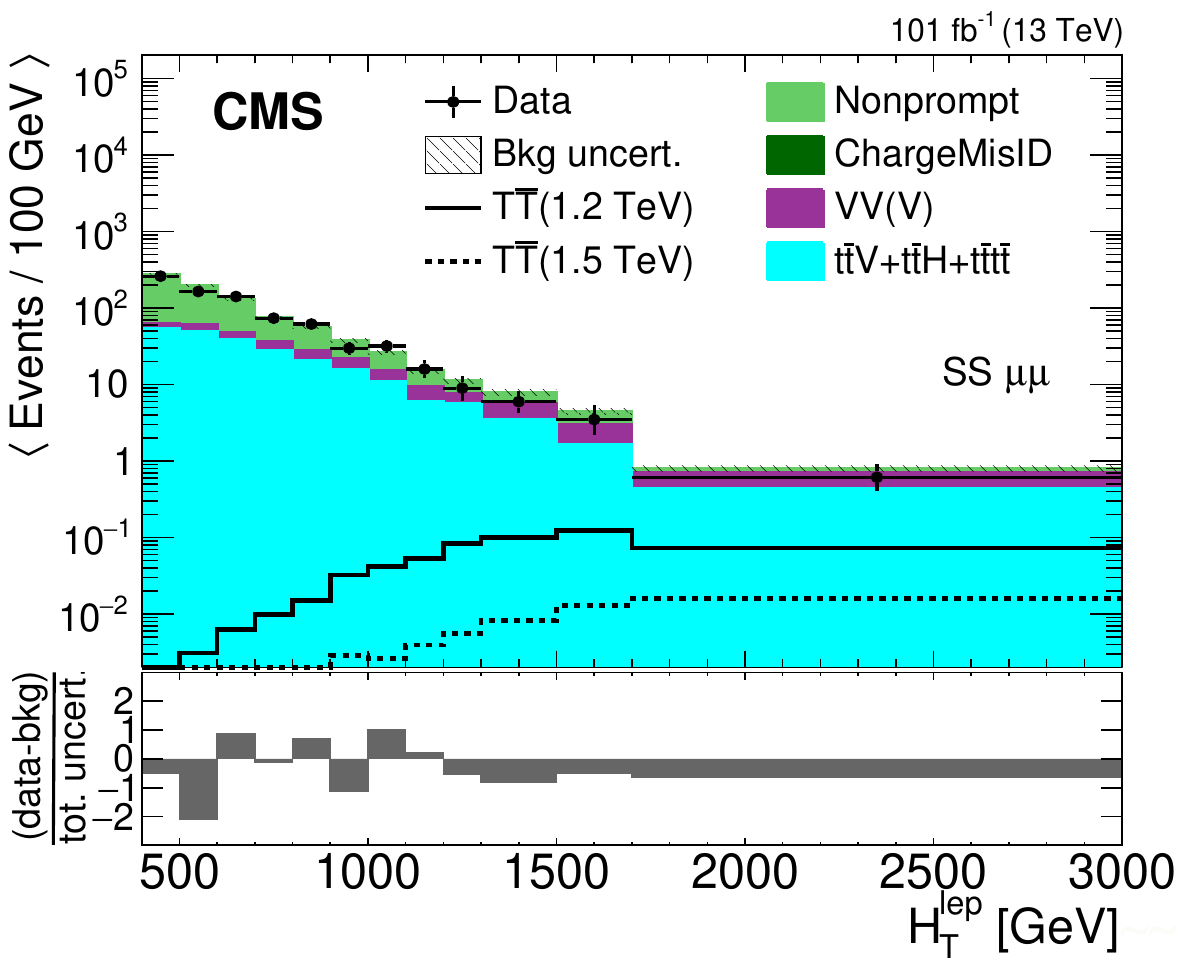}%
\hfill%
\includegraphics[width=0.48\textwidth]{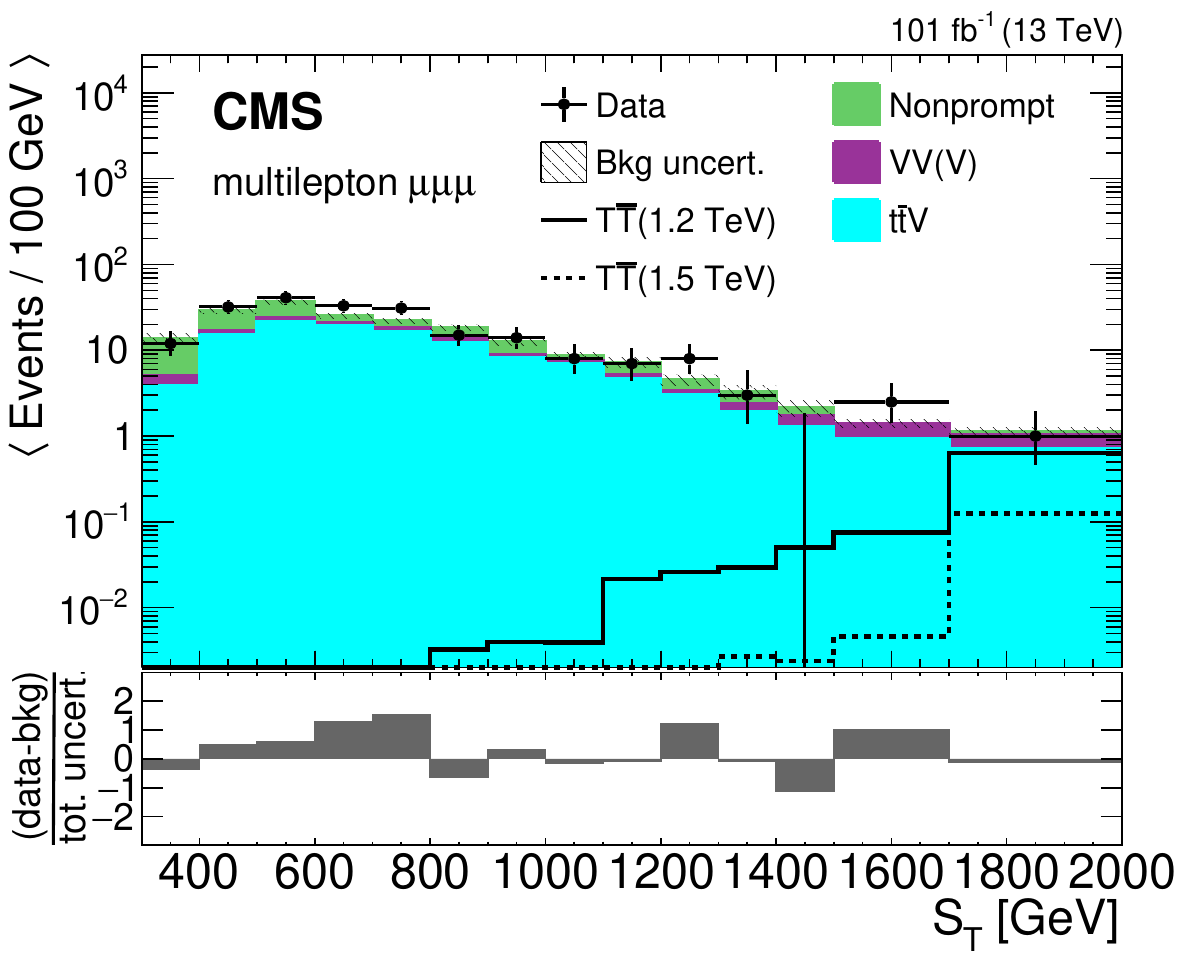}%
\caption{%
    Template histograms of \HTlep in the \mumu category of the SS dilepton channel (left) and \ST in the \mmm category of the multilepton channel (right).
    The observed data from 2017--2018 (combined for illustration) are shown using black markers, the predicted \TTbar signal for a mass of 1.2 (1.5)\TeV in the singlet scenario using solid (dashed) lines, and the postfit background estimates using filled histograms.
    Statistical and systematic uncertainties in the background estimate after performing the fit to the observed data are shown by the hatched region.
    The lower panels show the difference between the observed data and the background estimate as a multiple of the total uncertainty from both sources.
    Figures adapted from Ref.~\cite{B2G-20-011}.
}
\label{fig:2L3L_muons}
\end{figure}

A maximum likelihood fit combining multiple template histograms from the three leptonic 
channels is used to search for evidence of signal. Template histograms from a variety of 
kinematic observables are taken from the SRs of all three lepton channels, as well as some 
CRs. In the single-lepton channel, the \HT and \DeepAKeight jet tag CR distributions are 
included in the fit to constrain uncertainties in the background modeling. In the SR, the 
VLQ score from the NN is used to form template histograms for both high-purity events, in 
which both VLQ candidates contain the expected particle labels, and for low-purity events, 
which have at least one VLQ candidate without the expected particle labels.
The SR data are subdivided into 24 exclusive categories based both on the lepton flavor 
and the set of \DeepAKeight jet tags observed. In the SS dilepton channel, the \HTlep 
distribution is used to form template histograms in the three lepton flavor categories 
for 2017 and 2018 data. In the multilepton channel, the \ST distribution is fit in the 
four lepton flavor categories for all data-taking periods. Representative \HTlep (left) 
and \ST (right) distributions from the all-muon categories in these channels are shown in 
Fig.~\ref{fig:2L3L_muons}. In both of these channels the template histograms from 2016 
data are reproduced from Ref.~\cite{B2G-17-011}.
No significant excess of data over the SM background estimate is observed in any channel.

\begin{figure}[!t]
\centering
\includegraphics[width=0.48\textwidth]{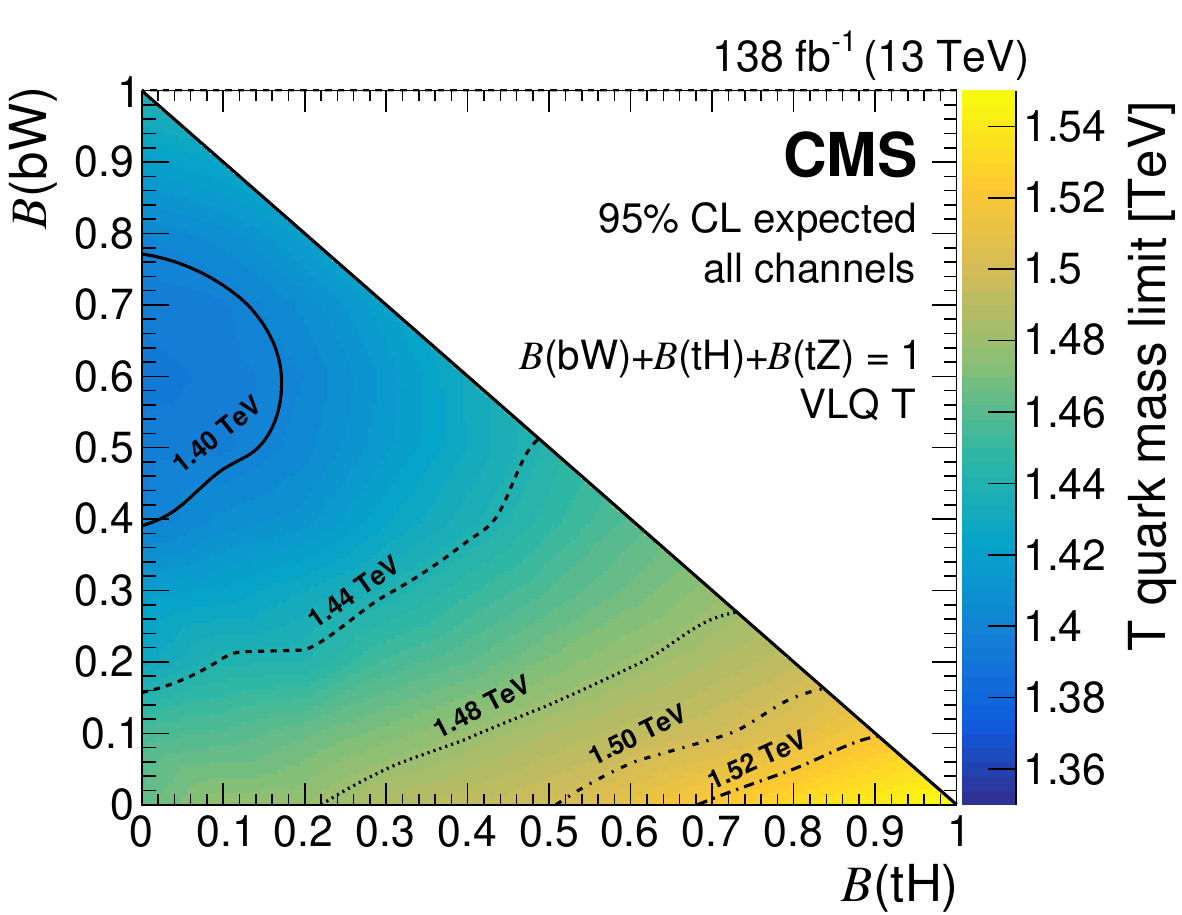}%
\hfill%
\includegraphics[width=0.48\textwidth]{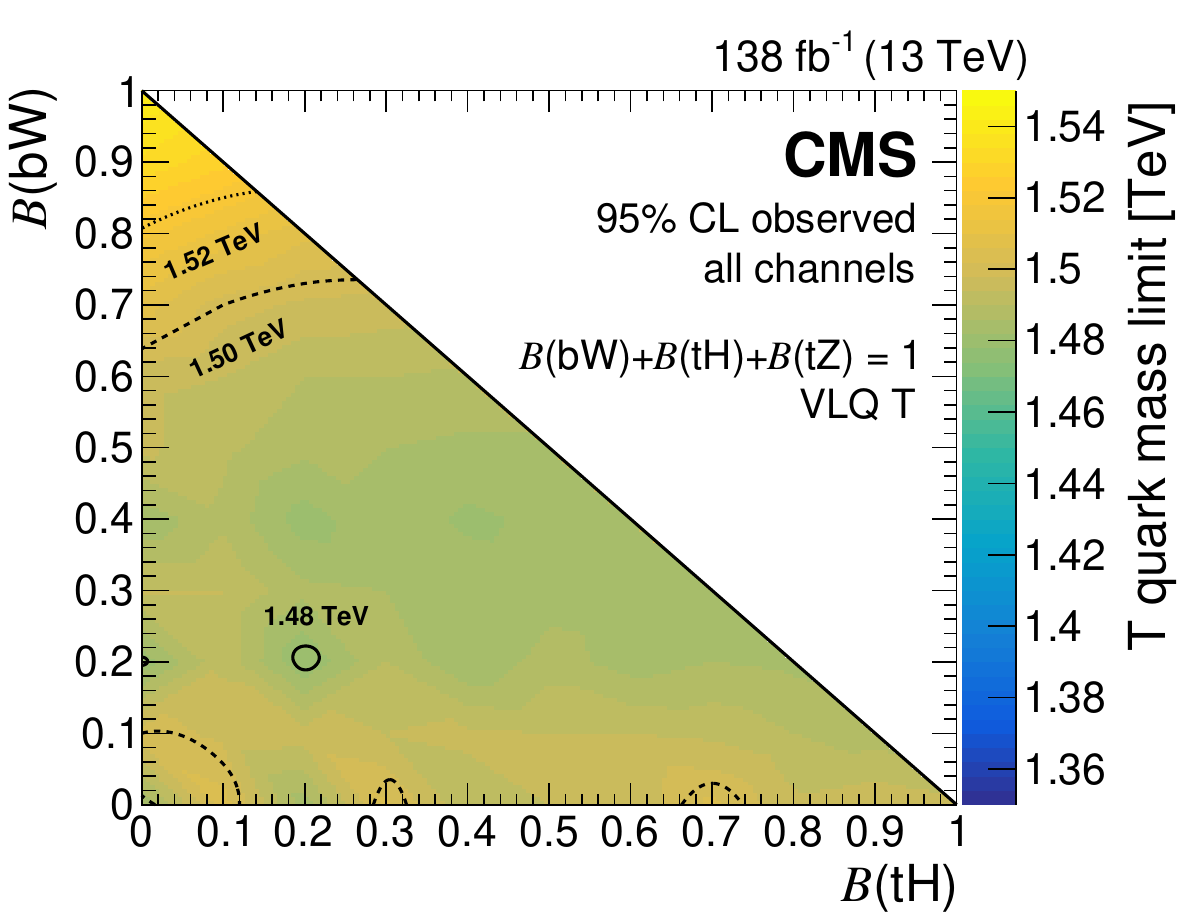}%
\caption{%
    The 95\% \CL expected (left) and observed (right) lower mass limits on pair-produced 
    \PQT quark masses, from the combined fit to the three leptonic channels, as functions 
    of their branching fractions to Higgs and \PW bosons.
    Mass contours are shown with lines of various styles.
    Figures adapted from Ref.~\cite{B2G-20-011}.
}
\label{fig:boxLimits}
\end{figure}

The dominant uncertainties in the single-lepton background predictions are the renormalization and factorization scale uncertainties, and the signal predictions are most sensitive to the \DeepAKeight heavy-particle misidentification uncertainties.
The dominant background uncertainties in the SS dilepton and multilepton channels are those affecting the nonprompt-lepton background estimation.
The searches for each VLQ flavor are independent, with only one flavor considered in the signal templates.
Figure~\ref{fig:boxLimits} shows the 95\% \CL expected (left) and observed (right) lower limits on the mass of pair-produced \PQT quarks for many possible branching fraction combinations, varying branching fractions in steps of 0.1 and requiring that $\BR(\qW)+\BR(\qH)+\BR(\qZ)=1$.
For branching fractions dominated by \PW boson decays, several single-lepton SR categories show a slight
deficit of data after the fit, but the expected and observed limits are consistent within two standard deviations.
From the scan, \PQT quarks with masses below 1.48--1.54\TeV are excluded at 95\% \CL, depending on the branching fraction.
From considering the 100\% \bW branching fraction limit, \yft quarks are excluded with masses below 1.54\TeV.
For both VLQs the strongest sensitivity is to decay modes with multiple \PQt quarks: $\TTbar\to\tH\tH$.
The sensitivity of the search is dominated by the single-lepton channel, with important contributions derived from the multilepton channel in branching fraction scenarios with significantly large \tZ decay rates.
The strength of the Run 2 \TTbar search compared to its predecessors shows the power of the expanded data set alongside
advances in NN jet classifiers.

\subsubsection{Searches for \texorpdfstring{\BBbar}{BBbar} production}
\label{sec:BBVLQ}

Many of the searches for \TTbar production presented in Section~\ref{sec:TTVLQ} also study the case of \BBbar production, using the same analysis strategy. 

\begin{figure}[!ht]
\centering
\includegraphics[width=0.48\textwidth]{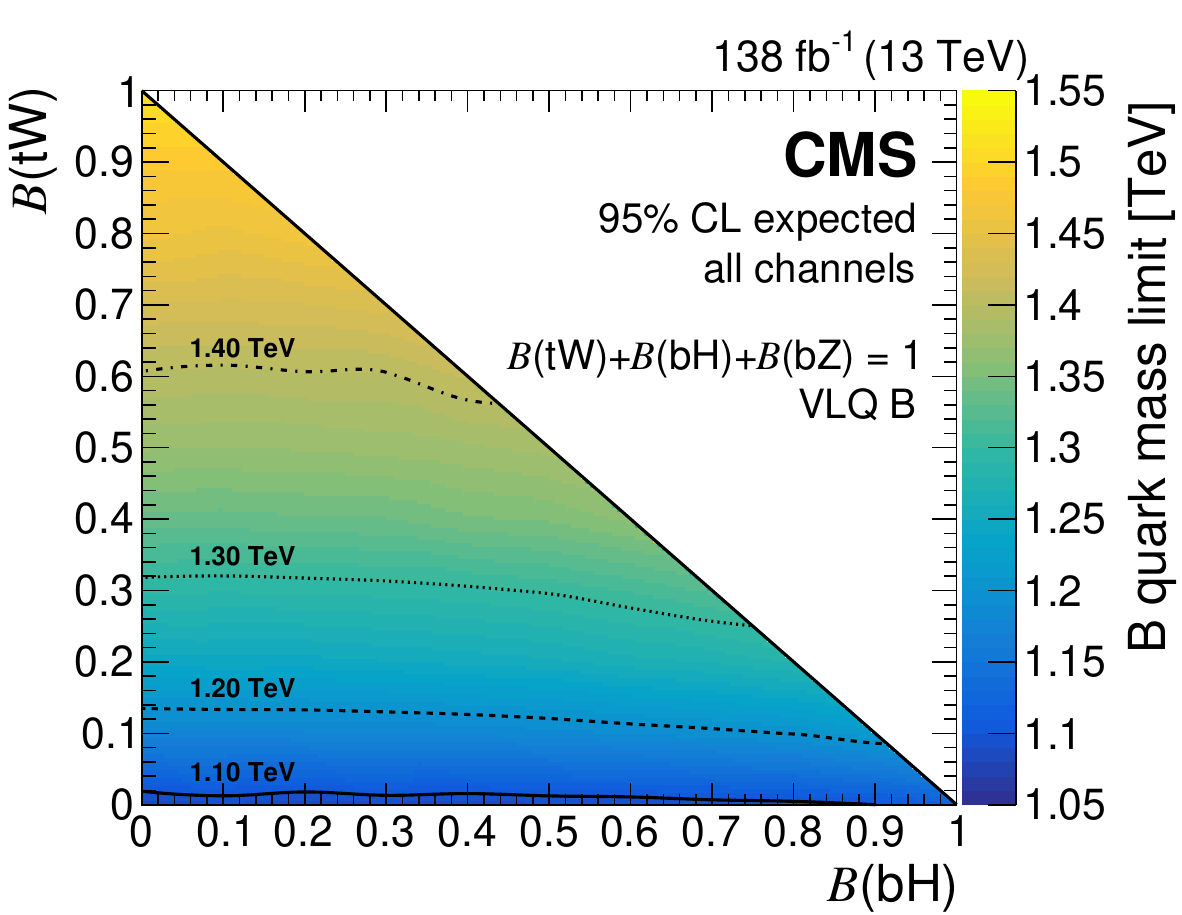}%
\hfill%
\includegraphics[width=0.48\textwidth]{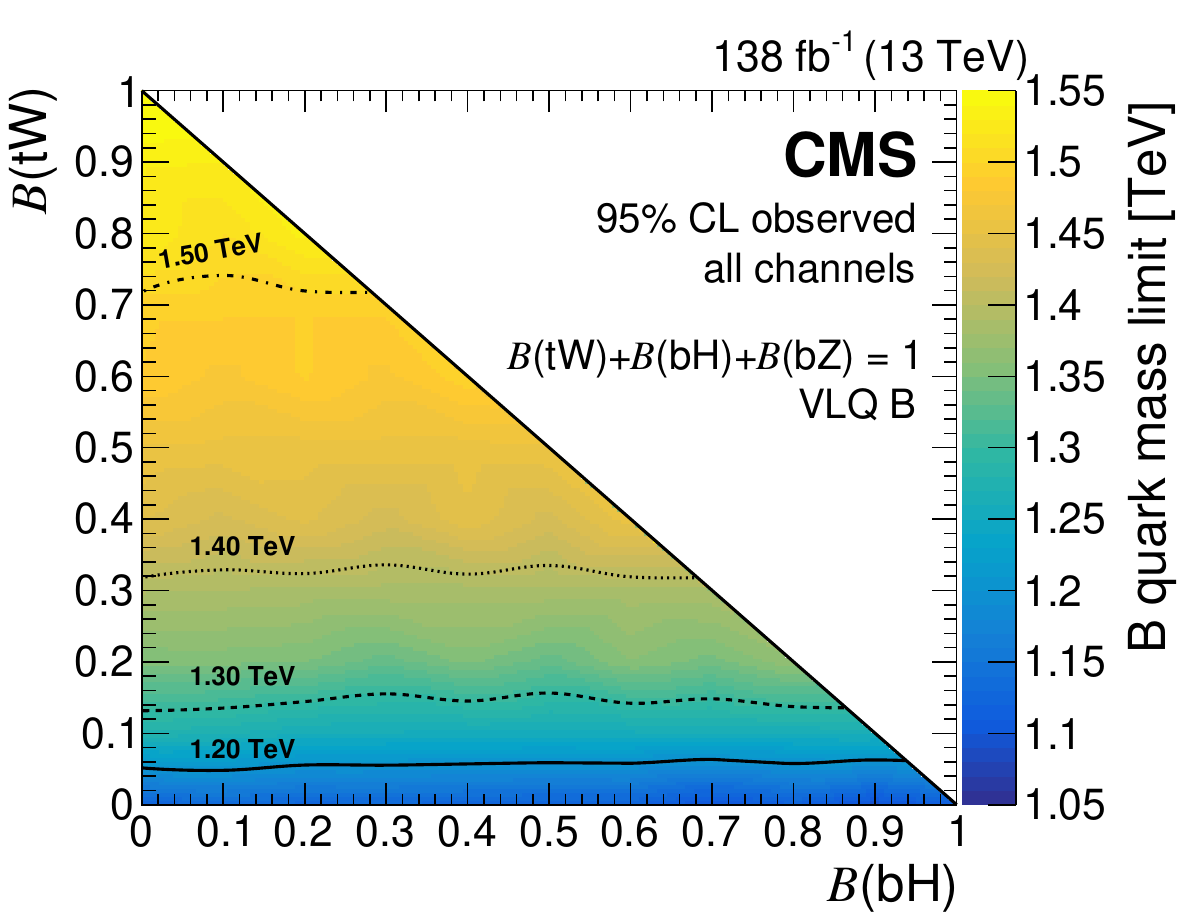}%
\caption{%
    The 95\% \CL expected (left) and observed (right) lower mass limits on pair-produced 
    \PQB quark masses, from the combined fit to the three leptonic channels, as functions 
    of branching fractions to Higgs and \PW bosons.
    Mass contours are shown with lines of various styles.
    Figures adapted from Ref.~\cite{B2G-20-011}.
}
\label{fig:boxLimitsBB}
\end{figure}

\cmsParagraph{$\BBbar\to$ leptons}
The OS dilepton analysis of Ref.~\cite{B2G-17-012} categorized events differently to search for \BBbar production with its different decay topology. That search excluded \PQB quarks with a mass below 1.13\TeV for $\BR(\bZ)=100\%$.
The leptonic search of Ref.~\cite{B2G-20-011} using the full Run 2 data set has strong sensitivity to \BtotW decays. In the single-lepton channel,
where events are categorized based on the set of \DeepAKeight jet tags observed, only 18 SR categories
are formed for the \BBbar interpretation, since \BBbar events with \BtobZ or \BtobH
decays have a low selection rate in this channel. Figure~\ref{fig:boxLimitsBB} shows the lower limits on
the \PQB quark mass derived for pair production as a function of the branching fractions to \bH and \tW. This search
excludes \PQB quarks with masses below 1.12--1.56\TeV, and offers the strongest sensitivity to
$\BBbar\to\tWtW$ decays. The \tWtW decay mode can also arise from \xft production, which is
excluded below 1.56\TeV.

\begin{figure}[!ht]
\centering
\includegraphics[width=0.48\textwidth]{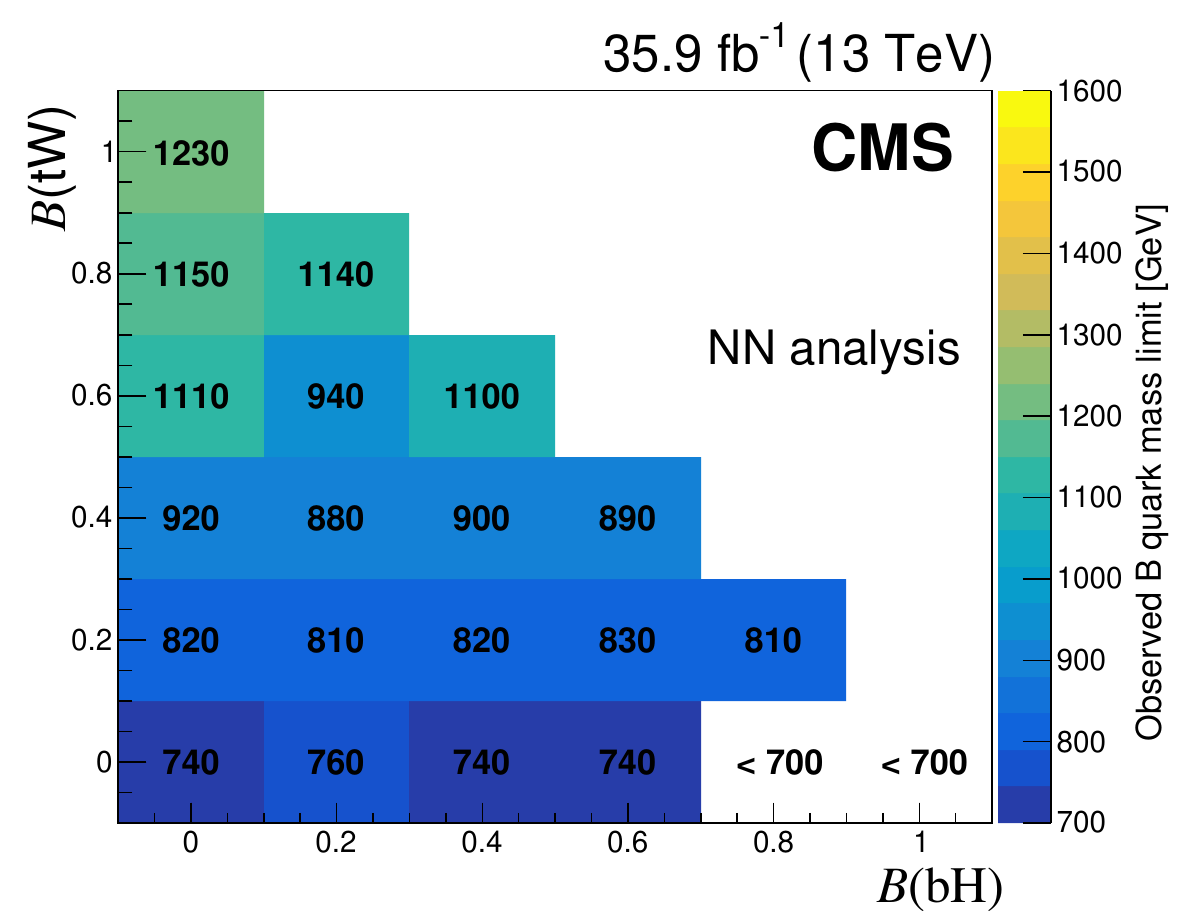}%
\hfill%
\includegraphics[width=0.48\textwidth]{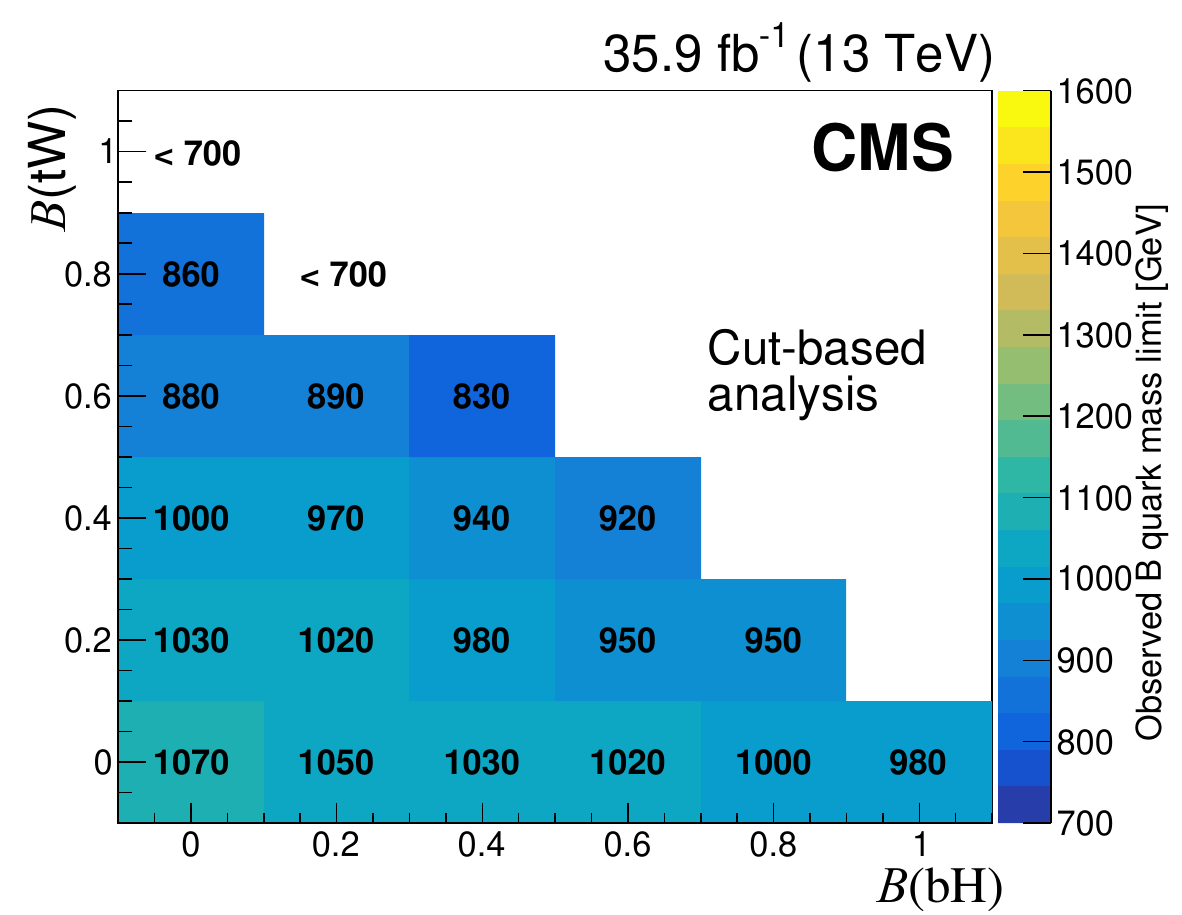}%
\caption{%
    Observed lower limit at 95\% \CL on \PQB quark masses as a function of the branching fractions to \bH and \tW, for the NN-based (left) and selection-based (right) approaches of the search for \BBbar production in the all-hadronic final state.
    Figures adapted from Ref.~\cite{B2G-18-005}.
}
\label{fig:BESTscanBB}
\end{figure}

\cmsParagraph{$\BBbar\to$ hadrons}
The NN- and selection-based analysis approaches in the hadronic pair production search using the 2016 data set~\cite{B2G-18-005} are interpretable in the \BBbar VLQ model. For the selection-based approach, the difference is simply in reinterpreting the \PW tagger as a \PZ boost tagger to target the \BtobZ channel. This analysis also benefits from using the \CSVvtwo algorithm to identify \PQb quarks in the \BtobZ and \BtobH channels.
The NN-based approach requires no modifications, though different
tag multiplicity categories are the most signal enriched compared with the \TTbar search. The fit is rerun over all six or 126 SRs for the selection-based and NN-based analyses, respectively. No significant excess above SM predictions is observed, and lower mass limits are calculated as functions of the \PQB quark branching fractions to \bH and \tW, as shown in Fig.~\ref{fig:BESTscanBB}. The NN-based analysis excludes \PQB quark masses below 1230\GeV for the \BtotW decay mode based on the strong performance of the \BEST algorithm for identifying \PQt quarks. The selection-based analysis with its stronger \PQb quark identification excludes \PQB quark masses below 1000\GeV for \BtobZ decays and below 980\GeV for \BtobH decays.

\cmsParagraph{$\BBbar\to 0\ell, \mathrm{OS}~2\ell$ }
Another search for \BBbar production was developed that specifically targets \BtobZ and \BtobH decays~\cite{CMS:2024xbc}, using the full Run 2 data set.
This analysis is based on the strategy of an earlier search
(Ref.~\cite{B2G-19-005}) that focused on \BtobH and \BtobZ decays in the hadronic final state.
However, the search using the full Run 2 data set expands on its predecessor analysis by considering
\BtotW decays of one \PQB quark in the pair, and by including an OS dilepton final state
optimized for \BtobZ decays to a pair of leptons.

The bosons produced in the \PQB quark decay often have a high Lorentz boost, so the two jets produced in a
hadronic \PW, \PZ, or Higgs boson decay may be reconstructed into a single merged jet. Similarly, in the
\tW decay mode the entire decay of the \PQt quark may be reconstructed as a single merged jet. As a result,
each decay mode may be represented by events with different jet multiplicities, depending on the number of
large-radius jets. In the leptonic category, also events with additional jets from initial-state radiation
(ISR) or final-state radiation (FSR) are included.
Events are divided into channels corresponding to the category, decay mode, and jet multiplicity, as
summarized in Table~\ref{tab:eventmodes}.

\begin{table}[!ht]
\centering
\topcaption{%
    Summary of channels considered for each category and jet multiplicity in the search for \BBbar production that specifically targets \BtobZ and \BtobH decays.
    Table adapted from Ref.~\cite{CMS:2024xbc}.
}
\label{tab:eventmodes}
\renewcommand{\arraystretch}{1.15}
\begin{tabular}{ccc}
    Jet & Leptonic & All-hadronic \\[-2pt]
    multiplicity & category & category \\
    \hline
    3 & \bHbZ, \bZbZ & \NA \\
    4 & \bHbZ, \bZbZ & \bHbH, \bHbZ, \bZbZ \\
    5 & \bHbZ, \bZbZ & \bHbH, \bHbZ, \bZbZ, \bHtW, \bZtW \\
    6 & \NA & \bHbH, \bHbZ, \bZbZ, \bHtW, \bZtW \\
\end{tabular}
\end{table}

The principal background in the leptonic category is DY dilepton production in association
with jets, whereas in the all-hadronic category the background is predominantly from QCD multijet events.
In all cases, potential signal events are distinguished from background events by requiring: the jets
are consistent with production from a \PW, \PZ, or Higgs boson; the dilepton pair in leptonic events
is consistent with a \PZ boson decay; the two reconstructed VLQs have consistent masses; and
some jets are \PQb tagged.

Small-radius jets are required to have $\pt>50\GeV$ and large-radius jets $\pt>200\GeV$, with
$\abs{\eta}<2.4$ in both cases. For the large-radius jets, the SD algorithm
is used to estimate the mass of the parent \PW, \PZ, or Higgs boson.
The \DeepJet algorithm is applied to small-radius jets to obtain single \PQb tags, whereas for merged
jets containing \bbbar pairs the double-\PQb algorithm is applied to large-radius jets.

In the leptonic channels, electrons or muons are selected with $\pt>50\GeV$, $\abs{\eta}<2.4$,
loose identification requirements, and a loose isolation requirement for muons.
Events are required to have two OSSF leptons with an invariant mass in the range
$80<\mellell<102\GeV$, and three to five small-radius jets, at least one of which must be \PQb tagged
according to the \DeepJet operating point with a 1\% misidentification rate.
In the hadronic channel,
events must have no leptons, four to six small-radius jets, and $\HT>1350\GeV$. Each channel carries a
different requirement for the number of single- and double-\PQb tags to be consistent with the decay mode
hypothesis. One double-\PQb tag is typically required in \bHbH or \bHbZ categories with only four or five
small-radius jets.
After the event selection, masses of VLQ candidates and their decay products are computed from the SD mass
of large-radius jets, and/or the invariant mass of combinations of small-radius jets.

The reconstruction of events and assignment of jets to parent particles is performed using a modified
\chisq metric, \chimod, that compares the mass of each reconstructed hadronically decaying \PQt quark
and \PW, \PZ, or Higgs boson with the average values found in simulation. The \chimod formula also includes
a term to compare the mass difference between VLQ candidates with the average value from simulation.
The decay mode hypothesis with the smallest \chimodndf value, where ndf is the number of degrees of
freedom, is used both to determine the assignment of jets and the overall event mode. In the leptonic
case, the \chimodndf value is also used to identify additional jets that are likely to be from ISR and FSR.
Since the \chimodndf values tend to be lower for signal events than background events, an upper threshold
is set for the minimum \chimodndf to provide background rejection. The threshold is optimized separately
for each category and channel to maximize the sensitivity of the analysis.

\begin{figure}[!ht]
\centering
\includegraphics[width=0.48\textwidth]{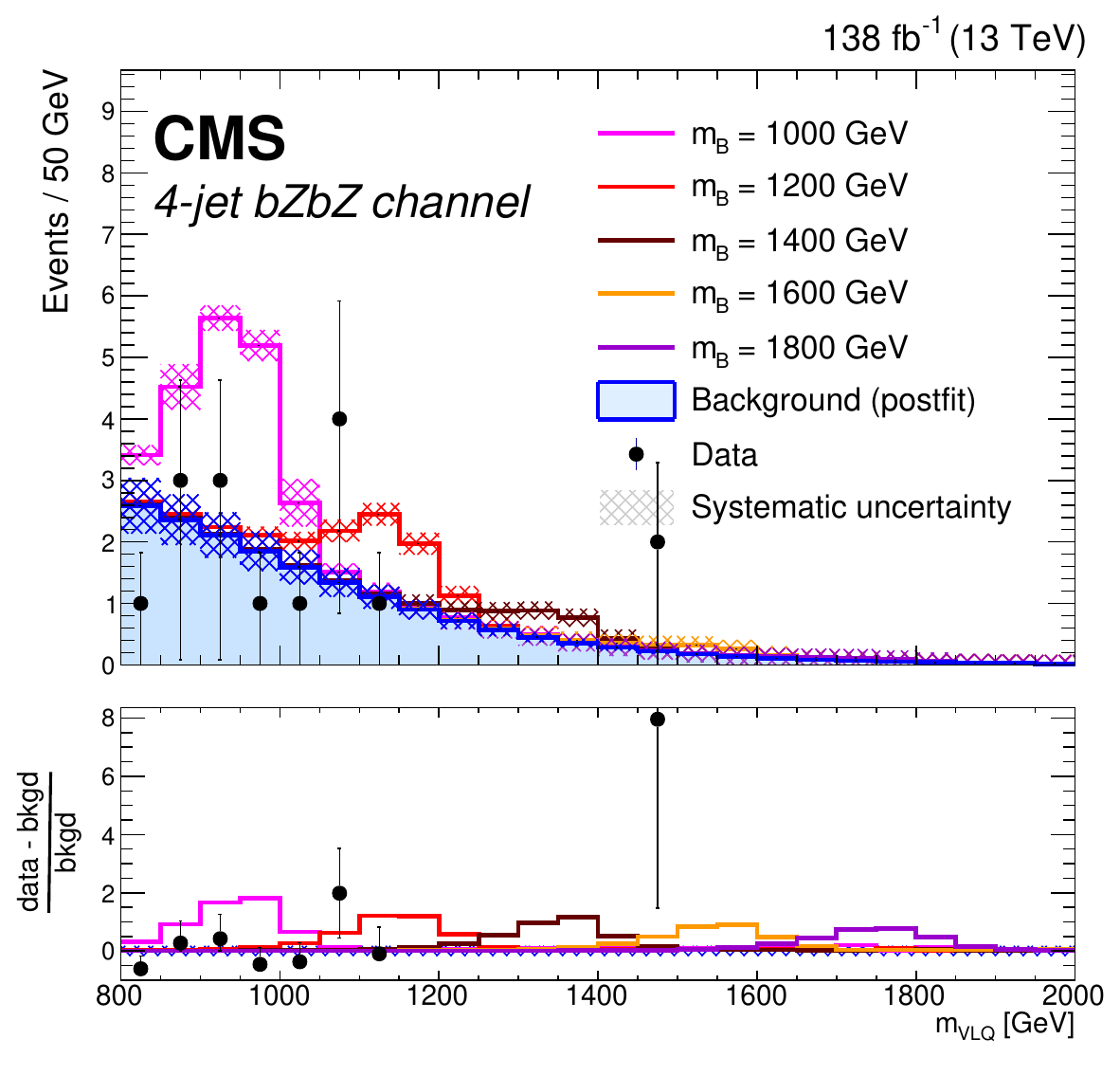}%
\hfill%
\includegraphics[width=0.48\textwidth]{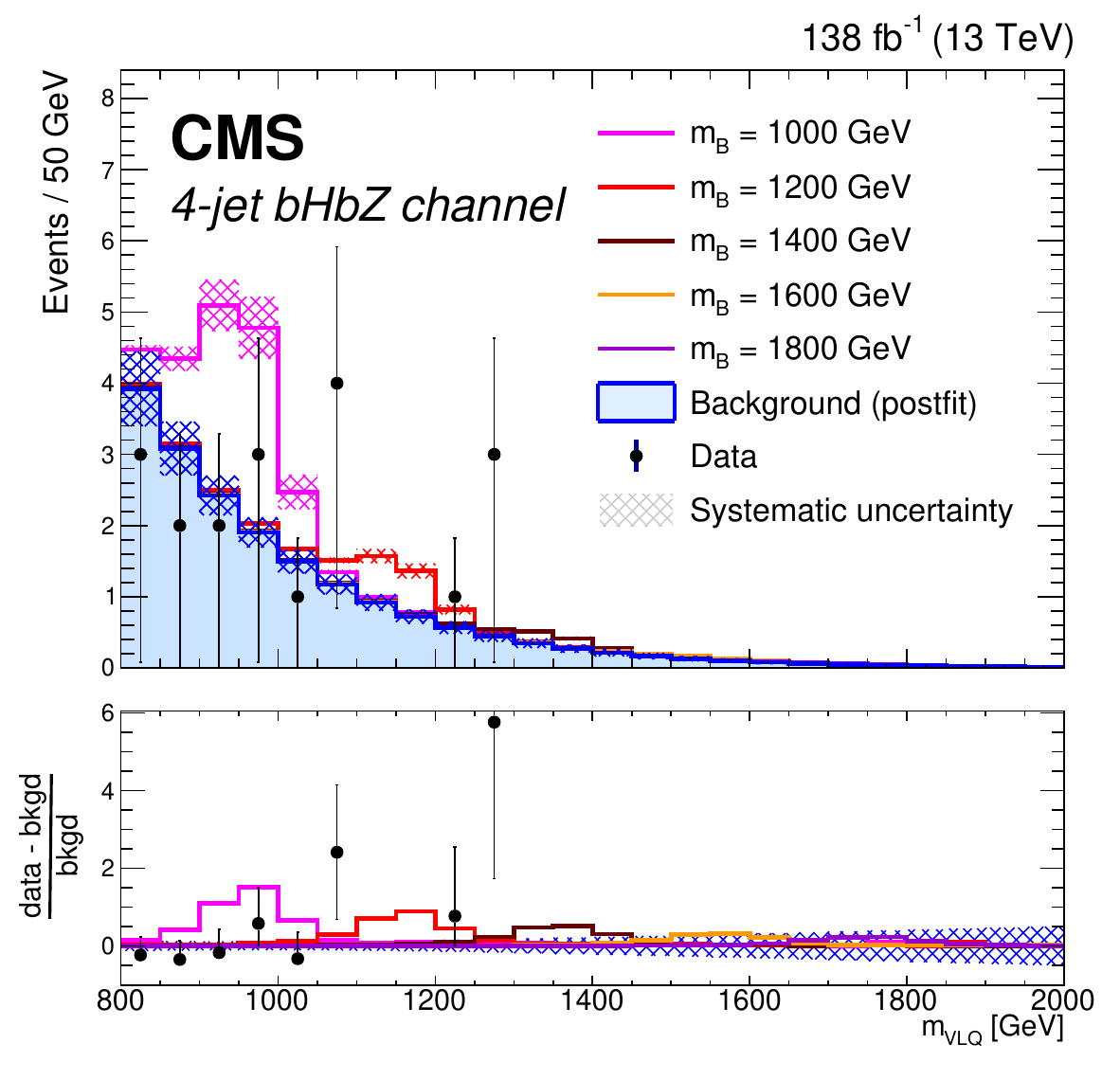}%
\caption{%
    Distributions of the reconstructed VLQ mass for expected background (blue histogram), signal plus background (colored lines), and observed data (black points) for events in the hadronic four-jet \bZbZ category (left) and the leptonic four-jet \bHbZ category (right) in the search for \BBbar production.
    Five signal masses are shown: 1000\GeV (pink), 1200\GeV (red), 1400\GeV (orange), 1600\GeV (yellow), and 1800\GeV (green).
    The signal distributions are normalized to the number of events determined by the expected VLQ production cross section.
    The hatched regions indicate the total systematic uncertainty in the background estimate.
    The lower panels show the difference between the observed data and the background estimate as a multiple of the background estimate.
    Figures taken from Ref.~\cite{CMS:2024xbc}.
}
\label{fig:sigbkgd}
\end{figure}

To estimate the background in the hadronic channel, a ``preselection'' sample of events passing all
selection requirements except for \PQb tagging is defined, and the distribution of the VLQ candidate mass
in this sample is fitted with an exponential function. The background estimate is constructed by
multiplying the fitted function by the ``background jet-tagged fraction'', which is determined in a
low-mass sideband region and corrected for any mass dependence using a region with large \chimodndf. In the leptonic category, the SR is defined
as events with $\chimodndf<5$ with a \PQb-tagged jet directly originating from the \PQB quark candidate, so
the CR consists of events that fail this \PQb tagging requirement. The VLQ mass distribution in the CR is
fitted with an exponential function over the mass range $800<\mvlq<2000\GeV$, and
this distribution is propagated
to the SR using a transfer factor: the ratio of \PQb-tagged events to \PQb-vetoed events with
$450<\mvlq<900\GeV$. For the estimated background distributions, uncertainties in each fit parameter and normalization factor
are propagated to the final background distribution.
Figure~\ref{fig:sigbkgd} shows the distribution of the reconstructed VLQ mass for the observed data,
expected background, and simulated signal events for one hadronic and one leptonic category.

\begin{figure}[!ht]
\centering
\includegraphics[width=0.48\textwidth]{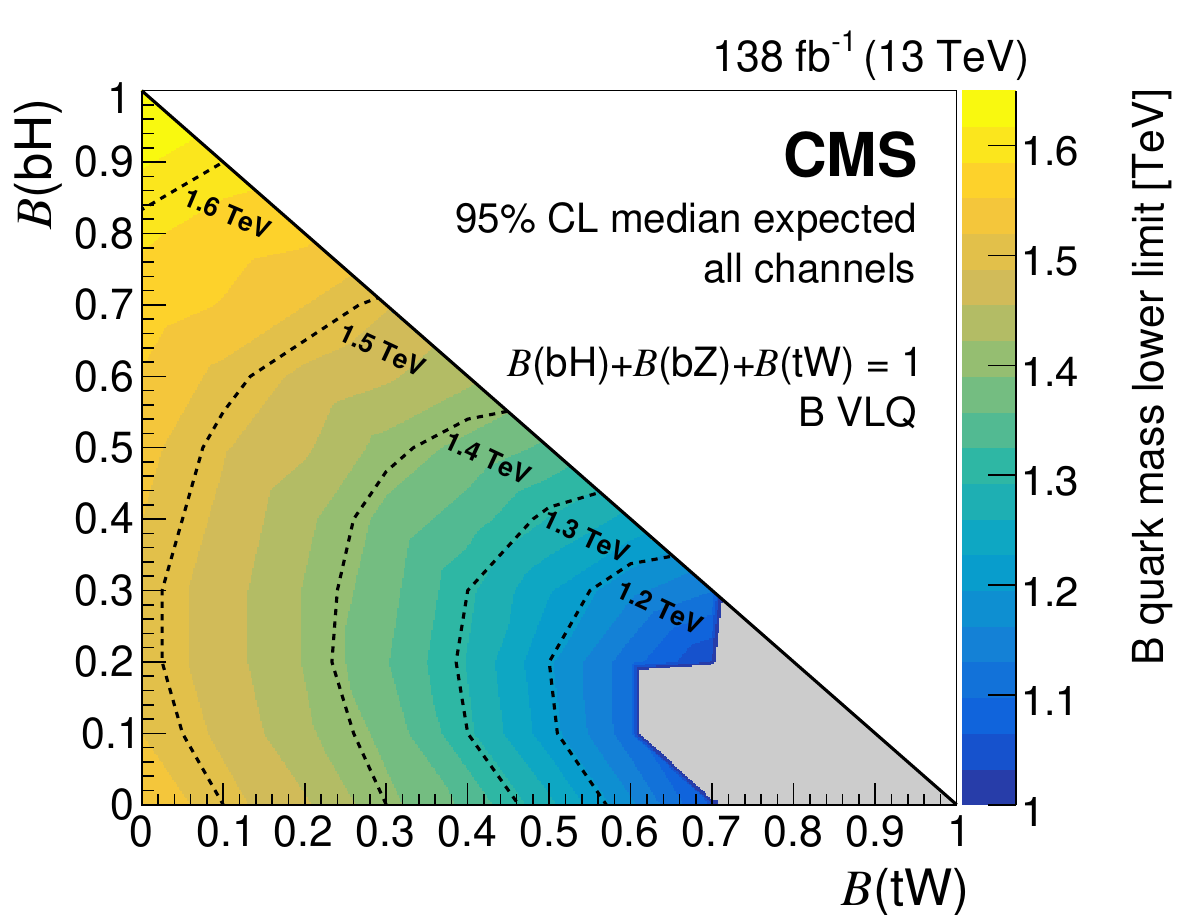}%
\hfill%
\includegraphics[width=0.48\textwidth]{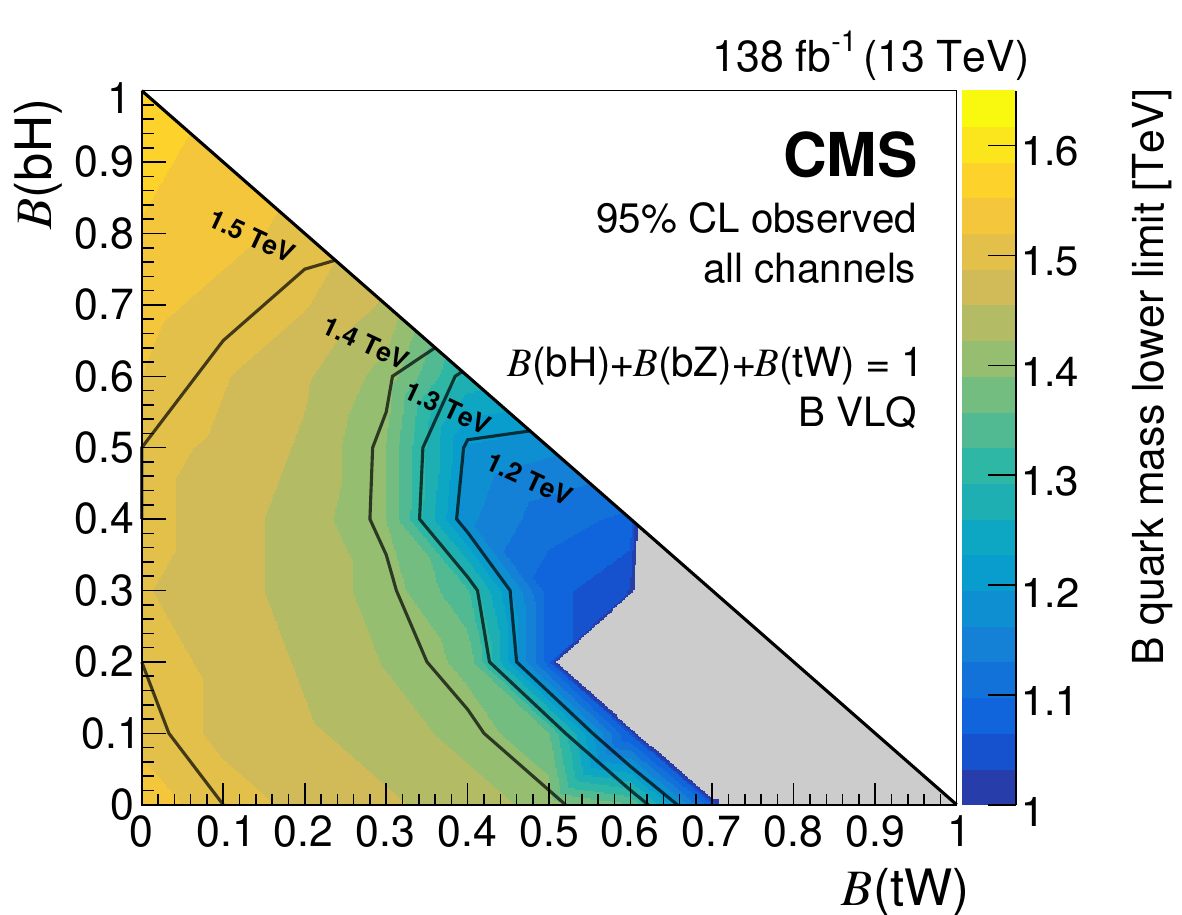}%
\caption{%
    Expected (left) and observed (right) lower limits on the \PQB quark mass at 95\% \CL 
    from the combination of the full Run 2 hadronic and OS dilepton channels, as a 
    function of the branching fractions \BrBbH and \BrBtW, with $\BrBtW=1-\BrBbH-\BrBbZ$.
    Results in the grey region, where the lower limit is less than 1.0\TeV, are omitted.
    Figures adapted from Ref.~\cite{CMS:2024xbc}.
}
\label{fig:results_triangle}
\end{figure}

No statistically significant excess of the observed data over the background expectation is observed,
and lower limits at 95\% \CL are set on the \PQB quark mass as a function of \PQB quark decay branching
fractions, combining the full Run 2 hadronic and OS dilepton channels.
Figure~\ref{fig:results_triangle} shows the expected and observed lower limits, respectively, on the
\PQB quark mass as a function of \BrBbH and \BrBtW, omitting scan points for which the exclusion limit is less
than 1000\GeV. Compared with the previous result by the CMS experiment~\cite{B2G-19-005},
the limits on the \PQB quark mass have been increased from 1570 to 1670\GeV, 1390 to
1560\GeV, and 1450 to 1560\GeV in the 100\% \BtobH, 100\% \BtobZ, and $\PQB\yft$ doublet cases, respectively.

\subsection{Single production}
\label{sec:SingleVLQ}

Analysis strategies for singly produced VLQs typically exploit the presence of a jet in the forward direction of the detector, originating from the quark produced in association with the VLQ, as seen in Fig.~\ref{fig:vlq_diagrams} (middle), as well as the high Lorentz boost of the decay products for high VLQ masses.
In the following, searches for singly produced \PQT quarks are discussed first, followed by searches for singly produced \yft, \xft, and \PQB quarks.

Single \PQT quark decays to \TtotZ and \TtotH are discussed in Sections~\ref{sec:TZlepVLQ} and ~\ref{sec:TZHbbVLQ}, respectively.
As mentioned in Section~\ref{subsubsec:vlq_crosssec}, the strength of the
EW production cross section depends on the coupling of the \PQT quark to third-generation
quarks, denoted by \kappaT. The value of this coupling may vary significantly depending on the mass and decay width
of the \PQT quark. The analyses are designed using different width approximations, including the NWA
and width approximations of 10, 20, and 30\%, while
considering different values for \kappaT.
Beyond the range of validity of the NWA,
the large width of the \PQT quark and the interference of single \PQT quark events with SM background events become significant
factors that must be considered in the analyses~\cite{Deandrea:2021vje, Banerjee:2021hfo}.

\subsubsection{\texorpdfstring{\TtotZ}{T to tZ}}
\label{sec:TZlepVLQ}

In this section, the studies are discussed in which the decay of the \PQT to a top quark and a \PZ boson were probed via various leptonic and hadronic decays of the \PZ boson and the \PQt quark.
Depending on the \PQT quark mass, the top quarks are produced with a high Lorentz boost and collimated decay products.
Thus, different reconstruction
strategies for the top quark are used for different mass ranges of the \PQT quark.
The quarks from the top quark decay tend to be reconstructed as individual, small-radius jets for
\PQT quark masses lower than $\approx$1\TeV.
However, for higher masses, the decay products of the boosted
top quark become highly collimated, producing overlapping jets. In this case, the analysis employs
substructure techniques to study large-radius jets and identify those originating from the top
quark or the \PW boson, which enhances the sensitivity of the analysis.
The top quark may be detected through
three distinct methods: fully merged, partially merged, or resolved, depending on whether a large-radius \PQt jet is identified; a large-radius \PW jet and a small-radius \PQb jet are identified; or three small-radius jets are reconstructed, respectively.
A particular feature of the direct production of a single \PQT quark is the presence
of an additional jet that is produced in the forward direction.
Below we describe analyses that have focused on searches in specific channels,
where the \PZ boson decays into two OS leptons, neutrinos, or undergoes a fully
hadronic decay.

\begin{figure}[!ht]
\centering
\includegraphics[width=0.48\textwidth]{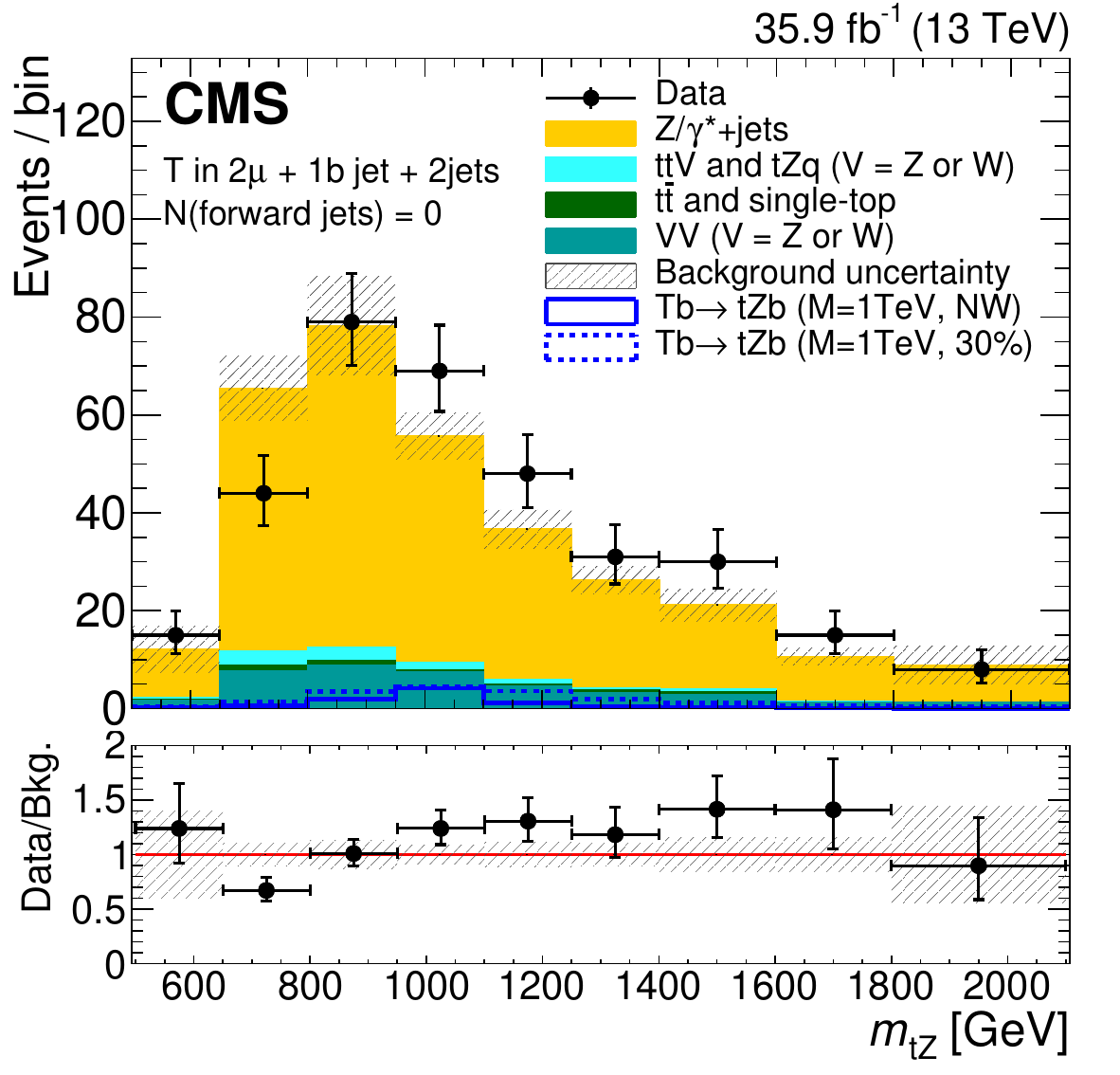}%
\hfill%
\includegraphics[width=0.48\textwidth]{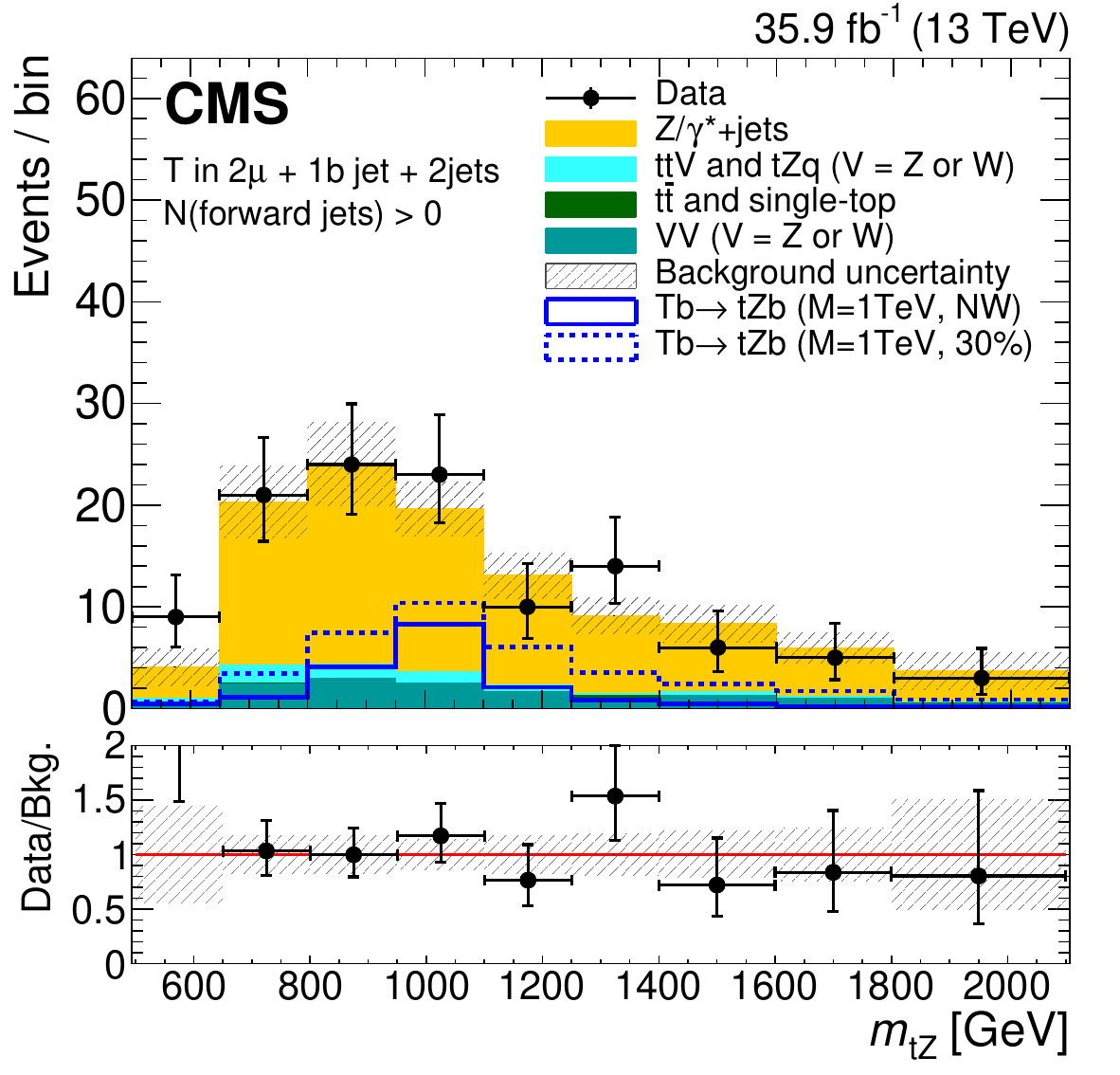}%
\caption{%
    Distributions of the reconstructed \PQT quark mass, \mtZ for the observed data, the background estimates, and the expected signal for the two categories where the singly produced \PQT quark is reconstructed in the resolved topology for events with the \PZ boson decaying into muons and no forward jets (left) and at least one forward jet (right).
    The background composition is taken from simulation.
    The expected signal is shown for two benchmark values of the width, for a \PQT quark produced in association with a \PQb quark: NWA and 30\% of the \PQT quark mass.
    The lower panel in each plot shows the ratio of the observed data to the background estimation, with the hatched band representing the uncertainties in the background estimate.
    Figures taken from Ref.~\cite{CMS:2017voh}.
}
\label{fig:Background3}
\end{figure}

\cmsParagraph{$\tZ\to\bqq\,\ellell$} The search in Ref.~\cite{CMS:2017voh} targets the detection of \TtotZ where the \PZ boson decays into a charged-lepton pair, using the 2016 data set.
The events are selected based on the presence
of two OS leptons, which can be either muons or electrons, with an invariant mass
within the range of 70 to 110\GeV. The forward jet is reconstructed as a small-radius
jet and is required to have $\pt>30$\GeV and $2.4<\abs{\eta}<5.0$.
Furthermore, in the partially merged and resolved categories, the presence of at least one \PQb-tagged jet, using the medium \DeepCSV working point, is required for the top quark reconstruction.
The two leptons from the \PZ boson decay must be spatially close to each other, with a
distance metric of $\DR<0.6$--1.4, depending on the category. Moreover, the leading-\pt
lepton, either a muon or an electron, must have a \pt greater than 120\GeV.
When more than one medium \PQb-tagged jet is present in the event, the one with the largest reconstructed top
quark \pt is selected for further reconstruction.
Additionally, in the resolved categories, the two jets with the lowest \PQb tagging discriminant
out of the three jets forming the top quark candidate must have a dijet invariant mass below 200\GeV. The signal is expected to accumulate as an excess over
the background events in the mass
spectrum of reconstructed \PQT quark candidates, \mtZ.
In this strategy, the main background process is $\PZ/\PGg^\ast+\text{jets}$ events, constituting
over 80\% of the total background. Smaller contributions originate from other sources like \ttV,
\tZq, \ttbar, single \PQt quark, and \VV diboson production.
Figure~\ref{fig:Background3} shows distributions of the reconstructed mass \mtZ of the \PQT quark for the observed data, the background estimates, and the expected signal for categories targeting the reconstructed \PQT quark in the resolved topology. The events
in this category involve the \PZ boson decaying into muons and no forward jet (left) and at least one forward jet (right).
Additional distributions of observed data, background estimates, and expected signal processes for various categories
based on different \PQT quark reconstruction topologies are reported in Ref.~\cite{CMS:2017voh}.

\begin{figure}[!ht]
\centering
\includegraphics[width=0.48\textwidth]{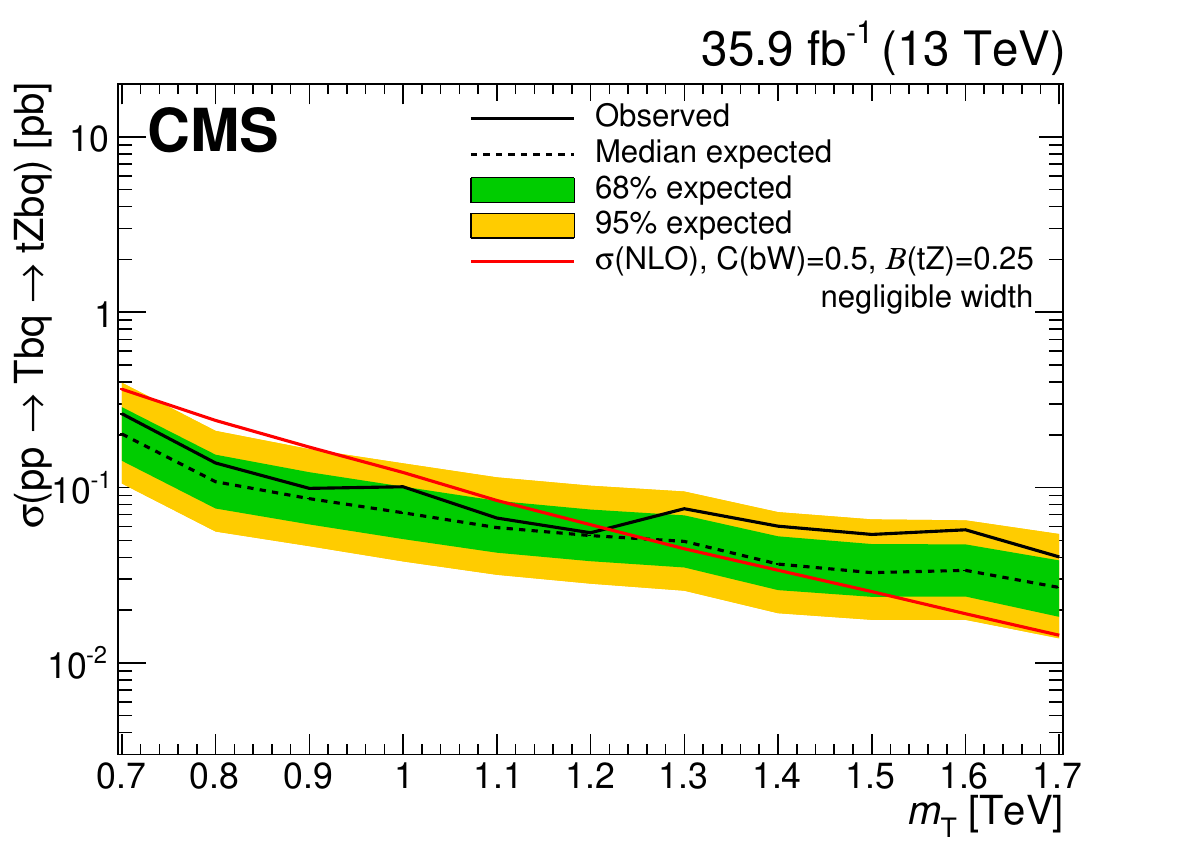}%
\hfill%
\includegraphics[width=0.48\textwidth]{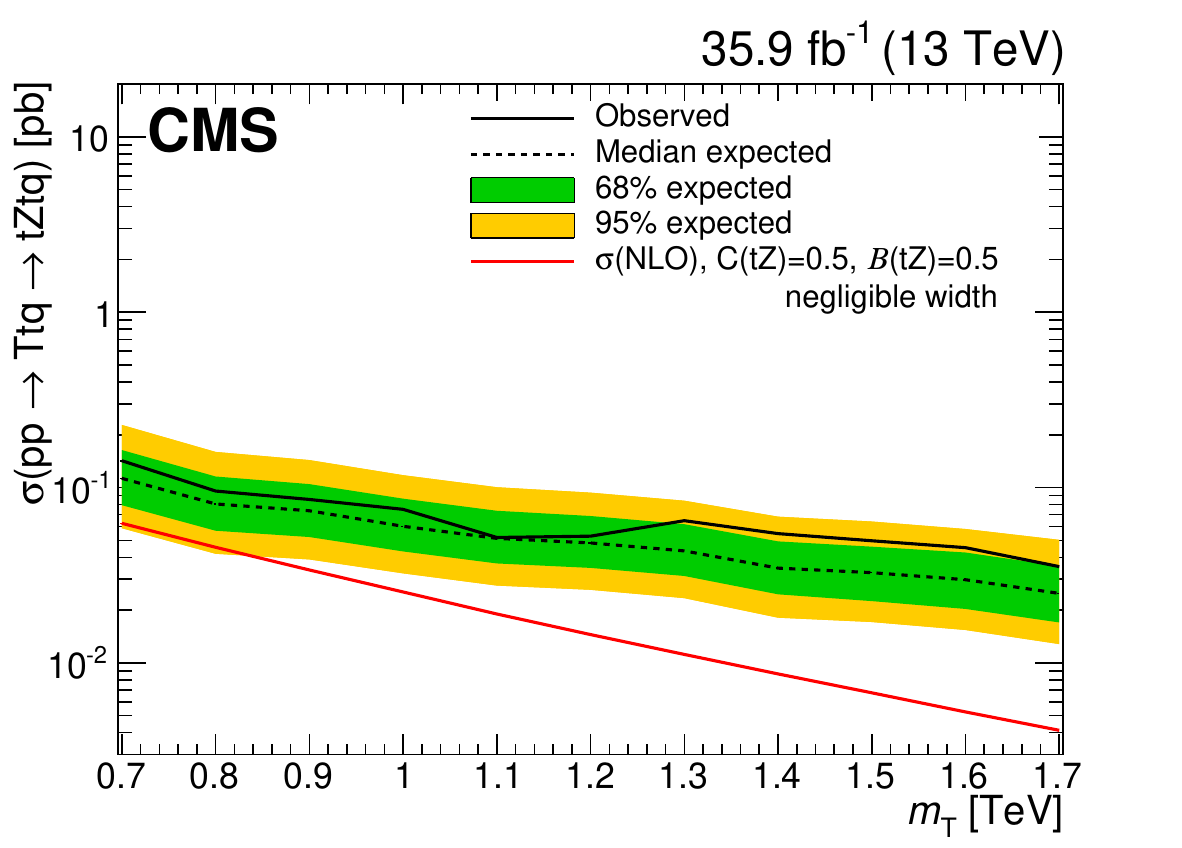}%
\caption{%
    Observed and expected upper limits on the product of the cross section and branching fraction for singlet LH \PQT quark (left) and doublet RH \PQT quark production (right) in association with a \PQb quark and a \PQt quark, respectively, in the NWA hypothesis.
    The \PQT quark decays to \tZ with a branching fraction $\BR(\TtotZ)$ of 0.25 (0.5) for the left (right) figure.
    The red lines represent theoretical cross sections calculated at NLO in perturbative QCD, whereas the inner (green) band and the outer (yellow) band indicate the regions containing 68 and 95\%, respectively, of the distribution of limits expected under the background-only hypothesis.
    Figures taken from Ref.~\cite{CMS:2017voh}.
}
\label{fig:limitTprime1}
\end{figure}

No significant deviation of the data from the expected background is observed in any of the search channels. Upper limits are set on the product of the cross section and
branching fraction of a \PQT quark decaying to \tZ.
In Fig.~\ref{fig:limitTprime1}, the observed and expected upper limits from the combined ten categories
in the search for singly produced \PQT quarks in the $\PZ\to\EE/\MM$ channels are shown
for the singlet LH \PQT quark production in association with a \PQb quark (left) and doublet RH \PQT quark
production in association with a \PQt quark (right) in the NWA hypothesis.
In this case, LH \PQT quarks produced in association with a \PQb quark and with $c_{\PW}=0.5$ are excluded for masses in the range of 0.7--1.2\TeV.
However, the limits on the production of a doublet RH \PQT quark in association with a \PQt quark do not impose constraints on the \PQT mass parameter.
Similar exclusion limit results are presented in Ref.~\cite{CMS:2017voh} as a function of width and mass of
\PQT in the ranges from 10 to 30\% and 0.8 to 1.6\TeV, respectively. The results are interpreted using
the model constructed in Refs.~\cite{Buchkremer:2013bha,Fuks:2016ftf,Oliveira:2014kla},
and an LH \PQT quark signal (in association with a \PQb quark) was excluded at 95\% \CL for masses below values in the range 1.34--1.42\TeV,
depending on the width, whereas an RH \PQT quark signal (in association with a \PQt quark) was excluded for masses below values
in the range 0.82--0.94\TeV.

\begin{figure}[htp!]
\centering
\includegraphics[width=0.48\textwidth]{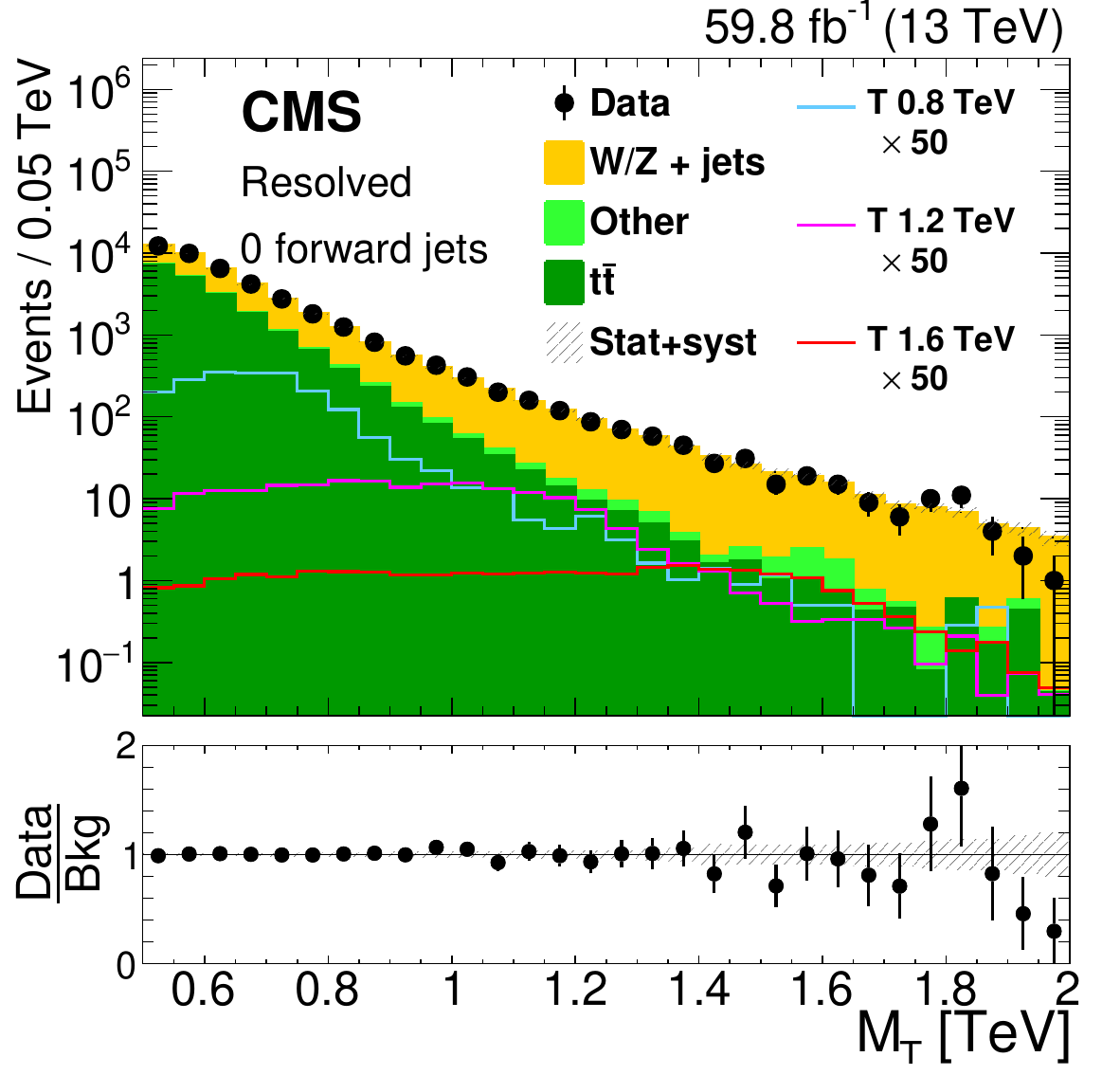}%
\hfill%
\includegraphics[width=0.48\textwidth]{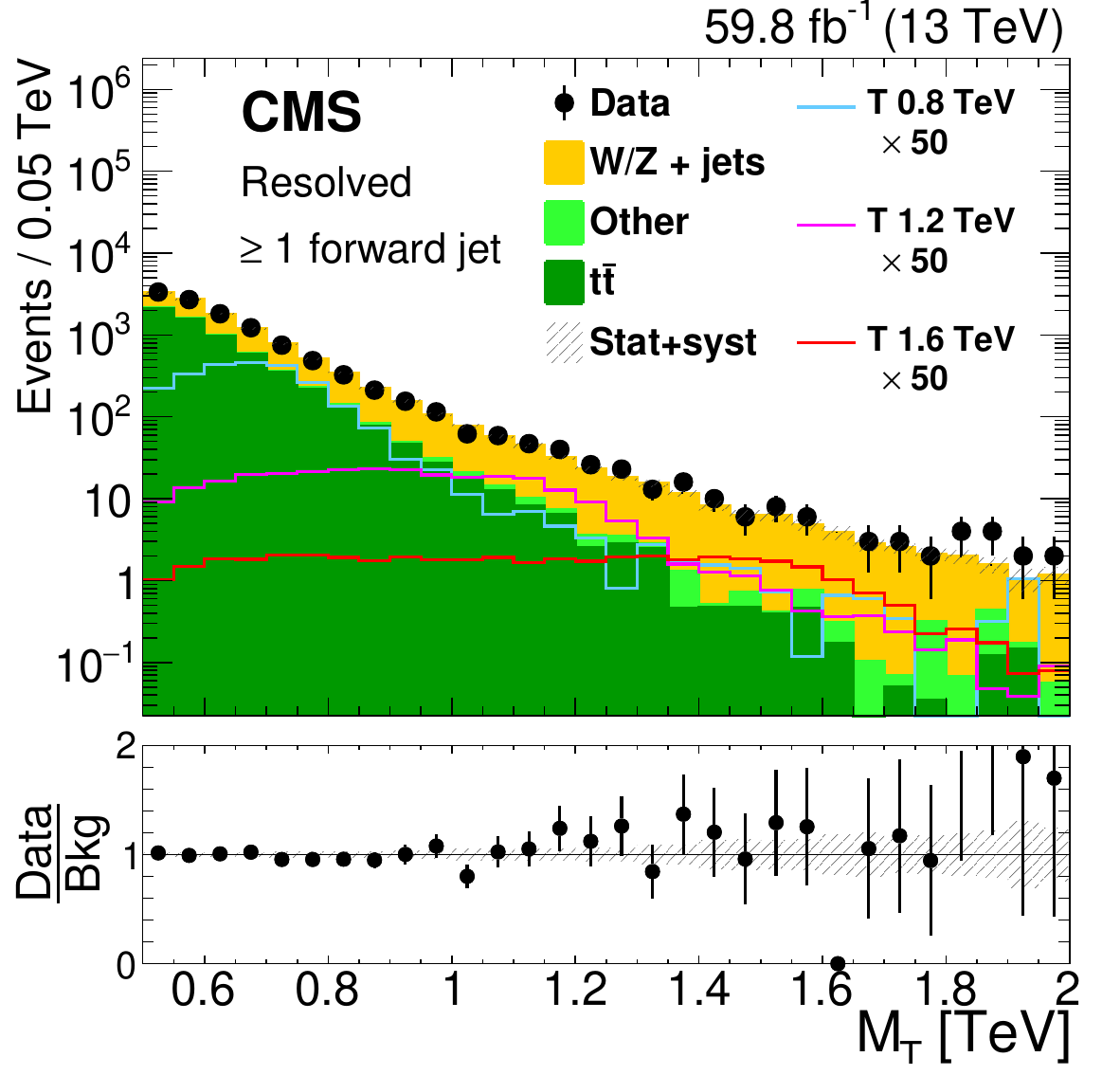}%
\caption{%
    Distributions from the 2018 data set of the transverse mass of the reconstructed top quark and \ptvecmiss system, for the selected events in the resolved categories, for events with no forward jet (left) and at least one forward jet (right).
    The distributions for the main background components have been determined in simulation with SFs extracted from CRs.
    All background processes and the respective uncertainties are derived from the fit to data, whereas the distributions of signal processes are represented according to the expectation before the fit.
    The lines show the signal predictions for three benchmark mass values (0.8, 1.2, and 1.6\TeV) for a \PQT quark of a narrow width.
    Figures taken from Ref.~\cite{B2G-19-004}.
}
\label{fig:MtResolved}
\end{figure}

\cmsParagraph{$\tZ\to\bqq\,\nunu$} The search for singly produced \PQT quarks in the mode \TtotZ with \PZ boson decays into neutrinos~\cite{B2G-19-004} 
also exploits the aforementioned three different top quark candidate reconstruction strategies. Events are selected from the full Run 2 data set with $\ptmiss > 200\GeV$, at least one
small-radius jet with $\pt > 30\GeV$ and $\abs{\eta} < 4.0$, and at least one top quark candidate constructed from 
small- or large-radius jets with $\abs{\eta} < 2.4$. To reduce the number of QCD multijet
background events, the angular separation $\Dphi_{\text{jet},\ptvecmiss}$ between each small-radius jet
and the \ptvecmiss vector must exceed 0.6. Events with identified electrons or muons are excluded.
Additional requirements are imposed on events in the resolved category to increase the
sensitivity of the analysis: $\pt>250\GeV$ for the resolved top quark candidate;
and the event \HT greater than 200\GeV. Each event that
passes the selection criteria is categorized into one of six groups based
on the type of reconstructed top quark candidate (merged, partially merged, or resolved)
and the presence or absence of at least one forward jet. When more than one type of top quark
candidate is reconstructed, the event is assigned to a single category based on a hierarchy
established through an optimization procedure aimed at obtaining the best expected exclusion
limit across the entire mass range: first the merged category, followed by the partially
merged category, and lastly the resolved category.
After the event selection, the major sources of background are {\ttbar}+jets, \wjets, and \zjets
events where the \PZ boson decays to neutrinos. Due to differences in the amount of observed data
in the CRs and to mismodeling corrections, different methods for determining
the correction factors are used for the resolved, merged, and partially merged categories.
The signal extraction is based on a simultaneous fit to the transverse mass of the top quark
candidate and \ptvecmiss system,
$\MT=\sqrt{\smash[b]{2\pt^{\PQt}\ptmiss(1-\cos\Dphi_{\PQt,\ptvecmiss})}}$,
in the six analysis categories.
Figure~\ref{fig:MtResolved} displays the $\MT$ distributions for the observed data
in 2018 and for the predicted backgrounds for events selected within the resolved categories
with no forward jet (left) and at least one forward jet (right). The distributions for the
merged and partially merged categories, as well as various data collection scenarios from different
years, are reported in Ref.~\cite{B2G-19-004}.

\begin{figure}[htp!]
\centering
\includegraphics[width=0.75\textwidth]{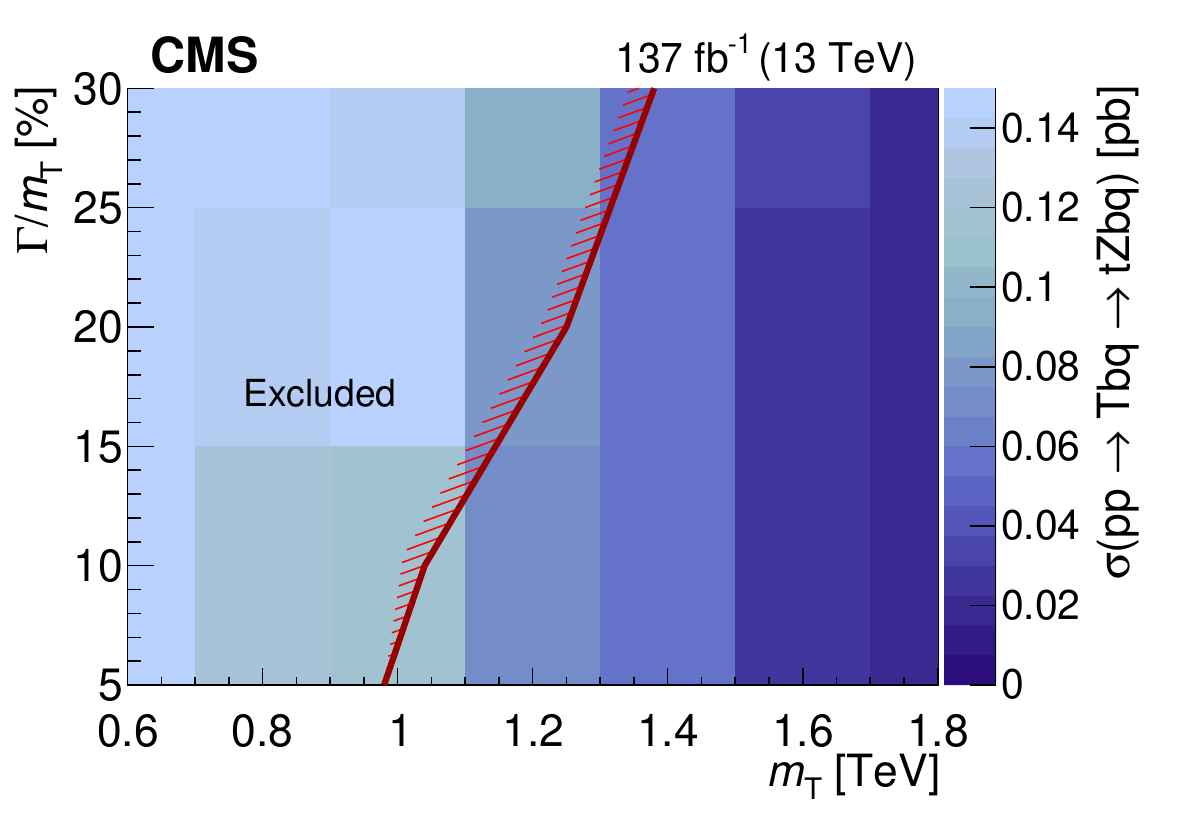}
\caption{%
    Observed 95\% \CL upper limit on the product of the single production cross section for a singlet VLQ \PQT quark and the \TtotZ branching fraction, as a function of the \PQT quark mass \mQT and width $\Gamma$, for widths from 5 to 30\% of the mass.
    A singlet \PQT quark that is produced in association with a bottom quark is assumed.
    The solid red line indicates the boundary of the excluded region (on the hatched side) of theoretical cross sections multiplied by the \PQT branching fraction.
    Figure taken from Ref.~\cite{B2G-19-004}.
}
\label{fig:combinedUL2D}
\end{figure}

Observed combined upper limits are derived for the product of the single production cross section
for the singlet \PQT quark and the \TtotZ branching fraction, for the six event
categories in the $\PZ\to\nunu$ channel combined. The result is presented as a function of the \PQT quark
mass \mQT, and for several width hypotheses, as shown in Fig.~\ref{fig:combinedUL2D}.
These results set the lower limits on the \PQT quark mass in the singlet model for various
resonance width hypotheses: values of \PQT quark mass lower than 0.98, 1.1, 1.3, and 1.4\TeV are
excluded for resonance widths 5, 10, 20, and 30\% of the mass, respectively.
As the cross section rises with assumed width according to Eq.~(\ref{eq:singleTxsec}), the constraints get stronger for larger relative width scenarios up to 30\%.

\cmsParagraph{$\tZ\to\bqq\,\bb$} {\tolerance=800
Analogous to the previously described analyses, search strategies for single \PQT in the all-hadronic channel with \PH/\Ztobb are defined based on the quarks
resulting from the top quark and Higgs/\PZ boson decays: a low-mass search and a high-mass search~\cite{B2G-19-001, B2G-18-003}.
The event selection criteria are based on the properties of the signal final state, specifically with $\PQt\to\bW$ and \PH/\Ztobb decays.
\par}

The low-mass search of Ref.~\cite{B2G-19-001} uses the full Run 2 data set. The final state comprises two jets from the \PW boson decay and three \PQb jets (two from the \PH or \PZ bosons and one from the top quark decay).
Events are selected if they contain at least six small-radius jets with $\pt>40\GeV$ and $\abs{\eta}<4.5$, in order to maintain high efficiency for selecting all the jets from the \PQT quark decay.
Tighter \pt thresholds are imposed on the first three leading jets to select objects consistent with the decay of a high-mass resonance.
Finally, three of the small-radius jets are required to lie in the central region of the detector and be \PQb-tagged using the tight \DeepCSV working point.
The main variable used in the low-mass search strategy is the reconstructed five-jet invariant mass from small-radius jets.
A multistep \chisq minimization algorithm is used to identify the jet combinations that reconstruct the best Higgs or \PZ boson, \PW boson, and top quark candidates.
Additional criteria are applied to optimize the signal reconstruction based on the individual \chisq scores, the angular separation between the selected objects, and the fraction of momentum carried by the decay products of the \PQT quark.
These criteria are described in detail in Ref.~\cite{B2G-19-001} and are defined to ensure that the five-jet invariant mass distribution is a smoothly falling spectrum.
The dominant background processes are QCD multijet production and top quark pair production, and their contributions are evaluated from the observed data using CRs with relaxed \PQb tagging requirements.

\begin{figure}[!ht]
\centering
\includegraphics[width=0.48\textwidth]{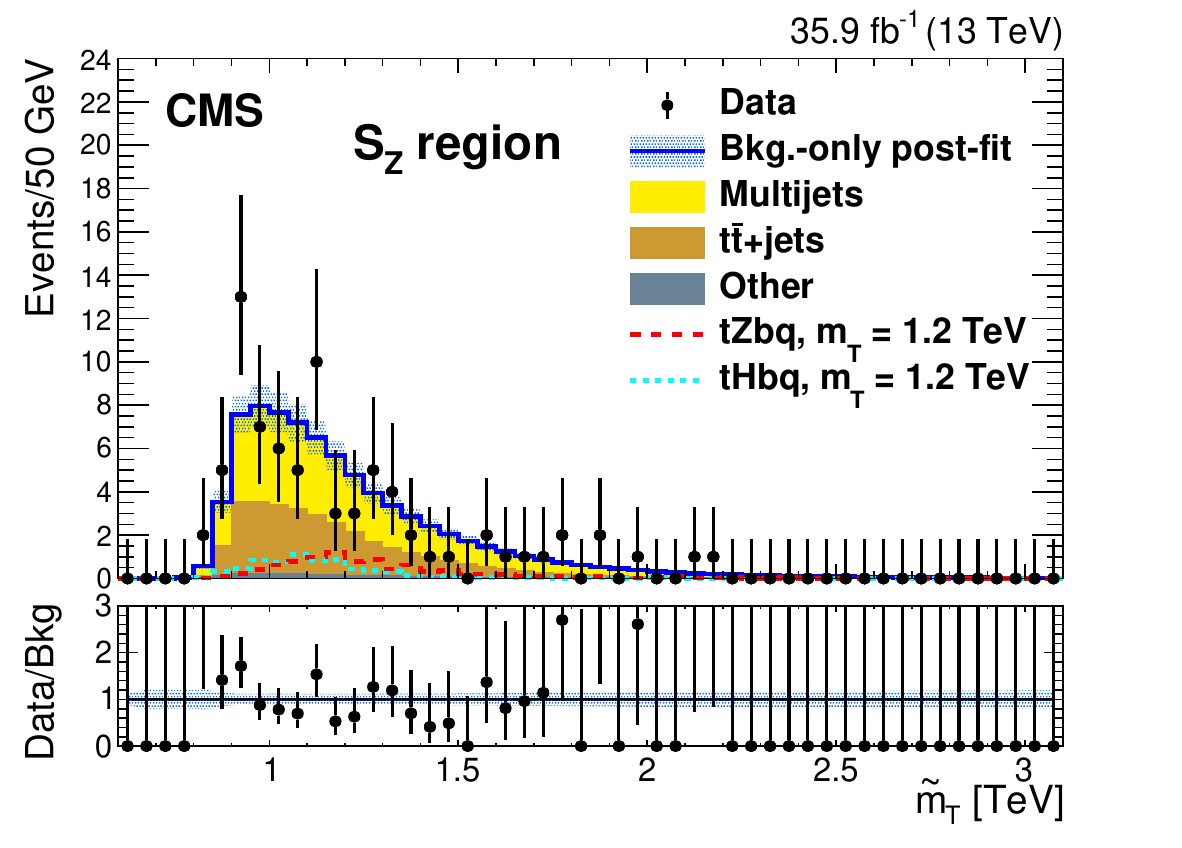}%
\hfill%
\includegraphics[width=0.48\textwidth]{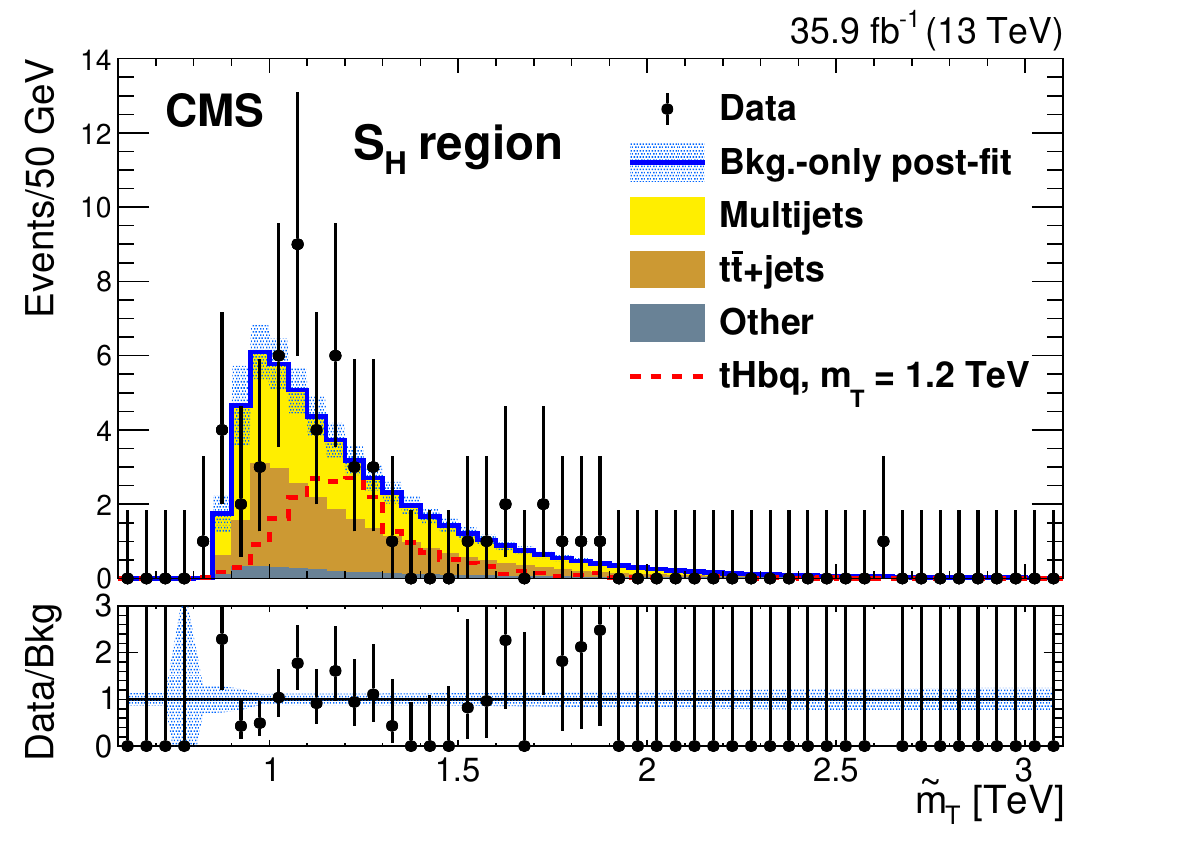}%
\caption{%
    Background-only postfit distributions of \mtilde, the adjusted \PQT mass sensitive observable defined in Ref.~\cite{B2G-18-003}, of the observed data for the SR of the \TtotZ (left) and \TtotH (right) channels, respectively, for the high-mass search.
    The dashed red histogram in each case represents an example signal for the \tZbq or \tHbq process with a \PQT quark mass of 1.2\TeV and a relative width of 30\%.
    The lower panels of the plots display the ratio of observed data to the fitted background for each bin.
    The error bars on the data points correspond to the 68\% \CL Poisson intervals, whereas the light blue band in each ratio panel represents the relative uncertainties in the fitted background.
    Figures taken from Ref.~\cite{B2G-18-003}.
}
\label{fig:postfit_tH_tZ}
\end{figure}

In Ref.~\cite{B2G-19-001}, the high-mass search strategy is also performed using the five-jet invariant mass as the main observable and hence focuses on \PQT masses below 1.2\TeV.
An earlier search performed by the CMS Collaboration \cite{B2G-18-003}, using the 2016 data set,
applies a different strategy for the high-mass regime, in which the invariant mass of the \TtotH and \TtotZ candidates is reconstructed using two large-radius jets.
This strategy is effective for \PQT masses of 1\TeV and above.
Jet substructure techniques are used to reconstruct the top quark and \PH/\PZ jets.
The presence of a forward jet is also required with a minimum \DR of 1.2 from the leading large-radius jets.
After event selection, the dominant background contributions, as in the low-mass search, are \ttbar and QCD multijet production.
The reconstructed mass of the \PQT candidate from the large-radius dijet system is adjusted for deviations of the individual
large-radius masses from the known top quark and an \PH or \PZ boson, and used as the main observable. Details on the adjusted
\PQT mass sensitive observable, \mtilde, can be found in Ref.~\cite{B2G-18-003}.
The resulting postfit \mtilde distribution of the observed data based on the background-only
hypothesis for \TtotZ is shown in Fig.~\ref{fig:postfit_tH_tZ} (left). Similar plots for other CRs may be found in Ref.~\cite{B2G-18-003}.

As the \Ztobb channel is merged with the \Htobb channel, the limits are presented jointly in the following Section~\ref{sec:TZHbbVLQ} in Figs.~\ref{fig:massLim_B2G19001} and \ref{fig:massLim_Tbq1}.
The exclusion limits derived from the \Ztobb channel alone is reported in Refs.~\cite{B2G-18-003} and \cite{B2G-19-001}.

\subsubsection{\texorpdfstring{\TtotH}{T to tH}}
\label{sec:TZHbbVLQ}

In this section, we discuss the searches for singly produced \PQT quarks, assumed to decay into a top quark
and a Higgs boson, considering various leptonic and hadronic channels for the top quark decay, and two photons
or \bbbar for the Higgs boson decay.

\begin{figure}[!ht]
\centering
\includegraphics[width=0.48\textwidth]{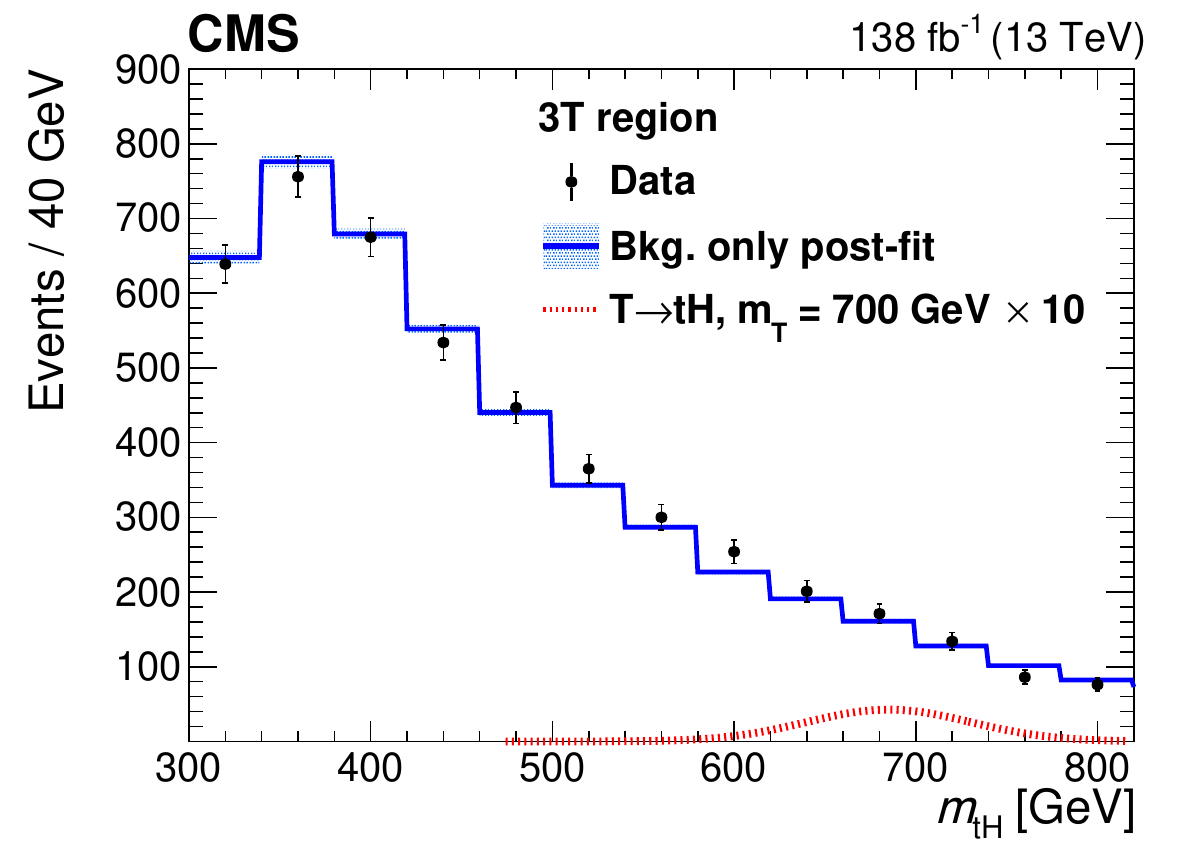}%
\hfill%
\includegraphics[width=0.48\textwidth]{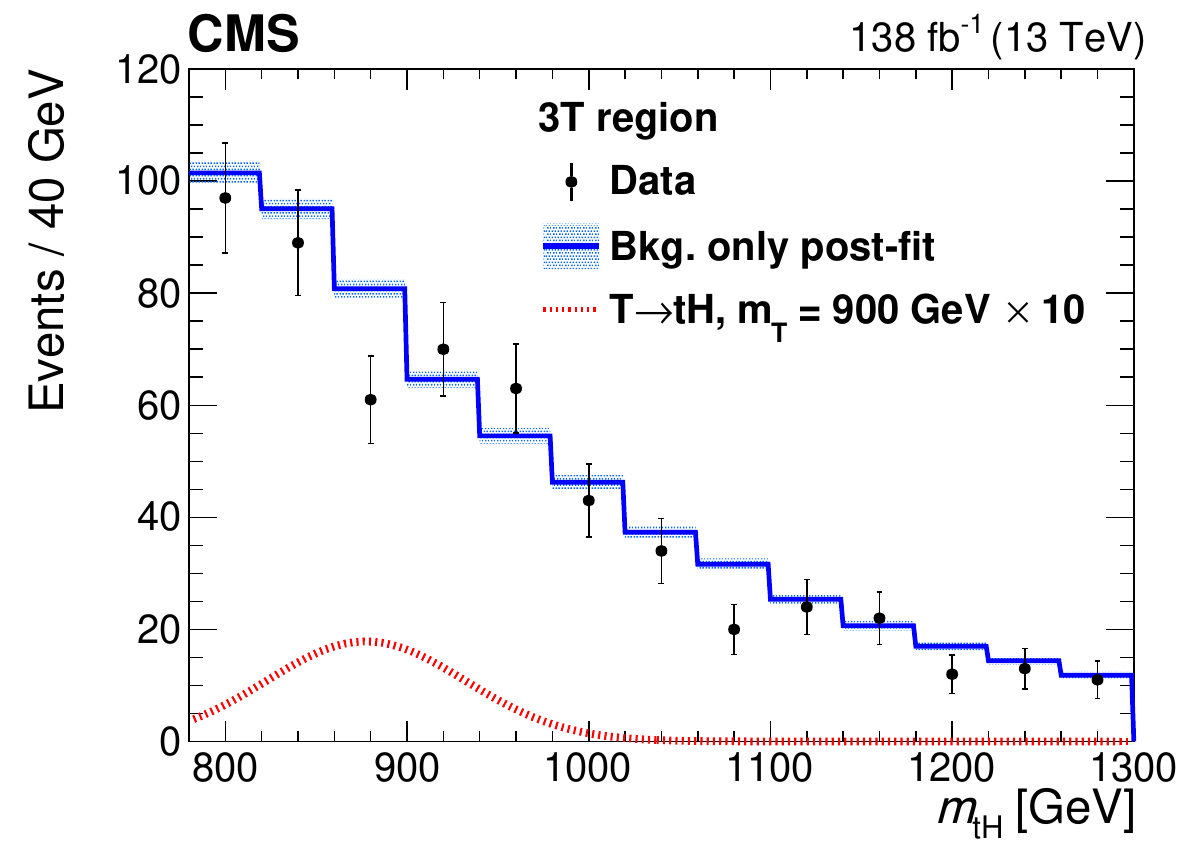}%
\caption{%
    Background-only postfit five-jet invariant mass distributions for the SR for the low-mass (left) and high-mass (right) selections.
    The shaded blue region represents the uncertainty in the fitted background estimate.
    The expected signal distributions (scaled for visibility) for a 700\GeV and a {900\GeV} \PQT quark are shown as red dashed lines for the low- and high-mass selections, respectively.
    Figures adapted from Ref.~\cite{B2G-19-001}.
}
\label{fig:postfit_tH}
\end{figure}

\cmsParagraph{$\tH\to\bqq\,\bb$} The full Run 2 search for singly produced \PQT quarks involving the Higgs boson decays to
\bbbar from Ref.~\cite{B2G-19-001} follows a similar strategy to that used for the \TtotZ, \Ztobb channel.
However, in the \TtotH channel, the reconstructed invariant mass of the \bbbar system must be greater
than 100\GeV, whereas for the \Ztobb channel, it must be less than 100\GeV.
The resulting postfit distributions of reconstructed \PQT mass sensitive observables in the observed
data based on the background-only hypothesis for \TtotH are shown in Fig.~\ref{fig:postfit_tH_tZ} (right), and in Fig.~\ref{fig:postfit_tH} for the low-mass (left) and high-mass (right) selections.

\begin{figure}[!ht]
\centering
\includegraphics[width=0.4\textwidth]{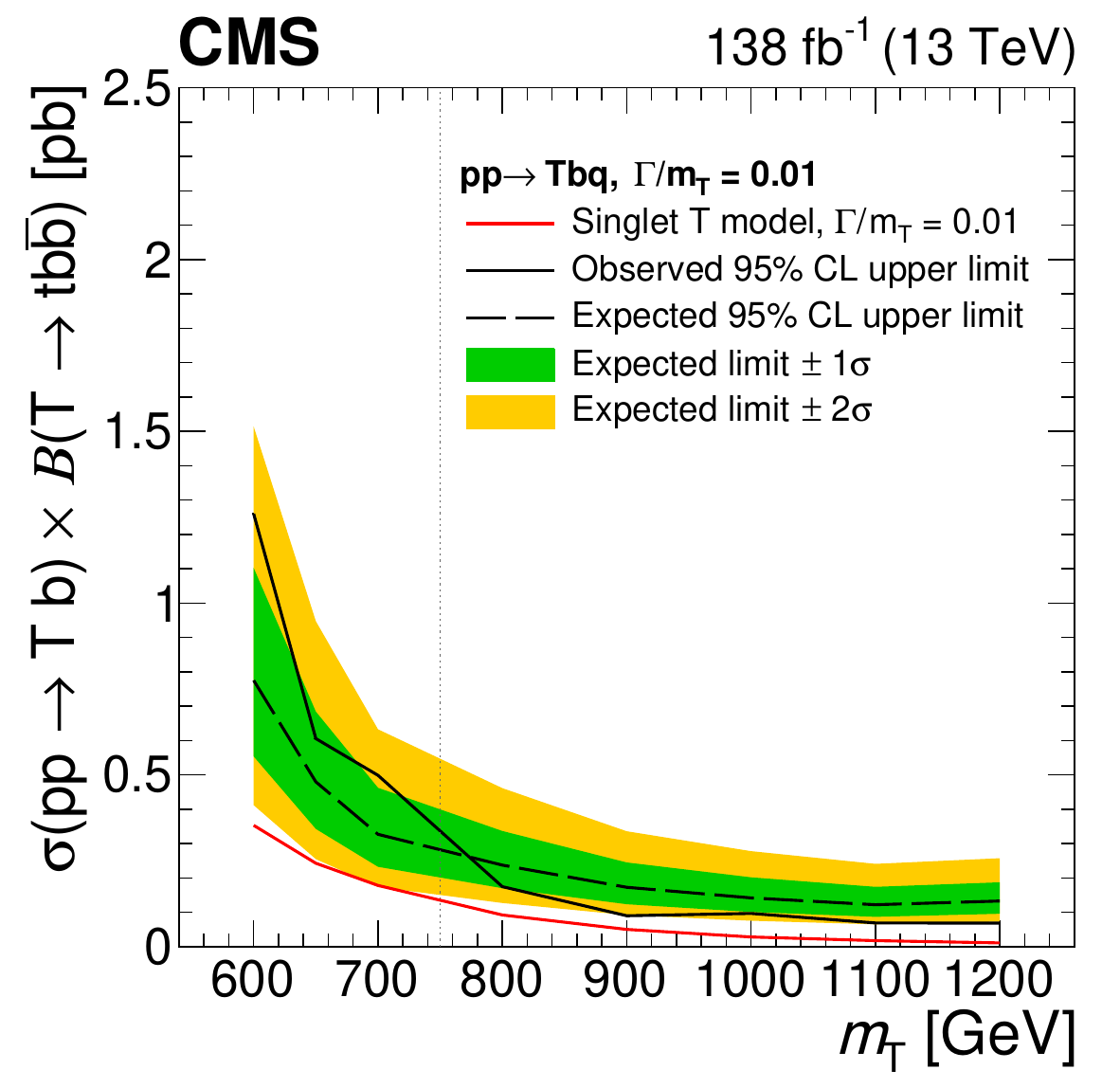}%
\hfill%
\includegraphics[width=0.56\textwidth]{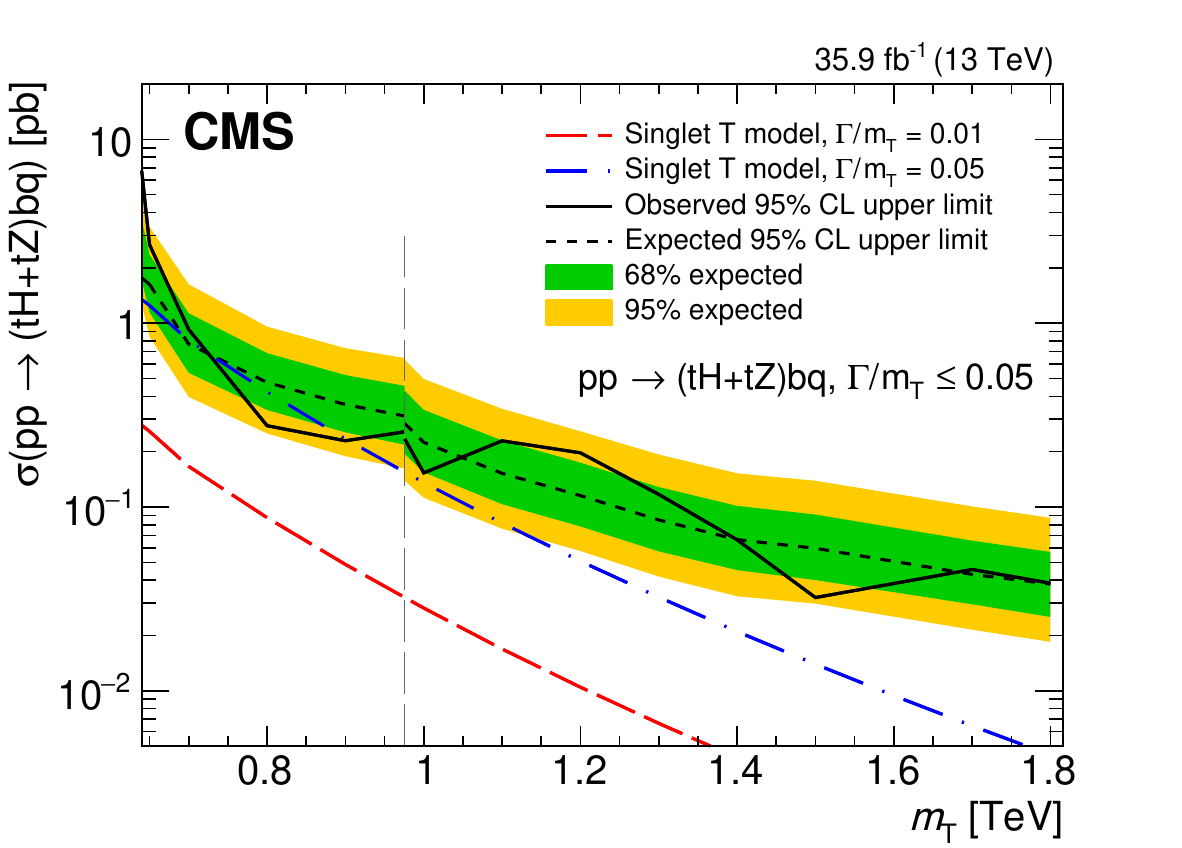}%
\caption{%
    Observed and median expected upper limits at 95\% \CL on the cross sections for single \PQT quark production associated with a \PQb quark, for the sum of \tHbq and \tZbq channels, as a function of the assumed values of the \PQT quark mass.
    The inner (green) band and the outer (yellow) band indicate the regions containing 68 and 95\%, respectively, of the distribution of limits expected under the background-only hypothesis.
    The left figure corresponds to the analysis strategy described in Ref.~\cite{B2G-19-001}, based on the five-jet invariant mass reconstruction of the \PQT.
    The figure on the right corresponds to the analysis strategy in Ref.~\cite{B2G-18-003}, which employs different reconstruction algorithms for the low- and high-mass searches.
    The vertical dashed lines represent the crossover point in sensitivity for the low-mass and high-mass selections.
    Figures adapted from Refs.~\cite{B2G-19-001, B2G-18-003}.
}
\label{fig:massLim_B2G19001}
\end{figure}

Upper limits are set on the cross section for the $\pp\to\Tbq$ production mode for the two decay channels (\tH and \tZ) individually as well as for their sum
({\tH}+\tZ). Both analysis strategies used in Refs.~\cite{B2G-18-003} and \cite{B2G-19-001}, described in Section~\ref{sec:TZlepVLQ}, derive the limits for a
singlet \PQT quark with a \GoM of 1\%. The low-mass search optimized in Ref.~\cite{B2G-19-001} gives a better sensitivity for \PQT quark masses below 1\TeV
whereas for higher masses the reconstruction based on large-radius jets used in Ref.~\cite{B2G-18-003} yields more stringent limits, as seen in Fig.~\ref{fig:massLim_B2G19001}.

\begin{figure}[!ht]
\centering
\includegraphics[width=0.48\textwidth]{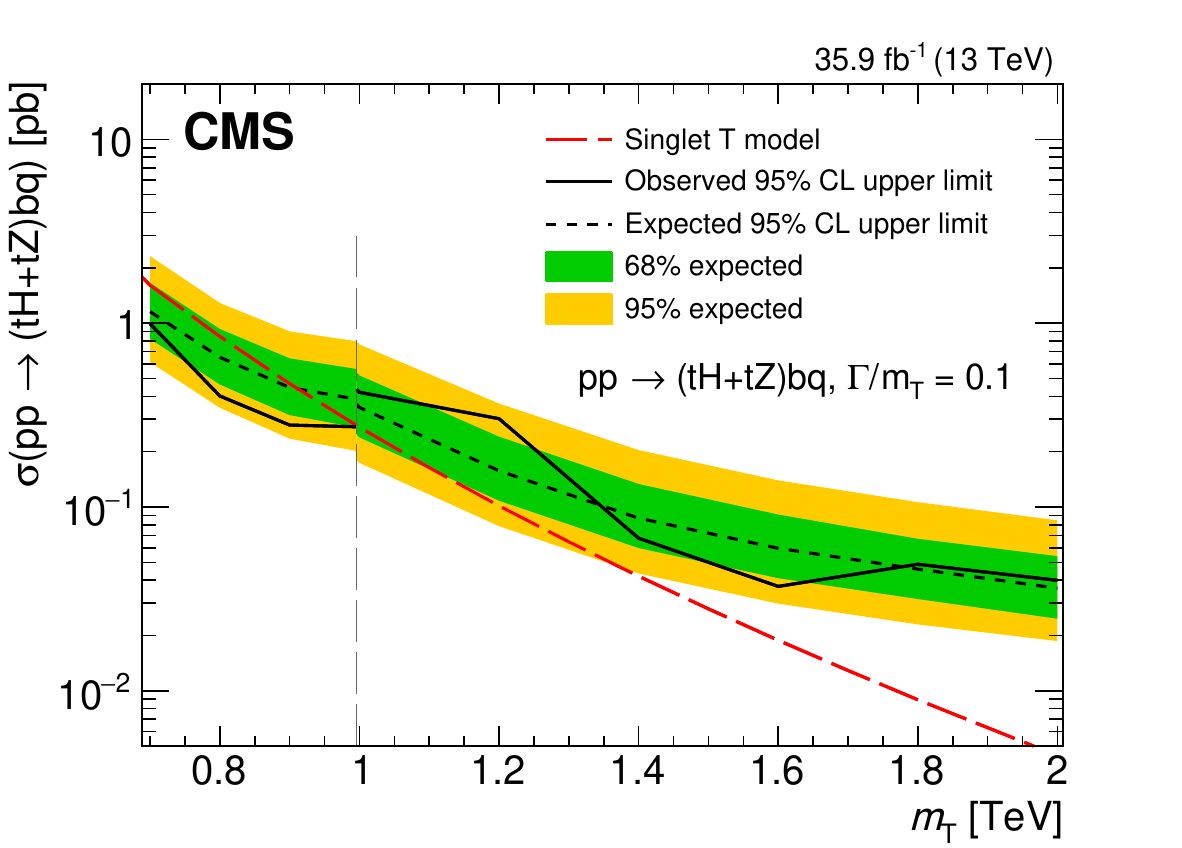}%
\hfill%
\includegraphics[width=0.48\textwidth]{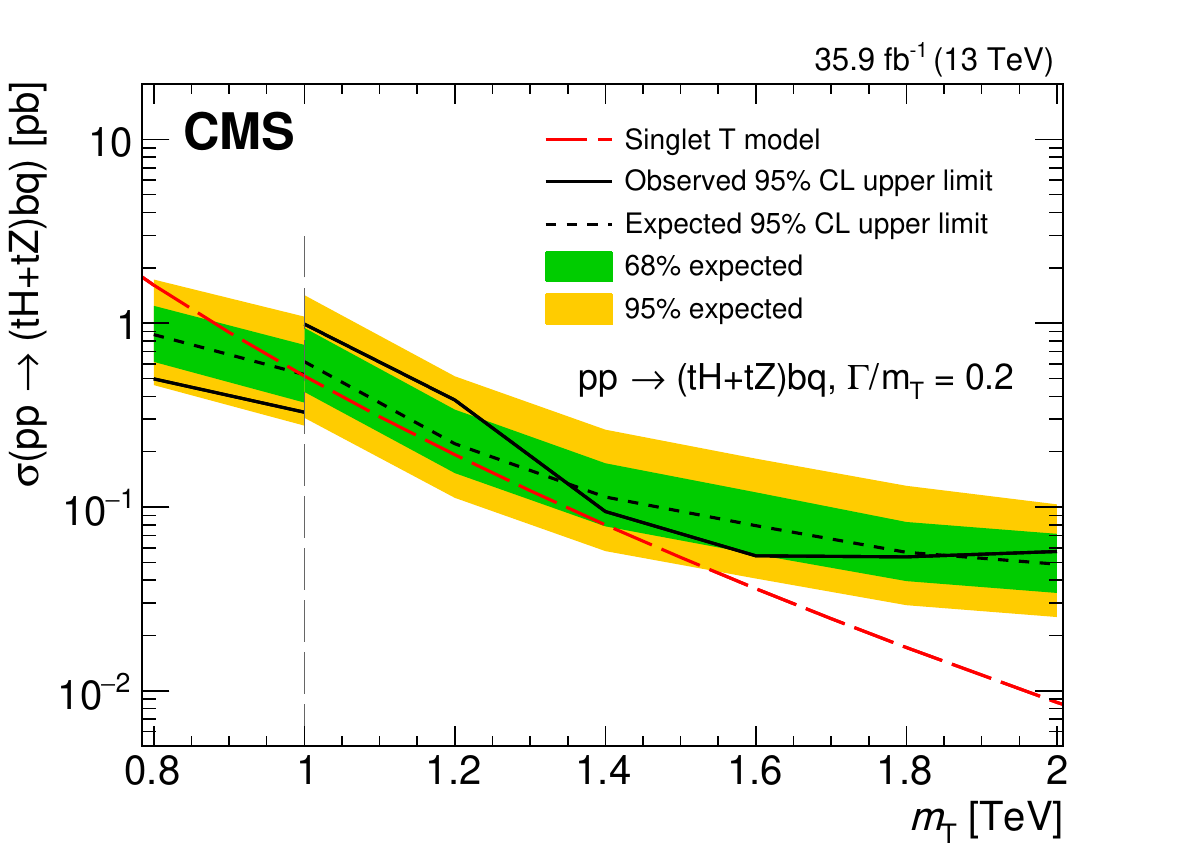} \\
\includegraphics[width=0.48\textwidth]{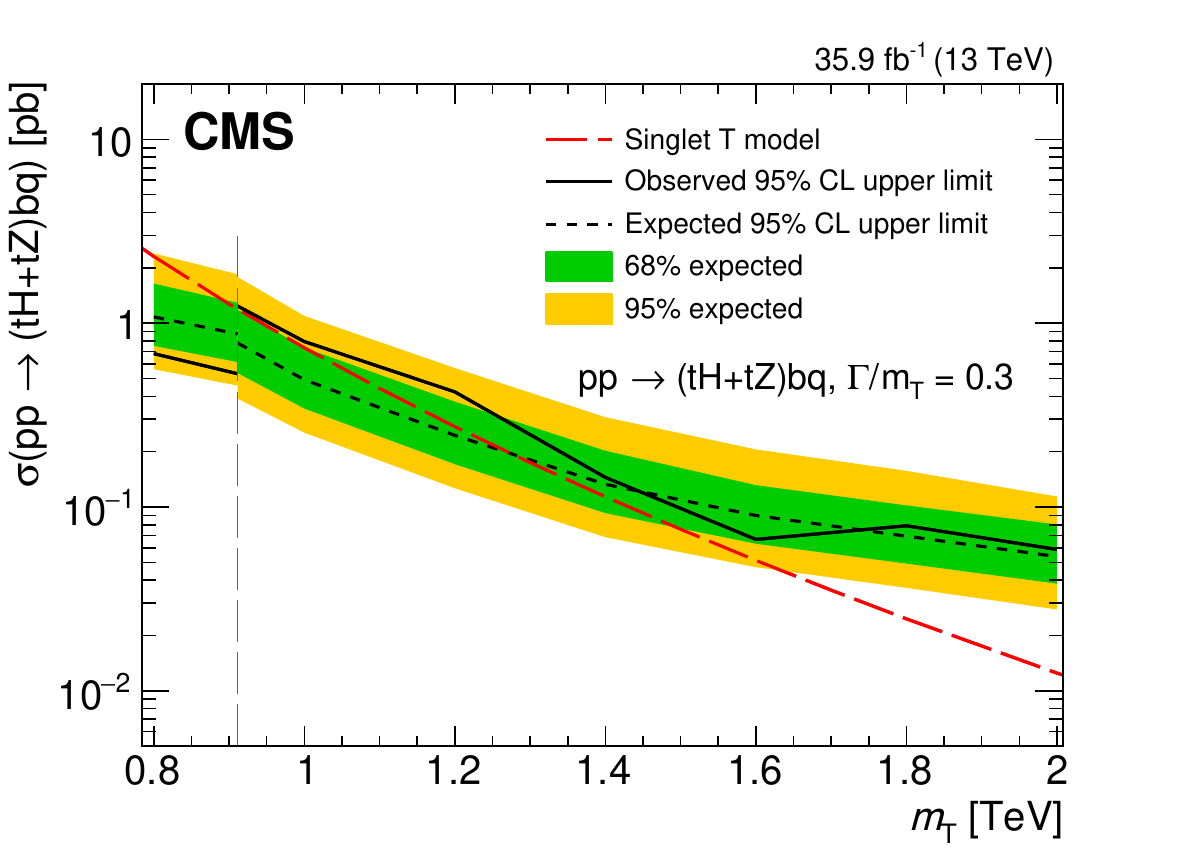}%
\caption{%
    Observed and median expected upper limits at 95\% \CL on the cross sections for single \PQT quark production associated with a \PQb quark, for the sum of \tHbq and \tZbq channels, as a function of the assumed values of the \PQT quark mass.
    The inner (green) band and the outer (yellow) band indicate the regions containing 68 and 95\%, respectively, of the distribution of limits expected under the background-only hypothesis.
    The results are given for relative widths of $\GoM=10$ (upper left), 20 (upper right), and 30\% (lower).
    The vertical dashed lines represent the crossover point in sensitivity for the low-mass and high-mass selections.
    Figures adapted from Ref.~\cite{B2G-18-003}.
}
\label{fig:massLim_Tbq1}
\end{figure}

Figure~\ref{fig:massLim_Tbq1} shows the upper limits on the cross section of \Tbq production after
combining the \tHbq and \tZbq channels.
The different figures correspond to different relative widths of 10, 20, and 30\%.
Similarly, in Ref.~\cite{B2G-18-003} analogous results for \Ttq production are presented
in the \tHtq and \tZtq channels, along with their sum. These results are also given for narrow relative
width $\GoM\leq5\%$ and relative widths of 10, 20, and 30\%.

For \PQT masses below 1\TeV, the models
describing $\pp\to\Tbq$ production are strongly constrained by the observed limits
from the low-mass search signature, which are generally more stringent than expected above 0.75\TeV;
for the \PQT singlet model, masses in the range 0.70 to 1\TeV are excluded at 95\% \CL for relative widths between 5 and 30\%.
The models corresponding to the associated production with a top quark
have lower cross sections with a median expected sensitivity for \PQT quark masses within
the ($\PQT\PQB$) doublet model of 0.82\TeV for the largest relative
width of 30\%. However, for this model, no range of masses is excluded at 95\% \CL for any
of the masses and relative widths considered here.

\cmsParagraph{$\tH\to\bqq\,\gammagamma$} All previously mentioned searches for singly produced \PQT quarks have primarily relied on the reconstructed \PQT quark invariant mass or transverse
mass as the primary observable to search for the presence of a signal.
However, the analysis in Ref.~\cite{B2G-21-007} is designed to utilize the high-resolution reconstruction of the Higgs
boson mass in the diphoton decay channel, with a precision of 1--2\%,
to search for a signal that exhibits a peak at the Higgs boson mass above the smoothly decreasing
diphoton mass background.
The analysis uses the full Run 2 data set, and is aimed to specifically target the detection of the two photons originating from
the decay of the Higgs boson resulting from the decay of the \PQT quark.
The event selection process involves the use of diphoton triggers, which require a minimum
of two photons with asymmetrical conditions on their transverse momenta: $\pTgone>30\GeV$
and $\pTgtwo>18$ or 22\GeV, depending on the data taking period.
Additionally, selection criteria~\cite{CMS:2021kom} are applied based on the shape of the
electromagnetic shower and based on the isolation in the calorimeter. The diphoton invariant mass (\mgg)
must be greater than 90\GeV to pass the HLT to ensure that both photons originate
from the PV.
For efficient selection of photons associated with the PV, a separate multivariate
analysis (MVA) known as the ``photon ID MVA''~\cite{CMS:2021kom}
is utilized in the offline event selection. This MVA
relies on observables such as the isolation and the shape of the photon shower in ECAL. The events
must have a minimum of two photons selected by the ID MVA within the ECAL and the
fiducial region of the tracker (with $\abs{\eta}<2.5$, excluding the
ECAL barrel-endcap transition region, $1.44<\abs{\eta}<1.57$). Additionally, the
photon pair has to meet the following criteria: the invariant mass of the photon pair must be between 100 and 180\GeV, the transverse momentum of the leading
photon (\pTgone) divided by \mgg must be greater than 1/3, and the transverse momentum
of the second photon (\pTgtwo) divided by \mgg must be greater than 1/4. If multiple
diphoton pairs exist, the pair with the highest $\pt(\gammagamma)$ is chosen~\cite{CMS:2021kom}.
The events are categorized based on the leptonic or hadronic decays of the top quark.
Events including a pair of photons, at least one electron or muon, and a \PQb-tagged jet, are categorized as the leptonic category. Those with zero leptons, a pair of photons and three jets, including at least one \PQb-tagged jet, are assigned to the hadronic channel.
Events with two leptons from the DY processes are rejected.

\begin{figure}[b!]
\centering
\includegraphics[width=0.48\textwidth]{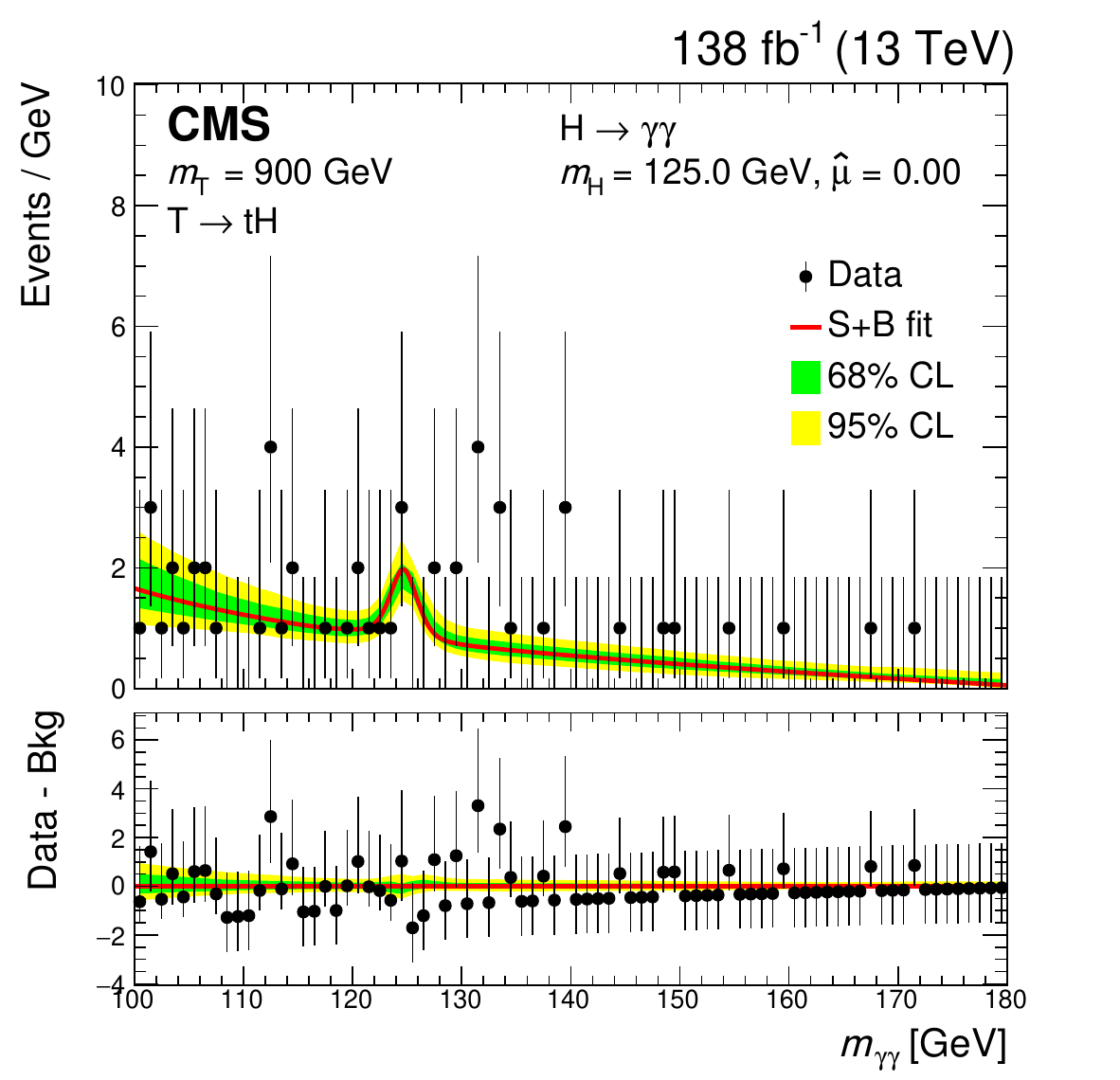}%
\hfill%
\includegraphics[width=0.48\textwidth]{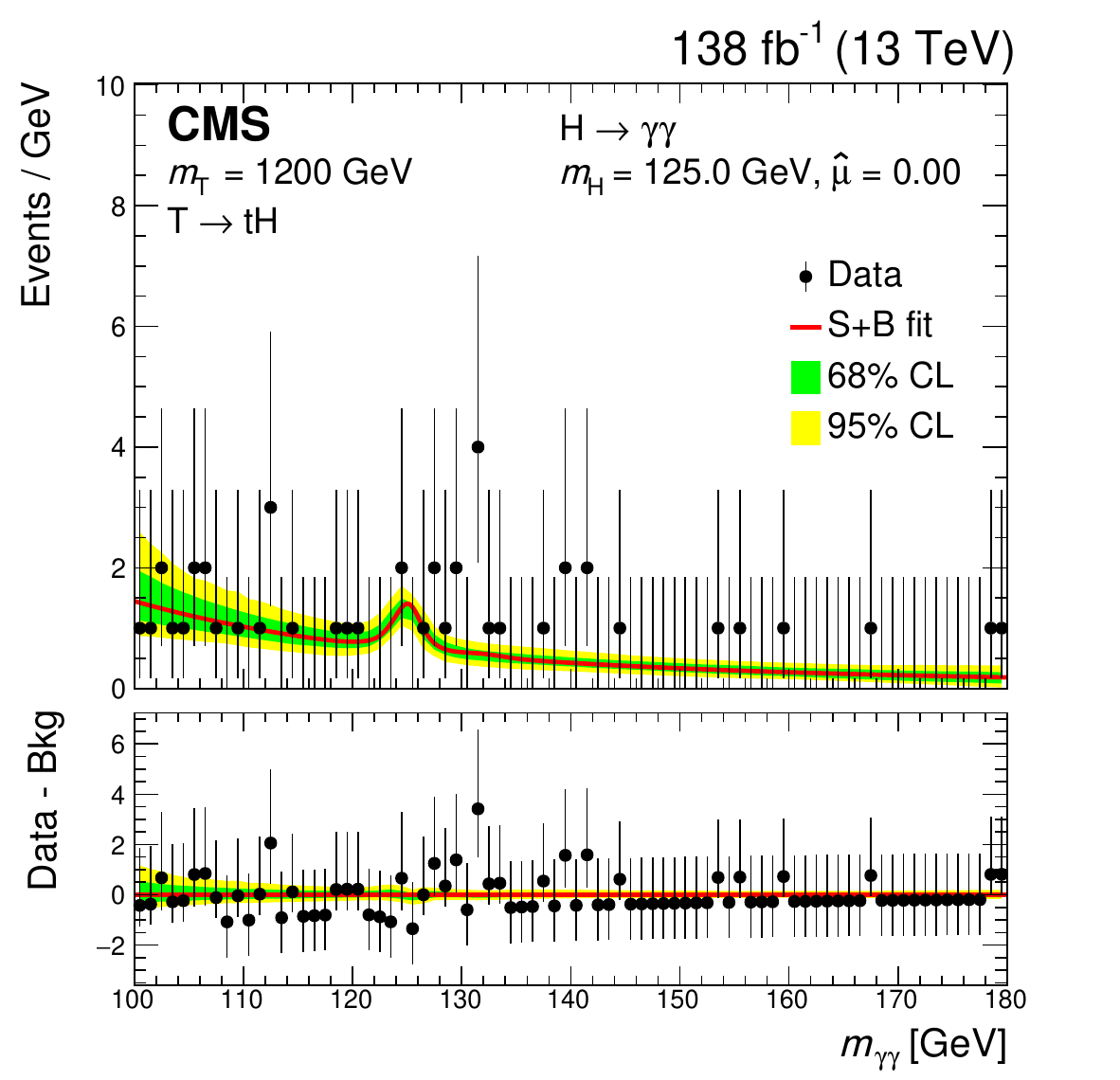}%
\caption{%
    Distributions of the observed data (black dots) and \mgg signal-plus-background model fits (red line) for a \PQT quark signal with \mQT of 900 (left) and 1200\GeV (right), combining the leptonic and hadronic channels.
    The green (yellow) band represents the 68 (95)\% \CL interval in the background component of the fit.
    The peak in the background component shows the considered irreducible SM Higgs boson contribution ($\Pg\Pg\PH$, VBF, $\PV\PH$, \ttH, and \tH).
    Here, $\hat{\mu}$ is the best fit value of the signal strength parameter $\mu$, which is zero for the two \mQT values considered.
    The lower panel shows the residuals after the subtraction of the background component.
    Figures adapted from Ref.~\cite{B2G-21-007}.
}
\label{fig:result_SplusB}
\end{figure}

At this level of the event selection, Higgs boson production associated with a top quark pair (\ttH) with \Htogg
is the dominant background among all SM Higgs boson production
processes, since it also leads to a peak in the \mgg spectrum at the Higgs boson mass.
The \mgg spectrum from the \PQT quark signal also peaks at \mH due to the \TtotH decay.
To separate the \PQT signal from the SM Higgs boson background processes,
MVA discriminants based on BDTs are implemented~\cite{hoecker_tmva_2009}
separately for each category (BDT-SMH). Furthermore, an additional BDT (BDT-NRB) is trained to
suppress the sizeable nonresonant background contributions (\tZ, \ttX, \Wgamma, QCD multijet, {\PGg}+jets,
and {\gammagamma}+jets and \Zgamma) in the hadronic category.
Higgs bosons from both SM processes and \PQT decays are expected
to peak on a smoothly falling \mgg distribution in the range $100<\mgg<180\GeV$. Models of the
signal and \ttH background processes are obtained by fitting the
\mgg distributions in simulation with a sum of at most five Gaussian functions,
separately for each category. The models used to describe the nonresonant background processes are extracted
from the observed \mgg spectrum in the region $\mgg\in[100,180]\GeV$ using a discrete profiling method~\cite{envelope}.
This technique estimates the systematic uncertainty in the background estimate associated
with choosing a particular analytic function to describe the \mgg spectrum.
The chosen functions are from a list of families of functions:
exponentials, power laws, polynomials, and Laurent series~\cite{envelope}.
However, the degrees of freedom for these functions are decided in each case using
a detailed $\mathcal{F}$-test~\cite{fisher_1922} with a loose requirement on the goodness of fit.
Figure~\ref{fig:result_SplusB} displays the observed data distributions, with the corresponding signal-plus-background model fit to the \mgg distribution, for \mQT values of 900 and 1200\GeV.

\begin{figure}[ht!]
\centering
\includegraphics[width=0.75\textwidth]{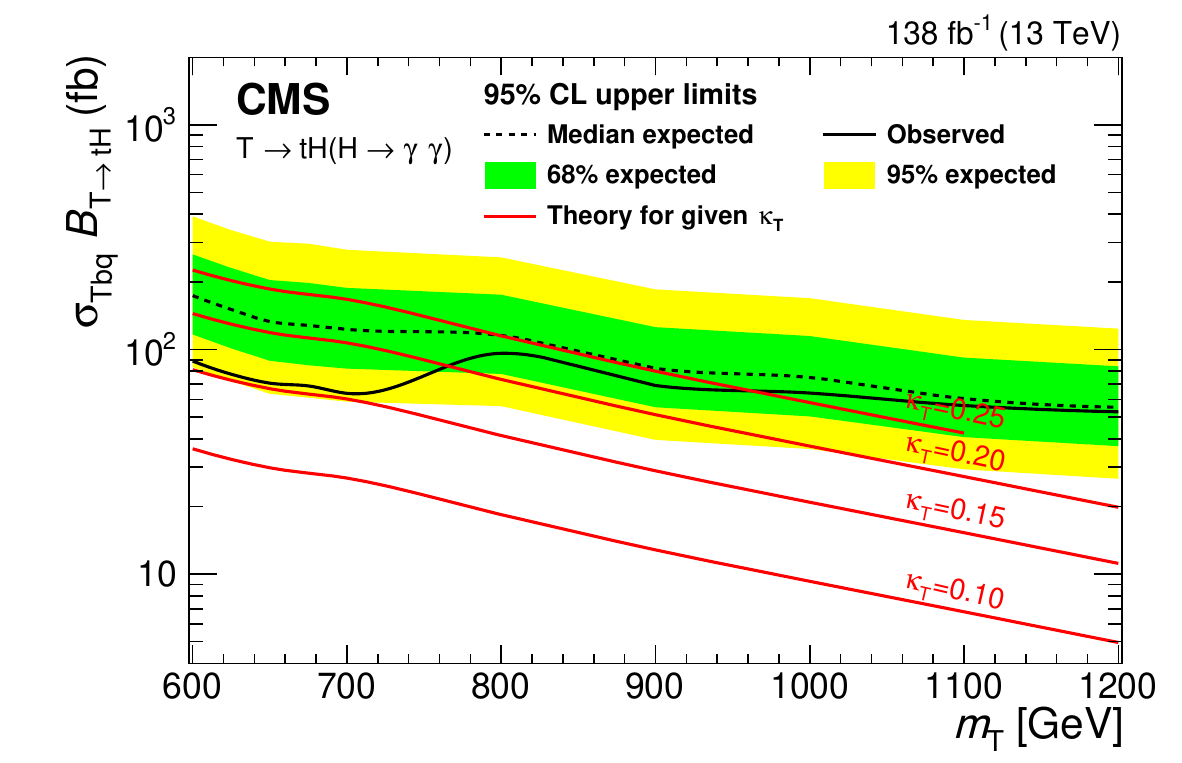}
\caption{%
    Expected (dotted black) and observed (solid black) upper limits at 95\% \CL on $\sigma_{\Tbq}\BR(\TtotH)$ are displayed as a function of \mQT, combining the leptonic and hadronic channels.
    The inner (green) band and the outer (yellow) band indicate the regions containing 68 and 95\%, respectively, of the distribution of limits expected under the background-only hypothesis.
    The theoretical cross sections for the singlet \PQT production with representative \kappaT values fixed at 0.1, 0.15, 0.2, and 0.25 (for $\GoM<5\%$) are shown as red lines.
    Figure adapted from Ref.~\cite{B2G-21-007}.
}
\label{fig:result_xsec_combined}
\end{figure}

No statistically significant excess above the SM backgrounds in any channels or mass ranges is
observed. Upper limits on the signal strength modifiers
$\mu_{\text{obs}} = \sigma_{\text{obs}}/\sigma_{\text{th}}$ and $\mu_{\text{exp}} = \sigma_{\text{exp}}/\sigma_{\text{th}}$, are derived for different assumed
\mQT values, using a maximum likelihood fit of the \mgg distributions, keeping the \mH parameter
of the model fixed at 125\GeV.
Finally, the upper limits on $\mu_{\text{obs}}$ and $\mu_{\text{exp}}$ are translated into upper limits on $\sigma_{\Tbq}\BR(\TtotH)$, as displayed in Fig.~\ref{fig:result_xsec_combined}
together with the theoretical cross sections for the singlet \PQT production process with
representative \kappaT values fixed at 0.1, 0.15, 0.2, and 0.25 (for $\GoM<5\%$).
Assuming a coupling to third-generation quarks of $\kappaT=0.25$ and a relative decay width of
$\GoM<5\%$, the EW production of a singlet \PQT quark is excluded up to a mass of 960\GeV at 95\% \CL.

\subsubsection{\texorpdfstring{$\PQT/\yft\to\bW$}{T/Y to bW}}
\label{sec:TYbW}

\cmsParagraph{$\bW\to\PQb\Pell\PGn$} A search for singly produced VLQs that decay to \bW, sensitive to both \PQT and \yft quarks, is performed in the \ljets channel using the 2015 data set~\cite{B2G-16-006}.
Events are selected if they contain exactly one charged lepton (electron or muon), with $\pt>30\GeV$ and $\abs{\eta}<2.1$.
The presence of at least two small-radius jets is required, one \PQb-tagged jet ($\pt>200\GeV$ and $\abs{\eta}<2.4$) and one forward jet ($\pt>30\GeV$ and $2.4<\abs{\eta}<5.0$), as well as $\ptmiss>50\GeV$ to account for the neutrino from the \PW boson decay.
Additional event selection criteria are imposed to suppress the contribution from the dominant background processes, \ttbar and \wjets; the transverse mass \MT of the lepton-\ptmiss system is required to satisfy $\MT<130\GeV$ whereas the scalar sum \ST of the transverse momenta of the lepton, \PQb jet, and \ptmiss is required to be $\ST>500\GeV$.
The VLQ signal is expected to show as an excess in the invariant mass \minv distribution reconstructed from the lepton, the \PQb jet, and the neutrino four-momenta.
A binned likelihood fit is performed to the observed \minv spectrum.
No significant deviations from the SM predictions are observed. Upper limits at 95\% \CL are set on the cross sections for the single production of \yft and \PQT in the mass range from 0.7 to 1.8\TeV, assuming a narrow decay width and $c_{\PW}=0.5$.
In this model, for a $\yft\to\bW$ branching fraction of 100\%, \yft masses between 0.85 and and 1.40\TeV are excluded at 95\% \CL.
Similar exclusion limits are achieved for \PQT also for a branching fraction \TtobW of 100\%.

\subsubsection{\texorpdfstring{\PQB}{B} quark production}
\label{sec:BVLQ}

\cmsParagraph{$\tW\to\bqq\,\lnu$ or $\blnu\,\qq$} A search for $\PQB/\xft\to\tW$ was carried out in the \ljets
channel~\cite{B2G-17-018} using the 2016 data set.
The charged lepton may originate either from
a $\PQt\to\Wb\to\lnu\PQb$ decay, or from the \PW boson from the $\PQB/\xft$ decay.
Leptons (electrons or muons) are selected with $\pt>55\GeV$, and are identified with a two-dimensional
isolation requirement in order to achieve a high selection efficiency for decays of high Lorentz boosted
\PQt quarks.
The analysis uses \PW and \PQt tagging, based on the SD jet mass, \tauTO,
\tauTT, and subjet \PQb tagging. Selected events are attributed to five
categories, defined by the presence of either a \PQt tag, a \PW tag, or two, one, or no
\PQb-tagged small-radius jets. 
Large-radius jets are used to identify the hadronic decays of highly boosted top quarks and \PW bosons. For top quark jets and \PW boson jets, a \pt greater than 400 GeV and 800 GeV is required, respectively.
In the \PQt tag category, the VLQ mass ($m_{\mathrm{reco}}$) is reconstructed
from the four-momenta of the \PQt-tagged jet, the charged lepton, and \ptvecmiss. In all
other categories, it is reconstructed using combinations of small-radius jets, where the
best combination is chosen based on a \chisq estimator.
The data sample is divided into an SR with a forward jet and a CR without one. The background distribution in the reconstructed VLQ mass
in the SR is estimated from the corresponding distribution in the
CR. This allows for a background estimation from data of all
SM backgrounds in this search. Residual differences in the shapes of these
distributions in the signal and CRs may arise from different
background compositions due to the presence of a forward jet. The observed
differences are small, with average values of 10\%, and are corrected
by using factors derived from simulation.

\begin{figure}[!ht]
\centering
\includegraphics[width=0.41\textwidth]{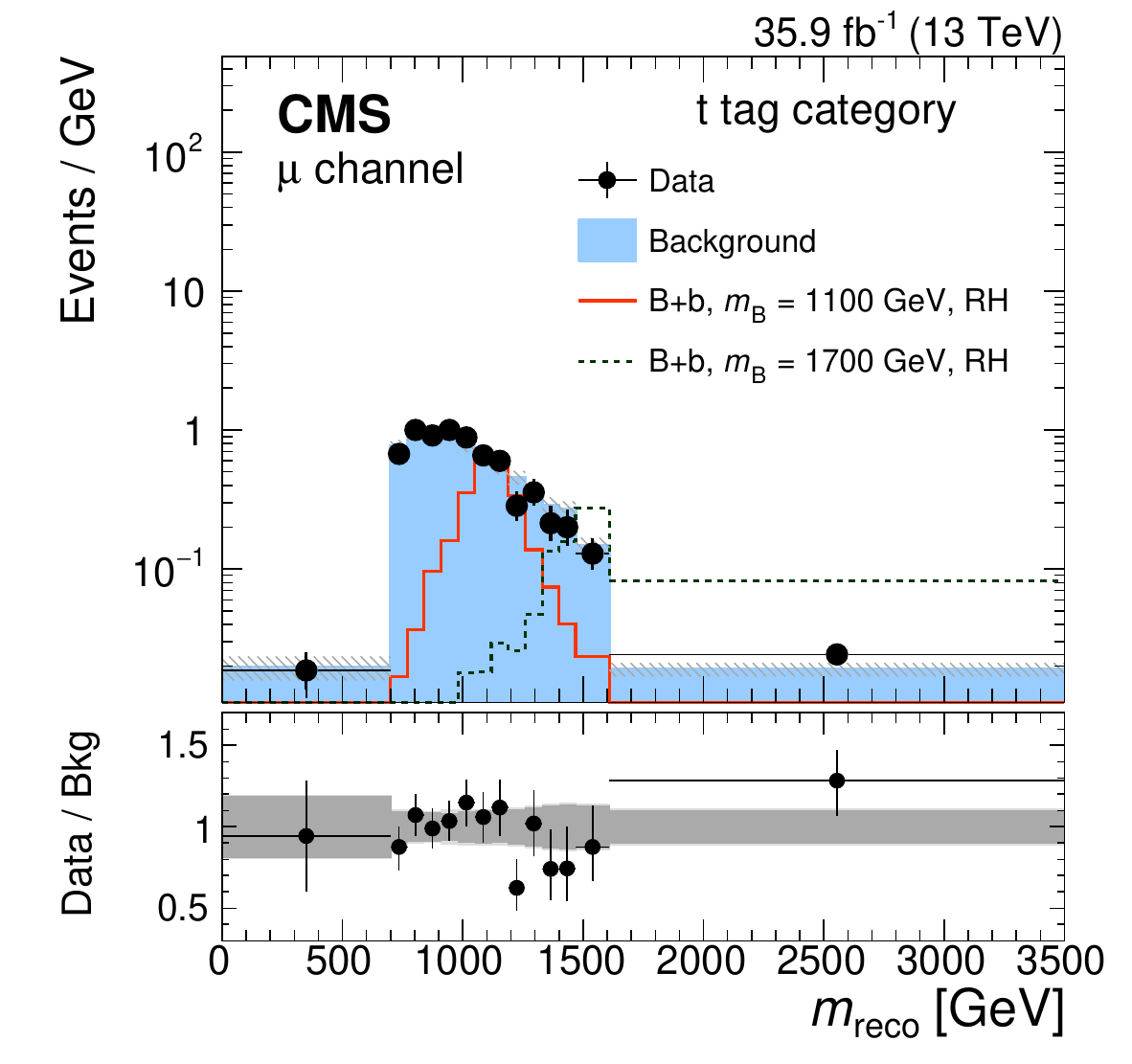}%
\hfill%
\includegraphics[width=0.55\textwidth]{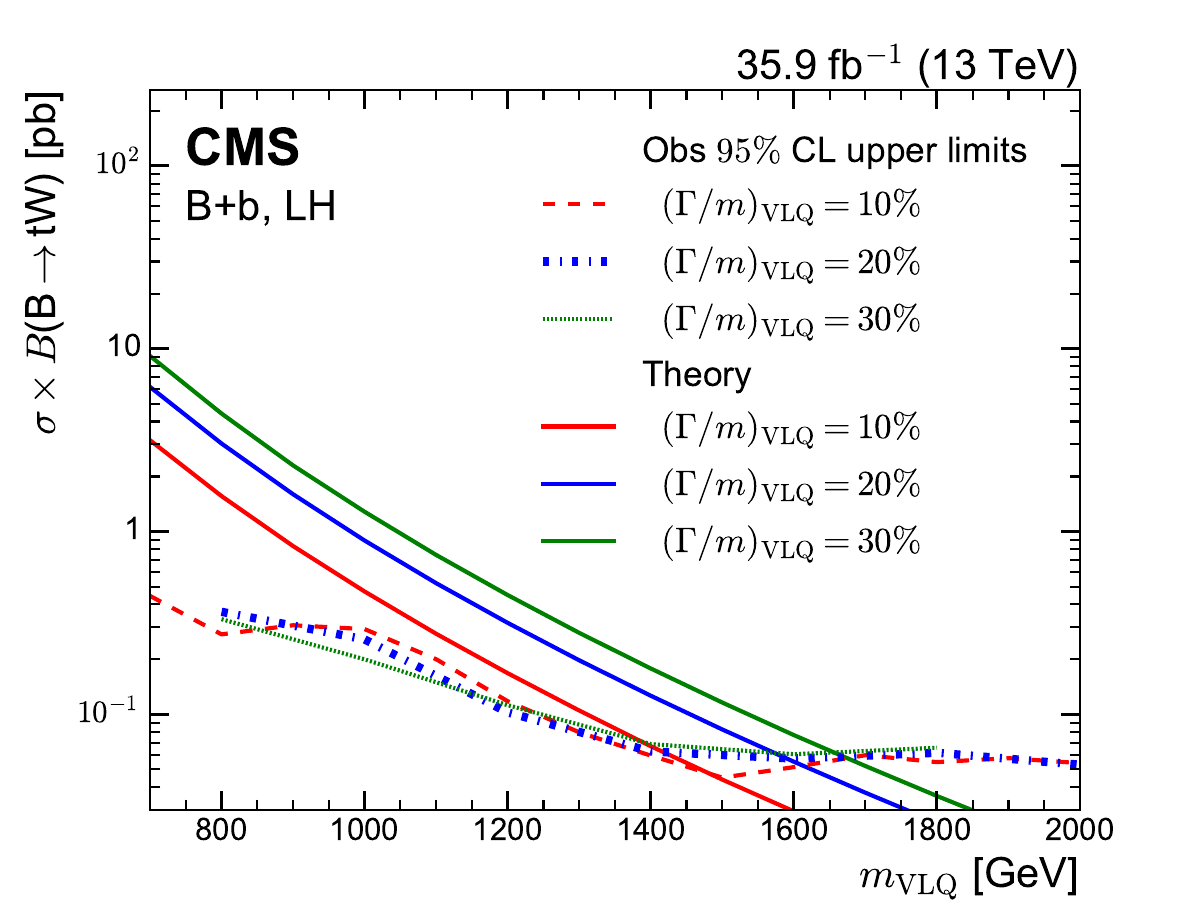}%
\caption{%
    The distribution in the reconstructed \PQB quark mass in events with one \PQt-tagged jet and a forward jet, where the SM background is obtained from a CR without a forward jet (left).
    The product of the observed upper limits on the cross section and \BrBtW as a function of \mvlq for different relative decay widths of the \PQB quark (right), for single \PQB quark production in association with a \PQb quark.
    Figures taken from Ref.~\cite{B2G-17-018}.
}
\label{fig:cms_BtW}
\end{figure}

The distribution of the reconstructed \PQB mass is shown in Fig.~\ref{fig:cms_BtW} (left)
for the \PQt tag category in the {\PGm}+jets channel. The signal distributions for a
\PQB quark with RH couplings, produced in association with a \PQb quark are shown as well,
for two different values of \mvlq with an assumed production cross section of 1\unit{pb}.
Upper limits on the product of the cross section and \BrBtW on the single--production processes \Bb and
$\xft\PQt$ are derived by combining the five categories measured in the muon and electron channels.
The observed (expected) upper limits for \Bb production with LH couplings
and in the NWA are between 0.04 (0.04)\unit{pb} and 0.3 (0.2)\unit{pb} for
$\mvlq=1.8$ and 1.0\TeV, respectively. A comparison of the observed exclusion limits of \Bb production with LH couplings for
relative VLQ widths of 10, 20, and 30\% is shown in Fig.~\ref{fig:cms_BtW} (right).
Similar exclusion limits are obtained for $\xft\PQt$ production.

A more recent search using data the full Run 2 data set
targets the single production of an excited \PQb quark (\PQbst)~\cite{B2G-20-010}. The dominant decay
through the weak interaction, $\PQbst\to\tW$, results in the same final state as for a \PQB quark.
However, because the \PQbst is predominantly produced through the strong force, no forward jet
is expected in this search, and thus the analysis is designed to be inclusive in the number of forward jets.
The signal is reconstructed with an isolated lepton from the \PW boson decay, and with a \PQt-tagged
jet. The analysis uses the \HOTVR algorithm~\cite{Lapsien:2016zor} for the reconstruction and
identification of the boosted \PQt quark. The variable size of the \HOTVR jets allows for an efficient
reconstruction of boosted \PQt quarks starting at $\pt>200\GeV$~\cite{CMS:2020poo}, such that a
single analysis strategy may cover mass range from 0.7 up to 4\TeV. Two signal categories
are defined, based on the number of \PQb-tagged small-radius jets. The 1\PQb category shows the
highest sensitivity, whereas the 2\PQb category serves to constrain the dominant background
from \ttbar production. The background from processes without \PQt quarks, originating from misidentified \PQt jets,
is obtained from data by an extrapolation from events with no \PQb-tagged jets.
The distribution in the reconstructed mass of the \tW system is used to set upper limits on the product of the cross section and branching fraction.

\cmsParagraph{$\tW\to\bqq\,\qq$} The $\PQbst\to\tW$ signal is also searched for in an analysis in the all-hadronic final state~\cite{CMS:2021iuw} using the full Run 2 data set.
The analysis uses two large-radius jets with $\pt>400\GeV$ and a $\Delta\phi > \pi/2$ to ensure a back-to-back topology
of the two jets. Because of the high jet \pt threshold, the analysis is only sensitive to masses
larger than 1.2\TeV.
The SR is defined by the presence of a \PW- and a \PQt-tagged jet~\cite{Larkoski:2014wba}.
The analysis is performed in two dimensions, where the distribution in the plane
$(\mQt,\mtW)$, with \mQt the reconstructed top quark mass, is probed for a potential signal.
This allows for the use
of a novel method to construct the multijet background template, which relies on
a parametrization of the pass-fail ratio as a function of \mQt.
The distribution in the $(\mQt,\mtW)$ plane of multijet events passing the \PQt-tagging requirement
is calculated by multiplying the distribution of multijet events failing the requirement by a pass-fail ratio.
The two-dimensional pass-fail ratio is obtained from data in the sideband regions, and forms a surface
parametrized by the product of a second-order polynomial in \mQt and a first-order polynomial in \mtW.
The advantage of this method is that it interpolates the pass-fail ratio into the SR from the
enclosing sidebands, such that the analysis may be fully tested and verified
before including the observed data in the SR.
Once the observed data in the SR is examined, the predicted pass-fail ratio may
be compared with the observed one to validate the multijet background estimation in the
SR. Besides the multijet background, \ttbar production is an important background
in this search as well. In order to validate the modeling of this background,
a dedicated CR is included.

\begin{figure}[!ht]
\centering
\includegraphics[width=0.48\textwidth]{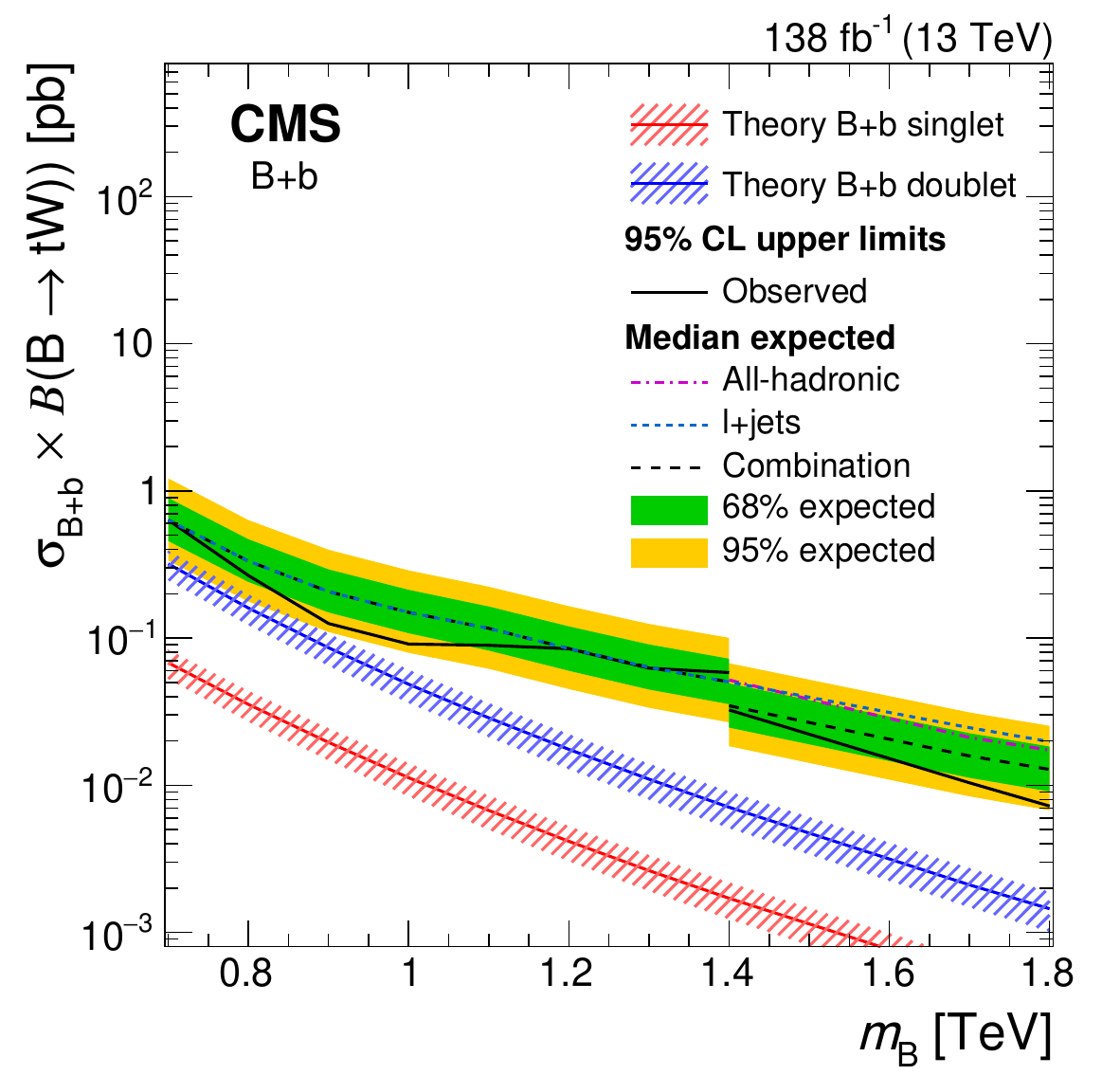}%
\hfill%
\includegraphics[width=0.48\textwidth]{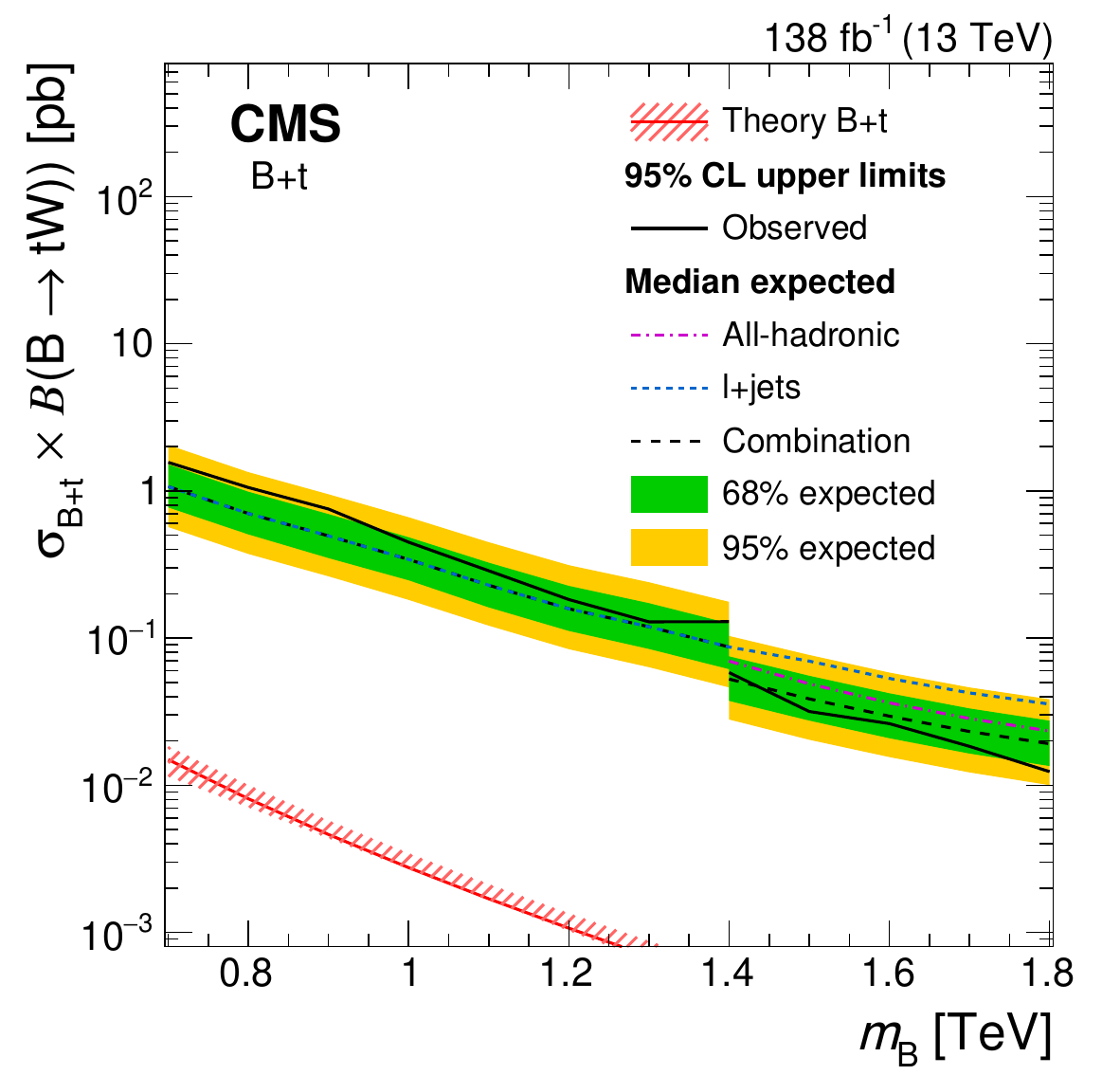}%
\caption{%
    Upper limits on the product of the production cross section and branching fraction to \tW of the \Bb (left) and \Bt (right) production modes at 95\% \CL.
    Colored lines show the expected limits from the \ljets (dotted) and all-hadronic (dash-dotted) channels, where the latter start at \PQB masses of 1.4\TeV.
    The observed and expected limits from the combination are shown as solid and dashed black lines, respectively.
    The inner (green) band and the outer (yellow) band indicate the regions containing 68 and 95\%, respectively, of the distribution of the limits expected under the background-only hypothesis.
    The theoretical cross sections are shown as the red and blue lines, where the uncertainties due to missing higher orders are depicted by shaded areas.
    Figures adapted from Ref.~\cite{B2G-20-010}.
}
\label{fig:bprime_limits}
\end{figure}

The analyses in the \ljets and all-hadronic final states are combined~\cite{B2G-20-010} to obtain upper limits on the product of production cross section and branching fraction for \BtotW, as shown in Fig.~\ref{fig:bprime_limits}.
In the mass range probed by both analyses, between 1.4 to 1.8\TeV, very
similar sensitivity is observed, resulting in a combined limit significantly stronger than
the limits from the individual analyses.
The limits in the range 0.7--1.4\TeV are obtained from the \ljets
analysis only.
Compared with the previous analysis in the \ljets channel~\cite{B2G-17-018}, the exclusion limits
in this mass range are about 10--30\% better for \Bb production. At high \mQB,
where both analyses contribute, the limits are up to 75\% better.
For \Bt production, the limits at low masses cannot be improved because of the
second \PQb jet, which results in most signal events being reconstructed in the 2\PQb
category of the \ljets search. At high \mQB, the combination improves the
previous limits by about 50\%.

\begin{figure}[!ht]
\centering
\includegraphics[width=0.75\textwidth]{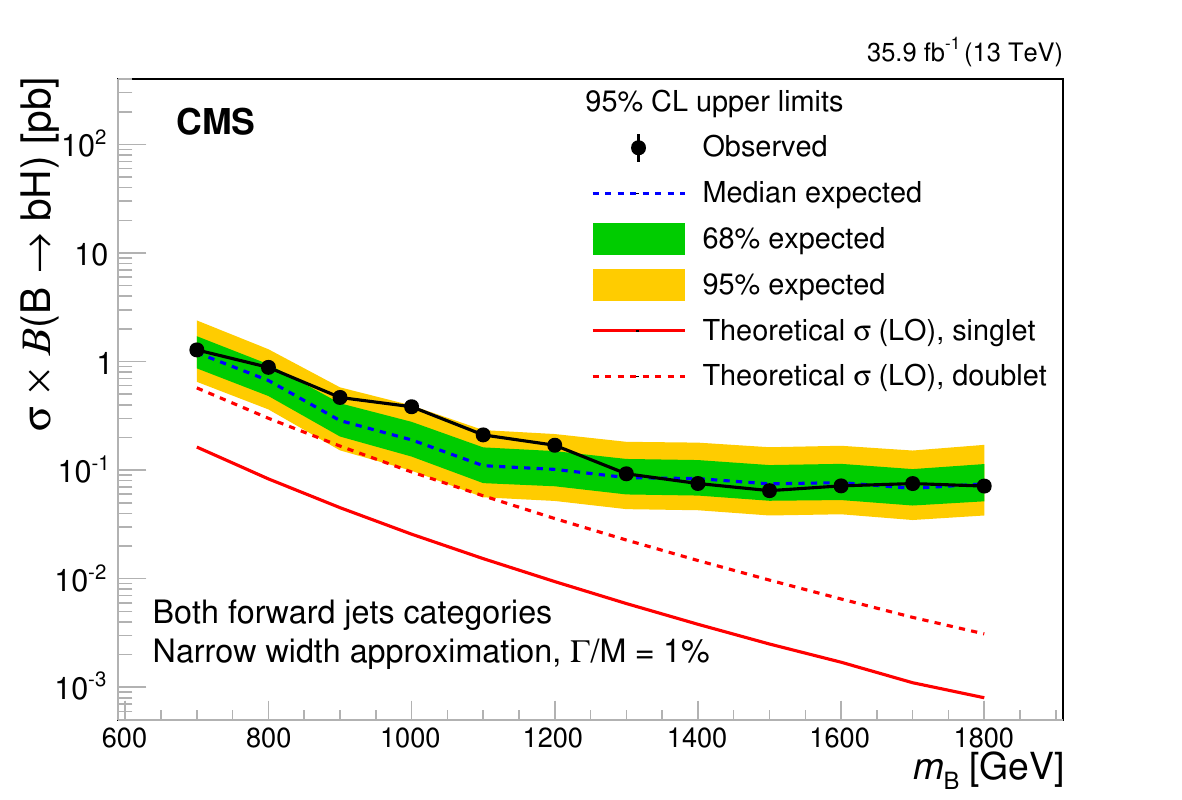}
\caption{%
    Observed and expected 95\% \CL upper limits on the product of the \PQB quark production cross section and branching fraction to \bH, as a function of the signal mass, under the NWA.
    The results are shown for the combination of 0 and $>$0 forward-jet categories.
    The continuous red curves correspond to the theoretical expectations for singlet and doublet models.
    Figure taken from Ref.~\cite{B2G-17-009}.
}
\label{fig:limits_ww}
\end{figure}

\cmsParagraph{$\bH\to\PQb\,\bb$} The CMS Collaboration has also performed a search targeting the single production of \PQB quarks with decays to \bH,
based on the 2016 data set~\cite{B2G-17-009}.
The search is carried out in the all-hadronic final state, and is optimized for the
\BtobH channel, where the \Htobb decay is reconstructed by an
\PH-tagged large-radius jet. The \PH jets are defined by the pruned jet
mass in the range 105--135\GeV and by two \PQb-tagged subjets, where the
subjets have been obtained with the SD algorithm.
Events in the SR require an \PH jet balanced by a high-\pt \PQb-tagged
small-radius jet.
Trigger requirements lead to a selection of $\HT>950\GeV$,
calculated from all small-radius jets with $\pt>30\GeV$.
Events are sorted into four categories, based on the presence of a forward jet and
the value of \HT.
The low-mass category with $\HT<1250\GeV$ shows higher sensitivity for
signals with $\mQB<1500\GeV$, whereas the multijet background is reduced in
the high-mass category by imposing $\HT>1250\GeV$, resulting in a better sensitivity
for signals with $\mQB>1500\GeV$. The main background in this search is
multijet production, with only 5--7\% from \ttbar production. Other SM processes
give negligible contributions. The multijet background is estimated from three
sideband regions, obtained by requiring only one \PQb-tagged subjet and/or
changing the SD jet mass to $75<\mjet<105\GeV$ or $\mjet>135\GeV$.
For a reliable extrapolation of the background to the SR,
the subjet \PQb-tagging has to be uncorrelated from the
SD jet mass, which has been verified using simulation.
The analysis excludes the products of cross sections and branching fractions above 0.07 and
0.4\unit{pb} for $\mQB=1.8$ and 1\TeV, respectively. The limits worsen by
factors between 1.3 for $\mQB=1\TeV$ and 2.1 for $\mQB=1.8\TeV$, when increasing the
relative decay width of the \PQB quark from 1\% (NWA) to 30\%.
The observed and expected 95\% \CL upper limits for the product of the production cross section of \PQB and the branching fraction to \bH in the NWA hypothesis, is shown in Fig.~\ref{fig:limits_ww}, whereas the
similar plots for the production of \PQB under the larger width approximations can be found
in Ref.~\cite{B2G-17-009}.

\subsection{Production through heavy resonance decays}
\label{sec:resVLQ}

Several models predict new heavy spin-1 resonances, generically denoted as the \PZpr and \PWpr bosons, that may decay into
VLQs.
In cases where the mass of the heavy resonances would be larger than 2\mvlq,
decays into pairs of VLQs are allowed. These may occur with branching fractions of 60\% and higher~\cite{Chala:2014mma}.
The additional decay channel results in large decay widths of the heavy resonances, such that
exclusion limits on pair production of VLQs may be reinterpreted to probe these models~\cite{Araque:2015cna}.
If the mass of the heavy resonance is smaller than 2\mvlq but larger than
$\mvlq+\mQq$, ``heavy-light'' decays into a VLQ and a SM quark with mass \mQq may have sizeable
branching fractions~\cite{Liu:2018hum}.
For the electrically charged resonance \PWpr, these heavy-light decays include
$\PWpr\to\Tbbar$ and $\PWpr\to\PQt\PAQB$,
complementing the ``light-light'' decays $\PWpr\to\PQt\PAQb$ and ``heavy-heavy'' decays
$\PWpr\to\PQT\PAQB$.
Although the light-light and heavy-heavy decays are covered by searches for $\PQt\PAQb$ resonances and VLQ pair production,
dedicated searches are needed for heavy-light decays.
For the electrically neutral \PZpr the situation is similar, where
the heavy-light decays $\PZpr\to\PQB\PAQb$ and $\PZpr\to\TTbar$
are not covered by searches for \ttbar resonances and VLQ pair production.

\subsubsection{Production via the \texorpdfstring{\PZpr}{Z'} boson}
\label{sec:resVLQZprime}

The case of a large coupling between the \PZpr boson and up-type quarks may be investigated
in searches for $\pp\to\PZpr\to\Tt$. A promising decay channel is
\WbWb~\cite{Deandrea:2017rqp}, which differs from the kinematics
of the $\PZpr\to\ttbar\to\WbWb$ resonance search in two important aspects.
The large mass of the \PQT quark results in very different boosts of the two \Wb systems.
The \PQt quark from the \PZpr boson decay may receive a large Lorentz boost if the mass
difference $\mZpr-\mQT$ is large, whereas the boost of the \PQT quark will be moderate
at most for \mZpr in the range 1.5--4.0\TeV and \mQT between 0.7 and 3.0\TeV. When considering
the constraint $\mZpr<2\mQT$, Lorentz factors not larger than $\gamma=1.5$ are realized
for the \PQT, such that the \Wb system from its decay cannot be reconstructed in a single jet.
However, the \PW boson and \PQb quark will be approximately back-to-back, with large \pt in the laboratory rest frame.
The second aspect is that the \Wb system from the \PQT quark decay will have a mass close
to $\mQT\gg\mQt$, such that the usual selection employed in \ttbar resonance searches
will result in a rejection of these events. Both aspects, the different boosts and
different masses of the two \Wb systems, result in an insensitivity of earlier
searches to this signal, despite the same final state.

A dedicated search has been carried out by the CMS Collaboration in the all-hadronic
final state using the 2015 data set~\cite{Sirunyan:2017bfa}.
The analysis selects events with a three-jet
topology, with one large-radius \PQt-tagged, one large-radius \PW-tagged, and one small-radius \PQb-tagged jet.
The \PQt and \PW tagging rely on \tauTT and \tauTO, respectively, in conjunction with \mSD.
The \PQb-tagged small-radius jet must not
overlap with the two large-radius jets. Two SRs are defined, depending on the
presence of a \PQb-tagged subjet in the identified \PQt jet. Both SRs have
approximately the same signal efficiency, with different background efficiencies and
compositions. The subjet \PQb tag reduces the multijet background by a factor of about four,
such that the corresponding SR has better sensitivity than the one without a
subjet \PQb tag. However, the latter still contributes to the overall sensitivity of this search
and validates the multijet background estimation, which is obtained from sideband regions
with vetoes on \PQb-tagged jets and subjets. The uniquely identified decay particles
of the signal decay chain allow for a reconstruction of \mZpr and \mQT, where both masses
can be determined in case a potential signal in the data is observed.

\begin{figure}[!ht]
\centering
\includegraphics[width=0.48\textwidth]{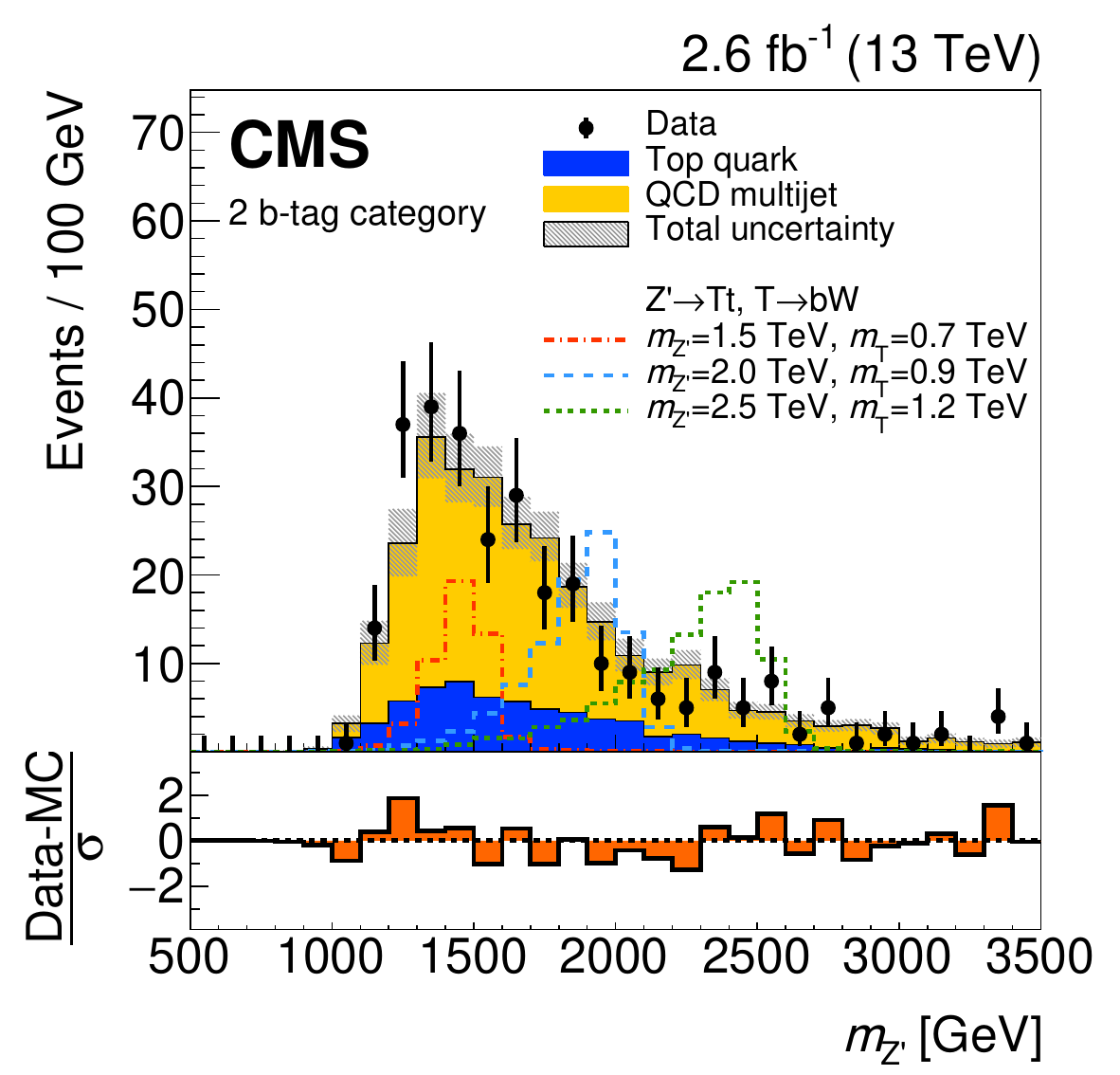}%
\hfill%
\includegraphics[width=0.48\textwidth]{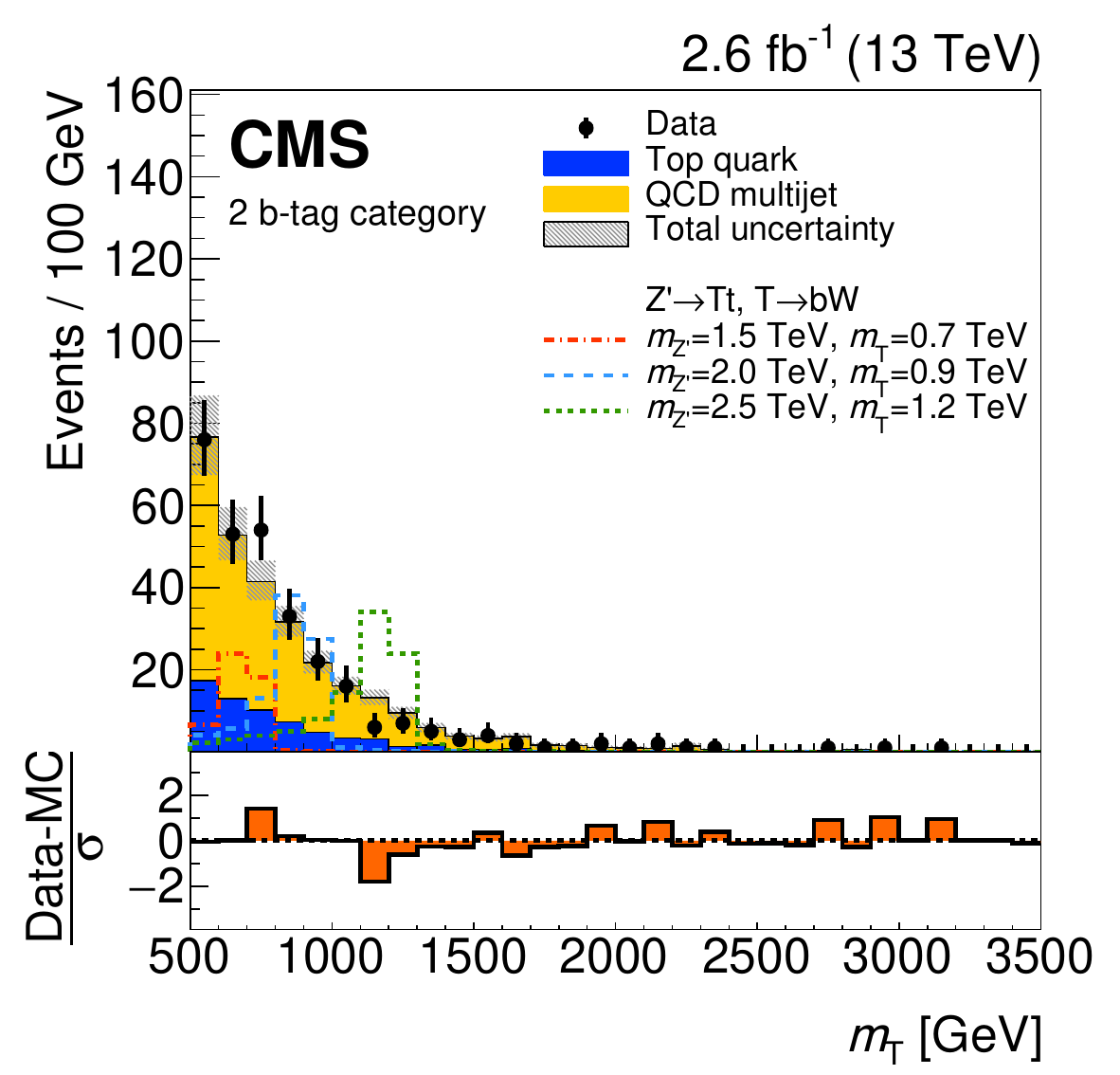}%
\caption{%
    Reconstructed \mZpr (left) and \mQT (right) distributions obtained in a search for $\pp\to\PZpr\to\TTbar$ in the all-hadronic final state.
    The \PZpr boson is reconstructed using a \PQt-, a \PW-, and a \PQb-tagged jet, whereas the \PQT quark is reconstructed using the latter two jets.
    The lower panels show the difference between the data and the estimated backgrounds divided by the sum in quadrature of the statistical uncertainties in data and backgrounds, and the systematic uncertainties in the estimated backgrounds.
    Figures adapted from Ref.~\cite{Sirunyan:2017bfa}.
}
\label{fig:Mzp_Mvlq_allhad}
\end{figure}

The distributions in \mZpr and \mQT for the
SR with a subjet \PQb tag are shown in Fig.~\ref{fig:Mzp_Mvlq_allhad}.
The relative mass resolution for signal events is about 15\%, such that
pronounced peaks on the falling background would be visible. Since both distributions are
obtained from the same events in the observed data, only the distribution in \mZpr
is used to extract upper cross section limits on a potential signal.
Upper limits on the cross section range from 0.13 to 10\unit{pb}, depending on the chosen hypotheses.

Since the all-hadronic search achieves high sensitivity for $\PQT\to\Wb$ decays,
the channels $\PQT\to\Ht$ and $\PQT\to\Zt$ have been targeted in a dedicated search in Ref.~\cite{Sirunyan:2018rfo}
optimized for $\pp\to\PZpr\to\Tt\to\Htt$ and \Ztt.
The search has been carried out by the CMS Collaboration in the \ljets final state
using the 2016 data set.
The presence of \PQt quarks, a Higgs boson or a \PZ boson, and \PW bosons in the decay chain makes this search
special in terms of single VLQ searches, where usually only two of these particles are
produced in a given channel. The analysis considers events with one high-\pt
lepton and a \PV- or \PH-tagged jet. In addition, events in the SR
are categorized depending on the presence of a \PQt-tagged jet. The substructure taggers rely
on the \mSD of large-radius jets, where the mass regions for \PV, \PH,
and \PQt tagging are 60--115, 100--150, and 150--220\GeV, respectively.
The \PV-tagged jets must fulfill $\tauTO<0.5$, \PQt-tagged jets $\tauTT<0.57$,
and \PH-tagged jets must have either one ($\PH_{1\PQb}$) or two ($\PH_{2\PQb}$) subjet \PQb tags.
The overlap between the \PV and \PH taggers is resolved by giving priority to the \PH tagger
for jets that fulfill both criteria, which results in an overall better sensitivity of this search.

One \PQt quark decay is reconstructed using the lepton, \ptvecmiss, and an additional jet.
The possibility to reconstruct the other \PQt quark decay with a \PQt jet depends on the boost of the
two \PQt quarks in the event, and thus on \mQT and the difference $\mZpr-\mQT$.
Events without a \PQt-tagged jet are reconstructed using a combination of small-radius jets,
not overlapping with the \PV- and \PH-tagged large-radius jet. All possible assignments
of jets to the leptonic and hadronic \PQt quark decay cascades are considered, and the hypothesis
with the smallest difference of reconstructed and expected \mQt is chosen.
In the \Htt and \Ztt channels there is an ambiguity concerning which \PQt quark is
emitted in the \PZpr boson decay, such that the \mQT observable cannot be reconstructed. The reconstruction
of \mZpr is achieved by summing the four-momenta of the chosen \ttbar
system and the tagged \PV or Higgs boson.
Six SRs are defined for each lepton flavor,
categorized by a \PV-, $\PH_{1\PQb}$-, or $\PH_{2\PQb}$-tagged jet and
the presence or absence of a \PQt-tagged jet,
resulting in a total of 12 SRs.
The background is estimated from simulation, necessitating the measurement of efficiencies
and misidentification rates of the three substructure taggers used. These measurements are
performed in samples enriched in \ttbar and multijet events. Differences in efficiencies
between data and simulation are used to derive correction factors, which
are generally found to be compatible with unity within the uncertainties.

\begin{figure}[!ht]
\centering
\includegraphics[width=0.48\textwidth]{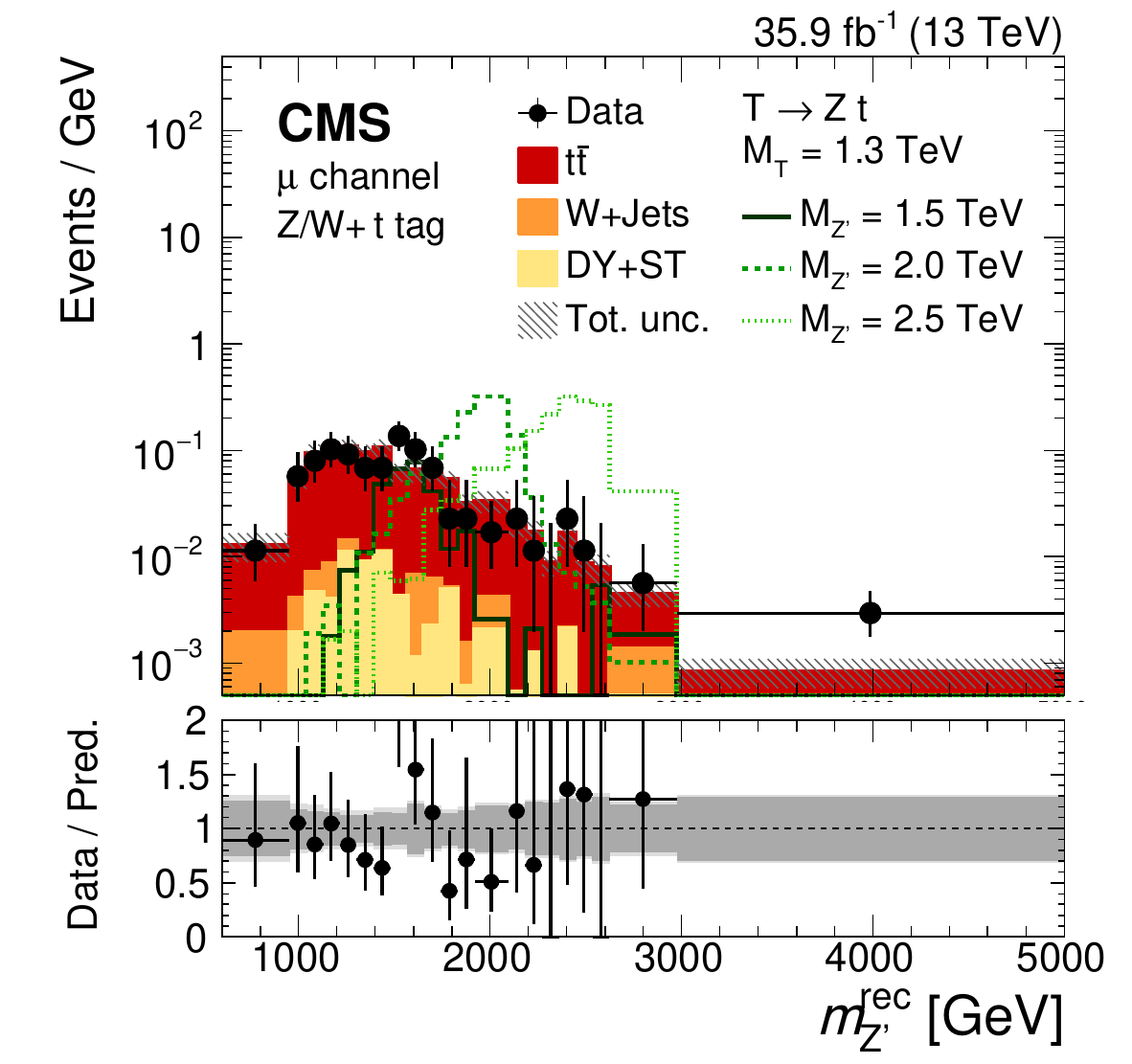}%
\hfill%
\includegraphics[width=0.48\textwidth]{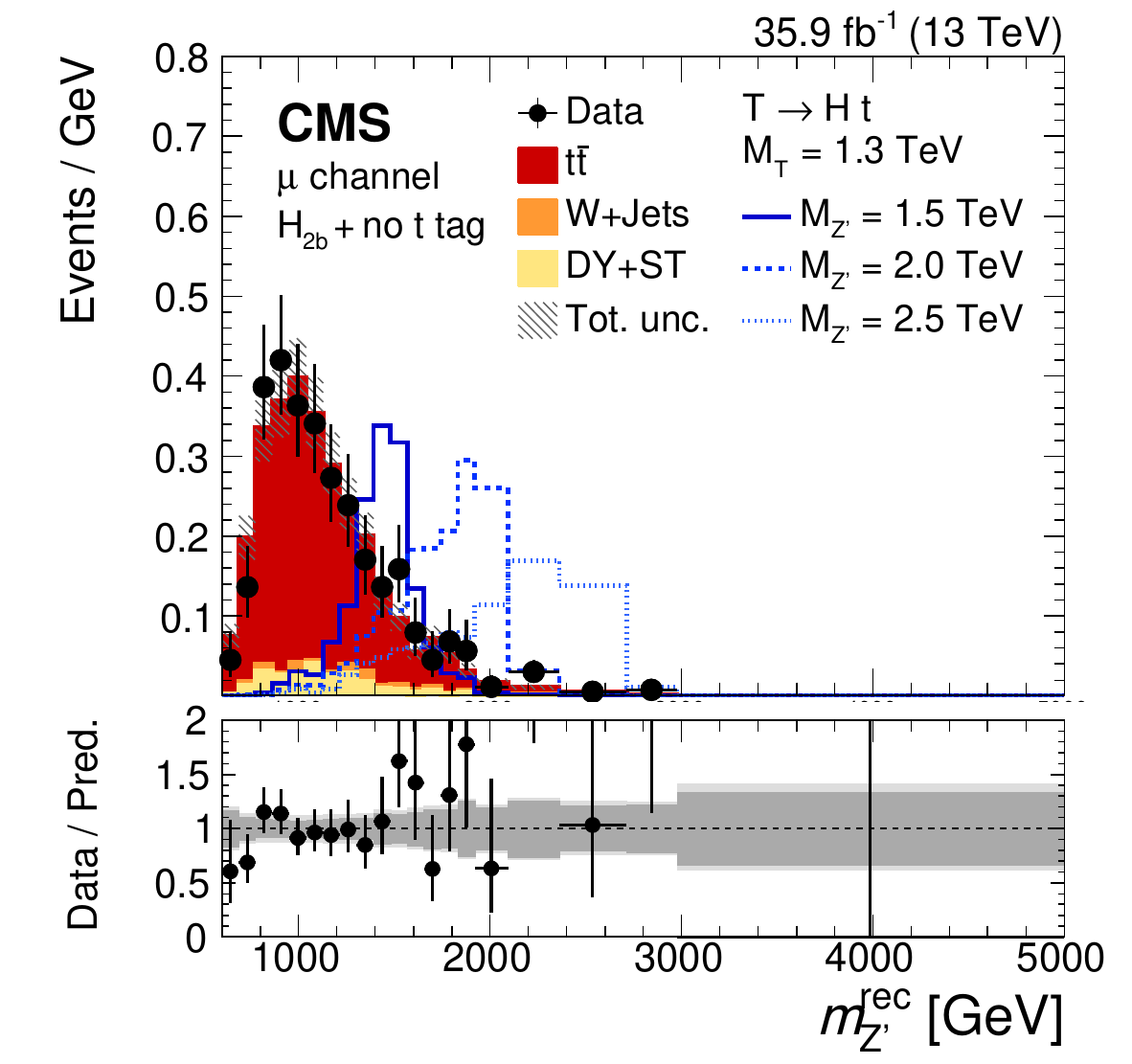}%
\caption{%
    Reconstructed \mZpr distributions obtained in a search for $\pp\to\PZpr\to\TTbar$ in the \ljets final state, in events with a \PV- and a \PQt-tagged jet (left) and in events with an \PH-tagged jet (right).
    The lower panels show the ratio of the observed data to the background prediction.
    Figures adapted from Ref.~\cite{Sirunyan:2018rfo}.
}
\label{fig:Zp}
\end{figure}

In addition to these measurements, CRs enriched with the two main backgrounds,
\ttbar and \wjets, are used to validate the simulation and constrain systematic uncertainties
in the modeling of these backgrounds.
Two reconstructed \mZpr distributions in the {\PGm}+jets channel are presented in
Fig.~\ref{fig:Zp}, where the signal predictions for $\mQT=1.3\TeV$ are overlaid.
The \Ztt channel with a \PV tag and a \PQt tag is shown (left), as well as
the \Htt channel with an $\PH_{2\PQb}$ tag without a \PQt tag (right).

In the \Ztt channel, the signal efficiency for $\mZpr=1.5\TeV$
is smaller than for signals with higher \PZpr boson masses because of the small mass difference $\mZpr-\mQT$.
This results in a \PQt quark emitted from the \PZpr boson decay nearly at rest, such that only the \PQt quark and \PZ boson
from the $\PQT\to\Zt$ decay receive a large boost. Compared with signals with $\mZpr=2.0$ and 2.5\TeV, there
is only one boosted \PQt quark instead of two, thus the selection efficiency is reduced by a factor of two in
this category.

In categories without a \PQt-tagged jet, the efficiency is comparable for \mZpr between
1.5 and 2.0\TeV. The efficiency for $\mZpr=2.5\TeV$ is smaller, because events with a \PQt-tagged jet
are more frequent and are reconstructed in the corresponding category.
This search achieves the best sensitivity to production of $\PQT\to\Ht$ in a resonance decay and similar
sensitivity to $\PQT\to\Zt$ as a nonresonant single VLQ search by the CMS Collaboration in the dilepton channel~\cite{CMS:2017voh}, discussed in Section~\ref{sec:TZlepVLQ},
which may be interpreted in this model as well.
Upper limits on the product of the cross section of the process $\pp\to\PZpr\to\Tt$ and the branching fraction
$\BR(\PQT\to\Ht,\Zt,\Wb)$ are derived. The simultaneous sensitivity to $\PQT\to\Ht$ and
$\PQT\to\Zt$ results in the most stringent constraints to date on models with a heavy gluon and on composite Higgs
models, predicting $\PZpr\to\Tt$ decays.

\subsubsection{Production via the \texorpdfstring{\PWpr}{W'} boson}
\label{sec:resVLQWprime}

A search for the heavy-light decay of a \PWpr boson has been performed by the CMS Collaboration
in the all-hadronic final state, using the 2016 data set~\cite{Sirunyan:2018fki}.
The search has been optimized for \WprtoBt and
\WprtoTb, which both result in the \tHb final state for
the decays $\PQB\to\Hb$ and $\PQT\to\Ht$~\cite{Vignaroli:2014bpa}. The analysis targets high \mWpr and \mvlq,
such that the Higgs boson and \PQt quark are produced with large Lorentz boost and may be reconstructed using large-radius jets.
The smallest mass difference considered in this search is $\mWpr-\mvlq=200\GeV$. In \WprtoTb decays,
even this mass difference results in high momenta for the \PQb quark, \PQt quark, and \PH boson.
In \WprtoBt decays, this mass difference will result in low-momentum \PQt quarks that cannot be
reconstructed in a single large-radius jet. In this regime, the analysis
loses sensitivity because the two decays \WprtoTb and \WprtoBt
are assumed to happen with the same frequency~\cite{Vignaroli:2014bpa}.
The \PH and \PQt tagging algorithms select jets with \mSD in the range
105--135 and 105--210\GeV, respectively. In addition, \PH jets have to pass a
selection based on the discriminator from the double-\PQb tagger and \PQt jets
must have $\tauTT<0.8$ and a subjet \PQb tag. The SR is defined by events
with an \PH-, \PQt-, and \PQb-tagged jet.

The distribution in the reconstructed \PWpr boson mass is used
to search for a signal, obtained from the four-vector sum of the three identified jets in the event.
Sideband and validation regions are used to estimate the multijet background.
These are obtained by inverting the \PQb tag discriminant, \PQt tag
\tauTT, or \PH tag jet mass requirements.
The multijet background in the SR is predicted from data in events passing the inverted \PH tag,
weighted by a transfer function that describes the ratio of probabilities for a jet to pass
the \PH- or inverted \PH-tagging selections. The transfer function is calculated using events
passing the inverted \PQt-tagging requirements.
This approach is validated in simulation and in a dedicated validation region in data. 
This method has the distinct advantage that it provides the background estimation for
any distribution, such that the background model may be checked thoroughly.
In the SR, the multijet background constitutes about 70\% of the total background,
the remaining part originating from \ttbar production.
No significant deviation from the background prediction is observed in the data
and cross section upper limits on \PWpr boson production in the \tHb
decay mode are reported as a function of \mWpr, for several \mvlq hypotheses, though no exclusion is found.

An extension of the heavy-light \PWpr boson analysis includes $\PQB\to\Zb$ and $\PQT\to\Zt$ decays,
such that the final state is complemented by the signature \tZb.
The analysis is carried out in the all-hadronic final state as well, and extends the earlier
analysis~\cite{Sirunyan:2018fki} by including additional sideband and validation regions, and
uses the full Run 2 data set~\cite{B2G-20-002}.
An increase in sensitivity is achieved by updating the \PQt tagging algorithm to use the jet-mass-decorrelated \ImageTop~\cite{CMS:2020poo} method, described in Section~\ref{sec:boosted}.
This allows for an unbiased selection in \mSD, which is important for the
employed background estimation method to work.
The \PH jets are selected using the double-\PQb tagger, and \PZ jets need to have $\tauTO<0.45$.
The SR is defined by the presence of a \PQt jet, an \PH or \PZ jet, and an additional
small-radius \PQb jet.

\begin{figure}[!ht]
\centering
\includegraphics[width=0.46\textwidth]{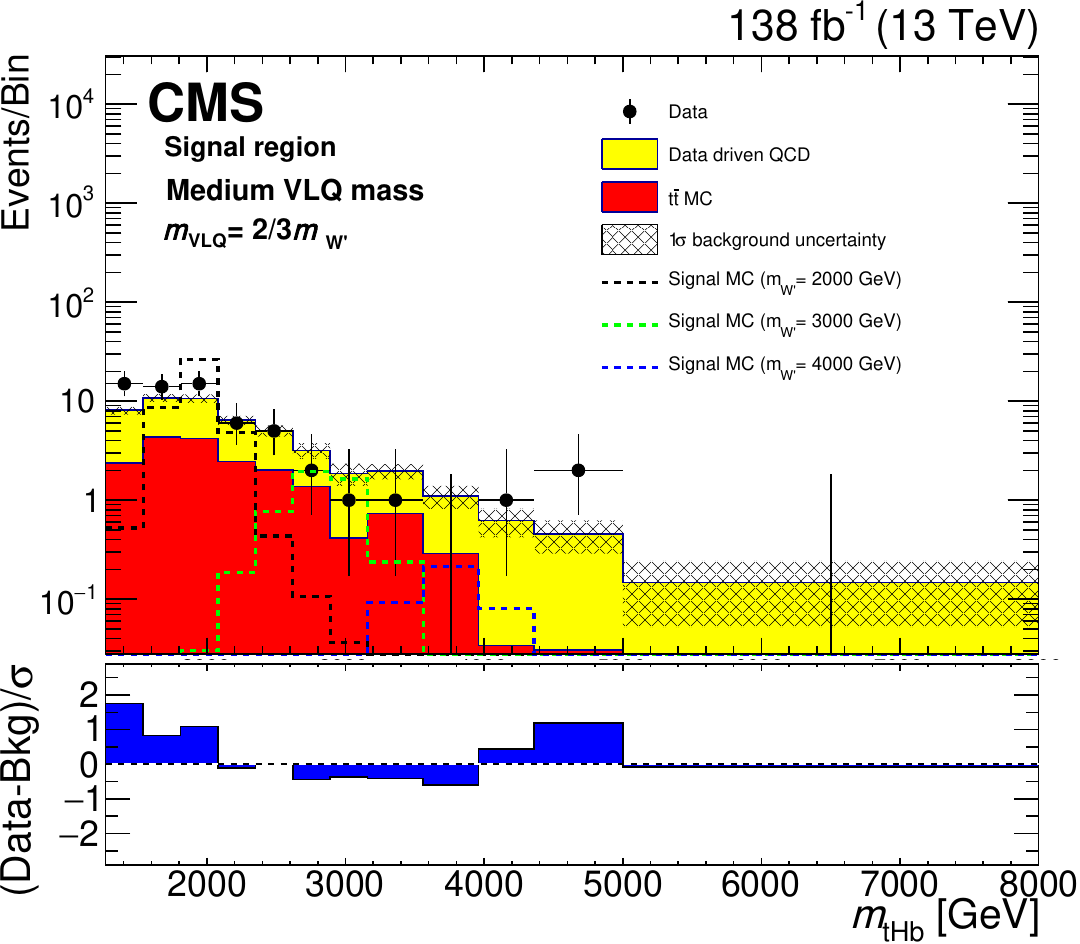}%
\hfill%
\includegraphics[width=0.50\textwidth]{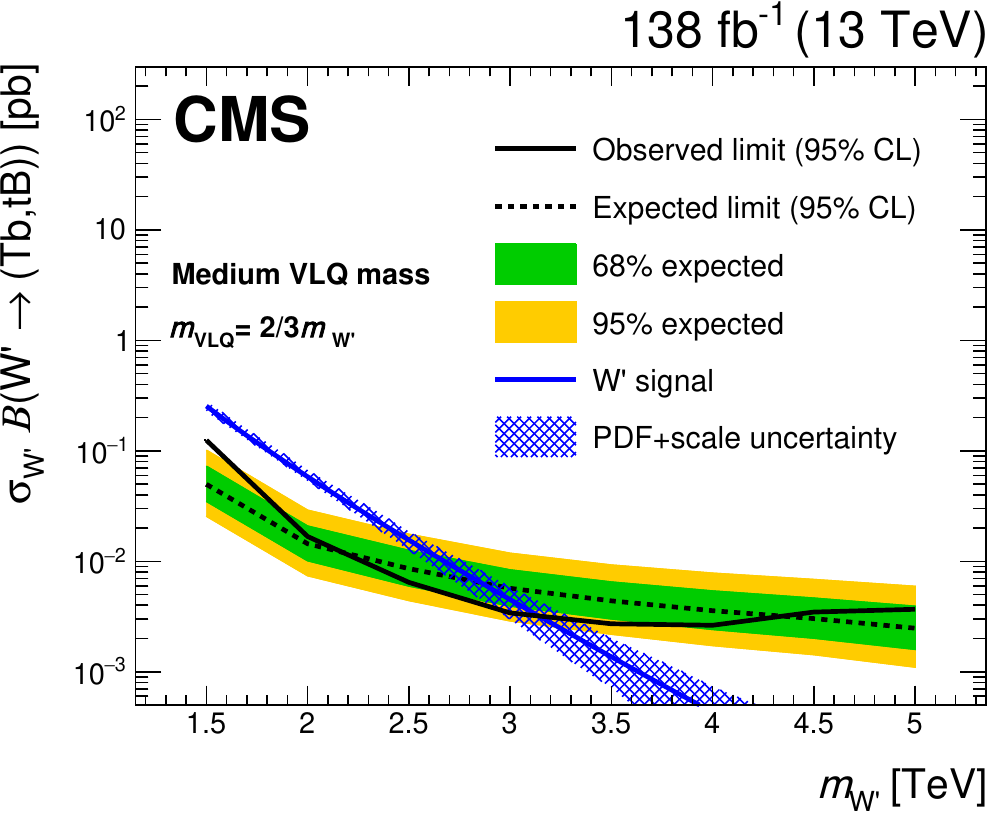}%
\caption{%
    Reconstructed \PWpr boson mass distributions obtained in a search for $\pp\to\PWpr\to\Tbbar/\PQB\PAQt$ in the all-hadronic final state, in events with a \PQt-, \PH- and \PQb-tagged jet (left).
    Upper limits at 95\% \CL on the product of the cross section and branching fraction for the production of a \PWpr boson with decays to \Tbbar and $\PQB\PAQt$ (right).
    Figures adapted from Ref.~\cite{B2G-20-002}.
}
\label{fig:WprimetHb}
\end{figure}

The \PQt, \PH, and \PZ jets are defined by a selection based on \mSD,
where mutually exclusive regions are defined such that two SRs, \tHb and
\tZb, are measured. The multijet background is estimated by an extension of the method used in the
previous analysis, where additional validation regions are used to confirm the validity of the method.
The reconstructed mass distribution in the \tHb SR is shown in Fig.~\ref{fig:WprimetHb} (left).

The background is composed of about 50\% multijet production, and 50\% \ttbar production.
Other SM processes have a negligible contribution in this SR.
The value of \mvlq is chosen to be $\mvlq=2/3\mWpr$ for the signals displayed here, but
choices of $\mvlq=1/2\mWpr$ and $\mvlq=3/4\mWpr$ have also been analyzed.
The resulting upper limits on the production of a \PWpr boson with consecutive decay to \Tbbar or
$\PQB\PAQt$ are shown in Fig.~\ref{fig:WprimetHb} (right) for $\mvlq=2/3\mWpr$.
Here, a branching fraction of 50\% into the \tHb and \tZb final states
is assumed. A relaxation of this assumption is also probed, where the limits are most stringent
for a pure \tHb final state, because of a better background suppression for \PH-tagged jets.
A \PWpr boson with a mass below 3.1\TeV is excluded at 95\% \CL for $\mvlq=2/3\mWpr$ and equal
branching fractions into \tHb and \tZb.
The analysis places the most stringent limits to date in this model.

\subsection{Combinations and summaries of search channels}
\label{sec:VLQcombinationAndsummary}

Collectively, the search program of the CMS experiment for VLQs provides a comprehensive look at VLQ decays to third generation quarks, with lower mass
limits generally extending above 1\TeV, and towards 1.6\TeV for some signal hypotheses. Here we present two combinations
of searches described earlier: in Section~\ref{sec:BBcombo}, we describe a combination of \BBbar searches, and in Section~\ref{sec:singleTcombo}, a combination of single \PQT quark searches is discussed.
Finally, in Section~\ref{sec:VLQsummary}, summary plots are presented showing comparisons of the results of many individual searches.

\subsubsection{Combination of \texorpdfstring{\BBbar}{BBbar} searches}
\label{sec:BBcombo}

\begin{figure}[ht]
\centering
\includegraphics[width=0.48\textwidth]{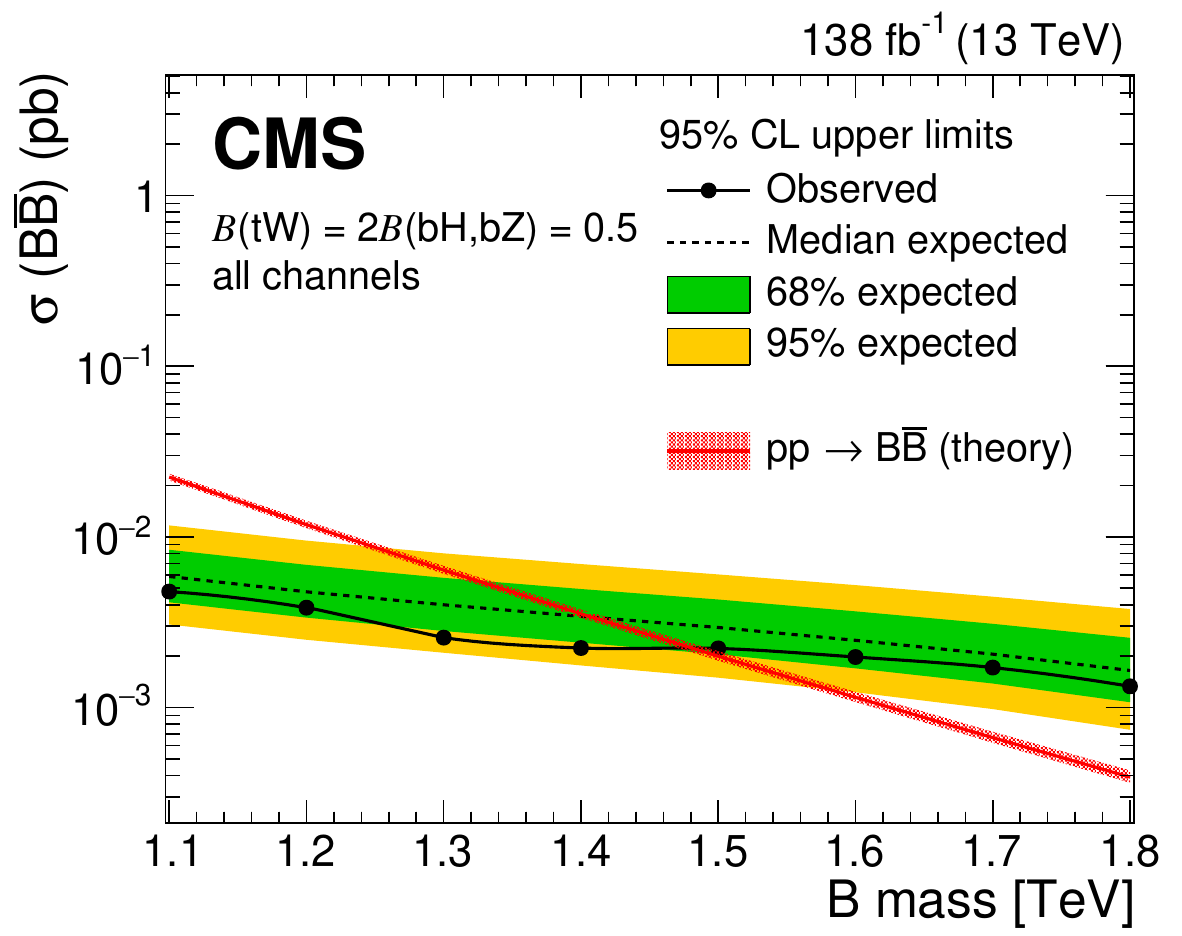}%
\hfill%
\includegraphics[width=0.48\textwidth]{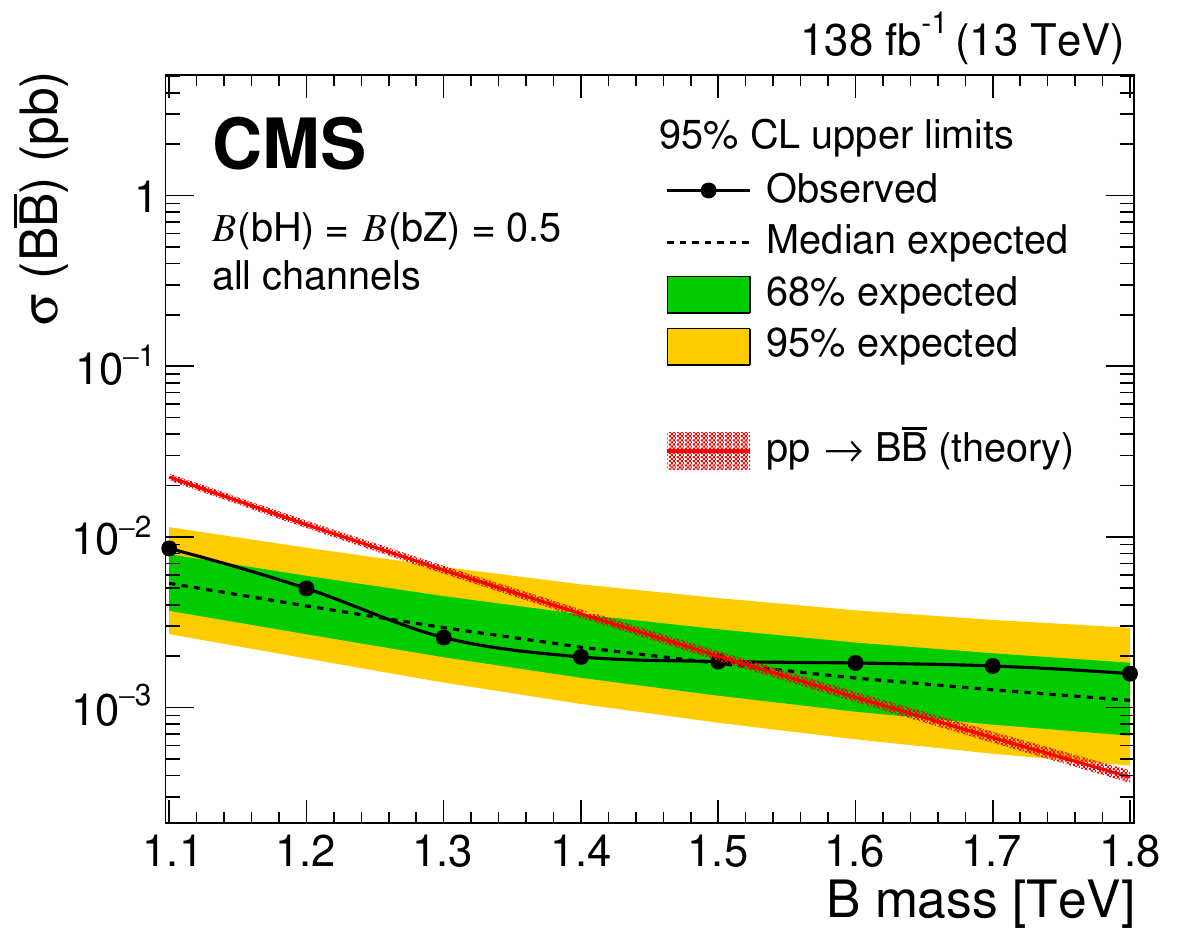}%
\caption{%
    Observed (solid lines) and expected (dashed lines) 95\% \CL upper limits on \BBbar production as a function of the \PQB quark mass for the singlet (left) and doublet (right) branching fraction scenarios, from the combination of two searches for \BBbar production.
    Predicted cross sections are shown by the red line surrounded by a band representing energy scale and PDF uncertainties in the calculation.
}
\label{fig:BBcombo1D}
\end{figure}

\begin{figure}[!ht]
\centering
\includegraphics[width=0.48\textwidth]{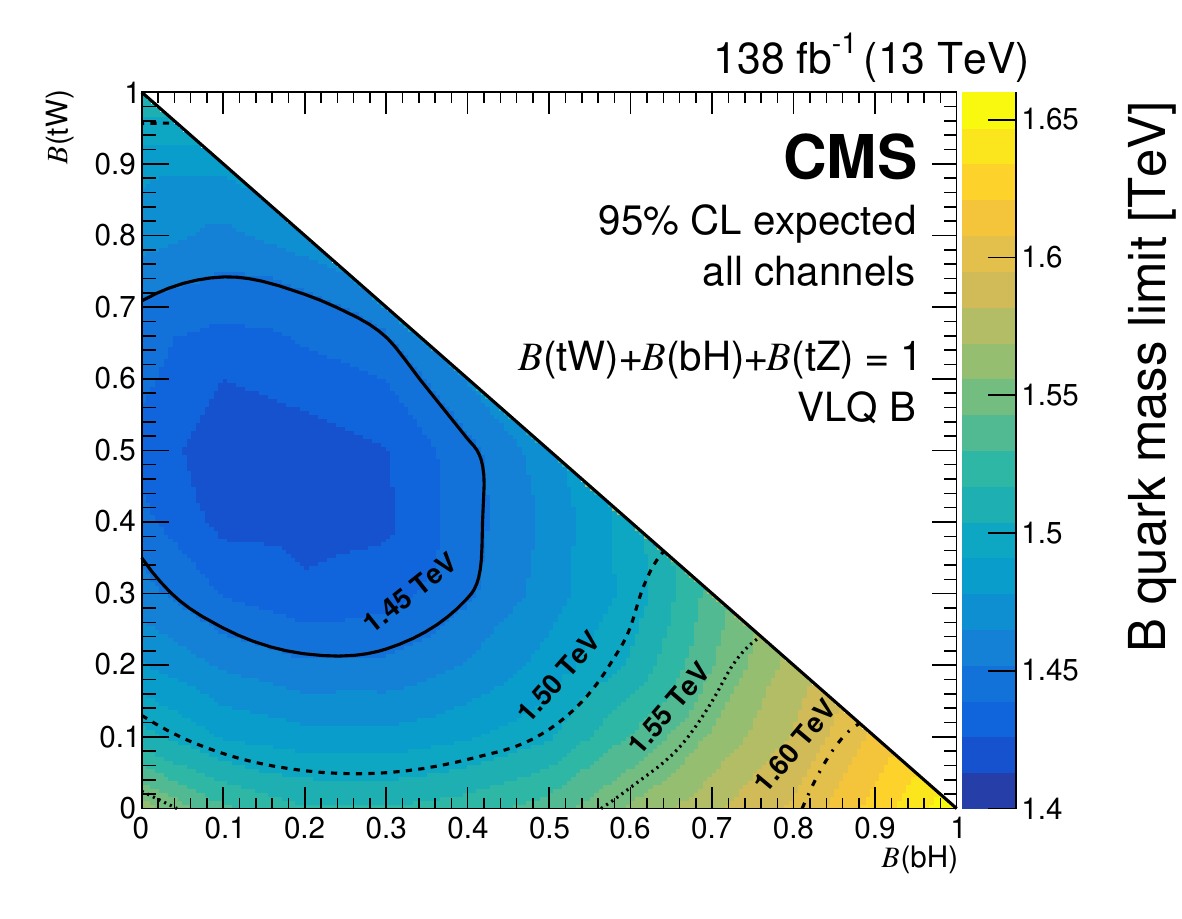}%
\hfill%
\includegraphics[width=0.48\textwidth]{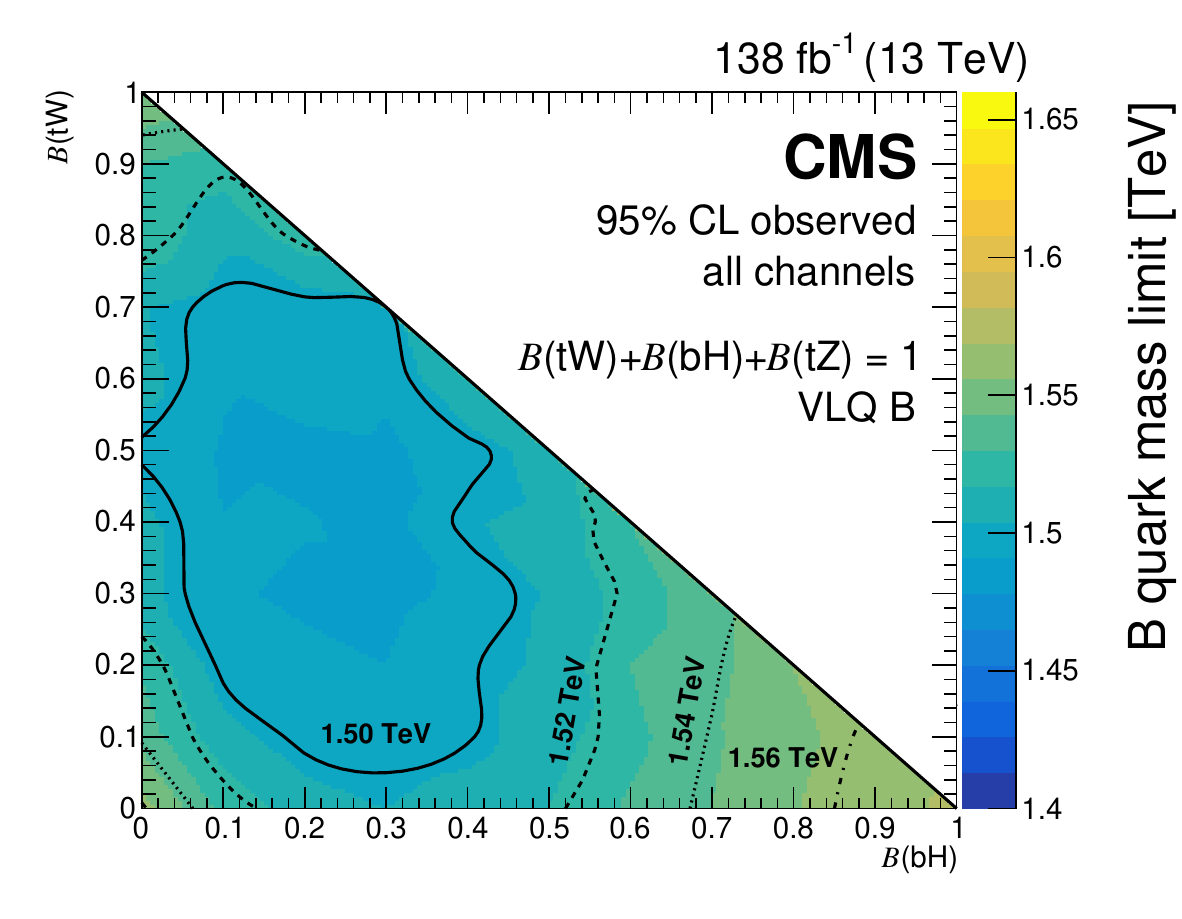}%
\caption{%
    Expected (left) and observed (right) lower limits on the \PQB quark mass at 95\% \CL from the combination of two searches for \BBbar production.
    The limits are shown as a function of the branching fractions \BrBbH and \BrBtW, with $\BrBtW=1-\BrBbH-\BrBbZ$.
}
\label{fig:BBcombo}
\end{figure}

The \BBbar searches that utilize the full Run 2 data set were designed to have mutually exclusive lepton selection criteria.
Combining the searches of Refs.~\cite{B2G-20-011} and~\cite{CMS:2024xbc}, both described in Section~\ref{sec:BBVLQ}, brings together hadronic, single-lepton, dilepton (SS and OS),
and multilepton final states. The hadronic and OS dilepton channels from Ref.~\cite{CMS:2024xbc} are sensitive to \BtobZ and \BtobH
decays, whereas the other leptonic channels from Ref.~\cite{B2G-20-011} are sensitive to \BtotW decays. The combination is performed by
simultaneously fitting all template distributions from the various individual final states to determine a common signal strength parameter.
Many of the background estimates in these channels are derived from the observed data and therefore have independent uncertainties.
Uncertainties in the signal hypothesis were correlated across all channels and all years of data collection, with the exception of the jet
energy scale and resolution uncertainties, which were left uncorrelated due to different treatments of data collection years across the
two searches. Limits on the production cross section for \BBbar reflect the separate strengths of the input searches in cases of 100\%
\BtobZ, \BtobH, or \BtotW branching fractions, and become stronger than either individual search in the case of heavily mixed branching fractions.
Figure~\ref{fig:BBcombo1D} shows the upper limits on the production cross section of \PQB quark pairs in the singlet and doublet branching
fraction scenarios. Figure~\ref{fig:BBcombo} shows the \PQB quark lower mass limits as a function of the \PQB quark branching fractions to \bH and \tW.
Pair production of \PQB quarks decaying to any third-generation quark is excluded for \PQB quark masses below 1.49\TeV, a significant
increase in the general lower mass limit compared to both the strongest Run-1 limit of 900\GeV, and any of
the individual Run 2 searches. The limits on \BBbar
production from this combination show similar sensitivity across all branching fractions to the limits
on \TTbar production from Ref.~\cite{B2G-20-011}, which excludes \PQT quarks below 1.48\TeV for all
third-generation decays.

\subsubsection{Combination of single \texorpdfstring{\PQT}{T} quark production searches}
\label{sec:singleTcombo}

A statistical combination has been performed of three searches for single \PQT quark production in different final states~\cite{B2G-21-007,B2G-19-001,B2G-19-004}, which have been described individually in Sections~\ref{sec:TZlepVLQ} and~\ref{sec:TZHbbVLQ}.
The decay modes that have been considered are \tH, with
Higgs boson decays into \bbbar or \gammagamma, and \tZ, with \PZ boson decays into \bbbar or $\nunu$.
All the final states combined here are defined as mutually exclusive such that they could be considered statistically independent observations.

The combined exclusion limit calculations include the full correlation of the systematic
uncertainties obtained for individual channels and for each year of data taking.
Nuisance parameters related to the same underlying effect, such as the uncertainty in the theoretical
prediction or the energy scale uncertainty in the final-state objects, are correlated across the different channels.
Uncertainties common to all input analyses in the combination, such as in the integrated luminosity, pileup, and PDF
uncertainties, are assumed to be correlated across all analyses. In contrast, all other
uncertainties are treated as uncorrelated in the combination, as the analysis channels are
assumed to be independent of each other.
The fit of the inclusive combined signal strength
($\mu$) involves a total of $\approx$600 nuisance parameters.
More details on the categories that are combined, which are created according to various criteria such
as signal-to-background ratios, mass resolutions, and multiplicities of physics objects, are reported in the references to the individual analyses.

\begin{figure}[ht!]
\centering
\includegraphics[width=0.48\textwidth]{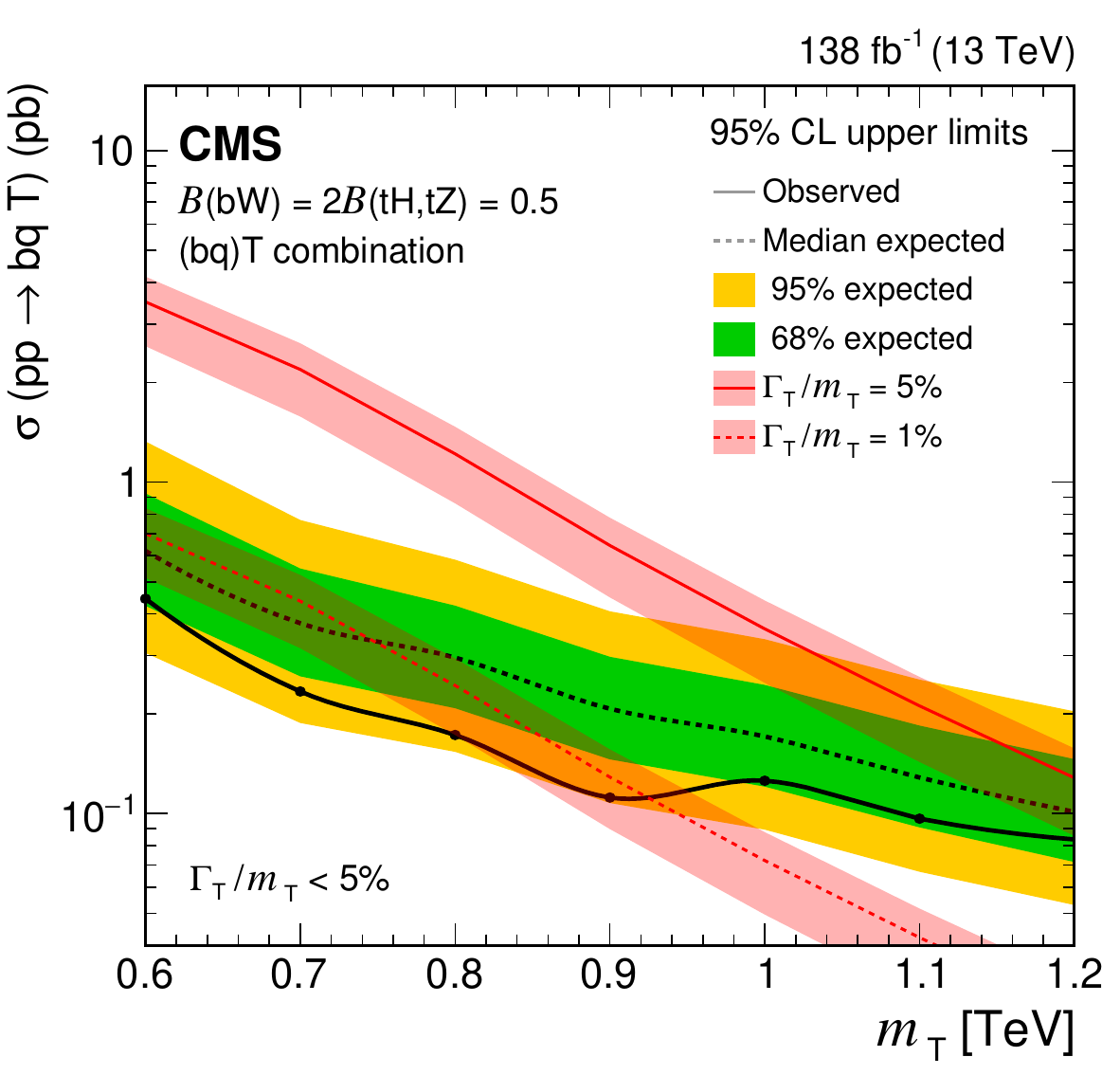}
\caption{%
    Observed and expected 95\% \CL upper limits on the production cross section of a single \PQT quark in association with a \PQb quark in a singlet scenario, versus the \PQT quark mass.
    Theoretical predictions for relative widths of 1 and 5\% of the mass are shown as red solid line and red dashed line, respectively.
}
\label{fig:limitSingleTcombo}
\end{figure}

\begin{figure}[ht!]
\centering
\includegraphics[width=0.48\textwidth]{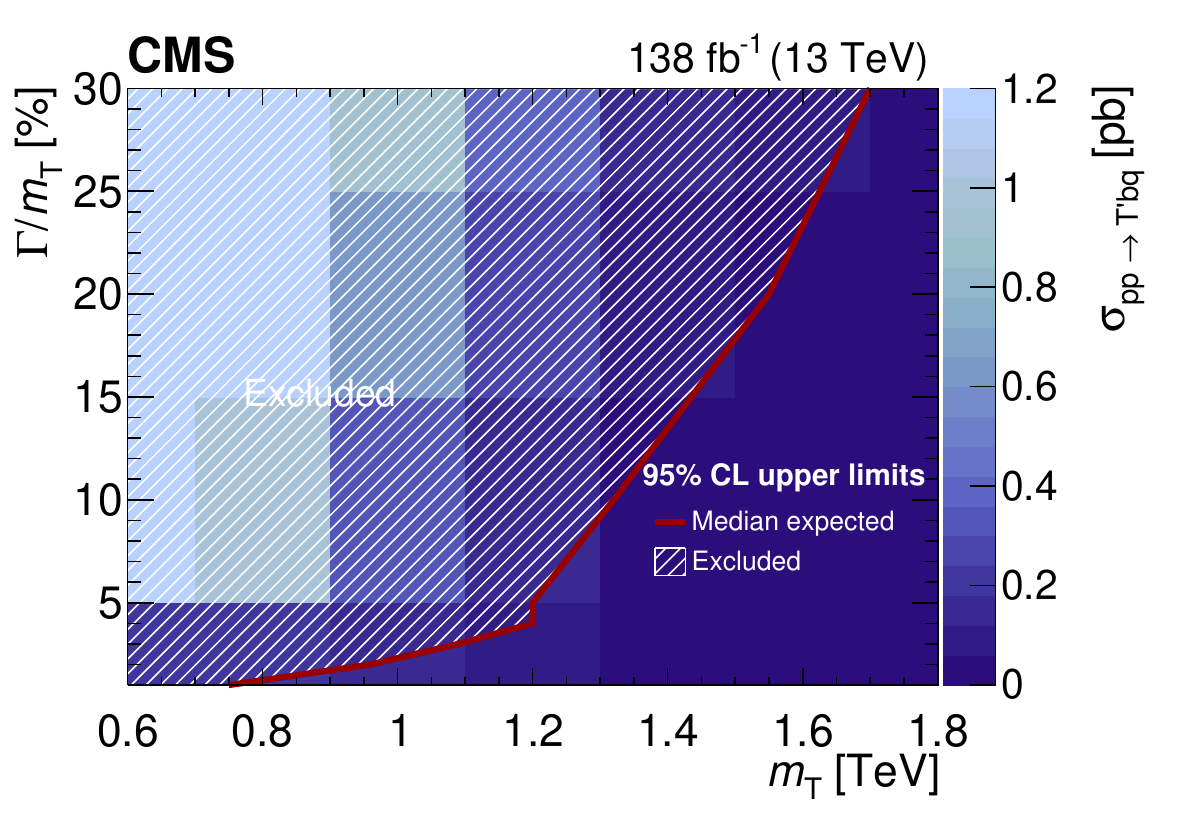}%
\hfill%
\includegraphics[width=0.48\textwidth]{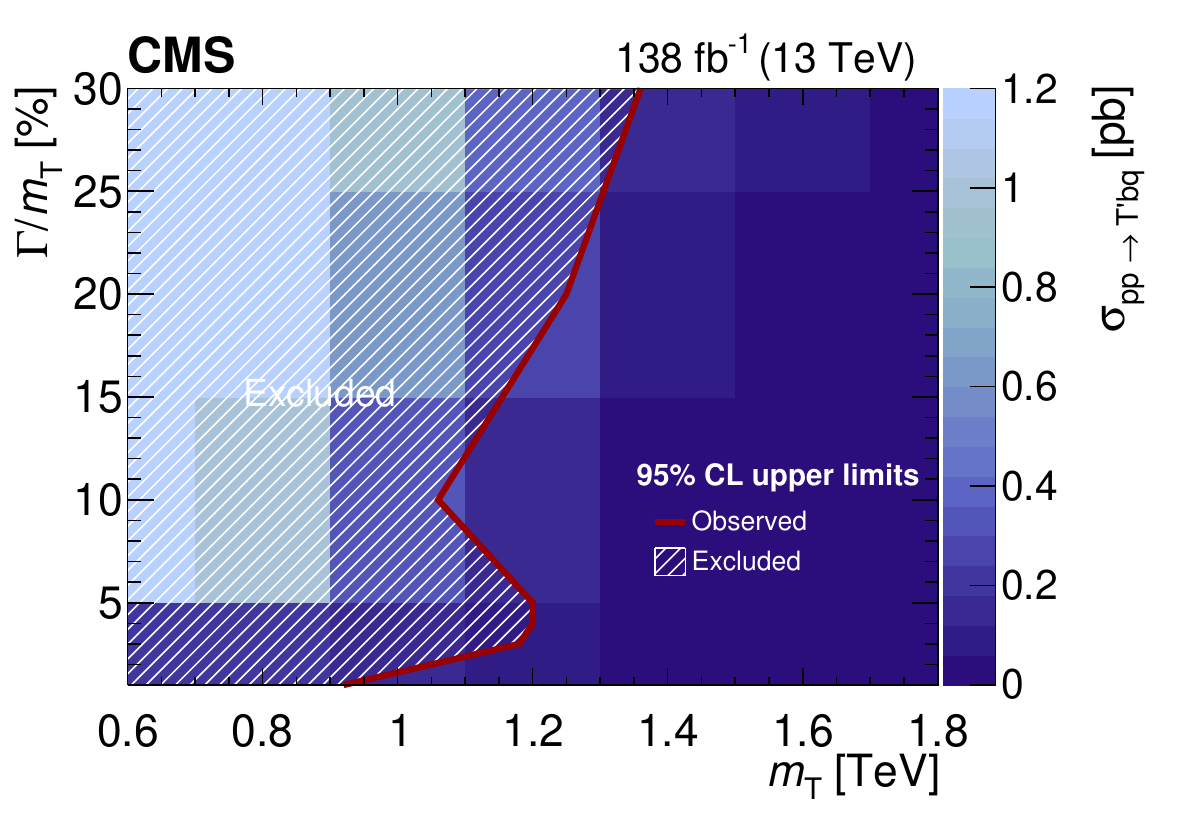}%
\caption{%
    Expected (left) and observed (right) 95\% \CL upper limits on the product of the single-production cross section and the $\PQT\to\PQt\PZ/\PH$ branching fraction for a singlet \PQT quark, as a function of the \PQT quark mass \mQT and width $\Gamma$, for relative widths from 1 to 30\% of the mass.
    A singlet \PQT quark that is produced in association with a \PQb quark is assumed.
    The solid red line indicates the boundary of the excluded region (hatched area) of theoretical cross sections.
}
\label{fig:summarySingleTcombo}
\end{figure}

As previously discussed in Section~\ref{sec:SingleVLQ}, in the searches for singly produced \PQT quarks, the analyses are designed
using different decay width approximations, including NWA and relative width approximations of 10, 20, and 30\%.
In case of the NWA, the combination of searches for single \PQT quark production
in the \Htogg~\cite{B2G-21-007}, \PZ/\Htobb~\cite{B2G-19-001}, and $\PZ\to\nunu$~\cite{B2G-19-004} channels,
using the full Run 2 data set,
could potentially result in the most stringent exclusion limits on the \PQT quark mass.
For the other width approximation scenarios, no combination is carried out and the existing
analysis in the $\PZ\to\nunu$ channel provides the most stringent exclusion to date.
Figure~\ref{fig:limitSingleTcombo}
shows the cross section exclusion limit as a function of the \PQT quark mass under the NWA, compared to theory predictions corresponding to width scenarios with $\GoM=1$ and 5\%.
Figure~\ref{fig:summarySingleTcombo}
illustrates the most stringent \PQT quark mass exclusion under various width approximations.
Assuming a relative decay width
of $\GoM=5$, 10, 20, and 30\%, the EW production of a singlet \PQT quark is excluded up to a
mass of 1.20, 1.06, 1.25, and 1.36\TeV, respectively, at 95\% \CL.
In the following section, we discuss how these limits translate to constraints on the coupling strength.

\subsubsection{Summary of searches}
\label{sec:VLQsummary}

We conclude the section on VLQ results by comparing the sensitivity of the various VLQ searches in terms of limits on cross sections in \pp collisions at $\sqrt{s}=13$\TeV and model parameters.
Since the sensitivity depends on assumptions on the production and decay of the VLQs, we compare
them in representative benchmark scenarios.
Overall, no significant deviation of the observed limits from the expected limit by more than two standard deviations was found.
We thus report the observed and median expected cross section limits, while omitting bands indicating the regions containing 68 and 95\% of the distribution of limits expected under the background-only hypothesis.

\begin{figure}[htp!]
\centering
\includegraphics[width=0.48\textwidth]{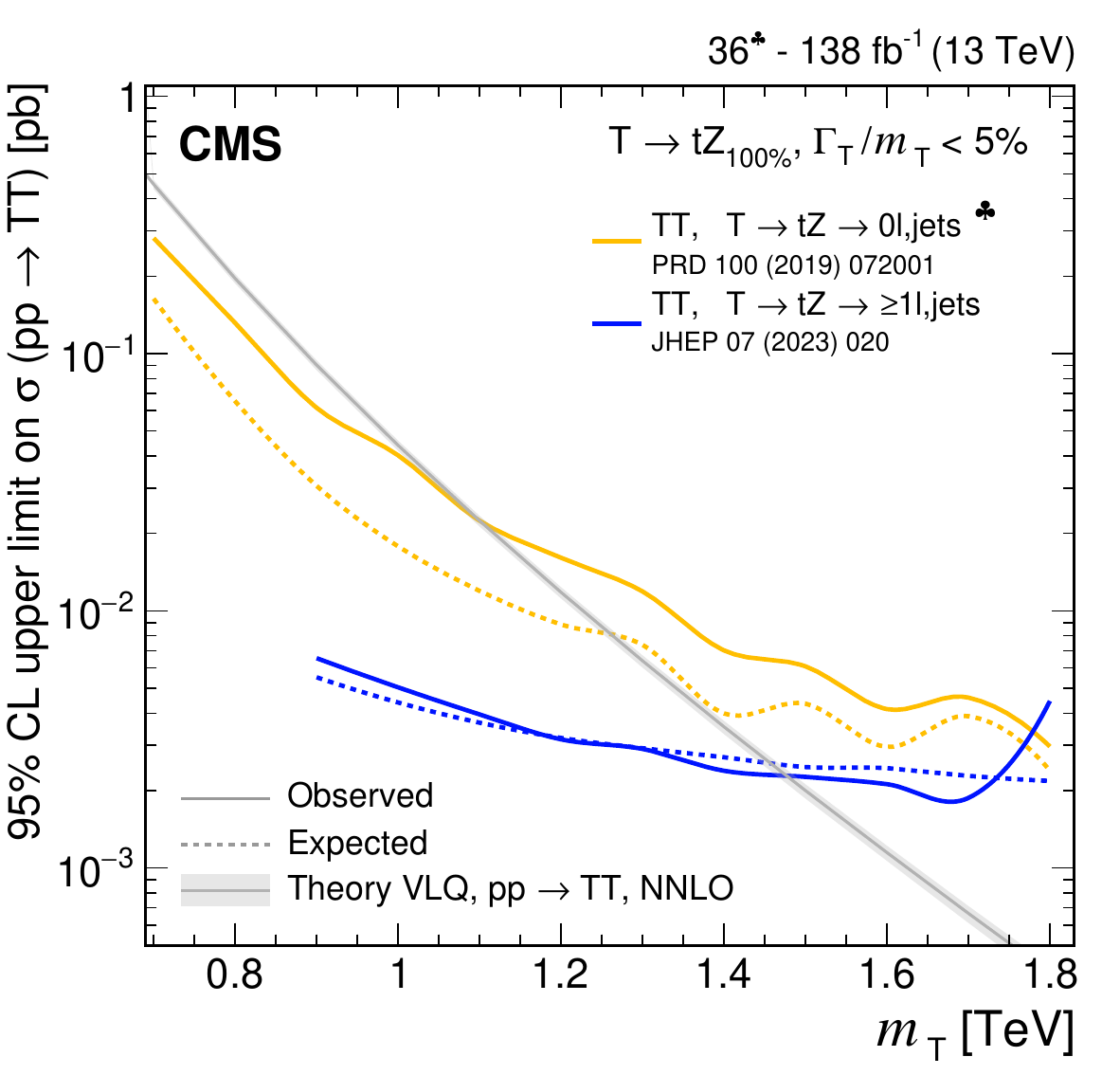}%
\hfill%
\includegraphics[width=0.48\textwidth]{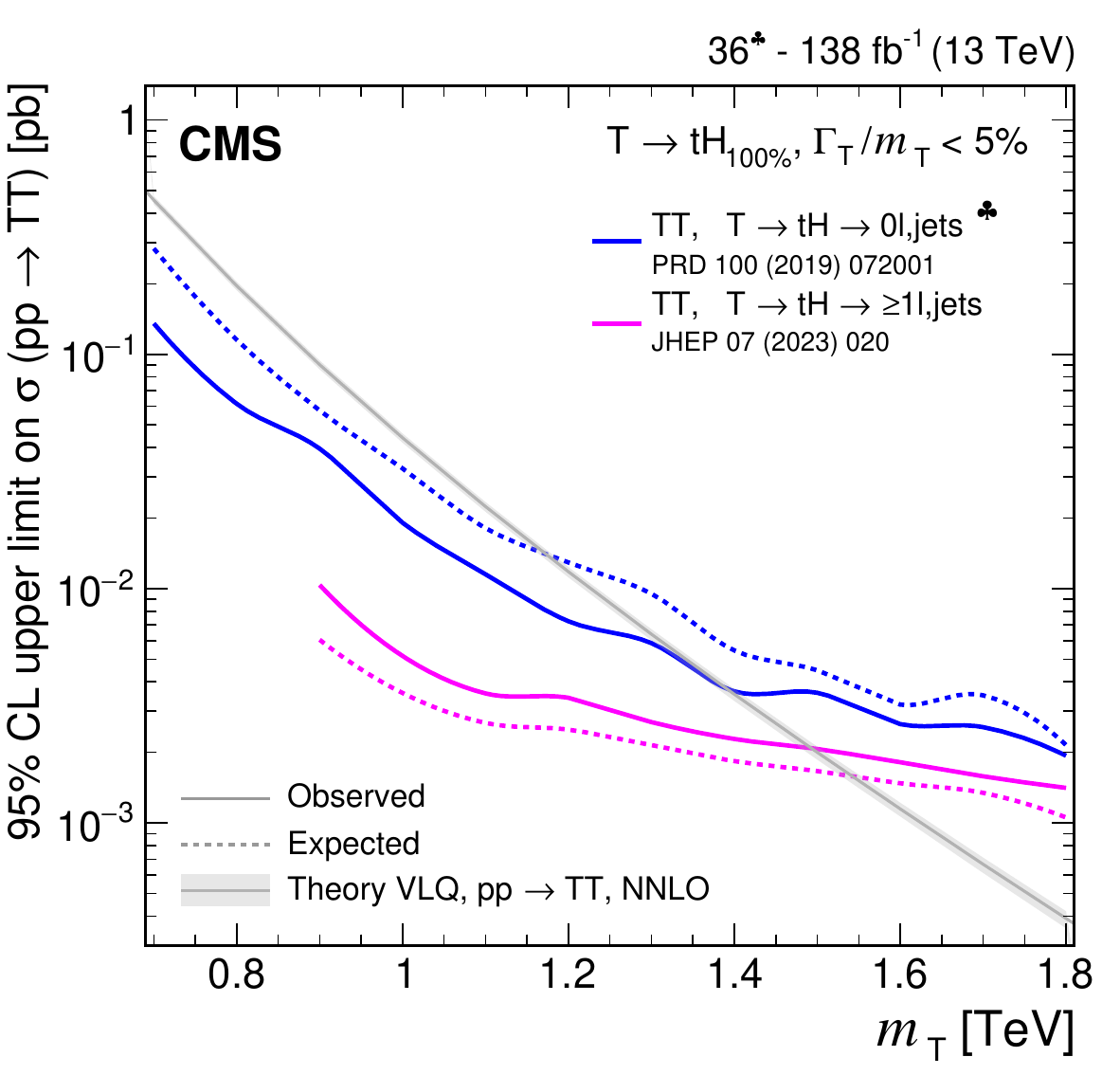} \\
\includegraphics[width=0.48\textwidth]{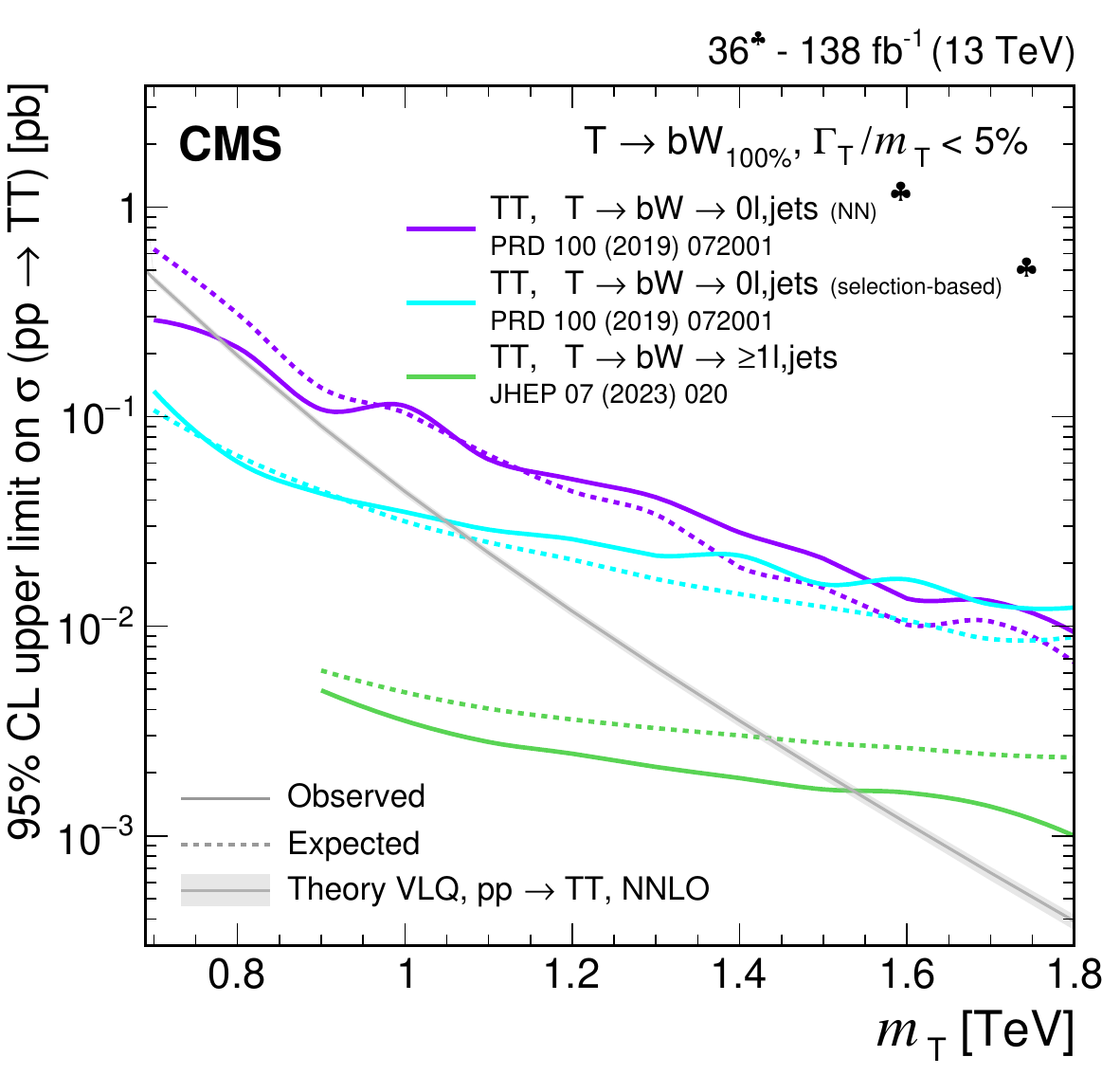}%
\caption{%
    Observed and expected 95\% \CL upper limits on the production cross section of a pair of vector-like \PQT quarks decaying to \tZ (upper left), \tH (upper right), and \bW (lower), as a function of the \PQT quark mass, obtained by different analyses: 0\ljets (NN, selection-based)~\cite{B2G-18-005}, and $\geq$1\ljets~\cite{B2G-20-011}.
    A theory prediction at NNLO in perturbative QCD of the pair production cross section in the NWA is superimposed.
    Searches using data corresponding to an integrated luminosity of 36\fbinv, rather than the full Run 2 integrated luminosity of 138\fbinv, are indicated with a spade symbol in the legend.
}
\label{fig:summaryPairTT100}
\end{figure}

\begin{figure}[htp!]
\centering
\includegraphics[width=0.48\textwidth]{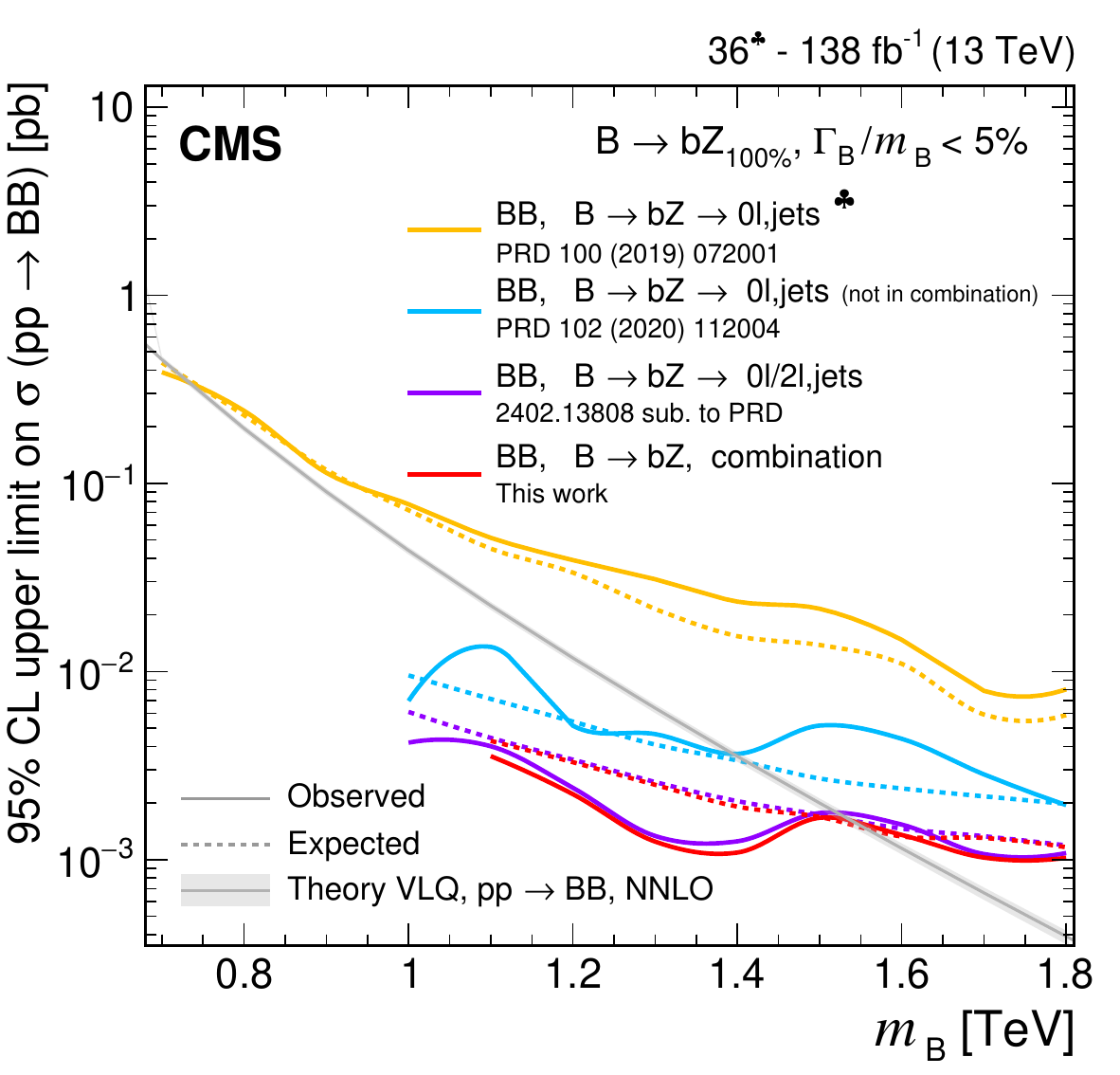}%
\hfill%
\includegraphics[width=0.48\textwidth]{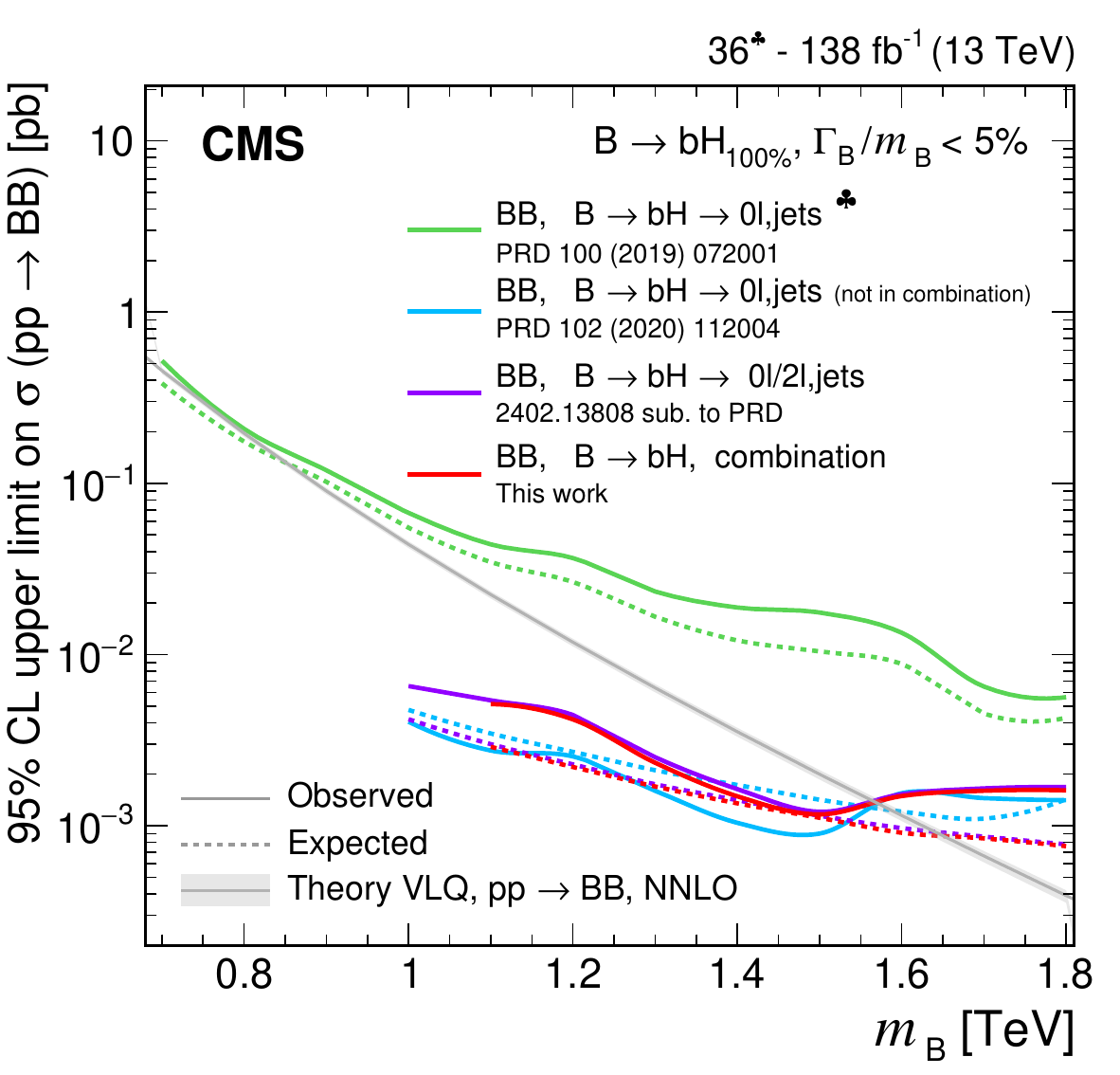} \\
\includegraphics[width=0.48\textwidth]{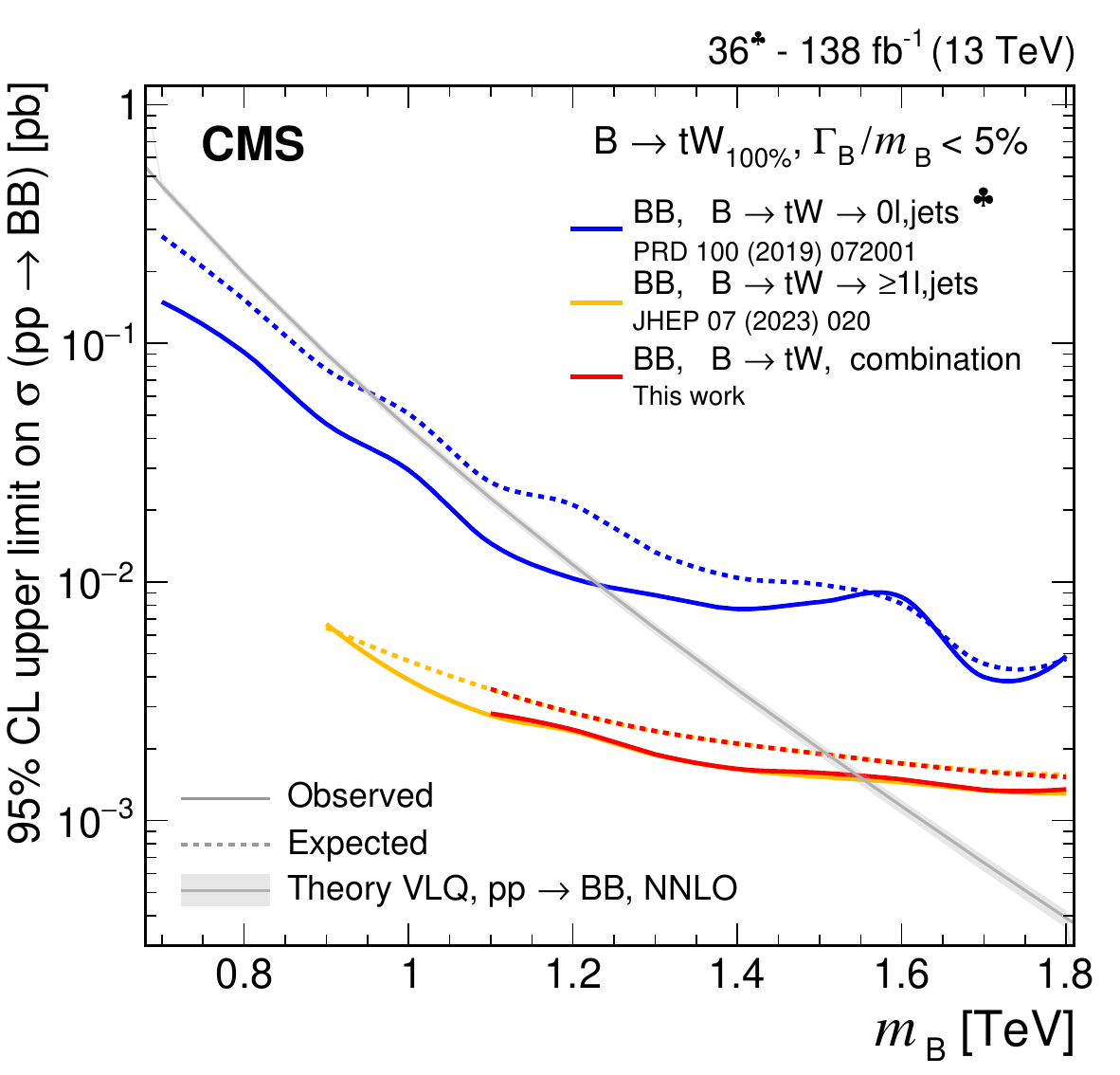}%
\caption{%
    Observed and expected 95\% \CL upper limits on the production cross section of a pair of vector-like \PQB quarks decaying to \bZ (upper left), \bH (upper right), and \tW (lower), as a function of the \PQB quark mass, obtained by different analyses: 0\ljets (NN)~\cite{B2G-18-005}, 0\ljets~\cite{B2G-19-005}, $\geq$1\ljets~\cite{B2G-20-011}, 0\Pell/2\ljets~\cite{CMS:2024xbc}, and the \BBbar combination of Section~\ref{sec:BBcombo}.
    A theory prediction at NNLO in perturbative QCD of the pair production cross section in the NWA is superimposed.
    Searches using data corresponding to an integrated luminosity of 36\fbinv, rather than the full Run 2 integrated luminosity of 138\fbinv, are indicated with a spade symbol in the legend.
}
\label{fig:summaryPairBB100}
\end{figure}

The upper limits on the cross section as functions of the VLQ mass for \TTbar and \BBbar
production via the strong interaction are presented for three benchmark scenarios.
In Figs.~\ref{fig:summaryPairTT100} and \ref{fig:summaryPairBB100}, a branching fraction of
100\% for all VLQ decay modes is assumed.
Although this scenario does not represent a realistic model, since branching fractions sum up
to more than 100\%, it allows us to compare the sensitivity of the searches within each decay mode.
The crossing point of the predicted cross section and the upper limit on the cross section
corresponds to the maximal mass excluded if the VLQ solely decayed through a single decay mode.
For all decay modes of \PQT, the $\geq$1\ljets search~\cite{B2G-20-011} contributes most
to the sensitivity.
This search is better because of the larger full Run 2 data set
and advanced NN-based \PQt tagging techniques, compared to the earlier 0\ljets
\TTbar search~\cite{B2G-18-005} using the 2016 data set.
Among the three decay modes, \tZ, \tH, and \bW, analyses targeting
\tH yield the strongest expected constraints. However, the strongest observed constraint
comes from \bW.
For the decay modes of \BtobZ and \bH, the sensitivity of the \BBbar
combination is dominated by the 0\Pell/2\ljets search~\cite{CMS:2024xbc}, whereas for
the \BtotW decay mode the $\geq$1\ljets~\cite{B2G-20-011} channel dominates.
Both searches benefit from the larger data set, compared to the 0\ljets search~\cite{B2G-18-005}.
Although the expected sensitivity of the \BBbar combination is always better than the
0\ljets \BBbar search~\cite{B2G-19-005}, the latter happens to give the most stringent observed
limits in the \bH decay mode. Among the three decay modes to \bZ, \bH, and
\tW, analyses targeting \bH yield the strongest constraints.
While the results for \xft and \yft pair production are not explicitly shown in these summaries,
the sensitivity to their production is solely driven by the searches targeting the \tW
and \bW decay modes, respectively, since none of these searches rely on charge
information to interpret the VLQ signal.

\begin{figure}[htp!]
\centering
\includegraphics[width=0.48\textwidth]{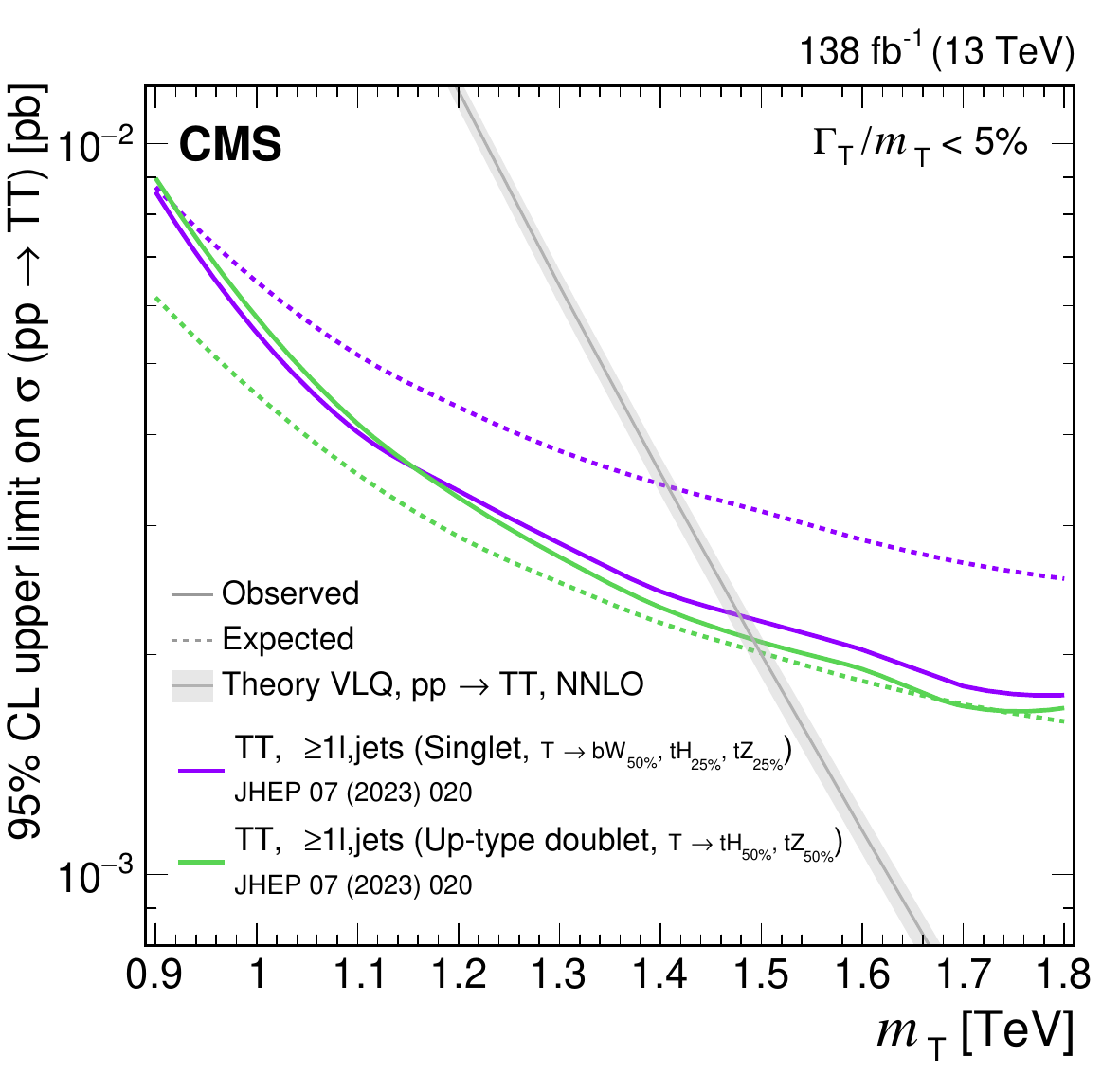} \\
\includegraphics[width=0.48\textwidth]{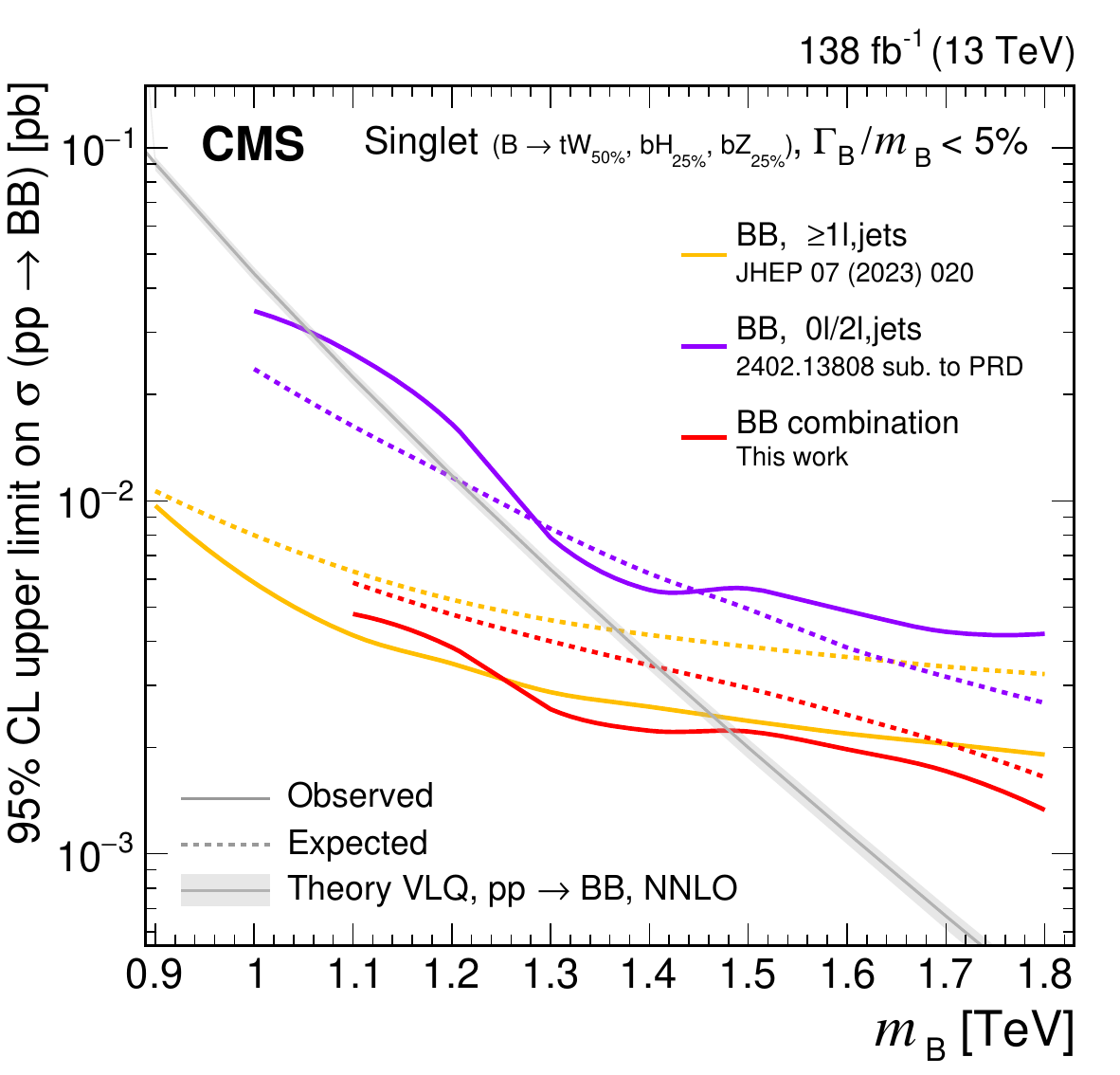}%
\hfill%
\includegraphics[width=0.48\textwidth]{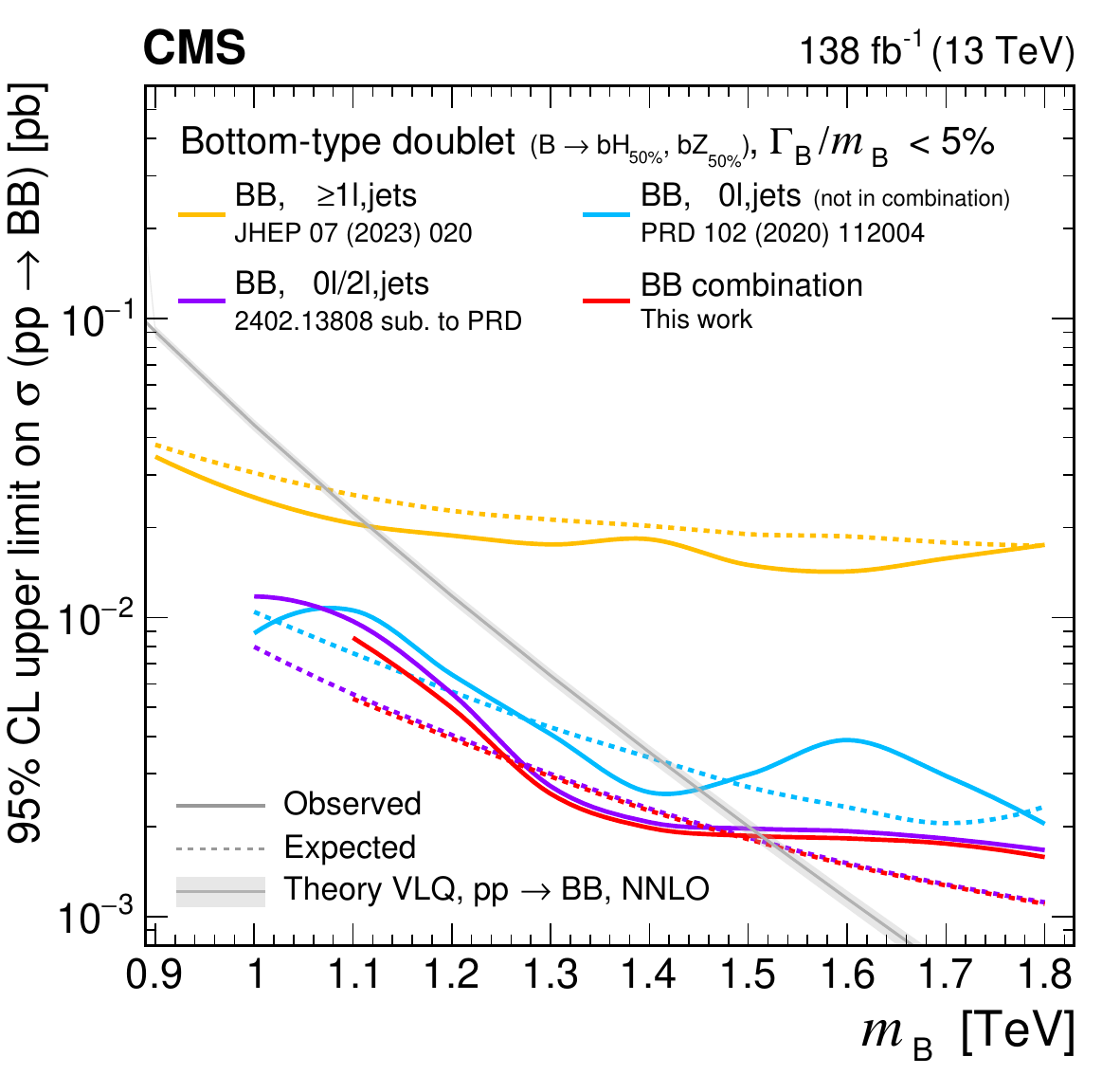}%
\caption{%
    Observed and expected 95\% \CL upper limits on the production cross section of a pair of vector-like \PQT or \PQB quarks, as functions of their mass, obtained by different analyses: 0\ljets~\cite{B2G-19-005}, $\geq$1\ljets~\cite{B2G-20-011}, 0\Pell/2\ljets~\cite{CMS:2024xbc}, and the \BBbar combination.
    A theory prediction at NNLO in perturbative QCD of the pair production cross section in the NWA is superimposed.
    Branching fractions of a singlet (upper and lower left panel) and doublet (upper and lower right panel) are assumed.
}
\label{fig:summaryPairSingletDoublet}
\end{figure}

{\tolerance=800
In Fig.~\ref{fig:summaryPairSingletDoublet}, branching fractions corresponding to a singlet and a doublet scenarios are assumed.
The crossing point of the predicted cross section and upper limit on the cross sections correspond to the lower limit on the mass in these scenarios.
In the singlet scenario, the $\geq$1\ljets~\cite{B2G-20-011} search contributes most to the sensitivity.
In the doublet scenario, the 0\Pell/2\ljets~\cite{CMS:2024xbc} contributes most to the sensitivity at high masses.
For \BBbar and \TTbar production, constraints on the doublet scenario are stronger than for the singlet scenario, since the experimentally better constrained decay modes to \bH and \tH have larger branching fractions.
\par}

\begin{figure}[htp!]
\centering
\includegraphics[width=0.76\textwidth]{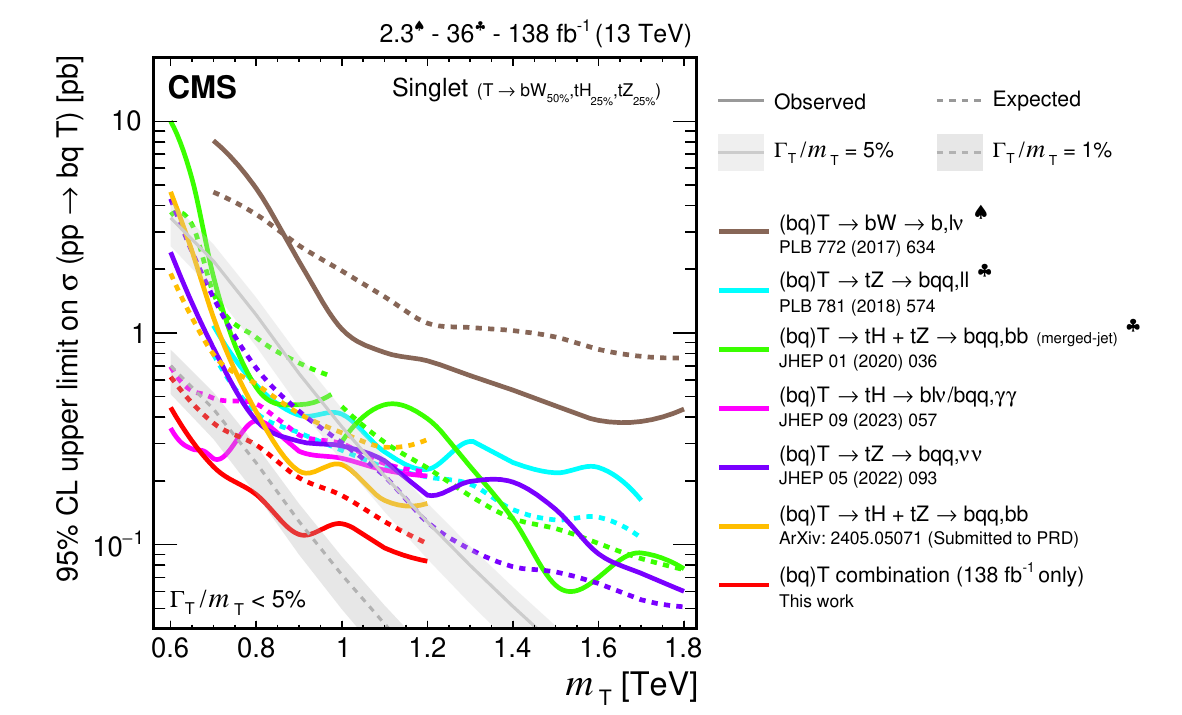} \\
\includegraphics[width=0.48\textwidth]{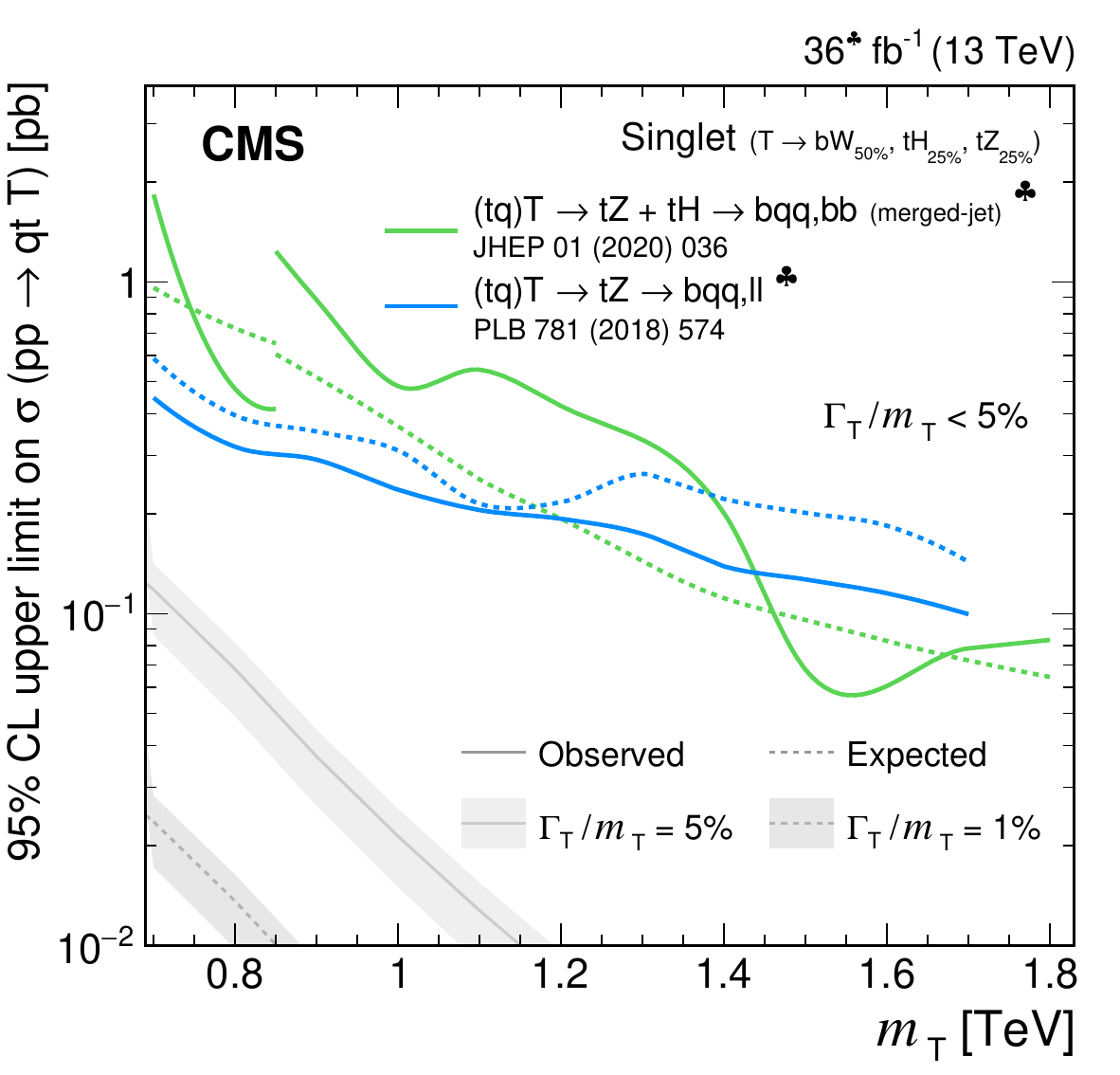}
\includegraphics[width=0.48\textwidth]{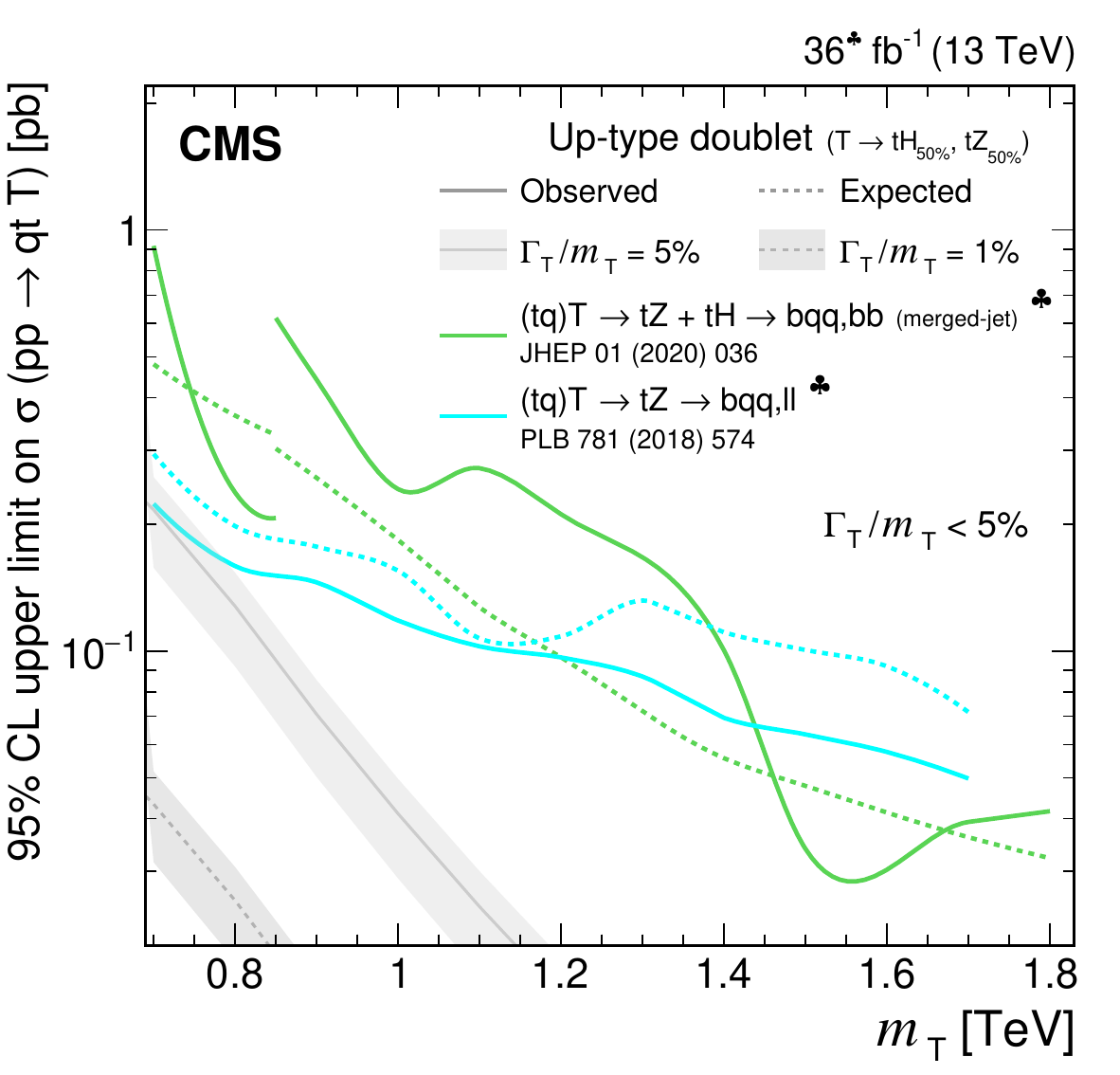}
\caption{%
    Observed and expected 95\% \CL upper limits on the production cross section of a single \PQT quark in association with a \PQb quark (upper) or a \PQt quark (lower row) in a singlet (upper and lower left) and doublet (lower right) scenario, versus the \PQT quark mass, obtained by different analyses: $\TtobW\to\bqq\,\lnu/\blnu\,\qq$~\cite{B2G-16-006}, $\TtotZ\to\bqq\,\ellell$~\cite{CMS:2017voh}, $\TtotH+\tZ\to\bqq\,\bb$ (merged-jet)~\cite{B2G-18-003}, $\TtotH\to\blnu/\bqq\,\gammagamma$~\cite{B2G-21-007}, $\TtotZ\to\bqq\,\nunu$~\cite{B2G-19-004}, $\TtotH+\tZ\to\bqq\,\bb$~\cite{B2G-19-001}, and the single \PQT quark combination of Section~\ref{sec:singleTcombo}.
    Only the three analyses using the full Run 2 data set are included in the single \PQT quark combination.
    Two theory predictions at LO in perturbative QCD are superimposed, corresponding to different VLQ widths.
    Searches using data corresponding to an integrated luminosity of 2.3\fbinv and 36\fbinv, rather than the full Run 2 integrated luminosity of 138\fbinv, are indicated with a heart and spade symbol, respectively, in the legend.
}
\label{fig:summarySingleT}
\end{figure}

\begin{figure}[htp!]
\centering
\includegraphics[width=0.48\textwidth]{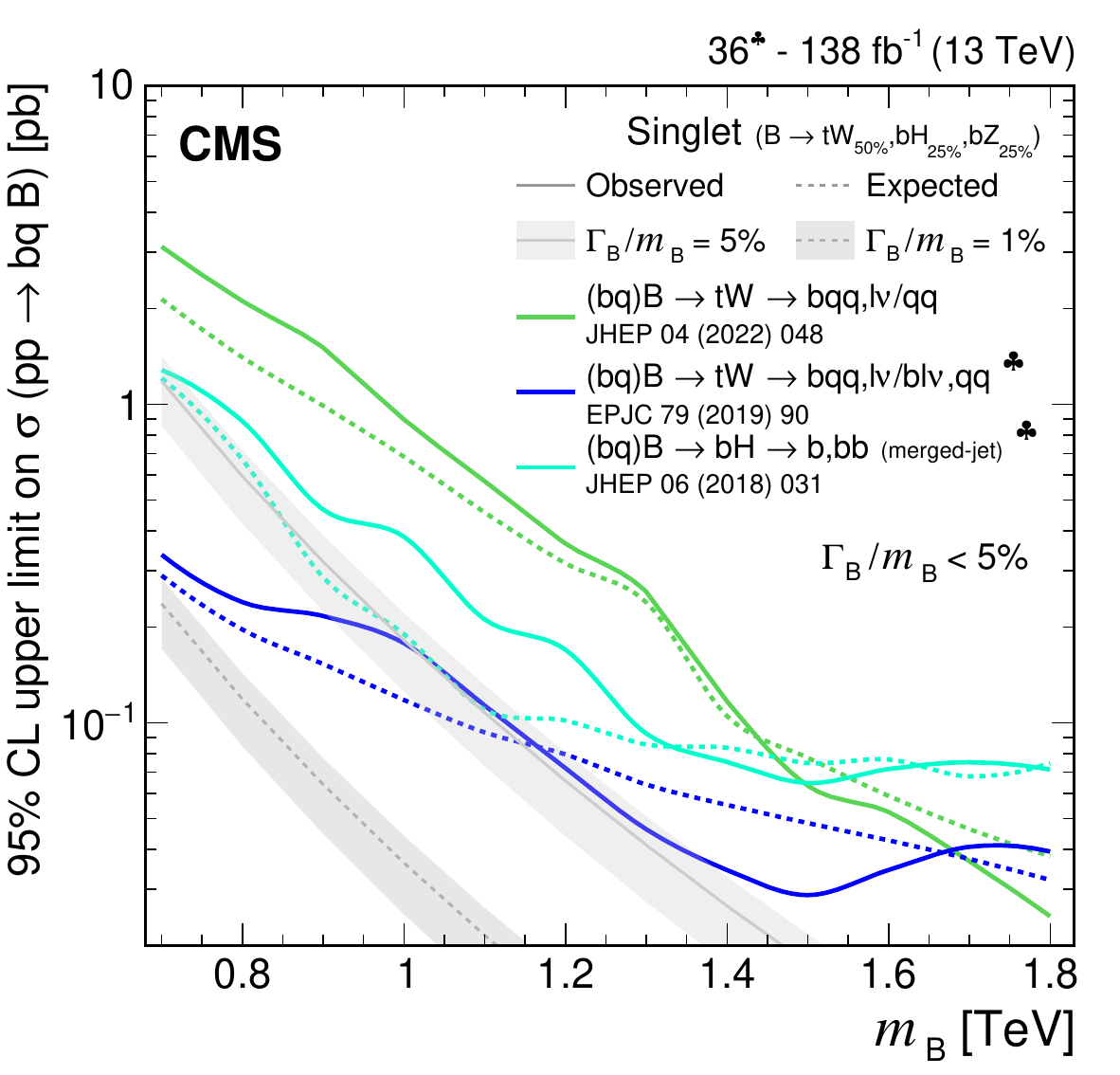}%
\hfill%
\includegraphics[width=0.48\textwidth]{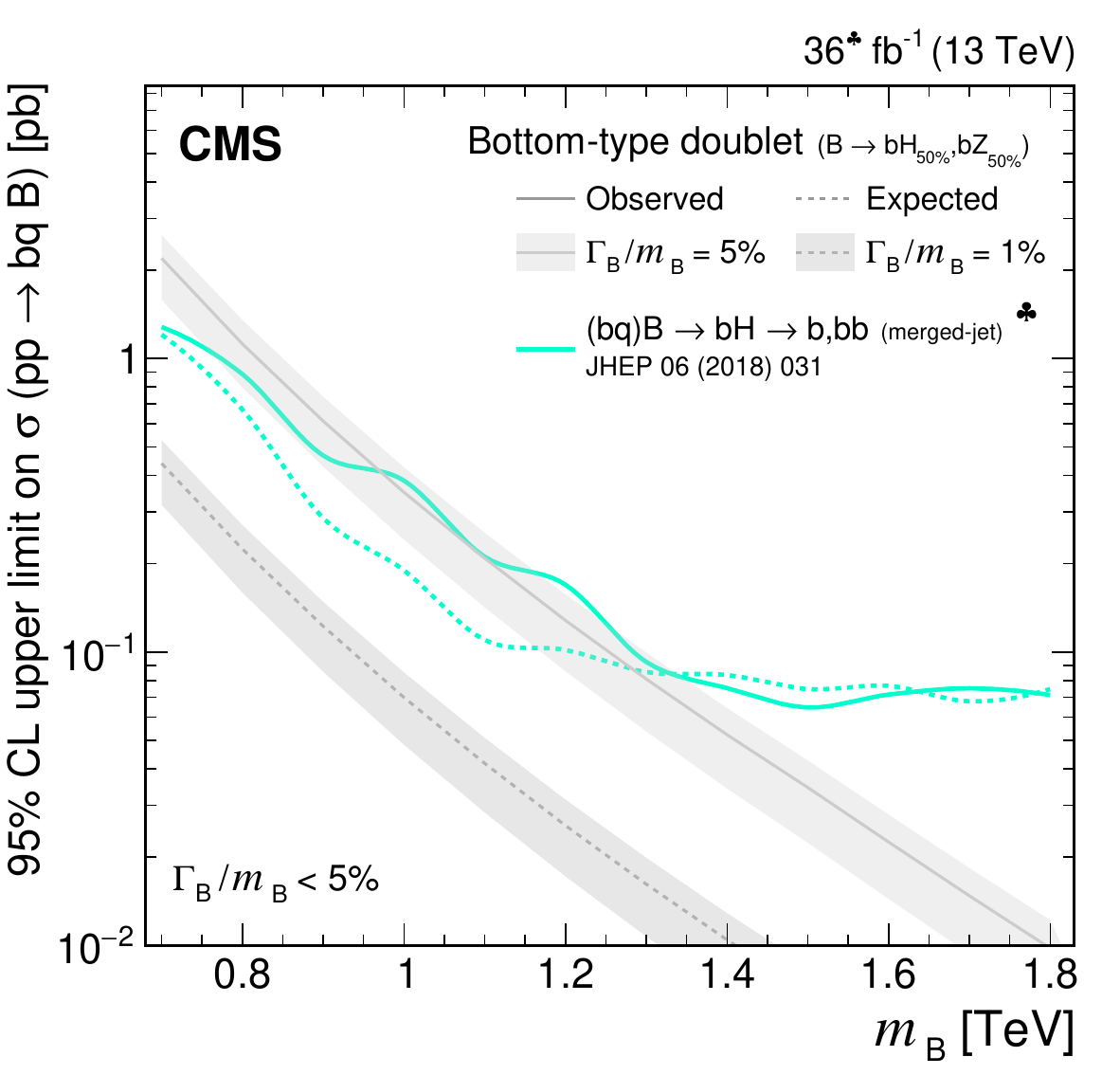} \\
\includegraphics[width=0.48\textwidth]{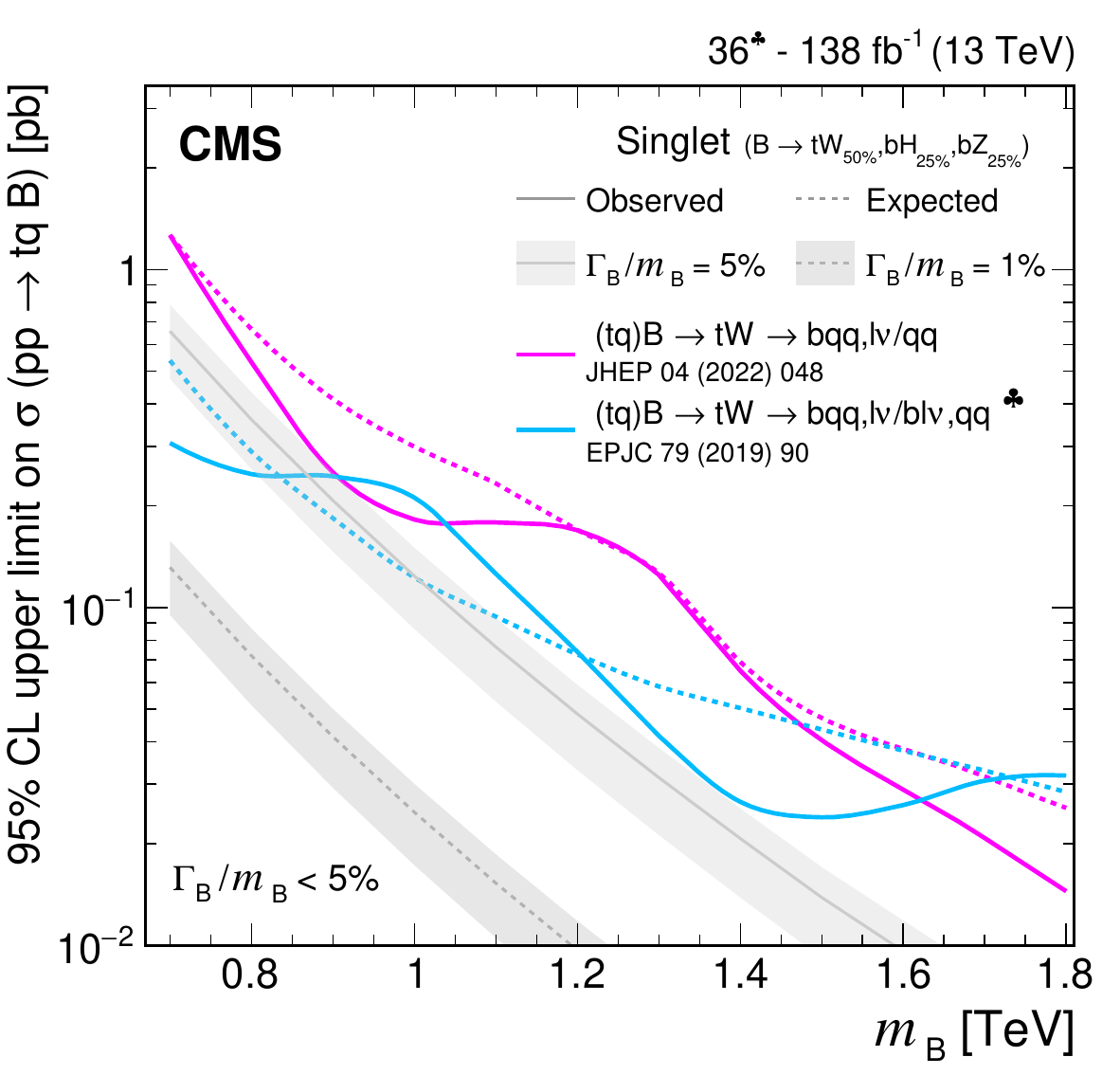}%
\caption{%
    Observed and expected 95\% \CL upper limits on the production cross section of a single \PQB quark in association with a \PQb quark (upper row) or a \PQt quark (lower) in a singlet (upper left and lower) and doublet (upper right) scenario, versus the \PQB quark mass, obtained by different analyses: $\BtotW\to\bqq\,\lnu/\qq$~\cite{B2G-20-010}, $\BtotW\to\bqq\,\lnu/\blnu\,\qq$~\cite{B2G-17-018}, and $\BtobH\to\PQb\,\bb$~\cite{B2G-17-009}.
    Two theory predictions at LO in perturbative QCD are superimposed, corresponding to different VLQ widths.
    Searches using data corresponding to an integrated luminosity of 36\fbinv, rather than the full Run 2 integrated luminosity of 138\fbinv, are indicated with a spade symbol in the legend.
}
\label{fig:summarySingleB}
\end{figure}

The upper limits on the cross section as a function of VLQ masses for single \PQT quark and \PQB quark production via the EW interaction are shown in Figs.~\ref{fig:summarySingleT} and \ref{fig:summarySingleB}.
In Fig.~\ref{fig:summarySingleT} (upper), the searches for a single \PQT quark in association with a \PQb quark are compared with cross section predictions in a singlet scenario assuming two different scenarios for the VLQ decay width.
For a relative width assumption of 5\%, multiple searches contribute with similar sensitivity to the mass exclusion limit.
The combination of single \PQT quark searches, described in Section~\ref{sec:singleTcombo}, thus significantly improves the sensitivity.
Figure~\ref{fig:summarySingleT} (lower) and Fig.~\ref{fig:summarySingleB} cover more single production modes for VLQs.

\begin{figure}[htp!]
\centering
\includegraphics[width=0.76\textwidth]{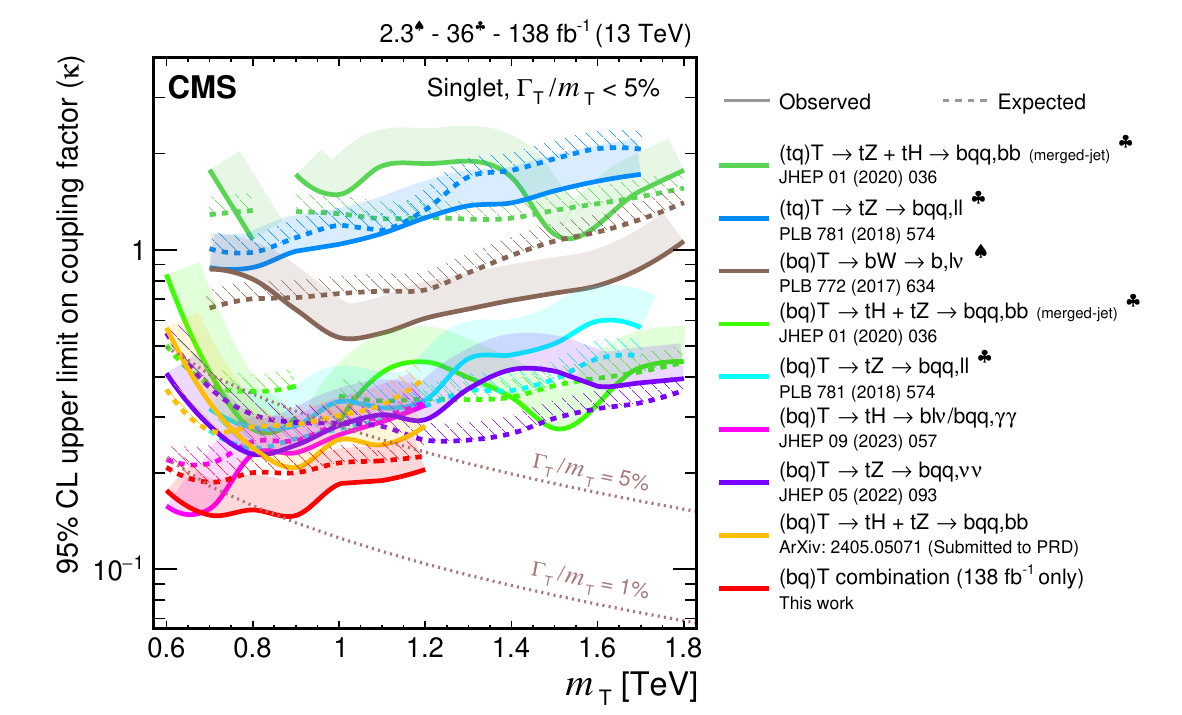} \\
\includegraphics[width=0.48\textwidth]{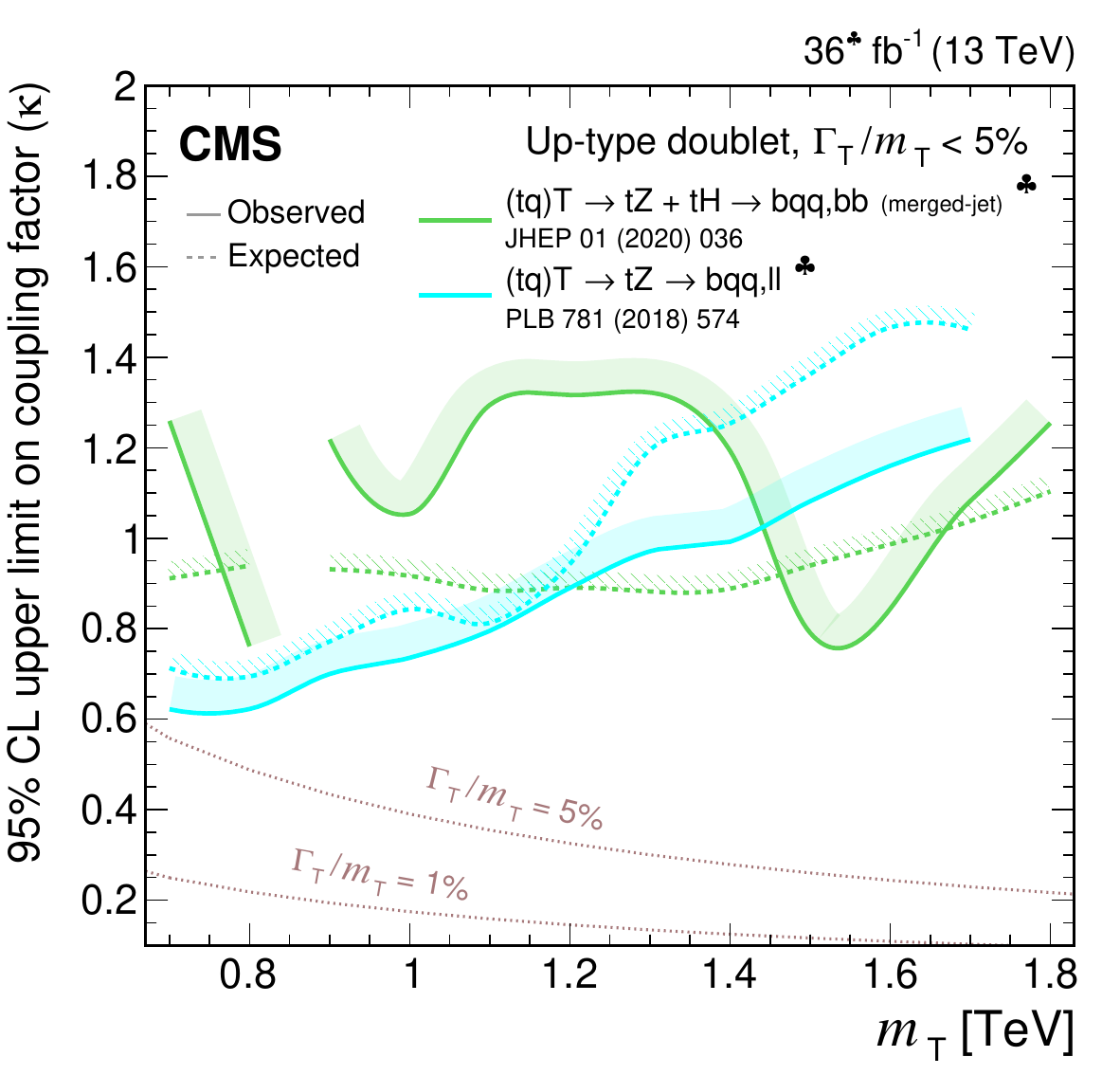}
\caption{%
    Observed and expected 95\% \CL upper limits on the coupling strength $\kappa$ for single \PQT quark production in a singlet (upper) and doublet (lower) scenarios as functions of the \PQT quark mass, obtained by different analyses: $\TtotH+\tZ\to\bqq\,\bb$ (merged-jet)~\cite{B2G-18-003}, $\TtotZ\to\bqq\,\ellell$~\cite{CMS:2017voh}, $\TtobW\to\bqq\,\lnu/\blnu\,\qq$~\cite{B2G-16-006}, $\TtotZ\to\bqq\,\ellell$~\cite{CMS:2017voh}, $\TtotH\to\blnu/\bqq\,\gammagamma$~\cite{B2G-21-007}, $\TtotZ\to\bqq\,\nunu$~\cite{B2G-19-004}, $\TtotH+\tZ\to\bqq\,\bb$~\cite{B2G-19-001}, and the single \PQT quark combination of Section~\ref{sec:singleTcombo}.
    Only the three analyses using the full Run 2 data set are included in the single \PQT quark combination.
    Two theory predictions at LO in perturbative QCD are superimposed, corresponding to different VLQ widths.
    Searches using data corresponding to an integrated luminosity of 2.3\fbinv and 36\fbinv, rather than the full Run 2 integrated luminosity of 138\fbinv, are indicated with a heart and spade symbol, respectively, in the legend.
}
\label{fig:summarySingleTCoupling}
\end{figure}

\begin{figure}[htp!]
\centering
\includegraphics[width=0.76\textwidth]{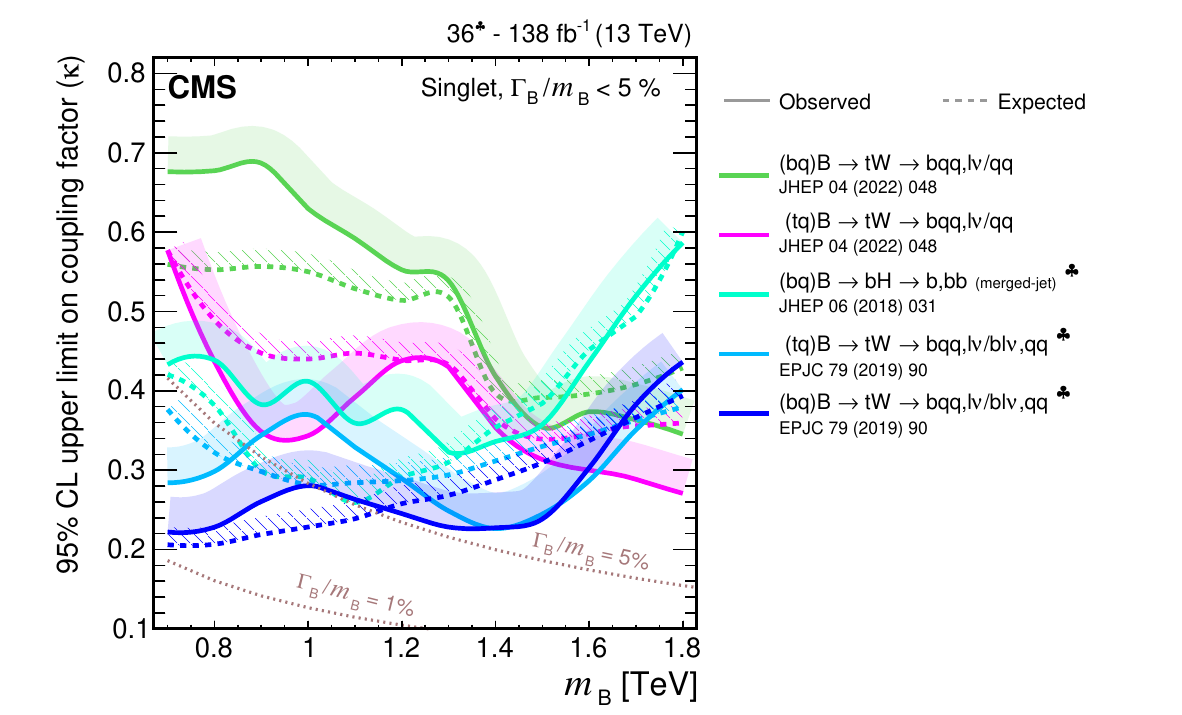} \\
\includegraphics[width=0.48\textwidth]{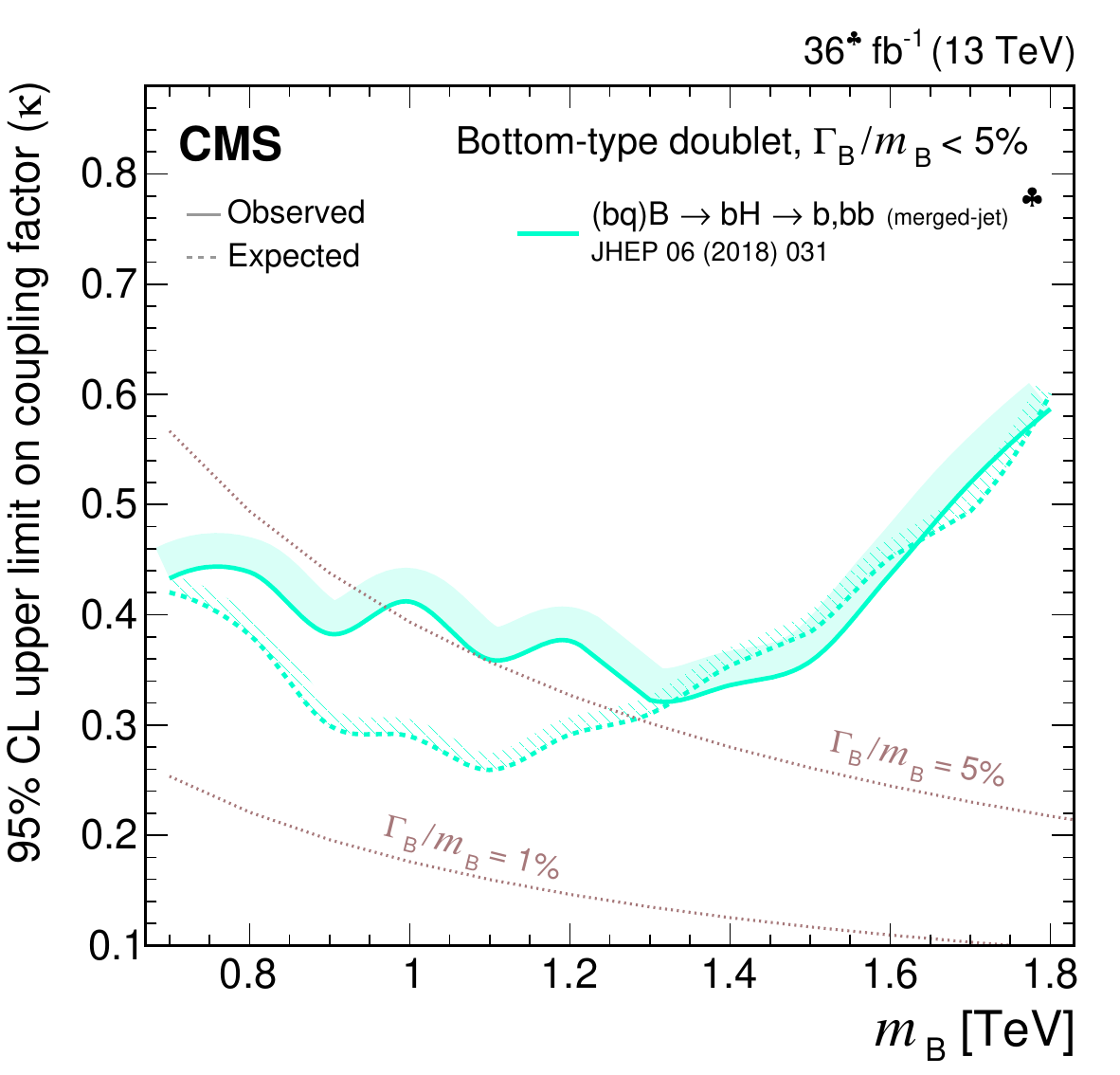}
\caption{%
    Observed and expected 95\% \CL upper limits on the coupling strength $\kappa$ for single \PQB quark production in a singlet (upper) and doublet (lower) scenarios as functions of the \PQB quark mass, obtained by different analyses: $\BtotW\to\bqq\,\lnu/\qq$~\cite{B2G-20-010}, $\BtobH\to\PQb\,\bb$~\cite{B2G-17-009}, and $\BtotW\to\bqq\,\lnu/\blnu\,\qq$~\cite{B2G-17-018}.
    Two theory predictions at LO in perturbative QCD are superimposed, corresponding to different VLQ widths.
    Searches using data corresponding to an integrated luminosity of 36\fbinv, rather than the full Run 2 integrated luminosity of 138\fbinv, are indicated with a spade symbol in the legend.
}
\label{fig:summarySingleBCoupling}
\end{figure}

\begin{figure}[htp!]
\centering
\includegraphics[width=0.48\textwidth]{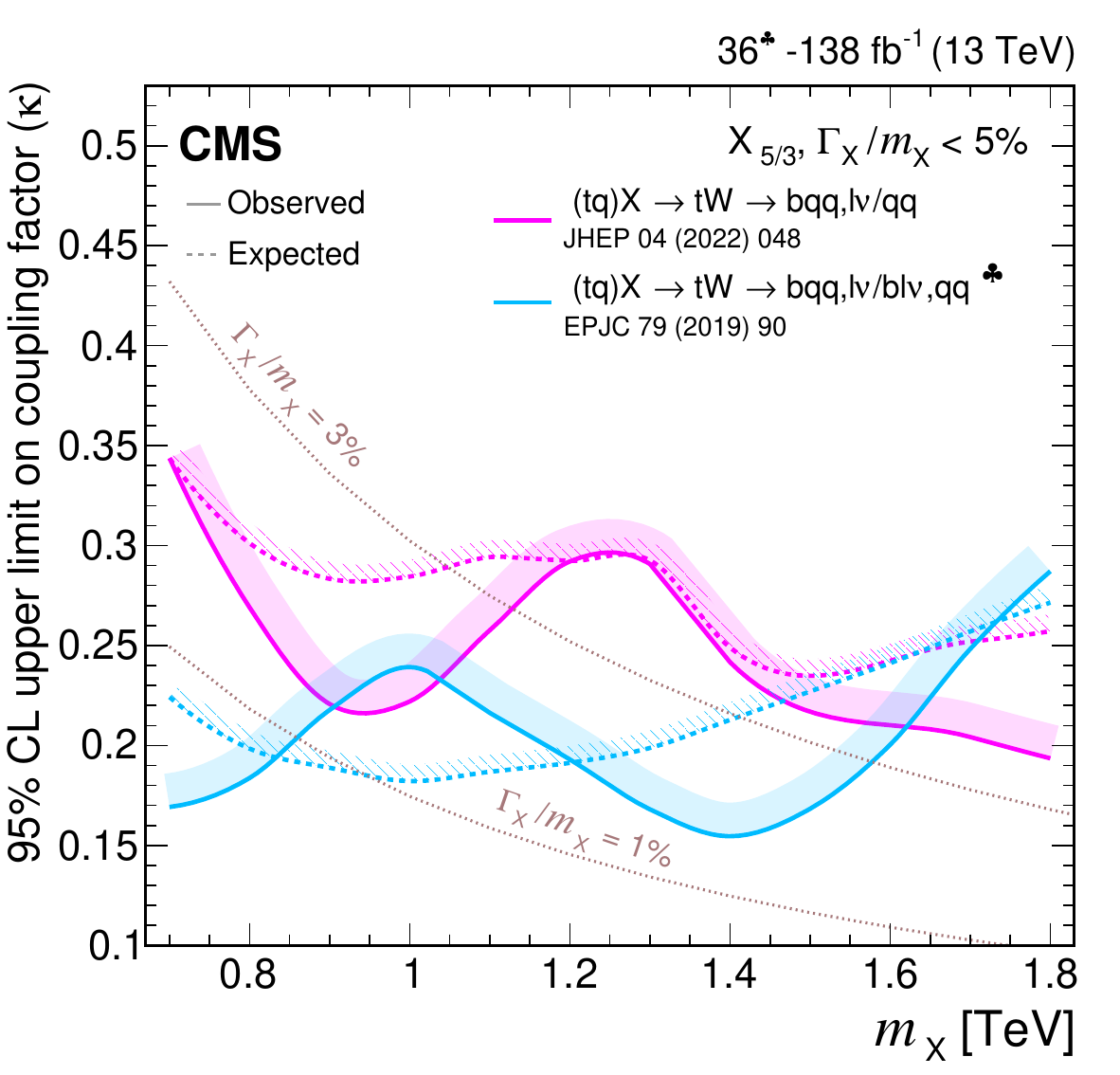}%
\hfill%
\includegraphics[width=0.48\textwidth]{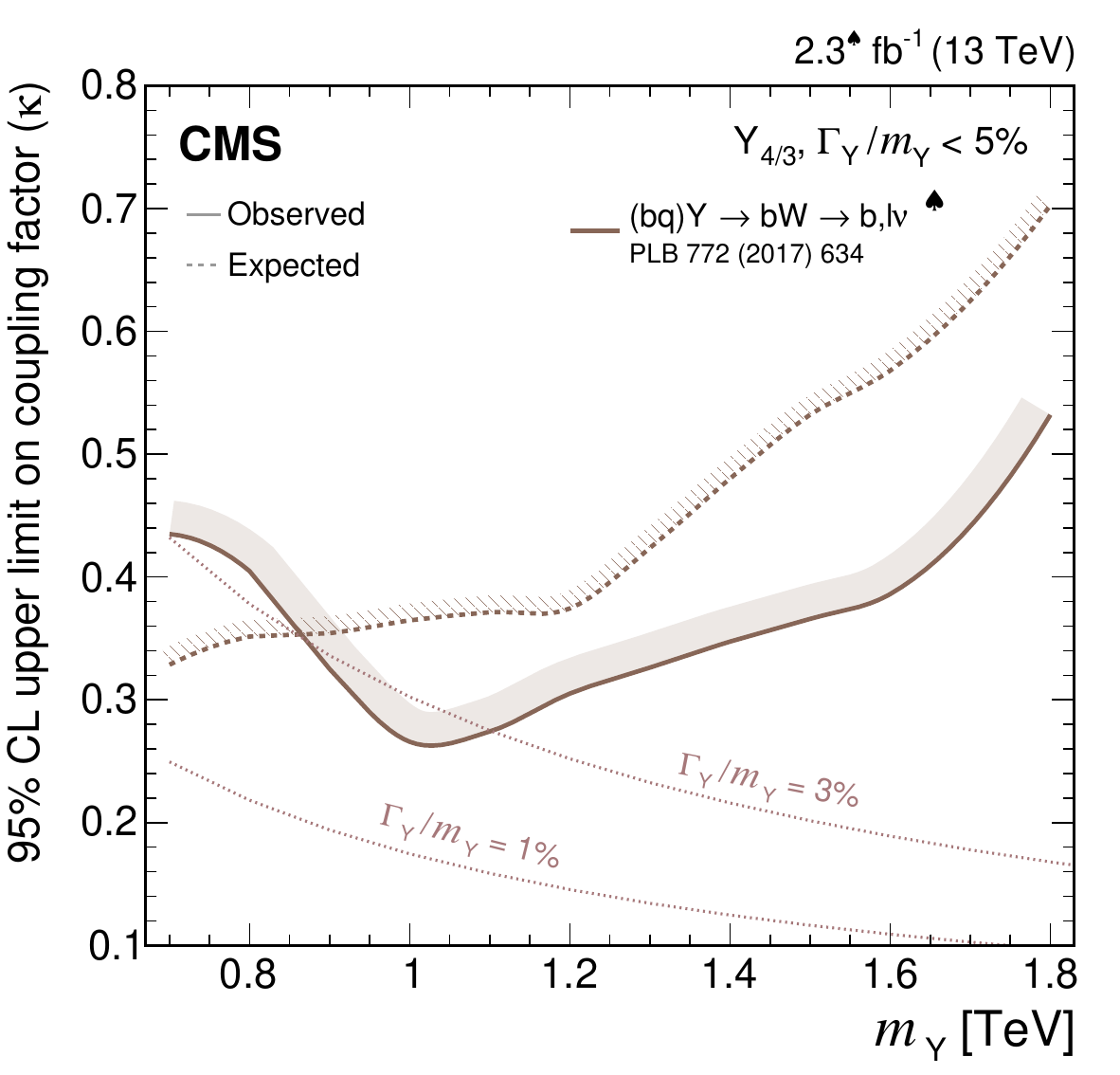}%
\caption{%
    Observed and expected 95\% \CL upper limits on the coupling strength $\kappa$ for single \xft (left) and \yft (right) production as functions of the VLQ mass, obtained by different analyses: $\xft\to\tW\to\bqq\,\lnu/\qq$~\cite{B2G-20-010}, $\xft\to\tW\to\bqq\,\lnu/\blnu\,\qq$~\cite{B2G-17-018}, and $\yft\to\PQb\PW\to\PQb\,\lnu$~\cite{B2G-16-006}.
    Searches using data corresponding to an integrated luminosity of 2.3\fbinv and 36\fbinv, rather than the full Run-s2 integrated luminosity of 138\fbinv, are indicated with a heart and spade symbol, respectively, in the legend.
}
\label{fig:summarySingleXCoupling}
\end{figure}

The upper limits on the cross section can be translated into upper limits on the coupling parameter $\kappa$ as  functions of the VLQ mass.
The results are shown in Figs.~\ref{fig:summarySingleTCoupling}--\ref{fig:summarySingleXCoupling} for singlet and doublet scenarios.
For single \PQT quark in the singlet scenario, couplings larger than 0.4 are excluded at 95\% \CL across the entire mass range, and at the lowest mass of 0.6\TeV, couplings as low as 0.15 are excluded.

In data collected from 2022--2025, known as Run 3, VLQ searches will benefit from a slight increase in the collision energy to 13.6\TeV,
which increases the cross section for VLQ production. Algorithms for identifying jets of various
flavors continue to improve, as do analysis techniques for reconstructing VLQ decays and the
theoretical models used to simulate VLQ production. For standard decays of VLQs to third
generation particles, Run 3 is expected to push the sensitivity to VLQs toward the 2\TeV range.
The CMS experiment is also beginning to explore exotic decays of VLQs, as introduced in Section~\ref{subsec:theory_vlq}.
Run 3 will offer a rich data set from which to extract new information about VLQs with a broad variety of decays.

\subsection{Future prospects for VLQ searches at the HL-LHC}
\label{sec:futurevlq}

The physics capabilities of the Phase-2 upgrade of CMS for the HL-LHC have been studied by projecting many existing searches to $\sqrt{s}=14\TeV$ \pp collisions, and assuming a final integrated luminosity of 3000\fbinv. Searches for VLQs typically rely on identifying \PQb quarks and hadronic decays of boosted particles within jets, all of which incorporate track and vertex information.
Future VLQ searches could benefit from the expanded coverage of the CMS tracking detector.

A single-lepton search for \PQT quarks originally performed using the 2016 data set~\cite{B2G-17-011} has been projected to HL-LHC conditions, considering the operational conditions of the CMS Phase-2 detector~\cite{CMS-PAS-FTR-22-002}.

{\tolerance=800
To study signal and background processes with HL-LHC conditions,
\pp collision events are simulated using the \MGvATNLO 2.6.5~\cite{Alwall:2014hca} event generator, interfaced with \PYTHIA8~\cite{Sjostrand:2007gs} for parton showering. 
Signal event samples are simulated assuming \PQT quark masses between 1000 to 3000\GeV, with equal branching fractions for each third-generation decay mode, such that all decay modes are well populated by simulated events.
The \delphes~\cite{deFavereau:2013fsa} program is used to simulate the Phase-2 CMS detector response. 
Since the decay mode could not be accessed in the \delphes simulations, this projection is performed using the assumption of equal branching fractions for each third-generation decay mode.
\par}

The \delphes program uses the anti-\kt algorithm~\cite{Cacciari:2008gp} to cluster hadronic jets,
with distance parameters of 0.4 and 0.8, similiar to the treatment used in the reconstruction in the
CMS experiment.
The \DeepJet \PQb tagging algorithm was emulated in \delphes by providing the efficiency for 
light, \PQc, or \PQb quark jets to be identified as \PQb-tagged jets. In this analysis, a small-radius 
jet is considered \PQb tagged if it passes the emulation of the medium \DeepJet working point.
A large-radius jet is identified as originating from the hadronic decay of a \PW boson if it has $\pt>200\GeV$, \mSD ranging from 60 to 110\GeV, and an $N$-subjettiness ratio of $\tauTO<0.55$.
Large-radius jets are identified as \PH tagged if they have $\pt>300\GeV$, \mSD ranging from 60 to 160\GeV, and at least one small-radius jet that overlaps with the large-radius jet is \PQb tagged.

Selected events must have exactly one high-quality, isolated electron or muon, zero additional looser-quality leptons, and
$\ptmiss>75\GeV$.
Three or more small-radius jets are required, with $\pt>300$, 150, and 100\GeV, and at least one jet must be \PQb tagged.
If no \PW- or \PH-tagged jets are found in the event, an additional small-radius jet is required with $\pt>30\GeV$.
Two or more large-radius jets are required, and the SR contains events in which the minimal angular separation between the highest \pt large-radius jet and another large-radius jet is $\DR<3$, excluding many background events with a back-to-back topology of $\DR=\pi$.

\begin{figure}[!htp]
\centering
\includegraphics[width=0.48\textwidth]{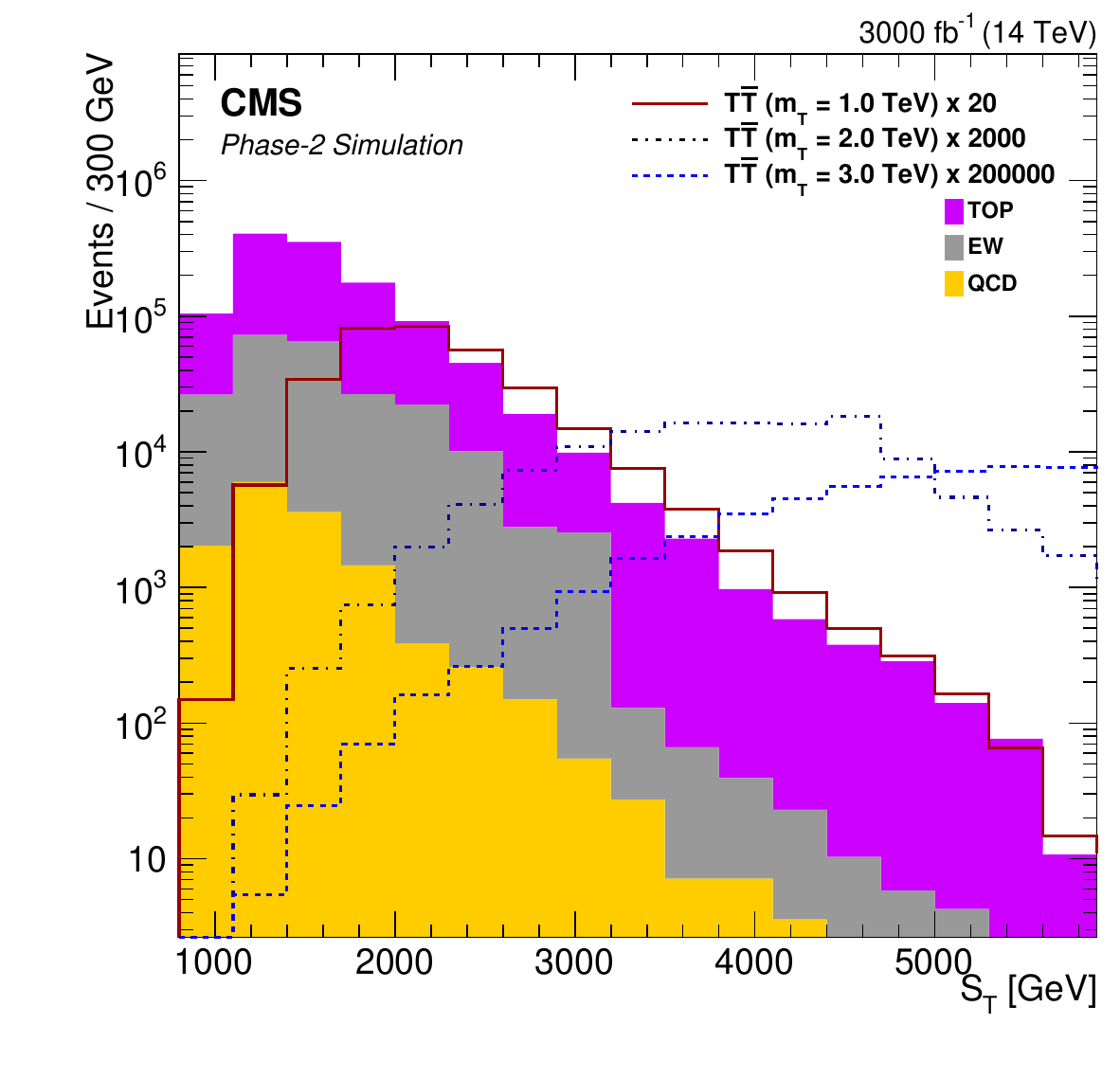}%
\hfill%
\includegraphics[width=0.48\textwidth]{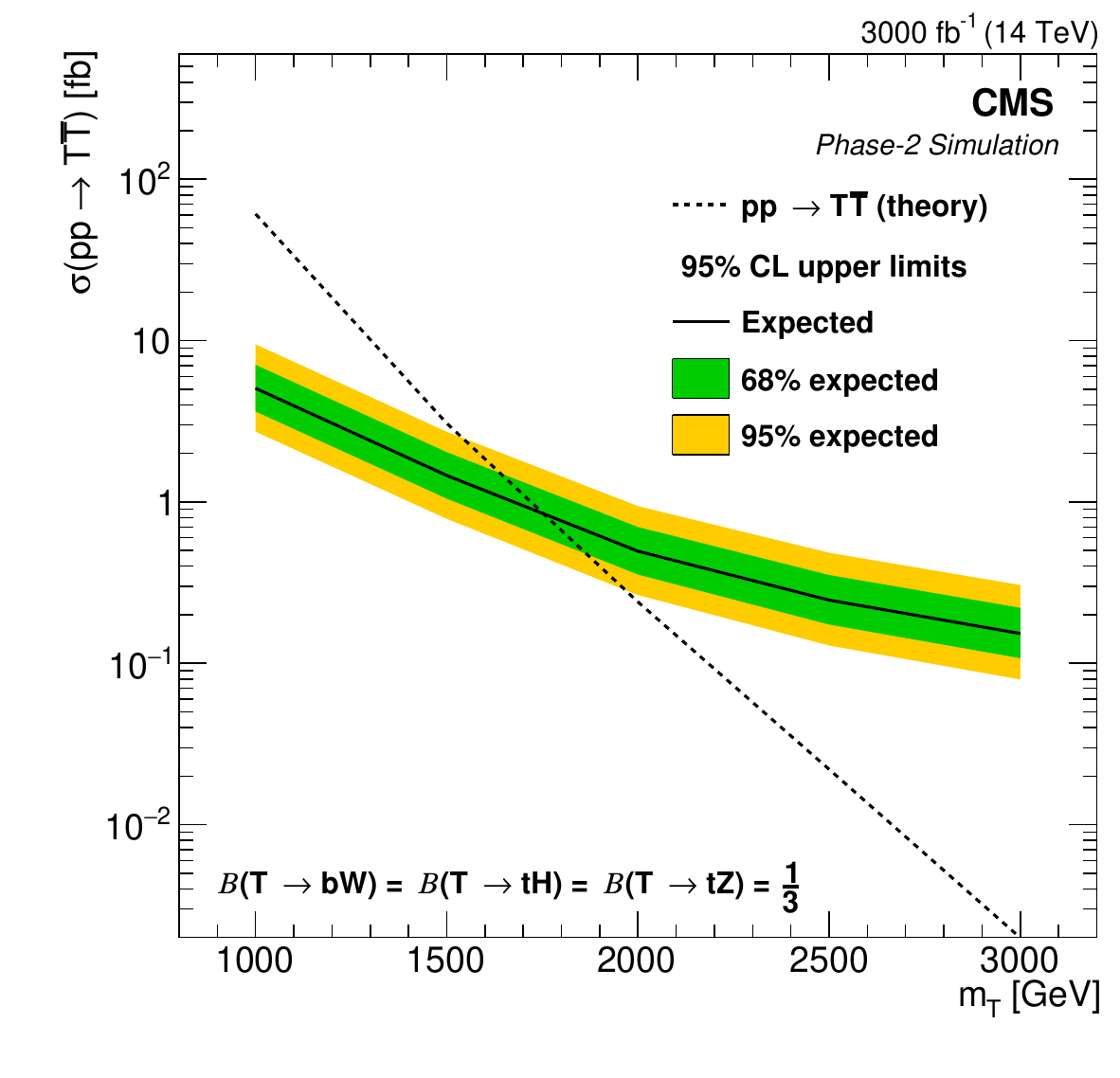}%
\caption{%
    Distributions of the \ST observable for signal and background processes (left), with signal distributions scaled by factors of 20, 2000, and 200\,000, depending on the \PQT quark mass, and expected upper limits at 95\% \CL on the \TTbar production cross section (right).
    The inner (green) and the outer (yellow) bands indicate the regions containing 68 and 95\%, respectively, of the distribution of limits expected under the background-only hypothesis.
    Figures adapted from Ref.~\cite{CMS-PAS-FTR-22-002}.
}
\label{fig:stdistsr}
\end{figure}

Events are categorized into eight different SRs based on the
number of \Pb-tagged, \PW-tagged, and \PH-tagged jets. Events with \PH-tagged jets are further separated
based on the number of \PQb-tagged subjets of the \PH-tagged jet.
Figure~\ref{fig:stdistsr} (left) displays
the \ST distributions for signal and background events in the eight SRs combined.

\begin{figure}[!htp]
\centering
\includegraphics[width=0.48\textwidth]{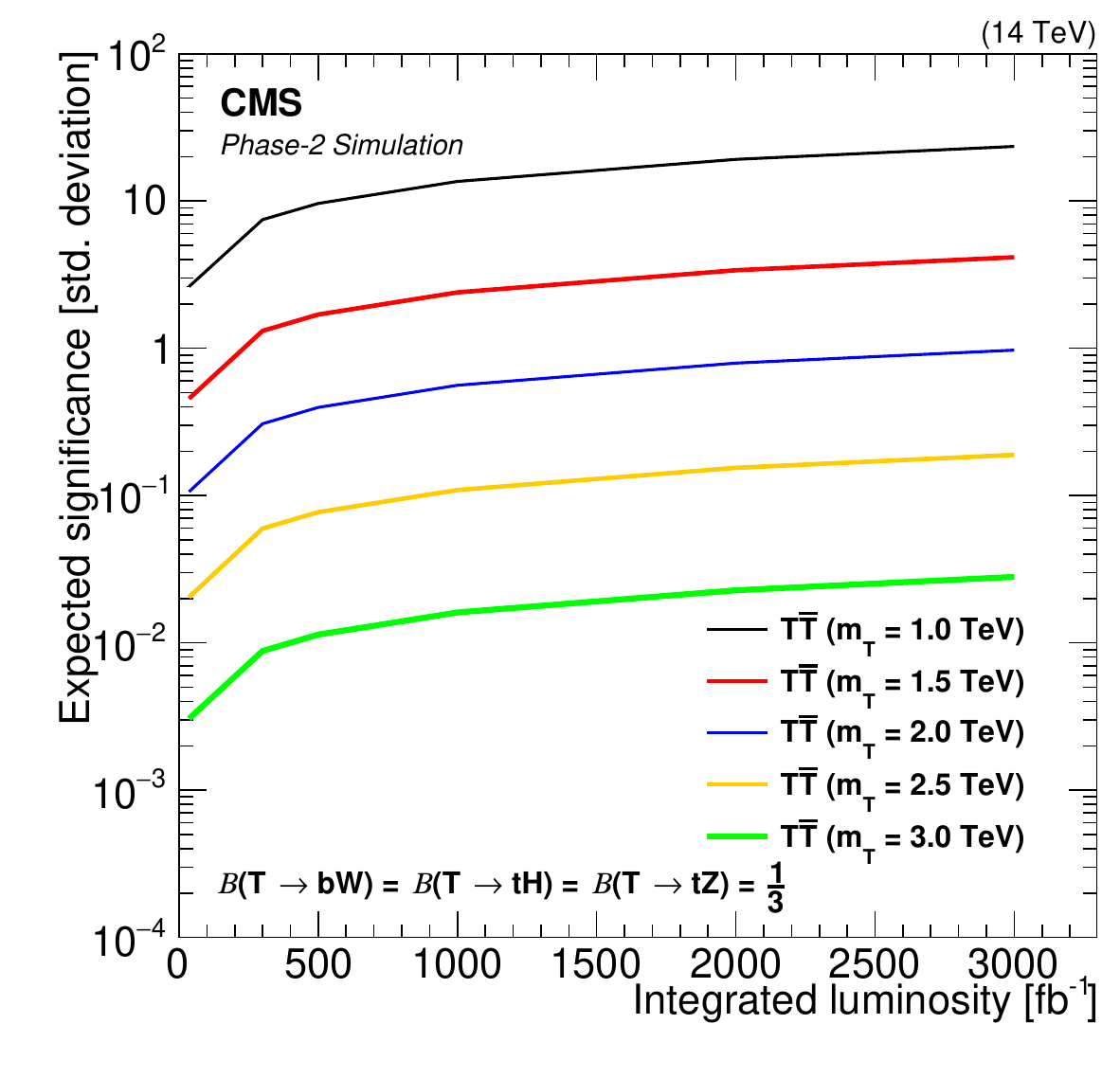}%
\hfill%
\includegraphics[width=0.48\textwidth]{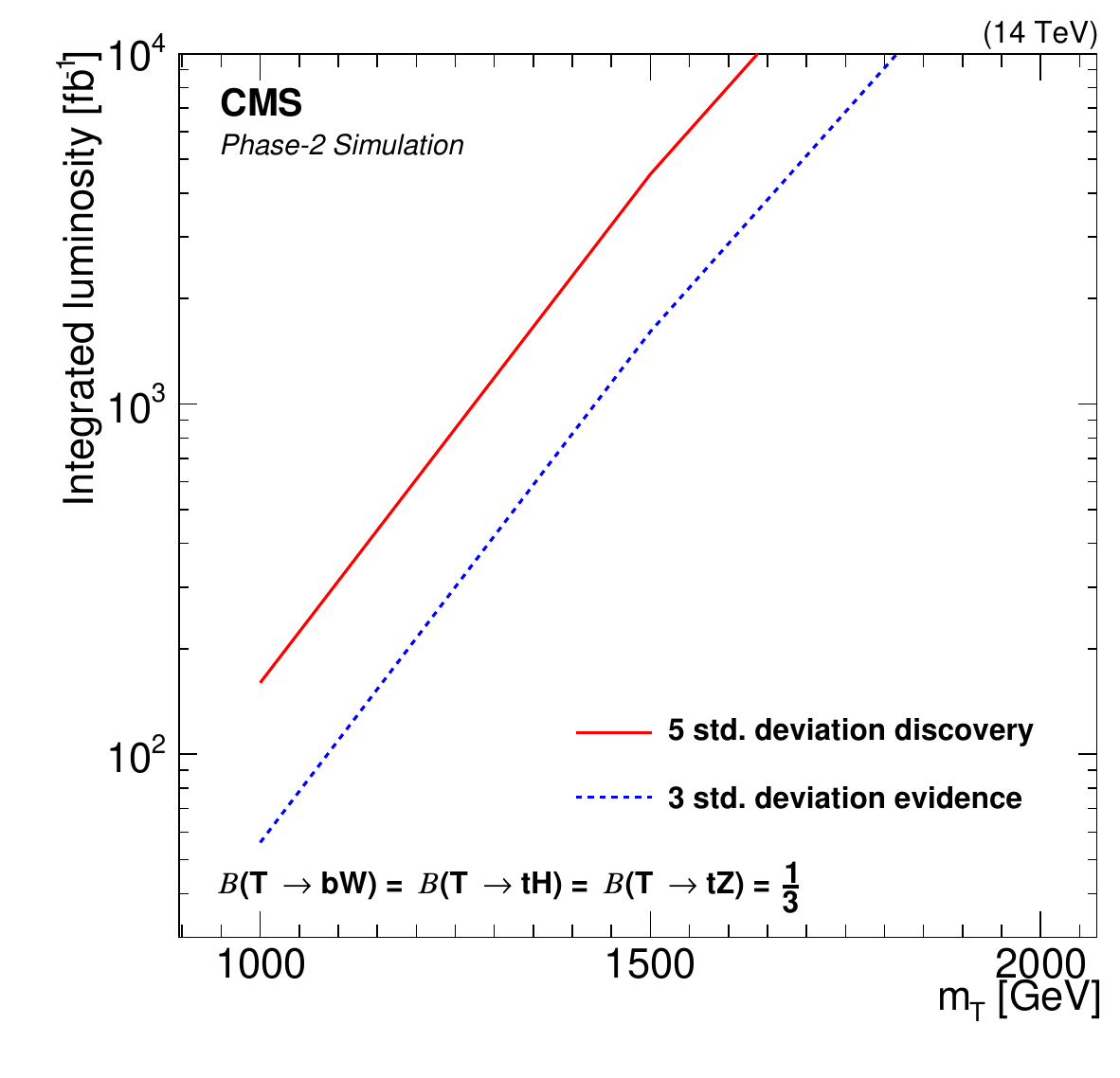}%
\caption{%
    Expected significances for \PQT quark pair production as a function of the integrated luminosity at the HL-LHC, assuming equal branching fractions for \TtobW, \tZ, \tH decays (left).
    Discovery potential at three and five standard deviations for \PQT quark pairs, as a function of the \PQT quark mass and the integrated luminosity (right).
    Figures adapted from Ref.~\cite{CMS-PAS-FTR-22-002}.
}
\label{fig:significancelumiproj}
\end{figure}

Following the Yellow Report of Ref.~\cite{yellow_report_2019}, experimental uncertainties for signal and
background yields are included.
Using a simultaneous maximum likelihood fit of the \ST distribution in the eight SRs
for each of the \PQT quark mass points under consideration, upper limits on the \TTbar production cross section are calculated.
The results are displayed in Fig.~\ref{fig:stdistsr} (right).
Figure~\ref{fig:significancelumiproj} displays the expected significance as a function of the
HL-LHC integrated luminosity, as well as the integrated luminosity required for the discovery of the
\PQT quark at an expected significance of three and five standard deviations.

At the HL-LHC, with 3000\fbinv of data, the discovery of the \PQT quark with a significance of five standard deviations significance
may be achieved for masses up to 1440\GeV. In the absence of signal, this study projects a 95\% \CL exclusion for \PQT quarks with masses below 1750\GeV.
Compared to the expected lower mass limits from the original 2016 search of Ref.~\cite{B2G-17-011}, the limits derived in this projection study are
more stringent by approximately 600\GeV. The \TTbar searches summarized in Section~\ref{sec:TTVLQ}, particularly the single-lepton search
of Ref.~\cite{B2G-20-011}, already show significant increases in sensitivity
due to jet tagging algorithms using ML techniques, as well as other analysis improvements developed during Run 2. This projection therefore
provides insight into how the reach of those searches might improve in the future with the HL-LHC data taking.

\clearpage
\section{Theoretical motivation for vector-like leptons}
\label{sec:vll}

{\tolerance=800
Vector-like leptons (VLLs) are color-singlet counterparts of VLQs, \ie, leptons with left- and right-handed components transforming in the same way under the EW gauge symmetry group.
Such new states arise in a wide variety of BSM scenarios, including but not limited to supersymmetric models~\cite{Martin:2009bg,Graham:2009gy,Endo:2011mc,Zheng:2019kqu,Aguilar-Saavedra:2009fxa}, models with extra spatial dimensions~\cite{Kong:2010qd,Huang:2012kz}, and grand unification~\cite{Nevzorov:2012hs,Dorsner:2014wva,Joglekar:2016yap}.
Expansions of the SM with one or more vector-like fermion families may provide a dark matter candidate~\cite{Schwaller:2013hqa,Halverson:2014nwa,Bahrami:2016has,Bhattacharya:2018fus}, and account for the mass hierarchy between the different generations of particles in the SM via their mixings with the SM fermions~\cite{Agashe:2008fe,Redi:2013pga,Falkowski:2013jya}.
Furthermore, VLLs are also among the proposed solutions~\cite{Endo:2011mc,Dermisek:2013gta, Megias:2017dzd,Kawamura:2019rth,Hiller:2020fbu} to the observed tensions between the experimental measurements and the SM prediction of the anomalous magnetic moment of the muon~\cite{Muong-2:2006rrc,Muong-2:2021ojo}.
\par}

Vector-like leptons may be \SU2 doublets $\VLLL=(\VLLE,\VLLN)$ or singlets \VLLE, where \VLLE and \VLLN denote the electrically charged and neutral states, respectively. These heavy lepton states are independent of possible neutrino mass generation mechanisms~\cite{Aguilar-Saavedra:2009fxa}.    
In the doublet models, the \VLLE and \VLLN are mass-degenerate at tree level and may be pair produced, $pp\to\VLLE\VLLEbar/\VLLN\VLLNbar$, or produced in association, $pp\to\VLLEbar\VLLN/\VLLE\VLLNbar$.
Only the $pp\to\VLLE\VLLEbar$ production mode is available for the singlet model.

Prior to direct searches at the LHC, a lower bound of about 100\GeV was placed by the L3 Collaboration at the CERN LEP collider on such additional heavy lepton states~\cite{Achard:2001qw}.

\subsection{Minimal models with VLLs}
\label{subsec:vll_minimal_theory}

In minimal extensions of the SM with VLLs, the newly introduced states are assumed to mix through Yukawa interactions with the leptons of the SM and decay into SM boson-lepton pairs~\cite{delAguila:1982fs,Fishbane:1985gu,Fishbane:1987tx,Montvay:1988av,Fujikawa:1994we,Kumar:2015tna,Bhattiprolu:2019vdu}. Electroweak precision data allows the mixing angle between VLLs and SM leptons to be ${\lesssim}10^{-2}$, permitting prompt decays for mass values in the neighborhood of the EW scale~\cite{Dermisek:2014cia,Dermisek:2014qca}.
In the doublet model, these decay modes are $\VLLE\to\PZ\Pell$ and $\PH\Pell$, and $\VLLN\to\PW\Pell$, with the branching fractions of \VLLE dependent on the mass \mVLLE.
Similarly, \VLLE in the singlet model may decay to $\PZ\Pell$, $\PH\Pell$, and $\PW\PGn$, with the branching fractions also governed by \mVLLE.

\begin{figure}[ht!]
\centering
\includegraphics[width=0.4\textwidth]{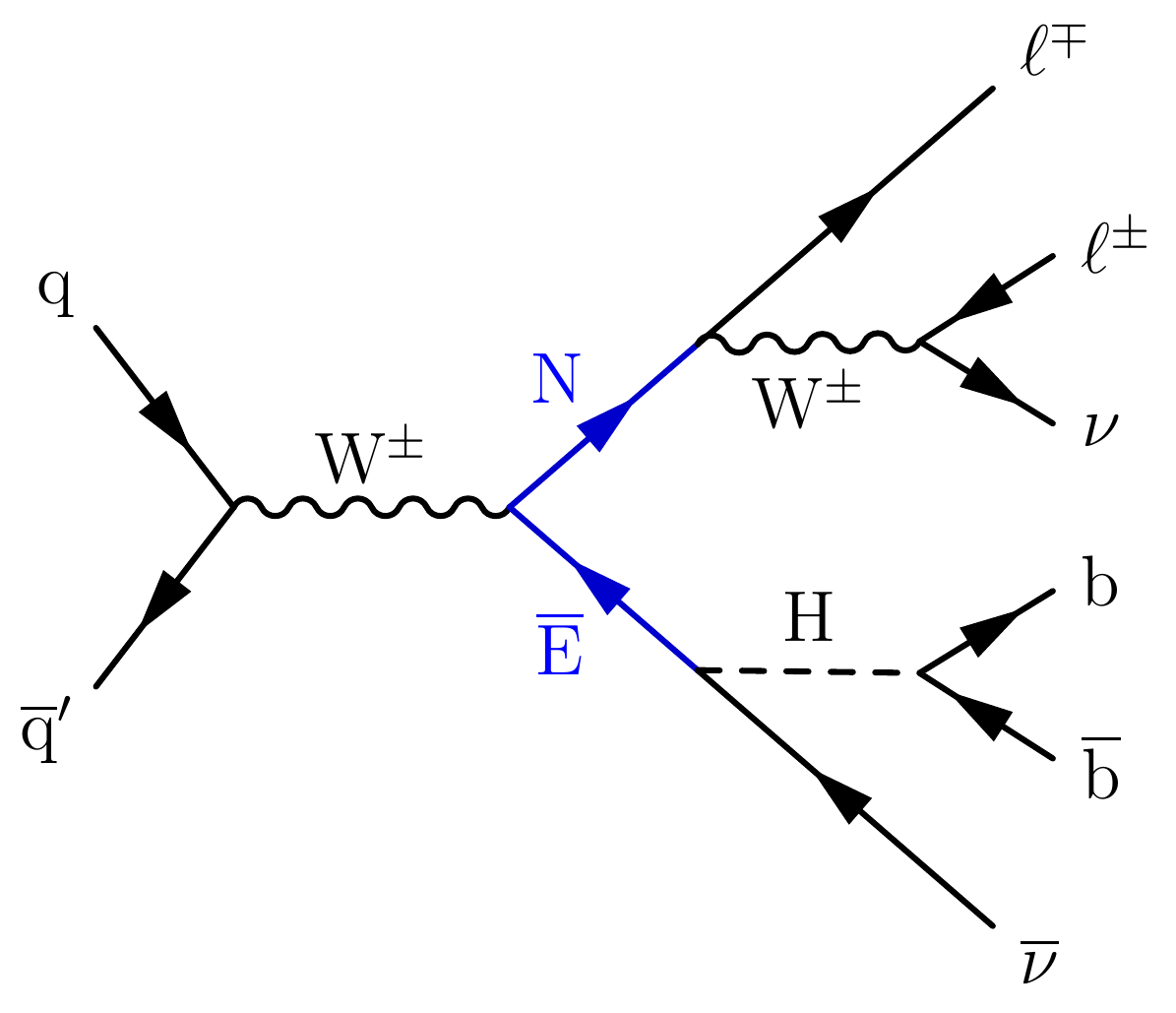}%
\hspace*{0.05\textwidth}%
\includegraphics[width=0.4\textwidth]{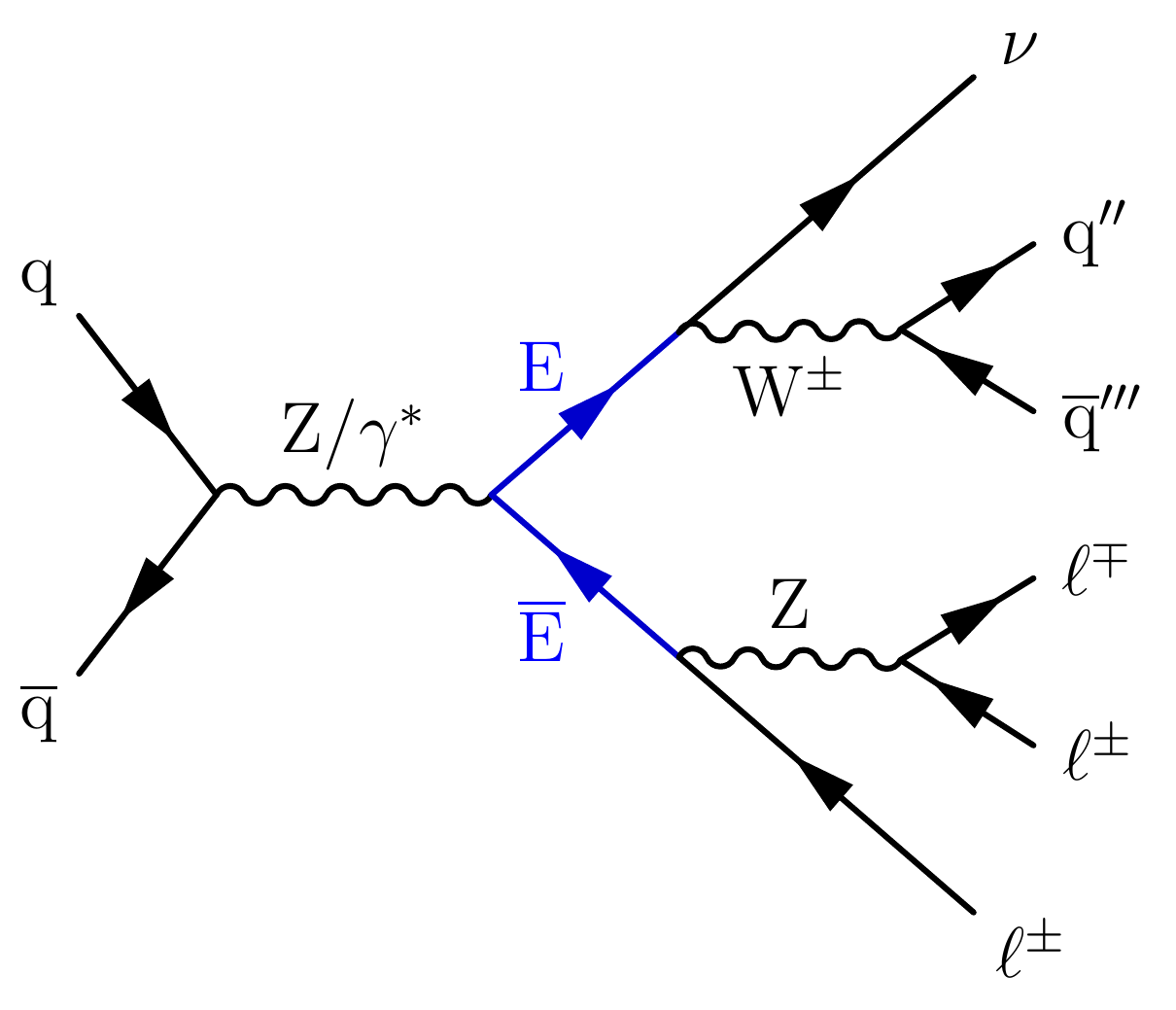}%
\caption{%
    Example processes illustrating production and decay of doublet (left) and singlet (right) VLL pairs at the LHC that result in multilepton final states.
}
\label{fig:VLLFeynman}
\end{figure}

{\tolerance=800
An example of a complete decay chain for the associated production in the doublet scenario is $\VLLN\VLLEbar\to\PWp\Pellm\PH\Pellp\to\Pellp\PGn\Pellm\bbbar\Pellp$ and for the pair production in the singlet scenario would be $\VLLE\VLLEbar\to\PWm\PGn\PZ\Pellp\to\PQqpr\PAQq\PGn\Pellp\Pellm\Pellp$.
Figure~\ref{fig:VLLFeynman} illustrates these two decay chains, which exemplify the production and decay of VLL pairs that result in multilepton final states.
\par}

\subsection{The 4321 model with VLLs}
\label{subsec:vll_4321_theory}

The 4321 model~\cite{4321_model,benchmark,threeSite_PS,nonuniversal4321} is an ultraviolet-complete (UV) model that extends the SM gauge groups to a larger $\SU4\times\SU3^\prime\times\SU2_{\mathrm{L}}\times\mathrm{U}(1)^\prime$ group, which then gives rise to the apparent groups in the SM after spontaneous symmetry breaking.
The lightest new particles in this model are the VLLs. The model furthermore contains additional heavier gauge boson \PZpr and vector leptoquark U states.

\begin{figure}[ht!]
\centering
\includegraphics[width=0.24\textwidth]{Figure_001-b.pdf}%
\hfill%
\includegraphics[width=0.24\textwidth]{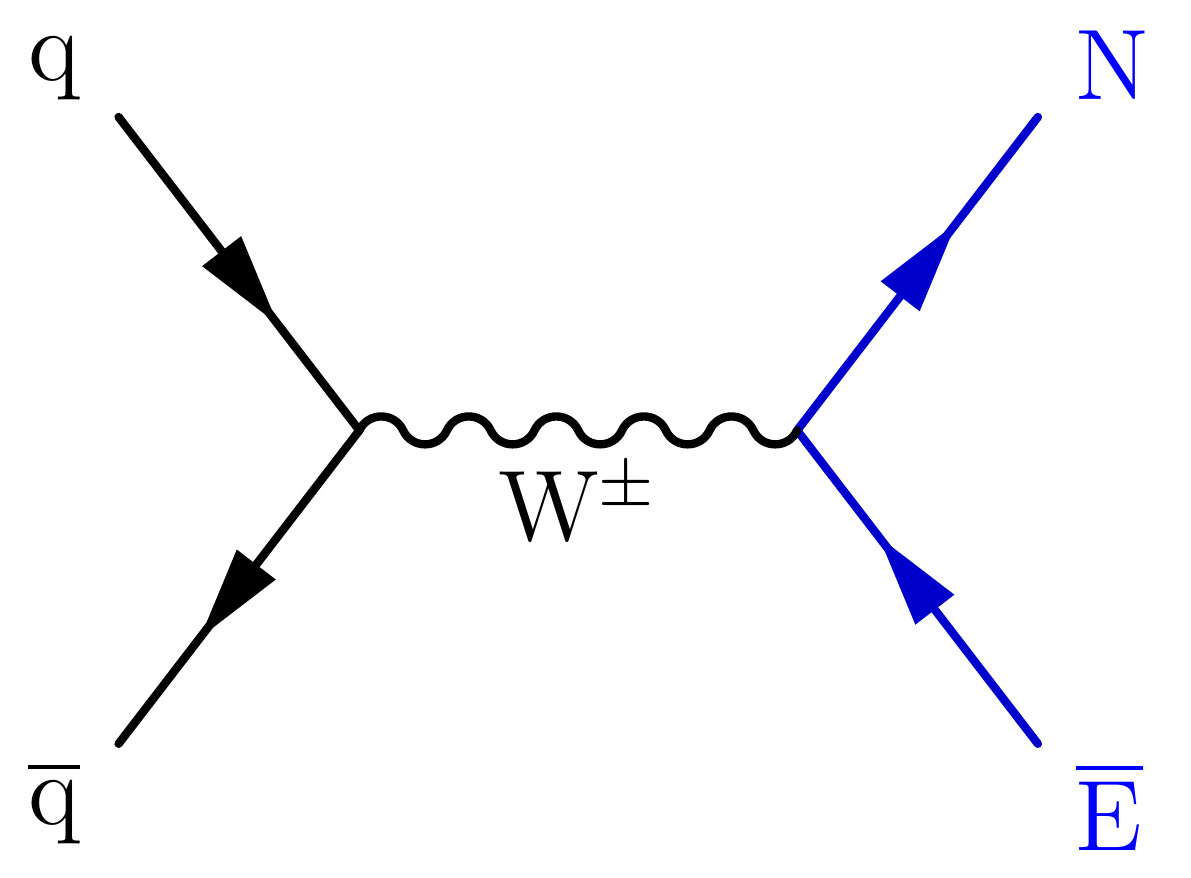}%
\hfill%
\includegraphics[width=0.24\textwidth]{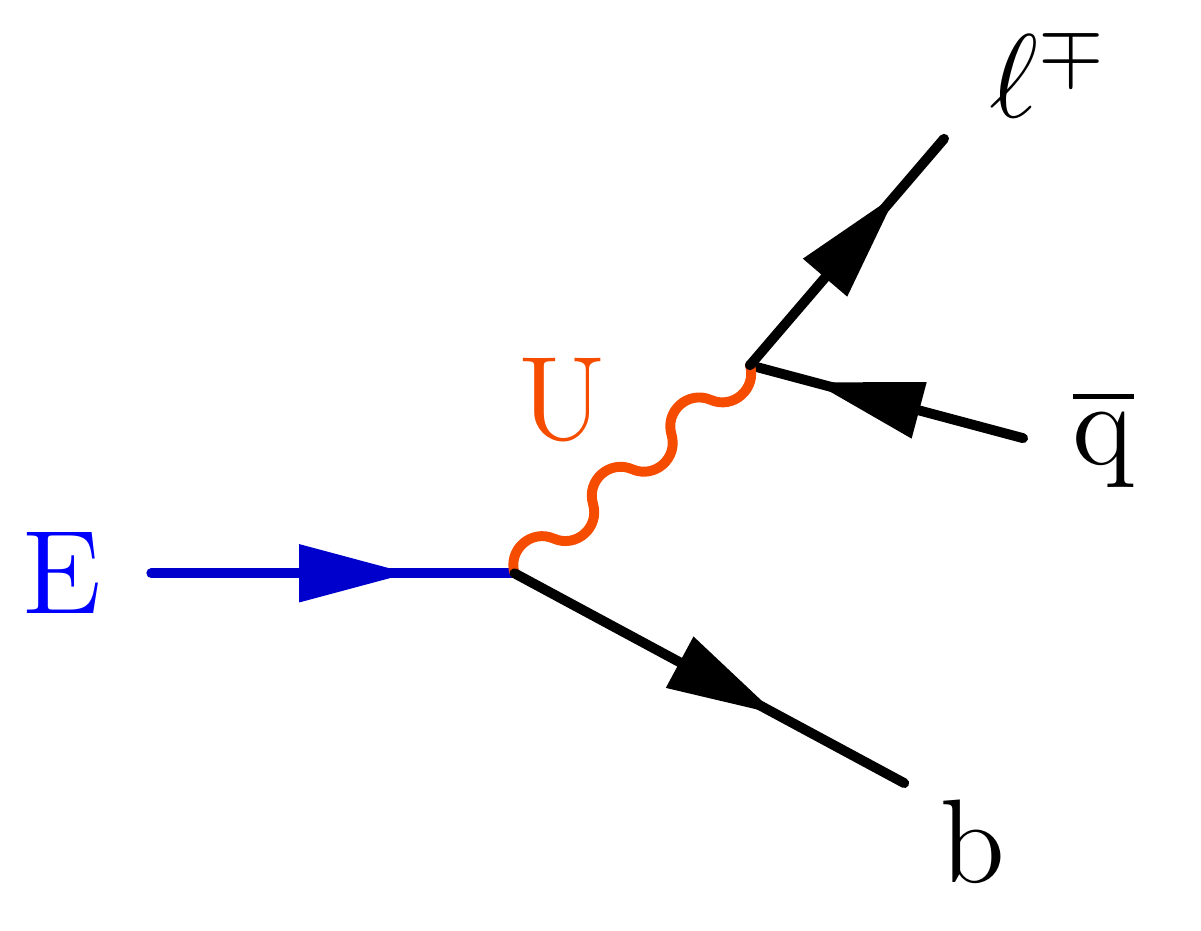}%
\hfill%
\includegraphics[width=0.24\textwidth]{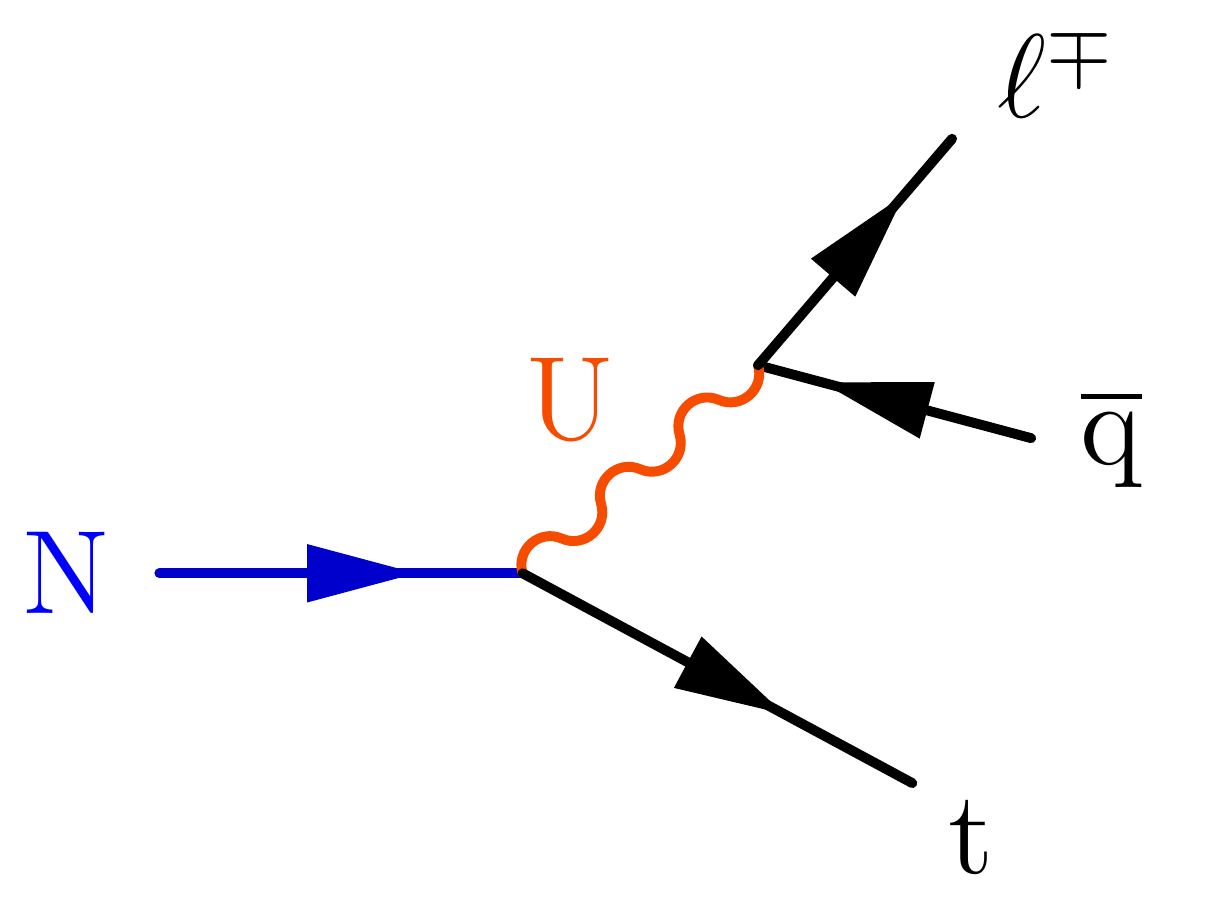}%
\caption{%
    Example diagrams showing $s$-channel EW production of VLL pairs through SM bosons, as expected at the LHC (left two diagrams).
    In these diagrams, \VLLL represents either the neutral VLL, \VLLN, or the charged VLL, \VLLE.
    The VLL decays are mediated by a vector leptoquark \PU (right two diagrams).
    In the 4321 model, these decays are primarily to third-generation leptons and quarks.
}
\label{fig:feynman_diagrams}
\end{figure}

In addition to the EW production modes through their couplings to the SM \PW and \PZ/\PGg bosons, the VLLs in the 4321 model may also be produced via interactions with the new \PZpr boson.
Conversely, these VLLs are expected to decay primarily through their interactions with the vector leptoquark in the model, to two SM quarks and one lepton.
Examples of diagrams showing the EW pair and associated production of VLLs, as well as diagrams of the VLL decays are shown in Fig.~\ref{fig:feynman_diagrams}.

The 4321 model provides a quark-lepton unification at the \TeVns scale, while simultaneously respecting many other measurements that are in agreement with SM expectations and lepton flavor universality~\cite{PhysRevLett.68.3398,Abbott:1999pk,Aaij:2016qqz,Aad:2020ayz,LHCb:2017rln,Belle:2017ilt,Aaij:2017uff,Belle:2019rba}. Additionally, the 4321 model can be used as a benchmark UV-complete model~\cite{benchmark,4321footprints} and allows one to fully explore the resulting phenomenology at the LHC.

\section{Review of vector-like lepton searches}

\subsection{Overview of the CMS search program}

The CMS Collaboration has carried out three direct searches targeting extensions of the SM with VLLs in the $\sqrt{s}=13\TeV$ \pp collision data set.
In the first of these efforts, multilepton final states with electrons and muons were probed using a data set collected during 2016 and 2017, and the first direct constraints were set on doublet models with vector-like leptons coupled to third-generation SM leptons in the mass range of 120--790\GeV~\cite{Sirunyan:2019ofn}.
This result has been superseded by a second search targeting both doublet and singlet models, conducted with the larger full Run 2 data set
with additional multilepton final states including hadronically decaying tau leptons (\tauh)~\cite{CMS:2022nty}.
The third search performed by the CMS Collaboration probes a non-minimal SM extension involving VLLs and other BSM states in the context of the 4321 model
in an all-hadronic final state involving multiple jets and hadronically decaying tau leptons~\cite{CMS:2022cpe}.
The latter two searches are detailed below.

\subsection{Search for VLLs in the minimal model in multileptonic signatures}\label{sec:exo21002}

Using the full Run 2 data set,
the CMS Collaboration has performed a search for an \SU2 doublet ($\VLLE{_3}$, $\VLLN_{3}$) and singlet ($\VLLE_{3}$) VLL extension of the SM with couplings to the third-generation SM leptons~\cite{CMS:2022nty}.
All charged and neutral lepton decay modes are considered, namely $\VLLE_{3}\to\PZ\PGt$, $\PH\PGt$, and $\VLLN_{3}\to\PW\PGt$ in the doublet model, and $\PZ\PGt$, $\PH\PGt$, $\PW\PGn$ in the singlet model.

The search probes multilepton events, which are categorized in seven orthogonal channels based on the number of light charged leptons (electrons or muons) and hadronically decaying tau leptons, defined as:
\begin{itemize}
\item at least four light leptons and any number of \tauh candidates (4\eormu),
\item exactly three light leptons and at least one \tauh candidates ($3\eormu1\tauh$),
\item exactly three light leptons and no \tauh candidates (3\eormu),
\item exactly two light leptons and at least two \tauh candidates ($2\eormu2\tauh$),
\item exactly two light leptons and exactly one \tauh candidates ($2\eormu1\tauh$),
\item exactly one light lepton and at least three \tauh candidates ($1\eormu3\tauh$), and
\item exactly one light lepton and exactly two \tauh candidates ($1\eormu2\tauh$).
\end{itemize}
In the 4\eormu channel, only the leading four light leptons in \pt are used in the subsequent analysis.
Likewise, in the $3\eormu1\tauh$, $2\eormu2\tauh$, and $1\eormu3\tauh$ channels, only the leading one, two, and three \tauh are used, respectively.

The SM background processes, such as \WZ, \ZZ, \ttZ, and \ttW production in which three or more reconstructed charged leptons originate from decays of SM bosons contribute mostly to the irreducible background in various channels of this search. A smaller background contribution arises from ISR or FSR photons that convert asymmetrically such that only one of the produced electrons is reconstructed in the detector, or from misidentifying on-shell photons as electrons. The dominant source of such backgrounds, collectively referred to as the conversion background, is DY events with an additional photon (\Zgamma). These backgrounds are estimated using simulation normalized to observed data in the dedicated CRs. Another important background component is the misidentified lepton background due to jets being misidentified as leptons, which is estimated using control samples in data via the matrix method (Section~\ref{sec:bkgest}).

\begin{figure}[ht!]
\centering
\includegraphics[width=0.48\textwidth]{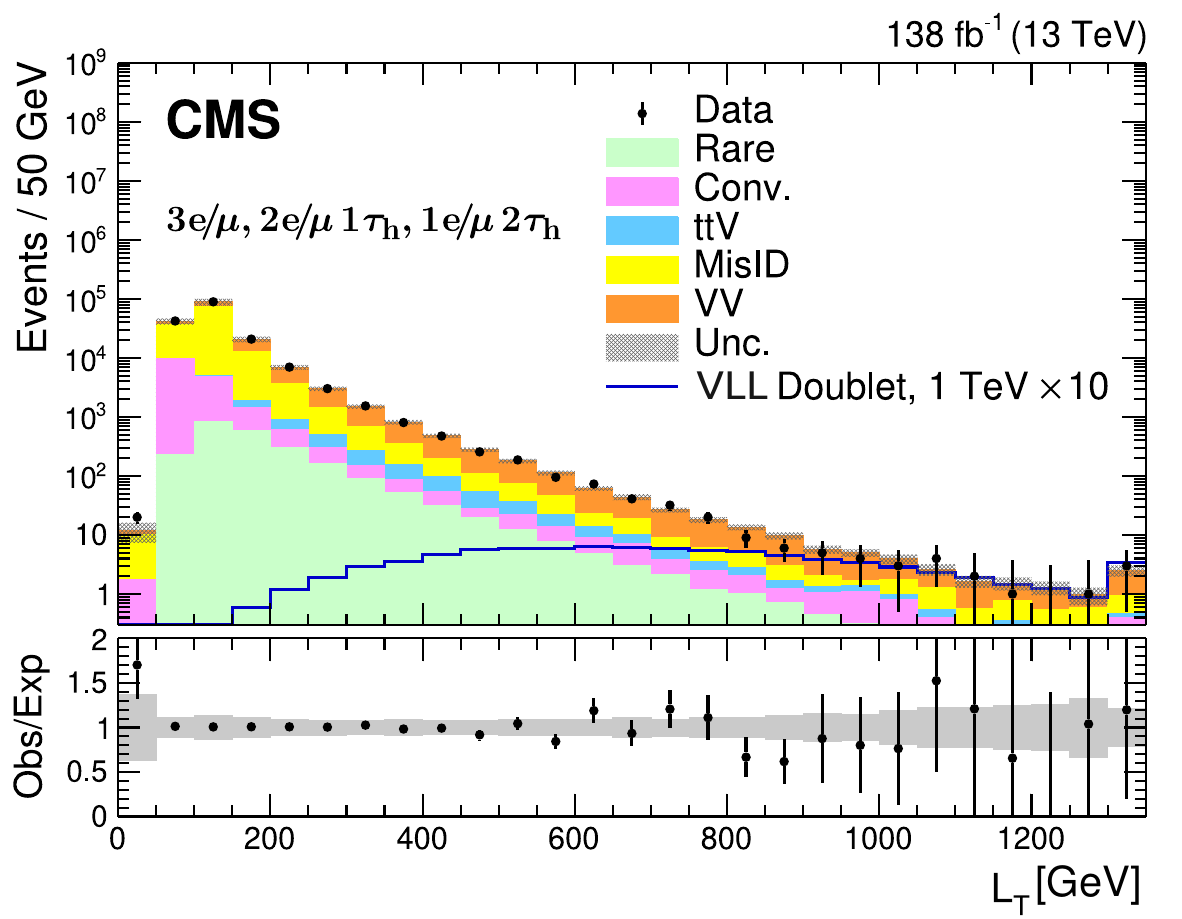}%
\hfill%
\includegraphics[width=0.48\textwidth]{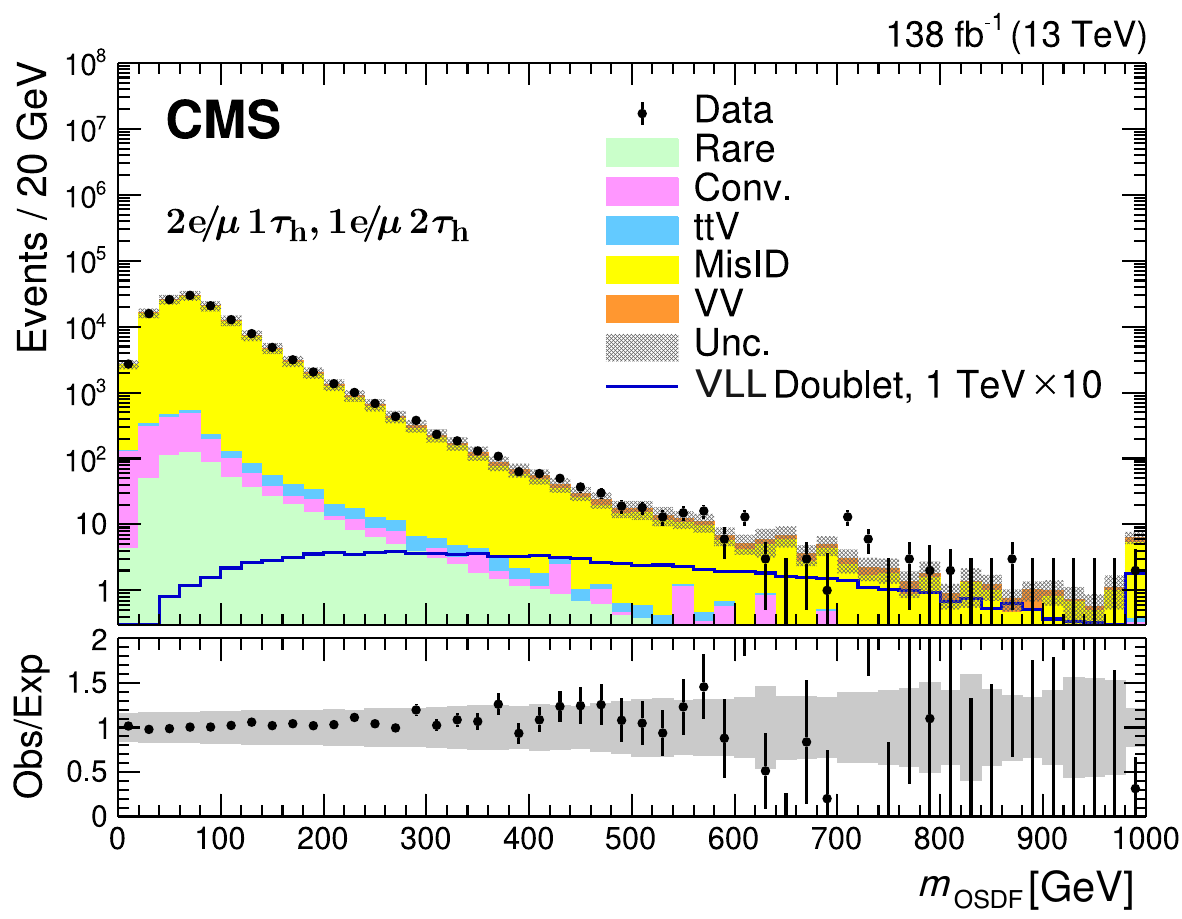}%
\caption{%
    The \LT distribution in 3\eormu, $2\eormu1\tauh$, and $1\eormu2\tauh$ events (left), and the invariant mass distribution of the OS different-flavor ($m_{\text{OSDF}}$) light lepton and tau lepton pair in $2\eormu1\tauh$ and $1\eormu2\tauh$ events (right).
    The rightmost bin contains the overflow events.
    The lower panel shows the ratio of observed events to the total expected background prediction.
    The gray band on the ratio represents the quadratic sum of statistical and systematic uncertainties in the SM background prediction.
    The expected SM background distributions and the uncertainties are shown after fitting the data under the background-only hypothesis.
    For illustration, an example signal hypothesis for the production of vector-like leptons coupled to third-generation SM leptons in the doublet scenario for a VLL mass of 1\TeV, before the fit, is overlaid. The signal yield is scaled by a factor of 10 for visualization purposes.
    Figures adapted from Ref.~\cite{CMS:2022nty}.
}
\label{fig:LT_VLL}
\end{figure}

Selected events in the seven channels are further categorized in a model-independent way, based on the characteristics of the SM backgrounds, or in a model-dependent way, based on the output of BDTs trained to identify the signal against the SM backgrounds, to define a number of SRs.
The model-independent SRs are defined by splitting the channels into various regions based on the charge, flavor, invariant mass of lepton pairs, and kinematic properties of leptons, jets, and \ptmiss, as well as multiplicity of \PQb-tagged jets. The observable \LT is defined as the scalar \pt sum of all charged leptons that constitute the channel. For example, in the 4\eormu channel, \LT is calculated from the four light leptons leading in \pt, while for the $3\eormu1\tauh$ channel, it is calculated from the three light leptons and the leading \tauh candidate. The observable \HT is defined as the scalar \pt sum of all jets. Additionally, the scalar sum of \LT, \HT, and \ptmiss is defined as \ST. In each region, the \ST distribution is probed as the VLL signals are expected to produce broad enhancements in the tails of this observable.
This scheme gives 805 independent SR bins in each year of data taking in Run 2, a detailed breakdown of which may be found in Ref.~\cite{CMS:2022nty}.
Figure~\ref{fig:LT_VLL} illustrates example \LT and dilepton invariant mass distributions, together with the expected signal distribution for a 1\TeV VLL mass.

\begin{figure}[ht!]
\centering
\includegraphics[width=\textwidth]{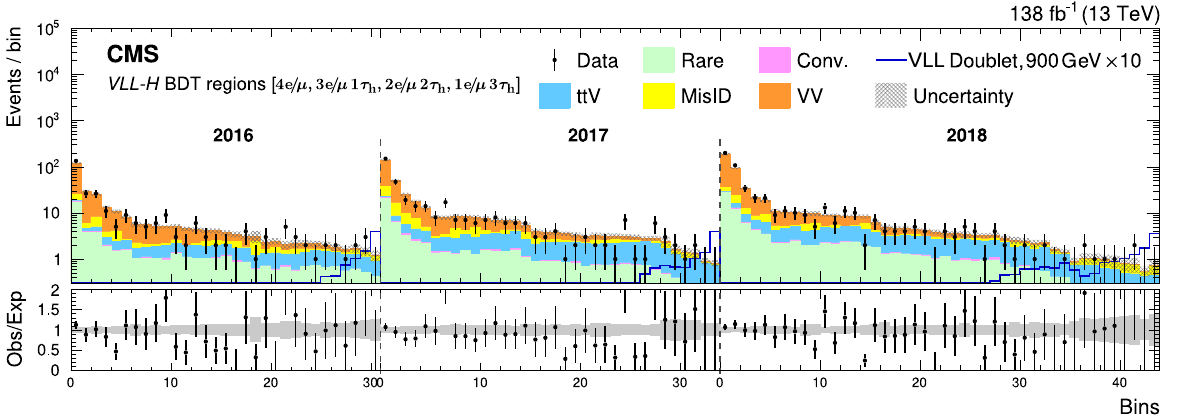}
\caption{%
    The \textit{VLL-H} BDT regions for the four-lepton channels for the full Run 2 data set. The lower panel shows the ratio of observed events to the total expected background prediction.
    The gray band on the ratio represents the quadratic sum of statistical and systematic uncertainties in the SM background prediction.
    The expected SM background distributions and the uncertainties are shown after fitting the data under the background-only hypothesis.
    For illustration, an example signal hypothesis for the production of the vector-like leptons coupled to the third generation SM leptons in the doublet scenario for a VLL mass of 900\GeV, before the fit, is overlaid. The signal yield is scaled by a factor of 10 for visualization purposes.
    Figure adapted from Ref.~\cite{CMS:2022nty}.
}
\label{fig:resultVLLHighMVAbins}
\end{figure}

In the model-dependent approach, a set of BDTs are trained using both doublet and singlet VLL scenarios targeting three signal mass windows (low, medium, high) exploiting up to 48 physics object- and event-level observables. Using the BDT score, a number of variable-width regions is defined for each of the combined three-lepton and four-lepton channels in each data-taking year. These define the BDT regions for all three signal mass windows in which an analysis is performed using only the number of observed events, \ie, not using shape information of the distributions of observables.
Example BDT region distributions for the high-mass BDTs (\textit{VLL-H} BDT) for the four-lepton channels (4\eormu, $3\eormu1\tauh$, $2\eormu2\tauh$, $1\eormu3\tauh$) with the full Run 2 data set for the doublet VLL model are shown in Fig.~\ref{fig:resultVLLHighMVAbins}.

\begin{figure}[hbt!]
\centering
\includegraphics[width=0.48\textwidth]{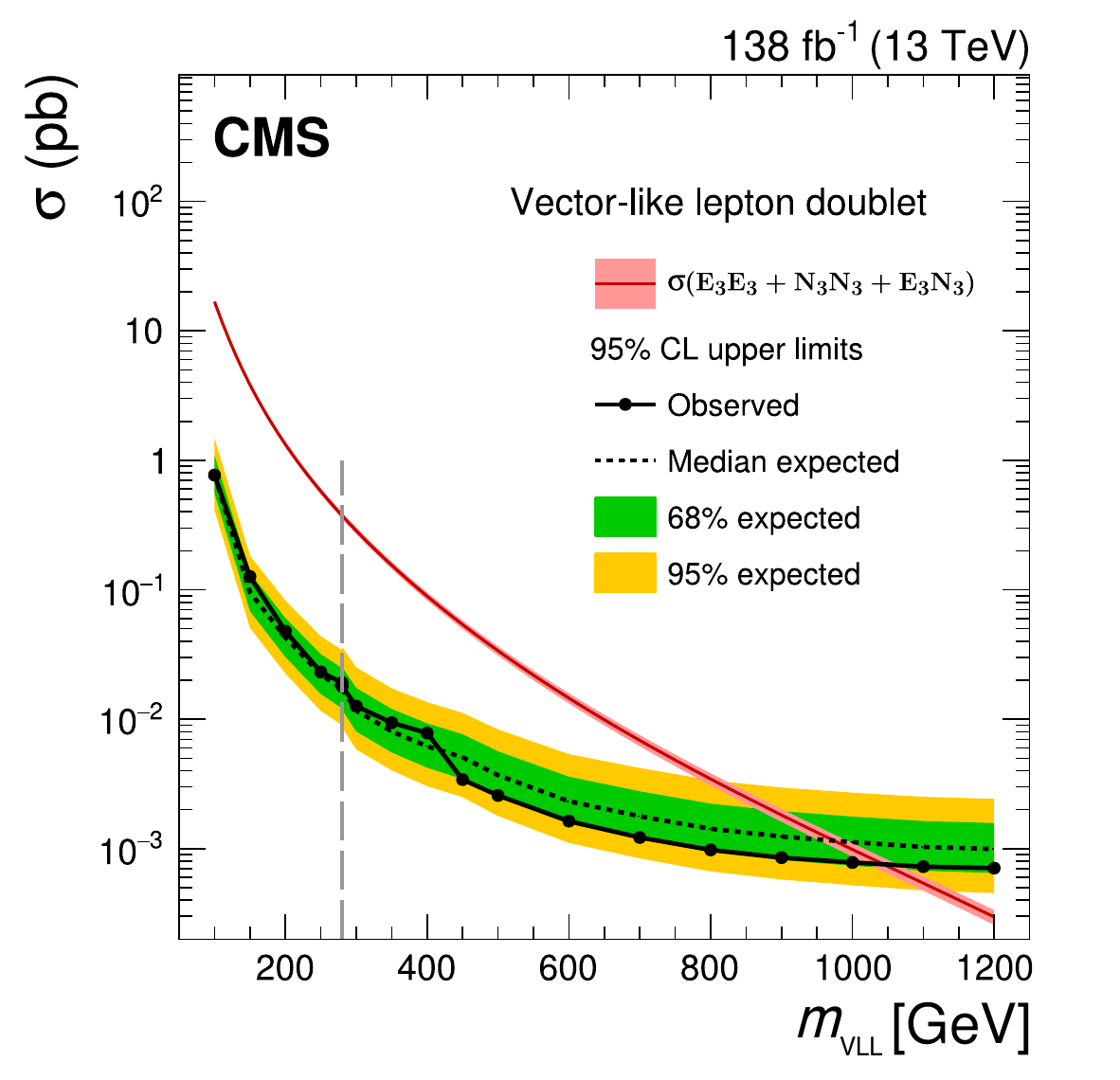}%
\hfill%
\includegraphics[width=0.48\textwidth]{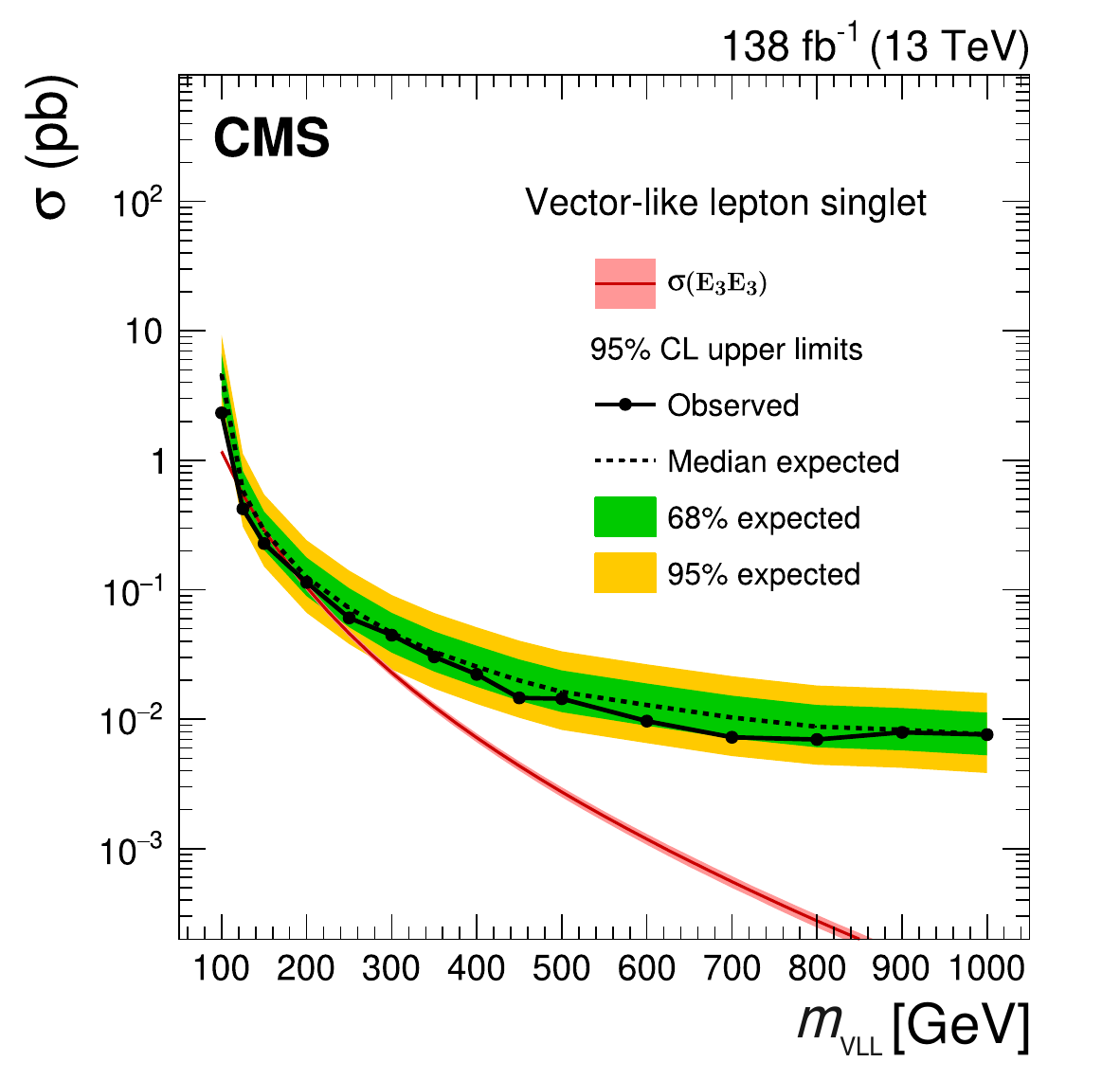}%
\caption{%
    Observed and expected upper limits at 95\% \CL on the production cross section for the vector-like leptons coupled to third-generation SM leptons in the doublet model (left) and singlet model (right).
    For the doublet vector-like lepton model, to the left of the vertical dashed gray line, the limits are shown from the model-independent scheme, while to the right the limits are shown from the model dependent BDT regions.
    For the singlet vector-like lepton model, the limit is shown from the model-independent scheme for all masses.
    Figures adapted from Ref.~\cite{CMS:2022nty}.
}
\label{fig:limitsVLL}
\end{figure}

Figure~\ref{fig:limitsVLL} shows the observed and expected cross section limits for the doublet and singlet VLL models.
In the doublet model, vector-like leptons coupled to third-generation SM leptons with masses up to 1040\GeV are excluded, while the expected mass exclusion is at 970\GeV.
The most stringent limit for VLL masses below below 280\GeV is given by the model-independent scheme, and by the model dependent BDT regions for VLL masses above 280\GeV.
In the singlet model, the most stringent limits are given by the model-independent scheme over the entire mass range.
The expected exclusion for the singlet model is only at a VLL mass $\approx$150\GeV, while the observed exclusion is in the VLL mass range of 125--170\GeV. These are the most stringent limits on this model from the LHC. The model-dependent SRs are typically more sensitive than their model-independent counterparts, except for the lowest signal masses. This is because at low masses, the BDT training process is degraded by the low signal yield.
The less stringent constraints observed in the singlet model arise from the notably lower cross section of VLL pair production, which proceeds exclusively through the $\pp\to\PZ/\PGg\to\VLLE\VLLEbar$ and involves a weaker gauge coupling strength compared to the doublet scenario. Additionally, the prevalent $\VLLE\to\PW\PGn$ decay mode in the singlet model might not result in energetic charged leptons in the final state.

These bounds apply to all vector-like leptons coupled to third-generation SM leptons that undergo prompt decays in the detector; and aside from this assumption, the analysis is insensitive to the precise values of the mixing angles.

\subsection{Search for VLLs in the 4321 model in hadronic signatures}
\label{sec:results_vll_4321AllHad}

The first search for VLLs in the context of the 4321 model was performed using the 2017 and 2018 data sets,
corresponding to an integrated luminosity of 96.5\fbinv~\cite{CMS:2022cpe}.
The analysis targets VLL decays via their couplings to SM fermions through leptoquark interactions, resulting in third-generation fermion signatures.
The primary signature for this model is a final state with four \PQb quarks and two charged third-generation SM leptons accompanying multiple light-flavor jets.
Events with at least three \PQb-tagged jets and varying \PGt lepton multiplicities are selected, where the hadronic tau lepton decay channels are targeted.
The 2016 data set is not considered in the analysis, as it predates the tracker upgrade~\cite{CMS-DP-2018-033} that significantly improves the \PQb jet identification performance crucial to the online event selection. In addition, it was subject to statistical limitations in the estimate of the misidentified tau lepton background from control samples in data, performed separately for each data-taking period. Therefore, this data set was deemed not to contribute significantly to the analysis compared to the 2017 and 2018 data. Similarly, VLL signal masses below 500\GeV are not targeted in this analysis to be compatible with the online event selection requirements.

Events are categorized into 0\tauh, 1\tauh, and 2\tauh categories, each with dedicated background estimation and signal optimization.
The data are also split by the data-taking year, either 2017 or 2018, to reflect the different accelerator and detector conditions.
A simultaneous binned maximum likelihood fit is performed over all categories to extract the VLL signal.

In all categories, \ttbar and QCD multijet production are the primary sources of background.
A number of CRs enriched in the backgrounds are used to help constraining the background estimates using data.

In the 0\tauh category, the primary background is from QCD multijet events, and is estimated using an ABCD method.
This method uses \ptmiss and a graph-based DNN~\cite{Mikuni:2020wpr} output trained to distinguish signal from QCD multijet background as the two independent axes.
In this category, \ttbar events, including those with an additional \PQb quark pair (\ttbb) or an additional boson (\ttX) form a smaller, but still relevant background.
The distribution of the number of jets, which tends to be higher for the signal, is then used in the fit for this category.

In the 1\tauh and 2\tauh categories, the primary background is \ttbar events.
In both categories, contributions from events, which include misidentified \tauh leptons are estimated separately for QCD multijet, and \ttbar events using control samples in data.

\begin{figure}[ht!]
\centering
\includegraphics[width=0.48\textwidth]{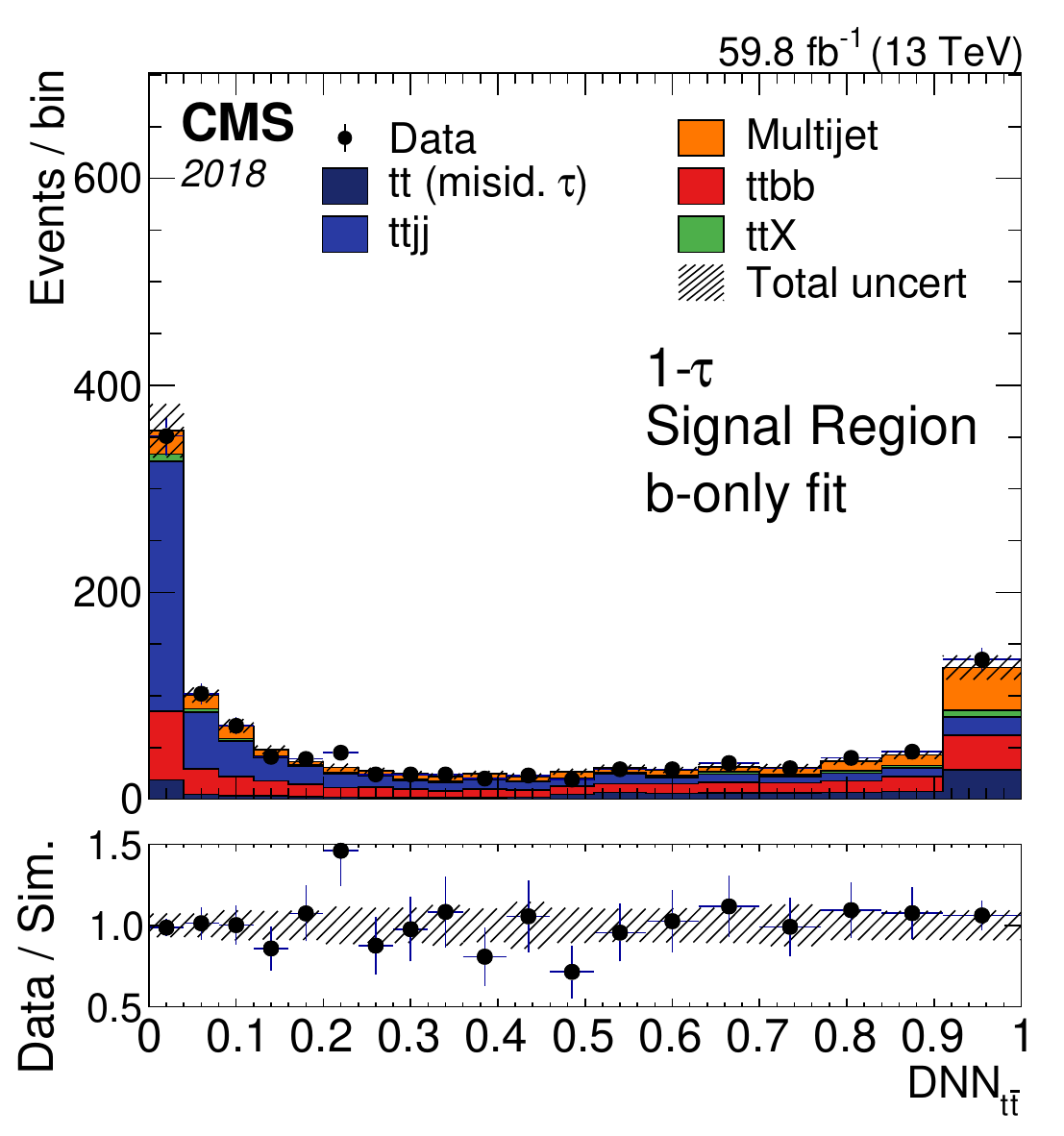}%
\hfill%
\includegraphics[width=0.48\textwidth]{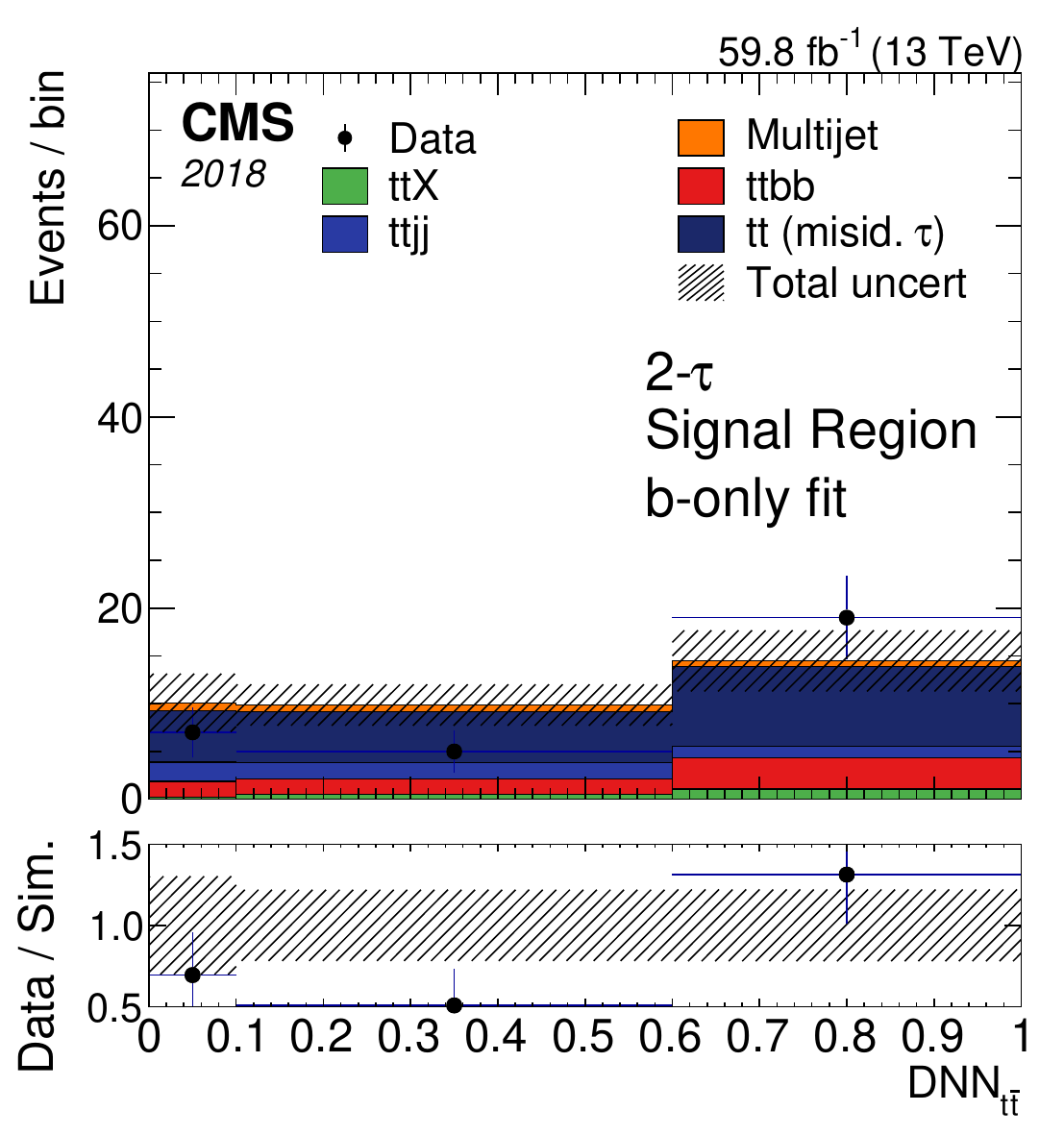}\\
\includegraphics[width=0.48\textwidth]{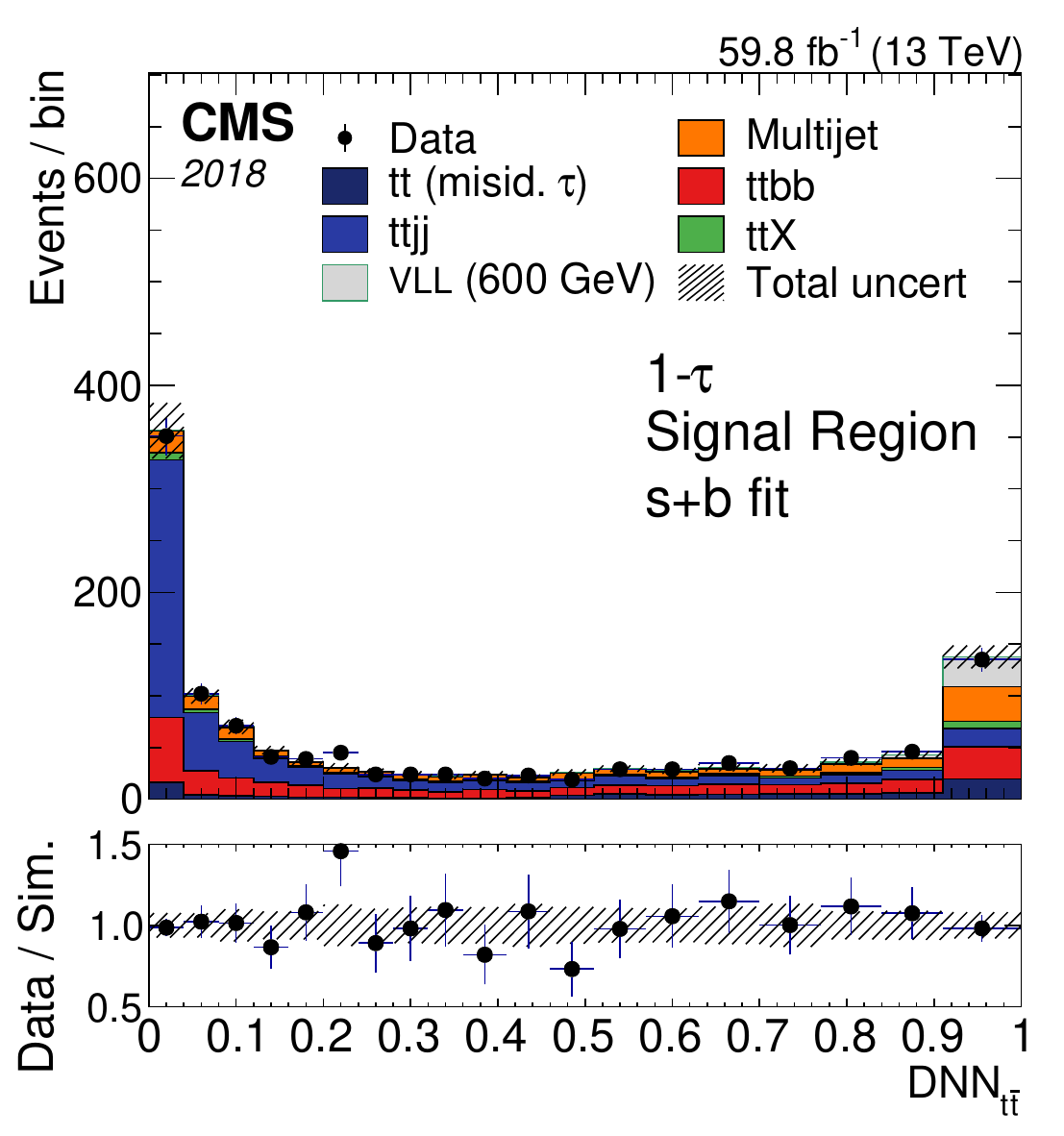}%
\hfill%
\includegraphics[width=0.48\textwidth]{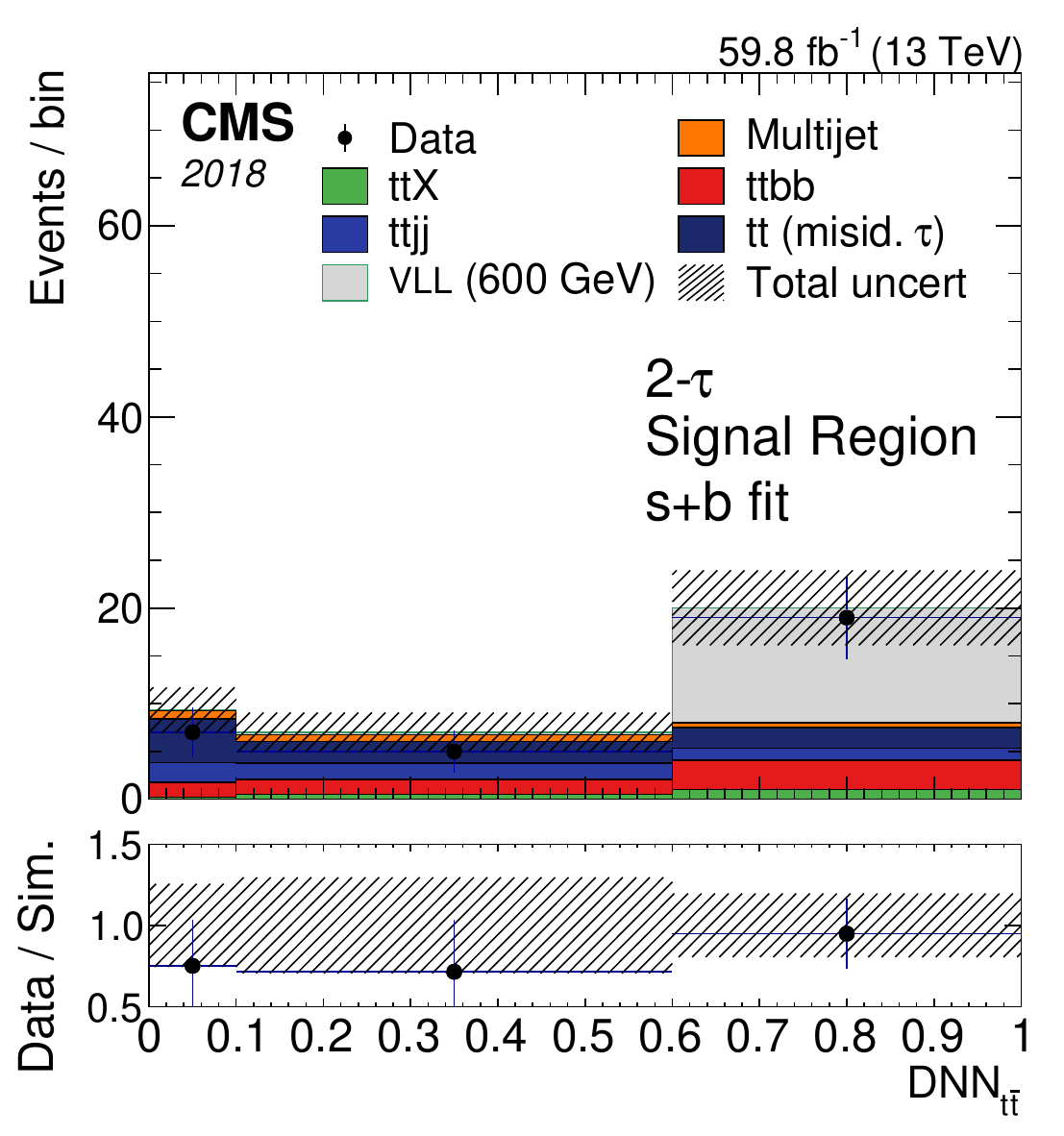}%
\caption{%
    Postfit distributions for the 2018 data set in the 1\tauh (left) and 2\tauh (right) channels.
    The upper row shows the background-only fit and the lower row shows the fit including the signal.
    Not shown here, but included in the fit, are the 2017 data and the 0\tauh channel.
    Figures taken from Ref.~\cite{CMS:2022cpe}.
}
\label{fig:4321vll_postfit}
\end{figure}

In the 1\tauh category, QCD multijet events with misidentified \tauh leptons, and \ttbar (including \ttbb and \ttX) production with real \tauh leptons are the primary backgrounds, while \ttbar events with misidentified \tauh leptons play a smaller role.
In the 2\tauh category, \ttbar events with misidentified \tauh leptons are the primary background, and \ttbar events with two real \tauh leptons also make an important contribution, whereas QCD multijet events do not contribute significantly.

The same graph-based DNN architecture used in the 0\tauh category is used in the 1\tauh and 2\tauh categories, but is trained to discriminate signal from \ttbar using all final state objects (jets and \tauh leptons) as inputs.
For each object, its $\eta$, $\phi$, \pt, mass, charge, and the value of its \DeepJet score~\cite{Bols_2020,CMS-DP-2018-058} are passed to the NN. For jets, the charge entry is always set to zero. For \tauh candidates, the \DeepJet score is always set to zero, which corresponds to an extremely low probability of being a \PQb jet.

The NN is trained to discriminate \ttbar from the signal hypothesis, using a mix of VLL signal masses in the range 500--1050\GeV, and its output distribution is used as input to the fit.
Data and postfit expectations for the 1\tauh and 2\tauh channels for the 2018 data set are shown in Fig.~\ref{fig:4321vll_postfit}. The 0\tauh channel, which is less sensitive, is not shown, nor is the 2017 data used in the fit, which shows very similar trends.

\begin{figure}[ht!]
\centering
\includegraphics[width=0.48\textwidth]{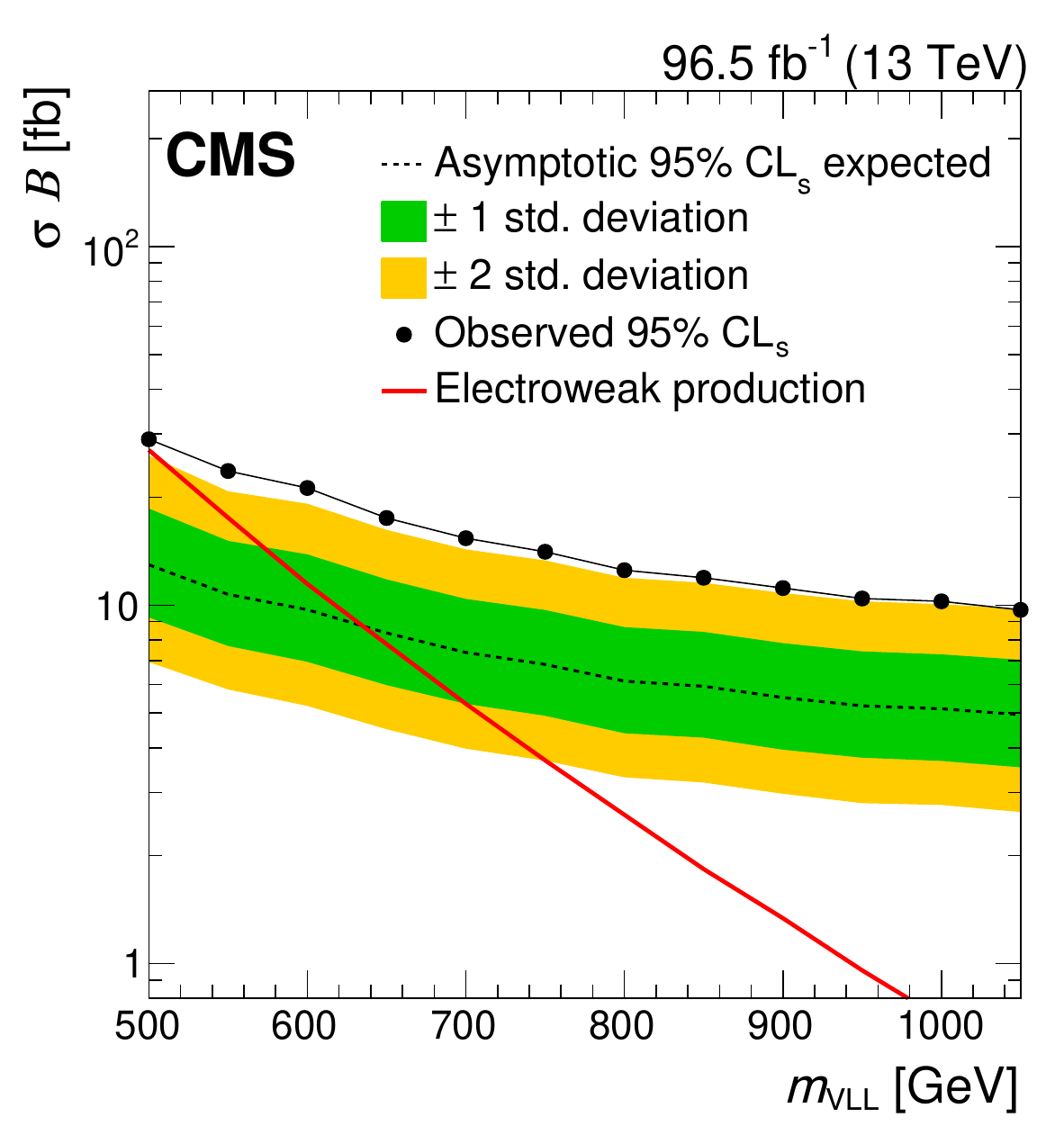}
\caption{%
    Expected and observed 95\% \CL upper limits on the product of the VLL pair production cross section and the branching fraction to third-generation quarks and leptons, combining the 2017 and 2018 data and all \tauh multiplicity channels.
    The theoretical prediction in the 4321 model for EW production of VLLs is also shown.
    Figure adapted from Ref.~\cite{CMS:2022cpe}.
}
\label{fig:vll_had_limits}
\end{figure}

The observed data show mild excesses in the highest DNN bins for the 1\tauh and 2\tauh categories for both 2017 and 2018 compared to the background-only hypothesis.
The magnitude of the combined excess is independent of the assumed mass hypothesis. Across the mass range, the largest local significance of 2.8 standard deviations is observed at the VLL mass point of 600\GeV.
No excess is visible in the less sensitive 0\tauh category for either year.
Accordingly, the observed exclusion limits on the signal cross section are above the expected limits, as shown in Fig.~\ref{fig:vll_had_limits}.
Upper limits at 95\% \CL on the product of the VLL pair production cross section and their branching fraction to third-generation fermions are set between 10 and 30\unit{fb}, depending on the assumed VLL mass hypothesis. This is the first direct search for VLLs in the context of the 4321 model at the LHC.

\subsection{Future prospects for VLL searches}

Vector-like leptons appear in a variety of BSM models, as described in Section~\ref{sec:vll}. While vector-like leptons coupled to third-generation SM leptons have already been probed, no CMS searches have probed vector-like leptons coupled to first- and second-generation SM leptons with the Run 1 or Run 2 data sets.
In the minimal model of VLLs, the newly introduced final states are assumed to mix through Yukawa interactions with the same generation of the SM leptons and decay into SM boson-lepton pairs.
Model independent SRs of the seven multilepton channels from the search discussed in Section~\ref{sec:exo21002} may be utilized to extrapolate the sensitivity for these three generations of VLL models (both singlet and doublet scenarios) to the HL-LHC with a total integrated luminosity of 3000\fbinv at $\sqrt{s}=14\TeV$. The physics observable \LTmet in the SR of all seven channels is used to project the sensitivity for these models.

In order to utilize these Run 2 analysis results to project the sensitivity to the HL-LHC,
individual background yields in each SR bin of the analysis using all of the Run 2 data are scaled to 3000\fbinv of integrated luminosity. In addition, prompt background (\WZ, \ZZ, \ttW, or \ttZ) yields already take into account the enhancement due to the higher center-of-mass energy.
The signal acceptance is estimated from simulated samples of all three coupling scenarios with Run 2 detector conditions. The cross section for the VLL signal is calculated at LO for $\sqrt{s}=14\TeV$, and is then corrected to NLO using a correction factor derived at 13\TeV.

\begin{figure}[p!]
\centering
\includegraphics[width=0.48\textwidth]{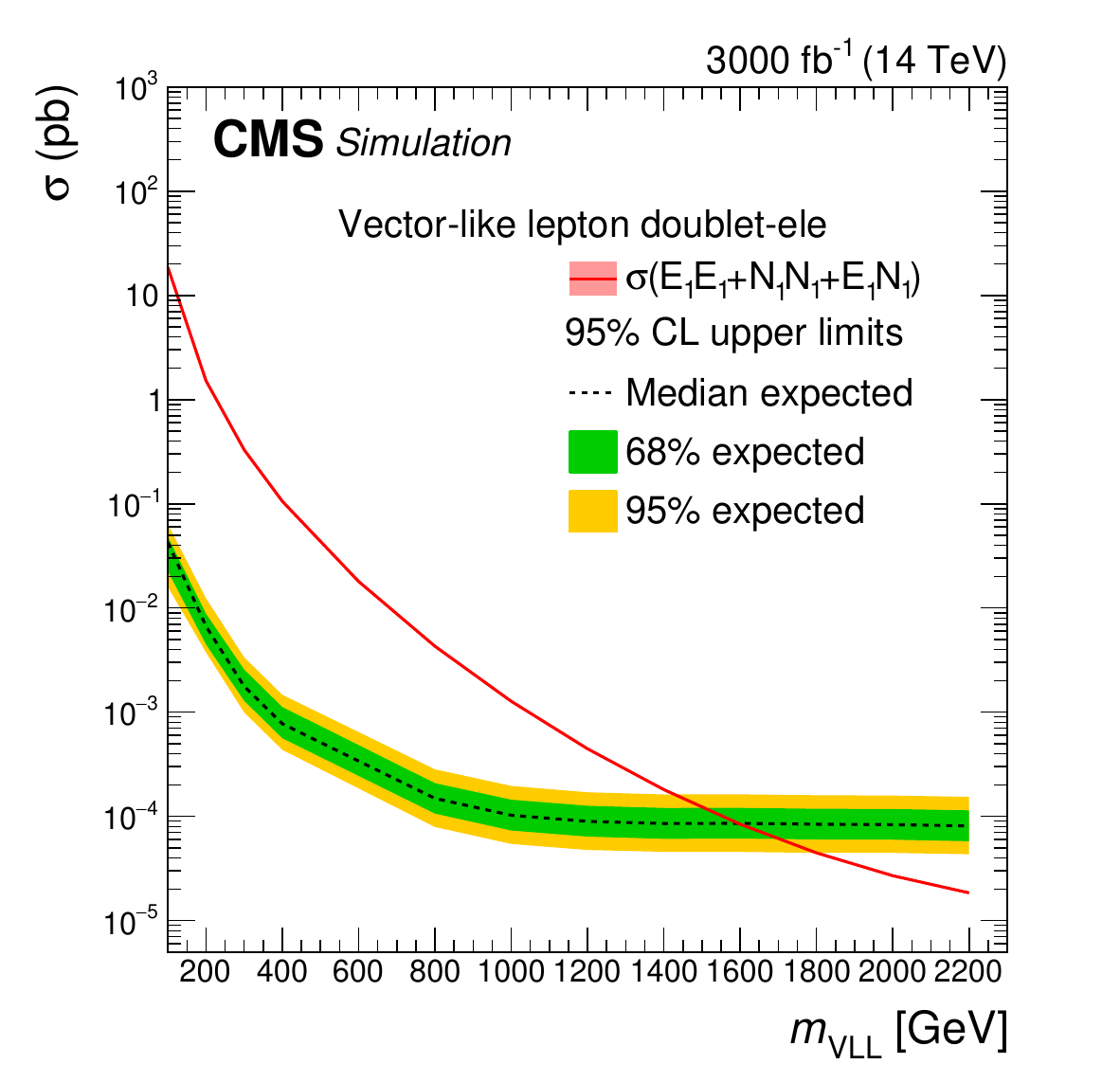}%
\hfill%
\includegraphics[width=0.48\textwidth]{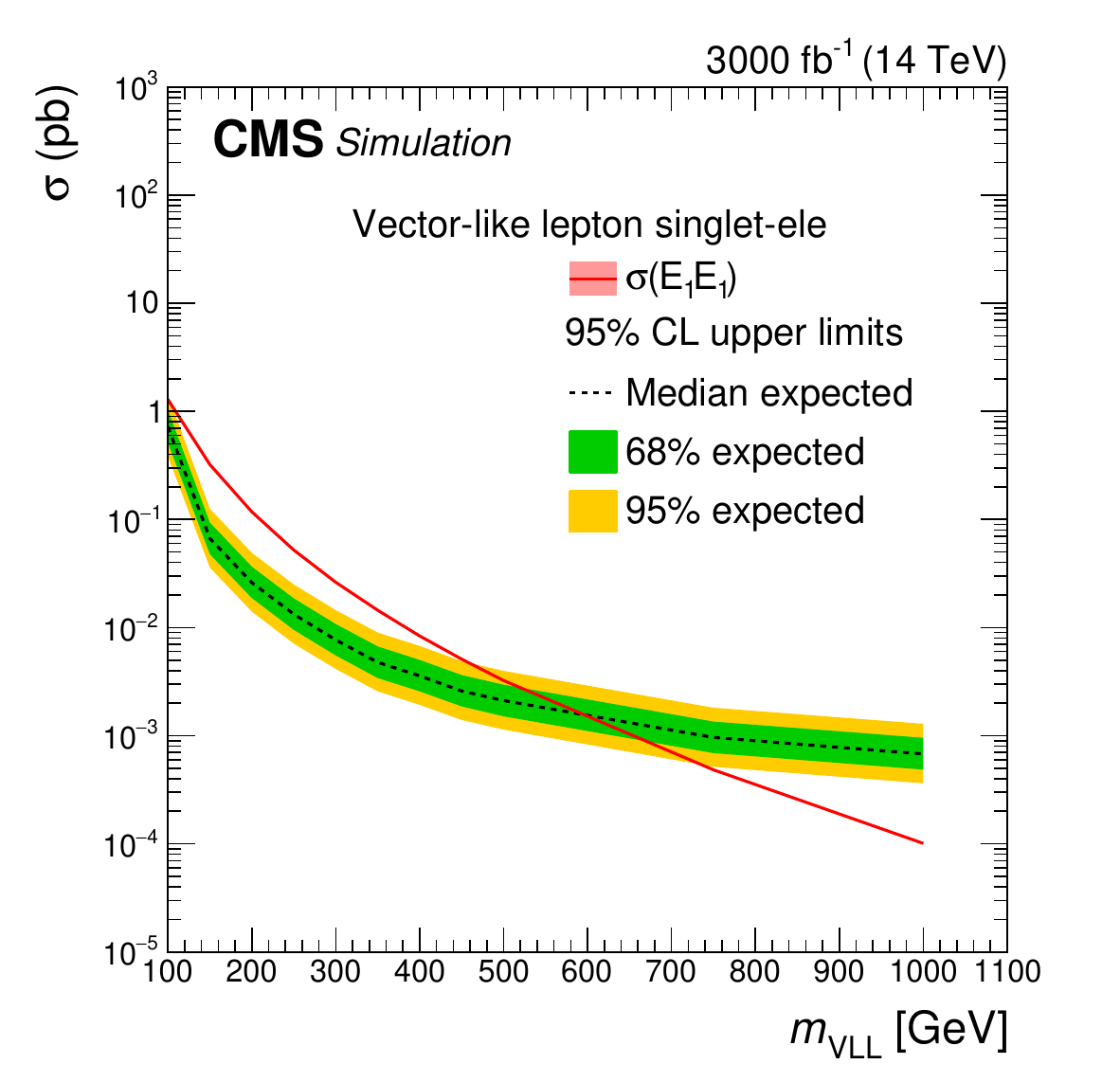} \\
\includegraphics[width=0.48\textwidth]{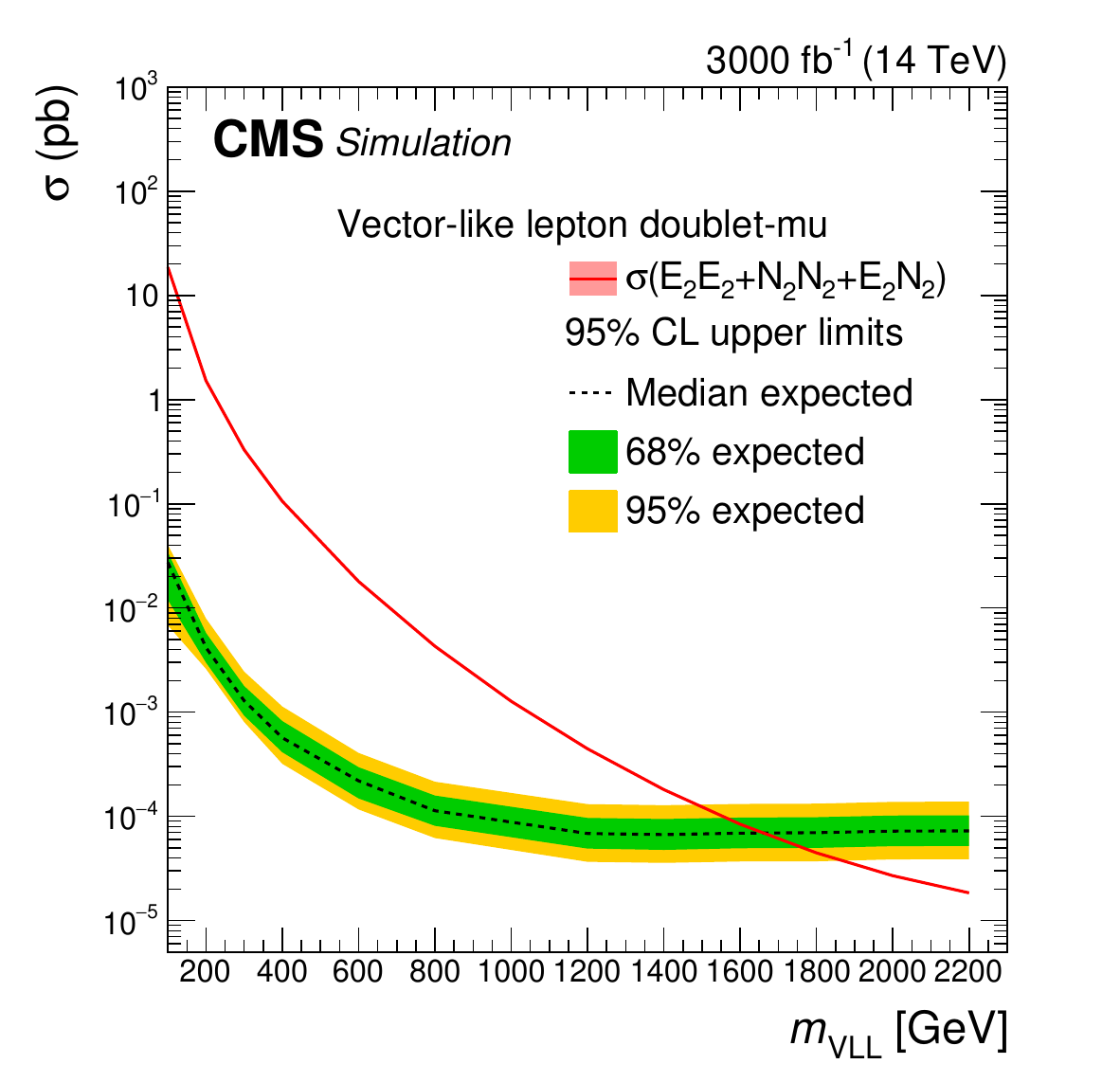}%
\hfill%
\includegraphics[width=0.48\textwidth]{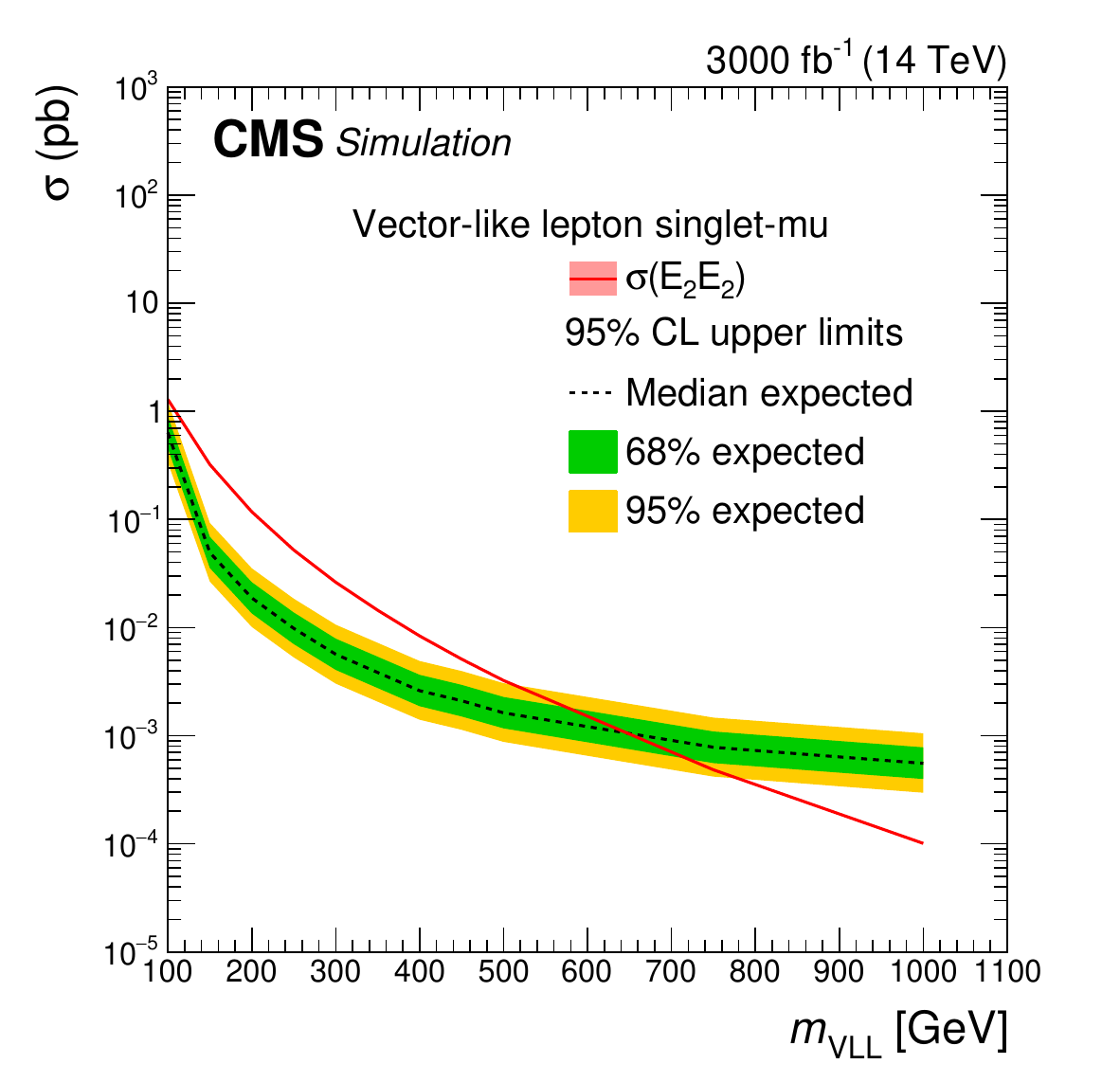} \\
\includegraphics[width=0.48\textwidth]{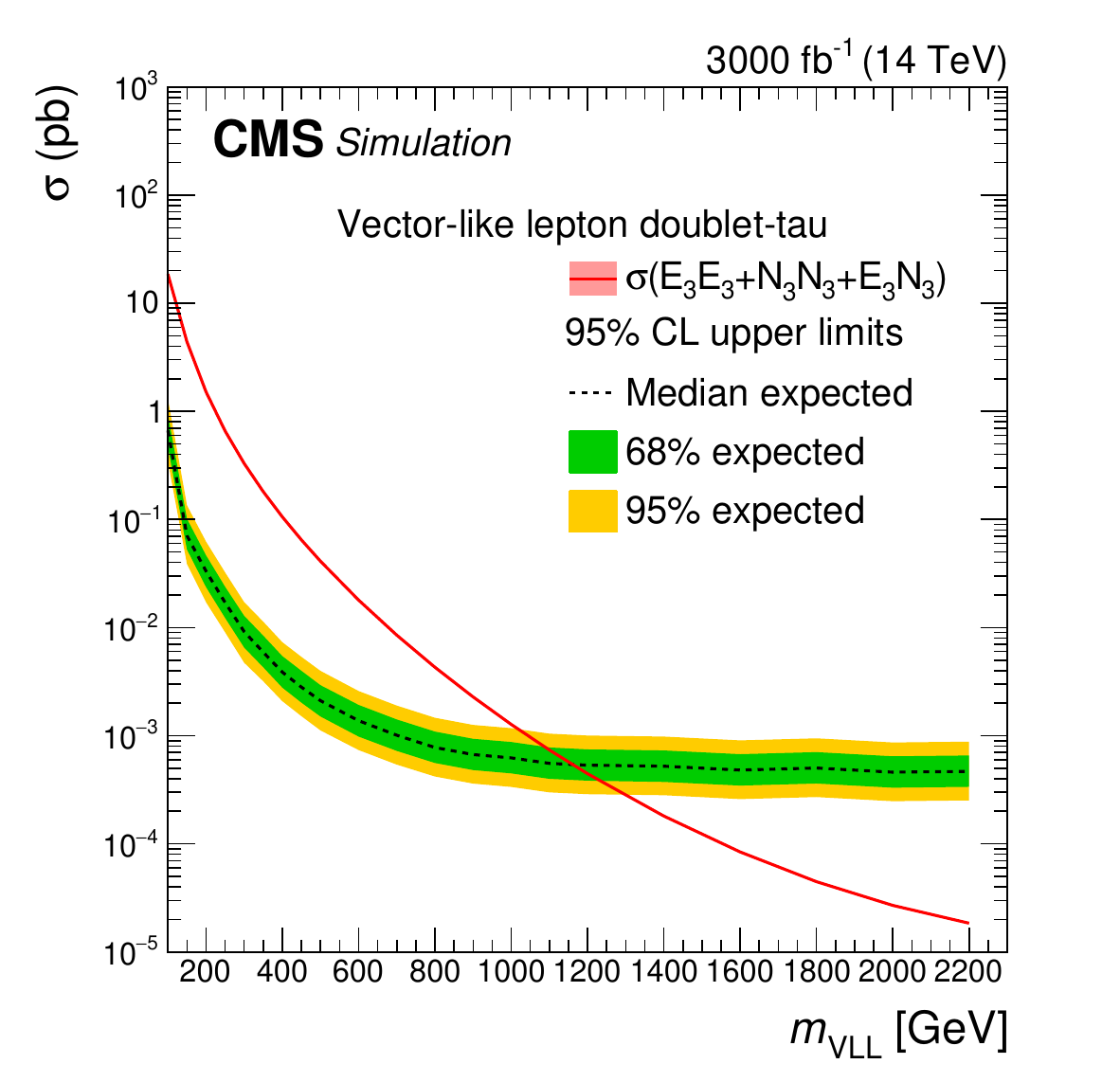}%
\hfill%
\includegraphics[width=0.48\textwidth]{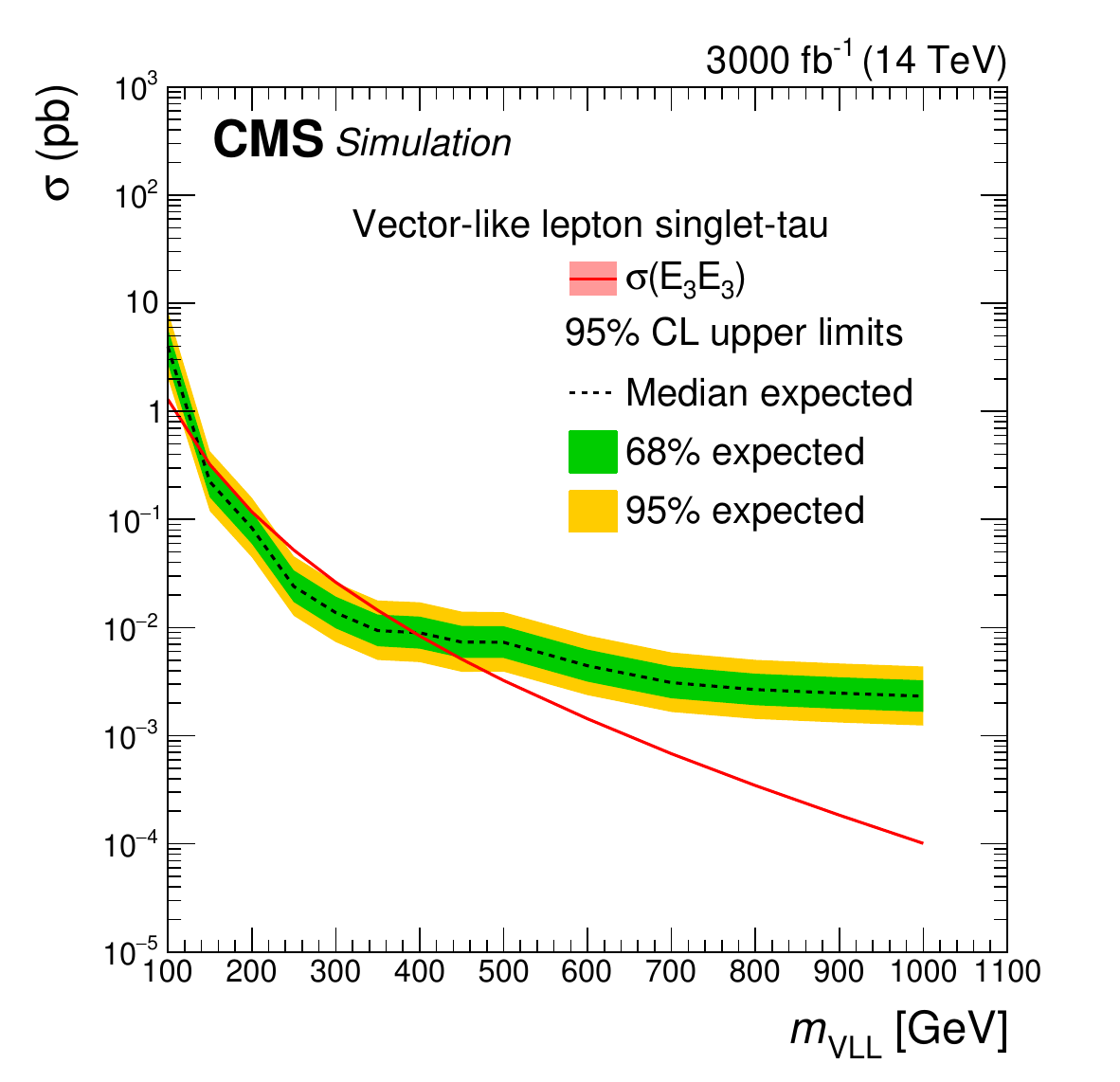}%
\caption{%
  Expected HL-LHC exclusion limits for vector-like leptons coupled to first-generation (upper row), second-generation (middle row), and third-generation SM leptons (lower row) in the doublet model (left) and the singlet model (right).
  For both models, limits are calculated using \LTmet from the model independent SRs for all masses.
}
\label{fig:vll_emutau}
\end{figure}

Following the Yellow Report on BSM physics at the HL-LHC and High-Energy LHC~\cite{yellow_report_2019}, experimental uncertainties for signal and background yields were taken into consideration.
Figure~\ref{fig:vll_emutau} shows the expected upper limits on the cross section at the HL-LHC for the production of vector-like leptons coupled to first-, second-, and third-generation SM leptons. Both the singlet ($\VLLE_{\textrm{i}}$) and the doublet ($\VLLE_{\textrm{i}}$, $\VLLN_{\textrm{i}}$) scenarios are considered, where $\mathrm{i}$ = 1, 2, 3 denotes VLLs coupled to first-, second-, and third-generation SM leptons, respectively. For the doublet model, the VLLs are expected to be excluded at 95\% \CL up to a mass of 1600\GeV ($\VLLE_{1}$, $\VLLN_{1}$), 1630\GeV ($\VLLE_{2}$, $\VLLN_{2}$), and 1150\GeV ($\VLLE_{3}$, $\VLLN_{3}$). The singlet VLLs are expected to be excluded up to a mass of 600\GeV ($\VLLE_{1}$), 640\GeV ($\VLLE_{2}$), and between a mass of 150 and 395\GeV ($\VLLE_{3}$). The weaker cross section limits obtained for the singlet model follow the same reasons mentioned in Section~\ref{sec:exo21002}. These projected sensitivities using model-independent SRs are better by a factor of approximately 2--3 in the upper limit of the cross section compared to the Run 2 results discussed in Section~\ref{sec:exo21002}. The Run 2 results used a BDT to enhance sensitivity, whereas the HL-LHC will provide a much larger data set and there will be the opportunity for more advanced ML techniques and optimization. Thus the eventual reach in terms of VLL mass is expected to be higher than the projected sensitivities here.

\clearpage
\section{Theoretical motivation for heavy neutral leptons}

Unlike charged leptons and quarks in the SM, which can have both LH and RH chiralities, neutrinos are observed as exclusively LH. Neutrinos were initially believed to be massless due to their chiral nature, since the interaction between a particle and the SM Higgs boson flips its chirality. However, the revelation of neutrino oscillations~\cite{Super-Kamiokande:1998kpq}, wherein neutrinos change flavor as they propagate, conclusively demonstrated that neutrinos possess mass, albeit on a scale remarkably smaller than other particles in the SM. In fact, neutrinos are more than 5 million times lighter than electrons. Physics beyond the minimal SM is required to explain the origin of neutrino masses.

The inability of the minimal SM to account for the observed neutrino masses and oscillations underscores its limitations. Heavy neutral leptons, also known as RH neutrinos, provide a possible extension to the SM. With an RH nature complementing the predominantly LH nature of SM neutrinos, HNLs emerge as an alternative or complementary mechanism to the SM Higgs boson Yukawa couplings, responsible for neutrino mass generation involving the mixing~\cite{Mohapatra:1979ia,Schechter:1980gr,Foot:1988aq} in most theoretical models.

The presence of matter in the universe today implies an asymmetry between matter and antimatter in the early universe, known as baryon asymmetry. The exact cause of this imbalance remains an open question in cosmology.  In the framework of baryogenesis via leptogenesis~\cite{Fukugita:1986hr,Davidson:2008bu}, HNLs could potentially contribute to the matter-antimatter asymmetry through $CP$-violating decays in the early universe.

In Sections~\ref{sec:HNL_th_type-I}--\ref{sec:th_hnl_composite}, we describe various mechanisms for HNL production. We explore various theoretical models aimed at explaining HNL and SM LH neutrinos mass patterns and the matter-antimatter asymmetry. These models serve as a benchmark, and include \TypeOne and \TypeThree seesaw mechanisms, the left-right symmetric model (LRSM), and composite models.

\subsection{Production of HNLs in the \TypeOne seesaw model}
\label{sec:HNL_th_type-I}

Neutrinos, being electrically neutral, may exhibit a Dirac or a Majorana nature, determining whether they are distinct from or the same as their antiparticles. The \TypeOne seesaw mechanism~\cite{Minkowski:1977sc,Magg:1980ut,Mohapatra:1980yp,Schechter:1981cv} involves adding predominantly weak-isospin singlet HNLs to the particle content of the standard model to explain the smallness of neutrino masses. In addition to Majorana mass terms for the RH fields, it includes Yukawa couplings between the LH neutrinos, the RH neutrinos, and the Higgs field, \ie, the Dirac mass terms. These couplings allow for a mixing between the light, LH neutrinos and the HNLs. As a result of these Yukawa couplings, the observed neutrino mass patterns emerge. The masses of the light, LH neutrinos are suppressed by the heaviness of HNLs, leading to small masses for the neutrinos.

A prominent theoretical framework is the neutrino minimal standard model (\nuMSM)~\cite{Asaka:2005an}. The \nuMSM incorporates three HNLs into the SM. It extends the SM Lagrangian density by the \TypeOne seesaw mechanism. In addition to providing a solution to the origin and smallness of neutrino mass, this model may also explain other enigmas in the universe, such as dark matter and the matter-antimatter asymmetry.

The search program of the CMS experiment focussing on the \TypeOne seesaw mechanism involves a comprehensive investigation of the HNL production. This includes the search for both Dirac and Majorana HNLs, leading to processes involving lepton number conservation (LNC), as well as those involving lepton number violation (LNV). Within these scenarios, production and decay, within a single generation and across generations, are probed. This approach allows for the exploration of lepton flavor conservation and the search for lepton flavor violation.

A crucial characteristic of HNLs, labeled as \PN, to consider is their lifetime, \tauhnl. Depending on two main factors, namely their masses and mixing with the three lepton generations, HNLs may exhibit a wide range of lifetimes, varying from short-lived to long-lived states. The proper lifetime of an HNL may be described by the following equation:
\begin{equation}\label{eqn:lifetime}
	\frac{1}{\tauhnl} = \Gammatot(\mhnl,\mixpareN,\mixparmN,\mixpartN) = \Gammae+\Gammam+\Gammat,
\end{equation}
where \Gammatot is the total decay width of an HNL; \Gammae, \Gammam, and \Gammat are the partial widths for the decay to an electron, a muon, and a tau lepton, or to their respective neutrino partners, respectively; \mhnl is the HNL mass; and \mixpareN, \mixparmN, and \mixpartN are the mixing matrix elements of the three lepton generations. The \Gammatot may be expressed as
\begin{equation}
	\Gammatot \propto \GF^2\mhnl^5\sum_{\Pell=\Pe,\PGm,\PGt}\mixparsqlN,
\end{equation}
with \GF being the Fermi coupling constant. The proper lifetime of the HNLs, measured in seconds, is inversely proportional to \Gammatot, as expressed in Eq.~\eqref{eqn:lifetime}. Specifically, the lifetime is proportional to $1/\mhnl^5\sum_{\Pell=\Pe,\PGm,\PGt}\mixparsqlN$. This means that for a fixed value of \mixparsqlN, smaller masses correspond to longer lifetimes.

By probing these different properties of the HNLs, the CMS experiment attempts to cover a broad spectrum of interactions and potential signatures, as discussed next.

\begin{figure}[!ht]
\centering
\includegraphics[width=0.48\textwidth]{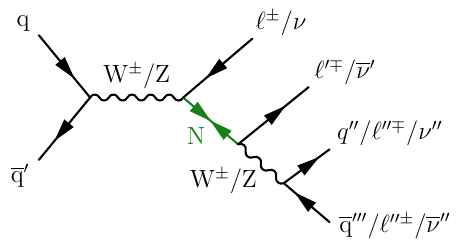}
\caption{%
    Representative Feynman diagram of a Majorana HNL, labeled as \PN, produced through the decay of a \PW or \PZ boson.
}
\label{fig:FD_HNLs}
\end{figure}

The primary production of HNLs considered is through the decay of a \PW boson due to its particularly high production cross section~\cite{Datta:1993nm,Han:2006ip,delAguila:2007qnc}. The decay of the \PW boson yields a charged lepton and a neutrino. The charged lepton arising from the \PW boson decay is an important component in the trigger strategy of various analyses. The final states considered in each analysis, depend on the HNL decay process. The Feynman diagram depicted in Fig.~\ref{fig:FD_HNLs} encapsulates the full spectrum of possible decay scenarios in the context of HNL production through the \PW and \PZ boson decays.

The decay channel $\PN\to\Pellpm\qqbarpr$, with \PN a Dirac HNL, is dominant with an approximate  branching fraction of 50\%, while the $\PN\to\Pellpm\Pellmp\PGn$ decay channel follows closely with a branching fraction of around 23\%. Another significant decay channel is $\PN\to\PGn\qqbarpr$, accounting for an approximate branching fraction of 18\%.

\begin{figure}[!ht]
\centering
\includegraphics[width=0.48\textwidth]{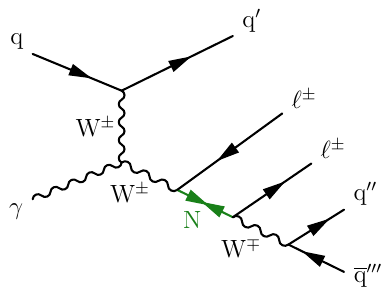}
\caption{%
    Representative Feynman diagram of a Majorana HNL, labeled as \PN, produced through the \Wgamma fusion process and with two charged leptons and jets in the final state.
}
\label{fig:FD_HNLs_Wgamma}
\end{figure}

Figure~\ref{fig:FD_HNLs_Wgamma} shows a hypothetical production mode of HNLs via \Wgamma fusion~\cite{Dev:2013wba,Alva:2014gxa}, which has been considered in one of the searches to enhance the sensitivity to HNLs with masses above several hundred \GeVns. This $t$-channel process is complementary to the search for HNLs in the $s$-channel shown in Fig.~\ref{fig:FD_HNLs}.

\begin{figure}[ht!]
\centering
\includegraphics[width=0.48\textwidth]{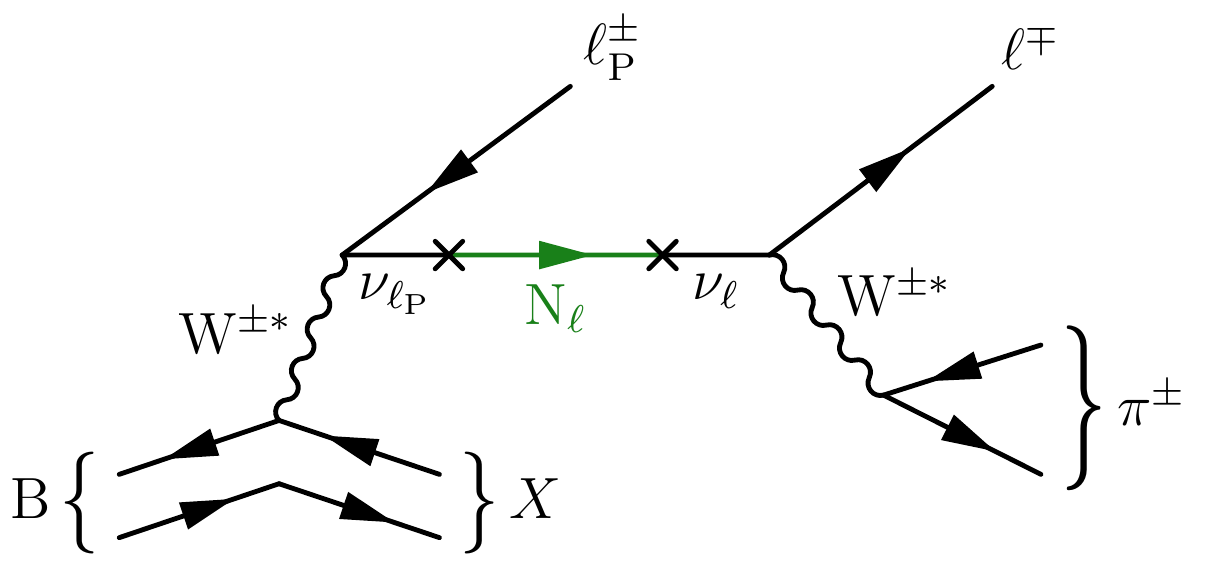}
\caption{%
    Representative Feynman diagram showing the semileptonic decay of a \PB meson into the primary lepton (\PellP), a hadronic system (\PX), and a neutrino, which contains the admixture of an HNL.
    The HNL propagates and decays weakly into a charged lepton \Pellpm and a charged pion \PGpmp.
}
\label{fig:feynMain}
\end{figure}

An additional HNL production process is searched for in the decays of \PB mesons~\cite{CMS-PAS-EXO-22-019}. This is interesting to probe as \PB mesons are produced in \pp collision events with a much higher rate than \PW bosons, and are therefore a more prominent source of neutrinos. A representative Feynman diagram of this process is shown in Fig.~\ref{fig:feynMain}.

\begin{figure}[!ht]
\centering
\includegraphics[width=0.38\textwidth]{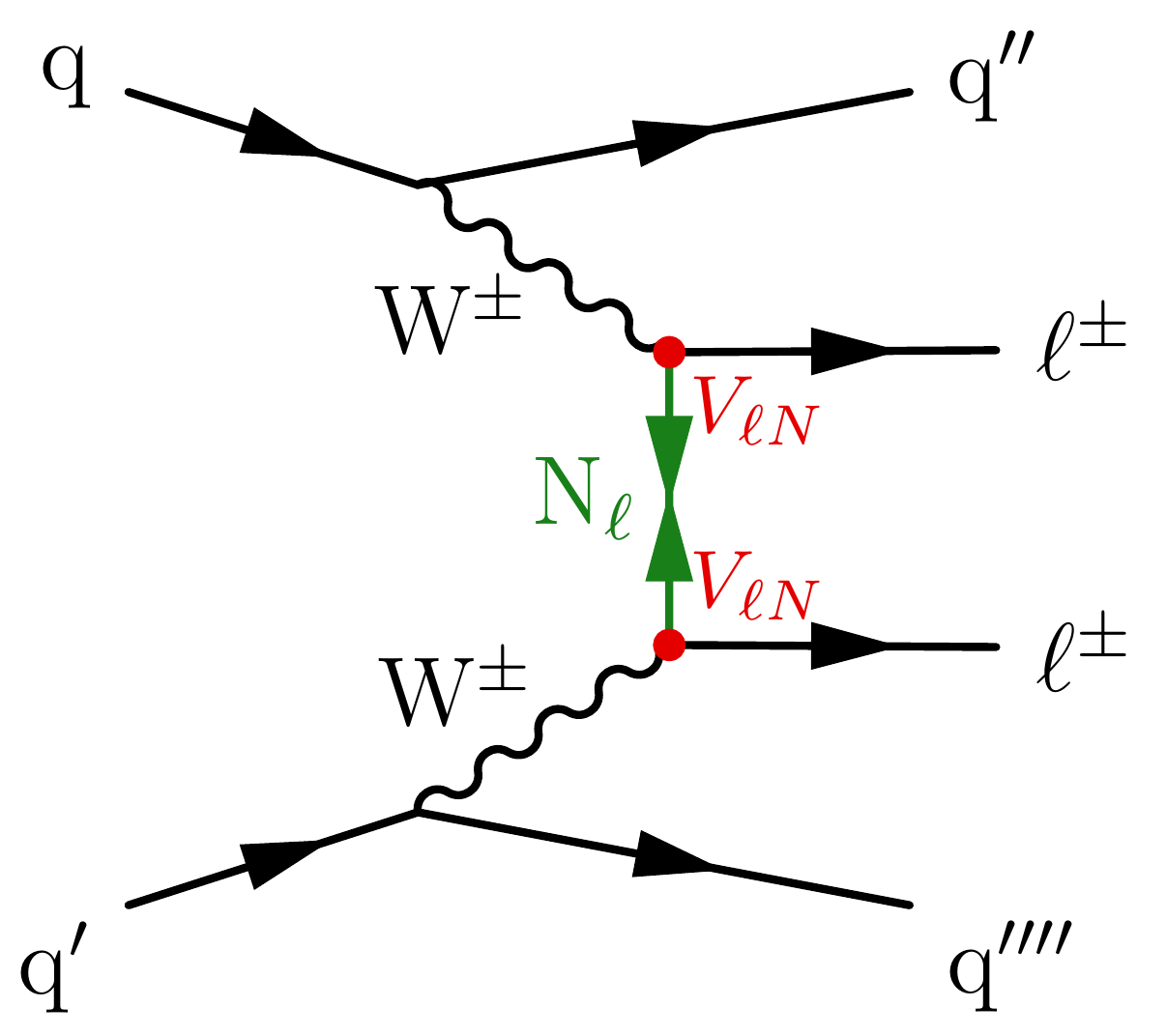}%
\hspace{0.05\textwidth}%
\includegraphics[width=0.38\textwidth]{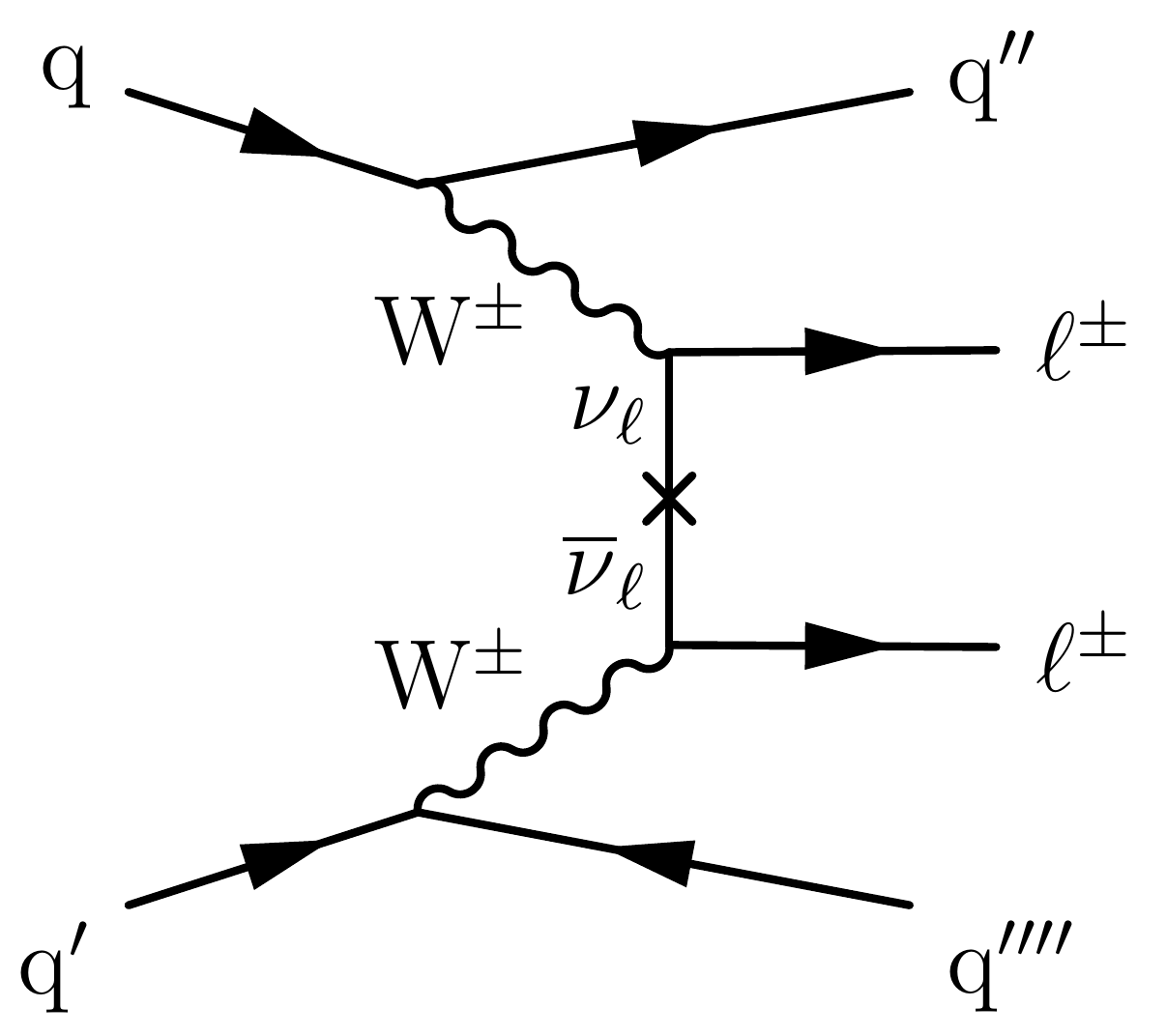}%
\caption{%
    Example Feynman diagrams of VBF processes with heavy Majorana neutrino production (left) and processes mediated by the Weinberg operator (right).
}
\label{fig:FD_HNLs_VBF}
\end{figure}

A Majorana HNL in the context of the \TypeOne seesaw model would also induce a process where two SS \PW bosons fuse and lead to the production of a pair of SS leptons~\cite{Datta:1993nm,Fuks:2020att}, notably with the absence of neutrinos in the final state as illustrated in the Feynman diagram in Fig.~\ref{fig:FD_HNLs_VBF} (left).

It is worth noting that the cross section of this kind of $t$-channel processes (processes characterized by the exchange of a virtual particle) is less sensitive to the mass of the intermediate particle compared with $s$-channel quark-antiquark annihilation processes discussed previously and shown in Fig.~\ref{fig:FD_HNLs}. The Vector Boson Fusion (VBF) processes, presented in Fig.~\ref{fig:FD_HNLs_VBF} (left), may complement searches for heavy Majorana neutrinos in the $t$-channel at the \TeVns mass scale.

Additionally, these VBF-type processes are analogous to the VBF processes induced by the dimension-5 Weinberg Operator~\cite{Fuks:2020zbm}. This operator is proposed~\cite{Weinberg79} to extend the SM Lagrangian with terms of the form
\begin{equation}
    \mathcal{L}_5 = \frac{\Cfivell}{\Lambda} \big[\Phi\cdot\overline{L}^c_{\Pell}\big] \big[L_{\Pellpr}\cdot\Phi\big],
\end{equation}
where \Pell and \Pellpr are different lepton flavors (electrons, muons, or tau leptons); $\Lambda$ is the energy scale at which the particles responsible for neutrino masses becomes a non negligible parameter; \Cfivell is a flavor-dependent Wilson coefficient; $L_{\Pell}=(\PGn_{\Pell},\Pell)$ is the LH lepton doublet; and $\Phi$ is the SM Higgs doublet with a vacuum expectation value $v=\sqrt{2}\langle\Phi\rangle\approx246\GeV$. The Weinberg operator provides a natural formalism for generating neutrino masses as shown in Eq.~\eqref{eqn:Majorana_N_mass}, where $m_{\ellellpr}$ represents the effective dilepton mass:
\begin{equation}\label{eqn:Majorana_N_mass}
    m_{\ellellpr}=\Cfivell v^2/\Lambda.
\end{equation}
This mechanism introduces LNV, as the Majorana neutrino is its own antiparticle. This mechanism shares similarities with the process of neutrino double beta decay.

\subsection{Production of HNLs in the \TypeThree seesaw model}
\label{subsec:typeIIIseesaw}

In contrast to the \TypeOne seesaw mechanism (Section~\ref{sec:HNL_th_type-I}), where heavy states of mass \mhnl involving weak-isospin singlets were introduced, the \TypeThree seesaw mechanism introduces an \SU2 triplet of heavy leptons~\cite{Foot:1988aq}. The neutrino masses are generically reduced relative to charged-fermion masses by a factor $v/\mhnl$, where $v$ is the vacuum expectation value of the Higgs field. For sufficiently large \mhnl (of the order of $10^{14}\GeV$), small neutrino masses are generated even for Yukawa couplings of $\approx$1. Smaller Yukawa couplings are required to obtain small neutrino masses while keeping \mhnl close to a few hundreds of \GeVns, such that these heavier lepton states may be produced at the LHC. These new triplet states may be produced through gauge interactions, such that the possible smallness of the Yukawa couplings does not affect the production cross section of the heavy states.

\begin{figure}[ht!]
\centering
\includegraphics[width=0.45\textwidth]{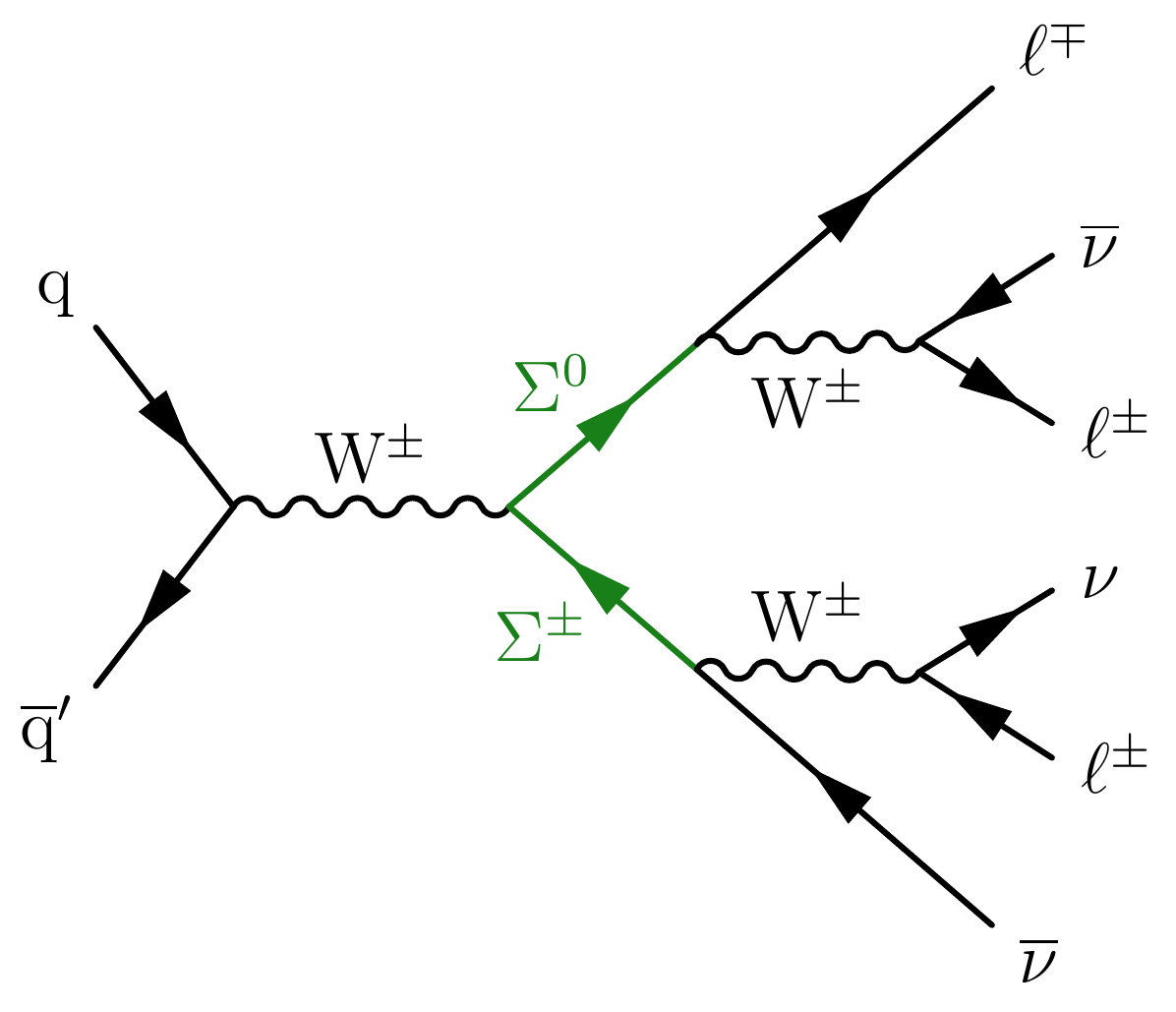}%
\hspace{0.05\textwidth}%
\includegraphics[width=0.45\textwidth]{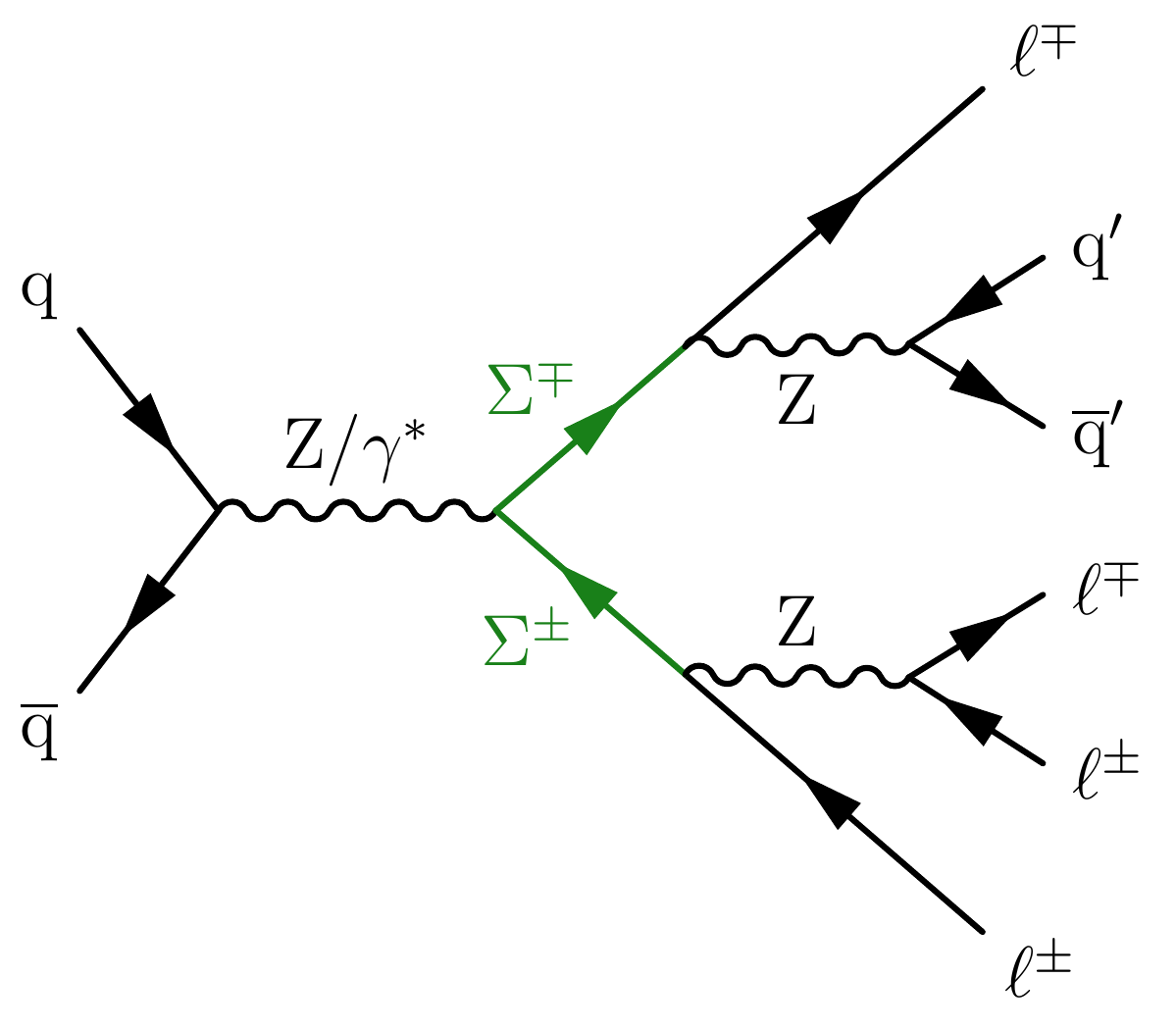}%
\caption{%
    Example Feynman diagrams illustrating production and decay of \TypeThree seesaw heavy lepton \PGS pairs at the LHC that may result in multilepton final states.
}
\label{fig:SeesawFeynman}
\end{figure}

Within the \TypeThree seesaw model, these massive leptons are two heavy Dirac charged leptons (\PGSpm) and a heavy Majorana neutral lepton (\PGSz). These heavy leptons may be pair-produced through LO EW interactions in charged-charged ($\PGSpm\PGSmp$) and charged-neutral ($\PGSp\PGSz$ or $\PGSm\PGSz$) modes. The seesaw leptons are assumed to mix with the SM leptons, and decay to a \PW, \PZ, or Higgs boson and an SM lepton (\PGn, or $\Pell=\Pe$, \PGm, \PGt). The three production modes, combined with the nine possible combinations of boson-SM lepton decay, yield 27 distinct signal production and decay modes. An example of the complete decay chain that may yield multiple leptons in the final state is $\PGSpm\PGSz\to\PWpm\PGn\PWpm\Pellmp\to\Pellpm\PGn\PGn\Pellpm\PGn\Pellmp$. Two diagrams exemplifying the production and decay process of \PGS pairs that may result in multilepton final states are shown in Fig.~\ref{fig:SeesawFeynman}.

The total width of these new heavy-lepton states and their decay branching fractions to SM leptons of flavor \Pell ($\BR_{\Pell}$) are proportional to $\mixparsq{\Pell}/(\mixparsq{\Pe}+\mixparsq{\PGm}+\mixparsq{\PGt})$, where \mixparl is the heavy-light lepton mixing angle. If all three \mixparl values are less than ${\approx}10^{-6}$, the \PGS states may have sufficiently long lifetimes to produce leptons at secondary vertices. Electroweak and low-energy precision measurements enforce an upper limit on the mixing angles of $10^{-4}$ across all lepton flavors~\cite{Biggio:2019eeo,Das:2020uer}. This bound allows for prompt decays of heavy leptons in the mass ranges accessible to collider experiments~\cite{Abada:2007ux,Abada:2008ea,Franceschini:2008pz,Cai:2017mow,Ashanujjaman:2021jhi,Ashanujjaman:2021zrh,Aguilar-Saavedra:2009fxa}. The heavy-lepton states are assumed to be degenerate in mass and their decays are assumed to be prompt in the corresponding analysis discussed later in Section~\ref{sec:type-III}. The \PGS decay branching fractions to different bosons are determined solely by their masses.

\subsection{Production of HNLs in the left-right symmetric model}
\label{subsec:LRSMtheory}

\begin{figure}[htb!]
\centering
\includegraphics[width=0.43\textwidth]{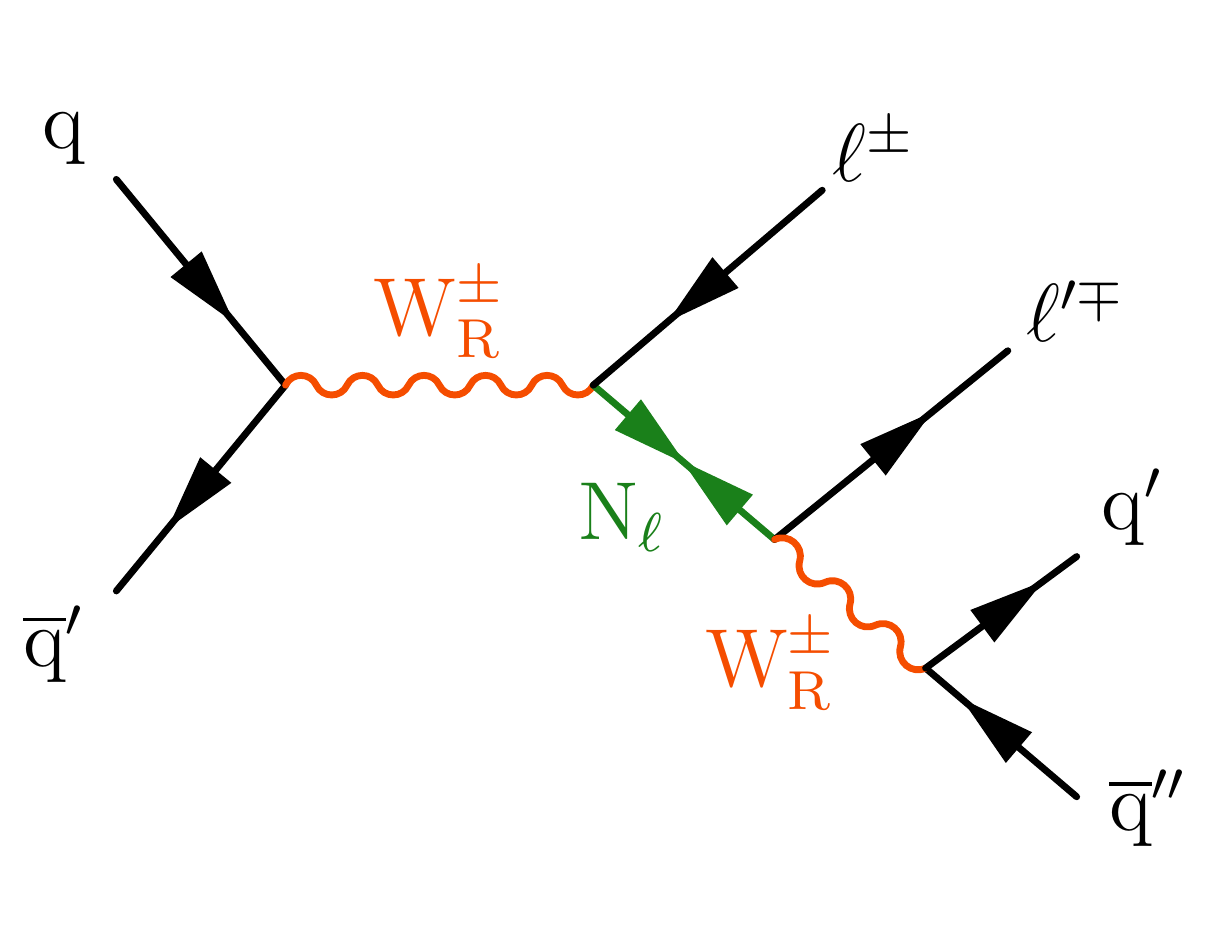}%
\hspace*{0.05\textwidth}%
\includegraphics[width=0.47\textwidth]{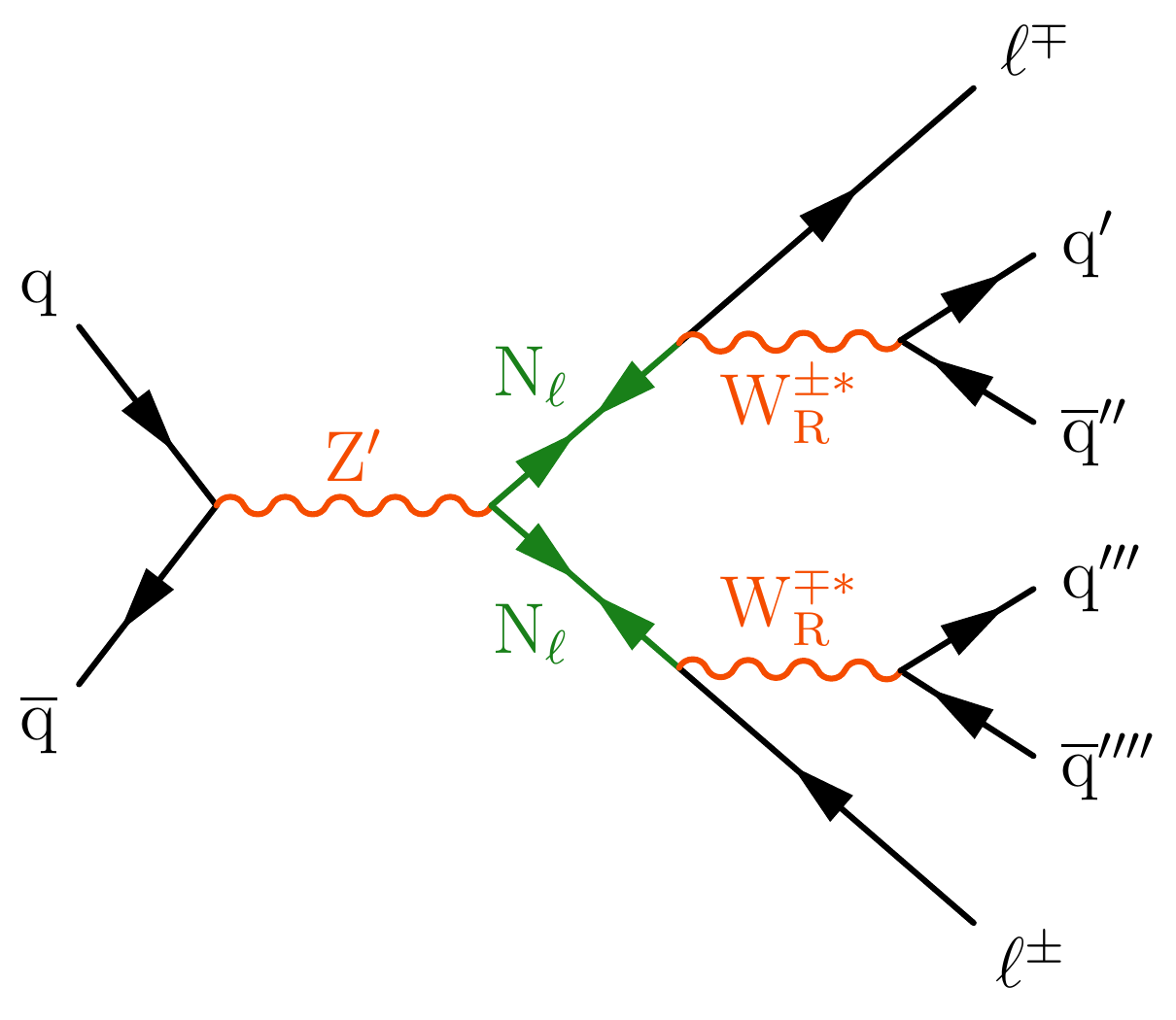}
\caption{%
    Representative Feynman diagrams for the production of a heavy Majorana neutrino, labeled as \PNell, via the decay of a \PWR (left) and \PZpr boson (right).
}
\label{fig:lrsm_feynman}
\end{figure}

The LRSM is a renormalizable framework that is constructed by adding the $\SU2_{\mathrm{R}}$ gauge group to the SM, introducing the heavy partners of the SM \PW and \PZ bosons, namely \PWRpm and \PZpr bosons~\cite{Pati:1974yy, Mohapatra:1974gc}. This naturally embeds the seesaw mechanism, providing answers for the small SM neutrino masses with a heavy Majorana-type particle \PN. The LRSM may be directly tested by searching for the postulated \PWRpm and \PZpr bosons at the LHC~\cite{PhysRevLett.50.1427,Maiezza_2010,Mattelaer:2016ynf}. The searches are focused on cases where \PN couples exclusively to leptons with one single flavor, assuming three RH neutrinos (\PNe, \PNGm, and \PNGt) with different masses. As a result of the nature of Majorana-type particles, the \PN allows for a lepton number violation by two units.
The production process for \PN at the LHC may either be mediated by the \PWR boson with a charged lepton, or a resonant \PZpr boson that produces a pair of HNLs, as shown in Fig.~\ref{fig:lrsm_feynman}. As the \PWR and \PZpr bosons are assumed to be heavy, such production channels of the \PN yield various interesting event topologies. For parameter points in the \mhnl--\mWR or \mhnl--\mZpr plane, large mass gaps between the mediating heavy gauge boson and the \PN (\ie, $\mhnl/\mWR\ll1$ or $\mhnl/\mZpr\ll1$) are likely to induce a large Lorentz boost for the \PN.
Because of the collimation of the decay products of the boosted HNL, jet substructure is a powerful tool to distinguish HNL decays from QCD multijet background in this channel.

\subsection{Production of HNLs in composite models}
\label{sec:th_hnl_composite}

\begin{figure}[ht!]
\centering
\includegraphics[width=\textwidth]{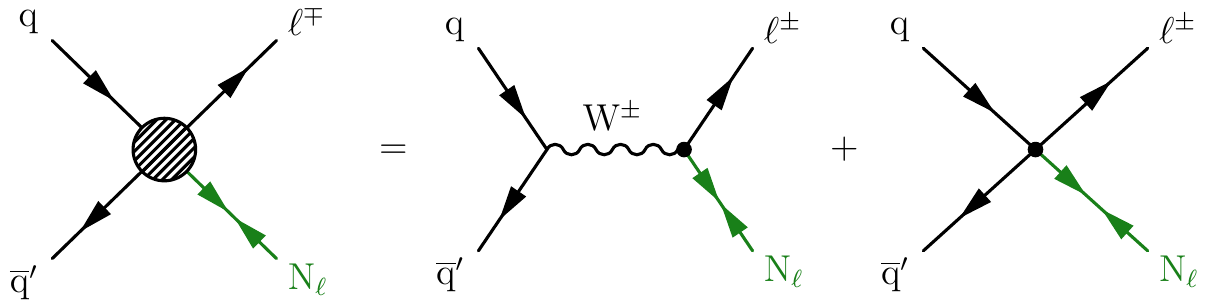}
\caption{%
    The fermion interaction as a sum of gauge (center) and contact (right) interaction contributions.
}
\label{fig:interactions}
\end{figure}

\begin{figure}[ht!]
\centering
\includegraphics[width=\textwidth]{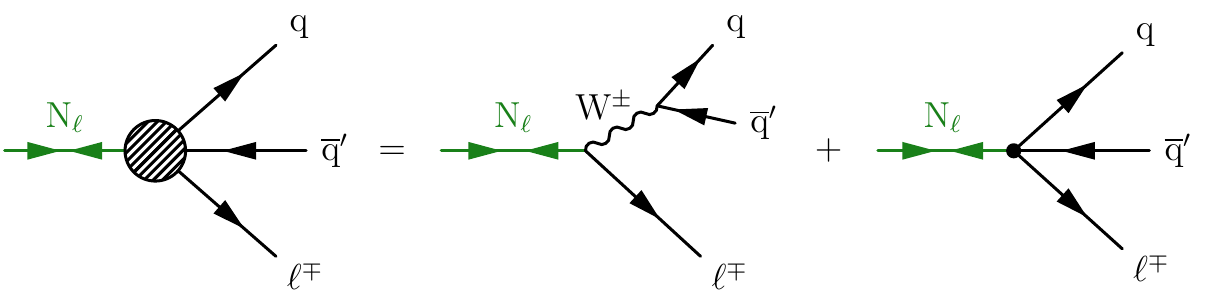}
\caption{%
    Example diagrams for the decay of a heavy composite Majorana neutrino to $\Pell\qqbarpr$.
}
\label{fig:decay}
\end{figure}

In the context of the composite-fermion theory~\cite{Pati:1975md,Greenberg:1974qb,Eichten:1979ah,Eichten:1983hw,Harari:1982xy}, quarks and leptons possess an internal substructure that manifests itself at a sufficiently high energy scale $\Lambda$, the compositeness scale.
A relevant phenomenological feature of the compositeness scenarios is the existence of excited states of quarks and leptons, with masses lower than or equal to $\Lambda$, interacting with SM fermions via effective field theory (EFT) interactions~\cite{Terazawa:1976xx,Cabibbo,Baur:1989kv,Baur:1987ga}.
A particular case of such excited states is an HNL (\PNell, $\Pell=\Pe$, \PGm, \PGt)~\cite{O3,O2,O1,Biondini:2021vip}: a composite Majorana fermion often associated with the observed baryon asymmetry in the universe.
Such composite Majorana neutrinos would also lead to detectable effects in neutrinoless double beta decay experiments \cite{O2,Biondini:2021vip}.
As a general phenomenological framework, we consider the composite neutrino model given in Ref.~\cite{Leonardi:2015qna}, in which two types of effective interactions, the gauge interaction (GI) and the contact interaction (CI), enter into both the production and decay mechanisms and are governed, respectively, by the Lagrangian densities:
\begin{equation}\begin{aligned}
  \mathcal{L}_{\text{GI}} &= \frac{gf}{\sqrt{2}\Lambda}\overline{\PN}\sigma^{\mu\nu}(\partial_\mu\PW_\nu)P_{\mathrm{L}}\Pell+\text{h.c.},
  \\
  \mathcal{L}_{\text{CI}} &= \frac{g_\ast^2\eta}{\Lambda^2}\PAQqpr\gamma^\mu P_{\mathrm{L}}\PQq\overline{\PN}\gamma_\mu P_{\mathrm{L}}\Pell+\text{h.c.}
\end{aligned}\end{equation}
Here \PN, \Pell, \PW, and \PQq are the \PNell, charged lepton, \PW boson, and quark fields, respectively; $P_{\mathrm{L}}$ is the LH chirality projection operator; and $g$ is the $\SU2_{\mathrm{L}}$ gauge coupling. The effective coupling for contact interactions, $g^2_\ast$, takes the value $4\pi$~\cite{Leonardi:2015qna}. The factors $f$ and $\eta$ are additional couplings in the composite model; they are taken here to be unity, a choice that is commonly adopted in phenomenological studies and experimental analyses of composite-fermion models.
The total amplitude for the production process is given by the coherent sum of the gauge and contact contributions, as shown in Fig.~\ref{fig:interactions}, together with the decay modes shown in Fig.~\ref{fig:decay}.
The production process is dominated by the CI mechanism for the entire parameter space probed at the LHC, while for the decay, the dominant interaction changes depending on $\Lambda$ and the \PNell mass~\cite{Leonardi:2015qna}.

\section{Review of heavy neutral lepton searches}
\label{sec:results_hnl}

In the preceding section, we outlined the theoretical foundations of HNL models and their potential role in addressing the enigma of neutrino masses within various theoretical frameworks. In this section, we review the experimental efforts at the CMS experiment aimed at uncovering the existence of these hypothetical HNL particles. The results from searches for HNLs in the context of the \TypeOne seesaw model are discussed first in Section~\ref{sec:type-I}. Afterwards a review of the searches for HNLs in the \TypeThree seesaw model is presented in Section~\ref{sec:type-III}, followed by a review of results in the LRSM model in Section~\ref{sec:LRSM}. Finally, the searches for heavy composite Majorana neutrinos are presented in Section~\ref{sec:HeavyComposite_results}.

\subsection{Searches for HNLs in the \TypeOne seesaw model}
\label{sec:type-I}

In this section, we summarize the experimental searches for HNLs within the framework of the \TypeOne seesaw model, taking into account their mass and lifetime parameters. As discussed in Section~\ref{sec:HNL_th_type-I}, HNLs may exhibit different lifetimes based on their mass and mixing parameters.

For the high-mass regime, typically above $\approx$15\GeV, HNLs are expected to have a relatively short lifetime. Searches for short-lived (prompt) HNLs benefit from standard detector reconstruction techniques to capture their signatures in the CMS detector. On the other hand, in the low-mass regime, typically below $\approx$15\GeV, HNLs are anticipated to be long-lived. In this scenario, searches for HNLs with long lifetimes rely on innovative techniques to reconstruct the displaced decays of the HNLs. The successful implementation of challenging reconstruction methods for the detection of long-lived HNLs have significantly increased their discovery potential at the LHC.

By presenting both prompt and long-lived HNL searches, we offer a comprehensive perspective on the distinct challenges and methods associated with each category, covering a complementary parameter space of searches for HNLs. First, the CMS experiment results on prompt HNLs are discussed, followed by an overview of the long-lived HNL searches. In this theoretical framework, the HNL mass and mixing are free model parameters, leading to the presentation of exclusion limits in two-dimensional planes of mass versus squares of mixing matrix elements.

\subsubsection{Searches for prompt HNLs}

Several prompt HNL analyses are reviewed. We organize the review by prompt HNL decay channel, starting with semihadronic decays $\PN\to\Pellpm\qqbarpr$, followed by the leptonic decays corresponding to $\PN\to\Pellpm\Pellmp\PGn$ and $\PN\to\Pellpm\Pellpm\PGn$~\cite{delAguila:2008hw}.

First, we review a search for HNL production through the decay of \PW boson as illustrated in Fig.~\ref{fig:FD_HNLs} where \PN decays into a lepton and a \PW boson~\cite{CMS:2018jxx}.
The \PW boson, in turn, further decays into two quarks. This series of production and decay processes results in the presence of two leptons and jets in the final state.
This search uses the 2016 data set.

This search considers cases where the \PN inclusively mixes with either electron (\Pe) or muon (\PGm) flavor SM neutrinos. Additionally, since the \PN is a Majorana particle, it allows for processes with LNV. Consequently, the analysis examines the final state consisting of an SS lepton pair in conjunction with jets. In addition to the the $s$-channel production mode of the \PN, this study introduces the \Wgamma VBF channel shown in Fig.~\ref{fig:FD_HNLs_Wgamma}. The inclusion of the \Wgamma VBF channel enhances the sensitivity of the search, particularly for a larger \PN mass range, where it becomes the dominant production mechanism for HNLs. In this analysis, the mass range covered extends from 20 up to 1600\GeV, providing an exploration of prompt HNLs across a broad spectrum of masses.

Compared to an earlier analysis at $\sqrt{s}=8\TeV$ performed at the CMS experiment~\cite{CMS:2015qur}, in which events were selected with two SS leptons and two small-radius jets, this search adds two additional SRs to compensate for the signal acceptance loss in both small and large \mhnl regions. For \mhnl below 80\GeV, an SR with one small-radius jet is added to increase the signal acceptance of the restricted phase space of jets. For larger masses, especially above 500\GeV, the signal acceptance is shown to be recovered by taking into account the event topology where the decay products of the \PW boson are merged into a large-radius jet.
Three potential sources of background events that are specifically related to the SS dilepton final state are considered: SM physics processes that are able to produce two prompt SS leptons; events that arise from misidentification of leptons; and lastly, OS lepton pairs in which one of the charges is mismeasured. This last source is shown to be negligible for the \mumu and \emu channels. Backgrounds with prompt SS leptons mainly arise from events with two vector bosons (\WpmWpm,\WZ, \ZZ), and these contributions are estimated using MC simulation. Events containing objects misidentified as prompt leptons constitute the most important background for low-mass signals, and they originate from \PQb hadron decays, light-quark or gluon jets. The simulation is not reliable in estimating the misidentified-lepton backgrounds, and we therefore estimate these backgrounds directly from collision data by calculating lepton misidentification probabilities. The last background category is from lepton sign mismeasurements in events with jets and two opposite-sign leptons. The probability of mismeasuring the sign of a prompt electron is obtained from simulated $\PZ\to e^{\pm}e^{\mp}$ events and is parametrized as a function of the electron \pt.

The exclusion limits are determined using a cut-and-count method, applied to different SR selections corresponding to various mass scenarios. The SRs are first categorized into two groups based on the mass scenario of the \PN, and a selection tailored to each group is applied: $\mhnl<80\GeV$ and $\mhnl\geq80\GeV$.
Furthermore, within each of these two mass scenario categories, the SRs are further divided into two separate regions, each targeting a different kinematic phase space. Regardless of the mass scenario, a first SR dedicated to the resolved kinematic phase space, is defined by requiring at least two small-radius jets and no large-radius jets.
The second SR, designed for the scenario involving high Lorentz boosts, has different requirements. For the low-mass scenario, exactly one small-radius jet and zero large-radius jets are required, while for the high-mass scenario, the large-radius jet multiplicity must be at least one.
In addition, the SRs are further categorised based on the flavor combination of the leptons in the lepton pair, leading to events with \mumu, \ee, and \emu pairs. With these three lepton flavor channels, the analysis has a total of 12 separate SRs.

\begin{figure}[ht!]
\centering
\includegraphics[width=0.48\textwidth]{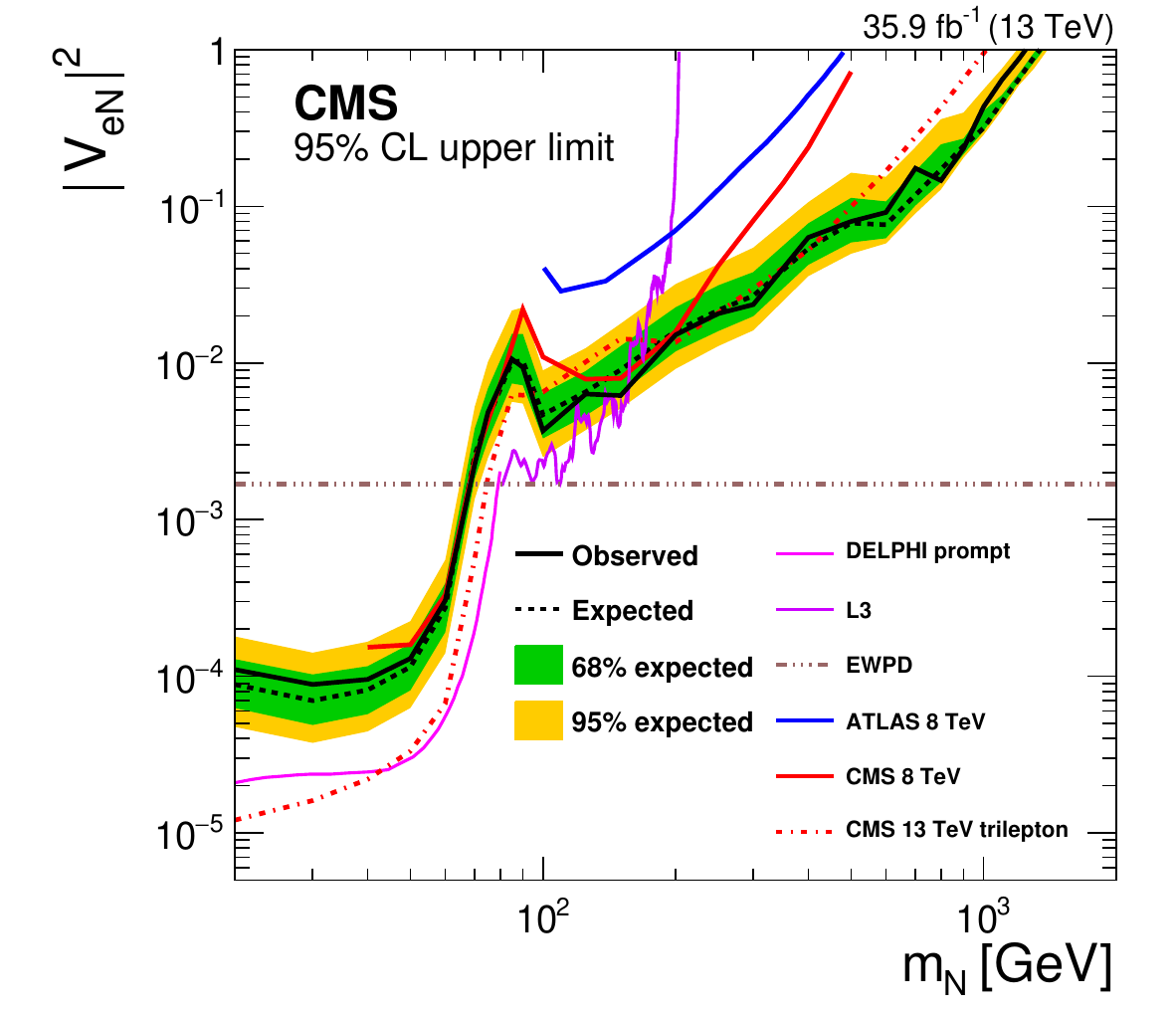}%
\hfill%
\includegraphics[width=0.48\textwidth]{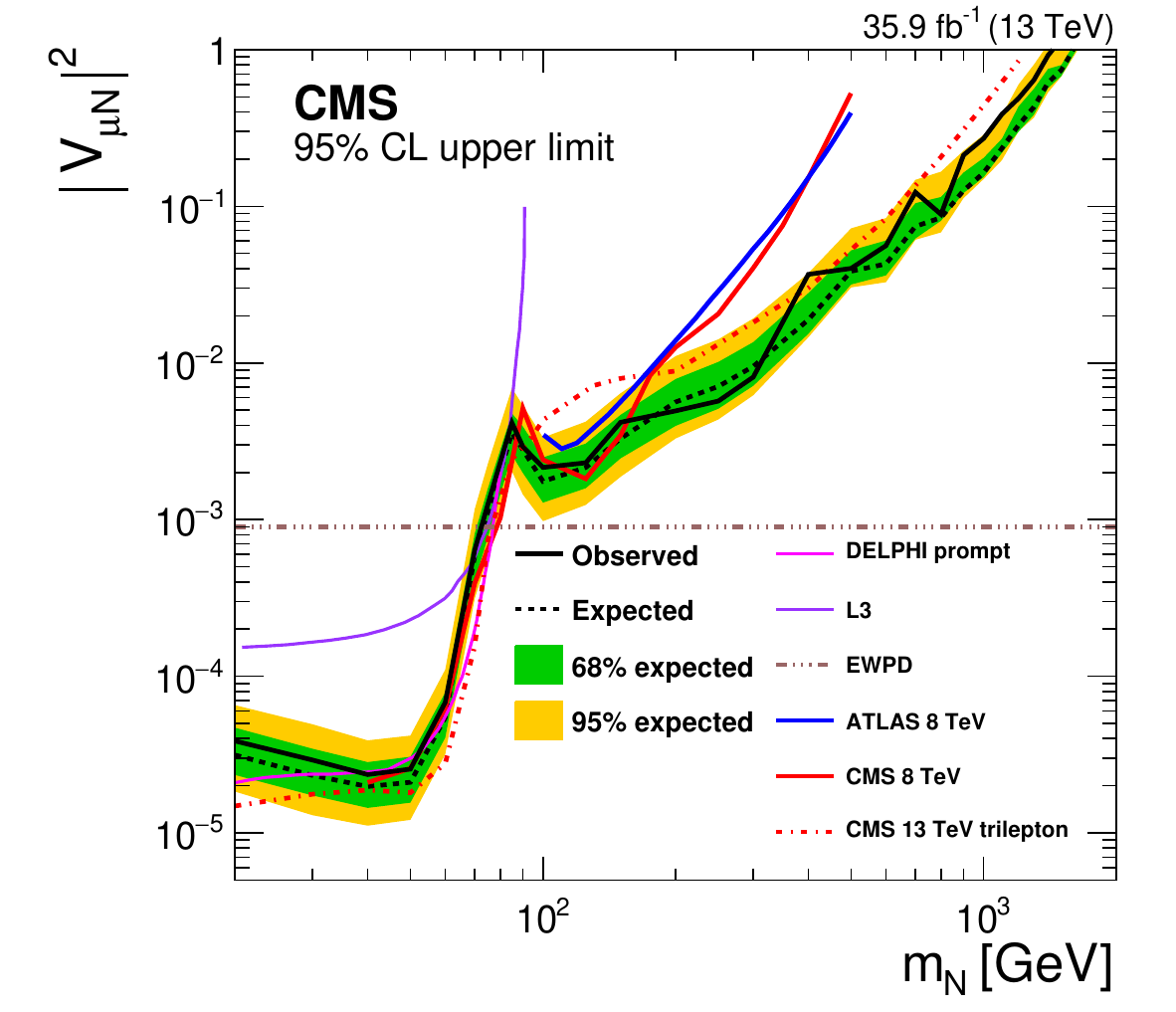} \\
\includegraphics[width=0.48\textwidth]{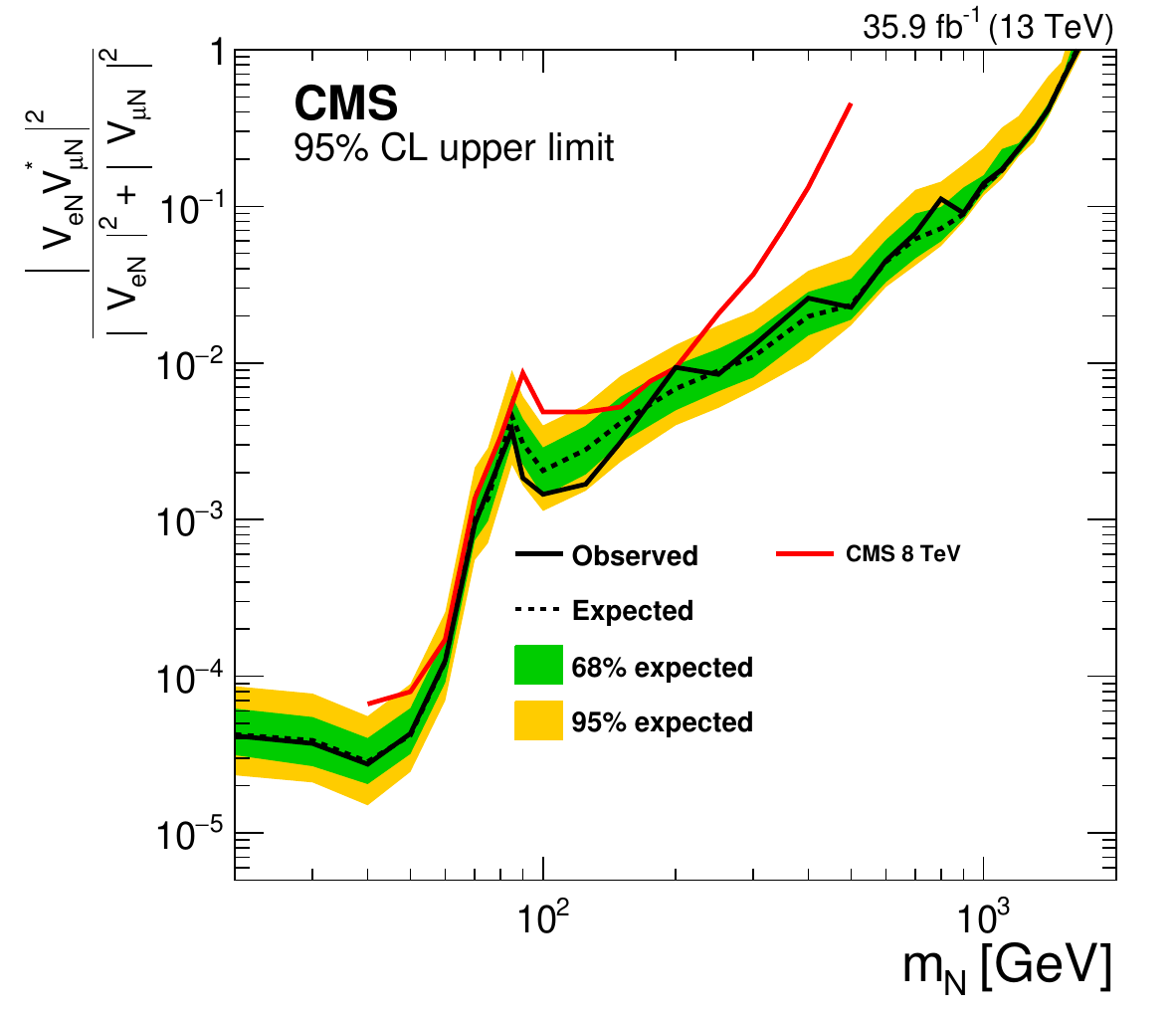}%
\caption{%
    Expected (observed) upper limits at 95\% \CL shown with a dashed (solid) black line, derived on heavy neutrino mixing parameters \mixparsqeN, \mixparsqmN, and \mixparsqemN as functions of the HNL mass.
    The dashed brown line in the upper two figures shows constraints from Electroweak Precision Data (EWPD)~\cite{deBlas:2013gla} on the \mixparsqeN and \mixparsqmN parameters. 
    The lower figure, reproduced from Ref.~\cite{CMS:2018jxx}, does not show the corresponding EWPD limits.
    The upper limits from other direct searches at the DELPHI experiment~\cite{DELPHI:1996qcc}, the L3 experiment~\cite{ADRIANI1992371, L3:2001zfe}, and the ATLAS experiment~\cite{ATLAS:2015gtp} are superimposed.
    Also shown are the upper limits from the CMS experiment at $\sqrt{s}=8\TeV$ using the 2012 data set~\cite{CMS:2015qur} with a solid red line, and the search in the trilepton final state~\cite{CMS:2018iaf} based on the same 2016 data set as used in this analysis with a dashed red line.
    Figures adapted from Ref.~\cite{CMS:2018jxx}.
}
\label{fig:EXO-17-028_Limits}
\end{figure}

Exclusion limits are derived on HNL masses in the range from 20 to 1600\GeV. Upper limits at 95\% \CL on mixing parameters are placed ranging up to 1240, 1430, and 1600\GeV for $\mixparsqeN=1$, $\mixparsqmN=1$, and $\mixparsqemN=1$ assumptions, respectively, as shown in Fig.~\ref{fig:EXO-17-028_Limits}.

The second search for HNLs we review is a search in which the same HNL production mode is probed, \ie, $s$-channel and \Wgamma VBF ($t$-channel) processes. Fully leptonic decays of HNLs are considered, where both $\PN\to\Pell\PW\to\Pell\Pellpr\PGn$ and $\PN\to\PGn\PZ\to\PGn\Pellpr\Pellpr$ result in final states with three leptons and a neutrino~\cite{EXO-22-011}.
The full Run 2 data set is analyzed.
Events are selected with three leptons, including for the first time up to one hadronically decaying tau lepton (\tauh).
The HNL signal models are considered with an \PN mixing exclusively to one of the SM neutrino generations: electron, muon, or tau neutrinos.
When only electron (muon) neutrino mixing is assumed, only \eee and \eem (\emm and \mmm) events are relevant for the search.
On the other hand, when only tau neutrino mixing is assumed, all combinations of three electrons and muons are important when both tau leptons in the final state decay leptonically, and events with $\ee\tauh$, $\emu\tauh$, and $\mumu\tauh$ are relevant when one tau lepton decays hadronically.

Selected events are divided into low- and high-mass regions by requiring the leading lepton \pt to be below or above 55\GeV, respectively.
The low-mass region targets HNL scenarios with masses below the \PW boson mass that result in a compressed lepton momentum spectrum, whereas the high-mass region targets HNL masses above the \PW boson mass.
In each region, events are further categorized based on whether or not they have an OSSF lepton pair, where events with an OSSF pair have larger background contributions, but events without are only possible for Majorana HNLs.
Events are then binned into search regions based on kinematic properties of the selected leptons and \ptmiss, to separate signal from background events, as well as potential HNL scenarios with different masses.
The search region categorization is similar to the previous trilepton search based on the 2016 data set~\cite{CMS:2018iaf}.

To further improve the separation between signal and background contributions, an alternative strategy with ML discriminants based on BDTs is applied as well.
The BDTs are trained in the five mass ranges for HNL masses between 10 and 400\GeV, separately for the three different mixing scenarios.
In addition to kinematic properties of the leptons in the events, information about reconstructed jets is provided as input to the BDTs as well.

The SM background contributions after the event selection arise mostly from diboson (\WZ, \ZZ, and \Vgamma) production, with smaller contributions from nonprompt leptons and rare top quark and Higgs boson production.
The nonprompt-lepton background is estimated with a misidentification-rate method from sidebands in the observed data. Precise estimations of these backgrounds are important for channels including hadronically decaying tau leptons.
All other backgrounds are estimated from simulation, and the three diboson processes are validated in dedicated CRs, from which additional simulation correction factors are derived.

\begin{figure}[ht!]
\centering
\includegraphics[width=0.48\textwidth]{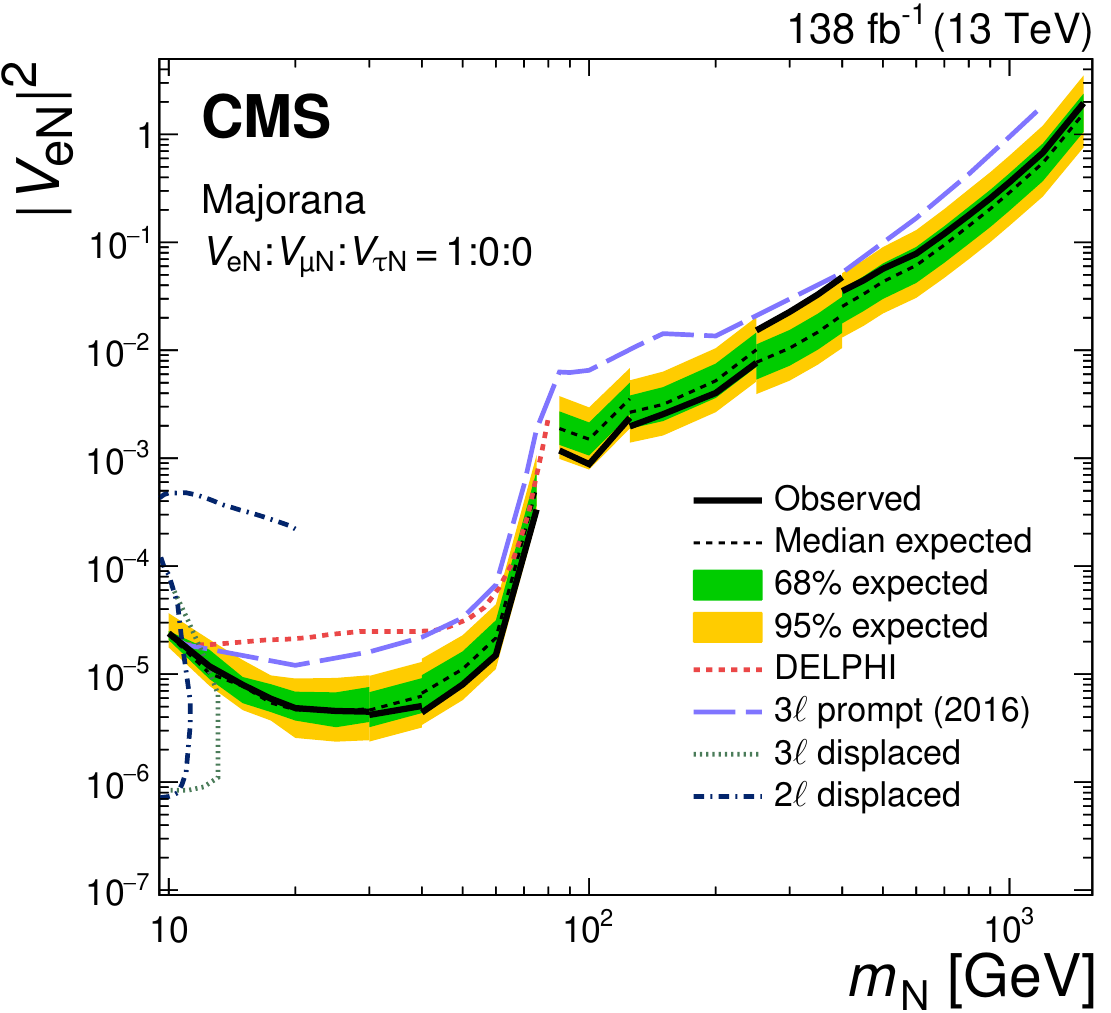}%
\hfill%
\includegraphics[width=0.48\textwidth]{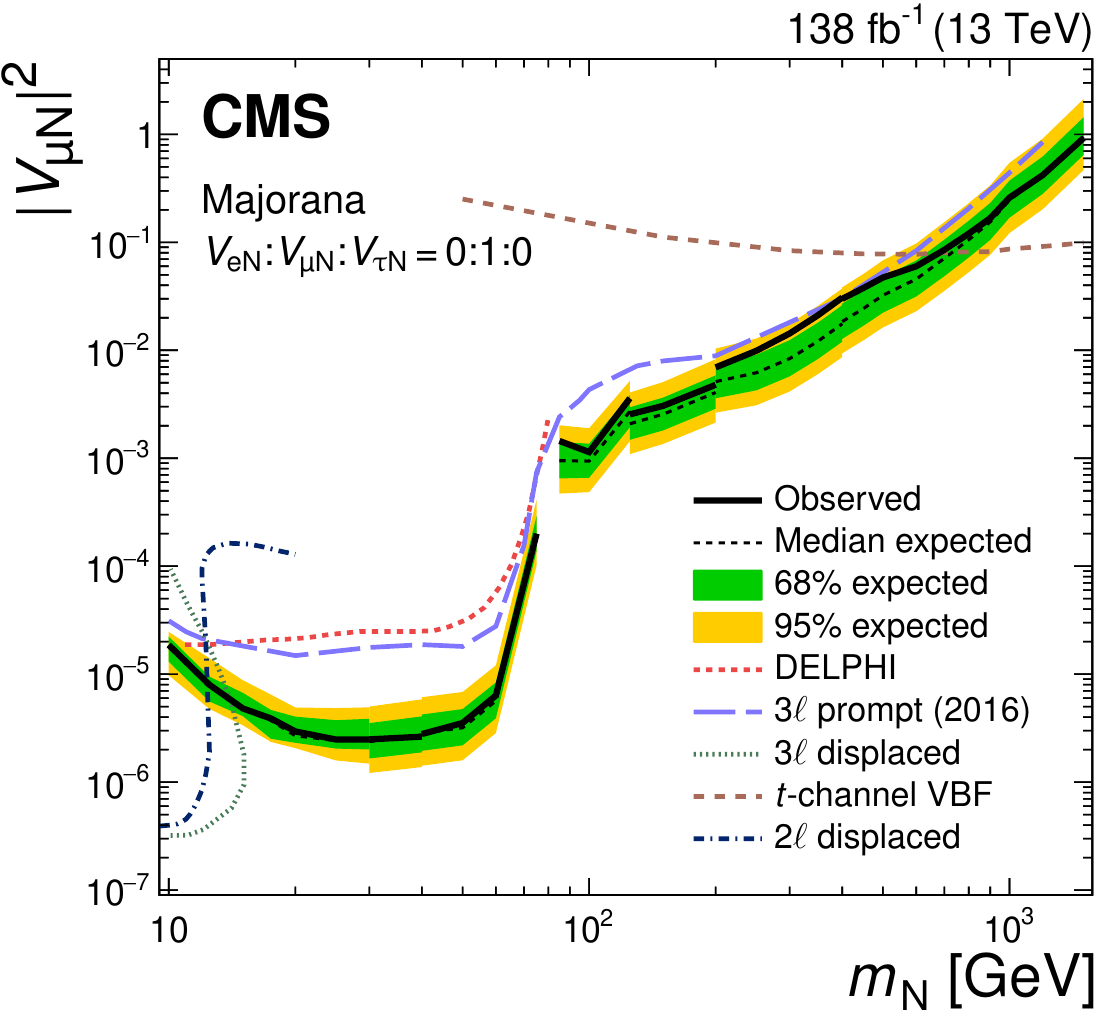} \\
\includegraphics[width=0.48\textwidth]{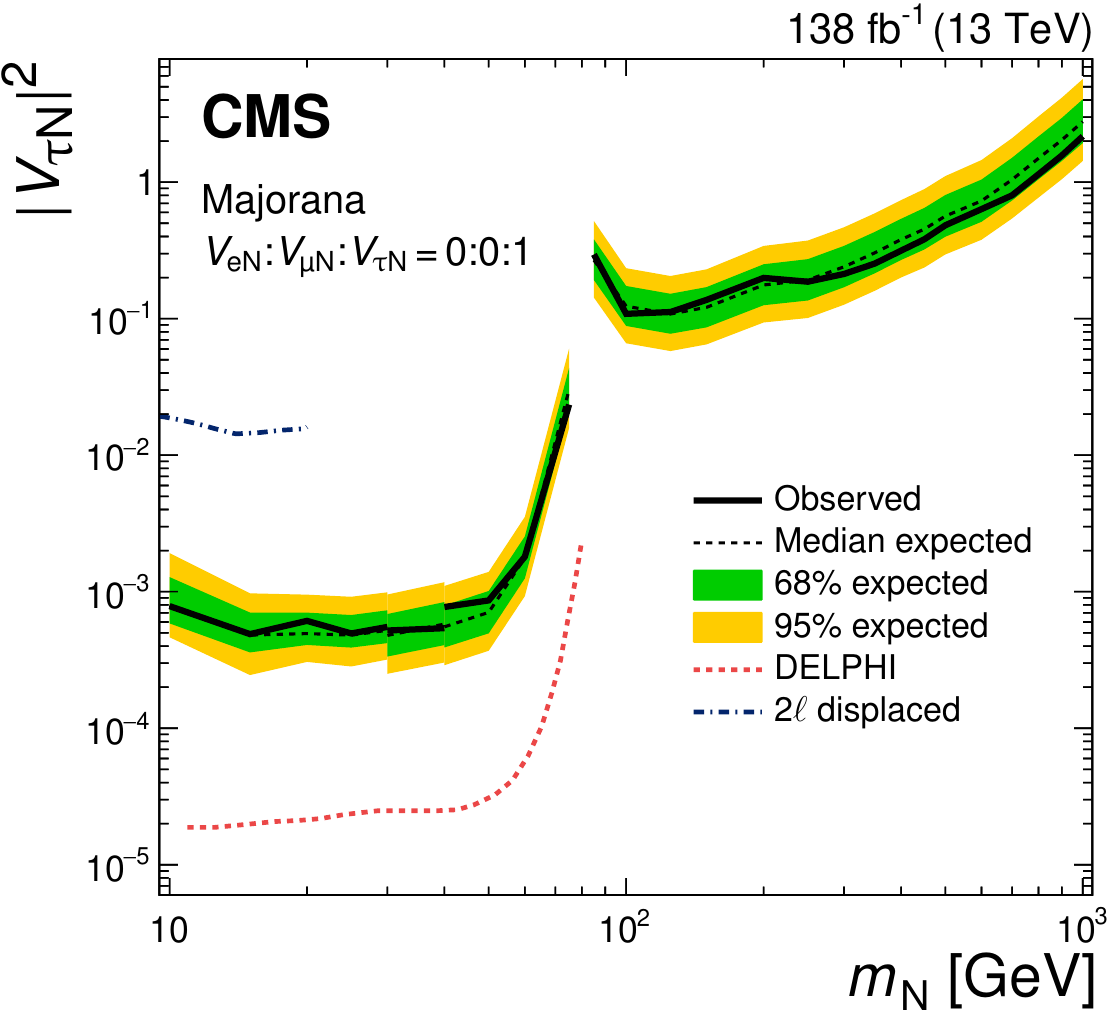}%
\caption{%
    Expected (observed) upper limits at 95\% \CL derived on heavy neutrino mixing parameters \mixparsqeN, \mixparsqmN, and \mixparsqtN as functions of the HNL mass \mhnl. No exclusion limit is evaluated for the range $75<\mhnl<85\GeV$, where HNL production through \PW boson decays has a resonance and the analysis strategy changes from using the low- or high-mass region.
    The area above the solid (dashed) black curve indicates the observed (expected) exclusion region.
    The upper limits from other direct searches at the DELPHI experiment~\cite{DELPHI:1996qcc} and the CMS experiment~\cite{CMS:2018iaf,CMS:2022fut,CMS:2022hvh,CMS:2023jqi} are superimposed.
    Figures taken from Ref.~\cite{EXO-22-011}.
}
\label{fig:EXO-22-011_Limits}
\end{figure}

For each HNL signal scenario (mixing scenario, mass, and Majorana or Dirac nature), a binned likelihood fit is performed to the BDT distributions in an optimized combination of SRs.
Based on the upper limit on the HNL production cross section derived from these fits, exclusion limits at 95\% \CL on the mixing strength as a function of the HNL mass are derived.
The exclusion limits for the case of Majorana HNL are shown in Fig.~\ref{fig:EXO-22-011_Limits}, and cover HNL masses between 10\GeV and 1.5\TeV.
Since the kinematic properties of HNL signal events are different below and above the \PW boson mass, no interpolation of limits between the highest evaluated \mhnl value with the low-mass strategy and the lowest evaluated \mhnl value with the high-mass strategy is performed, and exclusion limits are not evaluated for the range $75<\mhnl<85\GeV$.
For electron and muon mixing, the limits improve over the results from Ref.~\cite{CMS:2018iaf} over the full mass range by up to one order of magnitude.
For tau neutrino mixing, previous limits by the DELPHI experiment for HNL masses below the \PW boson mass are up to two orders of magnitude more stringent~\cite{DELPHI:1996qcc}, but for masses above the \PW boson mass, experimental limits are presented for the first time.

The CMS Collaboration performed another search~\cite{CMS:2022hvh} for Majorana neutrinos and for the signatures related to the Weinberg operator in the VBF processes~\cite{Fuks:2020zbm,Fuks:2020att}, using the full Run 2 \pp collision data set at $\sqrt{s}=13\TeV$. The considered process is the $t$ channel where SS \PW boson pairs decay into SS lepton pairs via a {\TeVns}-scale Majorana neutrino or through a Weinberg operator process, as discussed in Section~\ref{sec:HNL_th_type-I} and illustrated in Fig.~\ref{fig:FD_HNLs_VBF}.

In this search, exclusively muon neutrino mixing with the HNL is considered, hence only events with an SS dimuon pair in the final state are analyzed. The final state consists of two well-identified isolated SS muons and two jets with a large rapidity separation as well as a large dijet invariant mass.
To discriminate the signal from the SM EW \WpmWpm events, SRs for the HNL and Weinberg operator analyses are defined in bins of the azimuthal separation observable $\Dphi_{\ellell}$ and \ptmiss, respectively.
Events from the \ttbar process that have only one \PW boson decaying leptonically are the main source of the so-called nonprompt-lepton backgrounds, which originate from leptonic decays of heavy quarks or hadrons misidentified as leptons. The nonprompt-lepton background is estimated from a data sample by applying weights to events containing muon candidates that fail the nominal selection criteria while passing a less stringent requirement. To select event samples enriched in nonprompt leptons, a \PQb tag CR is defined requiring at least one \PQb-tagged jet in addition to the SR selection.
A \WZ CR, requiring the presence of three muons in an event, is used to estimate \WpmZ background contributions. The \PZ boson decay product is obtained from the OS dimuon combination with the invariant mass closest to the \PZ boson mass. Similarly, the \WZb CR is defined by requiring the same selection as for the \WZ CR, but requiring at least one \PQb-tagged jet. The dominant backgrounds in the SR are SM EW \WpmWpm production and the contribution from nonprompt leptons.

Two separate fits are performed: one for the heavy Majorana neutrino analysis using the $\Dphi_{\ellell}$ bins in the SR, and the \PQb-tagged, \WZ, and \WZb CRs; and a second for the Weinberg operator analysis with the \ptmiss bins in the SR, and the \PQb-tagged, \WZ, and \WZb CRs.
The normalization factors for the \WW, \WZ, and \tZq background processes, affecting both the SRs and CRs, are included as free parameters in the fit together with the signal strength. The bin boundaries are chosen to optimize the signal sensitivity.

\begin{figure}[!ht]
\centering
\includegraphics[width=0.48\textwidth]{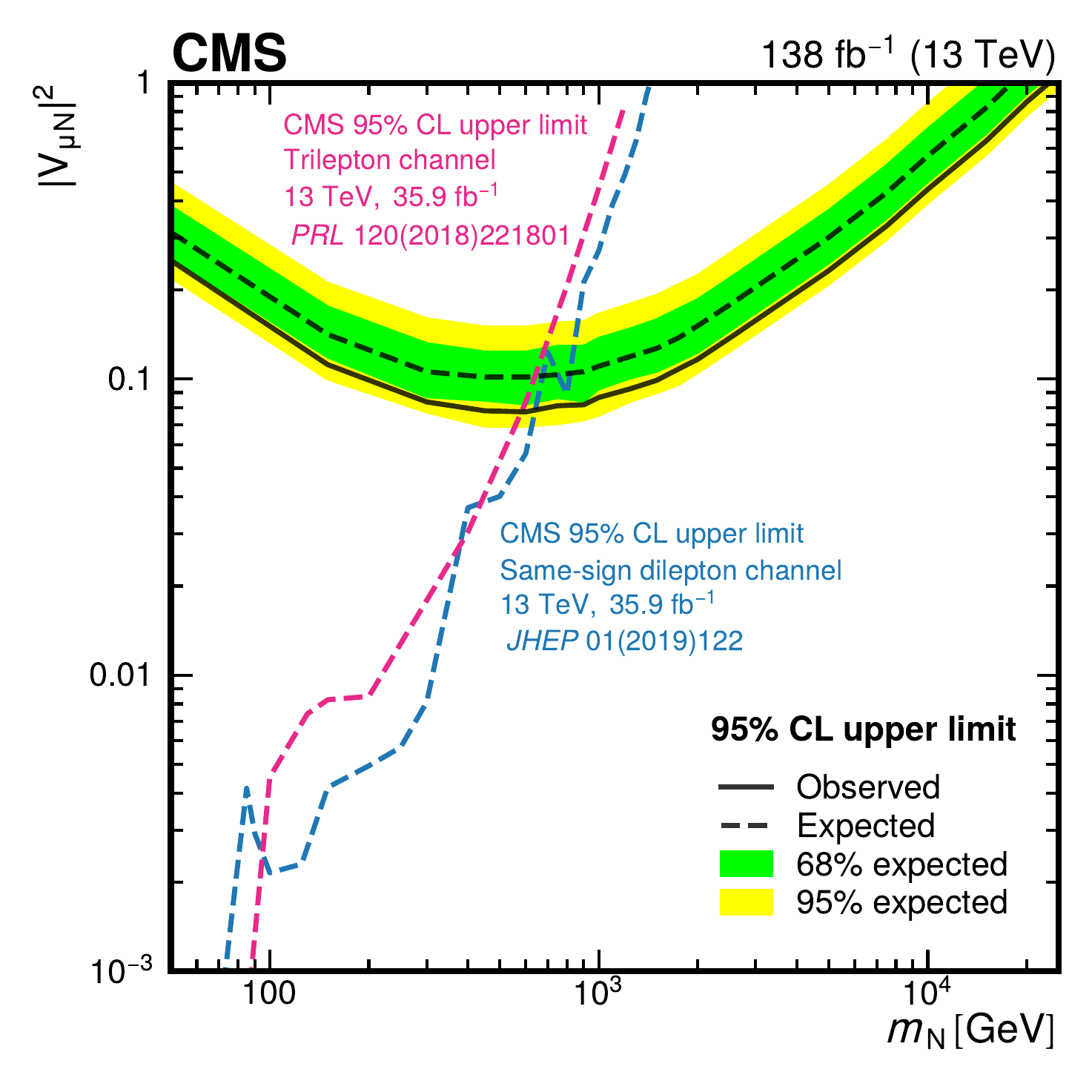}
\caption{%
    Upper limits on \mixparsqmN at 95\% \CL as a function of \mhnl.
    The black dashed curve shows the median expected upper limit, while the inner (green) band and the outer (yellow) band indicate the regions containing 68 and 95\%, respectively, of the distribution of limits expected under the background-only hypothesis.
    The solid black curve is the observed upper limit~\cite{CMS:2022hvh}.
    The red dashed curve displays the observed upper limits from Ref.~\cite{CMS:2018iaf}, while the blue dashed curve shows the observed upper limits from Ref.~\cite{CMS:2018jxx}.
    Figure adapted from Ref.~\cite{CMS:2022hvh}.
}
\label{fig:EXO-21-003_upper_limits}
 \end{figure}

The results are found to agree with the predictions of the SM. Using the relationship between the cross section and the squares of mixing matrix elements for the heavy Majorana neutrino analysis, upper limits at 95\% \CL are derived on \mixparsqmN, as shown in Fig.~\ref{fig:EXO-21-003_upper_limits}. These results surpass those obtained in previous searches by the ATLAS and CMS Collaborations~\cite{CMS:2018iaf,CMS:2018jxx,Aad:2019kiz,ATLAS2022} for $\mhnl\gtrsim650\GeV$, and set the first direct limits for $\mhnl>2\TeV$. According to Eq.~\eqref{eqn:Majorana_N_mass}, for the $\ellellpr=\mumu$ channel, a limit on the effective \mumu Majorana mass $\abs{\mmumu}=C_5^{\mumu}v^2/\Lambda$ is obtained from the limit on $\abs{\mathrm{C}_5^{\ellellpr}/\Lambda}^2$ in the Weinberg operator analysis. The observed (expected) 95\% \CL upper limit on $\abs{\mmumu}$ is found to be 10.8 (12.8)\GeV. This upper limit on \mmumu is the first obtained using a collider experiment, and it improves upon a previous limit set by the NA62 Collaboration~\cite{NA62:2019eax,Fuks:2020zbm}.

\subsubsection{Searches for long-lived HNLs}

In this section, we review the searches conducted by the CMS Collaboration for HNLs with displaced signatures, starting with searches for HNLs produced through \PW boson decays, specifically focusing on the $s$ channel. First we describe a search for the HNLs in semihadronic decays, followed by the search in fully leptonic decays. Next, a search for HNLs with inclusive decay in the muon system is discussed. Finally, we conclude this section with a search for HNLs produced from the decay of B mesons.

The first search considers a Majorana or Dirac HNL that is produced in association with one charged lepton (\Pell) and decays to a second charged lepton (\Pellpr) and jets~\cite{CMS:2023jqi}, as shown in Fig.~\ref{fig:FD_HNLs}. Scenarios in the mass range $2<\mhnl<20\GeV$, assuming inclusive coupling to all three lepton generations, are considered.
The full Run 2 data set
is analyzed, using events with two leptons (electrons or muons) and one to four jets.

Events are categorized based on the lepton flavor and charge combination, the significance of the impact parameter of the second lepton track \dxysig, and whether the HNL decay products overlap (``boosted'') or not (``resolved'').
By implementing this categorization scheme with a total of 48 SRs, a wide range of the HNL model parameter space is explored, enabling a comprehensive study for a variety of HNLs scenarios with distinct signatures.

This search employs a model-independent displaced jet tagging algorithm aimed at suppressing the larger number of background events.
The tagger is a DNN developed specifically for maximizing the efficiency for identifying displaced jets.
It utilizes a supervised ML technique to solve a multiclass classification problem. Various output classes are defined to differentiate between jets from SM processes or generic displaced ones.
Furthermore, a domain adaptation technique is used to ensure accurate performance of the resulting classifier in the observed data and the simulation. An initial version of this algorithm was developed in Ref.~\cite{CMS:2019dqq}.

\begin{figure}[!ht]
\centering
\includegraphics[width=0.48\textwidth]{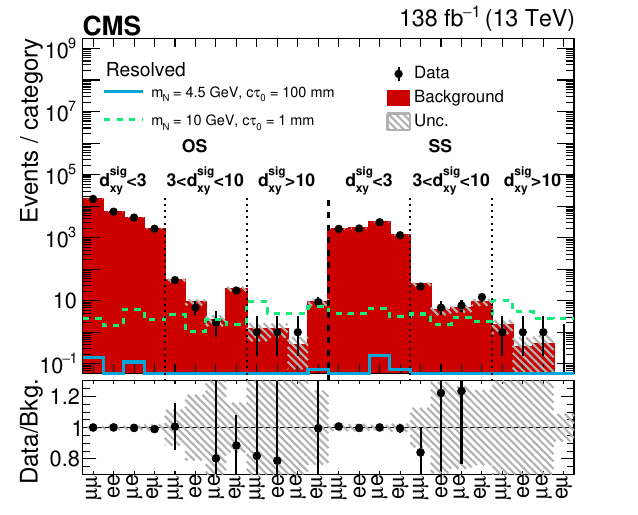}%
\hfill%
\includegraphics[width=0.48\textwidth]{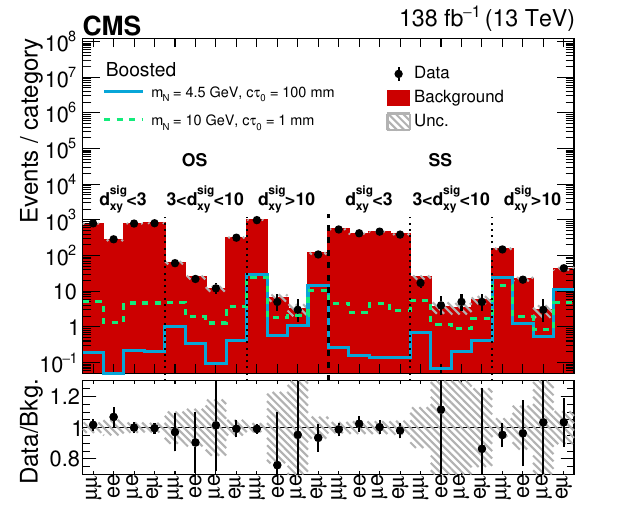}%
\caption{%
    Expected and observed background yields in 48 categories for resolved (left) and boosted (right) events.
    Two benchmark HNL scenarios are overlaid with masses of 4.5 and 10\GeV, and proper decay lengths of $\ctauhnl=100$ and 1\mm, respectively.
    The \dxysig quantity is the significance of the impact parameter of the second lepton track.
    Figures taken from Ref.~\cite{CMS:2023jqi}.
}
\label{fig:Yields_HNL_21-013}
\end{figure}

By requiring that the jet closest to the second lepton passes an optimized threshold on the tagger score, the background contribution is reduced by a factor of about $\mathcal{O}(10^{3})$ while 10--20\% of signal is retained.
The main sources of the remaining background arise from QCD multijet events and instrumentation effects, such as events with misreconstructed leptons or incorrectly reconstructed tracks.
As these backgrounds are not accurately modeled in simulation, an ABCD method (see Section~\ref{sec:bkgest}) is used to estimate the background yield from the observed data.
The two chosen uncorrelated observables in this method are the tagger output score and the invariant mass of the tagged jet and the two-lepton system, which approximates the mass of the on-shell \PW boson in signal events.
The estimated background yields per category are presented in Fig.~\ref{fig:Yields_HNL_21-013} and compared to the observed data.
The number of observed events agrees well with the expected background, indicating no evidence for an HNL signal.

\begin{figure}[!ht]
\centering
\includegraphics[width=0.48\textwidth]{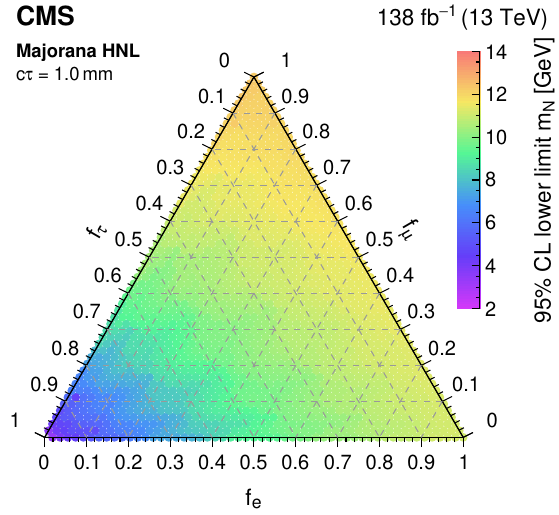}%
\hfill%
\includegraphics[width=0.48\textwidth]{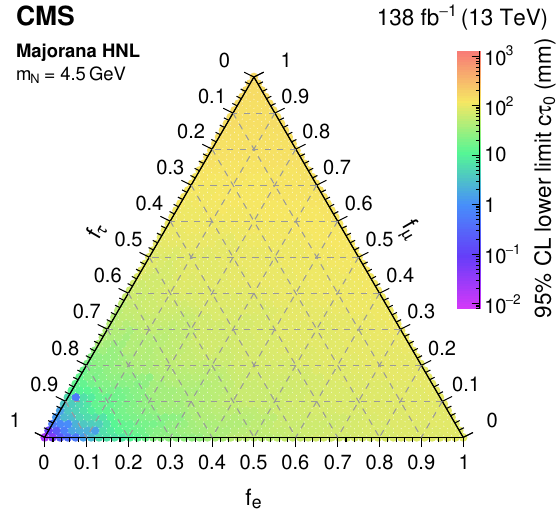}%
\caption{%
    Observed 95\% \CL lower limits on the mass (left) and the proper lifetime (right) for Majorana HNL production with $\ctauhnl=1\mm$ and $\mhnl=4.5\GeV$, respectively, as functions of the relative coupling strengths to electrons ($f_{\Pe}$), muons ($f_{\PGm}$), and tau leptons ($f_{\PGt}$).
    Figures adapted from Ref.~\cite{CMS:2023jqi}.
}
\label{fig:Triangle_HNL_21-013}
\end{figure}

The results are used to constrain the parameter space of HNL models by determining an upper limit on the HNL production cross section for each HNL mass and coupling hypothesis.
In Fig.~\ref{fig:Triangle_HNL_21-013}, the 95\% \CL lower limits on Majorana HNL production for either a fixed proper lifetime or a fixed mass are presented as a function of the relative coupling strengths to electrons, muons, and tau leptons.

Another search for long-lived HNLs produced in leptonic decays of \PW bosons~\cite{CMS:2022fut} has been performed, using the full Run 2 data set.
The postulated HNL may be either Dirac or Majorana in nature and mix exclusively to SM neutrinos from the \PW boson decays.
The considered signal process is illustrated in Fig.~\ref{fig:FD_HNLs}. The HNL subsequently decays into either a \PW boson and a charged lepton \Pell, or a \PZ boson and a neutrino \PGn. The EW gauge boson then decays leptonically, leading to a final state with three charged leptons (electrons or muons) and a neutrino ($\PN\to\Pell\Pellpr\Pellprpr\PGn$).

The HNL decay width in this final state is generally dominated by the \PWst-mediated diagrams.
This search focuses on scenarios where the HNL lifetime is such that its decay vertex may be resolved from the \pp interaction point. Therefore \Pellpr and \Pellprpr form an SV and have typically large impact parameters.
If the HNL is of Majorana nature, \Pell and \Pellpr (or \Pell and $\PGn_{\Pellpr}$) may either have the same chirality (LNV)
or opposite chirality (LNC).
In the case of an HNL decay mediated by a \PWst boson, an LNV decay may lead to final states with no OSSF lepton pairs, namely $\Pepm\Pepm\PGmmp$ or $\PGmpm\PGmpm\Pemp$.
Since the SM backgrounds in these final states are relatively small, these SRs are very sensitive to HNL signals.
In contrast, decays mediated by a \PZst boson and LNC decays are always accompanied by an OSSF lepton pair, resulting in final states such as $\Pepm\Pemp\PGmpm$ or
$\PGmpm\PGmmp\Pepm$.

Selected events must contain a prompt electron or muon, and two displaced OS leptons in any flavor combination. Prompt electrons are selected using a multivariate discriminant~\cite{CMS:2020uim}, and prompt muons must pass tight track quality requirements~\cite{CMS:2018rym}. Additional selections are applied on the maximum transverse and longitudinal impact parameters with respect to the PV and on the lepton isolation. Nonprompt electrons and muons (\Pellpr, \Pellprpr) are required to satisfy track quality and isolation requirements. These are optimized using standard sets of sequential requirements used in prompt lepton identification~\cite{CMS:2020uim,CMS:2018rym}, but removing those requirements that may affect the selection efficiency for leptons not emerging from the PV, such as a veto on photon conversions or the requirements on the number of tracker hits. Such ``displaced'' electrons and muons are required to have a transverse impact parameter $\abs{\dxy}>0.01\cm$.
Given the purity of the SR, the selection is a simple set of sequential requirements, exploiting the kinematic properties of a \PW boson decay and the small mass of the HNL. The tracks of \Pellpr and \Pellprpr are fitted to a common SV using a Kalman filter approach~\cite{Fruhwirth:1987fm}. The transverse distance between the PV and the SV (\DeltaTwoD) is used for the event categorization to maximize the sensitivity to different levels of displacement.

The main backgrounds in the SR originate from top quark,
DY, and \wjets production with misidentified hadrons or leptons from
light- and heavy-flavor hadron decays that pass the displaced-lepton
selection.
The contribution from these background leptons is estimated
with a ``tight-to-loose'' method using control samples in data.
Samples enriched with hadronic jets are used to measure the
probability for displaced leptons passing a ``loose'' isolation
criterion  to also satisfy the ``tight'' criterion for selecting
\Pellpr and \Pellprpr.
A weight based on this probability is then applied to the events found
in a second CR, obtained with a selection similar to that
of the SR, but inverting the isolation requirement on
either displaced lepton.
Two classes of background leptons are considered: single-background
(SB) leptons, \ie, single reconstructed leptons produced via one of the
aforementioned mechanisms; and double-background (DB) leptons, \ie,
pairs of reconstructed leptons produced in the decay chain of the same
hadron or from a quarkonium state.
Because the DB leptons are not independent, the selection probability is estimated for the whole system and not as a product of two uncorrelated probabilities.
The DB leptons from \PQb hadron decays are the dominant background
for dilepton masses $\mlprlprpr<4\GeV$, while
they are negligible above this threshold.

\begin{figure}[p!]
\centering
\includegraphics[width=0.9\textwidth]{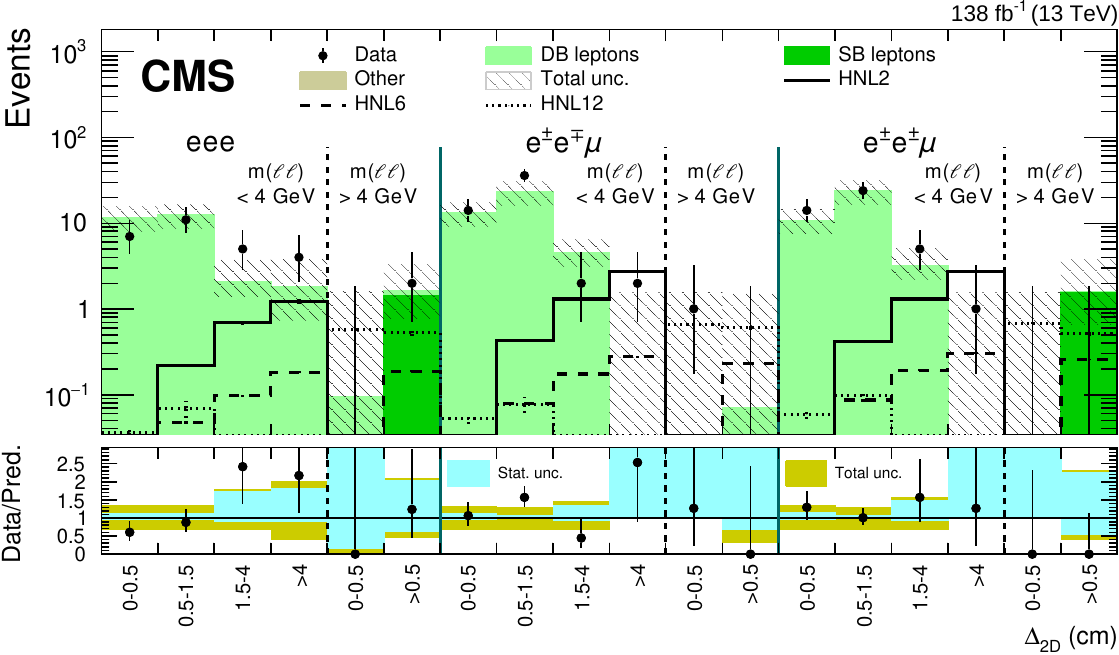} \\
\includegraphics[width=0.9\textwidth]{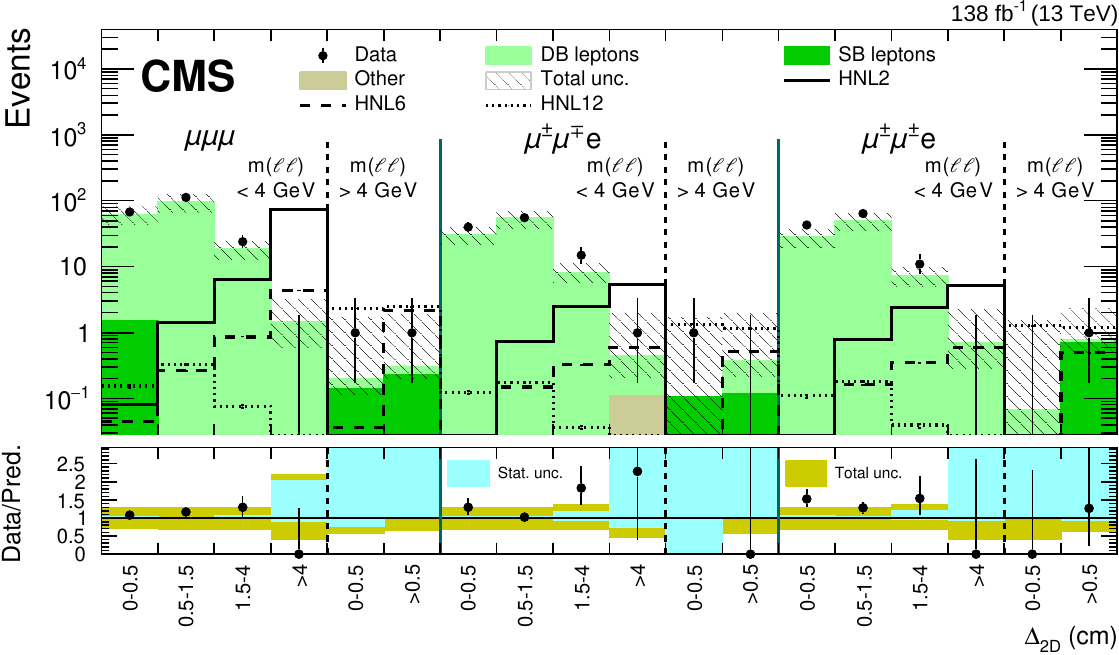}
\caption{%
    Comparison between the number of observed events in data and the background predictions (filled histograms) in the SR for \eeX
    (upper) and \mmX (lower) final states.
    The hatched band indicates the total systematic and statistical uncertainty in the background prediction.
    The lower panels indicate the ratio between the observed data and the prediction, where missing points indicate that the ratio lies outside the axis range.
    Predictions for signal events are shown for several benchmark hypotheses for Majorana HNL production:
    $\mhnl=2\GeV$ and $\mixparsqlN=0.8\times10^{-4}$ (HNL2),
    $\mhnl=6\GeV$ and $\mixparsqlN=1.3\times10^{-6}$ (HNL6),
    $\mhnl=12\GeV$ and $\mixparsqlN=1.0\times10^{-6}$ (HNL12).
    Small contributions from background processes that are estimated from simulation are collectively referred to as ``Other''.
    Figures taken from Ref.~\cite{CMS:2022fut}.
}
\label{fig:exo-20-009_shapes}
\end{figure}

\begin{figure}[t!]
\centering
\includegraphics[width=0.48\textwidth]{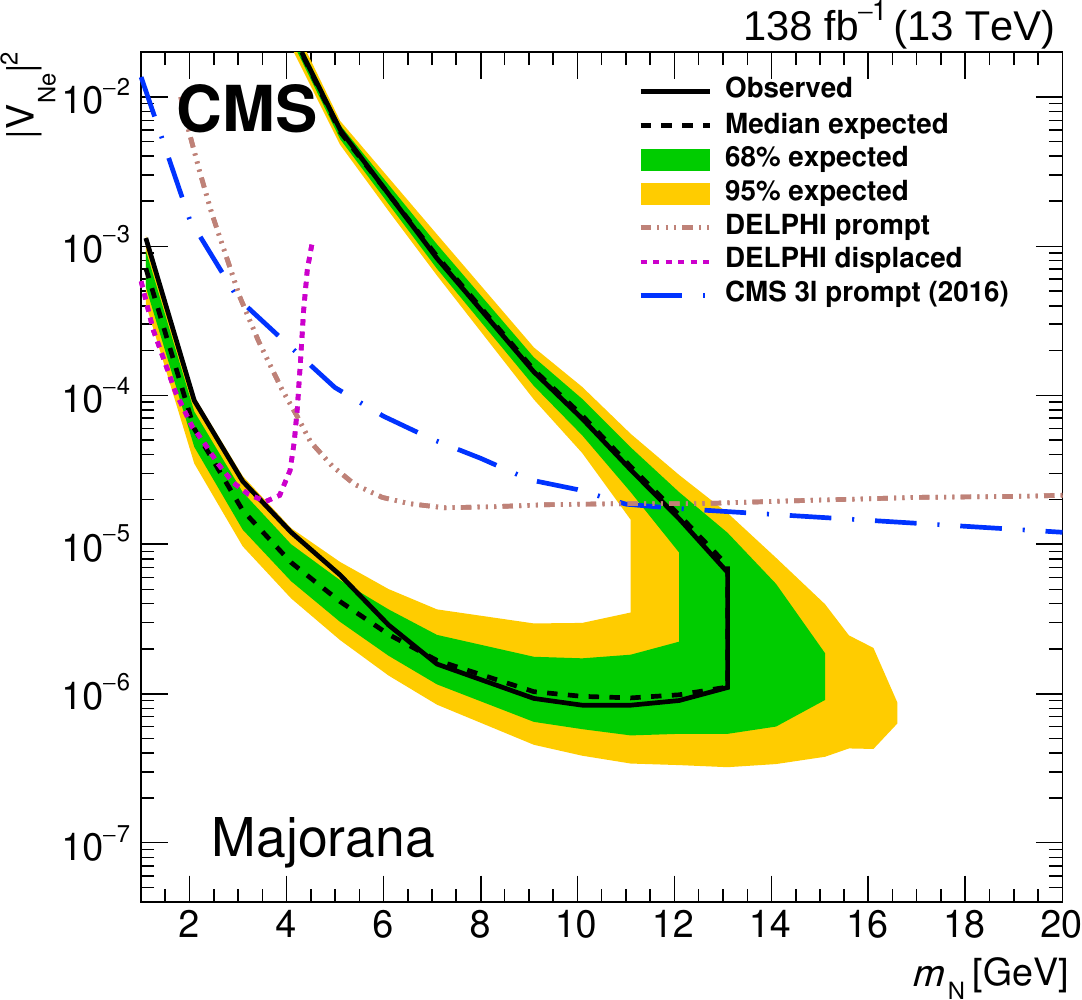}%
\hfill%
\includegraphics[width=0.48\textwidth]{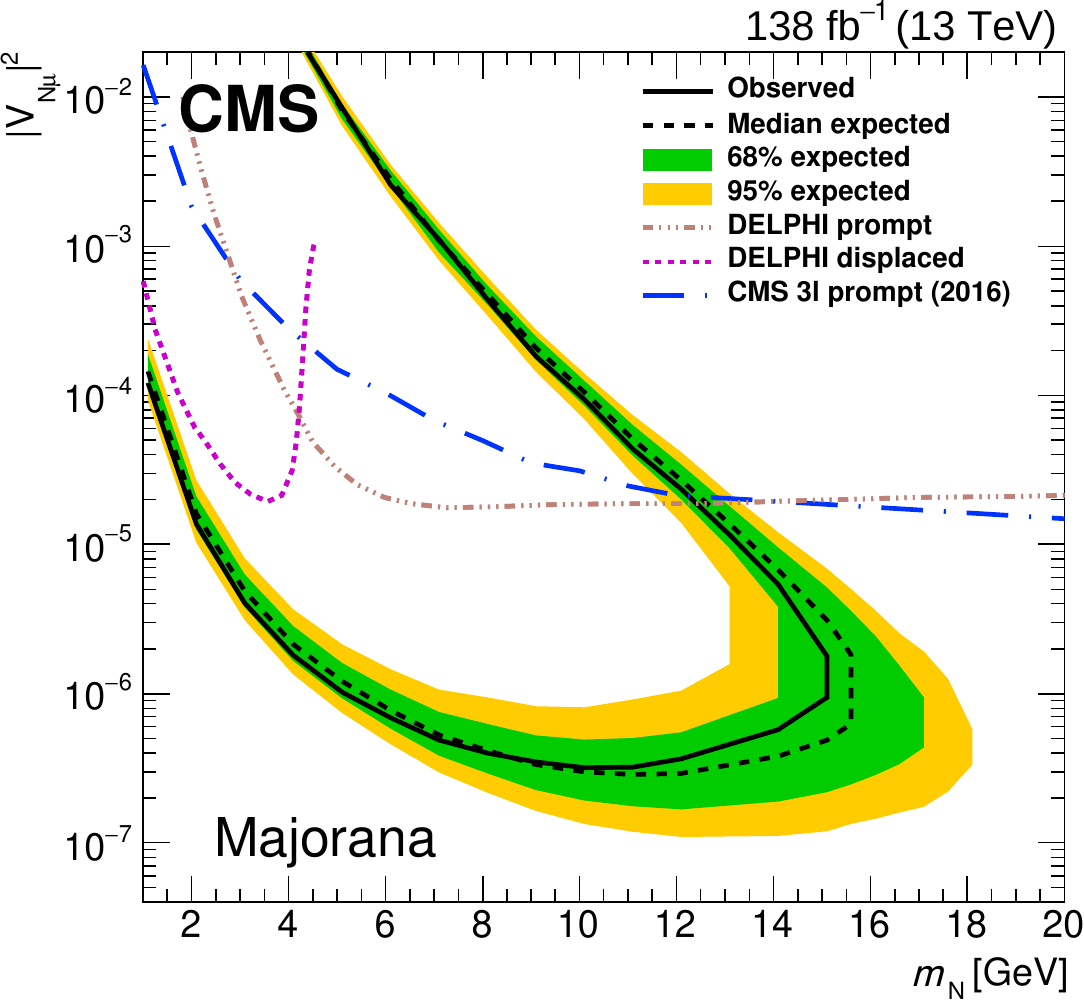} \\[1ex]
\includegraphics[width=0.48\textwidth]{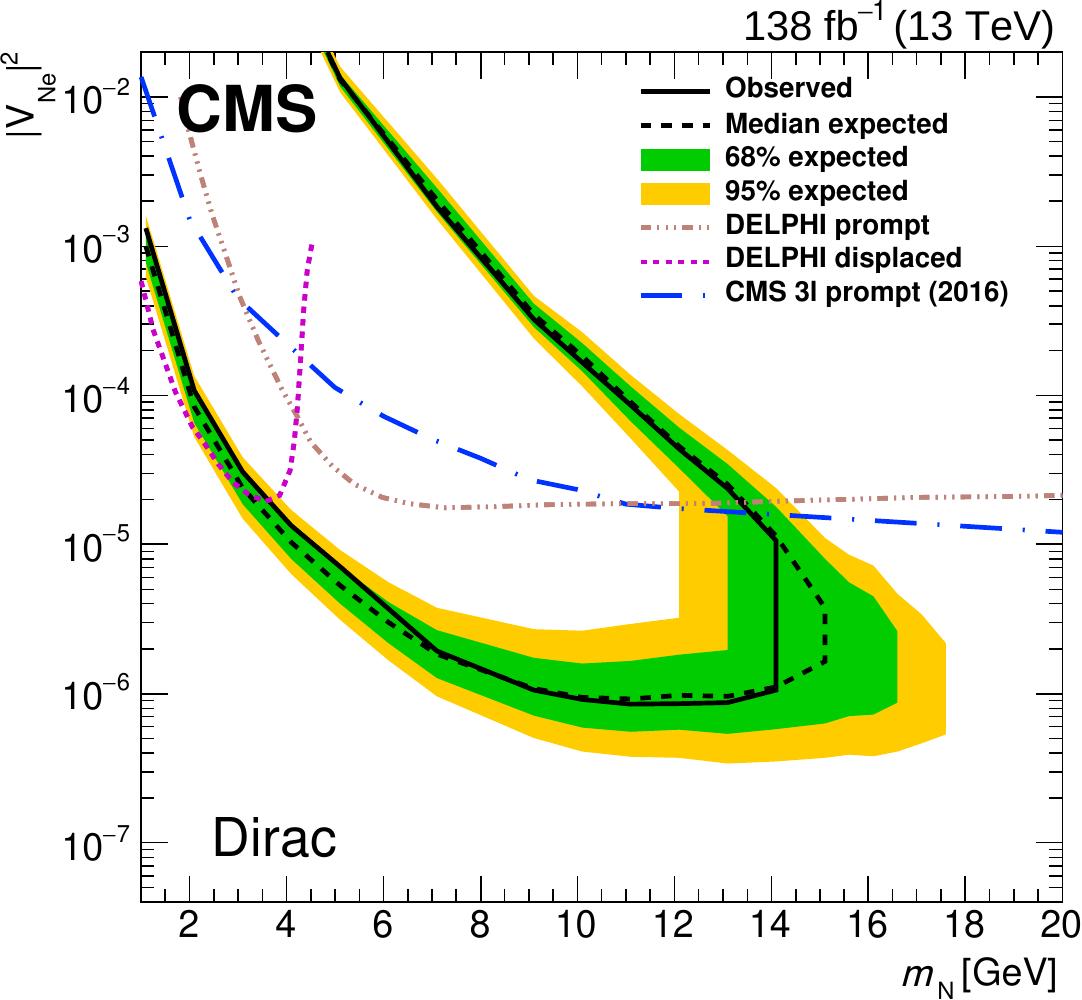}%
\hfill%
\includegraphics[width=0.48\textwidth]{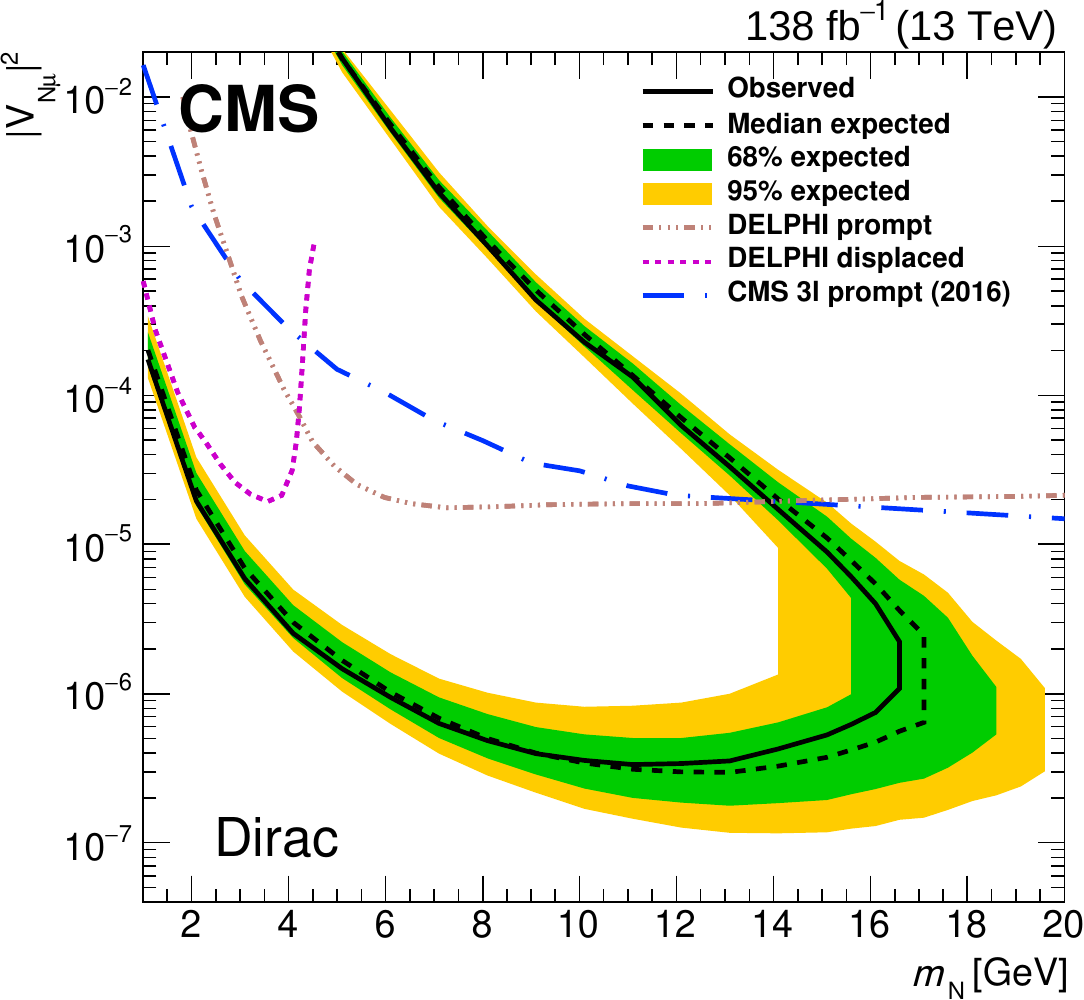}%
\caption{%
    The limits at 95\% \CL on \mixparsqeN (left) and \mixparsqmN (right) as functions of \mhnl for a Majorana (upper) or Dirac (lower) HNL.
    The area inside the solid (dashed) black curve indicates the observed (expected) exclusion region.
    Results from the DELPHI~\cite{DELPHI:1996qcc} and the CMS~\cite{CMS:2018iaf,CMS:2018jxx} Collaborations are shown as upper limits, \ie, the area above the curves indicates the respective observed exclusion region.
    Figures adapted from Ref.~\cite{CMS:2022fut}.
}
\label{fig:exo-20-009_limits}
\end{figure}

The two main observables used to categorize the selected events and
discriminate the hypothetical HNL signal from the background are
\DeltaTwoD and \mlprlprpr.
The expected and observed yields are shown in
Fig.~\ref{fig:exo-20-009_shapes}, split in four \DeltaTwoD
bins for $\mlprlprpr<4\GeV$ (where the DB
leptons dominate), and in two \DeltaTwoD bins for
$\mlprlprpr>4\GeV$ (dominated by SB leptons).
The expected HNL signal events for a selection of representative signal
scenarios are also shown.
Only scenarios in which the HNL mixes with a
single neutrino flavor are considered in this search, \ie, only one
of \mixparsqeN and \mixparsqmN is
nonzero. The \eeX channels (\eee, $\Pepm\Pemp\PGmpm$, and
$\Pepm\Pepm\PGmmp$), shown in Fig.~\ref{fig:exo-20-009_shapes} (upper), are sensitive to the \mixparsqeN mixing parameter,
while the \mmX channels (\mmm, $\PGmpm\PGmmp\Pepm$,
and $\PGmpm\PGmpm\Pemp$), shown in Fig.~\ref{fig:exo-20-009_shapes}
(lower), are sensitive to \mixparsqmN.
The number of observed events in data is in good agreement with the SM
background expectations within the statistical and systematic
uncertainties, and no significant excess is found for any final state
or SR bin.
Using these distributions, exclusion limits on
\mixparsqeN and \mixparsqmN are derived as a function of
\mhnl, separately for the cases of Majorana and Dirac HNLs.
The limits
are shown in
Fig.~\ref{fig:exo-20-009_limits}.

A search based on a novel detector signature, known as the muon detector shower (MDS), has been recently developed for detecting neutral long-lived particles (LLPs) with lifetimes in the range of 0.1 to 10\unit{m}~\cite{CMS:2021juv,EXO-22-017}. The considered process is shown in Fig.~\ref{fig:FD_HNLs} with an inclusive final state. This analysis is sensitive to HNLs with masses ranging from 1 to approximately 3.5\GeV. The full Run 2 \pp collision data set is analyzed, corresponding to an integrated luminosity of 137\fbinv collected at $\sqrt{s}=13\TeV$.

The MDS signature arises when LLPs decay within the muon system of the CMS detector, where the material in the iron return yoke structure induces a particle shower, creating a geometrically localized and isolated cluster of hits in the detectors. Because of the shielding in front of the muon system, MDSs are rarely produced by standard model background processes and can be a powerful signature to search for LLPs.
The analysis~\cite{EXO-22-017} utilizes the MDS signature to search for HNLs, which are reconstructed as an MDS.
Since the decay products of any hadronic decay modes of the HNL may be reconstructed as an MDS, the search is sensitive to HNL mixing to all three generations of leptons, including tau leptons.

Jets, with significant energy leakage beyond the calorimeter systems, and bremsstrahlung from muons, are the primary SM processes that may mimic MDS signatures. The MDS cluster selection is designed to reject such background events.
Those MDS clusters that are matched to jets or muons with sufficiently high \pt or specific detector patterns of hits and segments are vetoed.
The presence of the MDS signature along with the associated vetos suppresses SM background by a factor exceeding $10^7$.
In addition to the MDS object, the search selects events with one prompt, isolated lepton (electron or muon) passing the tight identification requirements.
Events are categorized based on the prompt lepton selection and the subdetector system in which the MDS cluster is reconstructed, namely the DTs (CSCs) in the barrel (endcap) region(s).

\begin{figure}[!ht]
\centering
\includegraphics[width=0.7\textwidth]{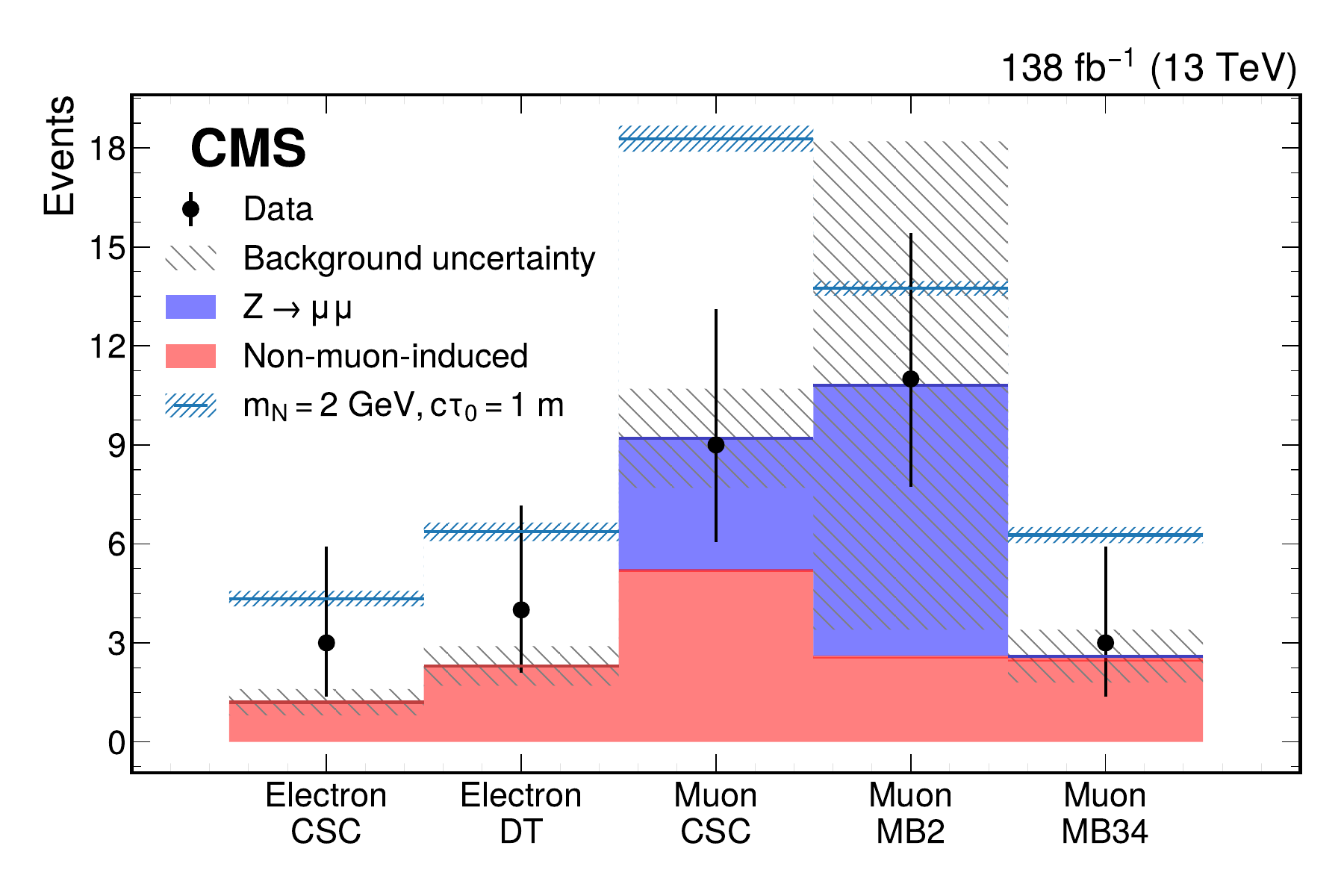}
\caption{%
    Expected and observed number of events in the SR of different event categories.
    Signal yields of a Majorana HNL with a mass of 2\GeV and with a proper decay length of 1\unit{m} are overlaid on top of the expected background estimated using the ABCD method.
    Figure taken from Ref.~\cite{EXO-22-017}.
}
\label{fig:Yields_HNL_22-017}
\end{figure}

The ABCD method is applied to estimate the background, using the number of hits in the MDS cluster and the azimuthal separation (\Dphi) between the MDS cluster and the prompt lepton as the two uncorrelated observables.
In the prompt-muon event categories, $\PZ\to\mumu$ events might mimic the back-to-back configuration between the MDS cluster and the prompt lepton, when one of the two muons from the \PZ boson decay undergoes bremsstrahlung in the muon detector and fails to be reconstructed as a muon object.
A dedicated leptonic \ttbar data sample is used to estimate the rate of such muons passing the SR selections.
Figure~\ref{fig:Yields_HNL_22-017} shows good agreement between the observed number of events in the SRs and the background predictions in the different event categories.

\begin{figure}[!ht]
\centering
\includegraphics[width=0.48\textwidth]{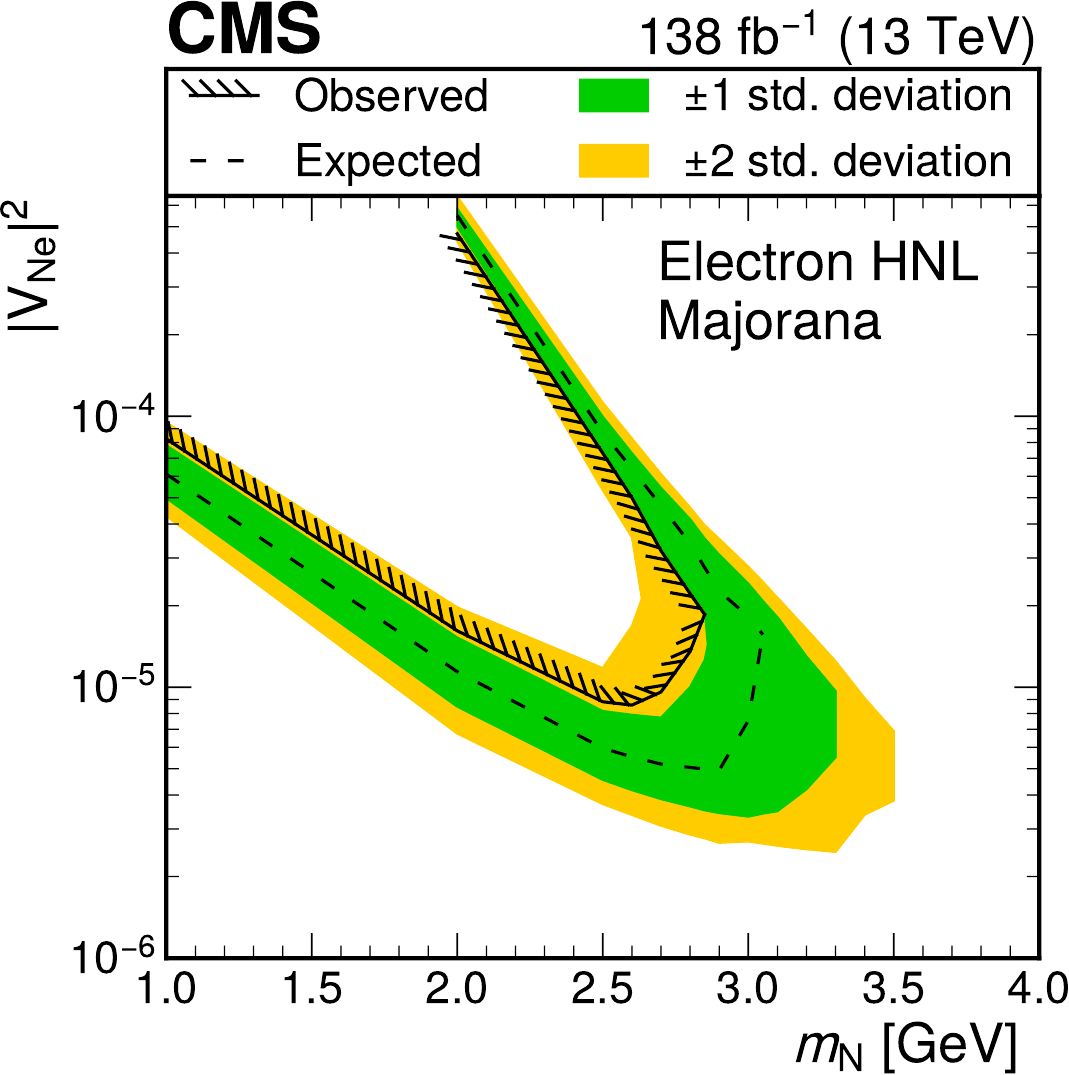}%
\hfill%
\includegraphics[width=0.48\textwidth]{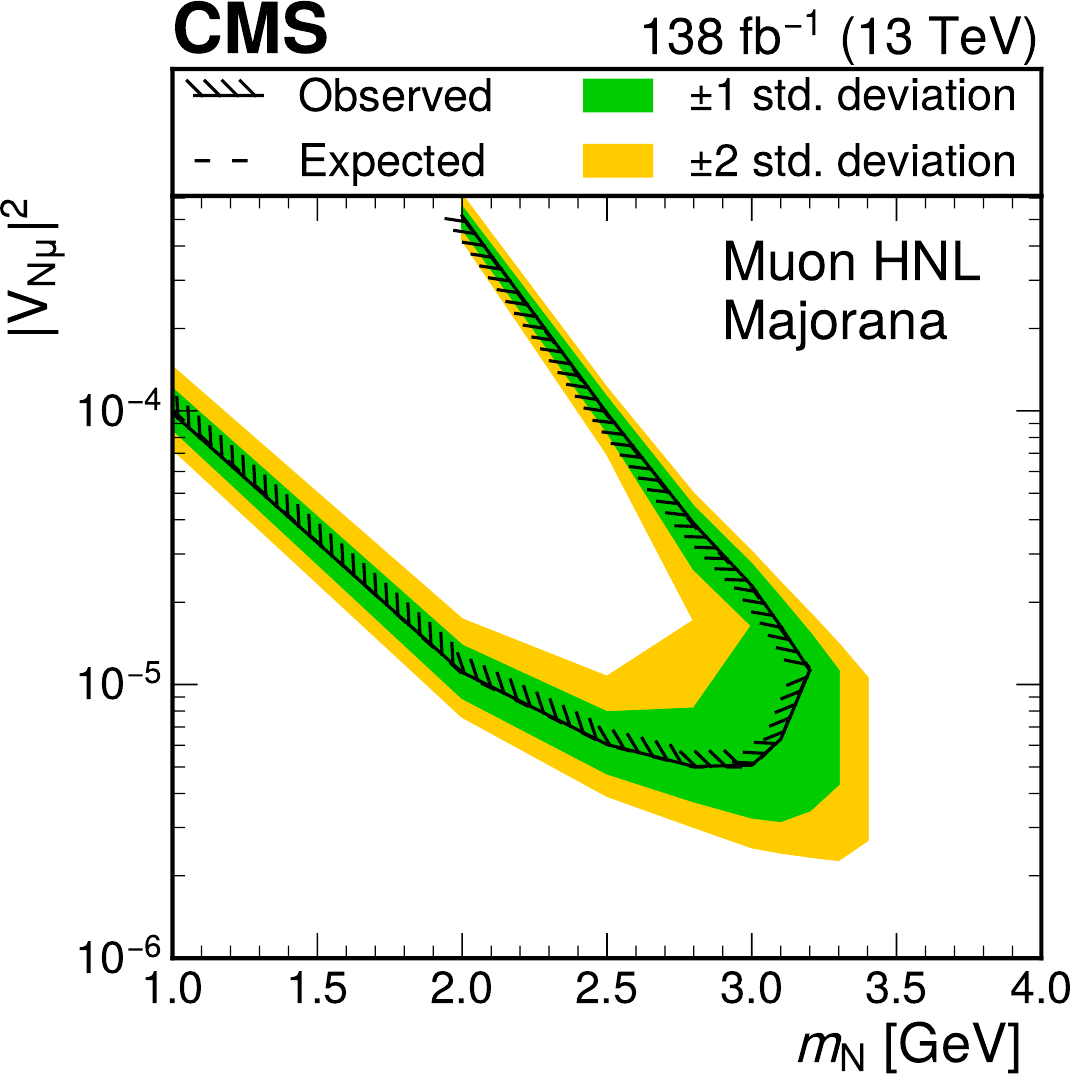} \\[1ex]
\includegraphics[width=0.48\textwidth]{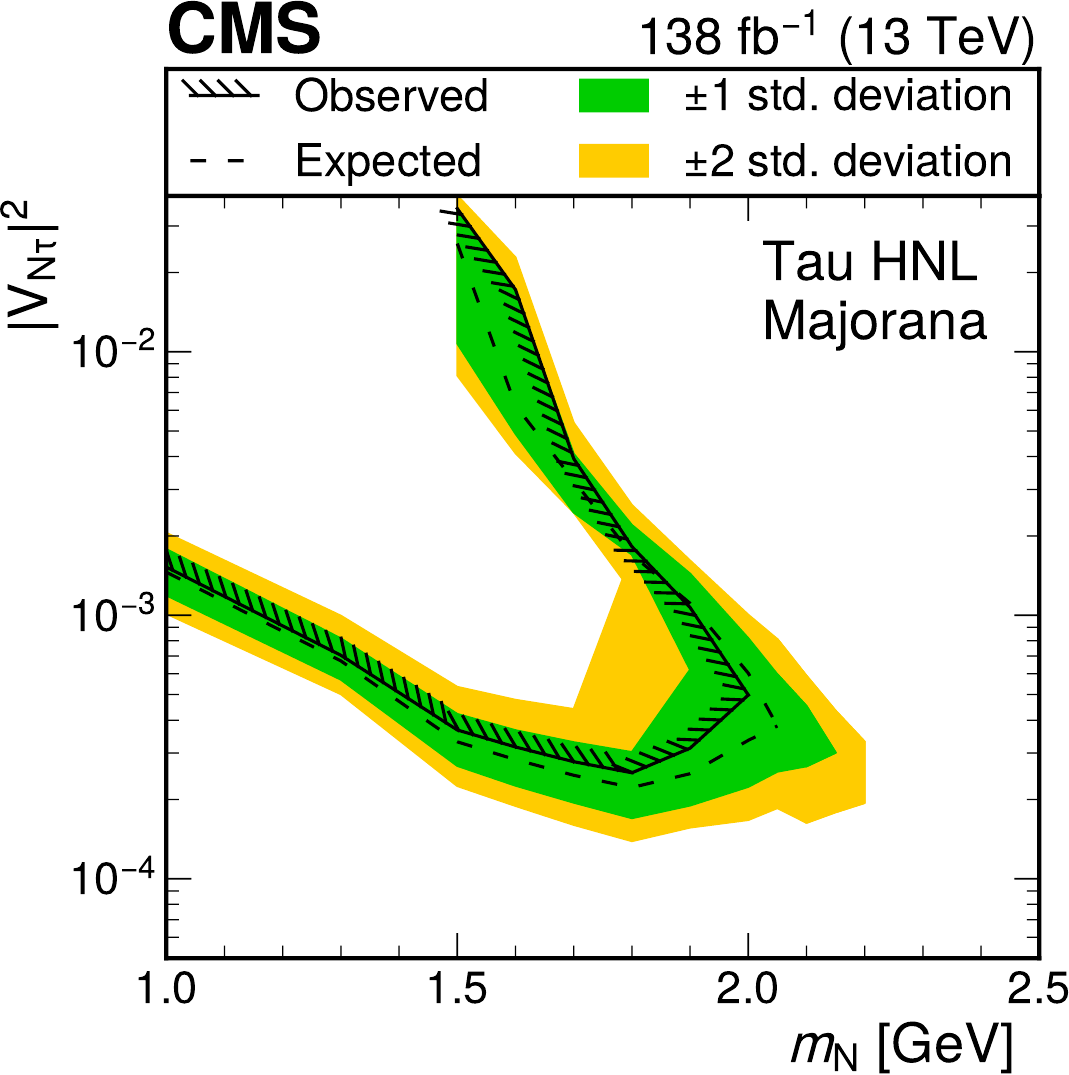}%
\caption{%
    Expected and observed upper limits at 95\% \CL on Majorana HNL production as functions of the HNL mass (\mhnl), and assuming mixing of the HNL with only one generation. The limits are shown for pure electron mixing (upper left), pure muon mixing (upper right), and pure tau neutrino mixing (lower). The limits in the tau neutrino mixing scenario are obtained by combining the results from the electron and muon decay channels of the tau lepton. Figures adapted from Ref.~\cite{EXO-22-017}.
}
\label{fig:limits_HNL_22-017}
\end{figure}

Upper limits on the HNL production cross section are determined as a function of the HNL mass for different mixing hypotheses.
Figure~\ref{fig:limits_HNL_22-017} shows the limits on the electron, muon, and tau neutrino mixing parameters as a function of the HNL mass, assuming mixing of the HNL with only one generation. 
This search extends the existing limits towards the parameter space with longer lifetimes or smaller \mixparsqlN.

Finally, we review a search for long-lived HNLs in the leptonic and semileptonic decays of \PB mesons~\cite{CMS-PAS-EXO-22-019}. The search probes HNLs with masses in the range $1<\mhnl<3\GeV$ and mean proper decay lengths in the range $10^{-2}<\ctauhnl<10^4\mm$, and targets HNLs decaying within the CMS tracker volume. A novel and key feature of the search is the use of a special \PQb hadron enriched data sample, referred to as the B-parking data sample~\cite{CMS-DP-2019-043,CMS-PAS-EXO-23-007, CMS:2024syx}, which corresponds to an integrated luminosity of 41.6\fbinv and contains of the order of $10^{10}$ \bbbar events.

For the first time at the CMS experiment, the decays of \PB mesons, with $\PB=(\PBu,\PBd,\PBs,\PBc)$, are considered as the source of signal. This production channel offers complementary sensitivity to that of the other searches for long-lived HNLs discussed above, which consider decays of \PW bosons as the source of the heavy neutrinos. Indeed, \PB mesons are significantly more abundant in \pp collisions than \PW bosons, and consequently, are potentially a more prominent source of signal. Moreover, because the \PB mesons have a lower mass than the \PW boson, the HNL states produced in \PB decays have lower momenta than those produced in \PW boson decays. For LLP signatures, the softer momentum spectrum is an advantage as it leads to a higher fraction of HNLs that decay within the CMS tracker volume.

The LO Feynman diagram for the process considered in this search is shown in Fig.~\ref{fig:feynMain}. The signature of the signal process is defined as $\PB\to\PellP\PN\PX$, $\PN\to\Pellpm\PGpmp$, where the leptons \PellP and \Pell may have same or opposite sign and may be either a muon or an electron, provided that one of them is a muon passing a \PB parking trigger. The hadronic recoil system, \PX, is treated inclusively and is not reconstructed. Because the HNL is long lived, the decay products \Pellpm and \PGpmp originate from a vertex displaced with respect to the \PB decay vertex and reconstructed using a kinematic vertex fit~\cite{Prokofiev:2005zz}. The neutral system $\Pellpm\PGpmp$ may be reconstructed to obtain the invariant mass \mellpmpimp. Thus, the strategy consists in searching for a peak consistent with the expected signal shape in the \mellpmpimp distribution. An advantage of this method is that a large number of mass hypotheses can be tested, in steps of the approximate detector resolution ($\sim$10\MeV). This provides the possibility to directly probe the mass of the HNL if a signal is present.

Events are classified according to 24 exclusive SRs based on (i) the significance of the transverse decay length of the $\Pellpm\PGpmp$ system; (ii) the relative sign of the charged leptons; (iii) the invariant mass of the $\PellP\Pellpm\PGpmp$ system; and (iv) the two lepton flavor channels, allowing for exactly zero or one electron in the final state, respectively. Backgrounds, which arise from strong-interaction processes, are suppressed using a parametric neural network~\cite{Baldi:2016fzo} that assesses a broad range of event properties. A search for HNL states is performed using simultaneous maximum likelihood fits to the \mellpmpimp distributions in the 24 categories. No significant excess of events over the SM background is observed in any of the fit regions.

\begin{figure}[p!]
\centering
\includegraphics[width=0.44\textwidth]{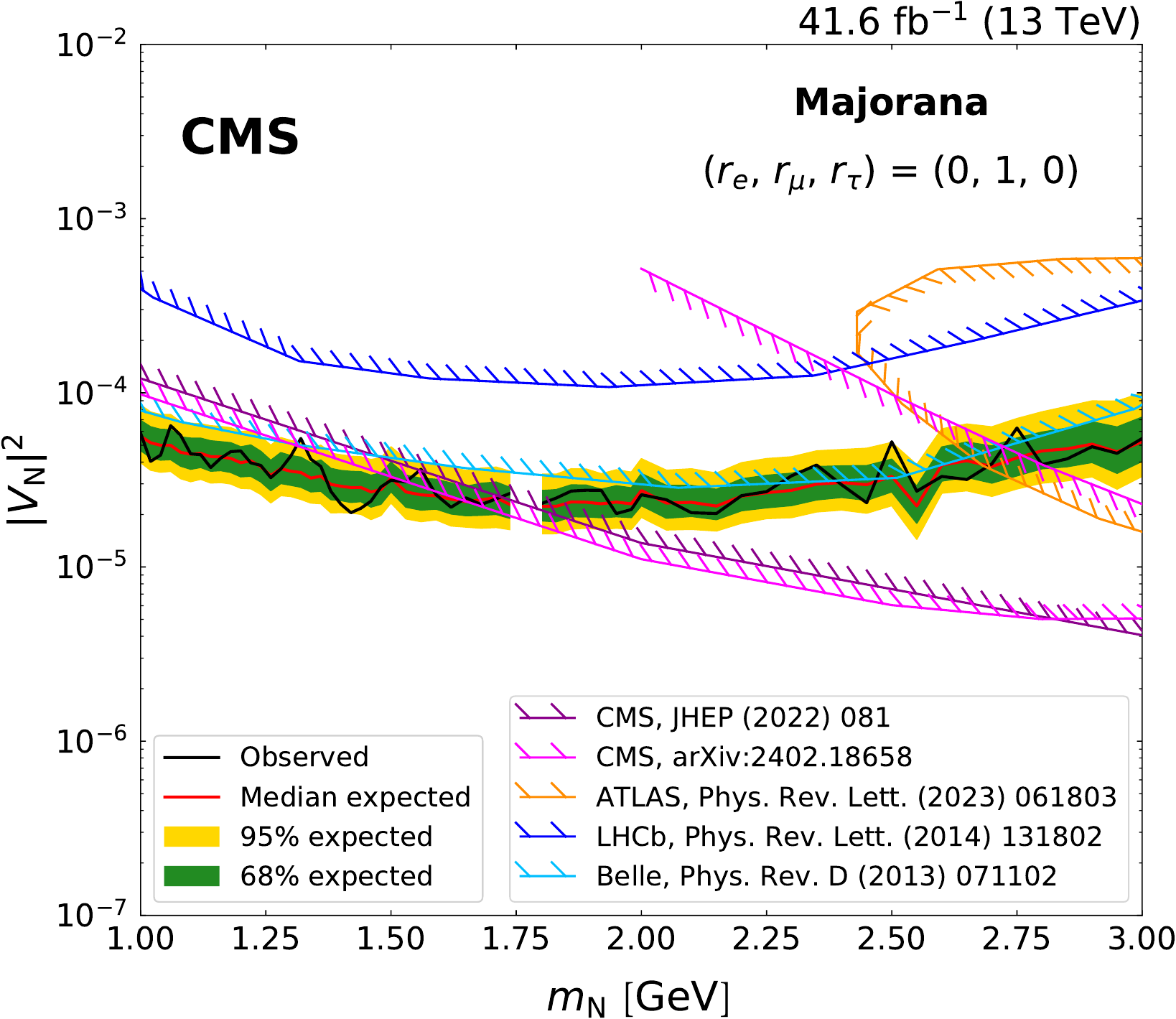}%
\hspace*{0.05\textwidth}%
\includegraphics[width=0.44\textwidth]{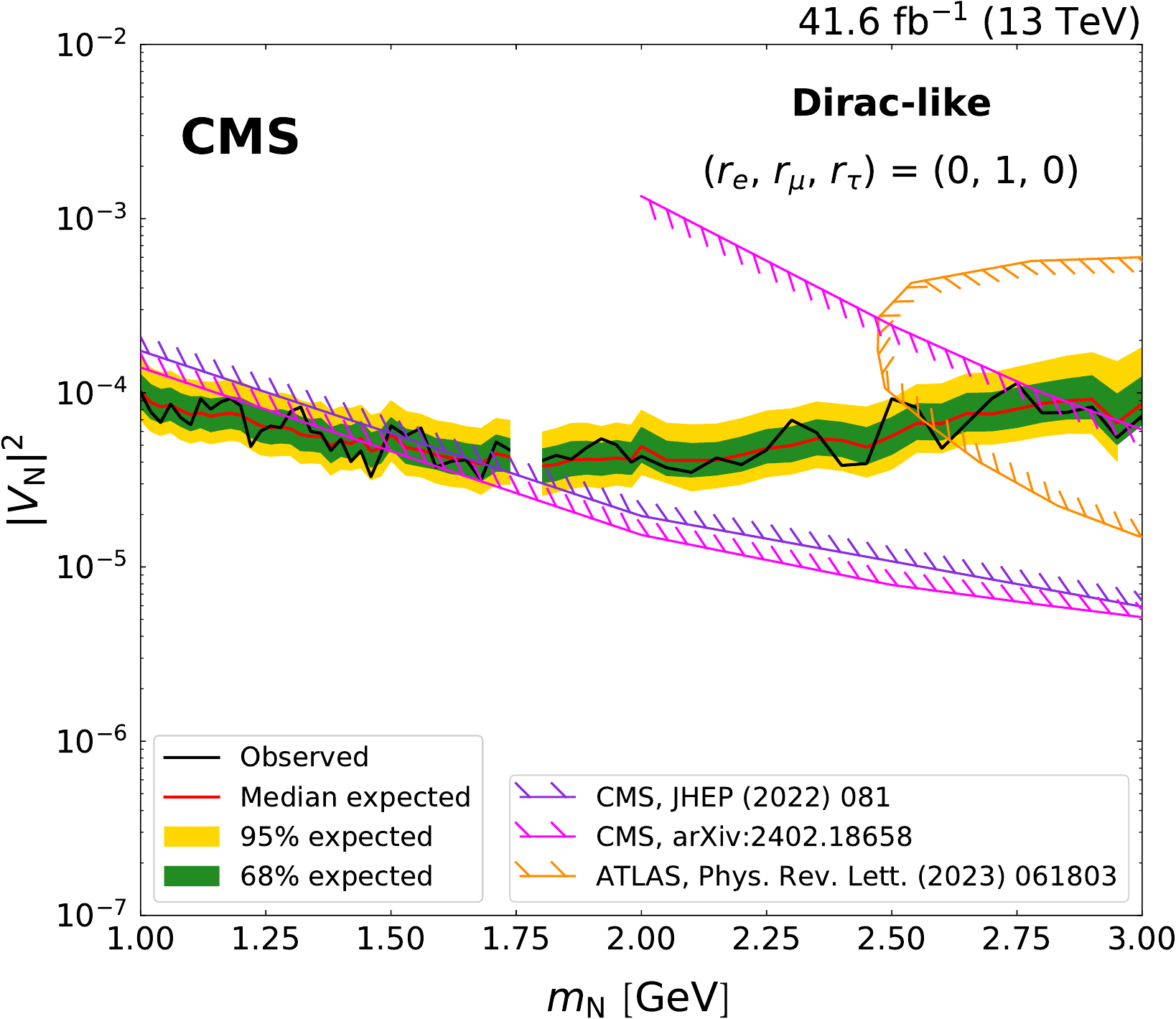} \\[1ex]
\includegraphics[width=0.44\textwidth]{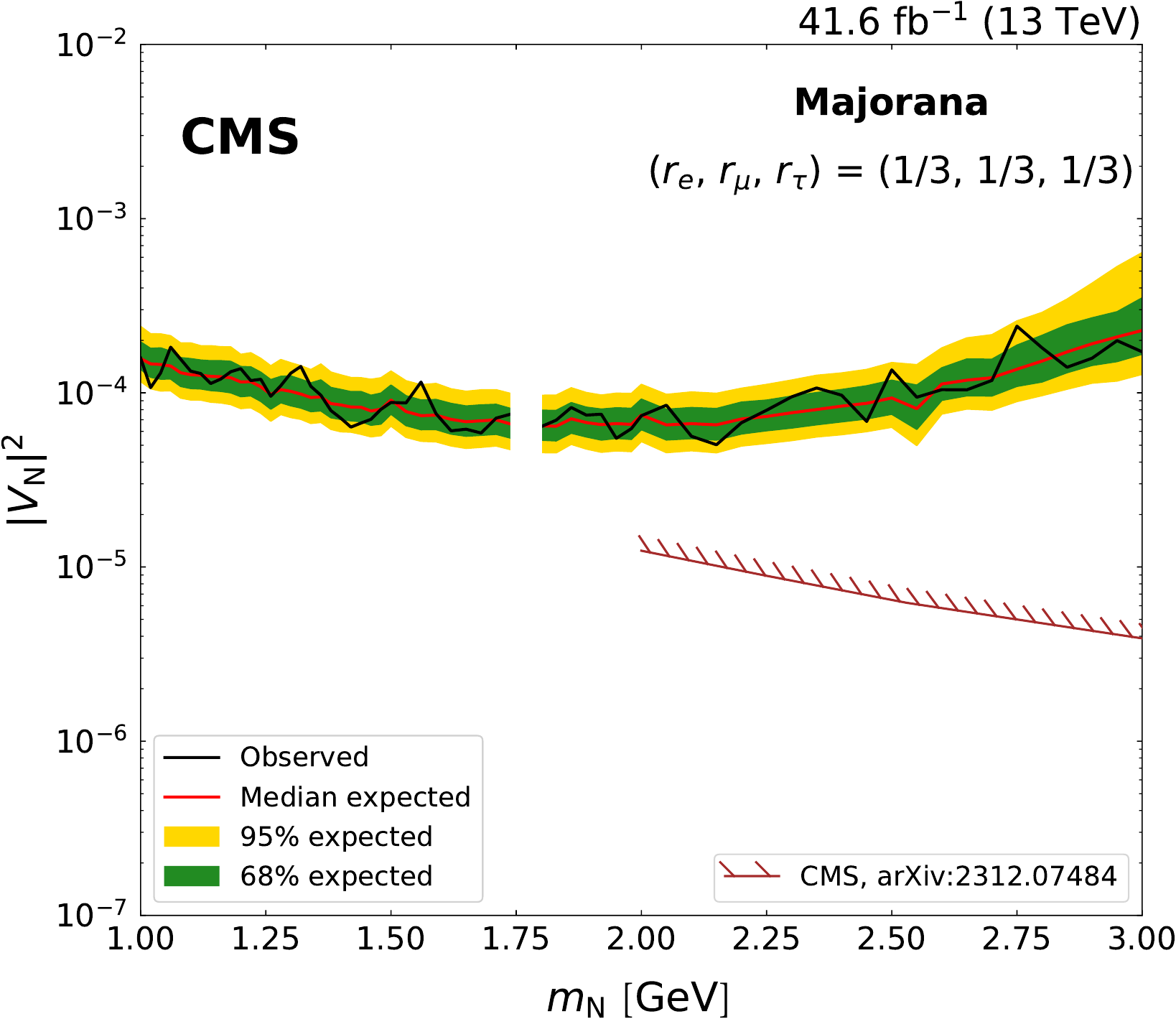}%
\hspace*{0.05\textwidth}%
\includegraphics[width=0.44\textwidth]{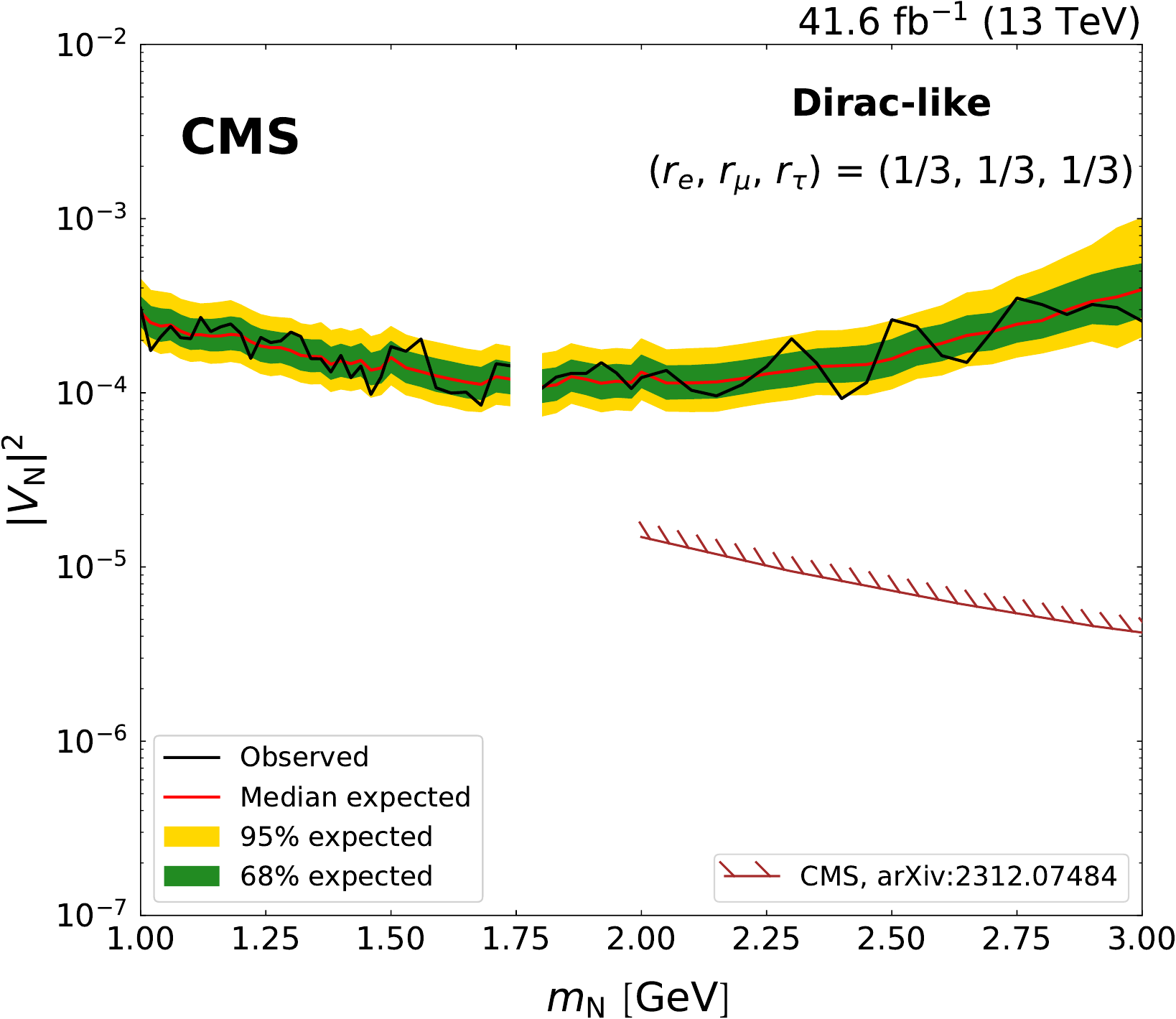}%
\caption{%
    Expected and observed limits at 95\% \CL on \mixparsqN as functions of \mhnl, in the Majorana (left column) and Dirac (right column) scenarios.
    The limits are shown for the mixing scenarios $(\rehnl,\ruhnl,\rthnl)=(0,1,0)$ (upper row) and $(\rehnl,\ruhnl,\rthnl)=(1/3,1/3,1/3)$ (lower row).
    Results from the CMS~\cite{CMS:2022fut,EXO-22-017,CMS:2023jqi}, ATLAS~\cite{ATLAS2022}, LHCb~\cite{LHCb:2014osd}, and Belle~\cite{Belle:2013ytx} Collaborations are superimposed for comparison.
    The mass range with no results shown corresponds to a vetoed region around the \PDz mass.
    Figures taken from Ref.~\cite{CMS-PAS-EXO-22-019}.
}
\label{fig:EXO-22-019}
\end{figure}

\begin{figure}[htp!]
\centering
\includegraphics[width=0.44\textwidth]{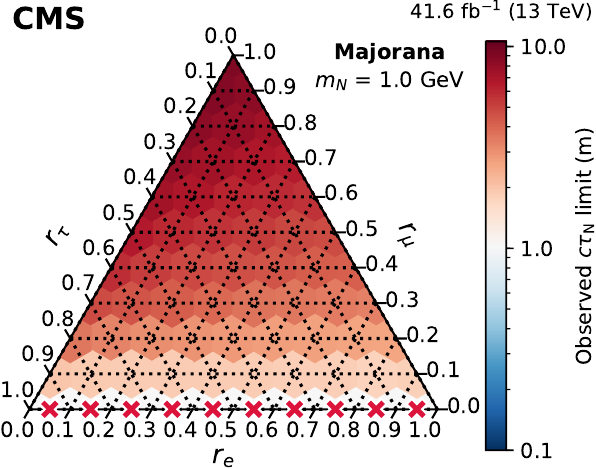}%
\hspace*{0.05\textwidth}%
\includegraphics[width=0.44\textwidth]{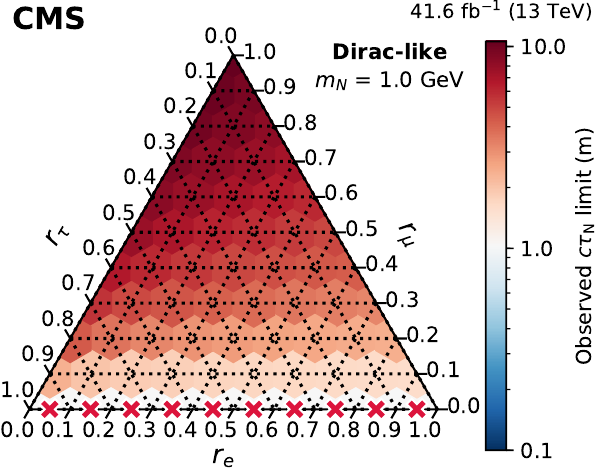}%
\caption{%
    Observed limits at 95\% \CL on \ctauhnl as a function of the mixing ratios $(\rehnl,\ruhnl,\rthnl)$ for $\mhnl=1\GeV$ in the Majorana (left) and Dirac (right) scenarios.
    The red crosses indicate that there is no exclusion found for that point.
    The orientation of the value markers on each axis identifies the associated internal lines on the plot.
    Figures taken from Ref.~\cite{CMS-PAS-EXO-22-019}.
}
\label{fig:EXO-22-019second}
\end{figure}

The results are interpreted as upper limits at 95\% \CL on $\mixparsqN=\sum_{\Pell=\Pe,\PGm,\PGt}\mixparsqlN$ as functions of \mhnl, shown for representative scenarios specified by different values of the mixing ratios $\rehnl\equiv\mixparsqeN/\mixparsqN$, $\ruhnl\equiv\mixparsqmN/\mixparsqN$, and $\rthnl\equiv\mixparsqtN/\mixparsqN$. In Fig.~\ref{fig:EXO-22-019}, limits are shown for the mixing scenarios $(\rehnl,\ruhnl,\rthnl)=(0,1,0)$ and $(\rehnl,\ruhnl,\rthnl)=(1/3,1/3,1/3)$, for the separate hypotheses of a Majorana or Dirac particle. The most stringent exclusion $\mixparsqN>2.0\times10^{-5}$ is obtained for a Majorana HNL of mass $\mhnl=1.95\GeV$ in the scenario $(\rehnl,\ruhnl,\rthnl)=(0,1,0)$. Furthermore, the most stringent exclusion limits to date are obtained on \mixparsqN for masses $1<\mhnl<1.7\GeV$ from a collider experiment. Finally, lower limits at 95\% \CL on \ctauhnl are also presented for 66 combinations of \rehnl, \ruhnl, and \rthnl. Figure~\ref{fig:EXO-22-019second} shows the limits for $\mhnl=1\GeV$ in both the Majorana and Dirac scenarios. The most stringent limit $\ctauhnl<10.5\unit{m}$ is obtained for a Dirac HNL in the scenario $(\rehnl,\ruhnl,\rthnl)=(0,1,0)$. It is the first time that lower limits on \ctauhnl in the form of ternary plots for masses $\mhnl<2.0\GeV$ are presented.

\subsubsection{Summary and complementarity of channels}

\begin{figure}[ht!]
\centering
\includegraphics[width=0.48\textwidth]{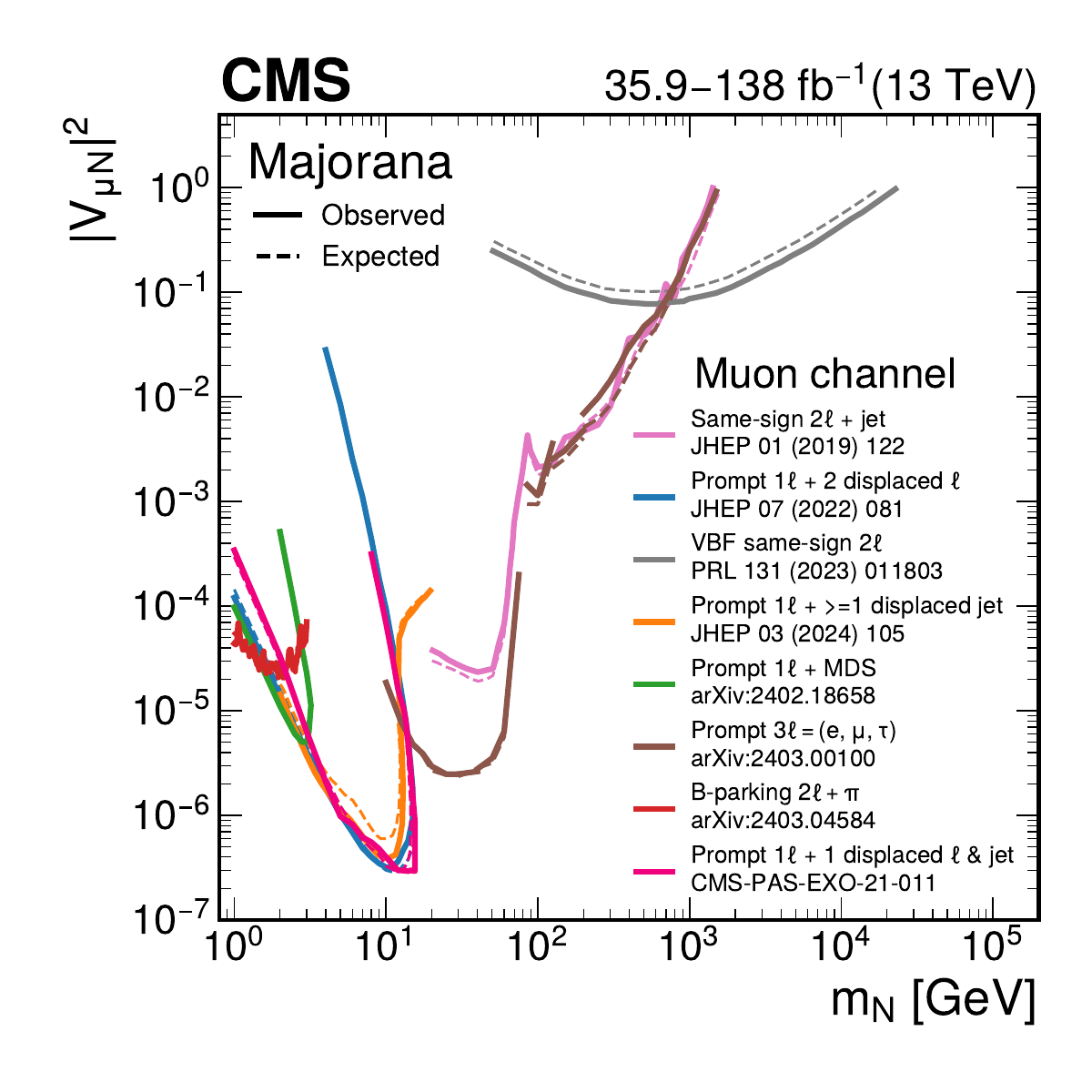}%
\hfill%
\includegraphics[width=0.48\textwidth]{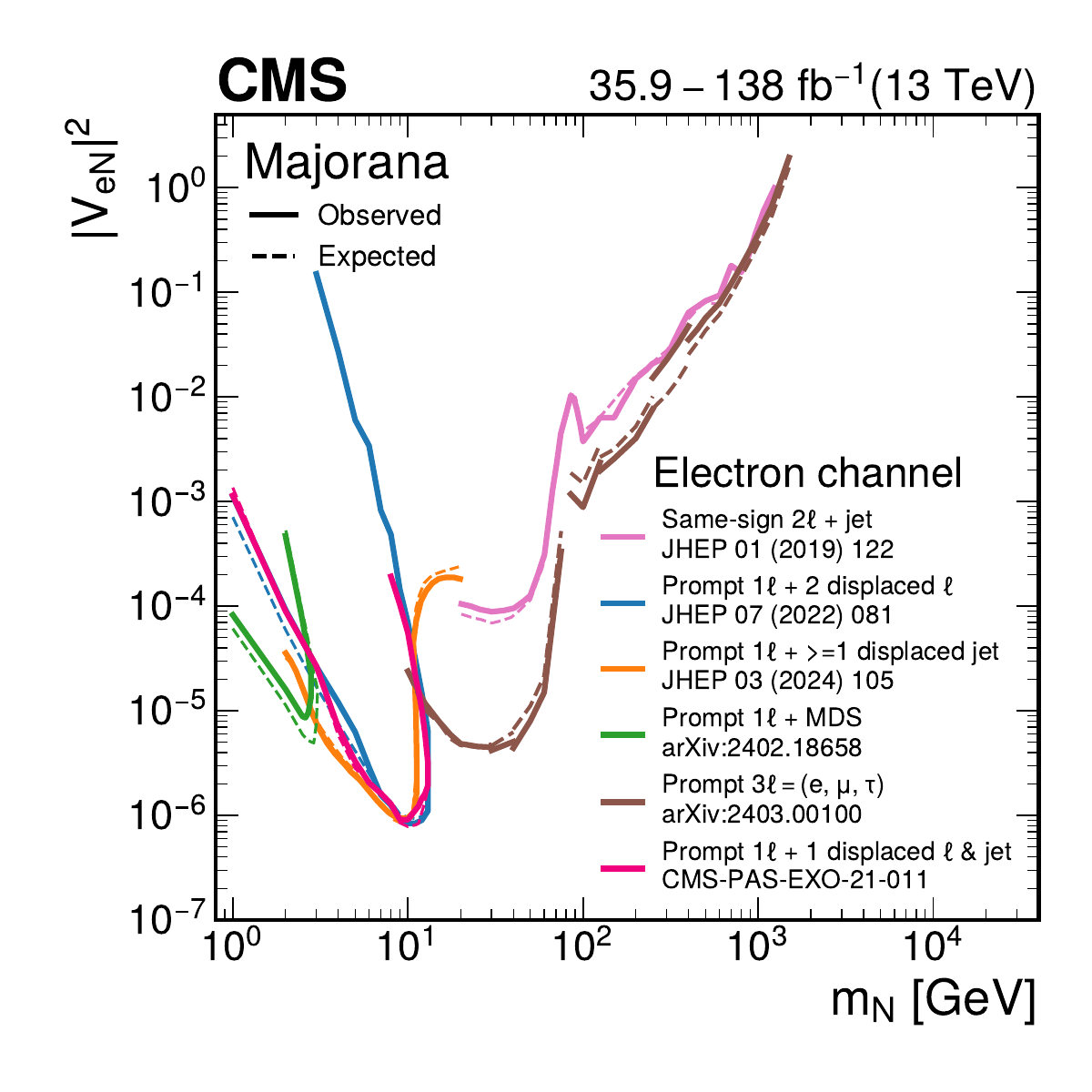} \\
\includegraphics[width=0.48\textwidth]{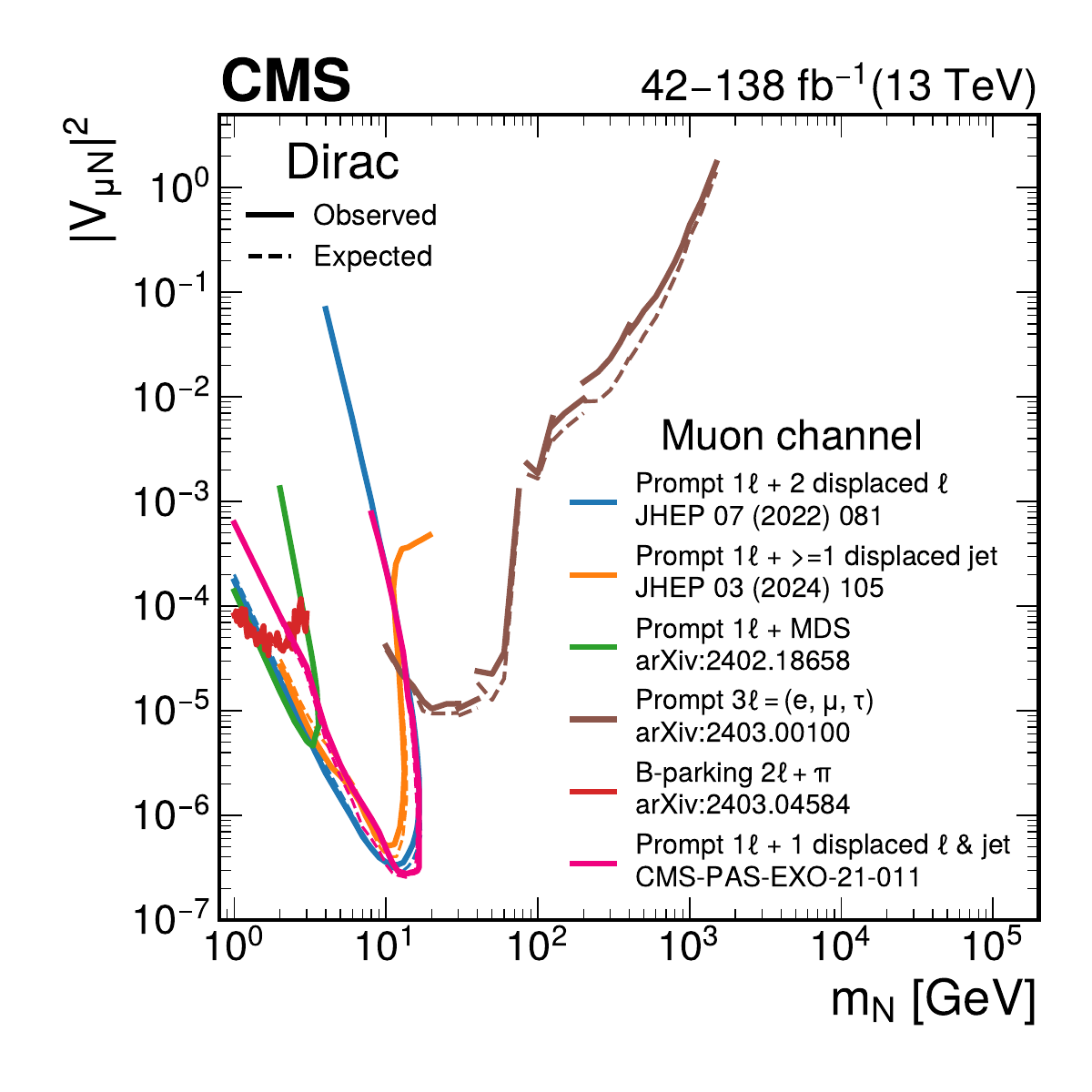}%
\hfill%
\includegraphics[width=0.48\textwidth]{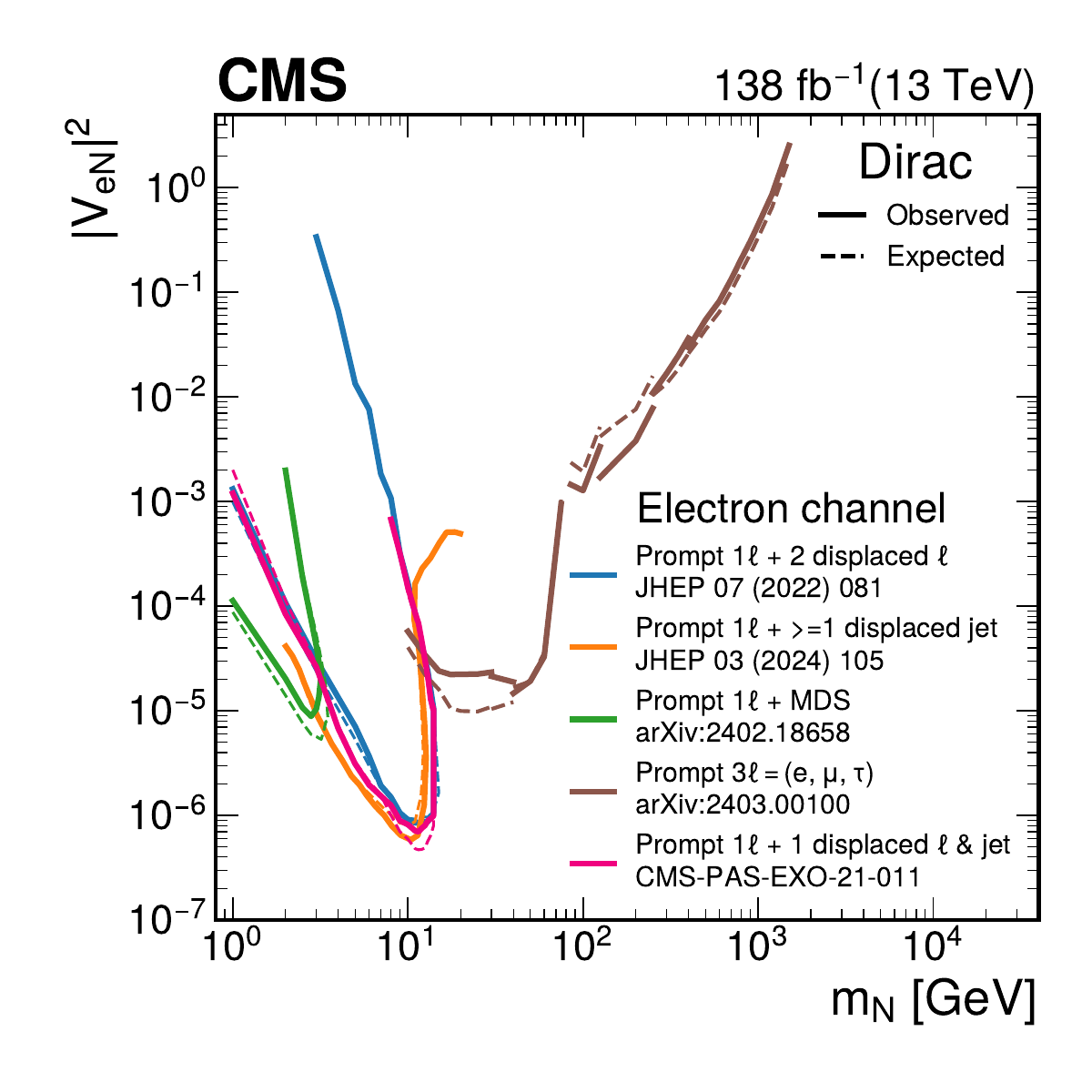}
\caption{%
    Summary of searches at the CMS experiment for long-lived HNLs in the \TypeOne seesaw model.
    The observed limits at 95\% \CL on the mixing parameter \mixparsqlN as a function of the HNL mass \mhnl are shown, for Majorana and Dirac HNLs (upper and lower row, respectively), and in the muon and electron channel (left and right column, respectively).
}
\label{fig:displacedHNLsummary}
\end{figure}

We conclude the \TypeOne seesaw section by summarizing and comparing these HNL searches. Figure~\ref{fig:displacedHNLsummary} presents a summary of the limits on the mixing parameter \mixparsqlN from prompt and long-lived \TypeOne seesaw HNL searches, covering a broad mass range from 1\GeV to 10\TeV for the pure muon and electron mixing scenarios for both Dirac and Majorana HNLs.

The exploration of long-lived HNLs covers masses \mhnl below 20\GeV. Techniques targeting displaced $\PN\to\Pell\Pell\PGn$ and $\PN\to\Pellpm\qqbarpr$ decays within the tracker volume such as displaced-vertex reconstruction and displaced jet tagging are used, leading to the most stringent limits in the mass range of 3 to 20\GeV.
Notably, for lower masses between 1 to 3\GeV, the search utilizing muon detector shower signatures results in the strongest bounds.
Between 1 to 2\GeV, the search using the \PB-parking data set provides the most stringent limits for muon-type HNLs, due to the large cross section of B meson production. These searches are complementary, with each dominating in a specific mass region. Therefore, a statistical combination does not provide a significant gain in sensitivity.

For \mhnl greater than 20\GeV, in scenarios where \PN exclusively mixes with muons, the search for prompt $\PN\to\Pell\Pell\PGn$ provides the most stringent limits in the \mhnl range from $\approx$20\GeV up to $\approx$100\GeV for both Dirac and Majorana HNLs. In the higher mass range, the search for prompt Majorana $\PN\to\Pellpm\qqbarpr$ exhibits comparable sensitivity despite relying solely on the 2016 data set. This suggests that expanding this search using the full Run 2 data set and probing Dirac HNLs could enhance the reach in this parameter space region. Lastly, the VBF search proves valuable in covering the very high mass region, where it attains the strongest constraints up to $\mhnl\approx10\TeV$.

\subsection{Search for HNLs in the \TypeThree seesaw model}
\label{sec:type-III}

The CMS Collaboration has carried out multiple direct searches for \TypeThree seesaw HNLs using the \pp collision data sets collected at $\sqrt{s}=7$ and 13\TeV.
In the first search at the CMS experiment for such heavy leptons, a $\sqrt{s}=7\TeV$ data set with an integrated luminosity of 4.9\fbinv was analyzed~\cite{CMS:2012ra}, probing the flavor-democratic scenario ($\BRe=\BRm=\BRt$), as well as electron- and muon-only scenarios, targeting final states with three electrons or muons.
Subsequent searches with the 2016 and full Run 2 data sets
expand the experimental final states to three and four electrons or muons~\cite{CMS:2017ybg,CMS:2019lwf}.
The latest $\sqrt{s}=13\TeV$ search by the CMS Collaboration on \TypeThree seesaw HNLs presented in Ref.~\cite{CMS:2022nty} improves on these previous results by utilizing three- and four-lepton signatures with electrons, muons, and hadronically decaying tau leptons, as well as improved analysis techniques, and sets the most stringent limits on such HNLs.

In particular, this most recent search with the CMS experiment on \TypeThree seesaw HNLs~\cite{CMS:2022nty} considers seven distinct multilepton final states based on the number of light charged leptons (electrons or muons) and hadronically decaying tau leptons. These orthogonal channels are defined as:
\begin{itemize}
\item $\geq$4 light leptons and any number of \tauh candidates (4\eormu),
\item exactly 3 light leptons and $\geq$1 \tauh candidates ($3\eormu1\tauh$),
\item exactly 3 light leptons and no \tauh candidates (3\eormu),
\item exactly 2 light leptons and $\geq$2 \tauh candidates ($2\eormu2\tauh$),
\item exactly 2 light leptons and exactly one \tauh candidates ($2\eormu1\tauh$),
\item exactly one light lepton and $\geq$3 \tauh candidates ($1\eormu3\tauh$), and
\item exactly one light lepton and exactly 2 \tauh candidates ($1\eormu2\tauh$).
\end{itemize}
In the 4\eormu channel, only the four light leptons leading in \pt are used in the subsequent analysis. Likewise, in the $3\eormu1\tauh$, $2\eormu2\tauh$, and $1\eormu3\tauh$ channels, only the leading one, two, and three \tauh are used, respectively.

The SM background processes, such as \WZ, \ZZ, \ttZ, and \ttW production in which three or more reconstructed charged leptons originate from decays of SM bosons, are the largest source of irreducible background in various channels of this search. A smaller background contribution arises from ISR or FSR photons that convert asymmetrically such that only one of the resultant electrons is reconstructed in the detector, or from the misidentification of on-shell photons as electrons. The dominant source of such backgrounds, collectively referred to as the conversion background, is DY events with an additional photon (\Zgamma). These backgrounds are estimated using simulation and normalized to observed data in the dedicated CRs. Another significant background component is the misidentified lepton background due to jets being misidentified as leptons, which is estimated using control samples in data via the matrix method (Section~\ref {sec:bkgest}).

Selected events in the seven channels are further categorized in a model-independent way depending on the dominant SM background processes, or in a model-dependent way, based on the output of BDTs trained to identify the \TypeThree seesaw signal against the SM backgrounds. The model-independent SRs are defined by splitting the channels into various regions based on the charge, flavor, invariant mass of lepton pairs, and kinematic properties of leptons, jets, and \ptmiss, as well as the multiplicity of \PQb-tagged jets. In each region, the \ST distribution is probed as the HNLs are expected to produce broad enhancements in the tails of \LT, \ptmiss, \HT, or dilepton mass observables.
This scheme gives 805 independent SR bins in each data-taking year in Run 2, a detailed breakdown of which may be found in Ref.~\cite{CMS:2022nty}.

In the model-dependent approach, separate BDTs are trained for the flavor-democratic scenario and for the $\BRt=1$ scenario. Since this analysis probes a wide mass range of \TypeThree seesaw signals, the training is performed in small signal mass windows (low, medium, and high). The training process considers a combination of up to 48 physics object- and event-level observables as input features. In the BDT evaluation, the BDT scores in the three-lepton (3\eormu, $2\eormu1\tauh$, $1\eormu2\tauh$) channels are combined into a single distribution. Similarly, the BDT scores in the four-lepton (4\eormu, $3\eormu1\tauh$, $2\eormu2\tauh$, $1\eormu3\tauh$) channels are combined into one distribution to further increase the signal sensitivity. Using the BDT score, a number of variable-width regions is defined for each of the combined three-lepton and four-lepton channels in each data-taking year. These define the BDT regions for all three signal mass windows in which an analysis is performed using only the number of observed events, \ie, not using shape information of the distributions of observables. A detailed breakdown of the BDT regions may be found in Ref.~\cite{CMS:2022nty}. The model-dependent SRs are typically more sensitive than their model-independent counterparts, except for the lowest signal masses. This is because at low masses, the BDT training process is degraded by the low signal yield.

\begin{figure}[ht!]
\centering
\includegraphics[width=0.48\textwidth]{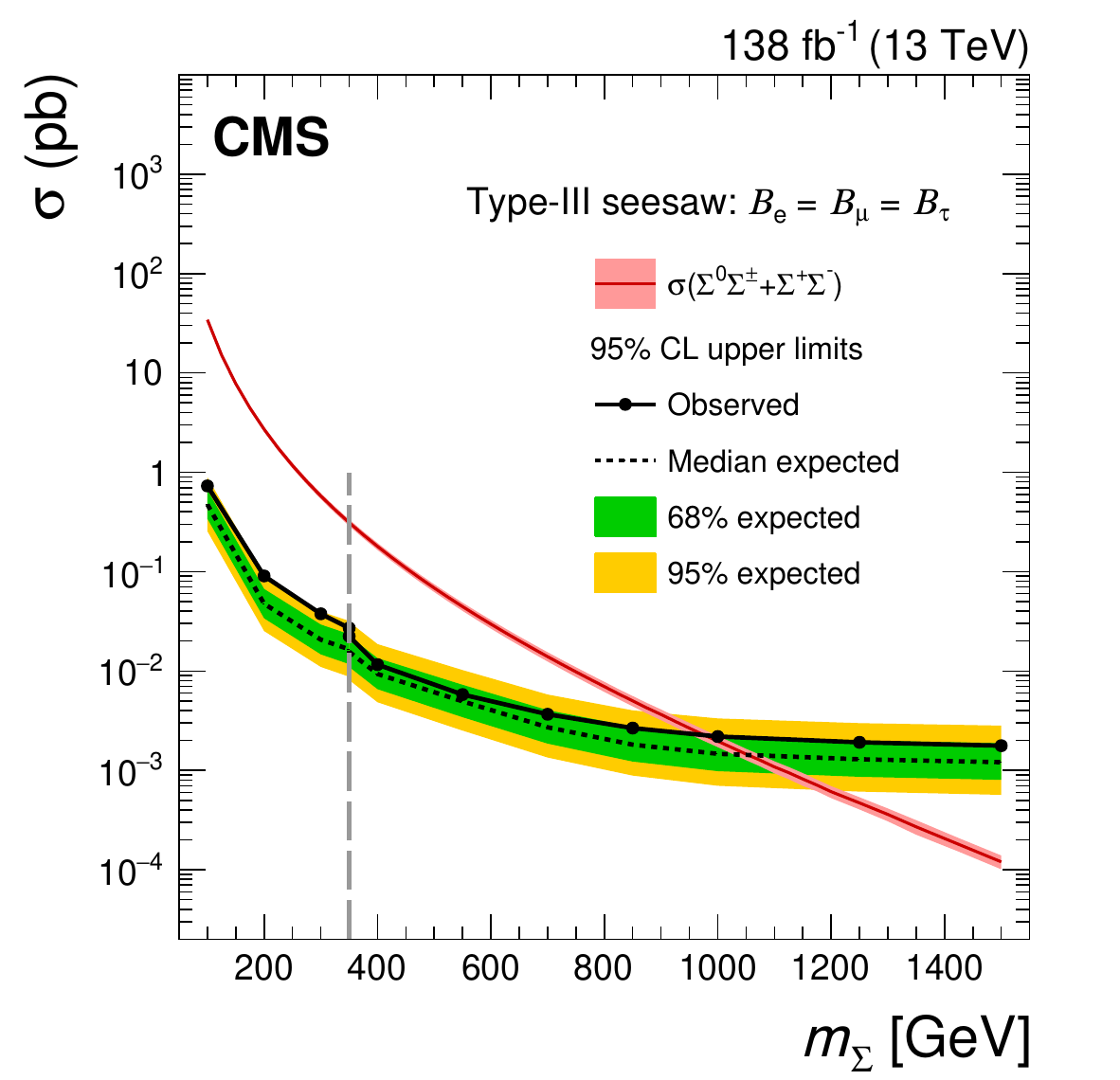}
\caption{%
    Observed and expected upper limits at 95\% \CL on the production cross section for \TypeThree seesaw HNLs in the flavor-democratic scenario using the model-independent schemes and the BDT regions.
    To the left of the vertical dashed gray line, the limits are shown from the model-independent SR, and to the right the limits are shown obtained using the BDT regions.
    Figure adapted from Ref.~\cite{CMS:2022nty}. Production cross sections for the signal model are calculated at NLO plus next-to-leading logarithmic precision, assuming that the heavy leptons are SU(2) triplet fermions~\cite{Fuks:2012qx,Fuks:2013vua}.
}
\label{fig:limitsSeesaw}
\end{figure}

\begin{figure}[ht!]
\centering
\includegraphics[width=0.48\textwidth]{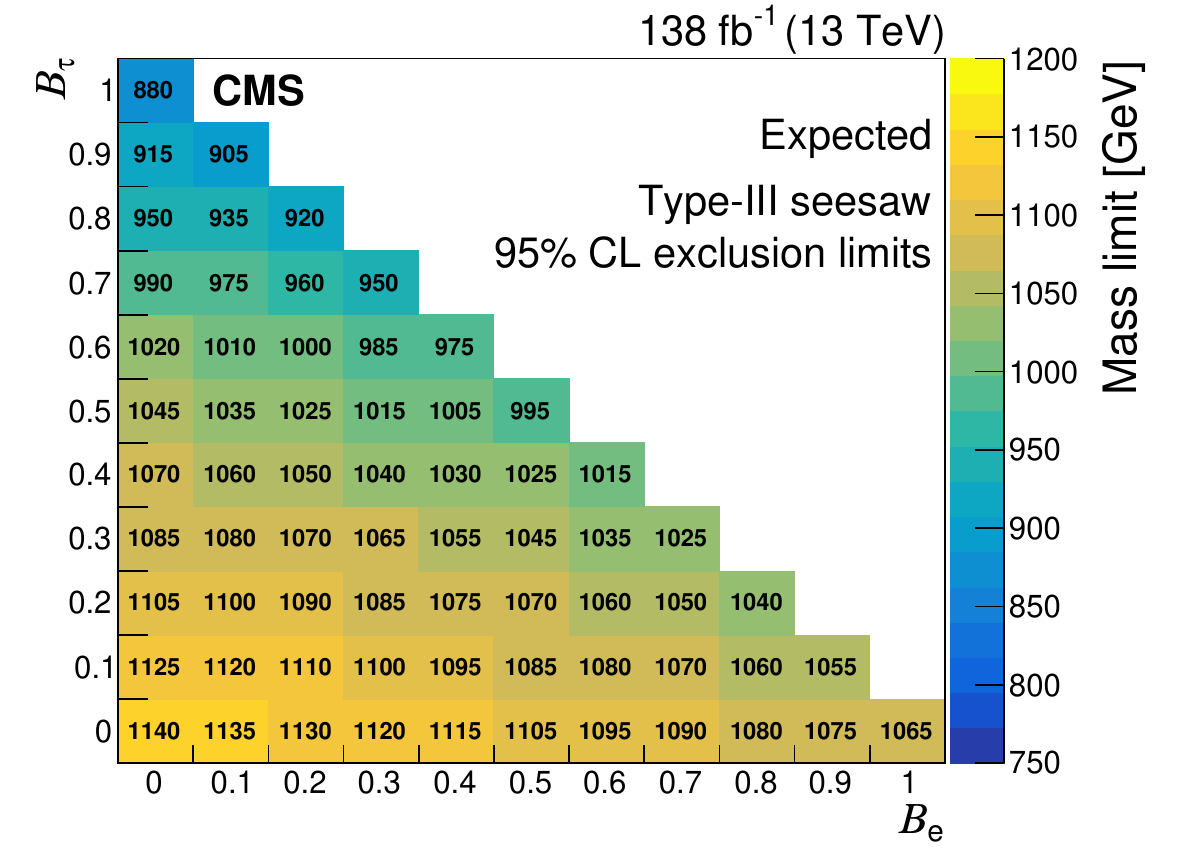}%
\hfill%
\includegraphics[width=0.48\textwidth]{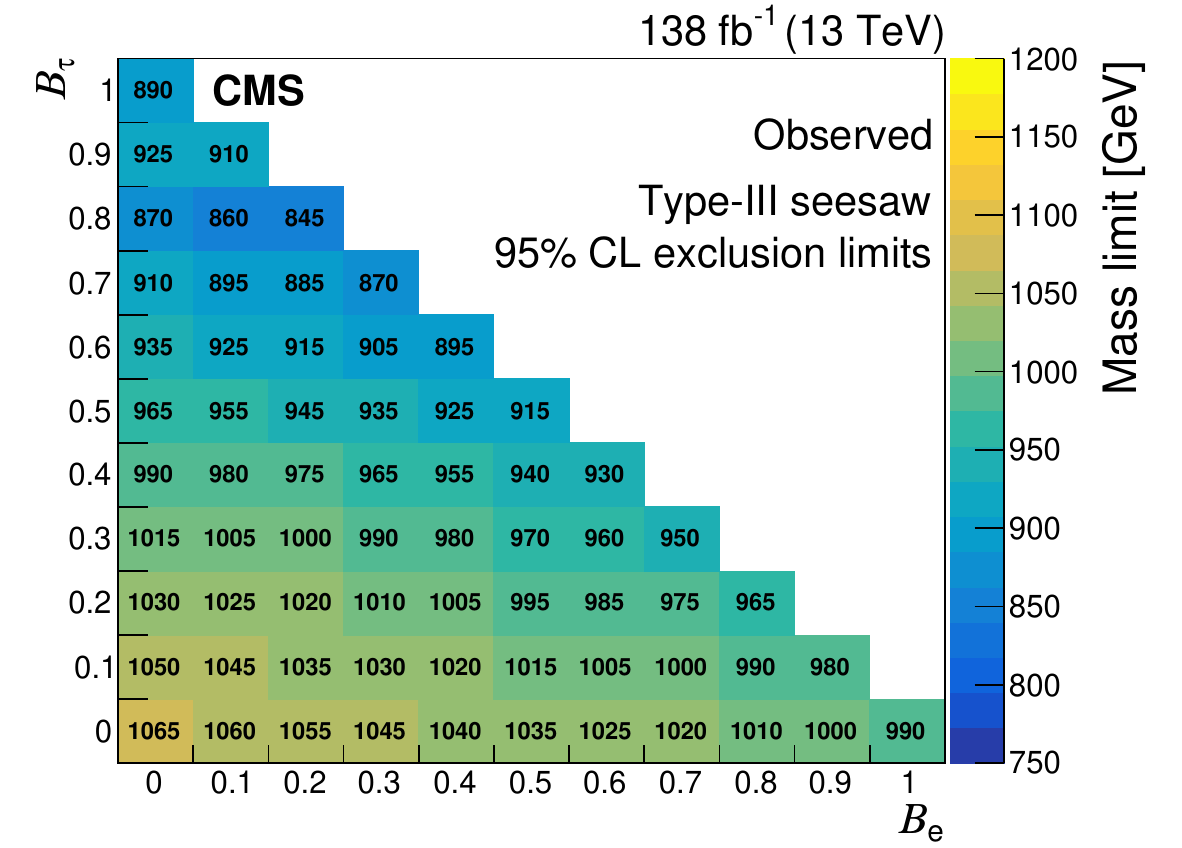}%
\caption{%
    Expected (left) and observed (right) lower limits at 95\% \CL on the mass of \TypeThree seesaw HNLs in the plane defined by \BRe and \BRt, with the constraint that $\BRe+\BRm+\BRt=1$.
    For $\BRt\geq0.9$, these limits are obtained using the high mass BDT trained assuming $\BRt=1$, and for the other decay branching fraction combinations, the limits use the $\BRe=\BRm=\BRt$ BDT.
    Figures adapted from Ref.~\cite{CMS:2022nty}.
}
\label{fig:limitsSeesawTri}
\end{figure}

Figure~\ref{fig:limitsSeesaw} shows the observed and expected upper limits at 95\% \CL on the cross section of \TypeThree seesaw HNL production in the
flavor-democratic scenario.
The observed (expected) lower limit on the mass \mGS of the heavy lepton \PGS in this scenario is 980 (1060)\GeV.
The most stringent expected limit for $\mGS<350\GeV$ is given by the model-independent scheme, and by the BDT regions for higher signal mass values.
The \PGS decay branching fractions to SM leptons of the different flavors are free parameters,
subject to the constraint that $\BRe+\BRm+\BRt=1$.
The observed and expected lower limits on \mGS in the plane defined by \BRe and \BRt are shown in
Fig.~\ref{fig:limitsSeesawTri}. For $\BRt\geq0.9$, these limits are obtained using the high mass BDT trained assuming $\BRt=1$, and for the other decay branching fraction combinations, the limits use the $\BRe=\BRm=\BRt$ BDT.
The strongest constraints are obtained assuming $\BRm=1$ ($\mGS>1070\GeV$), while the weakest are obtained assuming $\BRt=1$ ($\mGS>890\GeV$), which is due to the overall higher efficiency of reconstructing and identifying muons than \tauh in the experiment.

\subsection{Searches for HNLs in the left-right symmetric model}
\label{sec:LRSM}

In this section, we present the experimental outcomes from the CMS Collaboration in the searches for HNLs within the framework of the LRSM model. Searches considering HNLs produced through the decay of the right-handed \PW boson, denoted by \PWR, are discussed in Section~\ref{subsubsec:HNL_WR}, and searches for the right-handed \PZ boson, denoted by \PZpr, are discussed in Section~\ref{subsubsec:HNL_Zprime}. It is assumed that the gauge couplings associated with the left- and right-handed SU(2) groups are equal and the \PNGt decays are prompt.

\subsubsection{Searches for HNLs from right-handed \texorpdfstring{\PW}{W} bosons}
\label{subsubsec:HNL_WR}

A search has been performed for a heavy, RH gauge boson \PWR, which couples to RH fermions~\cite{CMS:2021dzb}, using the full Run 2 data set.
This search uses a final state consisting of two same-flavor leptons (\ee or \mumu) and two jets. The search covers two regions of phase space, one where the decay products of the HNL are collimated into a single large-radius jet (labeled ``\PQJ''), and one where the decay products are well separated. By including the regime where the decay products of the HNL are merged into a large-radius jet, this search probes areas of the phase space where the \PWR boson is heavy compared to the \PN (\ie, $\mWR/\mhnl\geq1$). The sensitivity in this regime is increased by identifying the HNL using the lepton subjet fraction ($\mathrm{LSF}_3$) algorithm~\cite{Brust:2014gia} to determine the consistency of the jet with three subjets, where one subjet is dominated by the four-momentum of the lepton. These events are referred to as ``boosted'' events, in contrast to ``resolved'' events where the two jets from the HNL decay are reconstructed separately.
The dominant SM processes that contribute to the backgrounds in this search are DY production of lepton pairs with additional jets in the final state, and leptonic decays of pair-produced top quarks.
These backgrounds are estimated from simulation and the modeling is corrected using dedicated CRs for both backgrounds.

\begin{figure}[!ht]
\centering
\includegraphics[width=0.48\textwidth]{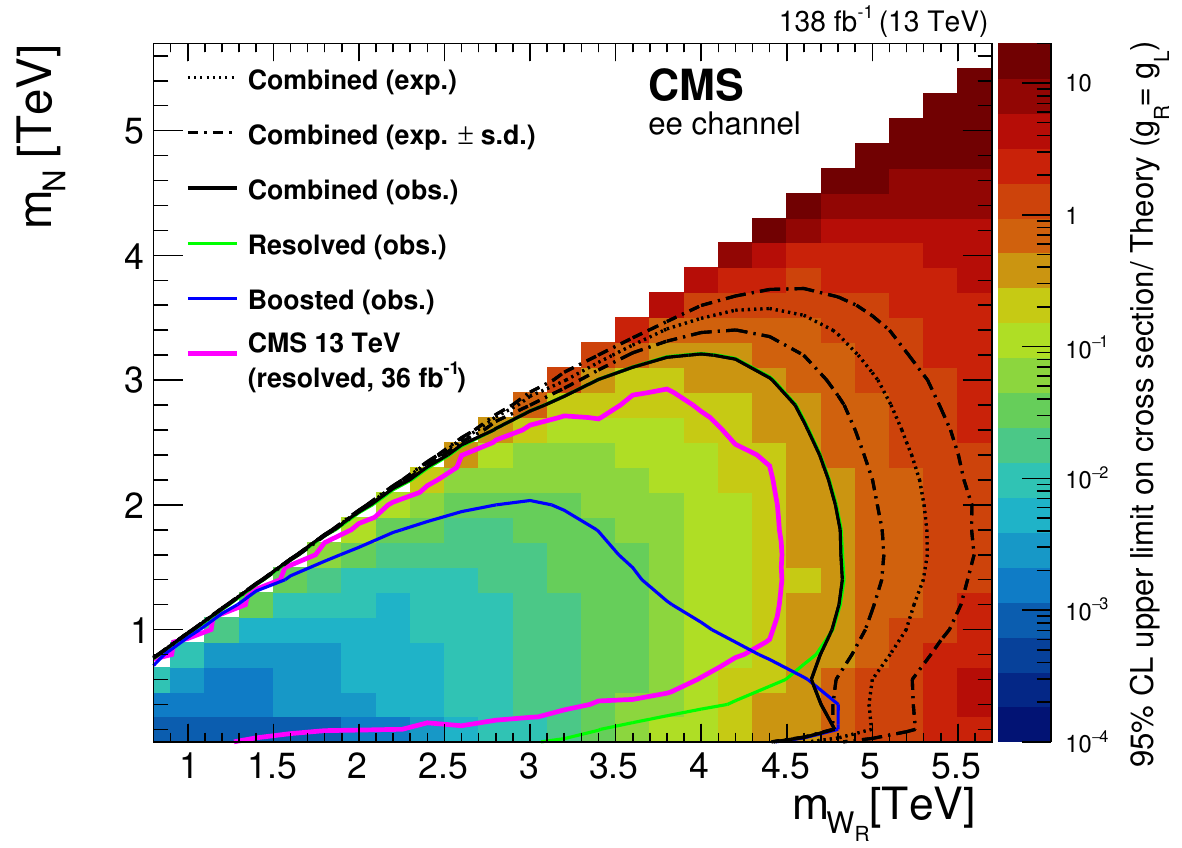}%
\hfill%
\includegraphics[width=0.48\textwidth]{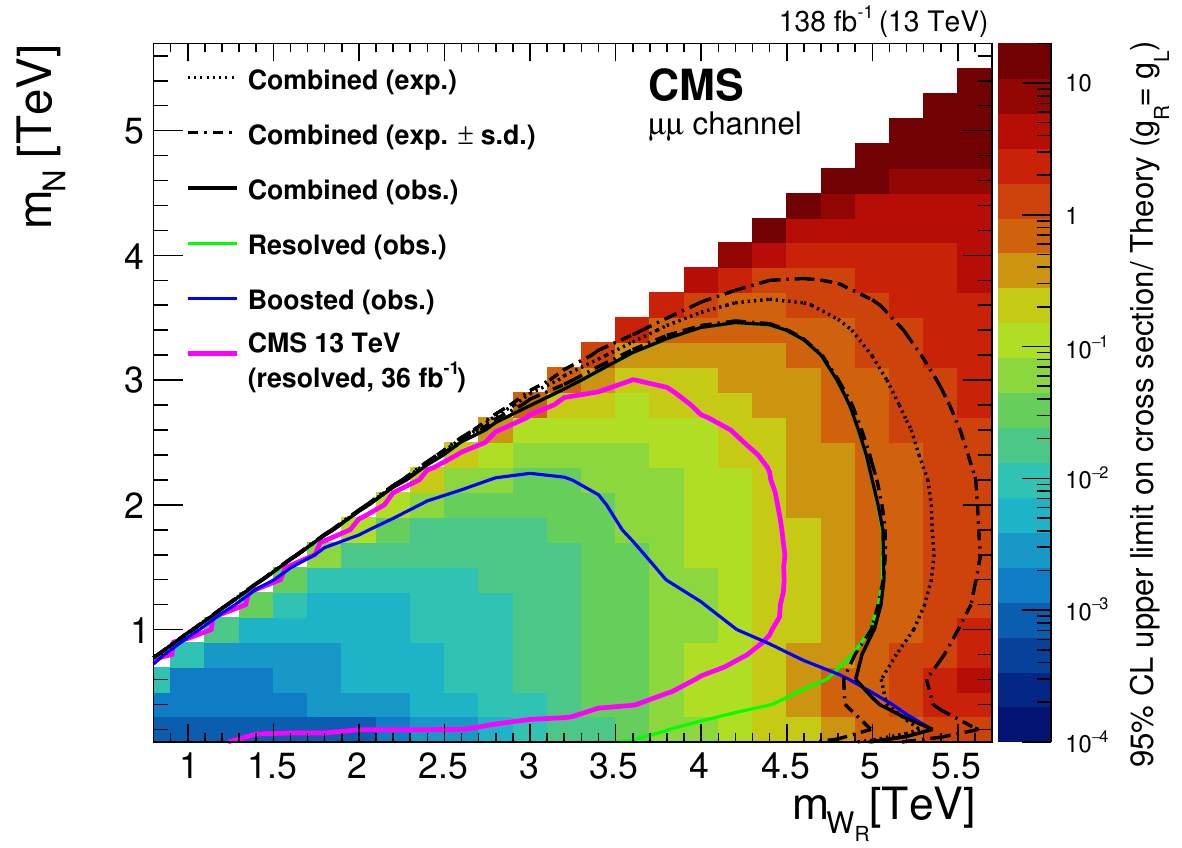}%
\caption{%
    The observed upper limits at 95\% \CL on the product of the production cross section and the branching fraction of a right-handed \PWR boson divided by the theory expectation for a coupling constant $g_{\text{R}}$ equal to the SM coupling of the \PWR boson ($g_{\text{L}}$), for the electron channel (left) and muon channel (right).
    The observed exclusion regions are shown for the resolved (solid green), boosted (solid blue), and combined (solid black) channels, together with the expected exclusion region for the combined result (dotted black).
    The dash-dotted lines represent the 68\% coverage of the boundaries of the expected exclusion regions.
    The observed exclusion regions obtained in the previous search performed by the CMS Collaboration~\cite{Sirunyan:2018pom} are bounded by the magenta lines.
    The biggest improvement may be seen in the $\mhnl<0.5\TeV$ region, where the new boosted category greatly improves the sensitivity with respect to the previous result.
    Figures adapted from Ref.~\cite{CMS:2021dzb}.
}
\label{fig:2DFullRun2Limits}
\end{figure}

A maximum likelihood fit is performed using the invariant mass distributions ($m_{\Pell\Pell\PQj\PQj}$ for the resolved region or $m_{\Pell\PQJ}$ for the boosted region), simultaneously in the SRs and the CRs.
Upper limits are derived on the product of the cross section for \PWR production and the branching fractions, $\sigma(\pp\to\PWR)\BR(\PWR\to\ee(\mumu)\qqbarpr)$, for various \mWR and \mhnl hypotheses.
The upper limits across the entire \mWR--\mhnl plane are shown in Fig.~\ref{fig:2DFullRun2Limits}. With $\mhnl=\mWR/2$, the observed (expected) lower limit at 95\% \CL on \mWR is 4.7 (5.2)\TeV and 5.0 (5.2)\TeV for the electron and muon channels, respectively. For $\mhnl=0.2\TeV$, the limits exclude the phase space up to $\mWR=4.8$ (5.0) and 5.4 (5.3)\TeV for the electron and muon channels, respectively. The local $p$-value of the signal strength, as a function of \mWR and \mhnl, is obtained from fits to the observed data with the signal strength at each point treated as a free parameter. The most extreme $p$-value is observed in the electron channel, at the $(\mWR,\mhnl)=(6.0,0.8)\TeV$ mass point, with a value $1.58\times10^{-3}$, corresponding to a local significance of 2.95 standard deviations. The look-elsewhere effect~\cite{Gross:2010qma} is taken into account by using pseudo-experiments to calculate the probability under the background-only hypothesis of observing a similar or larger excess in the electron channel across the full mass range considered in the analysis. This probability is $2.7\times10^{-3}$, corresponding to a global significance of 2.78 standard deviations.

The LRSM model is also investigated using events with final states with two \PGt leptons in two analyses that consider different decays of the \PGt pair: one leptonic and one hadronic~\cite{exo-16-023}, and both hadronic~\cite{CMS:2018iye}.
For both searches, models that predict the presence of leptoquarks are included in the comparison to the observed data, but are not explicitly reported here as they are not directly related to the scope of this section.

A final state with two \PGt leptons and two jets is studied in the context of the search for a \PWR boson, as described in Ref.~\cite{exo-16-023}. The search follows the decay chain $\PWR\to\PGt\PNGt$, where $\PNGt\to\PGt\PWR^\ast\to\PGt\qqbar$. One \PGt (denoted by $\PGt_{\Pell}$) decays into a light lepton (\Pe or \PGm) and a neutrino, leading to \ptmiss, while the other \PGt (denoted by \tauh) decays hadronically.
The search utilizes the 2015 data set and part of the 2016 data set,
corresponding to an integrated luminosity of 12.9\fbinv, which was the only available data set at the time of publication. The simulated \PWR signal samples cover a mass region ranging from 1000 to 4000\GeV in steps of 500\GeV.

Several SM processes mimic the signatures explored in this search.
The production of top quark pairs is the dominant background because of the presence of leptons, \ptmiss, and both light- and heavy-flavor jets.
Additionally, \PW or \PZ boson production in association with jets, diboson or single top quark production, and QCD multijet processes also contribute to the SM background.
The results of the analysis are obtained from a binned maximum likelihood fit to the \ST distributions in the $\Pe\tauh$ and $\PGm\tauh$ channels simultaneously.
The quantity \ST is defined as the scalar sum of the \ptmiss and the \pt of the electron or muon, the \tauh candidate, and the two jets.

\begin{figure}[ht!]
\centering
\includegraphics[width=0.50\textwidth]{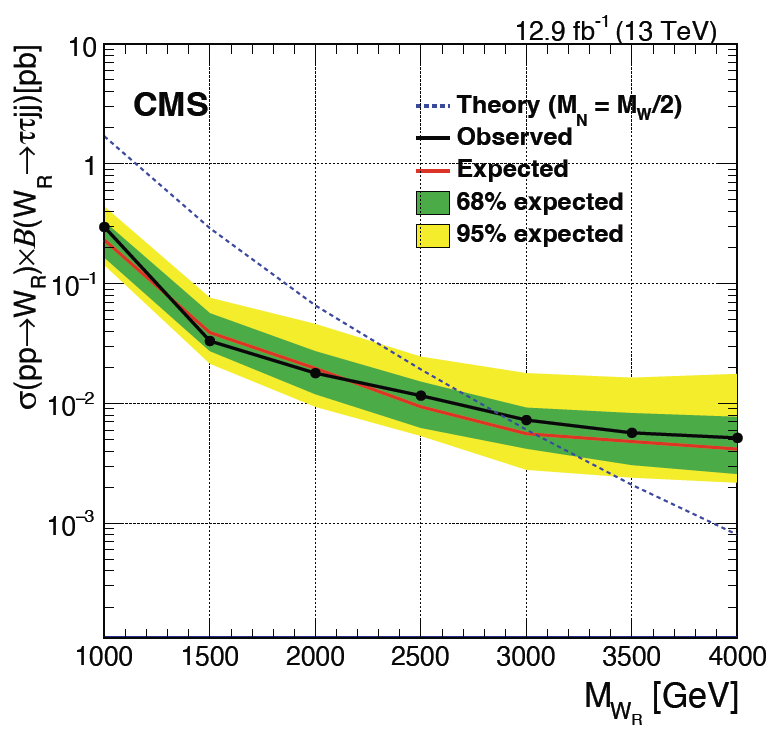}%
\hfill%
\includegraphics[width=0.46\textwidth]{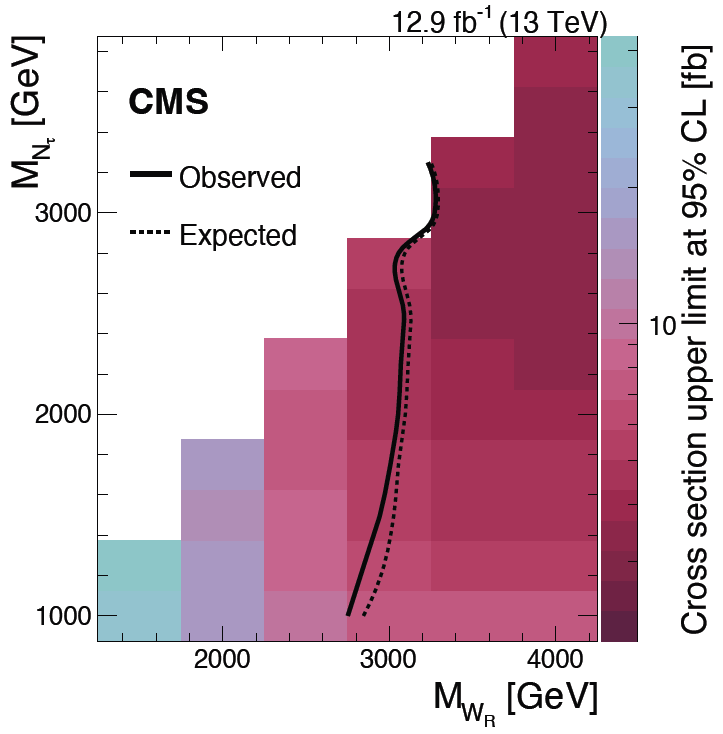}%
\caption{%
    Observed and expected limits at 95\% \CL on the product of cross section and branching fraction, obtained from the combination of the $\Pe\tauh$ and $\PGm\tauh$ channels (left), and the observed and expected upper limits at 95\% \CL on the production cross section as functions of the mass \mWR of the \PWR boson and the mass \mNGt of the HNL (right).
    The inner (green) band and the outer (yellow) band in the left figure indicate the regions containing 68 and 95\%, respectively, of the distribution of limits expected under the background-only hypothesis.
    The dashed dark blue curve in the left figure represents the theoretical prediction for the product of the \PWR boson production cross section and the branching fraction for decay to a \PGt lepton and RH neutrino, assuming the mass of the RH neutrino to be half the mass of the \PWR boson.
    Figures taken from Ref.~\cite{exo-16-023}.
}
\label{fig:CL_xsec_exo16_023}
\end{figure}

Upper limits on the product of the cross section and branching fraction are set at 95\% \CL
based on the binned distribution of the \ST observable.
Figure~\ref{fig:CL_xsec_exo16_023} (left) shows the observed and expected upper limits on the product of the cross section and branching fraction for the $\PWR\to\PGt\PNGt$ decay mode.
Assuming the mass of the HNL \mNGt to be half the mass of the \PWR boson \mWR, the observed (expected) lower limit at 95\% \CL on the mass of heavy right-handed \PWR bosons is determined to be 2.9 (3.0)\TeV.
Figure~\ref{fig:CL_xsec_exo16_023} (right) shows the observed and expected upper limits on the production cross section as functions of \mWR and \mNGt.

Another search in the LRSM context explored the process $\PWR\to\tauh\PNGt\to\tauh(\tauh\qqbar)$ where the two tau leptons decay hadronically~\cite{CMS:2018iye}.
The search is performed using the 2016 data set.

Signal samples are simulated with \PWR masses ranging from 1 to 4\TeV, in steps of 0.25\TeV.
Furthermore, in the considered mass range, it is assumed that the \PNe and \PNGm are too heavy to be decay products of \PWR, and thus $\PWR\to\PGt\PNGt$ and $\PWR\to\qqbarpr$ are the dominant decay modes.
The branching fraction for the $\PWR\to\PGt\PNGt$ decay is approximately 10--15\%, depending on the \PWR and \PNGt masses.
For the \PWR mass range of interest for this analysis, the $\PNGt\to\PGt\qqbar$ branching fraction is close to 100\%.

The dominant SM processes that contribute to the backgrounds in this search are, in order of importance: \ttbar production with a \PGt lepton pair in the final state, QCD multijet production, and \PZ or \PW boson production with additional jets.
All these backgrounds are estimated from simulation, except for QCD multijet production, which is estimated from control samples in data.

\begin{figure}[ht!]
\centering
\includegraphics[width=0.45\textwidth]{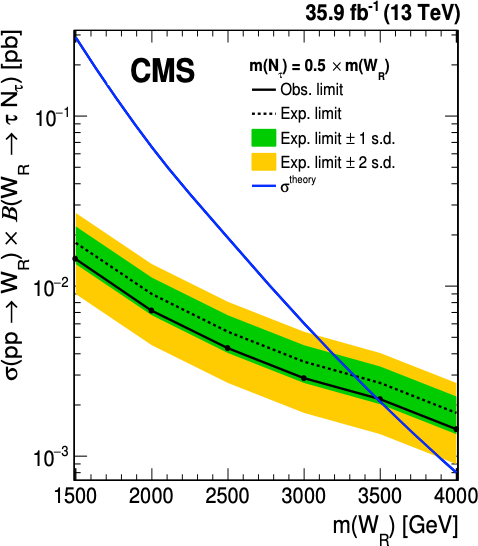}%
\hfill%
\includegraphics[width=0.51\textwidth]{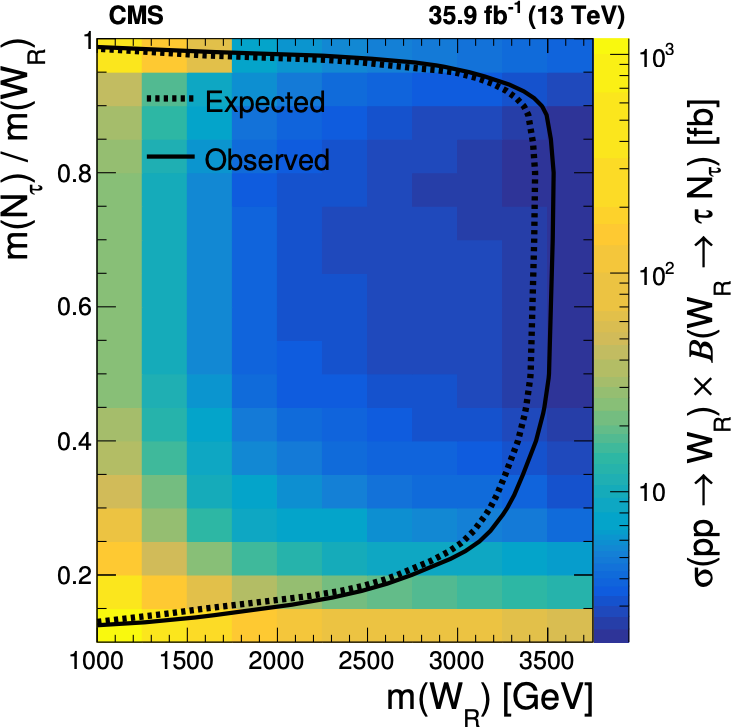}%
\caption{%
    Upper limits at 95\% \CL on the product of the cross section and the branching fraction for the production of \PWR bosons decaying to \PNGt as function of the \PWR boson mass (left).
    The observed (expected) limit is shown as solid (dashed) black lines, and the inner (green) band and the outer (yellow) band indicate the regions containing 68 and 95\%, respectively, of the distribution of limits expected under the background-only hypothesis.
    The theoretical cross section is indicated by the solid blue line.
    Expected and observed limits at 95\% \CL on the product of the cross section and the branching fraction for $\PWR\to\PNGt\PGt$ as a function of \mWR and $\mNGt/\mWR$ (right).
    Figures taken from Ref.~\cite{CMS:2018iye}.
}
\label{fig:CL_exo17_016}
\end{figure}

The results are presented as upper limits at 95\% \CL on the signal production cross sections.
Maximum likelihood fits are performed using the final $m(\PGt_{\text{h},1},\PGt_{\text{h},2},\PQj_1,\PQj_2,\ptmiss)$ observable---a partial mass observable constructed from the system formed by the two tau leptons, the two jets, and the \ptmiss---to derive the expected and observed limits.
Figure~\ref{fig:CL_exo17_016} (left) shows the expected and observed limits on the cross section, as well as the theoretical signal prediction~\cite{Huitu:1997,Barenboim:1997} as a function of \mWR.
For heavy neutrino models with strict left-right symmetry and assuming only the \PNGt flavor contributes significantly to the \PWR decay width and the \PNGt mass is 0.5\mWR, \PWR masses below 3.50 (3.35)\TeV are observed (expected) to be excluded at 95\% \CL.
This result is the most stringent limit to date in the considered model context.
Figure~\ref{fig:CL_exo17_016} (right) shows the upper limits on the product of the production cross section and branching fraction, as a function of \mWR and $x=\mNGt/\mWR$.
The signal acceptance and mass shape are evaluated for each $(\mWR,x)$ combination and used in the limit calculation procedure described above.
The lower limits on the \PWR mass depend on the \PNGt mass.
Masses below $\mWR=3.52$ (2.75)\TeV are excluded at 95\% \CL, for the benchmark scenarios assuming the \PNGt mass is 0.8 (0.2) times the mass of the \PWR boson.

\subsubsection{Search for HNLs from right-handed \texorpdfstring{\PZ}{Z} bosons}
\label{subsubsec:HNL_Zprime}

In this section, we review the search for pair-produced right-handed Majorana neutrinos from the decay of an additional heavy neutral gauge boson \PZpr introduced in the LRSM, using the full Run 2 data set~\cite{CMS:2023ooo}.
The selected events are first categorized by the flavor of the charged leptons (electrons and muons) mixing with \PN. In addition, two or more jets with various combinations of jet cone sizes are selected to exploit the full parameter phase space. These selected final-state objects are all used to reconstruct the \PZpr boson, with the aim of searching for an excess of events in its invariant mass distribution. In this search, the charge of leptons is inclusively considered with no separation between SS and OS dilepton signal event yields. The mixing of heavy neutrinos is assumed to only occur with SM neutrinos of the same flavor.

The event topology of this signal process strongly depends on the mass difference between the \PZpr boson and the HNL (\PN). If the ratio $2\mhnl/\mZpr$ is large enough, \ie, close to unity, the lepton and two quarks from the decay of the HNL are spatially well separated, and reconstructed as one lepton satisfying certain quality isolation criteria and two small-radius jets. In contrast, if the ratio is much smaller than unity, due to the high Lorentz boost of the HNL its decay particles are collimated, allowing it to be detected as one large-radius jet. In such cases, leptons may be considered either encompassed in the large-radius jet or as an isolated lepton outside the jet.

Considering all possible event topologies discussed above, three different SRs are defined based on the multiplicity of large and small-radius jets. The first signal region, SR1, consists of having no large-radius jet present in the event, but requires the presence of two isolated leptons and at least four small-radius jets. In the second signal region, SR2, one large-radius jet is required along with at least one isolated lepton and two small-radius jets. The third signal region, SR3, requires at least two large-radius jets to be present in the event.
For all SRs, the analysis is restricted to regions in (\mhnl, \mZpr) where the background is low, namely $\mhnl>80\GeV$ and $\mZpr>300\GeV$.

\begin{figure}[ht!]
\centering
\includegraphics[width=0.48\textwidth]{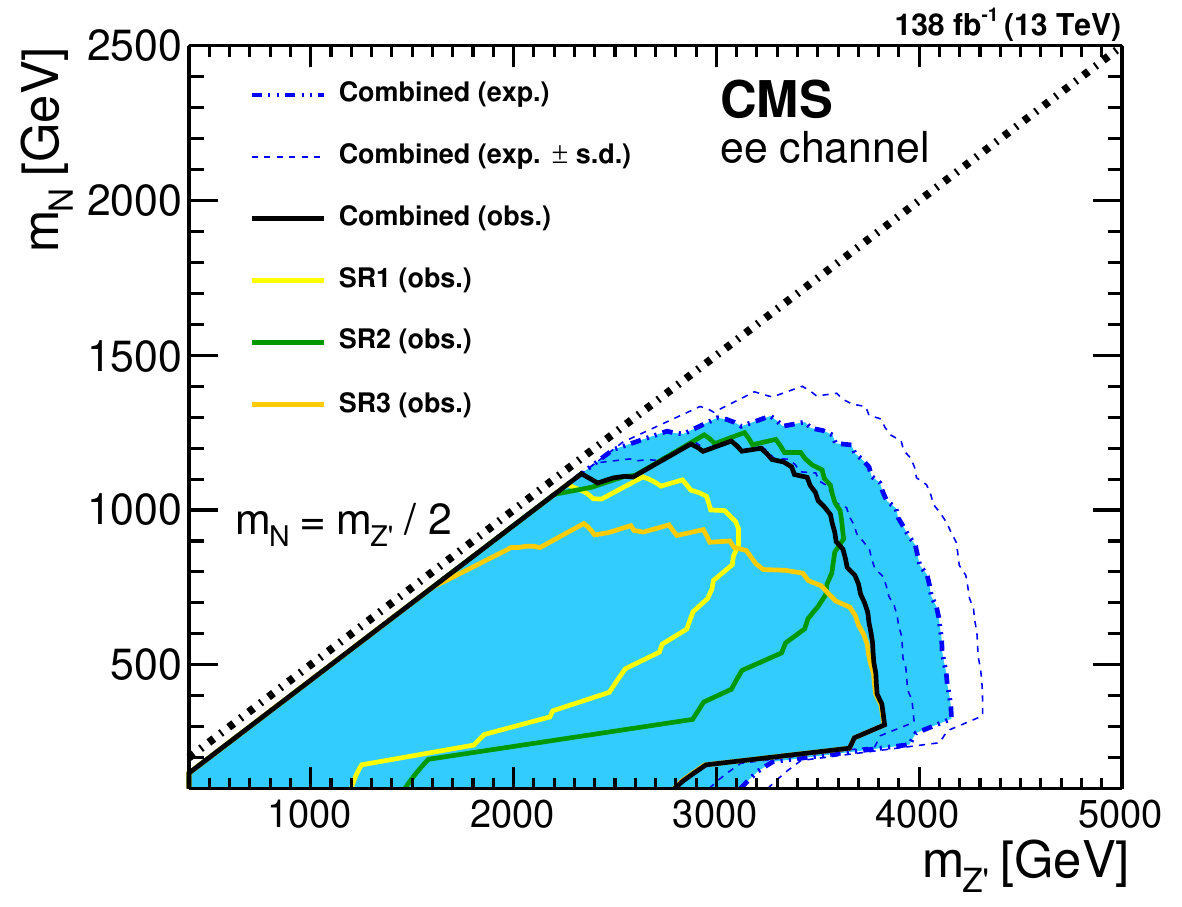}%
\hfill%
\includegraphics[width=0.48\textwidth]{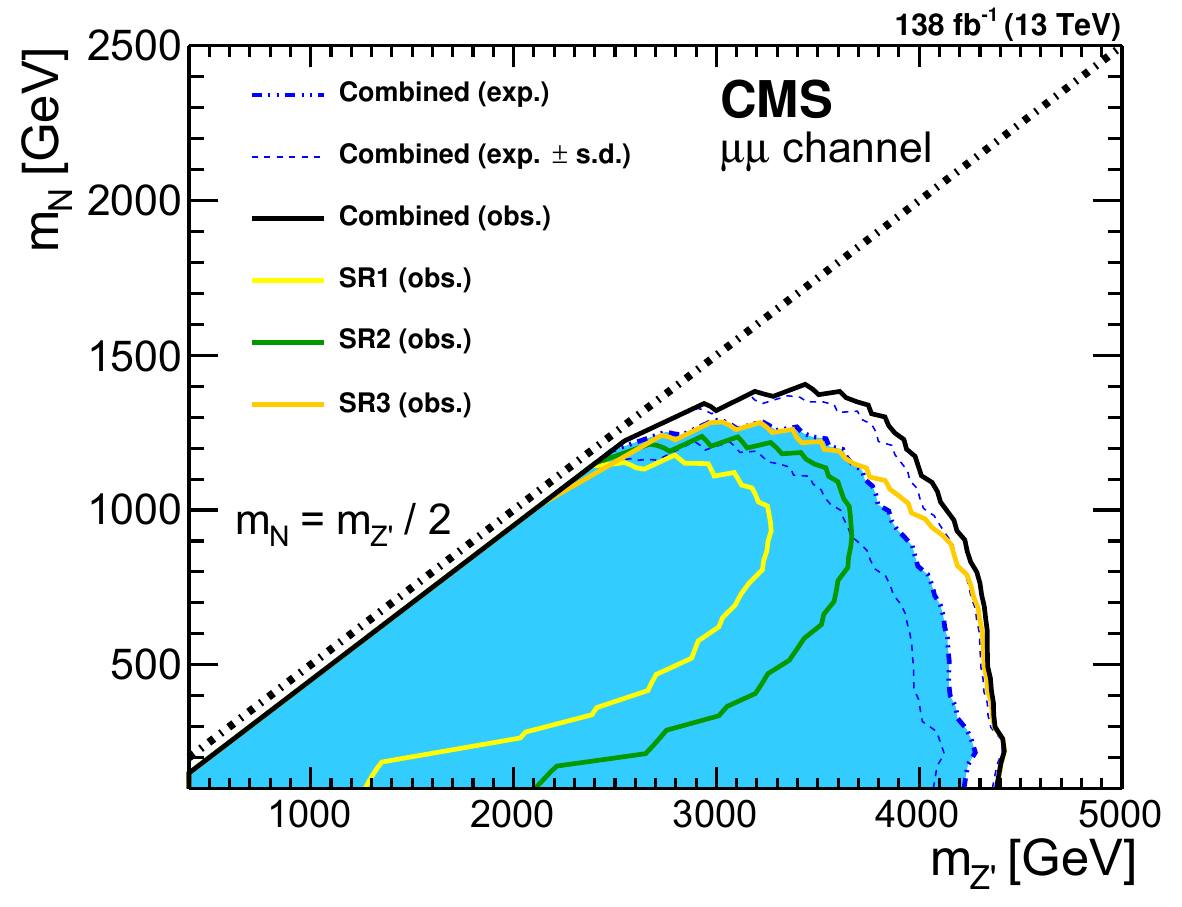}%
\caption{%
    The observed and expected exclusion limits in the \mhnl--\mZpr parameter space, in the dielectron channel (left) and the dimuon channel (right).
    Figures adapted from Ref.~\cite{CMS:2023ooo}.
}
\label{fig:exo-20-006-exclusion}
\end{figure}

A binned maximum likelihood fit is performed using the \mZpr distribution to extract the signal process.
No significant excess in data with respect to the SM prediction is observed. The upper limits are shown in Fig.~\ref{fig:exo-20-006-exclusion} in the two-dimensional \mhnl--\mZpr plane for the dielectron and the dimuon channel separately. The signal parameter point $(\mZpr,\mhnl)=(4.6,0.1)\TeV$ in the dielectron channel corresponds to the maximum local significance of 3.32 standard deviations. The observed (expected) lower limit at 95\% \CL on the mass of the \PZpr boson is 3.59 (3.90)\TeV and 4.10 (3.86)\TeV in the dielectron and dimuon channel, respectively, for scenarios where $\mhnl=\mZpr/4$.
For a small mass $\mhnl=100\GeV$, an example point in the parameter space in which the HNL is highly Lorentz boosted, the observed (expected) lower limits are $\mZpr=2.79$ (3.12)\TeV in the dielectron channel and 4.38 (4.22)\TeV in the dimuon channel. The use of a dedicated SR for boosted \PN event topologies provides a significant improvement in the sensitivity.
Scenarios assuming $\mWR\gg\mhnl$ in the \ee channel show weak sensitivity. This is because a requirement is placed on the energy fraction between HCAL and ECAL deposits as part of the trigger used for the electromagnetic objects, to improve selectivity.

\subsubsection{Summary and complementarity of channels}

\begin{figure}[ht!]

\centering
\includegraphics[width=0.48\textwidth]{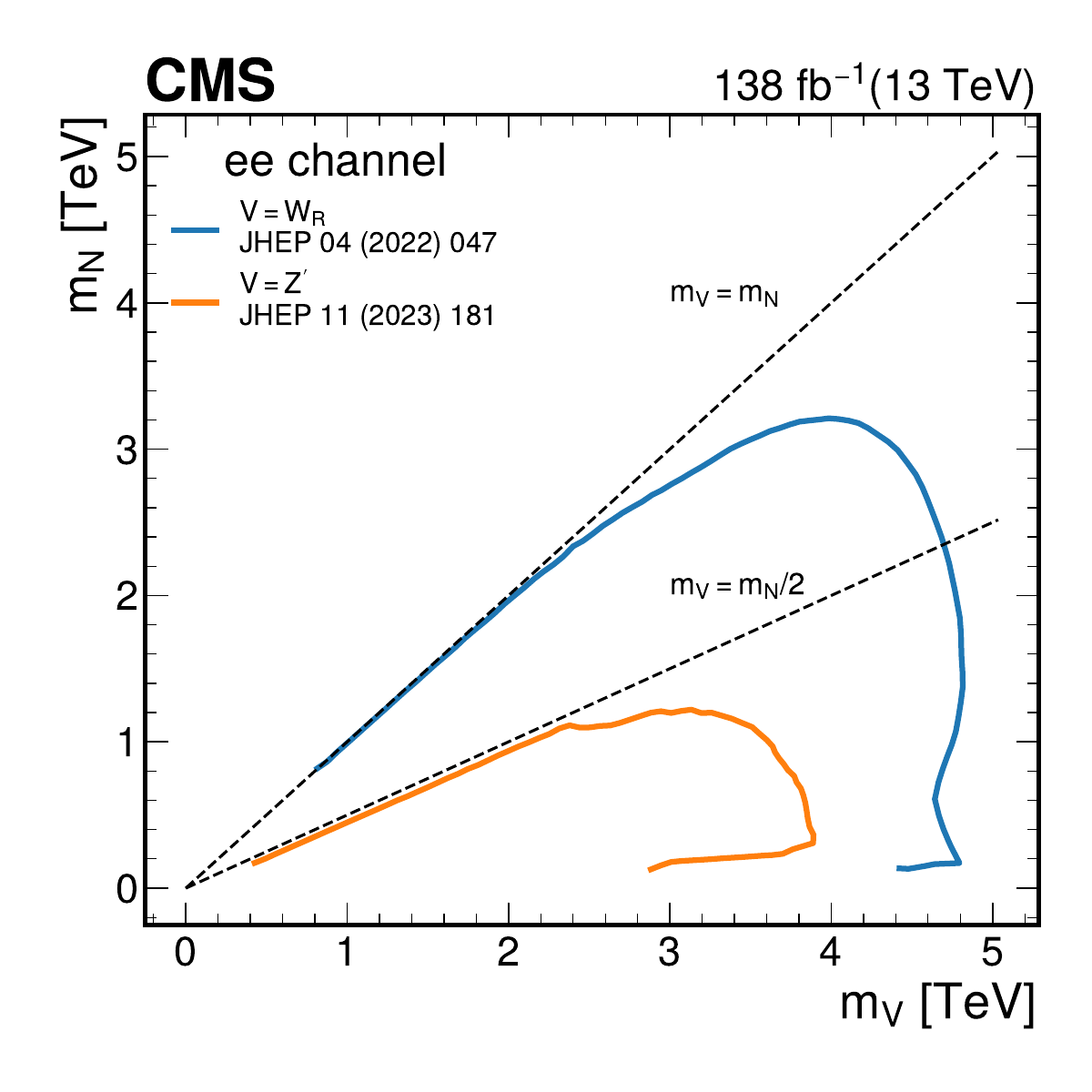}%
\hfill%
\includegraphics[width=0.48\textwidth]{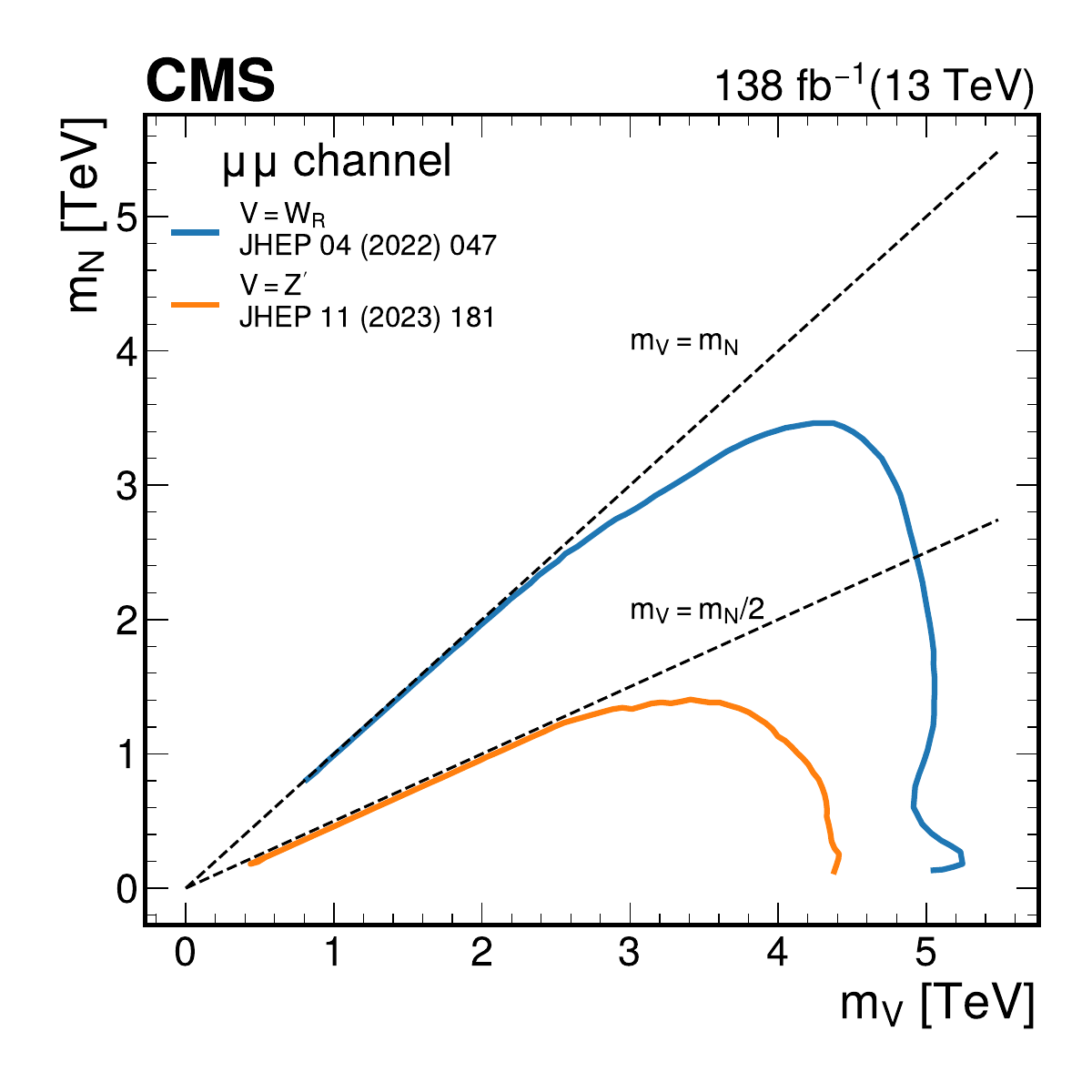}%
\caption{%
    Summary of searches at the CMS experiment for Majorana HNLs in the context of the LRSM model. The observed limits at 95\% \CL in the two-dimensional \mhnl--$m_{\PV}$ plane are shown in the electron and muon channel (left and right, respectively).
}
\label{fig:LRSM_summary}
\end{figure}

Two searches for LRSM HNLs are summarized in Fig.~\ref{fig:LRSM_summary}: the searches for $\PWR\to\Pell\PN$ and $\PZpr\to\PN\PN$. The \PWR search sets the most stringent limits on \mhnl as a function of \mWR thanks to the larger production cross sections.
Diagonal lines represent mass constraints in both channels as the searches are performed under the assumption that \PN is produced on-shell.
A weak sensitivity for the parameter space where $\mWR\gg\mhnl$ in the \ee channel for the \PZpr search is observed due to trigger requirements.

\subsection{Heavy composite Majorana neutrinos}
\label{sec:HeavyComposite_results}

In this section, we review a search for heavy composite neutrinos \PNell, introduced in Section~\ref{sec:th_hnl_composite}. These HNLs are produced in association with a charged lepton and decay to a charged lepton and a pair of quarks, leading to the experimental signature $\Pell\Pell\qqbarpr$, where \Pell is either an electron or a muon~\cite{CMS:2022chw}. Because the \PNell is a Majorana lepton at the \TeVns scale, the expected signal is characterized by two leptons \Pell that may be of the same or opposite sign, but are of the same flavor.
The analysis focuses only on the cases in which these leptons are both electrons or both muons, and the quark pair is detected as a large-radius jet.
For gauge-mediated decays of the \PNell, the fragmentation products of the two quarks from the \PW boson decay typically form at least one large-radius jet.
In the case of contact-mediated decays, the two quarks are well separated, but at least one of them will be contained within a large-radius jet.
The signal simulation shows that the efficiency for capturing one or both quarks in the jet is 98\% for the CI-dominated case $\mNell = 5$~TeV, and 95\% overall for the gauge or contact interaction with $\mNell >1$~TeV.

Events are selected with two same-flavor leptons, and jets are reconstructed as large-radius jets (labeled as ``\PQJ'').
The large-radius jets are required to have $\pt>190\GeV$, $\abs{\eta}<2.4$, and to be separated from leptons
by $\DR>0.8$.
Studies in simulation demonstrate that requiring one or more large-radius jets guarantees high signal efficiency for events with two leptons and is suitable for \PNell decays through both the gauge and the contact interactions.
The SR for the search for heavy composite Majorana neutrinos is defined by requiring two leptons, selected without specifying the sign, with the invariant mass $\mellell>300\GeV$ and at least one large-radius jet.
Restricting to the high-mass region allows for reducing the contributions from DY and part of the \ttbar background processes without affecting the signal acceptance.
Evidence of a signal is searched for by considering the distribution of the mass of the two leptons and the leading-\pt large-radius jet, \mellellJ. This observable provides good discrimination between the signal and the SM background contributions, but is also naturally correlated with the mass of the heavy composite neutrino.
In addition, two SM-dominated CRs are included in the maximum likelihood fit that help constrain the major background sources.
The DY background contribution is estimated from simulation, corrected by scale factors that adjust the simulated \mellellJ shape to match the observed data. An SF for each \mellellJ bin and each data-taking year is taken from the DY-dominated \mellell region around the \PZ boson mass peak, $60<\mellell<120\GeV$.
A DY CR is then defined that adopts the same criteria as the SR except in an \mellell sideband, $150<\mellell<300\GeV$, that lies adjacent to the SR. This CR provides the validation of the corrected simulation and improves the precision of the background prediction.
The second most important background arises from the leptonic decays of top quarks from \ttbar and single top quark production. The \mellellJ shape of this background is taken from the simulation, with a free normalization parameter in the fit for each year of data taking. To constrain these parameters, a top quark enriched CR is included in the fit, selecting SR events but requiring one muon and one electron in the final state.
After performing the fit, the data and expected SM background contributions agree well, and no excess of events above the expected background is observed.

\begin{figure}[!ht]
\centering
\includegraphics[width=0.48\textwidth]{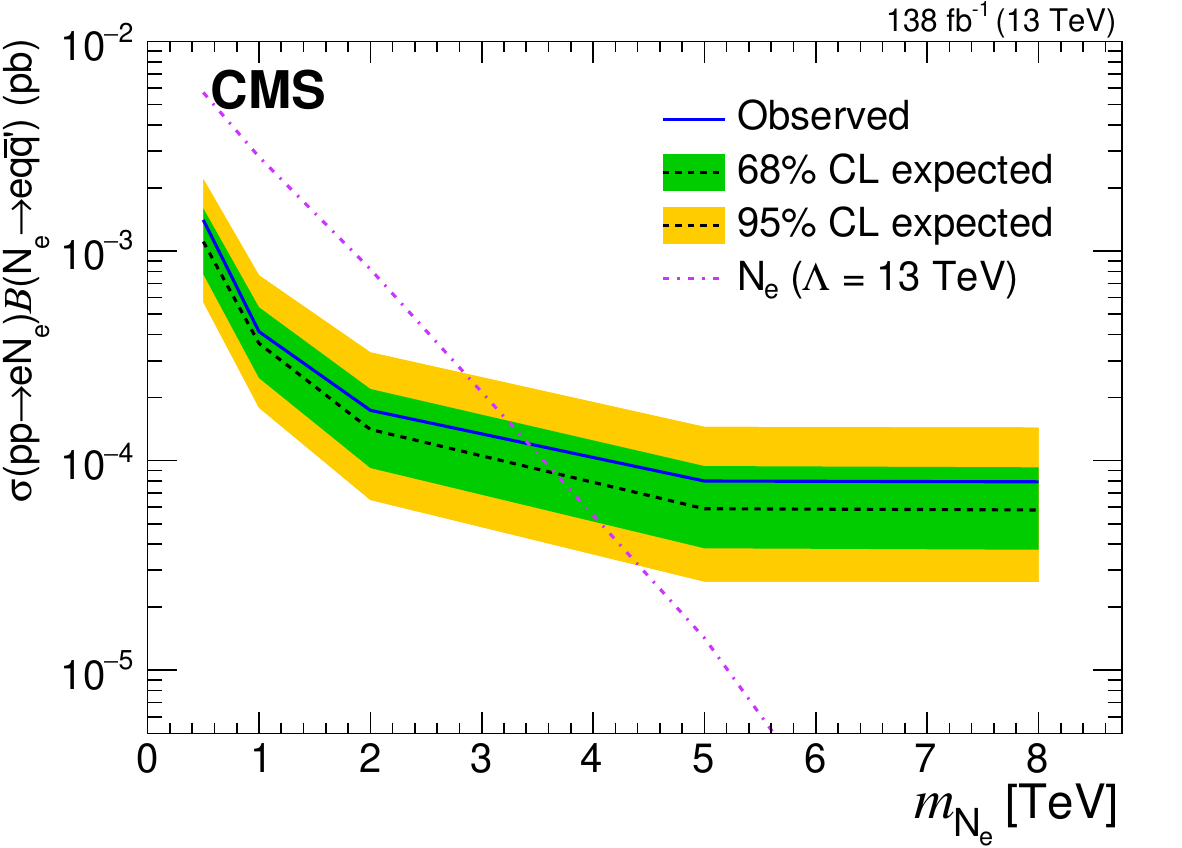}%
\hfill%
\includegraphics[width=0.48\textwidth]{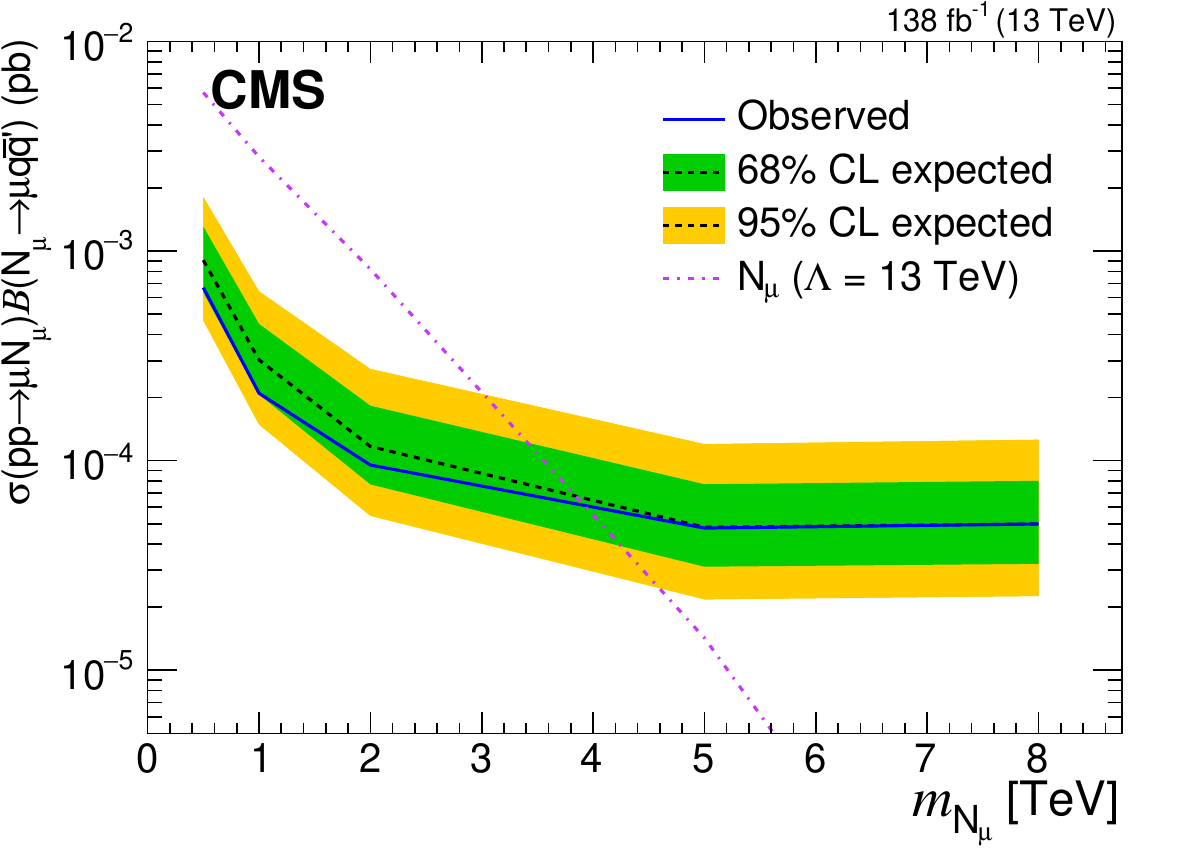}%
\caption{%
    Expected (dashed black) and observed (blue solid) exclusion limits for the $\ee\qqbarpr$ (left) and $\mumu\qqbarpr$ (right) channels in the search for heavy composite Majorana neutrinos.
    The inner (green) band and the outer (yellow) band indicate the regions containing 68 and 95\%, respectively, of the distribution of limits expected under the background-only hypothesis.
    Figures taken from Ref.~\cite{CMS:2022chw}.
}
\label{fig:Limit_comb_with_toys}
\end{figure}

\begin{figure}[!ht]
\centering
\includegraphics[width=0.48\textwidth]{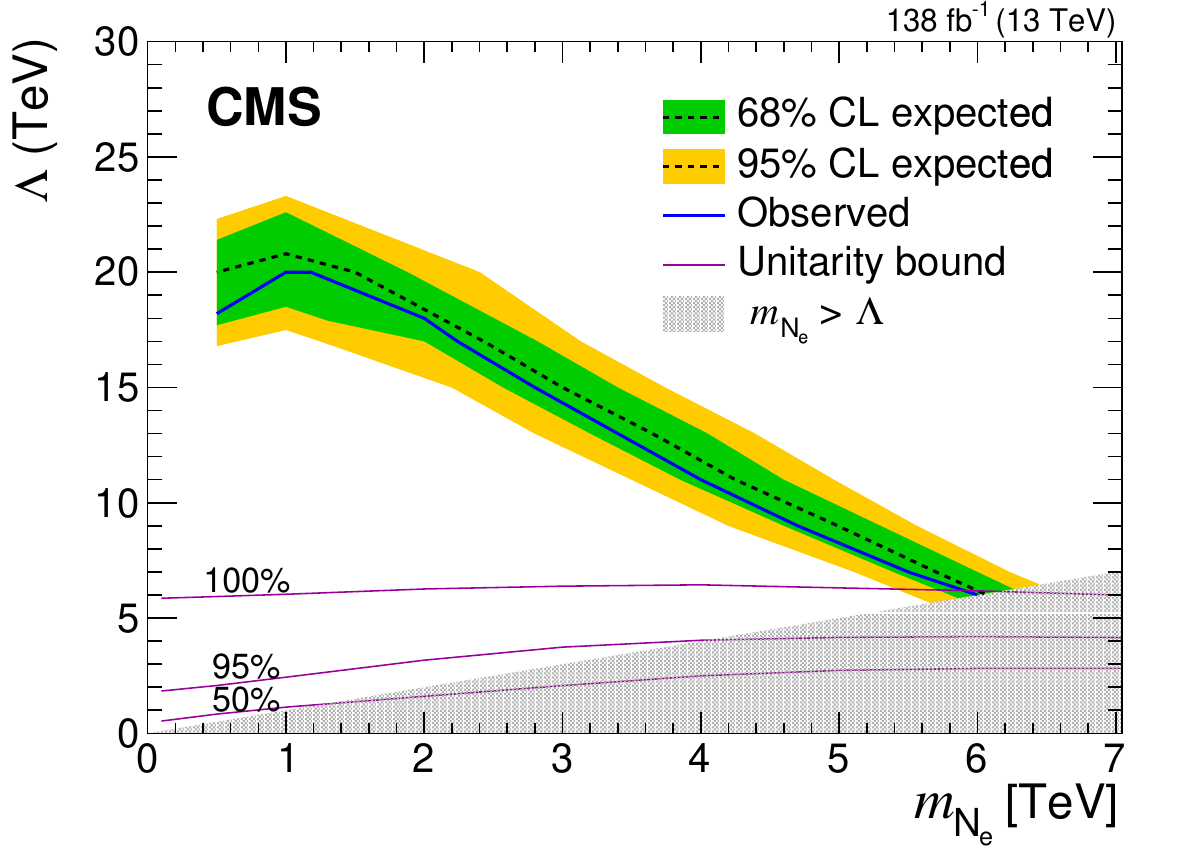}%
\hfill%
\includegraphics[width=0.48\textwidth]{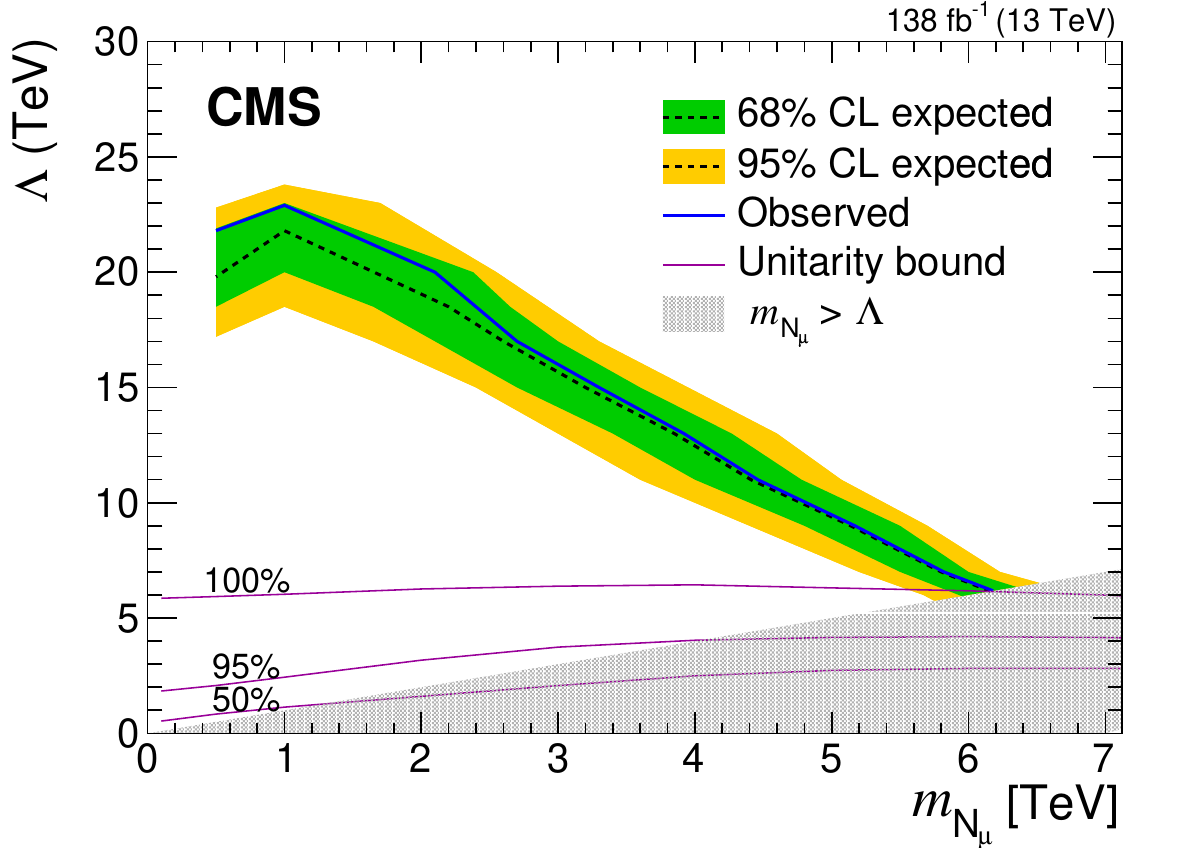}%
\caption{%
    Expected (dashed black) and observed (blue solid) exclusion limits for the $\ee\qqbarpr$ (left) and $\mumu\qqbarpr$ (right) channel in the two-dimensional plane \mNell--$\Lambda$.
    The solid violet lines represent the fraction of simulated events that satisfy the unitarity condition in the EFT approximation~\cite{Biondini:2019tcc} with the various percentages considered.
    The inner (green) band and the outer (yellow) band indicate the regions containing 68 and 95\%, respectively, of the distribution of limits expected under the background-only hypothesis.
    Figures taken from Ref.~\cite{CMS:2022chw}.
}
\label{fig:2D_Limit_eejj}
\end{figure}

The upper limits at 95\% \CL on the \PNell mediated process cross section are derived using the \mellellJ mass distributions of the estimated backgrounds, the expected signal, and the observed data.
The expected limits for the $\ee\qqbarpr$ channel and the $\mumu\qqbarpr$ channel, displayed in Fig.~\ref{fig:Limit_comb_with_toys}, provide an exclusion down to cross sections of the order of $10^{-4}$\unit{pb} for a vast range of \PNell signal hypotheses.
The limits on the process cross section may also be presented in the two-dimensional plane \mNell--$\Lambda$ for a more practical comparison with the unitarity restrictions on the parameter space. The results are shown in Fig.~\ref{fig:2D_Limit_eejj} for both channels.
For the case of $\Lambda=\mNell$, the existence of \PNe (\PNGm) is excluded for masses up to 6.0 (6.1)\TeV at 95\% \CL, improving by more than 1\TeV the world's most stringent limit on this kind of resonance~\cite{2017315}. These results are safe from a potential violation of the underlying EFT assumptions. Moreover, the accessible range of $\Lambda$ is almost twice the one reached in the previous search, extending the sensitivity to $\approx$20\TeV at lower \PNell masses.

\subsection{Future prospects for HNLs at the LHC}

The HNL search program of the CMS experiment offers a comprehensive insight into HNL production, decay, and the associated experimental constraints. Various theoretical models and several different and novel experimental methodologies are considered in these searches. In this report, we have reviewed all HNL results using data collected by the CMS detector during Run 2.

A straightforward and natural way to enhance the reach of HNL searches is to utilize the Run-3 data set collected by the CMS detector. Combining data from both Run 2 and Run 3 allows for the exploration of a significantly larger parameter space for prompt HNLs, particularly in searches for prompt $\PN\to\Pell\qqbarpr$ decays, given that the existing analysis has been conducted using the 2016 data set only. Additionally, during Run 3, the search for HNLs in the \WW VBF $t$ channel, analogous to double beta decay, can be expanded using newly developed trigger techniques, called VBF parking~\cite{CMS-PAS-EXO-23-007}, to maximize the efficiency in triggering on VBF events.

Furthermore, HNL production through the decay of \PZ bosons to active neutrinos, where at least one active neutrino mixes with the HNL in the $\PN\to\Pell\Pell\PGn$ channel within the \TypeOne seesaw model, remains unexplored. For high masses, where HNL decays are prompt, missing transverse momentum and lepton triggers can be utilized for this search. Similar searches exist in the context of the LRSM model, but they probe very high-mass HNLs, starting from 400\GeV, and in the semihadronic decay channel only.

\begin{figure}[!ht]
\centering
\includegraphics[width=0.75\textwidth]{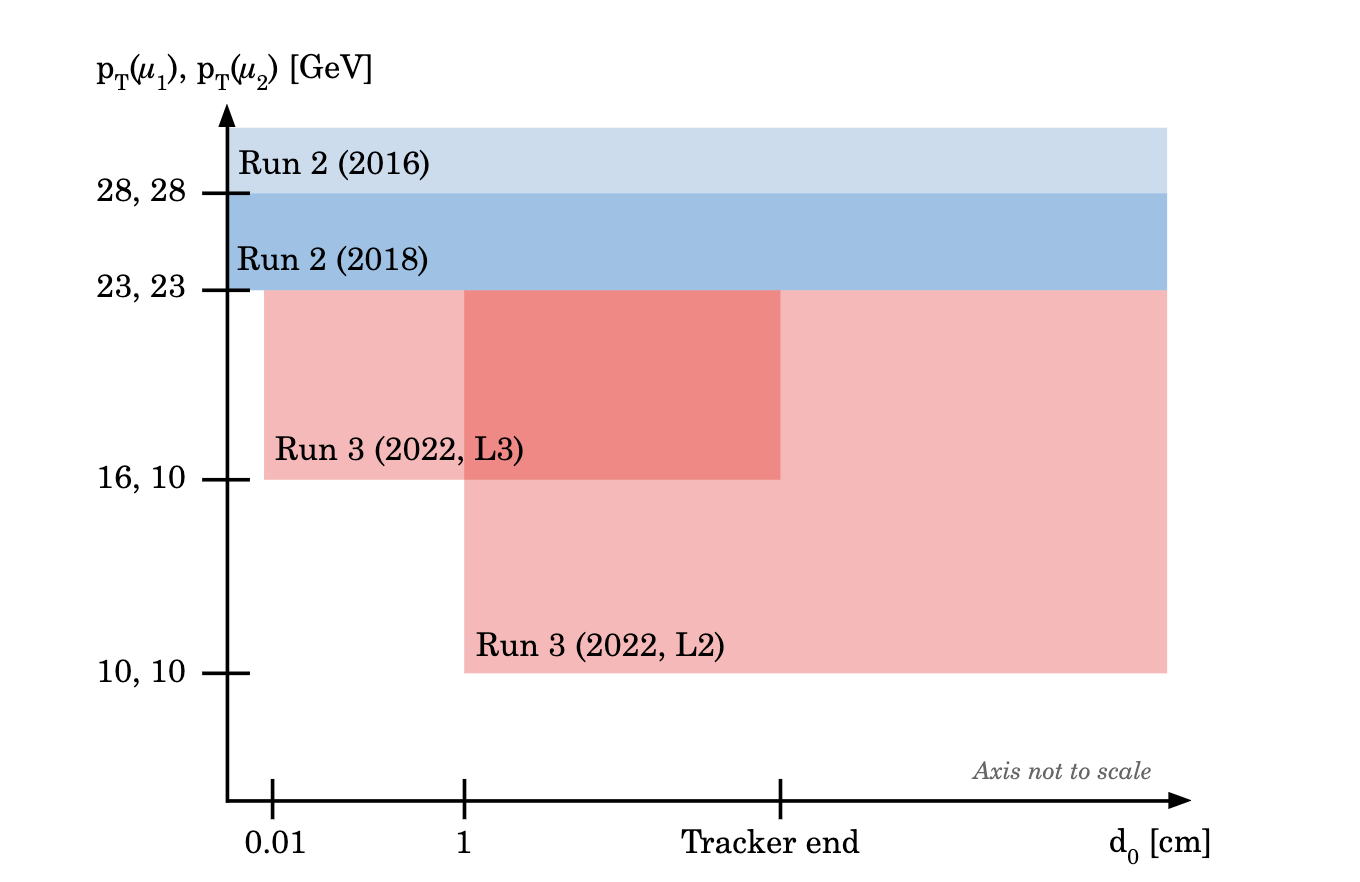}
\caption{%
    Coverage in the (\pt, \dzero) plane for displaced leptons with the 2016 and 2018 triggers, and the new Run-3 triggers, indicated in light blue, dark blue, and red, respectively~\cite{CMS-PAS-EXO-23-014}.
    Here, \dzero is the impact parameter of the charged lepton track with respect to the PV in the transverse plane.
}
\label{fig:displaced_leptons_trigger}
\end{figure}

For the long-lived HNL searches, the \PZ boson decay channel may yield more stringent limits. Despite the lower production cross section of \PZ bosons compared to \PW bosons, this approach has the potential to improve the sensitivity to HNL production, given the clean signature in the absence of QCD multijet background. This may be achieved by utilizing dedicated triggers for displaced leptons, as illustrated in Fig.~\ref{fig:displaced_leptons_trigger}.

Finally, a similar signature can be probed through the decay of the Higgs boson to active neutrinos. The Yukawa coupling within the framework of the \TypeOne seesaw model is very low, of the order of $10^{-5}$, making this channel challenging to probe. However, in models where Yukawa couplings are of the order of 1, such as in the inverse seesaw mechanism~\cite{Das:2017rsu}, the sensitivity to such decay channels increases significantly. This specific channel not only allows for the search for HNLs but also provides more insight into the relationship between the neutrino and the Higgs boson in general.

\clearpage
\section{Summary}
\label{sec:summary}

In this report, the physics program of the CMS experiment has been summarized for searches for physics beyond the standard model (SM) in the
context of models that introduce vector-like quarks (VLQs), vector-like leptons (VLLs), and heavy neutral leptons (HNLs).
Each of these three model classes provides a complementary perspective on the origin of mass of fundamental particles.
The VLQs extend the SM with nonchiral partners of SM quarks, and the searches focus on VLQs that couple to the third-generation quarks. The VLLs, introduced in
a class of models that can be particularly sensitive to leptonic anomalies, correspond to an analogous extension of the leptonic sector of the SM.
These searches target charged-lepton partners. The HNLs provide yet another perspective on the interplay between chirality and neutrino mass-generating
mechanisms, and produce distinct prompt and displaced signatures in the detector.

These searches probe unexplored areas of parameter space in several models beyond the SM, using Run 2 proton-proton collision data sets collected by the CMS detector during the years
2015 to 2018 corresponding to an integrated luminosity of up to 138\fbinv.
Two new statistical combinations of searches for VLQs have been performed.
Pair production of \PQB quarks with mass below 1.49\TeV is excluded at 95\% confidence level for any third-generation decay of the \PQB quark.
Single production of \PQT quarks in the narrow-width approximation is excluded at 95\% confidence level for \PQT quark masses below 1.20\TeV.
No evidence for physics beyond the SM has been observed, and stringent exclusion limits on new fermion masses and couplings have been placed.
One search for VLLs, detailed in Section~\ref{sec:results_vll_4321AllHad}, shows a modest excess of the observed data over the background-only prediction that requires further investigation using more data. No VLQ and HNL searches report excesses.

Using projections in the context of the future High-Luminosity LHC (HL-LHC) and the corresponding upgrades to the CMS detector, an increased discovery reach of new fermions
well into the \TeVns energy domain is expected. Although the environment of the HL-LHC with many simultaneous collisions will present new challenges for
particle reconstruction and identification, searches for new fermions will benefit from the increased collision energy, unprecedented integrated luminosity,
and the planned detector upgrades. Many of the searches presented in this report rely on identifying jets from the decays of massive SM particles, or feature
high-pseudorapidity jets from $t$-channel or vector boson fusion production modes. The expansion of the tracker volume and significant upgrades of the endcap calorimeter
and muon detectors will provide improved jet reconstruction and identification at high pseudorapidity in the HL-LHC era.

There are still unexplored regions of parameter space in various models beyond the SM involving VLQs, VLLs, and HNLs within reach of the LHC, that can yield a first glimpse of new physics in the near or longer term.
This includes considering nonminimal VLQ extensions such as decays of VLQs to scalar or pseudoscalar bosons, exploring VLQ production modes such as electroweak pair production, and expanding the searches for VLQs assuming a finite decay width.
Manifestations of VLLs in other models and final states than currently probed may also be considered, involving final states with muon detector shower signatures, final states with highly Lorentz-boosted decay products, or vector boson fusion modes of VLL pair production.
Future runs of the LHC will bring great opportunities to explore new model phase spaces, detector upgrades will provide improved particle reconstruction, and continued efforts in innovating analysis techniques will further enhance the potential to discover new physics.

\begin{acknowledgments}
\hyphenation{Bundes-ministerium Forschungs-gemeinschaft Forschungs-zentren Rachada-pisek} We congratulate our colleagues in the CERN accelerator departments for the excellent performance of the LHC and thank the technical and administrative staffs at CERN and at other CMS institutes for their contributions to the success of the CMS effort. In addition, we gratefully acknowledge the computing centers and personnel of the Worldwide LHC Computing Grid and other centers for delivering so effectively the computing infrastructure essential to our analyses. Finally, we acknowledge the enduring support for the construction and operation of the LHC, the CMS detector, and the supporting computing infrastructure provided by the following funding agencies: the Armenian Science Committee, project no. 22rl-037; the Austrian Federal Ministry of Education, Science and Research and the Austrian Science Fund; the Belgian Fonds de la Recherche Scientifique, and Fonds voor Wetenschappelijk Onderzoek; the Brazilian Funding Agencies (CNPq, CAPES, FAPERJ, FAPERGS, and FAPESP); the Bulgarian Ministry of Education and Science, and the Bulgarian National Science Fund; CERN; the Chinese Academy of Sciences, Ministry of Science and Technology, the National Natural Science Foundation of China, and Fundamental Research Funds for the Central Universities; the Ministerio de Ciencia Tecnolog\'ia e Innovaci\'on (MINCIENCIAS), Colombia; the Croatian Ministry of Science, Education and Sport, and the Croatian Science Foundation; the Research and Innovation Foundation, Cyprus; the Secretariat for Higher Education, Science, Technology and Innovation, Ecuador; the Estonian Research Council via PRG780, PRG803, RVTT3 and the Ministry of Education and Research TK202; the Academy of Finland, Finnish Ministry of Education and Culture, and Helsinki Institute of Physics; the Institut National de Physique Nucl\'eaire et de Physique des Particules~/~CNRS, and Commissariat \`a l'\'Energie Atomique et aux \'Energies Alternatives~/~CEA, France; the Shota Rustaveli National Science Foundation, Georgia; the Bundesministerium f\"ur Bildung und Forschung, the Deutsche Forschungsgemeinschaft (DFG), under Germany's Excellence Strategy -- EXC 2121 ``Quantum Universe" -- 390833306, and under project number 400140256 - GRK2497, and Helmholtz-Gemeinschaft Deutscher Forschungszentren, Germany; the General Secretariat for Research and Innovation and the Hellenic Foundation for Research and Innovation (HFRI), Project Number 2288, Greece; the National Research, Development and Innovation Office (NKFIH), Hungary; the Department of Atomic Energy and the Department of Science and Technology, India; the Institute for Studies in Theoretical Physics and Mathematics, Iran; the Science Foundation, Ireland; the Istituto Nazionale di Fisica Nucleare, Italy; the Ministry of Science, ICT and Future Planning, and National Research Foundation (NRF), Republic of Korea; the Ministry of Education and Science of the Republic of Latvia; the Research Council of Lithuania, agreement No.\ VS-19 (LMTLT); the Ministry of Education, and University of Malaya (Malaysia); the Ministry of Science of Montenegro; the Mexican Funding Agencies (BUAP, CINVESTAV, CONACYT, LNS, SEP, and UASLP-FAI); the Ministry of Business, Innovation and Employment, New Zealand; the Pakistan Atomic Energy Commission; the Ministry of Education and Science and the National Science Center, Poland; the Funda\c{c}\~ao para a Ci\^encia e a Tecnologia, grants CERN/FIS-PAR/0025/2019 and CERN/FIS-INS/0032/2019, Portugal; the Ministry of Education, Science and Technological Development of Serbia; MCIN/AEI/10.13039/501100011033, ERDF ``a way of making Europe", Programa Estatal de Fomento de la Investigaci{\'o}n Cient{\'i}fica y T{\'e}cnica de Excelencia Mar\'{\i}a de Maeztu, grant MDM-2017-0765, projects PID2020-113705RB, PID2020-113304RB, PID2020-116262RB and PID2020-113341RB-I00, and Plan de Ciencia, Tecnolog{\'i}a e Innovaci{\'o}n de Asturias, Spain; the Ministry of Science, Technology and Research, Sri Lanka; the Swiss Funding Agencies (ETH Board, ETH Zurich, PSI, SNF, UniZH, Canton Zurich, and SER); the Ministry of Science and Technology, Taipei; the Ministry of Higher Education, Science, Research and Innovation, and the National Science and Technology Development Agency of Thailand; the Scientific and Technical Research Council of Turkey, and Turkish Energy, Nuclear and Mineral Research Agency; the National Academy of Sciences of Ukraine; the Science and Technology Facilities Council, UK; the US Department of Energy, and the US National Science Foundation.

Individuals have received support from the Marie-Curie program and the European Research Council and Horizon 2020 Grant, contract Nos.\ 675440, 724704, 752730, 758316, 765710, 824093, 101115353, 101002207, and COST Action CA16108 (European Union) the Leventis Foundation; the Alfred P.\ Sloan Foundation; the Alexander von Humboldt Foundation; the Belgian Federal Science Policy Office; the Fonds pour la Formation \`a la Recherche dans l'Industrie et dans l'Agriculture (FRIA-Belgium); the Agentschap voor Innovatie door Wetenschap en Technologie (IWT-Belgium); the F.R.S.-FNRS and FWO (Belgium) under the ``Excellence of Science -- EOS" -- be.h project n.\ 30820817; the Beijing Municipal Science \& Technology Commission, No. Z191100007219010; the Ministry of Education, Youth and Sports (MEYS) of the Czech Republic; the Shota Rustaveli National Science Foundation, grant FR-22-985 (Georgia); the Hungarian Academy of Sciences, the New National Excellence Program - \'UNKP, the NKFIH research grants K 131991, K 133046, K 138136, K 143460, K 143477, K 146913, K 146914, K 147048, 2020-2.2.1-ED-2021-00181, and TKP2021-NKTA-64 (Hungary); the Council of Scientific and Industrial Research, India; ICSC -- National Research Center for High Performance Computing, Big Data and Quantum Computing and FAIR -- Future Artificial Intelligence Research, funded by the NextGenerationEU program (Italy); the Latvian Council of Science; the Ministry of Education and Science, project no. 2022/WK/14, and the National Science Center, contracts Opus 2021/41/B/ST2/01369 and 2021/43/B/ST2/01552 (Poland); the Funda\c{c}\~ao para a Ci\^encia e a Tecnologia, grant FCT CEECIND/01334/2018; the National Priorities Research Program by Qatar National Research Fund; the Programa Estatal de Fomento de la Investigaci{\'o}n Cient{\'i}fica y T{\'e}cnica de Excelencia Mar\'{\i}a de Maeztu, grant MDM-2017-0765 and projects PID2020-113705RB, PID2020-113304RB, PID2020-116262RB and PID2020-113341RB-I00, and Programa Severo Ochoa del Principado de Asturias (Spain); the Chulalongkorn Academic into Its 2nd Century Project Advancement Project, and the National Science, Research and Innovation Fund via the Program Management Unit for Human Resources \& Institutional Development, Research and Innovation, grant B37G660013 (Thailand); the Kavli Foundation; the Nvidia Corporation; the SuperMicro Corporation; the Welch Foundation, contract C-1845; and the Weston Havens Foundation (USA).
\end{acknowledgments}

\bibliography{auto_generated}

\clearpage
\appendix
\section{Glossary of acronyms}
\label{sec:glossary}
\begin{longtable}[l]{@{}ll}
BDT & Boosted decision tree \\
BSM & Beyond the standard model \\
CHS & Charged hadron subtraction \\
CI & Contact interaction \\
\CL & Confidence level \\
CMS & Compact Muon Solenoid \\
$CP$ & Charge conjugation parity \\
CSC & Cathode strip chamber \\
CR & Control region \\
DDT & Designed decorrelated tagger \\
DNN & Deep neural network \\
DT & Drift tube \\
ECAL & Electromagnetic calorimeter \\
EFT & Effective field theory \\
EW & Electroweak \\
FSR & Final-state radiation \\
HCAL & Hadronic calorimeter \\
HL-LHC & High-Luminosity Large Hadron Collider \\
HNL & Heavy neutral lepton \\
ISR & Initial-state radiation \\
JER & Jet energy resolution \\
JES & Jet energy scale \\
LH & Left handed \\
LHC & Large Hadron Collider \\
LLP & Long-lived particle \\
LNC & Lepton number conservation (or conserving) \\
LNV & Lepton number violation (or violating) \\
LO & Leading order \\
LRSM & Left-right symmetric model \\
MC & Monte Carlo \\
MDS & Muon detector shower \\
ML & Machine learning \\
MLP & Multilayer perceptron \\
NLO & Next-to-leading order \\
NN & Neural network \\
NNLO & Next-to-next-to-leading order \\
NWA & Narrow-width approximation \\
OS & Opposite sign \\
OSSF & Opposite-sign same flavor \\
PDF & Parton distribution function \\
PF & Particle flow \\
PUPPI & Pileup-per-particle identification \\
PV & Primary vertex \\
QCD & Quantum chromodynamics \\
RH & Right handed \\
RPC & Resistive plate chamber \\
SF & Scale factor \\
SM & Standard model \\
SR & Signal region \\
SS & Same sign \\
SV & Secondary vertex \\
VBF & Vector boson fusion \\
VLL & Vector-like lepton \\
VLQ & Vector-like quark \\
2D & Two-dimensional \\
\end{longtable}
\cleardoublepage \section{The CMS Collaboration \label{app:collab}}\begin{sloppypar}\hyphenpenalty=5000\widowpenalty=500\clubpenalty=5000
\cmsinstitute{Yerevan Physics Institute, Yerevan, Armenia}
{\tolerance=6000
A.~Hayrapetyan, A.~Tumasyan\cmsAuthorMark{1}\cmsorcid{0009-0000-0684-6742}
\par}
\cmsinstitute{Institut f\"{u}r Hochenergiephysik, Vienna, Austria}
{\tolerance=6000
W.~Adam\cmsorcid{0000-0001-9099-4341}, J.W.~Andrejkovic, T.~Bergauer\cmsorcid{0000-0002-5786-0293}, S.~Chatterjee\cmsorcid{0000-0003-2660-0349}, K.~Damanakis\cmsorcid{0000-0001-5389-2872}, M.~Dragicevic\cmsorcid{0000-0003-1967-6783}, P.S.~Hussain\cmsorcid{0000-0002-4825-5278}, M.~Jeitler\cmsAuthorMark{2}\cmsorcid{0000-0002-5141-9560}, N.~Krammer\cmsorcid{0000-0002-0548-0985}, A.~Li\cmsorcid{0000-0002-4547-116X}, D.~Liko\cmsorcid{0000-0002-3380-473X}, I.~Mikulec\cmsorcid{0000-0003-0385-2746}, J.~Schieck\cmsAuthorMark{2}\cmsorcid{0000-0002-1058-8093}, R.~Sch\"{o}fbeck\cmsorcid{0000-0002-2332-8784}, D.~Schwarz\cmsorcid{0000-0002-3821-7331}, M.~Sonawane\cmsorcid{0000-0003-0510-7010}, S.~Templ\cmsorcid{0000-0003-3137-5692}, W.~Waltenberger\cmsorcid{0000-0002-6215-7228}, C.-E.~Wulz\cmsAuthorMark{2}\cmsorcid{0000-0001-9226-5812}
% ...
\par}
\vskip\cmsinstskip
\dag:~Deceased\\
$^{1}$Also at Yerevan State University, Yerevan, Armenia\\
$^{2}$Also at TU Wien, Vienna, Austria\\
% ...
\end{sloppypar}
\end{document}